\documentclass[11pt, a4paper]{article}

\usepackage[utf8]{inputenc}
\usepackage[T1]{fontenc}
\usepackage[english]{babel}
\usepackage{amsmath, amssymb, amsthm, mathtools, mathrsfs}
\usepackage{enumitem}
\usepackage{geometry}
\usepackage[colorlinks=true, allcolors=blue]{hyperref}
\usepackage[capitalize]{cleveref}
\usepackage[numbers, sort&compress]{natbib}

\newtheorem{theorem}{Theorem}[section]
\newtheorem{proposition}[theorem]{Proposition}
\newtheorem{lemma}[theorem]{Lemma}
\newtheorem{corollary}[theorem]{Corollary}
\newtheorem{alphthm}{Theorem}

\theoremstyle{remark}\newtheorem{remark}[theorem]{Remark}

\newcommand{\cF}{\mathcal{F}}
\newcommand{\cN}{\mathcal{N}}
\newcommand{\cK}{\mathcal{K}}
\newcommand{\cV}{\mathcal{V}}
\newcommand{\ad}{\operatorname{ad}\!}
\renewcommand{\d}{\operatorname{d}\!}
\newcommand{\norm}[1]{\lVert#1\rVert}
\newcommand{\ip}[2]{\langle#1,#2\rangle}
\newcommand{\dGamma}{\operatorname{d}\!\Gamma}

\title{Quantum Fluctuations in Bose--Einstein Condensates:\\
Microscopic Correlations and Moment Bounds}
\author{
	\begin{minipage}[t]{0.31\textwidth}
		\centering
		Mauris Chueng\thanks{These authors contributed equally to this work.}~\thanks{Corresponding author.}\\
		\small School of Statistics and Data Science, Jilin University of Finance and Economics, Changchun, 130117, China\\[2pt]
		\small \texttt{maurischueng@gmail.com}
	\end{minipage}%
	\hfill
	\begin{minipage}[t]{0.31\textwidth}
		\centering
		Keheng Zhu\footnotemark[1]\\
		\small Academy for Multidisciplinary Studies, Capital Normal University, Beijing, 100048, China\\[2pt]
		\small \texttt{hexistartop@gmail.com}
	\end{minipage}
	\hfill
	\begin{minipage}[t]{0.31\textwidth}
		\centering
		Hexiang Wang\footnotemark[1]\\
		\small School of Mathematical Sciences, Nankai University, Tianjin, 300071, China\\[2pt]
		\small \texttt{Kui6539@outlook.com}
	\end{minipage}%
}
\date{}
\begin{document}
\maketitle
\begin{abstract}
We study higher moments of quantum fluctuations and microscopic correlation energy in the three-dimensional Gross--Pitaevskii regime. For a smooth, compactly supported, nonnegative radial interaction and an $N$-independent normalized initial condensate wave function in $H^6$, we propagate the third spectral moment of a positive correlated energy along the exact many-body evolution. A uniform initial bound controls the third moment of the depletion number and sixth moments of collective one-particle fluctuations on every compact time interval, independently of $N$. For a nonzero interaction and a smooth, compactly supported initial condensate, we also identify the excitation-energy measure at scale $N^2$, weighted by the first energy. This limit holds for norm-convergent pure-state families with a uniform bound on the initial third correlated-energy moment. Its positive-energy part is a spectral measure of the relative scattering operator $-2\Delta+v$, multiplied by a time-dependent dressed contact. We compute the radial density and prove strict positivity of the contact for nonzero references with finite kinetic energy and finite mean excitation number. The same evolving states have bounded physical fluctuation moments and excitation-energy moments of order $p>1$ bounded below by a positive multiple of $N^{2p-2}$, in the extended spectral sense. Inverse powers of the positive initial energy construct admissible pure and mixed families. For the norm-convergent pure-state class, we determine the scaled interaction-energy moments for $1\le p<2$ and bound the second moment by $O(N^2)$. A finite cubic approximation has norm error $O(N^{-3/16})$ on the energy form domain and $O(N^{-5/4})$ on a stronger initial domain of its generator.
\end{abstract}

\clearpage
\setcounter{tocdepth}{2}
\tableofcontents
\clearpage

\section{Introduction and summary}
\label{sec:introduction}

The Gross--Pitaevskii description originates in the work of
Gross and Pitaevskii \cite{Gross1961,Pitaevskii1961}. It describes
the condensate, while fluctuations around the condensate
retain the short-distance correlations generated by the interaction.
In three dimensions, these correlations enter both the effective
evolution and its comparison with the many-body evolution
\cite{BS,COS}. Controlling higher moments of the fluctuations
requires estimates adapted to the observables and their dependence
on the particle number. We seek bounds that are uniform in the
Gross--Pitaevskii regime.

We address this question for the Gross--Pitaevskii-scaled Hamiltonian
\cite{ESY2010,BS}
\[
	H_N^{\rm mb}=\sum_{j=1}^N-\Delta_{x_j}+\sum_{1\le i<j\le N}N^2v(N(x_i-x_j))
\]
on the symmetric $N$-particle space. The interaction is smooth,
compactly supported, nonnegative and radial. The condensate evolves
from an $N$-independent normalized initial wave function in $H^6$.
Our main result concerns the exact
many-body evolution of states with a bounded third moment of a
positive correlated energy. On every compact time interval, this
initial bound controls the third moment of the depletion and the
sixth moment of collective one-particle fluctuations at scale
$\sqrt N$, with constants independent of $N$. Taking the one-particle
observable to be the indicator of a measurable region gives, in
particular, a sixth-moment bound for its particle-number fluctuation.

\subsection{Setup and notation}

We use the bosonic Fock-space conventions of
\cite[Section 2]{BS} and \cite[Chapters 3 and 9]{DerezinskiGerard}.
The one-particle space is $L^2(\mathbb R^3)$, and
\[
	\cF=\bigoplus_{n\ge0}\cF^{(n)},\qquad
	\cF^{(n)}=L^2_s(\mathbb R^{3n}),\qquad
	\Omega=(1,0,\ldots).
\]
Inner products are antilinear in the first argument and linear
in the second. We write $T^*$ for the Hilbert-space adjoint.
For an integral kernel $k$, the transpose and adjoint kernels are
$k^{\mathsf T}(x,y)=k(y,x)$ and
$k^*(x,y)=\overline{k(y,x)}$, respectively. A symmetric pair
kernel satisfies $k^{\mathsf T}=k$.
The creation and annihilation fields obey
\[
	a^*(f)=\int f(x)a_x^*\d x,\qquad
	a(f)=\int\overline{f(x)}a_x\d x,\qquad
	[a(f),a^*(g)]=\langle f,g\rangle.
\]
For a one-particle operator $h$, its second quantization is
$\dGamma(h)|_{\cF^{(n)}}=\sum_{j=1}^n h_j$.
For a one-particle contraction $u$, the Fock-space lift is
$\Gamma(u)=\bigoplus_{n\ge0}u^{\otimes_s n}$, with identity
on the vacuum sector. In particular,
\[
	\cN=\dGamma(1),\qquad \cK=\dGamma(-\Delta),\qquad
	\mathbf1_n=\mathbf1_{\{n\}}(\cN),\qquad
	\Pi_L=\mathbf1_{[0,L]}(\cN).
\]
Here $\mathbf1_B(T)$ denotes the spectral projection of a
self-adjoint operator $T$ onto a Borel set $B$.
We write $\cF^{\le N}=\Pi_N\cF$. The parameter $N$ is the
total particle number; $\cN$ counts particles in Fock space
and excitations on the condensate-orthogonal space.

For functions and integral kernels, $\|\cdot\|_p$ denotes
the $L^p$ norm. The unadorned norm is the Hilbert-space norm
for vectors and the operator norm for bounded operators.
We also write $\|\cdot\|_{\mathrm{op}}$ for the latter.
The Schatten norms are denoted by $\|\cdot\|_{\mathfrak S_p}$;
in particular, $\|\cdot\|_{\mathrm{HS}}=\|\cdot\|_{\mathfrak S_2}$
\cite[Section 2.2.6]{DerezinskiGerard}.
For a square-integrable kernel $k$, $\|k\|_{\mathrm{HS}}=\|k\|_2$.
For a positive self-adjoint operator $H$, we use the form notation
\[
	H[u,v]=\langle H^{1/2}u,H^{1/2}v\rangle,\qquad
	H[u]=H[u,u],\qquad u,v\in D(H^{1/2}).
\]
Inequalities between unbounded positive operators are understood
in the sense of quadratic forms. The notation $H_1\asymp H_2$
means two-sided form bounds, with constants as specified in the
corresponding statement.

Set $M=\cN+1$, $A=\cK+M$, and $Q_N=A+\cV_N$, where
\[
	\cV_N=\frac12\int N^2v(N(x-y))a_x^*a_y^*a_ya_x\d x\d y.
\]
For the model application assume that $v$ is nonnegative, radial
and in $C_c^\infty(\mathbb R^3)$. The one-particle and Fock-space excitation projections are
$q_t=1-|\phi_t\rangle\langle\phi_t|$ and $P_t=\Gamma(q_t)$.
The physical excitation space is $P_t\cF^{\le N}$.
At fixed time we suppress $t$ and write $P_n=q^{\otimes_s n}$
for the restriction of $P$ to sector $n$.
Restricted operators are defined by restricting their closed forms,
using the representation theorem for semibounded forms
\cite{KatoBook,SimonForms}. We denote the restricted kinetic
operator by $K_{P_t}$ and set $A_{P_t}=K_{P_t}+M$.
These operators commute with $M$. In the abstract argument we suppress $P_t$
and work at a fixed time. Assume
\begin{equation}\label{eq:comparison}
	M\le A\le Q_N\le BMA.
\end{equation}
This follows for the GP potential by the scaled Sobolev inequality;
it also holds on the restricted forms. We use the Sobolev, Hardy
and convolution inequalities in the conventions of \cite{LiebLoss}.

\subsection{Exact evolution and microscopic energy}

We assume a nonnegative, radial $C_c^\infty$ potential and an
$N$-independent normalized initial condensate wave function in $H^6$.
We use the
modified GP background and fixed quadratic scattering radius of
Section~\ref{sec:number-compatibility}, together with the original cubic
phase at $m=N^{-1/2}$ and its cutoff $\Theta$.
The model statements concern all sufficiently large $N$ for which
these original cutoff parameters are defined.
Write $Q_t=Q_{N,P_t}$ and use the common connection of
Section~\ref{sec:covariant-generator}. More precisely, with
$\mathcal D_CG=GC-CG$ and $X=\nabla_tC$, set
\[
	\mathfrak J=e^{-C}\mathfrak Ge^C -i\int_0^1e^{-aC}Xe^{aC}\d a, \qquad \mathfrak J^{[3]}=\sum_{j=0}^3\frac{\mathcal D_C^j\mathfrak G}{j!}-iX.
\]
The exact logarithmic derivative is included in the first
generator; the second has the specified linear time term.
Let $U_{\rm ex},U_3$ denote their propagators. All constants
below are independent of $N$ on a fixed compact time interval $I$.

\begin{alphthm}
\label{thm:D}
Let $v\in C_c^\infty(\mathbb R^3)$ be nonnegative and radial.
Let $\phi_t$ solve the modified GP equation with an $N$-independent
normalized initial datum in $H^6(\mathbb R^3)$. Fix a sufficiently
small quadratic scattering radius $\ell\in(0,1/2)$, independent of $N$.
Use the cubic phase \eqref{spec:eq:actual-phase} at this background,
with $m=N^{-1/2}$ and the original COS cutoff $0\le\Theta_N\le1$.
It equals one for $n\le N^{1/2}/2+10$ and zero for
$n\ge N^{1/2}-10$, for sufficiently large $N$.
Fix a compact time interval $I$ and $s\in I$.
For a sufficiently large fixed $b_I$, put
$F_{\rm ex}(t)=\mathfrak J(t)+b_IM\asymp Q_t$ and let
$\xi_N(t)=U_{\rm ex}(t,s)z_N$, where $\|z_N\|=1$ and
\[
	E_{*,N}=\|F_{\rm ex}(s)^{3/2}z_N\|^2<\infty.
\]
Then
\[
	\sup_{t\in I}\left\{ \|F_{\rm ex}(t)^{3/2}\xi_N(t)\|^2 +\|M Q_t^{1/2}\xi_N(t)\|^2\right\} \le C_{I,s} E_{*,N}.
\]
Let $\mathcal U_{N,t}$ extract the condensate $\phi_t$,
and set $\Psi_N(t)=\mathcal U_{N,t}^*e^{B_t}e^{C_t}\xi_N(t)$. Up to a scalar phase, this state is the
exact $N$-body evolution of $\Psi_N(s)=\mathcal U_{N,s}^*e^{B_s}e^{C_s}z_N$. With
$\mathcal N_{\perp,N}(t)=\sum_{j=1}^N(q_t)_j$, one has
\[
	\sup_{t\in I} \langle\Psi_N(t),(\mathcal N_{\perp,N}(t)+1)^3\Psi_N(t)\rangle \le C_{I,s}E_{*,N}.
\]
For every bounded self-adjoint one-particle operator $O$, put
\[
	\mathcal X_{O,N}(t)=N^{-1/2}\sum_{j=1}^N \bigl(O_j-\langle\phi_t,O\phi_t\rangle\bigr).
\]
Its sixth moment obeys
\[
	\sup_{t\in I}\langle\Psi_N(t), \mathcal X_{O,N}(t)^6\Psi_N(t)\rangle \le C_{I,s}\|O\|^6 E_{*,N}.
\]
All constants are independent of $N$. Thus a family with
$\sup_N E_{*,N}<\infty$ has uniform bounds for all three moments.
Proposition~\ref{prop:exact-moment-preparation} gives such a family
from every uniformly first-energy-bounded reference vector. In particular, the normalized vectors
$F_{\rm ex}(s)^{-3/2}\Omega/
\|F_{\rm ex}(s)^{-3/2}\Omega\|$ have uniformly bounded $E_{*,N}$.
\end{alphthm}
\begin{proof}
See Theorem~\ref{thm:exact-third-moment},
Proposition~\ref{prop:exact-moment-preparation}, and
Theorem~\ref{thm:physical-sixth-moment}.
\end{proof}

\paragraph{Microscopic energy distribution.}
For this part assume also $v\ne0$ and take a common normalized
$\phi_0\in C_c^\infty(\mathbb R^3)$. Write
$H_N=\widehat F_{N,b}$ for the centered positive initial energy
in \eqref{eq:common-centered-energy} and $H=F_{\infty,b}$ for
its quadratic limit. Fix $b$ as in
Theorem~\ref{thm:arbitrary-reference-preparation}.
The initial class consists of normalized physical excitation
vectors with
\begin{equation}\label{eq:intro-joint-initial-class}
	z_N\longrightarrow\psi,\qquad
	\sup_{N\ge N_0}\|H_N^{3/2}z_N\|<\infty .
\end{equation}
Theorem~\ref{thm:admissible-initial-class} proves that such a
family exists exactly when $\|\psi\|=1$ and
$\psi\in D(H^{3/2})$. It constructs a recovery family whose
initial third moment converges to $\|H^{3/2}\psi\|^2$.
Every family in \eqref{eq:intro-joint-initial-class} has the
physical moment bounds of Theorem~\ref{thm:D}.

Let $\xi_N(t)=U_{\rm ex}(t,0)z_N$ and reconstruct $\Psi_N(t)$
as in Theorem~\ref{thm:D}. On the symmetric $N$-particle space,
the transformed excitation-energy observable is
\[
	\mathscr Q_N(t)=\mathcal U_{N,t}^*e^{B_t}
	Q_N(t)e^{-B_t}\mathcal U_{N,t}.
\]
The operator and its domain are defined by this unitary
conjugation of the physical excitation-space operator.
Its spectral probability and first-energy-weighted measure are
\begin{align*}
	\mathbb P_{N,t}(B)&=\langle\Psi_N(t),
	\mathbf1_B(\mathscr Q_N(t)/N^2)\Psi_N(t)\rangle,\\
	\nu_{N,t}(B)&=\langle\mathscr Q_N(t)^{1/2}\Psi_N(t),
	\mathbf1_B(\mathscr Q_N(t)/N^2)
	\mathscr Q_N(t)^{1/2}\Psi_N(t)\rangle .
\end{align*}
Thus $\nu_{N,t}(B)=N^2\int_B\lambda\,\d\mathbb P_{N,t}(\lambda)$.
Put $\psi(t)=\mathcal U_{2,\infty}(t,0)\psi$ and
$E_\infty(t)=A_{\infty,t}[\psi(t)]$, where
$A_{\infty,t}=K_{P_{\infty,t}}+M$ is the limiting kinetic
energy plus $M$. For every fixed $g\in C_b([0,\infty))$ and
compact interval $I$, the same theorem gives
\begin{equation}\label{eq:intro-joint-spectral-limit}
	\sup_{t\in I}\left|\int g\,\d\nu_{N,t}
	-E_\infty(t)g(0)-2c_\psi(t)\int g\,\d\mu_v\right|
	\longrightarrow0 .
\end{equation}
Here $c_\psi(t)$ is the dressed contact in
\eqref{eq:evolved-contact}, evaluated in $\psi(t)$, and
\[
	L=-2\Delta+v,\qquad
	\mu_v(B)=\langle L^{-1/2}v,\mathbf1_B(L)L^{-1/2}v\rangle .
\]
Section~\ref{sec:microscopic-spectrum} constructs this finite
measure. Its density is computed in
Proposition~\ref{prop:radial-microscopic-density}.
The contact is continuous and $\min_{t\in I}c_\psi(t)>0$.
For every fixed window $B=[b_0,d_0]\subset(0,\infty)$ with
$b_0<d_0$, Proposition~\ref{prop:physical-energy-comparison} gives
\begin{equation}\label{eq:intro-joint-window-law}
	\sup_{t\in I}\left|N^2\mathbb P_{N,t}(B)
	-2c_\psi(t)\int_B\lambda^{-1}\d\mu_v(\lambda)\right|
	\longrightarrow0 .
\end{equation}
The probability is of order $N^{-2}$, and the window contributes
order $N^{2p-2}$ to each fixed moment of order $p\ge0$.
For $p>1$, this gives the corresponding lower bound on the
full moment, interpreted in the extended spectral sense.
Corollary~\ref{cor:evolved-energy-probability} also gives the
conditional energy distribution within a fixed window.
Theorem~\ref{thm:admissible-interaction-spectrum} determines
the corresponding scaled interaction-energy moments for
$1\le p<2$ and gives second-moment bounds of order $N^2$.

More explicit initial-state constructions are given in
Theorems~\ref{thm:common-preparation},
\ref{thm:form-reference-preparation} and
\ref{thm:arbitrary-reference-preparation}. They use one or three
positive form equations, according to the regularity of the
reference vector. The form-domain construction includes
finite-excitation references with $H^1$ orbitals and the pure
quasi-free references of
Corollary~\ref{cor:quasifree-form-preparation}.
Theorem~\ref{thm:mixed-reference-preparation} treats reference
density matrices, including free Bose Gibbs references.

These are preparations of the physical initial state.
Theorem~\ref{thm:correlated-ground-state} also treats minimizers
of the positive correlated energy, which includes the fixed
number correction $bM$.
Under the Neumann overlap condition in
Theorem~\ref{thm:vacuum-third-obstruction}, the unprepared
vacuum lies outside the limiting third-moment domain.
Corollary~\ref{cor:finite-excitation-obstruction} gives the
corresponding statement for finite-excitation limits in $D(A)$.
The preparation and domain results specify which initial states
satisfy the hypothesis of Theorem~\ref{thm:D}.

\subsection{Finite cubic approximation}

The same energy estimates also control a finite cubic expansion.
The reference generator $\mathfrak J^{[3]}$ retains the interaction
energy. Its initial spectral norms determine the two error rates
in Theorem~\ref{thm:C}. We first state the block assumptions used
for the expansion.

Let $C=C_N$ be anti-self-adjoint and supported on particle numbers
$n\le L_N$. Assume that it changes the excitation number by at most
three and satisfies
\begin{align}
	\norm{C_{mn}z_n}
	&\le B\epsilon(n+1)^{3/2}\norm{z_n},\notag\\
	\norm{A_m^{1/2}C_{mn}z_n}
	&\le B(n+1)^{3/2}
	\big(\epsilon\norm{A_n^{1/2}z_n}+\norm{z_n}\big),
	\label{eq:blocks}\\
	0<\epsilon\le1,\qquad
	\epsilon\sqrt{L_N+2}&\le D_1. \label{eq:weak-budget}
\end{align}
All constants in these assumptions are independent of $N$.
For each fixed $N$ the phase is bounded both on Fock space and
on $D(Q_N^{1/2})$. Thus its form conjugation is analytic there.

In the standard cubic phase of \cite[Eq.~(3.2)]{COS},
$m=N^{-1/2}$, $\epsilon=N^{-3/4}$ and $L_N=O(\sqrt N)$.
The relative kernel is
$\sqrt N\,w^{\rm Neu}_{Nm}(N(x-y))\phi_t(y)$, up to its common
sign. Its relative $L^2$ norm is $O(\epsilon)$ and its gradient
norm is $O(1)$. The identity part of the annihilation field gives
the one-to-two blocks in \eqref{eq:blocks}. For the remaining
terms the three-variable kernel is
$\sqrt N\,w^{\rm Neu}_{Nm}(N(x-y))\phi_t(y)h_t(x,z)$, where
$h_t=\gamma_t-I$ or $\sigma_t$. Fubini bounds its norm by
$C\epsilon\norm{h_t}_2$ and its gradient by
$C(\norm{h_t}_2+\epsilon\norm{\nabla h_t}_2)$.
The bounds of \cite[Appendix A]{COS} give
$\norm{h_t}_2\le C_T$ and
$\norm{\nabla h_t}_2\le C_T\sqrt N$.
The creation and annihilation estimates on bosonic sectors
\cite[Section 3.4]{DerezinskiGerard} give contraction factors at most
$C(n+1)^{3/2}$. Projection is bounded in the energy form norm;
generalized $b$ factors are scalar contractions on each block.
These facts verify \eqref{eq:blocks} on compact time intervals.

\begin{alphthm}\label{thm:A}
Let $v\in C_c^\infty(\mathbb R^3)$ be nonnegative and radial.
Let $\phi_t$ solve the modified GP equation with an $N$-independent
normalized initial datum in $H^6(\mathbb R^3)$. Fix a sufficiently
small quadratic scattering radius $\ell\in(0,1/2)$, independent of $N$.
Use the cubic phase \eqref{spec:eq:actual-phase} at this background,
with $m=N^{-1/2}$ and the original COS cutoff $0\le\Theta_N\le1$.
It equals one for $n\le N^{1/2}/2+10$ and zero for
$n\ge N^{1/2}-10$, for sufficiently large $N$.
Fix a compact time interval $I$ and $s\in I$.
For a sufficiently large fixed $b_I$,
$F_3(t)=\mathfrak J^{[3]}(t)+b_IM\asymp Q_t$, and
\[
	\sup_{t\in I}\|Q_t^{1/2}U_3(t,s)z\| \le C_{I,s}\|Q_s^{1/2}z\|, \qquad z\in D(Q_s^{1/2}).
\]
For $z\in D(F_3(s)^{3/2})$ one also has
\[
	\sup_{t\in I}\|Q_t^{1/2}F_3(t)U_3(t,s)z\| \le C_{I,s}\|F_3(s)^{3/2}z\|.
\]
\end{alphthm}
\begin{proof}
These are Theorem~\ref{thm:Taylor-first-energy} and
Corollary~\ref{cor:Taylor-three-halves}, respectively.
Their proofs verify the needed form bounds and time
regularity for this same Taylor generator.
\end{proof}

\begin{alphthm}\label{thm:B}
Let $v\in C_c^\infty(\mathbb R^3)$ be nonnegative and radial.
Let $\phi_t$ solve the modified GP equation with an $N$-independent
normalized initial datum in $H^6(\mathbb R^3)$. Fix a sufficiently
small quadratic scattering radius $\ell\in(0,1/2)$, independent of $N$.
Use the cubic phase \eqref{spec:eq:actual-phase} at this background,
with $m=N^{-1/2}$ and the original COS cutoff $0\le\Theta_N\le1$.
It equals one for $n\le N^{1/2}/2+10$ and zero for
$n\ge N^{1/2}-10$, for sufficiently large $N$.
Fix a compact time interval $I$ and $s\in I$.
Put $R=\mathfrak J-\mathfrak J^{[3]}$.
For $T=R,\nabla_tR,i[M,R]$,
\[
	|T[u,v]|\le C_I N^{-3/2} \|M Q_t^{1/2}u\|\,\|Q_t^{1/2}v\|.
\]
This is an asymmetric energy-form estimate on its displayed
domain. The exact fourth integral remainder of the static
conjugation and the full time-term difference are retained.
\end{alphthm}
\begin{proof}
The model decomposition in
Theorem~\ref{thm:full-model-comparison} verifies
\eqref{eq:one-weight-static} for every low-degree group
and \eqref{eq:geometric-tail-static} for the exact tail.
Drop the spectral projections in the two bounding norms
and sum the summable coefficients. The tail with $K=2$
is smaller since $M\ge1$ and $N^{-2}\le N^{-3/2}$.
Proposition~\ref{prop:time-remainder-sharp} includes the
cubic time term and the two indicated derivatives.
\end{proof}

\begin{alphthm}\label{thm:C}
Let $v\in C_c^\infty(\mathbb R^3)$ be nonnegative and radial.
Let $\phi_t$ solve the modified GP equation with an $N$-independent
normalized initial datum in $H^6(\mathbb R^3)$. Fix a sufficiently
small quadratic scattering radius $\ell\in(0,1/2)$, independent of $N$.
Use the cubic phase \eqref{spec:eq:actual-phase} at this background,
with $m=N^{-1/2}$ and the original COS cutoff $0\le\Theta_N\le1$.
It equals one for $n\le N^{1/2}/2+10$ and zero for
$n\ge N^{1/2}-10$, for sufficiently large $N$.
Fix a compact time interval $I$ and $s\in I$.
For $z\in D(F_3(s)^{3/2})$,
\[
	\sup_{t\in I}\|(U_{\rm ex}(t,s)-U_3(t,s))z\| \le C_{I,s}N^{-5/4}\|F_3(s)^{3/2}z\|.
\]
The normalized preparations obtained from the seventh power of
the resolvent in \eqref{eq:Taylor-resolvent-data} have a uniformly
bounded initial $F_3(s)^{3/2}$ norm. On the larger energy form domain,
\[
	\sup_{t\in I}\|(U_{\rm ex}(t,s)-U_3(t,s))z\| \le C_{I,s}N^{-3/16}\|Q_s^{1/2}z\|.
\]
\end{alphthm}
\begin{proof}
The first assertion is
Theorem~\ref{thm:full-model-comparison}, using the low-degree support and inverse-energy argument of
Theorem~\ref{thm:inverse-comparison}. The preparation is proved in
Proposition~\ref{prop:Taylor-resolvent-preparation}.
The last assertion is
Theorem~\ref{thm:energy-class-comparison}.
\end{proof}

For a fixed bounded mean-field potential,
\cite[Theorem 1]{BPPS} constructs corrections to arbitrary order.
The expansion in \cite[Theorem 1]{BPPaS2020} also allows
$(N-1)^{-1}N^{3\beta}v(N^\beta x)$ in three dimensions for
$0\le\beta<1/12$. Theorems~\ref{thm:A}--\ref{thm:C} concern
the GP endpoint and the specified interacting reference generator.

\subsection{Relation to earlier work}

The experimental observation of condensation in dilute atomic gases
\cite{Anderson1995,Davis1995} provides the physical setting for this
problem. The review \cite{DGPS} discusses the condensate evolution,
collective excitations and the role of interactions. On the mathematical
side, the Gross--Pitaevskii energy functional for trapped gases was
derived in \cite{LSY2000}, and condensation in this regime was proved
in \cite{LS2002}. Complete condensation and quantitative depletion
bounds in the translation-invariant Gross--Pitaevskii regime were
established in \cite{BBCS2018,BBCS2020}.

The derivation of the time-dependent equation must retain the short-distance two-particle
correlations. The hierarchy approach of Erd\H{o}s, Schlein and Yau
\cite{ESY2006,ESY2010} identifies their contribution to the effective
coupling. Quantitative derivations using correlated fluctuation states
were obtained in \cite{BOS2015,BS}. These works supply the microscopic
motivation for the correlation transformations used below.
Renormalized excitation number operators provide another approach
to the Gross--Pitaevskii evolution and quantitative depletion bounds
\cite[Theorem 1.1 and Proposition 2.1]{BK2025}.
The introduction of that work also motivates stronger estimates
for excitation observables beyond norm approximation.
The energy-weighted comparison in
Theorem~\ref{thm:admissible-initial-class} supplies such control
for the initial class \eqref{eq:intro-joint-initial-class}.

Bogoliubov's approximation \cite{Bogoliubov1947} describes excitations
through a quadratic Hamiltonian. The mathematical theory of canonical
Bogoliubov transformations includes the implementability criterion of
Shale \cite{Shale} and the diagonalization results of \cite{NNS2016}.
Rigorous excitation-spectrum results were obtained for weakly
interacting bosons in \cite{Seiringer2011,GrechSeiringer2013}, for the
translation-invariant Gross--Pitaevskii gas in \cite{BBCS2019}, and for
trapped Gross--Pitaevskii gases in \cite{BSSSpectrum,NamTriay}.
The recent work \cite[Theorem 2.2]{ReichmannTriay2026}
extends the static low-energy theory on the three-dimensional
torus to interactions without spherical symmetry or finite range.
These spectral results concern low-energy states of a stationary
Hamiltonian. The moments studied here belong to a time-dependent
correlated energy.

The fluctuation approach to effective evolution has its origins in
the classical-field analysis of Hepp and Ginibre--Velo
\cite{Hepp1974,GinibreVelo1979}. Quantitative estimates on Fock space
were developed in \cite{RodnianskiSchlein2009}. Pair-correlation
corrections to mean-field evolution were studied in \cite{GMM2010,GMM2011}.
Condensate extraction and norm approximations in the Hartree and
nonlinear Schr\"odinger regimes were developed further in
\cite{LNS2015,NN2017,BCS2017,BNNS}. In the Gross--Pitaevskii regime,
Caraci, Oldenburg and Schlein \cite{COS} obtain a norm approximation
by quadratic fluctuation evolution after incorporating microscopic
correlations, including a cubic transformation. We use their cubic
phase and the quadratically transformed generator of \cite{BS}.
For the full fluctuation evolution, \cite[Proposition 3.2]{COS}
controls the first energy moment. For the quadratic evolution,
\cite[Proposition 4.3]{COS} also controls arbitrary moments of the
number operator and the second kinetic-energy moment. Theorem~\ref{thm:D} propagates the third spectral moment of the
exact correlated energy. A uniform initial bound then controls
the third depletion moment and sixth moments of collective
fluctuations along the exact evolution.

The microscopic spectral question also uses the cubic phase of
\cite[Section 3]{COS}. The surviving terms in its Eq.~(8.24)
already exhibit correlation-energy contributions of order one.
Here we resolve that contribution through bounded tests of
$Q_N(t)/N^2$, with the first energy as weight.
Theorems~\ref{thm:uniform-evolved-energy-spectrum}
and~\ref{thm:admissible-initial-class} identify the
limiting relative scattering measure and control its coefficient
uniformly on compact time intervals.
The contact in this statement is the explicitly defined form
\eqref{spec:eq:dressed-contact} in the chosen excitation
representation.
Contact relations for zero-range Bose gases link momentum tails,
short-distance pair correlations and scattering-length derivatives
\cite[Table I]{WernerCastin2012}. Radio-frequency spectroscopy of identical bosons also has a
two-body contact contribution proportional to $\omega^{-3/2}$
\cite[Eq.~(6)]{BKP2011}. The operator product expansion gives
contact-dependent scaling functions for the Bragg density response
\cite[Section II]{HofmannZwerger2017}.
Our observable is a one-time spectral measurement of $Q_N(t)$.
Moreover, at the $N^2$ energy scale the measure $\mu_v$ retains
the full relative Hamiltonian $-2\Delta+v$.
The threshold limit below takes $N\to\infty$ first and then
lowers the positive rescaled energy.
Proposition~\ref{prop:radial-microscopic-density} computes
its density from a radial scattering equation. The leading
threshold behavior is
$\rho_v(\lambda)\sim4\sqrt2\,a^2\lambda^{-1/2}$.
This calculation uses the standard half-line Green function
and Stone's formula \cite{Teschl2009}.
Corollary~\ref{cor:evolved-energy-probability} then gives the
$N^{-2}$ probability of every fixed positive microscopic
energy window along the exact evolution. Conditioned on
that window, the limiting density is proportional to
$\rho_v(\lambda)/\lambda$. Its shape depends on the potential;
the evolving contact sets the overall probability.
Proposition~\ref{prop:evolved-energy-defect} compares this
microscopic contribution with the unscaled measurement law.
That law converges to the spectral law of $A_{\infty,t}$ in
the quadratic reference state, but its first moment has the
positive excess $2D_vc(t)$. The energy at scale $N^2$
accounts for this excess. The statement concerns the same
prepared states as the physical-moment bounds.

A related short-distance question is correlation formation from
the product state $\phi^{\otimes N}$.
Erd\H{o}s, Michelangeli and Schlein \cite[Theorem 1.1]{EMS}
measure the developing zero-energy scattering structure in a
local relative-coordinate window during an initial time layer.
The present spectral statement follows inverse-energy preparations
on macroscopic time intervals.

Collective observables also connect the fluctuation evolution to
probabilistic limit theorems. Central limit theorems for mean-field
evolution were proved in \cite{BAKS,BSSCLT}; the corresponding
Gross--Pitaevskii results include the ground-state theorem of
\cite{RSCLT} and the time-dependent result of \cite{COS}.
Exponential bounds on the excitation number are available for
low-temperature equilibrium states on the torus \cite{NR} and for
low-energy trapped states \cite{BBR}. Since a spectral subspace is
preserved by its Hamiltonian, the low-energy bounds also control
evolution under the same stationary Hamiltonian. Here we propagate the third correlated-energy moment relative
to a moving condensate, under the initial bound in Theorem~\ref{thm:D}.
Uniform-in-time Bogoliubov approximation is available in the
fixed-potential mean-field regime under dispersive assumptions
\cite[Theorem 1.2]{NRS2026}. That result starts from product states
and uses a different interaction scaling. Our constants may
depend on the compact observation interval.

\subsection{Proof strategy and sources}

The main result for the exact evolution is Theorem~\ref{thm:D}.
We use the full
fluctuation generator after the quadratic and cubic correlation
transformations, with a linear number correction that makes it
positive. Its third spectral moment is the higher-order energy
propagated along the exact evolution.

We use the correlation transformations and scattering cancellations
of \cite{BS,COS}. The main additional model estimates are the
second time-derivative and number-commutator bounds in
Proposition~\ref{prop:full-second-form}. They preserve the cancellations
between singular kernels and include the time-dependent transformation
terms, with constants uniform in particle number.
Lemma~\ref{lem:exact-moment-coercivity} then relates the correlated
energy to number-weighted kinetic and interaction energy. For this
step, the positive form comparison and the first number-commutator
bound suffice, through the standard inverse-commutator identity.

The second derivative requires more than a repetition of the
first-energy estimate. In the quadratic logarithmic derivative
$-ie^{-B_t}\nabla_t(e^{B_t})$, it produces a third time derivative
of the pair kernel. The singular spatial terms must also be combined
before differentiation. For example, the kinetic and interaction
pair coefficients reduce, by the Neumann scattering equation, to
\[
	N^3\bigl\{(\Delta w_\ell)(Nr)+\tfrac12v(Nr)(1-w_\ell(Nr))\bigr\}
	=N^3\lambda_\ell f_\ell(Nr)\mathbf1_{|r|\le\ell}.
\]
Here $r=x-y$ and the Neumann function is extended by $1$ outside
its ball. The right-hand side is uniformly bounded, whereas the
separate interaction coefficient has a growing $L^2$ norm.
This cancellation is inherited from \cite[Section 5]{BS}.
The additional step is to control the cancelled kernel series on
a common $H^2$ neighborhood of the condensate and to justify two
time derivatives of that series. Lemma~\ref{lem:phase-second-regularity}
and Proposition~\ref{prop:full-second-form} provide these bounds,
including the third kernel derivatives in the logarithmic terms.
The $H^6$ assumption concerns the condensate; the propagated
fluctuation norm is defined by the correlated energy.

The derivative bounds allow us to apply Kisy\'nski's form-space
propagation theorem \cite{Kisynski}, in the formulation of
\cite[Theorem 3.10(iv)]{BLP}. Lemma~\ref{lem:three-halves-propagation}
records the dependence of its constants on the form bounds.
The number-moment estimates for the correlation transformations
\cite{BS,COS} and the condensate-extraction formalism \cite{LNS2015}
then give the depletion and collective-observable bounds. These are
physical consequences of the same energy estimate.

The microscopic spectral argument has two steps.
Pair dilation identifies the relative scattering measure in
Theorem~\ref{spec:thm:actual-COS-profile}.
Theorem~\ref{thm:uniform-evolved-energy-spectrum} transfers it to
the exact evolution by comparison in the energy form norm,
including the different condensate projections.
The completion argument in
Theorem~\ref{thm:admissible-initial-class} then reaches the full
norm-convergent initial class.

The contact is defined by the source form
\eqref{spec:eq:dressed-contact}.
Proposition~\ref{prop:contact-positive-kinetic} proves its strict
positivity at finite kinetic energy and finite mean excitation
number. Proposition~\ref{prop:contact-energy-coercivity} gives a
quantitative lower bound from the high-frequency tail of the
pair kernel and the canonical commutation relation.
Thus the spectral contribution and the physical moment bounds
hold for the same evolving family.
Proposition~\ref{prop:physical-energy-comparison} expresses both
energies on the physical $N$-particle space.

The source of each main input is summarized below.
The right column gives the additional estimate or application
proved in this paper.
\begin{center}
\small
\begin{tabular}{@{}p{0.22\linewidth}p{0.31\linewidth}p{0.39\linewidth}@{}}
\hline
Input & Established result & Present argument\\
\hline
Correlation transformations &
Quadratic and cubic phases, scattering cancellations
\cite{BS,COS} &
Number-weighted kernel bounds on a common parameter
neighborhood; two time derivatives in
Proposition~\ref{prop:full-second-form}.\\[3pt]
Positive form evolution &
Kisy\'nski's theorem and its higher-domain formulation
\cite{Kisynski,BLP} &
Uniform model hypotheses, tracked constants and the
physical moments in Theorem~\ref{thm:D}.\\[3pt]
Weighted coercivity &
The inverse-commutator identity \cite{GoleniaJecko2013} &
The specific number weight in
Lemma~\ref{lem:exact-moment-coercivity}.\\[3pt]
Microscopic correlations &
Correlation formation \cite{EMS} and the surviving energy
terms of \cite[Eq.~(8.24)]{COS} &
The relative spectral measure and its transfer to the
same exact evolving states.\\[3pt]
Spectral and probability tools &
Stone's formula \cite{Teschl2009} and the
L\'evy--Khintchine representation \cite{Bertoin2000} &
The model density, contact coefficient and independent
energy-sampling law.\\
\hline
\end{tabular}
\end{center}

Equation and lemma numbers from \cite{BS,COS} refer to the
arXiv versions specified in the bibliography.

\subsection{Plan of the paper}

The proofs have three parts, with common correlation and form
estimates. Section~\ref{sec:generators} defines the transformations
and the generators. Section~\ref{sec:taylor-energy} proves the
first-energy and weighted inverse estimates.
Proposition~\ref{prop:full-second-form} in
Section~\ref{sec:second-regularity} supplies two time derivatives
and the number commutators, with constants uniform in $N$.
These model estimates are the input to the propagation theorem.

The first part proves Theorem~\ref{thm:D}.
In Section~\ref{sec:exact-moment-stability},
Lemma~\ref{lem:exact-moment-coercivity} gives weighted coercivity
for the exact correlated energy.
Theorem~\ref{thm:exact-third-moment} then propagates its third
moment. The reconstruction estimates in
Theorem~\ref{thm:physical-sixth-moment} convert this bound into
the stated physical fluctuation bounds.

The second part connects these states to microscopic energy
measurements. Section~\ref{sec:microscopic-spectrum} uses pair
dilation and the relative scattering operator to compute the
initial spectral measure. Sections~\ref{sec:evolved-energy}
and \ref{sec:evolving-spectrum} combine the quadratic norm
approximation of \cite{COS} with exact third-moment propagation
to obtain first-energy convergence and transfer this measure
to the exact evolving state. Lemma~\ref{lem:third-energy-completion}
and Theorem~\ref{thm:admissible-initial-class} extend the argument
to the full class of norm-convergent families with the initial
third-energy bound. The interaction-energy measure and its
scaled moments follow in
Theorem~\ref{thm:admissible-interaction-spectrum}.

The third part proves the Taylor approximation in
Theorems~\ref{thm:A}--\ref{thm:C}.
Section~\ref{sec:inverse-comparison} gives the comparison principle.
The cubic source estimates of Section~\ref{sec:cubic-estimates}
and the generator expansion in Section~\ref{sec:model-expansion}
control the fourth integral remainder and the time term.
These estimates determine the two approximation rates.
Section~\ref{sec:conditional-comparisons} treats direct Duhamel
bounds with stronger trajectory norms.
Section~\ref{sec:discussion} discusses the results and possible extensions.

\section{Correlation transformations and finite cubic expansion}
\label{sec:generators}

We formulate the finite conjugation estimate and then verify its
number-weight assumptions for the Gross--Pitaevskii generator.
Covariant differentiation accounts for the moving excitation space
and the time dependence of the cubic transformation. The final
step constructs the Taylor evolution at fixed particle number.
The commutator identities follow the exponential-operator calculus
of \cite{Magnus,Wilcox}. We use a finite integral remainder and
prove its weighted form bounds explicitly.

\subsection{Asymmetric form remainder}
\label{sec:asymmetric-remainder}

\begin{lemma}\label{lem:words}
Let $M=\cN+1$ on a graded Hilbert space. Let $A,Q_N$ be
positive self-adjoint operators that strongly commute with $M$ and satisfy
$M\le A\le Q_N\le BMA$ as forms.
Let $C^*=-C$, with $C_{mn}=0$ if $|m-n|>3$ or
$\max(m,n)>L_N$. Suppose, for $z_n\in D(A_n^{1/2})$, that
\[
	\|C_{mn}z_n\|\le B\epsilon(n+1)^{3/2}\|z_n\|,\qquad
	\|A_m^{1/2}C_{mn}z_n\|
	\le B(n+1)^{3/2}\bigl(\epsilon\|A_n^{1/2}z_n\|+\|z_n\|\bigr).
\]
Assume $0<\epsilon\le1$ and $\epsilon\sqrt{L_N+2}\le D_1$,
with $B,D_1$ independent of $N$.
For fixed integers $j\ge1$, real $r$, and $z\in D(M^{r+3j/2}Q_N^{1/2})$,
\begin{equation}\label{eq:words}
	\norm{M^rQ_N^{1/2}C^jz} \le C_{j,r,D_1}\epsilon^{j-1} \norm{M^{r+3j/2}Q_N^{1/2}z}.
\end{equation}
For $j=0$ use the identity, with no factor $\epsilon^{-1}$.
\end{lemma}
\begin{proof}
There are finitely many number paths of a fixed length $j$.
Iteration of \eqref{eq:blocks} gives, with
$a_n=\norm{A_n^{1/2}z_n}$,
\[
	\norm{A_m^{1/2}(C^j)_{mn}z_n} \le B_j(n+1)^{3j/2} \big(\epsilon^j a_n+j\epsilon^{j-1}\norm{z_n}\big).
\]
Here $|m-n|\le3j$. Using \eqref{eq:comparison} and $A_n\ge n+1$
bounds the $Q_N$ energy norm of the block by
\[
	C_j\epsilon^{j-1}(n+1)^{3j/2} \big(\epsilon\sqrt{n+1}+j\big)a_n.
\]
All active inputs are within the cutoff, so the last parenthesis
is uniformly bounded by \eqref{eq:weak-budget}. The shifted weights
$(m+1)^r$ and $(n+1)^r$ are comparable for fixed $j,r$.
Finite-band Cauchy--Schwarz, followed by $A\le Q_N$, proves the
claim. Closure extends it from finite-number form-core vectors.
\end{proof}

For a Hermitian $Q_N$-continuous form $G$, define
$T_N=Q_N^{-1/2}GQ_N^{-1/2}$.
For an integer $s\ge2$ suppose
\begin{equation}\label{eq:number-compatible}
	\norm{M^\alpha T_NM^{-\alpha}}\le h_s,\qquad \alpha=0,\tfrac32,3,\ldots,\tfrac{3s}{2}.
\end{equation}
Products are first defined on finite-number vectors and are
required to have the displayed bounded extensions.

\begin{proposition}\label{prop:asymmetric-comm}
Let $M=\cN+1$ on a graded Hilbert space. Let $A,Q_N$ be
positive self-adjoint operators that strongly commute with $M$ and satisfy
$M\le A\le Q_N\le BMA$ as forms.
Let $C^*=-C$, with $C_{mn}=0$ if $|m-n|>3$ or
$\max(m,n)>L_N$. Suppose, for $z_n\in D(A_n^{1/2})$, that
\[
	\|C_{mn}z_n\|\le B\epsilon(n+1)^{3/2}\|z_n\|,\qquad
	\|A_m^{1/2}C_{mn}z_n\|
	\le B(n+1)^{3/2}\bigl(\epsilon\|A_n^{1/2}z_n\|+\|z_n\|\bigr).
\]
Assume $0<\epsilon\le1$ and $\epsilon\sqrt{L_N+2}\le D_1$,
with $B,D_1$ independent of $N$.
Fix an integer $s\ge2$. Let $G$ be a Hermitian form continuous
on $D(Q_N^{1/2})$, and put $T_N=Q_N^{-1/2}GQ_N^{-1/2}$.
Assume that the products initially defined on finite-number vectors
extend to bounded operators satisfying
\[
	\|M^\alpha T_NM^{-\alpha}\|\le h_s,
	\qquad \alpha=0,\tfrac32,3,\ldots,\tfrac{3s}{2},
\]
with $h_s$ independent of $N$.
For $u\in D(Q_N^{1/2})$ and $v\in D(M^{3s/2}Q_N^{1/2})$,
\[
	|(\mathcal D_C^sG)[u,v]| \le C_{s,D_1}h_s\epsilon^{s-2} \norm{Q_N^{1/2}u}\norm{M^{3s/2}Q_N^{1/2}v}.
\]
Here $\mathcal D_CG=GC-CG$ as a form.
\end{proposition}
\begin{proof}
Differentiation of $G[e^{tC}u,e^{tC}v]$ gives
\[
	(\mathcal D_C^sG)[u,v] =\sum_{j=0}^s\binom{s}{j}G[C^ju,C^{s-j}v].
\]
For $a=3j/2$, \eqref{eq:number-compatible} bounds the $j$th
pairing by
\[
	h_s\norm{M^{-a}Q_N^{1/2}C^ju} \norm{M^aQ_N^{1/2}C^{s-j}v}.
\]
If $1\le j\le s-1$, Lemma~\ref{lem:words} cancels all number
weights on $u$, puts their sum $3s/2$ on $v$, and supplies
$\epsilon^{s-2}$. For $j=0,s$ it supplies the stronger
$\epsilon^{s-1}$, which is at most $\epsilon^{s-2}$.
Sum the finitely many terms and extend by continuity.
\end{proof}

The phase satisfies the additional estimate
\begin{equation}\label{eq:phase-energy}
	\norm{M^rQ_N^{1/2}e^{tC}z} \le E_r\norm{M^rQ_N^{1/2}z}, \quad 0\le t\le1,\quad r=0,3s/2.
\end{equation}
For the standard phase this is
\cite[Lemma 3.1 and its proof in Section 7]{COS}, with number
weights $0$ and $3s$.

\begin{theorem}\label{thm:remainder}
Let $M=\cN+1$ on a graded Hilbert space. Let $A,Q_N$ be
positive self-adjoint operators that strongly commute with $M$ and satisfy
$M\le A\le Q_N\le BMA$ as forms.
Let $C^*=-C$, with $C_{mn}=0$ if $|m-n|>3$ or
$\max(m,n)>L_N$. Suppose, for $z_n\in D(A_n^{1/2})$, that
\[
	\|C_{mn}z_n\|\le B\epsilon(n+1)^{3/2}\|z_n\|,\qquad
	\|A_m^{1/2}C_{mn}z_n\|
	\le B(n+1)^{3/2}\bigl(\epsilon\|A_n^{1/2}z_n\|+\|z_n\|\bigr).
\]
Assume $0<\epsilon\le1$ and $\epsilon\sqrt{L_N+2}\le D_1$,
with $B,D_1$ independent of $N$.
Fix an integer $s\ge2$. Let $G$ be a Hermitian form continuous
on $D(Q_N^{1/2})$, and put $T_N=Q_N^{-1/2}GQ_N^{-1/2}$.
Assume that the products initially defined on finite-number vectors
extend to bounded operators satisfying
\[
	\|M^\alpha T_NM^{-\alpha}\|\le h_s,
	\qquad \alpha=0,\tfrac32,3,\ldots,\tfrac{3s}{2},
\]
with $h_s$ independent of $N$.
Suppose also that, with constants $E_0,E_{3s/2}$ independent of $N$,
\[
	\|M^rQ_N^{1/2}e^{tC}z\|\le E_r\|M^rQ_N^{1/2}z\|,
	\qquad 0\le t\le1,\quad r=0,3s/2.
\]
For $u\in D(Q_N^{1/2})$ and $v\in D(M^{3s/2}Q_N^{1/2})$,
the remainder after degree $s-1$ satisfies
\begin{align}
	R_{s,N}&=e^{-C}Ge^C-\sum_{j=0}^{s-1}
	\frac{\mathcal D_C^jG}{j!},\notag\\
	|R_{s,N}[u,v]|
	&\le C_{s,D_1}h_sE_0E_{3s/2}\epsilon^{s-2}
	\norm{Q_N^{1/2}u}\norm{M^{3s/2}Q_N^{1/2}v}.
	\label{eq:remainder}
\end{align}
\end{theorem}
\begin{proof}
Taylor's formula applied to the conjugation map gives the exact
integral remainder; see \cite{Wilcox} for the operator-exponential
identities underlying this calculation:
\[
	R_{s,N}[u,v]=\frac1{(s-1)!}\int_0^1(1-t)^{s-1} (\mathcal D_C^sG)[e^{tC}u,e^{tC}v]\d t.
\]
Apply Proposition~\ref{prop:asymmetric-comm} and
\eqref{eq:phase-energy}, and integrate. Fixed-$N$ analyticity
justifies the formula; the resulting bound gives the asserted
asymmetric domain extension.
\end{proof}

\subsection{Number-weight bounds for the full generator}
\label{sec:number-compatibility}

The weighted estimate \eqref{eq:number-compatible} requires
more information than the ordinary energy bound
$|G[u,v]|\le C\norm{Q_N^{1/2}u}\norm{Q_N^{1/2}v}$,
even when a first number-commutator bound is available. For example on $\ell^2(\mathbb N_0)$,
let $Me_n=(n+1)e_n$, $Q=M$, and
\[
	T_N=I+N^{-1}(|e_N\rangle\langle e_0| +|e_0\rangle\langle e_N|).
\]
Then $\norm{T_N}\le3/2$ for $N\ge2$ and $\norm{[M,T_N]}=1$,
but $\norm{M^\alpha T_NM^{-\alpha}}\ge(N+1)^\alpha/N$.
Additional model structure is necessary.

\begin{lemma}\label{lem:shifts}
Let $Q$ commute with $M$. Suppose
$G=G_0+\sum_\iota E_\iota$ as continuous energy forms, where
$G_0$ preserves particle number and
$T_\iota=Q^{-1/2}E_\iota Q^{-1/2}$ has number shift $d_\iota$.
If
\[
	\sum_\iota(1+|d_\iota|)^a\norm{T_\iota}<\infty,
\]
then
\[
	\norm{M^\alpha Q^{-1/2}GQ^{-1/2}M^{-\alpha}} \le \norm{Q^{-1/2}G_0Q^{-1/2}} +\sum_\iota(1+|d_\iota|)^a\norm{T_\iota}
\]
for $|\alpha|\le a$.
\end{lemma}
\begin{proof}
On an active block $m=n+d$, both $(m+1)/(n+1)$ and its reciprocal
are at most $1+|d|$. A single-shift operator is an orthogonal
direct sum of its blocks. Its weighted norm is therefore bounded
by $(1+|d|)^{|\alpha|}$ times its norm. Sum in operator norm.
\end{proof}

\subsubsection{Fixed-kernel field estimates}
\label{sec:fixed-kernel-estimates}

We first establish the number-weighted estimates used in both the
static generator bound and the time-regularity argument.
They depend on the field expansion of \cite[Lemmas 3.2 and 5.2]{BS}.
All operators in this subsection act on $\cF^{\le N}$.
For a symmetric Hilbert--Schmidt kernel $\eta$, use
\[
	B(\eta)=\frac12\int\bigl(\eta(x,y)b_x^*b_y^*
	-\overline{\eta(x,y)}b_xb_y\bigr)\d x\d y,
	\qquad b_x=\sqrt{1-\cN/N}\,a_x .
\]
The generalized fields include the physical number cutoff.
Set $\mathcal D_BX=[X,B]$ and
$\mathcal L_j(g)=\mathcal D_B^jb(g)/j!$ for $g\in L^2$.
The constants below are independent of $N$.

For symmetric
Hilbert--Schmidt kernels $\zeta,\beta$, set
\begin{equation}\label{eq:time-homogeneous-map}
	\mathcal T_J(\zeta,\beta)
	=-\frac{i}{(J+1)!}\mathcal D_{B(\zeta)}^J B(\beta),
	\qquad \mathcal D_B X=[X,B].
\end{equation}
This is a real homogeneous polynomial of degree $J$ in
$\zeta$ and is real linear in $\beta$.
Write $r_\beta(x)=\|\beta(x,\cdot)\|_2$ and
$\rho=\|\zeta\|_{\mathrm{HS}}$. The two terminal cases in
\cite[Lemma 5.2(iii), Eq.~(5.17)]{BS}, used in
\cite[Eqs.~(5.128)--(5.129)]{BS}, have bounds
\[
	C_0^J\rho^{J-1}r_\beta(x)r_\zeta(x)\|M^{1/2}z\|,
	\qquad C_0^J\rho^Jr_\beta(x)\|a_xz\|.
\]
The first case occurs only when $J\ge1$.
The factor $\nabla\eta$ in that calculation can be replaced
by any symmetric HS kernel $\beta$: the argument uses only
its row function and the integrals
\[
	\int r_\beta r_\zeta\d x\le\|\beta\|_{\mathrm{HS}}\rho,
	\qquad
	\int r_\beta(x)\|a_xz\|\d x
	\le\|\beta\|_{\mathrm{HS}}\|\cN^{1/2}z\|.
\]
For each $n+k=J$, the field expansion has $2^Jn!k!$
terms and coefficient $1/[2n!k!(J+1)]$.
There are $J+1$ such index pairs, and the adjoint terms
give only another factor two. Thus, for one fixed constant $c_{\rm time}$,
\begin{equation}\label{eq:time-homogeneous-form}
	|\mathcal T_J(\zeta,\beta)[z,z]|
	\le C(c_{\rm time}\rho)^J\|\beta\|_{\mathrm{HS}}\|M^{1/2}z\|^2 .
\end{equation}
For $J=0$ the same estimate follows directly from the pair
creation and annihilation bounds. This also checks the
low orders without a spatial derivative of $\beta$.

Let $\|T\|_{M,r}=\|M^{r-1/2}TM^{-r-1/2}\|$.
The numerical-radius inequality turns
\eqref{eq:time-homogeneous-form} into a bounded form
operator. Its particle-number bandwidth is at most $2J+2$.
To check the norm of a fixed particle-number shift, write
$A=M^{-1/2}\mathcal T_JM^{-1/2}$ and let $\Pi_n$ be the
$n$-particle projection. Orthogonality gives
\[
	\left\|\sum_n\Pi_{n+s}A\Pi_nz\right\|^2
	=\sum_n\|\Pi_{n+s}A\Pi_nz\|^2
	\le\|A\|^2\sum_n\|\Pi_nz\|^2 .
\]
The sum contains only $0\le n,n+s\le N$; all other
blocks vanish. Thus each fixed-shift part has norm at
most $\|A\|$. The preceding numerical-radius step uses
$\|A\|\le2\sup_{\|z\|=1}|\langle z,Az\rangle|$.
There are at most $4J+5$ shifts. On each nonzero block,
conjugation by $M^r$ multiplies the block by
$((n+s+1)/(n+1))^r$, whose modulus is at most
$(2J+3)^{|r|}$. Orthogonality gives the same bound for
each shifted operator. Hence
\begin{equation}\label{eq:time-homogeneous-weight}
	\|\mathcal T_J(\zeta,\beta)\|_{M,r}
	\le C_r(1+J)^{b_r}(c_{\rm time}\rho)^J\|\beta\|_{\mathrm{HS}} .
\end{equation}
One or two number commutators add only the factors
$2J+2$ or $(2J+2)^2$.

For the remaining estimates take $\zeta=\eta$ and
$\rho=\|\eta\|_{\mathrm{HS}}$. Choose the strict small-kernel
radius so that the geometric factors in these estimates are
less than one.

We also retain the two expansion indices in the conjugate
of $N^{-1/2}Mb(g)$.
Set $\mathcal S_0=M$ and
$\mathcal S_l=\mathcal D_B^lM/l!$ for $l\ge1$.
The identity $[M,B(\eta)]=-2iB(i\eta)$ gives
\[
	\mathcal S_l=2\mathcal T_{l-1}(\eta,i\eta),\qquad l\ge1,
\]
where $\mathcal T_j$ is defined in
\eqref{eq:time-homogeneous-map}.
Equation \eqref{eq:time-homogeneous-weight} therefore gives,
for every fixed $r\in\mathbb R$,
\[
	\|M^{r-1/2}\mathcal S_lM^{-r-1/2}\|
	\le C_r(1+l)^{b_r}(C_1\rho)^l,
	\qquad l\ge0 .
\]
The case $l=0$ is the identity after balancing by $M$.
For the transformed field, \cite[Lemma 5.2(i)]{BS} bounds
$M^{-1/2}$ times each order-$j$ term by
$C_0^j\rho^j\|g\|_2$ in operator norm.
Each term has a definite number shift $|d|\le2j+1$.
On a block from $n$ to $n+d$, moving the fixed number
weights introduces a power of $(n+d+1)/(n+1)$.
This ratio and its reciprocal are at most $2j+2$.
After summing the terms and dividing by $j!$, we obtain
\[
	\|M^{r+1/2}\mathcal L_j(g)M^{-r-1}\|
	\le C_r(1+j)^{b_r}(C_1\rho)^j\|g\|_2 .
\]
The ordered product factors as
\begin{align*}
	M^{r-1/2}\mathcal S_l\mathcal L_j(g)M^{-r-1/2}
	={}&\bigl(M^{r-1/2}\mathcal S_lM^{-r-1/2}\bigr)\\
	&\times\bigl(M^{r+1/2}\mathcal L_j(g)M^{-r-1}\bigr)M^{1/2} .
\end{align*}
Since $M\le N+1$ on the physical Fock space, this proves
\begin{equation}\label{eq:linear-number-product-degree}
	N^{-1/2}\|M^{r-1/2}\mathcal S_l\mathcal L_j(g)M^{-r-1/2}\|
	\le C_r(1+l+j)^{b_r}(C_1\rho)^{l+j}\|g\|_2 .
\end{equation}
For fixed $J=l+j$, summing over the $J+1$ pairs preserves
geometric decay. The expansions of $e^{-B}Me^B$ and
$e^{-B}b(g)e^B$ converge in their displayed weighted
norms. At fixed $N$ their product equals
$e^{-B}Mb(g)e^B$, and the displayed product bound is
uniform in $N$.

For the conjugate of $M^2/N$, write the ordered product as
\begin{align*}
	M^{r-1/2}\mathcal S_l\mathcal S_jM^{-r-1/2}
	={}&\bigl(M^{r-1/2}\mathcal S_lM^{-r-1/2}\bigr)\\
	&\times\bigl(M^{r+1/2}\mathcal S_jM^{-r-3/2}\bigr)M .
\end{align*}
The second factor uses the preceding estimate with weight
$r+1$. Since $M/N\le2$, we obtain
\begin{equation}\label{eq:number-square-degree}
	\frac1N\sum_{l+j=J}\|\mathcal S_l\mathcal S_j\|_{M,r}
	\le C_r(1+J)^{b_r}(C_1\rho)^J .
\end{equation}
The $J+1$ choices of $l,j$ only change the polynomial.
At fixed $N$, multiplication of the two convergent series
gives $e^{-B}M^2e^B/N$.
After the scalar subtraction, the number polynomial in
\cite[Section 5.2]{BS} is a linear combination of
$1,M,M/N,M^2/N$ with uniformly bounded coefficients.
Thus \eqref{eq:number-square-degree} covers this group.

For a bounded self-adjoint one-particle operator $h$, put
$\mathcal X_j(h)=\mathcal D_B^j\dGamma(h)/j!$ and
$\beta_h=(h\otimes1+1\otimes h)\eta$.
The kernel $\beta_h$ is symmetric and
$\|\beta_h\|_{\mathrm{HS}}\le2\|h\|\rho$.
Since $\dGamma(h)$ commutes with $M$, the number factors
in the generalized fields commute with it as well.
The canonical commutation relations give
$[\dGamma(h),B(\eta)]=-iB(i\beta_h)$.
The factor $-i$ turns the anti-Hermitian pair operator into
the required Hermitian pair sum. Therefore
\begin{align}
	\mathcal X_j(h)&=\mathcal T_{j-1}(\eta,i\beta_h),
	&&j\ge1,\notag\\
	\|\mathcal X_j(h)\|_{M,r}
	&\le C_r(1+j)^{b_r}(C_1\rho)^j\|h\|,
	&&j\ge0 .\label{eq:one-body-degree}
\end{align}
Here $j=0$ follows from $\|\dGamma(h)|_{\cF^{(n)}}\|\le n\|h\|$.
For $j\ge1$ we used \eqref{eq:time-homogeneous-weight}.
The resulting series equals $e^{-B}\dGamma(h)e^B$.

The normal quadratic terms contain generalized fields.
Define $\mathcal B_h=\int h(x,y)b_x^*b_y\d x\d y$,
with the same kernel convention as for $\dGamma(h)$.
For a bounded $h$ without a function kernel, this notation
denotes the corresponding one-particle quadratic form.
On the $n$-particle sector, the intermediate number in
$b_x^*b_y$ is $n-1$. Consequently
\begin{equation}\label{eq:normal-one-body-number}
	\mathcal B_h-N^{-1}\dGamma(h)
	=\left(1-\frac{\cN}{N}\right)\dGamma(h)
	=\left(1+\frac1N\right)\dGamma(h)-\frac1N M\dGamma(h).
\end{equation}
The vacuum identity follows from $\dGamma(h)\Omega=0$.
Conjugating the last product gives
$N^{-1}\sum_{l,j}\mathcal S_l\mathcal X_j(h)$.
The factorization used for \eqref{eq:number-square-degree}
applies with $\mathcal X_j(h)$ in place of $\mathcal S_j$.
Its sum at degree $J=l+j$ is bounded by
$C_r(1+J)^{b_r}(C_1\rho)^J\|h\|$.
This proves the bound for the first two terms of
\cite[Eq.~(5.88)]{BS}, with their number factors retained.

\subsubsection{The full generator}

We now use the precise model of \cite{BS}. Its condensate
$\phi_t=\widetilde\varphi_{N,t}$ solves the modified GP equation
with normalized $H^4$ initial datum. The fixed scattering radius
$\ell$ is sufficiently small as in that paper. Let $G_{N,t}$ be
its exact quadratically transformed generator, Eq.~(4.24), and
$c_{N,t}$ its scalar, Eq.~(4.25). On $P_t\cF^{\le N}$ write
\[
	\mathfrak G_{N,t}=P_t(G_{N,t}-c_{N,t})P_t.
\]
The parameter choices in the cubic phase are those of \cite{COS}
on this same background.

\begin{proposition}\label{prop:GP-weight}
Let $v\in C_c^\infty(\mathbb R^3)$ be nonnegative and radial.
Let $\phi_t$ solve the modified GP equation with a fixed normalized
initial datum in $H^4(\mathbb R^3)$. Fix a sufficiently small
quadratic scattering radius $\ell\in(0,1/2)$, independent of $N$.
Let $\mathfrak G_{N,t}=P_t(G_{N,t}-c_{N,t})P_t$, with
$G_{N,t},c_{N,t}$ as in \cite[Eqs.~(4.24)--(4.25)]{BS}.
For every fixed real $\alpha$ and $T<\infty$,
\begin{equation}\label{eq:GP-weight}
	\sup_{N,\,|t|\le T} \norm{M^\alpha Q_{N,P_t}^{-1/2} \mathfrak G_{N,t}Q_{N,P_t}^{-1/2}M^{-\alpha}} <\infty.
\end{equation}
\end{proposition}
\begin{proof}
We use the termwise proof in \cite[Section 5]{BS} to
establish the number-weighted extension of its Theorem 4.4.

\textit{Number degree and cancellations.}
In the expansion of \cite[Lemma 3.2]{BS}, an order-$j$ transformed
field is a sum of $2^j j!$ monomials. Each monomial has a definite
number shift, bounded in absolute value by $2j+1$.
Indeed, its number factors have shift zero, each pair factor has
shift $0$ or $\pm2$, and the terminal field contributes $\pm1$.
The counting identity in that lemma bounds the number of pair
factors by $j$. Products arising from up to four fields consequently
have shift at most $2J+4$, where $J$ is the sum of the expansion
indices. Kernel contractions and integrations by parts preserve
the shift.

First carry out the subtractions in \cite[Eqs.~(5.45), (5.63),
(5.89), (5.98), (5.115), (5.130)]{BS}. The large linear terms
cancel with their identical counterparts. Scalars have shift zero;
the singular pair terms are combined using the scattering equation
before estimating them. Their remaining Neumann-eigenvalue pair
kernel has uniformly bounded Hilbert--Schmidt norm and shift
$\pm2$. Thus these cancellations are compatible with number
weights.

\textit{Termwise majorants.}
For the nontrivial energy terms, the local estimates in that proof
give geometric majorants after the factorial coefficients and
the $2^j j!$ multiplicities have been accounted for. The relevant
locations are as follows:
\[
	\begin{array}{c|l}
	\text{family}&\text{termwise estimates in \cite{BS}}\\ \hline
	\text{linear remainder}&(5.52)\text{--}(5.58)\\
	\text{kinetic remainder}&(5.78)\text{--}(5.87)
	\text{ and the summary after }(5.87)\\
	\text{pair remainder}&(5.93)\text{--}(5.97)\\
	\text{cubic remainder}&(5.102)\text{--}(5.114)\\
	\text{quartic remainder}&(5.119)\text{--}(5.126)\\
	\text{quadratic time term}&(5.128)\text{--}(5.129)
	\end{array}
\]
More precisely, group a normal-ordered contribution, together
with any subtraction of the same shift needed at its lowest
orders, as $E_\iota$. If its indices have total $J$, the sum of
its numerical-radius bounds after inserting the series
coefficients is at most
\begin{equation}\label{eq:degree-majorant}
	C_T(1+J)^b\vartheta^{\max(J-b_0,0)} \norm{Q_N^{1/2}z}^2,\qquad 0<\vartheta<1,
\end{equation}
for fixed integers $b,b_0$. There are finitely many families,
with at most four expansion indices. Increasing $b$ absorbs the
number of index tuples of a given total.
For example, the kinetic summary after (5.87) retains
$\norm{\eta_t}^{\max(J-3,0)}$ for every elementary term.
The extra $n$ or $n+1$ factors in the pair and quartic estimates
are polynomial in $J$, not new factorials.
These estimates hold for each individual monomial.

One estimate requires retaining an expansion index that was dropped in the
published bound.
In (5.103) the power of $\norm{\eta_t}$ shown after simplification
contains $k+r-1$ but not $n$. Use instead (5.102) and the two
bounds on $P_2$ immediately preceding (5.103).
Put $\rho=\|\eta_t\|_{\mathrm{HS}}$,
$a(x)=\|\eta_t(x,\cdot)\|_2$ and
$F_N(x)=N^{5/2}|v(Nx)|$. Define the nonnegative function
$H_N=F_N*(|\phi_t|a)$.
Young's inequality gives
\begin{equation}\label{eq:cubic-convolution-majorant}
	\|H_N\|_2\le C_TN^{-1/2}\rho,\qquad
	\|H_N\|_\infty\le C_TN^{-1/2},\qquad
	\|a\|_2=\rho,\quad \|a\|_\infty\le C_T .
\end{equation}
Here $\|F_N\|_1=N^{-1/2}\|v\|_1$ and
$\|\phi_t\|_\infty\le C_T$.
Write $L=\|M^{1/2}z\|$, $L_1=\|Mz\|$,
$Z(x)=\|a_xz\|$ and $Y(x)=\|a_xM^{1/2}z\|$.
Then $\|Z\|_2\le L$ and $\|Y\|_2\le L_1$.
The two estimates for $P_1$ are
$C_0^n\rho^{n-1}a(x)L$ and $C_0^n\rho^nZ(x)$.
Those for $P_2$ are
$C_0^{k+r}\rho^{k+r-2}a(x)a(y)L_1$ and
$C_0^{k+r}\rho^{k+r-1}a(y)Y(x)$.
For $J=n+k+r$, their four products therefore give
\begin{align}
	\rho^{J-3}LL_1\int a(x)^2H_N(x)\d x
	&\le C_TN^{-1/2}\rho^{J-1}LL_1,\notag\\
	\rho^{J-2}L\int a(x)H_N(x)Y(x)\d x
	&\le C_TN^{-1/2}\rho^{J-1}LL_1,\notag\\
	\rho^{J-2}L_1\int a(x)H_N(x)Z(x)\d x
	&\le C_TN^{-1/2}\rho^{J-1}LL_1,\notag\\
	\rho^{J-1}\int H_N(x)Z(x)Y(x)\d x
	&\le C_TN^{-1/2}\rho^{J-1}LL_1 .
	\label{eq:cubic-four-products}
\end{align}
The first line uses $\|H_N\|_\infty\|a\|_2^2$;
the next two use $\|a\|_\infty\|H_N\|_2$ and
Cauchy--Schwarz. The last uses $\|H_N\|_\infty$.
The first line occurs only when $n\ge1$ and $k+r\ge2$.
The middle lines require $J\ge2$, and the last requires
$J\ge1$ since $k\ge1$. All powers are nonnegative on
their respective index sets. At $\rho=0$ use continuity.
Since $L_1\le\sqrt{N+1}L$ on $\cF^{\le N}$, each
elementary term is bounded by $C_TC_0^J\rho^{J-1}L^2$.
Its multiplicity is $2^Jn!k!r!$, while the series
coefficient is $1/[n!k!r!(k+r+1)]$.
There are at most $(J+1)(J+2)/2$ index triples.
The resulting bound is summable under the small-kernel
condition and proves \eqref{eq:degree-majorant} for this
family, retaining decay in every expansion index.

We next estimate the terms $W_2,W_3,W_4$ in
\cite[Eq.~(5.118)]{BS}. Set $W_N=N^2v(N\cdot)$ and
$\eta=\eta_\psi$, with $\psi$ in a bounded normalized
$H^2$ neighborhood. Write
\[
	R(x)=\|\eta(x,\cdot)\|_2,\qquad
	Z(x)=\|a_xz\|,\qquad Z_2(x,y)=\|a_xa_yz\|,
	\qquad L=\|M^{1/2}z\| .
\]
Here $\|R\|_\infty\le C_R$, $\|R\|_2=\rho\le1$,
$\|Z\|_2\le L$ and $\|W_N\|_1=N^{-1}\|v\|_1$.
The pointwise kernel bound is
$|\eta(x,y)|\le C_RN|\psi(x)\psi(y)|$.
For $\mathcal D_B X=[X,B]$ and $a=n+k$, define
\[
	F_{n,k}(x,y)=\frac{1}{n!k!}
	\|M^{1/2}(\mathcal D_B^k b_y)(\mathcal D_B^n b_x)z\| .
\]
For $a\ge2$, \cite[Lemma 5.2(iv), Eq.~(5.122)]{BS}
and $M\le N+1$ give
\begin{align}
	F_{n,k}(x,y)&\le C^{a+1}(1+a)
	\rho^{\max(a-2,0)}\mathcal E(x,y),\notag\\
	\mathcal E(x,y)&=NR(x)R(y)L+NR(x)Z(y)+NR(y)Z(x)
	+|\eta(x,y)|L+\sqrt N Z_2(x,y).
	\label{eq:quartic-regular-left}
\end{align}
The word multiplicity $2^a n!k!$ is included in $C^a$.
The two lowest orders satisfy the same estimate.
For $a=0$, $M^{1/2}b_yb_x$ is bounded by
$C\sqrt N Z_2(x,y)$ on each input sector.
For $a=1$, use the exact identity
\[
	[b_x,B]=\left(1-\frac{\cN}{N}\right)b^*(\eta_x)
	-\frac1N\int\eta(u,v)b_u^*a_v^*a_x\d u\d v .
\]
In $b_y[b_x,B]$, normal ordering the first term gives
the scalar $\eta(x,y)$ and the field $a^*(\eta_x)a_y$,
with bounded number coefficients. Normal ordering the
second term gives a pair creation operator followed by
$a_ya_x$, and a field $a^*(\eta_y)a_x$.
Their $M^{1/2}$-weighted bounds are, respectively,
\[
	C|\eta(x,y)|L,quad CNR(x)Z(y),quad
	C\rho\sqrt N Z_2(x,y),quad CR(y)Z(x).
\]
The product $[b_y,B]b_x$ has the same bounds with $x,y$
interchanged. Since $N\ge1$ and $\rho\le1$, each term
is bounded by $C\mathcal E(x,y)$. This proves \eqref{eq:quartic-regular-left}
also for $a=1$, including the boundary sectors.

After pairing with the two left fields, the right-hand
norms in $W_2,W_3,W_4$ are bounded by
\[
	\begin{array}{c|c}
	\text{term}&\text{right-hand bound}\\ \hline
	W_2&U(x,y):=R(x)Z(y)\\
	W_3&(C\rho)^{m+r}\{U(x,y)+V(x,y)\}\\
	W_4&(C\rho)^{m+r}V(x,y)
	\end{array}
	\qquad V(x,y):=R(x)R(y)L .
\]
For $W_2$ this is \cite[Eq.~(5.119)]{BS}.
For $W_3,W_4$ we divided by $m!r!$ in
\cite[Eq.~(5.123) and the following bound]{BS}.
All four indices remain in these estimates.

We give the spatial integrals for both $U$ and $V$.
The row bounds and Young's inequality imply
\begin{align*}
	\int W_N(x-y)R(x)^2Z(y)^2\d x\d y&\le C_RN^{-1}L^2,\\
	\int W_N(x-y)R(x)^2R(y)^2\d x\d y&\le C_RN^{-1},\\
	\int W_N(x-y)R(x)R(y)Z(x)Z(y)\d x\d y&\le C_RN^{-1}L^2.
\end{align*}
For the mixed terms, Cauchy--Schwarz in $y$ gives
\[
	\int W_N(x-y)R(x)^2R(y)Z(y)\d x\d y
	\le \|W_N\|_1\|R\|_\infty^2\|R\|_2\|Z\|_2
	\le C_RN^{-1}L .
\]
Thus the first three terms of $\mathcal E$, multiplied
by either $U$ or $V$, integrate to at most $C_RL^2$.
For the fourth term, use the pointwise bound on $\eta$:
\begin{align*}
	\int W_N|\eta|R(x)Z(y)\d x\d y
	&\le C_RN\|W_N\|_1\|\psi R\|_\infty
	\|\psi\|_2\|Z\|_2\le C_RL,\\
	\int W_N|\eta|R(x)R(y)\d x\d y
	&\le C_RN\|W_N\|_1\|\psi R\|_2^2\le C_R .
\end{align*}
Multiplying by the factors $L$ and $L^2$ in
$\mathcal E U$ and $\mathcal E V$ gives $C_RL^2$ in both cases.
Finally, $\int W_N Z_2^2=2\|\cV_N^{1/2}z\|^2$.
Cauchy--Schwarz with the first two spatial integrals gives
\begin{equation}\label{eq:quartic-regular-integrals}
	\int W_N\mathcal E(U+V)\d x\d y
	\le C_RL\bigl(L+\|\cV_N^{1/2}z\|\bigr)
	\le C_R Q_N[z] .
\end{equation}
For $W_3,W_4$, put $J=n+k+m+r$.
The power $\max(n+k-2,0)+m+r$ is at least
$\max(J-2,0)$. The remaining denominators in
\cite[Eq.~(5.118)]{BS} are at most one, and there are
$\binom{J+3}{3}$ index tuples. For $W_2$, set $J=n+k$.
Equations \eqref{eq:quartic-regular-left} and
\eqref{eq:quartic-regular-integrals} therefore yield
\eqref{eq:degree-majorant}, with $b_0=2$, for all three
terms and their adjoints. These bounds first hold on
smooth finite-particle vectors. Polarization and form
closure extend them to $D(Q_N^{1/2})$.
The coefficients have degree $O(J+1)$ in
$\psi,\overline\psi$, and every constant above is uniform
on the same $H^2$ neighborhood.

The quartic group has one further normalization that must
be retained before integrating its terminal kernels.
For the word in \cite[Eq.~(5.126)]{BS} with both terminal
fields unsmoothed and $J=n+k\ge1$, use
\cite[Lemma 5.2(iv), Eq.~(5.23)]{BS} with $\alpha=0$.
In that estimate $D$ is the norm of $M^{-1/2}$ times
the ordered product of the two field words.
The terminal lengths are $\ell_1=\ell_2=0$; the excluded
case is precisely $n=k=0$. With
$r_\eta(x)=\|\eta(x,\cdot)\|_2$ and $\rho=\|\eta\|_{\mathrm{HS}}$,
Eq.~(5.23) gives the following bound, where we have only
replaced its numerator $n$ by $J=n+k$:
\[
	C_0^J\rho^{J-1}
	\left\{\frac{J}{N}r_\eta(y)\|a_x z\|
	+\frac{\rho}{\sqrt N}\|a_xa_y z\|\right\}.
\]
The $N^{-1/2}$ in the second term is essential.
Write $W_N=N^2v(N\cdot)$ and let $\psi$ denote a
normalized condensate with $\|\psi\|_{H^2}\le R$.
The scattering and projection bounds give
\[
	\|\sqrt{W_N}\eta\|_2\le C_R\sqrt N,
	\qquad
	\left\|\int W_N(x-y)|\eta(x,y)|r_\eta(y)\d y\right\|_{L_x^2}
	\le C_R .
\]
For the second estimate split $\eta=k+\mu$.
For $k$, Young's inequality gives the explicit bound
\[
	\left\|\int W_N(x-y)|k(x,y)|r_\eta(y)\d y\right\|_2
	\le CN\|r_\eta\|_\infty\|\psi\|_2
	\|W_N*|\psi|\|_\infty
	\le C_R .
\]
For $\mu$, put $S_N(x)=Nw_\ell(Nx)$.
The scattering bound gives $\|S_N\|_1\le C$ for the
fixed radius $\ell$. Put $r_N=S_N*|\psi|^2$ and
$c_N=\int|\psi|^2r_N$. Since $S_N$ is even, the four terms
in $q_\psi\otimes q_\psi$ give
\[
	\mu(x,y)=\psi(x)\psi(y)\{r_N(x)+r_N(y)-c_N\},
	\qquad \|r_N\|_\infty+|c_N|\le C_R .
\]
Hence $|\mu(x,y)|\le C_R|\psi(x)\psi(y)|$.
The same displayed estimate, with $CN$ replaced by $C_R$,
bounds its contribution.
Consequently the last terminal term contributes at most
\begin{equation}\label{eq:quartic-bare-pair-normalization}
	N^{-1/2}\int W_N(x-y)|\eta(x,y)|
	\|a_xa_y z\|\d x\d y
	\le C_R\|\cV_N^{1/2}z\|.
\end{equation}
The other term uses the displayed $L_x^2$ bound and
$\int\|a_x z\|^2\d x=\|\cN^{1/2}z\|^2$.
The exterior factor is $\|M^{1/2}z\|$, paired with
the $M^{-1/2}$ in $D$. Thus the two contributions together
are bounded by
\[
	C_RC_0^J\rho^{J-1}\|M^{1/2}z\|
	\left\{\frac{J}{N}\|\cN^{1/2}z\|
	+\rho\|\cV_N^{1/2}z\|\right\}
	\le C_RC_0^J\left(\frac{J}{N}\rho^{J-1}+\rho^J\right)Q_N[z].
\]
This is the required form bound with geometric decay in $J$.

For terminal lengths $(0,1)$, one has $\ell_2\le n$.
Excluding the scalar term $(n,k)=(1,0)$ therefore leaves
$J\ge2$. The improved bound
\cite[Eq.~(5.21)]{BS} supplies $N^{-1}$ on the singular
contact. Its integral is bounded by
$N^{-1}\int W_N|\eta|^2\le C_R$.
The other terminal cases use the row and HS bounds in
\cite[Eqs.~(5.18)--(5.19)]{BS}.
This verifies the two singular endpoints of the quartic
majorant. These calculations start
on the smooth core in every sector $0\le n\le N$.
They extend to $D(Q_N^{1/2})$ by the displayed form bound;
this extension is independent of the cubic cutoff $\Theta$.
All constants just used are uniform on a bounded $H^2$
neighborhood of the condensate.

We now compute the two orders extracted in
\cite[Eq.~(5.124)]{BS}. For a symmetric kernel $f$, write
\[
	\mathcal P^*(f)=\int f(x,y)a_x^*a_y^*\d x\d y,
	\qquad \mathcal P(f)=\mathcal P^*(f)^* .
\]
There is no factor $1/2$ in this notation. Put
\begin{align*}
	g_\eta&=\int W_N(x-y)|\eta(x,y)|^2\d x\d y,
	&g_k&=\int W_N(x-y)|k(x,y)|^2\d x\d y,\\
	h_\eta(y,z)&=\int W_N(x-y)\eta(x,y)\overline{\eta(x,z)}\d x,
	&\mathcal H_\eta&=\dGamma(h_\eta),\\
	\mathcal C_\eta&=\mathcal P^*(W_N\eta)\mathcal P(\eta).
\end{align*}
The oriented first-order term is
\[
	\mathcal T_N=\int W_N(x-y)\eta(x,y)[b_x^*,B]b_y^*\d x\d y .
\]
On the input $n$-particle sector, set
$\theta_n=1-n/N$ and $p_n=\theta_n(\theta_n-N^{-1})$.
The exact commutator used above gives
\begin{equation}\label{eq:quartic-extracted-sector}
	\mathcal T_N\big|_{\cF^{(n)}}
	=p_ng_\eta I+\left(p_n-\frac{2\theta_n}{N}\right)\mathcal H_\eta
	-\frac{\theta_n}{N}\mathcal C_\eta .
\end{equation}
Indeed, the adjoint commutator begins with
$b(\eta_x)(1-\cN/N)$.
After $b_y^*$ acts, the number in this middle factor is
$n+1$. The two generalized fields contribute $\theta_n$.
Normal ordering therefore gives $p_n(g_\eta I+\mathcal H_\eta)$.
The remaining commutator term has coefficient $-\theta_n/N$.
Its normal-ordered part is $\mathcal C_\eta$.
The two contractions both equal $\mathcal H_\eta$, since
$W_N$ is even and $\eta$ is symmetric.
This proves \eqref{eq:quartic-extracted-sector}, including
the vacuum and the top sector.

The projection correction displayed above gives
\[
	\|\sqrt{W_N}\mu\|_2\le C_RN^{-1/2},\qquad
	\|\sqrt{W_N}k\|_2\le C_R\sqrt N,
	\qquad |g_\eta-g_k|\le C_R .
\]
For the first inequality, integrate
$|\mu(x,y)|^2\le C_R|\psi(x)\psi(y)|^2$
against $W_N$ and use $\|W_N\|_1=C/N$.
The last inequality follows by expanding
$|k+\mu|^2-|k|^2$ and applying Cauchy--Schwarz.
Also $g_\eta\le C_RN$ and
$|p_n-1|\le3(n+1)/N$. Hence
$|p_ng_\eta-g_k|\le C_R(n+1)$.
For the one-body term, Minkowski's inequality gives
\[
	\|h_\eta\|_{\mathrm{HS}}
	\le\left\|\int W_N(x-y)|\eta(x,y)|R(x)\d x\right\|_{L_y^2}
	\le C_R .
\]
The last bound is the preceding convolution estimate
with $x,y$ interchanged. Thus
$\|\mathcal H_\eta|_{\cF^{(n)}}\|\le C_Rn$.
Its coefficient in \eqref{eq:quartic-extracted-sector}
is uniformly bounded.

For the four-field term, pair the two creation fields
with the potential energy. Since
$\|\mathcal P(\eta)z_n\|\le\rho\sqrt{n(n-1)}\|z_n\|$,
\begin{align*}
	\frac{\theta_n}{N}|\langle z_n,\mathcal C_\eta z_n\rangle|
	&\le\frac{\sqrt{2g_\eta}}{N}
	\|\cV_N^{1/2}z_n\|\|\mathcal P(\eta)z_n\|\\
	&\le C_R\frac{n}{\sqrt N}\|z_n\|\|\cV_N^{1/2}z_n\|
	\le C_R\|M^{1/2}z_n\|\|\cV_N^{1/2}z_n\| .
\end{align*}
Here $n/\sqrt N\le\sqrt n$. Summing the sectors gives
$|(\mathcal T_N-g_kI)[z,z]|\le C_R Q_N[z]$.

The zero-order pair leaves
$P_\mu=\tfrac12\int W_N\mu b_x^*b_y^*\d x\d y+\mathrm{h.c.}$
after subtracting the bare pair with kernel $k$.
Cauchy--Schwarz between sectors whose particle numbers
differ by two yields
$|P_\mu[z,z]|\le C_RN^{-1/2}\|z\|\|\cV_N^{1/2}z\|$.
The coefficients $1/2$ and $1/4$ in
\cite[Eq.~(5.124)]{BS}, together with their adjoints,
therefore leave the bounded form
\begin{equation}\label{eq:quartic-extracted-remainder}
	P_\mu+\frac14\bigl(\mathcal T_N+\mathcal T_N^*-2g_kI\bigr)
\end{equation}
after the original pair and scalar subtractions.
All these terms have number shift $0$ or $\pm2$.
Their fixed-degree coefficient maps and bounds hold on
the same $H^2$ neighborhood. Polarization and closure
give the required forms and fixed number weights.

The remaining number polynomials, bounded one-body terms
and $N^{-1/2}Mb(h)$ are bounded by $M$ after conjugation.
Their bounds at each kernel degree are given in
\eqref{eq:linear-number-product-degree}--\eqref{eq:normal-one-body-number}.
The quadratic phase $B$ has bandwidth two
and block norm $C(n+1)$. Differentiation of number moments gives
$\norm{M^r e^{\pm B}M^{-r}}\le C_r$ for $r\in\mathbb R$.
For positive $r$ this follows from the bounded shifted difference
in $[M^{2r},B]$ and Gronwall; negative $r$ follows by duality.
An $M$-bounded finite-band form therefore stays compatible with
all number weights after conjugation with $e^B$.
On $\cF^{\le N}$, both $M^2/N$ and $N^{-1/2}Mb(h)$ have this
property. This also provides a direct alternative for the
quadratic logarithmic derivative via its exact integral formula.

\textit{Summation.}
A bound $|E_\iota[z,z]|\le b_\iota
\norm{Q_N^{1/2}z}^2$ gives
$\norm{Q_N^{-1/2}E_\iota Q_N^{-1/2}}\le2b_\iota$
by the numerical-radius inequality, also for a non-Hermitian
monomial. Multiplication by $(2J+5)^{|\alpha|}$ leaves the
series \eqref{eq:degree-majorant} summable. Apply
Lemma~\ref{lem:shifts}. The principal $\cK+\cV_N$ preserves
number and is bounded by $Q_N$. Finally, compression by $P_t$
preserves number shifts and leaves the form bounds valid.
This proves \eqref{eq:GP-weight}.
\end{proof}

\begin{remark}
The spatial estimates and scattering cancellations in \cite{BS}
yield the bound for each fixed $\alpha$, with a constant depending
on $\alpha$.
\end{remark}

\subsection{Covariant differentiation and the cubic generator}
\label{sec:covariant-generator}

We use Kato's unitary transport of projections
\cite{KatoTransport}. Define $Z_t=[\dot q_t,q_t]$,
$\mathscr Z_t=\dGamma(Z_t)$ and
$\nabla_t C=\partial_tC-[\mathscr Z_t,C]$.
The unitary transport $\dot W_t=Z_tW_t$ satisfies
$W_tq_0W_t^*=q_t$ and commutes with number after Fock lifting.
In particular $\nabla_tP_t=0$. Put $X_t=\nabla_tC_t$.
For the standard cubic kernel, its time derivative has the same
$L^2$ block bound as $C_t$:
\[
	\norm{(X_t)_{mn}}\le C_T\epsilon(n+1)^{3/2}.
\]
To check this, differentiate $\phi_t,\gamma_t-I,\sigma_t$ in
the kernels; their needed $L^\infty$ and Hilbert--Schmidt bounds
are in \cite[Appendix A]{COS}. The scattering profile is fixed.
The commutator with $\dGamma(Z_t)$ acts separately on the three
field operators. Its norm is at most $3\norm{Z_t}$ times the norm
of the one-to-two map or the three-variable kernel.
The sector bound retains its original power of particle number. Number cutoffs commute with the transport.

\begin{lemma}\label{lem:log}
For $\mathcal L_{C,t}=-ie^{-C_t}\nabla_t(e^{C_t})$,
\[
	|(\mathcal L_{C,t}+iX_t)[u,v]| \le C_T\epsilon^2\norm u\norm{M^3v}.
\]
\end{lemma}
\begin{proof}
At fixed $N$, the parameter-differentiation formula for bounded
exponentials \cite{Wilcox} gives
\[
	\mathcal L_C+iX =-i\int_0^1(1-r)e^{-rC}[X,C]e^{rC}\d r.
\]
The commutator has bandwidth six and block norm
$C_T\epsilon^2(n+1)^3$, so
$\norm{[X,C]M^{-3}}\le C_T\epsilon^2$.
The number-moment argument used above, now with cubic block
growth, gives
$\norm{M^3e^{rC}M^{-3}}\le \exp(C_T\epsilon\sqrt{L_N+4})$.
This is uniformly bounded. Use unitarity on the left argument
and integrate. The common transport preserves these norms.
\end{proof}

\begin{theorem}\label{thm:full}
Let $v\in C_c^\infty(\mathbb R^3)$ be nonnegative and radial.
Let $\phi_t$ solve the modified GP equation with a fixed normalized
initial datum in $H^4(\mathbb R^3)$. Fix a sufficiently small
quadratic scattering radius $\ell\in(0,1/2)$, independent of $N$.
Use the cubic phase \eqref{spec:eq:actual-phase} at this background,
with $m=N^{-1/2}$ and the original COS cutoff $0\le\Theta_N\le1$.
It equals one for $n\le N^{1/2}/2+10$ and zero for
$n\ge N^{1/2}-10$, for sufficiently large $N$.
Use $\mathfrak G_t=P_t(G_{N,t}-c_{N,t})P_t$, with the scalar
$c_{N,t}$ of \cite[Eq.~(4.25)]{BS}, and define
\[
	\mathfrak J_t=e^{-C_t}\mathfrak G_t e^{C_t}+\mathcal L_{C,t}, \qquad \mathfrak J_t^{[3]}=\sum_{j=0}^3 \frac{\mathcal D_{C_t}^j\mathfrak G_t}{j!}-iX_t.
\]
Then, for $|t|\le T$,
\begin{equation}\label{eq:full}
	|(\mathfrak J_t-\mathfrak J_t^{[3]})[u,v]| \le C_TN^{-3/2} \norm{Q_{N,P_t}^{1/2}u} \norm{M^6Q_{N,P_t}^{1/2}v}.
\end{equation}
The ambient-coordinate equation retains the common connection
$i\mathscr Z_t$.
\end{theorem}
\begin{proof}
Proposition~\ref{prop:GP-weight} supplies
\eqref{eq:number-compatible} for $s=4$.
Use the phase input \eqref{eq:phase-energy} and
Theorem~\ref{thm:remainder}, with $\epsilon=N^{-3/4}$.
Lemma~\ref{lem:log} bounds the time term since $Q_N\ge M$.
The covariant product rule gives the displayed exact generator;
the quadratic time derivative is already in $\mathfrak G_t$.
\end{proof}

\subsection{Construction at fixed particle number}
\label{sec:fixed-N-realization}

We first construct the Taylor evolution at fixed particle number;
the constants in this construction may depend on $N$. The uniform
estimates will be established in Section~\ref{sec:taylor-energy}.
On a fixed excitation space, let $K_P$ be the operator associated
with the restricted kinetic form and set $V=D((K_P+1)^{1/2})$.

\begin{lemma}\label{lem:domain}
Let $v\in C_c^\infty(\mathbb R^3)$ be nonnegative and radial.
Let $\phi_t$ solve the modified GP equation with a fixed normalized
initial datum in $H^4(\mathbb R^3)$. Fix a sufficiently small
quadratic scattering radius $\ell\in(0,1/2)$, independent of $N$.
Use the cubic phase \eqref{spec:eq:actual-phase} at this background,
with $m=N^{-1/2}$ and the original COS cutoff $0\le\Theta_N\le1$.
It equals one for $n\le N^{1/2}/2+10$ and zero for
$n\ge N^{1/2}-10$, for sufficiently large $N$.
At every fixed $N$ and time $t$:
\begin{enumerate}
 \item $\mathfrak G=K_P+R_0$, with $R_0$ bounded on the physical
 excitation space;
 \item $C$ preserves $D(K_P)$ and $V$, and
 $[K_P,C]:V\to\cF^{\le N}_\perp$ is bounded.
\end{enumerate}
\end{lemma}
\begin{proof}
The Neumann profile and its constant extension are defined in
\cite[Eq.~(4.1), Lemma 4.1]{BS}; the same construction at the cubic
radius is given in \cite[Lemma A.1]{COS}. The extension has continuous
value and first normal derivative. Its weak second derivatives are
piecewise bounded at fixed $N$; hence its compactly supported
defect is in $H^2$. Multiplication by the condensate and
projection preserve the resulting two-variable $H^2$ kernels.
The nonidentity hyperbolic series defining $\gamma,\sigma$
\cite[Eq.~(2.11)]{COS} converge in that space at fixed $N$: derivatives on the two
outer variables fall on the first and last kernels, and
factorials dominate the remaining operator norms.

Before quadratic conjugation the excitation generator is
$\cK$ plus a bounded operator at fixed $N$. This follows directly
from its terms of degree at most four: $v_N$ is bounded, pair
kernels are Hilbert--Schmidt, and the one-to-two maps have
finite operator norm. For the quadratic phase $B$,
$[\cK,B]$ is bounded because its pair kernel is
$(-\Delta_x-\Delta_y)\eta\in L^2$. It follows, by integrating
$e^{-rB}[\cK,B]e^{rB}$, that quadratic conjugation changes
$\cK$ only by a bounded operator. The quadratic time term is
also bounded on $\cF^{\le N}$. Compression changes the kinetic
operator by bounded off-diagonal pieces: $[\cK,P]$ is bounded
at fixed $N$ since $\phi\in H^2$. This proves the first assertion.

For the identity part of the cubic field, a typical sector map
is a finite sum of
\[
	z(X_n)\longmapsto k_N(x_i-x_j)\phi(x_j) z(X_{n+1}\setminus x_j).
\]
Subtract its action on the input kinetic operator from the
output kinetic operator. Second derivatives of $z$ cancel.
The remaining terms have either at most two derivatives on
$k_N\phi$, or one derivative there and one on $z$.
Fubini bounds the former by $C_N\norm z$ and the latter by
$C_N\norm{(\cK+1)^{1/2}z}$. The adjoint follows by integration
by parts. In the dressed terms all three variables are integrated
against an $H^2$ kernel; the same calculation is bounded, with
derivatives on annihilated variables transferred to that kernel.
For a product kernel $k_N(x-y)\phi(y)h(x,z)$ a cross derivative
is controlled by
$C_\phi\norm{\nabla k_N}_2\norm{\nabla h}_2$;
the other terms use $\norm{k_N}_{H^2}\norm h_{H^2}$.
All these norms are finite at fixed $N$.

The same product rules show $C:H^2\to H^2$ and $C:H^1\to H^1$
on the finitely many sectors. The generalized number factors and
the number cutoffs commute with kinetic energy; projection contributes
only bounded commutators. Thus the claims restrict to $K_P$.
\end{proof}

\begin{theorem}\label{thm:selfadjoint}
Let $v\in C_c^\infty(\mathbb R^3)$ be nonnegative and radial.
Let $\phi_t$ solve the modified GP equation with a fixed normalized
initial datum in $H^4(\mathbb R^3)$. Fix a sufficiently small
quadratic scattering radius $\ell\in(0,1/2)$, independent of $N$.
Use the cubic phase \eqref{spec:eq:actual-phase} at this background,
with $m=N^{-1/2}$ and the original COS cutoff $0\le\Theta_N\le1$.
It equals one for $n\le N^{1/2}/2+10$ and zero for
$n\ge N^{1/2}-10$, for sufficiently large $N$.
At fixed $N,t$, $\mathfrak J_t^{[3]}$ has a self-adjoint,
bounded-below realization on $D(K_{P_t})$.
For the time-dependent assertion, use parallel transport to a fixed
Hilbert space $\mathcal H_N=P_s\cF^{\le N}$ and put
$V=D((K_{P_s}+1)^{1/2})$. Suppose that $R_0(t)$ and
$X_t=\nabla_tC_t$ are $C^1$ in $\mathcal B(\mathcal H_N)$,
$C_t$ is $C^1$ in both $\mathcal B(\mathcal H_N)$ and $\mathcal B(V)$,
and $[K_{P_t},C_t]$ is $C^1$ in $\mathcal B(V,\mathcal H_N)$.
Require also that the transported $K_{P_t}-K_{P_s}$ is $C^1$
in $\mathcal B(\mathcal H_N)$. Then this family
has a unique unitary propagator preserving the common operator domain.
A sufficient background assumption is
$\phi\in C^1([-T,T];H^4)\cap C^2([-T,T];L^2)$.
This holds on intervals of regular existence for smooth
modified GP solutions, for example with $H^6$ initial data.
\end{theorem}
\begin{proof}
Write $\mathfrak G=K_P+R_0$. Every positive-order commutator
of $K_P$ with $C$ is a sum of $C^j[K_P,C]C^k$.
It is bounded $V\to\cF^{\le N}_\perp$ by Lemma~\ref{lem:domain}.
Commutators of $R_0$ and the term $X$ are bounded on the
Hilbert space. Therefore
\[
	\mathfrak J^{[3]}=K_P+R_N,\qquad \norm{R_Nz}\le C_{N,T}\norm{(K_P+1)^{1/2}z}.
\]
The remainder is symmetric on $D(K_P)$, by the commutator
identities and anti-self-adjointness of $C,X$.
The spectral inequality
$\sqrt{\lambda+1}\le\delta\lambda+c_\delta$
gives infinitesimal $K_P$ boundedness. The
Kato--Rellich theorem \cite{ReedSimonII} proves self-adjointness
on $D(K_P)$;
the analogous form inequality proves a lower bound.

The transport preserves $H^2$ and changes its kinetic operator
by a bounded term, since its one-body commutator with $-\Delta$
is bounded. On the fixed space the domain is thus constant.
Under the stated background regularity, differentiating the
kernel products above gives the required $C^1$ bounds. The
second time derivative is used only in differentiating the
quadratic logarithmic derivative, where its kernel is needed
in $L^2$. For fixed $N$ the relative kernel and its first two
spatial derivatives have finite norms.
The assumed $C^1$ dependence and the preceding graph-norm
equivalence verify conditions C1--C4 of
\cite[Section 1, Theorem 4]{Kato}, applied to
$-i\mathfrak J^{[3]}_t$. This supplies the propagator.
Application also in the reverse time direction and uniqueness
give unitarity. The constants in this construction may depend
on $N$.
\end{proof}

\section{Energy estimates for the Taylor evolution}
\label{sec:taylor-energy}

This section develops the uniform estimates for the Taylor evolution.
The operator and coercivity bounds lead to the weighted inverse
estimate and the first-energy propagation theorem. The comparison
on initial data in the energy form domain is Theorem~\ref{thm:energy-class-comparison}.
The conditional estimates distinguish the trajectory norms required
by direct Duhamel integration from the initial norms in
Theorem~\ref{thm:C}. The stronger initial-data result uses the
inverse-energy argument of Section~\ref{sec:inverse-comparison}.

\subsection{Operator bounds for the remainder}
\label{sec:operator-remainder}

The smoothness of $v$ also gives an operator estimate for the
remainder, complementing the preceding form bound. We write $K_P$
for the restricted kinetic operator, $A_P=K_P+M$, and
$Q=Q_{N,P_t}$. For each fixed number-weight exponent, the constants
below are uniform in $N$ and $|t|\le T$.

\begin{lemma}\label{lem:op-inputs}
Let $v\in C_c^\infty(\mathbb R^3)$ be nonnegative and radial.
Let $\phi_t$ solve the modified GP equation with a fixed normalized
initial datum in $H^4(\mathbb R^3)$. Fix a sufficiently small
quadratic scattering radius $\ell\in(0,1/2)$, independent of $N$.
Use the cubic phase \eqref{spec:eq:actual-phase} at this background,
with $m=N^{-1/2}$ and the original COS cutoff $0\le\Theta_N\le1$.
It equals one for $n\le N^{1/2}/2+10$ and zero for
$n\ge N^{1/2}-10$, for sufficiently large $N$.
Write $\mathfrak G=K_P+R_0$, retaining the scalar subtraction
$c_{N,t}$ of \cite[Eq.~(4.25)]{BS}. For every fixed $r\ge0$,
\begin{align}
	\norm{M^rR_0z}&\le C_{r,T}N^2\norm{M^{r+2}z},
	\label{eq:R0-op}\\
	\norm{M^r[K_P,C]z}
	&\le C_{r,T}\left(
	N\norm{M^{r+3/2}z}
	+\norm{M^{r+3/2}A_P^{1/2}z}\right).
	\label{eq:KC-op}
\end{align}
The commutator in \eqref{eq:KC-op} extends from $D(K_P)$
to the displayed weighted first-energy domain.
\end{lemma}
\begin{proof}
Write $K=\cK$ for the kinetic operator on the full Fock space.
We first keep track of the constants suppressed in
Lemma~\ref{lem:domain}. Before quadratic conjugation, the
nonkinetic excitation generator has number bandwidth at most four.
Its quartic term on the $n$ sector has norm at most
$C N^2(n+1)^2$, since $v$ is bounded. Pair kernels have
Hilbert--Schmidt norm at most $C_TN^{3/2}$, and cubic
one-to-two kernels have norm at most $C_TN$.
The remaining scalar, linear and one-body terms are bounded by
the same coarse $N^2M^2$ estimate. These statements follow
directly by scaling $N^3v(N(x-y))\phi(x)\phi(y)$ and
$N^{5/2}v(N(x-y))\phi(y)$ in $L^2$; convolution coefficients
use $\norm{N^3v(N\cdot)}_1=\norm v_1$.
Number factors and projections preserve these upper bounds.
Finite bandwidth gives the same bound with $M^r$ on the left.

For the quadratic kernel, the scattering equation and its
constant extension give
$\norm{\Delta_j\eta_t}_2\le C_TN^{3/2}$.
Indeed, the potential part of $\Delta(Nw_\ell(N\cdot))$
has this norm. The Neumann-eigenvalue part is smaller, and
derivatives on the condensate or the projections are controlled
by its $H^4$ norm. There is no boundary measure, because the
first normal derivative matches at the extension boundary.
Consequently
$\norm{M^r[K,B]z}\le C_{r,T}N^{3/2}\norm{M^{r+1}z}$.
The number-weight bounds for $e^{sB}$, uniformly for
$|s|\le1$, transfer these estimates through quadratic conjugation.
The quadratic logarithmic derivative is bounded by $C_{r,T}M$
with the corresponding weights. Integrate the commutator
$e^{-sB}[K,B]e^{sB}$, and then compress to the excitation space.
This proves \eqref{eq:R0-op}.

For \eqref{eq:KC-op}, let
$k_N(x)=\sqrt N\,w^{\rm Neu}_{N^{1/2}}(Nx)$.
Starting from the scattering equation and profile estimates in
\cite[Lemma A.1]{COS}, scaling and the matched extension give
\[
	\norm{k_N}_2\le C\epsilon,\qquad \norm{\nabla k_N}_2\le C,\qquad \norm{\nabla^2 k_N}_2\le CN,\qquad \epsilon=N^{-3/4}.
\]
The second-derivative bound follows also from the Laplacian
bound by the Fourier transform on $\mathbb R^3$.
For $h=p_t$ or $\sigma_t$, the hyperbolic series and the
quadratic-kernel estimate above imply
\[
	\norm h_2\le C_T,\quad \norm{\nabla h}_2\le C_TN^{1/2},\quad \norm{\nabla^2h}_2\le C_TN^{3/2}.
\]
For a pure second derivative of a kernel power, only its outer
factor is differentiated. A mixed derivative uses first
derivatives on both outer factors; their product is
$O(N)$, which is smaller. The factorial coefficients
make both series summable.

For the identity-field term, the product rule in Lemma~\ref{lem:domain}
gives a factor $CN$ multiplying the input
$L^2$ norm and a factor $C$ multiplying
its first kinetic norm.
For a dressed kernel $k_N(x-y)\phi(y)h(x,z)$,
its first-derivative norm is bounded uniformly, while
its second-derivative norm is at most
\[
	C_T\big(N+\norm{\nabla k_N}_2\norm{\nabla h}_2 +\epsilon\norm{\nabla^2h}_2\big) \le C_TN.
\]
The same estimates hold for projected kernels. Differentiating a
projection in any of the three variables gives derivatives of
$\phi$, whose norms are uniformly bounded. Writing the
operator in terms of projected fields before taking the commutator avoids
derivatives of the spectator projections. On the excitation space
$[K_P,C]=P[K,\widehat C]P$, where $\widehat C$ is this
projected-field expression and commutes with $P$.

The creation, annihilation and contraction factors are at most
$C(n+1)^{3/2}$. Applying the product rule also to the adjoints,
with an integration by parts on an annihilated variable, gives
\[
	\norm{([K_P,C])_{mn}z_n} \le C_T(n+1)^{3/2} \big(N\norm{z_n}+\norm{A_{P,n}^{1/2}z_n}\big).
\]
Its bandwidth is three. Multiplying by the shifted weights
and summing the blocks proves \eqref{eq:KC-op}.
\end{proof}

\begin{lemma}\label{lem:A-phase}
Let $v\in C_c^\infty(\mathbb R^3)$ be nonnegative and radial.
Let $\phi_t$ solve the modified GP equation with a fixed normalized
initial datum in $H^4(\mathbb R^3)$. Fix a sufficiently small
quadratic scattering radius $\ell\in(0,1/2)$, independent of $N$.
Use the cubic phase \eqref{spec:eq:actual-phase} at this background,
with $m=N^{-1/2}$ and the original COS cutoff $0\le\Theta_N\le1$.
It equals one for $n\le N^{1/2}/2+10$ and zero for
$n\ge N^{1/2}-10$, for sufficiently large $N$.
Let $\epsilon=N^{-3/4}$ and let $\Lambda_N$ contain both endpoints
of every nonzero block of $C$. Thus
$\epsilon(\Lambda_N+1)^{3/2}\le D$, with $D$ independent of $N$.
For every fixed real $r$ and $|a|\le1$,
\begin{align}
	\norm{M^r e^{aC}z}&\le C_{r,T}\norm{M^rz},\notag\\
	\norm{M^rA_P^{1/2}e^{aC}z}
	&\le C_{r,T}\left(\norm{M^rA_P^{1/2}z}
	+\norm{M^{r+3/2}z}\right).
	\label{eq:A-phase}
\end{align}
Suppose $X=\nabla_tC$ has number shifts at most three,
is supported in $n\le\Lambda_N$, and satisfies
\[
	\|X_{mn}z_n\|\le C_T\epsilon(n+1)^{3/2}\|z_n\|,\qquad
	\|A_{P,m}^{1/2}X_{mn}z_n\|
	\le C_T(n+1)^{3/2}
	\bigl(\epsilon\|A_{P,n}^{1/2}z_n\|+\|z_n\|\bigr).
\]
Then both estimates hold with $e^{aC}$ replaced by
$\nabla_t(e^{aC})$.
In the model application these bounds for $X$ follow, for
example, from a background bounded in $H^6$ on compact intervals.
\end{lemma}
\begin{proof}
Finite bandwidth, \eqref{eq:blocks}, and the stronger cutoff condition give
\[
	\norm{M^rCz}\le d_r\norm{M^rz},\qquad \norm{M^rA_P^{1/2}Cz} \le a_r\norm{M^rA_P^{1/2}z}+b_r\norm{M^{r+3/2}z},
\]
with constants independent of $N$. The notation $\Lambda_N$
denotes a cutoff containing every active input and output
sector of $C$, with $\Lambda_N=O(\sqrt N)$.
For $u_j=\norm{M^rA_P^{1/2}C^jz}$ and
$v=\norm{M^{r+3/2}z}$, iteration yields
\[
	u_j\le a_r^j u_0+ b_rj\max(a_r,d_{r+3/2})^{j-1}v .
\]
The exponential series proves \eqref{eq:A-phase}; its
$L^2$ version follows by summing $d_r^j$.
For the derivative use the exact integral
\[
	\nabla_t(e^{aC})=\int_0^a e^{(a-b)C}Xe^{bC}\d b.
\]
The analogous two bounds for $X$, followed by the phase
bounds, give the same energy form norm and number loss.
For $a<0$ reverse the integration direction.
The required blocks of $X$ follow by differentiating the
kernels once in time, including the projections on all three variables.
Under the stated regularity, $\dot\phi$ is bounded in $H^4$,
and the differentiated quadratic kernels satisfy the same
$L^2$ and first-spatial-derivative orders as the original ones.
This also follows directly from their scattering profile and
hyperbolic series; the relative profile itself is time-independent.
\end{proof}

\begin{theorem}\label{thm:operator}
For every fixed integer $s\ge2$, the conjugation remainder
$R_{s,N}$ of Theorem~\ref{thm:remainder}, with
$G=\mathfrak G_t$, has an $L^2$-valued extension satisfying
\begin{equation}\label{eq:operator-general}
	\norm{R_{s,N}z} \le C_{s,T}\left( N^{2-3s/4}\norm{M^{3s/2+2}z} +N^{-3(s-1)/4}\norm{M^{3s/2}A_{P_t}^{1/2}z}\right).
\end{equation}
Its domain is the intersection of the two displayed weighted
domains.
In particular, for the same exact and Taylor generators as in
Theorem~\ref{thm:full}, their form difference has an $L^2$-valued
extension satisfying
\begin{equation}\label{eq:operator}
	\norm{(\mathfrak J_t-\mathfrak J_t^{[3]})z} \le C_T\left( N^{-1}\norm{M^8z} +N^{-9/4}\norm{M^6A_{P_t}^{1/2}z}\right).
\end{equation}
The right-hand domain is
$D(M^8)\cap D(M^6A_{P_t}^{1/2})$.
Here the difference denotes the $L^2$-valued extension
of the form difference to this domain.
\end{theorem}
\begin{proof}
Finite-band iteration of the original blocks gives, for $j\ge1$,
\begin{align}
	\norm{M^r C^jz}
	&\le C_{j,r}\epsilon^j\norm{M^{r+3j/2}z},\notag\\
	\norm{M^r A_P^{1/2}C^jz}
	&\le C_{j,r}\left(
	\epsilon^j\norm{M^{r+3j/2}A_P^{1/2}z}
	+\epsilon^{j-1}\norm{M^{r+3j/2}z}\right).
	\label{eq:op-words}
\end{align}
For $j=0$ use the identity and omit the second term.
Using \eqref{eq:R0-op} between two words of total length $s$ gives
\[
	\norm{\mathcal D_C^sR_0z} \le C_{s,T}N^2\epsilon^s\norm{M^{3s/2+2}z}.
\]
For the kinetic term first write
$\mathcal D_C^sK_P=\mathcal D_C^{s-1}([K_P,C])$.
Every summand is $C^j[K_P,C]C^k$, with $j+k=s-1$.
Equations \eqref{eq:KC-op}--\eqref{eq:op-words} imply
\[
	\norm{\mathcal D_C^sK_Pz} \le C_{s,T}\left( (N\epsilon^{s-1}+\epsilon^{s-2})\norm{M^{3s/2}z} +\epsilon^{s-1}\norm{M^{3s/2}A_P^{1/2}z}\right).
\]
The input enters through its first kinetic norm; the
scattering kernels enter through derivatives of order at most two.

Use the integral Taylor remainder in operator form, initially
on $D(K_P)$. Its outer exponential is unitary. For the inner
exponential use Lemma~\ref{lem:A-phase}, which gives
\[
	\norm{M^{3s/2}A_P^{1/2}e^{rC}z} \le C_{s,T}\left(\norm{M^{3s/2}A_P^{1/2}z} +\norm{M^{3s/2+3/2}z}\right).
\]
The extra number term, multiplied by $\epsilon^{s-1}$,
is bounded by $N^2\epsilon^s\norm{M^{3s/2+2}z}$.
Since $N\epsilon^{s-1}+\epsilon^{s-2}\le
2N^2\epsilon^s$ for $\epsilon=N^{-3/4}$ and $N\ge1$,
this proves \eqref{eq:operator-general}.
Now specialize to $s=4$.
The logarithmic-derivative difference in Lemma~\ref{lem:log}
has operator norm bound $C_T\epsilon^2\norm{M^3z}$.
Finally
$N^2\epsilon^4=N^{-1}$,
$N\epsilon^3=N^{-5/4}$ and $\epsilon^3=N^{-9/4}$.
Absorb the lower number weights into $M^8$.
Density and the displayed bounds identify the extension with
the original form difference.
\end{proof}

\begin{remark}
The first-order time approximation in $\mathfrak J^{[3]}$
is sufficient for \eqref{eq:operator}. For a consistent
higher-order full generator, one may instead use
\[
	\mathfrak J^{(s)} =\sum_{j=0}^{s-1}\frac{\mathcal D_C^j\mathfrak G}{j!} -i\sum_{j=0}^{s-2}\frac{\mathcal D_C^jX}{(j+1)!}.
\]
The exact integral for $\mathcal L_C$ and the $L^2$ block
bounds of $C,X$ give a logarithmic-derivative remainder at most
$C_{s,T}\epsilon^s\norm{M^{3s/2}z}$.
It is absorbed by the right side of
\eqref{eq:operator-general}. Thus the same fixed-order estimate
holds for $\mathfrak J-\mathfrak J^{(s)}$.
The constants in these weighted operator estimates depend on $s$.
\end{remark}

\begin{corollary}
\label{cor:quadratic-test}
Let $v\in C_c^\infty(\mathbb R^3)$ be nonnegative and radial.
Let $\phi_t$ solve the modified GP equation with an $N$-independent
normalized initial datum in $H^6(\mathbb R^3)$. Fix a sufficiently
small quadratic scattering radius $\ell\in(0,1/2)$, independent of $N$.
Use the cubic phase \eqref{spec:eq:actual-phase} at this background,
with $m=N^{-1/2}$ and the original COS cutoff $0\le\Theta_N\le1$.
It equals one for $n\le N^{1/2}/2+10$ and zero for
$n\ge N^{1/2}-10$, for sufficiently large $N$.
Let $U_{2,N}(t;0)$ be the quadratic
reference evolution, with the same moving excitation spaces.
For initial data in the indicated domains,
\[
	\sup_{|t|\le T} \norm{(\mathfrak J_t-\mathfrak J_t^{[3]})U_{2,N}(t;0)\xi} \le C_TN^{-1} \norm{(\cK^2+M^{24})^{1/2}\xi}.
\]
The estimate evaluates the forcing term along the quadratic
reference evolution.
\end{corollary}
\begin{proof}
Since $A_P=K_P+M$ and $K_P$ commutes with $M$,
$M^{12}A_P=M^{12}K_P+M^{13} \le C(K_P^2+M^{24})$
as forms on $D(K_P)\cap D(M^{12})$.
Moreover $\norm{K_Pz}\le\norm{\cK z}$ for
$z\in P D(\cK)$, by compression. Proposition 4.3 of
\cite{COS} propagates every fixed number moment and the
second kinetic moment along $U_{2,N}$.
Apply these bounds, and $M^{16}\le M^{24}$,
in Theorem~\ref{thm:operator}.
\end{proof}

\subsection{Coercivity with number penalties}
\label{sec:penalty-coercivity}

We next quantify the lower bound of Theorem~\ref{thm:selfadjoint}
by introducing a number penalty. The penalty is an eighth moment
at the level of quadratic forms, whereas \eqref{eq:operator}
contains a sixteenth moment in an operator estimate. These bounds
give additional control of the remainder. First-energy propagation
follows from a linear number correction in
Section~\ref{sec:taylor-first-energy}.

\begin{lemma}\label{lem:inf-form}
For every $\delta>0$ there is $C_{\delta,T}$, independent of $N$,
such that
\begin{equation}\label{eq:inf-form}
	|(\mathfrak J_t-\mathfrak J_t^{[3]})[z,z]| \le \delta\,A_{P_t}[z] +C_{\delta,T}N^{-1}\norm{M^4z}^2 .
\end{equation}
The form domain is
$D(A_{P_t}^{1/2})\cap D(M^4)$.
\end{lemma}
\begin{proof}
Write the difference as $R_K+R_0^{\rm rem}+R_{\rm time}$,
where the first two terms are the degree-three conjugation
remainders of $K_P$ and $R_0$, respectively.
The argument of Lemma~\ref{lem:op-inputs} applies with every
fixed real weight $r$, including negative weights.
Indeed, all finite-band estimates compare two shifted positive
number weights; the quadratic and cubic exponentials act
boundedly on both positive and negative number-weight spaces.
In particular the proof of Theorem~\ref{thm:operator}, with
weights kept on the outer exponential as well, gives
\begin{align}
	\norm{M^{-4}R_0^{\rm rem}z}
	&\le C_TN^{-1}\norm{M^4z}, \label{eq:R0-balanced}\\
	\norm{M^{-3}R_Kz}
	&\le C_T\left(N^{-5/4}\norm{M^3z}
	+N^{-9/4}\norm{M^3A_P^{1/2}z}\right).
	\label{eq:RK-balanced}
\end{align}
In \eqref{eq:RK-balanced}, the term $\epsilon^2=N^{-3/2}$
has been absorbed into $N^{-5/4}$. The number loss from
Lemma~\ref{lem:A-phase} is absorbed as well, since this kinetic
remainder is supported on $\mathcal N\le\Lambda_N$ and
$N^{-9/4}(\Lambda_N+1)^{3/2}\le CN^{-3/2}$.
The same weighted argument
in Lemma~\ref{lem:log} yields
\[
	\norm{M^{-3/2}R_{\rm time}z} \le C_TN^{-3/2}\norm{M^{3/2}z}.
\]
The first and third contributions therefore have diagonal
bounds $C_TN^{-1}\norm{M^4z}^2$.

The particle-number cutoff is essential to the kinetic estimate.
Choose $\Lambda_N\le C\sqrt N$ containing both endpoints
of every active block of $C$, and put
$E_N=\mathbf1_{\{\mathcal N\le\Lambda_N\}}$.
Then $C=E_NCE_N$. Since $K_P$ commutes with $E_N$,
also $R_K=E_NR_KE_N$. Applying
\eqref{eq:RK-balanced} to $E_Nz$ gives
\begin{align*}
	|R_K[z,z]|
	&\le C_TN^{-5/4}\norm{M^3E_Nz}^2\\
	&\quad+C_TN^{-9/4}\norm{M^3E_Nz}
	\norm{M^3A_P^{1/2}E_Nz}\\
	&\le C_TN^{-5/4}\norm{M^3z}^2
	+C_TN^{-3/4}\norm{M^3z}\norm{A_P^{1/2}z}\\
	&\le \delta A_P[z]+C_{\delta,T}N^{-5/4}\norm{M^3z}^2.
\end{align*}
Here the cutoff gives $(\Lambda_N+1)^3\le CN^{3/2}$.
The error from Young's inequality is bounded by
$C_{\delta,T}N^{-3/2}\norm{M^3z}^2$.
Finally, use
$N^{-5/4}M^6\le N^{-1}M^8$ for $N\ge1$.
Sum the contributions and extend from the common core.
\end{proof}

\begin{theorem}
\label{thm:coercivity}
On each compact time interval there are constants
$b_T,c_T,C_T>0$, independent of $N$, such that the closed form
\[
	\mathcal E_{N,t} =\mathfrak J_t^{[3]}+b_TM+b_TN^{-1}M^8
\]
satisfies
\begin{equation}\label{eq:coercivity}
	c_T(Q_{N,P_t}+N^{-1}M^8) \le\mathcal E_{N,t} \le C_T(Q_{N,P_t}+N^{-1}M^8).
\end{equation}
The comparison holds at each fixed time.
\end{theorem}
\begin{proof}
The first inequality in \cite[Eq.~(4.26)]{BS} gives
$\mathfrak G_t\ge \frac12 Q_{N,P_t}-C_TM$; its upper bound
gives $|\mathfrak G_t[z,z]|\le C_TQ_{N,P_t}[z]$.
The phase-energy comparison holds for both $e^{C_t}$ and
$e^{-C_t}$: the proof of \cite[Lemma 3.1]{COS} applies with
the auxiliary phase parameter of either sign.
Consequently
\[
	e^{-C_t}Q_{N,P_t}e^{C_t}\ge c_TQ_{N,P_t}, \qquad e^{-C_t}Me^{C_t}\le C_TM.
\]
The operator $X_t$ has uniformly bounded Hilbert-space norm,
because $\epsilon(\Lambda_N+1)^{3/2}\le C$ for the standard
cutoff. The exact integral defining $\mathcal L_{C,t}$ then
implies $\norm{\mathcal L_{C,t}}\le C_T$.
We obtain
\[
	c_TQ_{N,P_t}-C_TM\le\mathfrak J_t,\qquad |\mathfrak J_t[z,z]|\le C_TQ_{N,P_t}[z].
\]
Use Lemma~\ref{lem:inf-form} with $\delta$ less than half the
positive lower coefficient and use $A_P\le Q_{N,P_t}$.
It follows that
\[
	\mathfrak J_t^{[3]} \ge c'_TQ_{N,P_t}-C'_TM-C''_TN^{-1}M^8, \qquad \mathfrak J_t^{[3]}\le C'''_T(Q_{N,P_t}+N^{-1}M^8).
\]
Choose $b_T$ larger than both negative constants and use
$M\le Q_{N,P_t}$. At fixed $N$ the added number operators
are bounded, so this is the closed form of the self-adjoint
operator in Theorem~\ref{thm:selfadjoint} plus bounded terms.
Its form domain agrees with that of the right side.
\end{proof}

\begin{proposition}
\label{prop:partial-derivative}
Let $v\in C_c^\infty(\mathbb R^3)$ be nonnegative and radial.
Let $\phi_t$ solve the modified GP equation with a fixed normalized
initial datum in $H^4(\mathbb R^3)$. Fix a sufficiently small
quadratic scattering radius $\ell\in(0,1/2)$, independent of $N$.
Use the cubic phase \eqref{spec:eq:actual-phase} at this background,
with $m=N^{-1/2}$ and the original COS cutoff $0\le\Theta_N\le1$.
It equals one for $n\le N^{1/2}/2+10$ and zero for
$n\ge N^{1/2}-10$, for sufficiently large $N$.
Assume that $\phi\in C^1([-T,T];H^4)\cap C^2([-T,T];H^2)$,
with both norms bounded independently of $N$.
Use the pieces $R_K,R_{\rm time}$ defined in
Lemma~\ref{lem:inf-form}. For every fixed real $r$,
\begin{align}
	\norm{M^r\nabla_t R_Kz}
	&\le C_{r,T}\left(
	N^{-5/4}\norm{M^{r+6}E_Nz}
	+N^{-9/4}\norm{M^{r+6}A_{P_t}^{1/2}E_Nz}\right),
	\label{eq:dt-RK}\\
	\norm{M^r\nabla_t R_{\rm time}z}
	&\le C_{r,T}N^{-3/2}\norm{M^{r+3}z}.
	\label{eq:dt-time}
\end{align}
In particular, for every $\delta>0$,
\[
	|(\nabla_tR_K+\nabla_tR_{\rm time})[z,z]| \le \delta A_{P_t}[z]+C_{\delta,T}N^{-1}\norm{M^4z}^2.
\]
\end{proposition}
\begin{proof}
The first time derivative $X=\nabla_tC$ has the same
$L^2$ and first-energy block bounds as $C$.
Repeating the second-spatial-derivative calculation in
Lemma~\ref{lem:op-inputs}, with one time derivative on a
coefficient, also gives
\[
	\norm{([K_P,X])_{mn}z_n} \le C_T(n+1)^{3/2} \left(N\norm{z_n}+\norm{A_{P,n}^{1/2}z_n}\right).
\]
Here $\dot\phi\in H^4$ controls the differentiated product
kernels; their relative profile is unchanged. The second
covariant derivative $X_1=\nabla_tX$ only needs an $L^2$
block estimate, which is
$\norm{(X_1)_{mn}}\le C_T\epsilon(n+1)^{3/2}$.
It uses $\ddot\phi\in H^2\subset L^\infty$ and the twice
time-differentiated Hilbert--Schmidt kernels. Derivatives of the connection act on the same three field operators.

After transporting to a fixed excitation space, let
$S_t=\nabla_tK_{P_t}$. Since $\nabla_tP_t=0$,
$S_t=P_t[\cK,\mathscr Z_t]P_t$.
Its one-body kernel is the compression of $[-\Delta,Z_t]$.
The assumption $\phi,\dot\phi\in H^2$ gives a uniform bound.
Thus $S_t$ is the second quantization of a bounded one-body
operator. Its commutator with a cubic monomial acts on each of the three field
operators. Thus $[S_t,C]$ has block bound
$C_T\epsilon(n+1)^{3/2}$. In particular
$\nabla_t[K_P,C]=[S_t,C]+[K_P,X]$
has the same coarse block estimate as $[K_P,C]$.

Differentiate the integral representation
\[
	R_K=\frac1{3!}\int_0^1(1-a)^3 e^{-aC}\mathcal D_C^3([K_P,C])e^{aC}\d a.
\]
In its middle factor, differentiation replaces one $C$ by
$X$, or replaces $[K_P,C]$ by the preceding differentiated
commutator. The same finite-product estimate gives a coefficient $N\epsilon^3+\epsilon^2$
for the norm of $M^6z$, and $\epsilon^3$ for the
norm of $M^6A_P^{1/2}z$. Differentiating either exponential uses
Lemma~\ref{lem:A-phase}. Its extra $M^{3/2}$ loss is absorbed
by $\epsilon^{s-2}$ at $s=4$, since every kinetic term and
its derivative is supported on $E_N$, and
$\epsilon(\Lambda_N+1)^{3/2}\le C$.
Keep $M^r$ throughout to obtain \eqref{eq:dt-RK}.

For the time part differentiate the integral in
Lemma~\ref{lem:log}. Its middle commutator has derivative
$\nabla_t[X,C]=[X_1,C]+[X,X]=[X_1,C]$.
This still contains two $L^2$ blocks and is bounded by
$C_T\epsilon^2M^3$ with every fixed number weight.
The derivatives of the two exponentials are bounded on those
number-weight spaces by Lemma~\ref{lem:A-phase}.
This proves \eqref{eq:dt-time}.
Finally use the balanced weights and the support argument
from Lemma~\ref{lem:inf-form}.
\end{proof}

We next control the remaining part of the differentiated remainder.
We differentiate the quadratic conjugation explicitly.

\begin{lemma}\label{lem:R0-dot}
Let $v\in C_c^\infty(\mathbb R^3)$ be nonnegative and radial.
Let $\phi_t$ solve the modified GP equation with a fixed normalized
initial datum in $H^4(\mathbb R^3)$. Fix a sufficiently small
quadratic scattering radius $\ell\in(0,1/2)$, independent of $N$.
Assume that $\phi\in C^1([-T,T];H^4)\cap C^2([-T,T];H^2)$,
with both norms bounded independently of $N$.
For every fixed real $r$,
\begin{equation}\label{eq:R0-dot}
	\norm{M^r\nabla_tR_0z}\le C_{r,T}N^2\norm{M^{r+2}z}.
\end{equation}
Here $R_0=\mathfrak G-K_P$ includes the scalar subtraction
and the quadratic logarithmic derivative.
\end{lemma}
\begin{proof}
We supply the quadratic-conjugation step explicitly. Let
$B=B(\eta)$ be the generalized quadratic phase, with $\eta$
uniformly small in Hilbert--Schmidt norm as in \cite{BS}.
For a bounded symmetric two-body operator $W$, write
$\mathbb W=\sum_{i<j}W_{ij}$ on each symmetric sector.
If $W,\eta$ are $C^1$ in their operator and Hilbert--Schmidt
norms, respectively, then
\begin{equation}\label{eq:quartic-differentiation}
	\norm{M^r\partial_t(e^{-B}\mathbb W e^B)z} \le C_r\big(\norm{\dot W}+\norm W\norm{\dot\eta}_2\big) \norm{M^{r+2}z}.
\end{equation}
The constant is uniform on a fixed sufficiently small ball for
$\eta$. The same assertion holds with covariant derivatives
after one-body transport.

To prove summability in \eqref{eq:quartic-differentiation}, expand
$\mathbb W+\int_0^1e^{-sB}[\mathbb W,B]e^{sB}\d s$.
The commutator contains a pair term and terms with two generalized
fields and one ordinary density. In the multiplication case
$W(x-y)$ its creation part is
\[
	\frac12\int W(u-v)\eta(u,v)b_u^*b_v^*\d u\d v +\frac12\int\big(W(x-u)+W(x-v)\big)\eta(u,v) b_u^*b_v^*a_x^*a_x\d x\d u\d v.
\]
The annihilation part is its adjoint. This identity follows by
commuting the ordinary quartic through a pair; it commutes with
the number factors in that pair. For general bounded $W$ the
same contractions are compositions of $W$ with the pair vector
$\eta$. Their few-particle map norms are at most
$C\norm W\norm\eta_2$.

Conjugate an ordinary density by first using its commutator
with $B$, which consists of two generalized fields. For example,
\[
	[a_x^*a_y,B]=\int\eta(y,z)b_x^*b_z^*\d z +\int\overline{\eta(x,z)}b_zb_y\d z.
\]
Thus only products of at most four conjugated generalized
fields, or two such fields and an ordinary density, remain.
Expand the generalized fields with
\cite[Lemmas 3.2--3.3]{BS}. Retain the $N^{-p}$ in each
$\Pi^{(2)}$ factor and the $N^{-k}$ in each terminal
$\Pi^{(1)}$ factor. Each additional field pair is accompanied
by $1/N$; on the physical sectors $M/N\le2$. The four terminal fields require a weight $M^2$, rather
than $M^3$.
Non-normal-ordered traces contain at least two
Hilbert--Schmidt factors; the trace-ideal Cauchy--Schwarz inequality
\cite{SimonTrace} bounds their traces.
For an arbitrary bounded $W$, each insertion contributes
at most $\norm W$ on the corresponding two-particle space.

More precisely, after the factorial series coefficients and
all multiplicities have been summed at total kernel degree
$J\ge1$, these rules bound the weighted operator norm
$\norm{M^rT_JM^{-r-2}}$ by $ C_0^J(1+J)^{b_r}\norm W\norm\eta_2^J $.
There are at most four series indices. Their number of
partitions, the four terminal-field shifts and the overall
number shift $|d|\le2J+4$ contribute only to the fixed
polynomial $(1+J)^{b_r}$. The remaining normalized
$\Pi$ factors have the bounds used in the proof of
\cite[Lemma 3.3]{BS}. This also establishes the claim for
every fixed negative $r$.
Differentiate a term: either $W$ is differentiated, or one
of its $J$ copies of $\eta$ is replaced by $\dot\eta$.
Its majorant is therefore
\[
	C_0^J(1+J)^{b_r} \left(\norm{\dot W}\norm\eta_2^J+ J\norm W\norm{\dot\eta}_2\norm\eta_2^{J-1}\right).
\]
Both series converge uniformly when $C_0\norm\eta_2<1$.
The degree-zero term is $\dot{\mathbb W}$.
This proves differentiation in the displayed weighted
operator norm, and \eqref{eq:quartic-differentiation}.
One can first use finite-rank pair kernels and finite-particle
core vectors and then pass to their Hilbert--Schmidt limits;
the majorants are unchanged.

Apply this to the physical potential,
$\norm W\le N^2\norm v_\infty$. Its covariant derivative is
the commutator with $Z_t\otimes1+1\otimes Z_t$, so its
two-body norm is at most $4\norm{Z_t}\norm W$.
The covariant derivative of $\eta$ is uniformly
Hilbert--Schmidt. Compression introduces no derivative term because $\nabla_tP_t=0$. Thus the differentiated conjugated quartic
has the bound \eqref{eq:R0-dot}.

For the remaining terms, the estimates allow one additional number factor.
Before quadratic conjugation, the nonquartic, nonkinetic
terms and their covariant derivatives are bounded, with
every fixed weight, by
$C_TN^{3/2}M^{3/2}$. This follows from the pair bound
$C_TN^{3/2}M$, the cubic bound $C_TNM^{3/2}$,
the number polynomials and the convolution bounds in
Lemma~\ref{lem:op-inputs}; the same kernel scaling applies
after replacing one condensate by its time derivative.
Moreover,
\[
	\norm{M^r\nabla_t(e^{sB})z} \le C_{r,T}\norm{M^{r+1}z},\qquad |s|\le1,
\]
by the exact derivative integral and the number bounds for
the quadratic phase. The derivatives of these nonquartic conjugates are consequently bounded by
$C_TN^{3/2}M^{5/2}\le C_TN^2M^2$ on $\cF^{\le N}$.

The change in kinetic energy is
$\int_0^1e^{-sB}[K_P,B]e^{sB}\d s$.
Both $[K_P,B]$ and its covariant derivative have weighted
bound $C_TN^{3/2}M$: the derivative of the pair kernel uses
at most two spatial derivatives and $\dot\phi\in H^4$.
The contribution of $\nabla_tK_P$ is a bounded one-body
commutator and obeys the same bound. The terms obtained by differentiating the two exponentials are therefore
bounded by $C_TN^{3/2}M^2$.
Finally the quadratic logarithmic derivative has an integral
with one $\nabla_tB$. Its derivative uses
$\nabla_t^2B$, of order $M$, and products of two
order-$M$ factors; hence it is bounded by $C_TM^2$.
Here $\ddot\phi\in H^2$ suffices for the required
Hilbert--Schmidt kernels. The same gradient and bounded-potential estimates bound the derivative of
the scalar in \cite[Eq.~(4.25)]{BS} by $C_TN^2$.
These bounds prove \eqref{eq:R0-dot}.
\end{proof}

\begin{corollary}\label{cor:full-dot}
Let $v\in C_c^\infty(\mathbb R^3)$ be nonnegative and radial.
Let $\phi_t$ solve the modified GP equation with a fixed normalized
initial datum in $H^4(\mathbb R^3)$. Fix a sufficiently small
quadratic scattering radius $\ell\in(0,1/2)$, independent of $N$.
Use the cubic phase \eqref{spec:eq:actual-phase} at this background,
with $m=N^{-1/2}$ and the original COS cutoff $0\le\Theta_N\le1$.
It equals one for $n\le N^{1/2}/2+10$ and zero for
$n\ge N^{1/2}-10$, for sufficiently large $N$.
Assume that $\phi\in C^1([-T,T];H^4)\cap C^2([-T,T];H^2)$,
with both norms bounded independently of $N$.
Let $R=\mathfrak J-\mathfrak J^{[3]}$ and $r\in\mathbb R$.
Then
\begin{align}
	\norm{M^r\nabla_tRz}
	&\le C_{r,T}\left(
	N^{-1}\norm{M^{r+8}z}
	+N^{-9/4}\norm{M^{r+6}A_P^{1/2}z}\right),
	\label{eq:full-dot}\\
	|(\nabla_tR)[z,z]|
	&\le\delta A_P[z]+C_{\delta,T}N^{-1}
	\norm{M^4z}^2,\qquad\delta>0 .
	\label{eq:full-dot-form}
\end{align}
\end{corollary}
\begin{proof}
Differentiate the integral Taylor remainder of $R_0$.
Its middle factor is a sum of products containing four factors of
$C$, distributed on either side of
$R_0$. The derivative replaces one $C$ by $X$,
or $R_0$ by $\nabla_tR_0$.
Equations \eqref{eq:R0-op} and \eqref{eq:R0-dot} give
$C_{r,T}N^2\epsilon^4M^{r+8}$ for each term.
The derivatives of the exterior phases are bounded on
every fixed number-weighted space by Lemma~\ref{lem:A-phase}.
Consequently
\[
	\norm{M^r\nabla_tR_0^{\rm rem}z} \le C_{r,T}N^{-1}\norm{M^{r+8}z}.
\]
Combine this with Proposition~\ref{prop:partial-derivative}
for \eqref{eq:full-dot}. Taking $r=-4$ in the last display and pairing with
$M^4z$ gives the bound $C_TN^{-1}\norm{M^4z}^2$.
The diagonal estimate for the two other parts is already
in that proposition. This proves \eqref{eq:full-dot-form}.
\end{proof}

For a common-domain solution $y$ of the Taylor equation,
the covariant energy identity reads
\begin{equation}\label{eq:energy-balance}
	\frac{\d}{\d t}\mathcal E_{N,t}[y,y] = \left(\nabla_t\mathfrak J_t^{[3]} +ib_T[\mathfrak J_t^{[3]},M] +ib_TN^{-1}[\mathfrak J_t^{[3]},M^8]\right)[y,y].
\end{equation}
At fixed $N$, differentiate on the common kinetic domain or
regularize there. The number operators are bounded and preserve
that domain. The next lemma estimates the moving-generator
derivative. Theorem~\ref{thm:Taylor-first-energy} proves the
first-energy bound using a comparison energy with a linear
number correction.

\begin{lemma}\label{lem:moving-G}
On the actual excitation space, uniformly on compact time intervals,
\[
	|(\nabla_t\mathfrak G_t)[z,z]| +|i[\mathfrak G_t,M][z,z]| \le C_T Q_{N,P_t}[z].
\]
The time derivative here includes the moving projections.
\end{lemma}
\begin{proof}
The number commutator is the compression of the second
estimate in \cite[Eq.~(4.26)]{BS}, because $P_t$ commutes
with $M$. For the derivative use $\nabla_tP_t=0$ and
$P_t\mathscr Z_tP_t=0$. On vectors in $P_t\cF^{\le N}$,
the product rule for $P_t(G_{N,t}-c_{N,t})P_t$ bounds its
covariant derivative by the sum of
\[
	\big|(\partial_t(G_{N,t}-c_{N,t}))[z,z]\big|, \qquad 2\big|\operatorname{Re} \ip{z}{[a^*(q_t\dot\phi_t)a(\phi_t),G_{N,t}]z}\big|.
\]
These are precisely the derivative and projection terms in
the proof of \cite[Proposition 6.1, following Eq.~(6.3)]{BS}.
The last two estimates of its Eq.~(4.26) bound them by
$C_TQ_{N,P_t}[z]$. Replacing $\dot\phi_t$ there by
$q_t\dot\phi_t$ changes only a multiple of
$a^*(\phi_t)a(\phi_t)$, whose compressed commutator
has zero quadratic form. This proves the assertion.
The estimate therefore includes the derivatives of the excitation
projection.
\end{proof}

\begin{proposition}
\label{prop:bounded-corrections}
Set $U=e^C$, $\mathcal M=U^*MU$,
$D=\mathcal M-M$ and $L=\mathcal L_C$. For every fixed
real $r$, both $T=D$ and $T=L$ satisfy
\begin{align}
	\norm{M^rTz}&\le C_{r,T}\norm{M^rz},\notag\\
	\norm{M^rA_P^{1/2}Tz}
	&\le C_{r,T}\big(\norm{M^rA_P^{1/2}E_Nz}
	+\norm{M^{r+3/2}E_Nz}\big).
	\label{eq:bounded-correction-graph}
\end{align}
They are self-adjoint and satisfy $T=E_NTE_N$.
For $R=\mathfrak J-\mathfrak J^{[3]}$ and every $\delta>0$,
\begin{equation}\label{eq:remainder-correction-comm}
	|i[R,M][z,z]|+|i[R,D][z,z]|+|i[R,L][z,z]| \le\delta A_P[z]+C_{\delta,T}N^{-1}\norm{M^4z}^2 .
\end{equation}
\end{proposition}
\begin{proof}
The middle factors in the exact formulas
\[
	D=\int_0^1e^{-sC}[M,C]e^{sC}\d s,\qquad L=-i\int_0^1e^{-sC}Xe^{sC}\d s
\]
have the same $L^2$ and first-energy blocks as $C$.
For $[M,C]$ this follows by multiplying a block by its
number shift, of absolute value at most three.
The strong cutoff condition and Lemma~\ref{lem:A-phase}
give \eqref{eq:bounded-correction-graph}, including every
fixed negative number weight. The support statements follow
because $C,X,[M,C]$ vanish on both sides outside $E_N$.
Self-adjointness follows directly from the definitions.

First treat the commutator with $M$. Introduce the real
number-gauge parameter $\theta$ and conjugate all operators
by $e^{i\theta M}$. Its derivatives of $C$ and $X$ have
the same blocks as the originals. For the nonkinetic
input, the proof of Lemma~\ref{lem:R0-dot} applies with
$\eta$ replaced by $e^{2i\theta}\eta$ and with the bare
nonquartic fields multiplied by their number-gauge phases.
The quartic potential and $K_P$ are unchanged. Thus
$\norm{M^r[M,R_0]z}\le C_{r,T}N^2\norm{M^{r+2}z}$.
The same differentiated-word argument as in
Proposition~\ref{prop:partial-derivative} and
Corollary~\ref{cor:full-dot} now applies.
For the time remainder, the differentiated middle factor is
$[i[M,X],C]+[X,i[M,C]]$. Each term contains two $L^2$
blocks, giving a bound of order $\epsilon^2M^3$.
For the kinetic remainder its two-sided support on $E_N$
is unchanged. In particular,
\[
	|i[R,M][z,z]| \le\delta A_P[z]+C_{\delta,T}N^{-1}\norm{M^4z}^2 .
\]

For $T=D,L$, keep the decomposition
$R=R_K+R_0^{\rm rem}+R_{\rm time}$.
The balanced estimate for $R_0^{\rm rem}$ and the first
line of \eqref{eq:bounded-correction-graph} give
$|R_0^{\rm rem}[z,Tz]| \le C_TN^{-1}\norm{M^4z}^2$.
For the kinetic part, the balanced bound in the proof of
Lemma~\ref{lem:inf-form} and the second line of
\eqref{eq:bounded-correction-graph} give, with $z_E=E_Nz$,
\begin{align*}
	|R_K[z,Tz]|\le C_T\big(&N^{-5/4}\norm{M^3z_E}^2\\
	&+N^{-9/4}\norm{M^3z_E}
	(\norm{M^3A_P^{1/2}z_E}+\norm{M^{9/2}z_E})\big).
\end{align*}
Use $M^3E_N\le C N^{3/2}E_N$ on the energy factor.
Young's inequality bounds that product by
$\delta A_P[z]+C_{\delta,T}N^{-3/2}\norm{M^3z}^2$.
The extra graph-loss term has the same latter bound since
$M^{3/2}E_N\le C N^{3/4}E_N$.
All the number terms are bounded by
$C_{\delta,T}N^{-1}\norm{M^4z}^2$.
The time remainder is bounded by $C_TN^{-3/2}\norm{M^{3/2}z}^2$.
Hermiticity bounds each diagonal commutator by twice
the corresponding mixed pairing. Choosing the intermediate
Young constants smaller proves \eqref{eq:remainder-correction-comm}.
\end{proof}

\begin{theorem}\label{thm:adapted-energy}
There are constants $b,d>0$, depending on the compact time
interval but not on $N$, such that
\[
	\widetilde{\mathcal E}_{N,t} =\mathfrak J_t^{[3]}-\mathcal L_{C,t} +b\,e^{-C_t}Me^{C_t}+dN^{-1}M^8
\]
is a closed positive form with
\begin{equation}\label{eq:adapted-comparison}
	c_T(Q_{N,P_t}+N^{-1}M^8) \le\widetilde{\mathcal E}_{N,t} \le C_T(Q_{N,P_t}+N^{-1}M^8).
\end{equation}
Along a common-domain solution of the same Taylor evolution,
\begin{equation}\label{eq:adapted-balance}
	\left|\frac{\d}{\d t}\widetilde{\mathcal E}_{N,t}[y,y] -dN^{-1}i[\mathfrak J_t^{[3]},M^8][y,y]\right| \le C_T\widetilde{\mathcal E}_{N,t}[y,y].
\end{equation}
The energy derivative is therefore bounded up to the
displayed penalty commutator.
\end{theorem}
\begin{proof}
Write $H=U^*\mathfrak G U$, so
$\mathfrak J=H+L$ and $\mathfrak J^{[3]}=H+L-R$.
The lower bound for $\mathfrak G$, the phase-energy comparison
and the number comparison give
$H\ge c_TQ-C_T\mathcal M$.
Choose $b$ large, absorb the arbitrarily small $A_P$ term
from Lemma~\ref{lem:inf-form}, and then choose $d$ to absorb
its $N^{-1}M^8$ term. This proves both bounds in
\eqref{eq:adapted-comparison}. At fixed $N$, $L$ and
$\mathcal M$ are bounded and the lower bound gives the
closed positive form on the common first-energy domain.

Since $U^*\nabla_tU=iL$, the product rule gives
\[
	\nabla_tH=U^*(\nabla_t\mathfrak G)U+i[H,L], \qquad \nabla_t\mathcal M=i[\mathcal M,L].
\]
Insert these identities into
$\nabla_t\widetilde{\mathcal E}
     +i[\mathfrak J^{[3]},\widetilde{\mathcal E}]$.
The connection terms cancel exactly, leaving
\begin{align}\label{eq:adapted-source}
	&U^*\big(\nabla_t\mathfrak G+i b[\mathfrak G,M]\big)U
	-\nabla_tR+i[R,L]-i b[R,\mathcal M]\notag\\
	&\hspace{35mm}{}+dN^{-1}i[\mathfrak J^{[3]},M^8].
\end{align}
Lemma~\ref{lem:moving-G} and the phase-energy bound control
the first term by $C_TQ$. Corollary~\ref{cor:full-dot}
controls the second. For the next two use
$\mathcal M=M+D$ and
Proposition~\ref{prop:bounded-corrections}.
The sum of their absolute diagonal forms is bounded by
$C_T(Q+N^{-1}M^8)$, which is bounded by
$C_T\widetilde{\mathcal E}$.
This proves \eqref{eq:adapted-balance}.
All identities hold first on the fixed-$N$ common kinetic
domain and then as the corresponding form identities.

\end{proof}

The preceding bounds can be sharpened by retaining the bare
potential before taking operator norms. We give the details because
the time derivative requires more than a static estimate.

\begin{lemma}\label{lem:bare-split}
Let $v\in C_c^\infty(\mathbb R^3)$ be nonnegative and radial.
Let $\phi_t$ solve the modified GP equation with a fixed normalized
initial datum in $H^4(\mathbb R^3)$. Fix a sufficiently small
quadratic scattering radius $\ell\in(0,1/2)$, independent of $N$.
Use the cubic phase \eqref{spec:eq:actual-phase} at this background,
with $m=N^{-1/2}$ and the original COS cutoff $0\le\Theta_N\le1$.
It equals one for $n\le N^{1/2}/2+10$ and zero for
$n\ge N^{1/2}-10$, for sufficiently large $N$.
Assume that $\phi\in C^1([-T,T];H^4)\cap C^2([-T,T];H^2)$,
with both norms bounded independently of $N$.
Write
\[
	\mathfrak G=K_P+V_P+S,\qquad V_P=P\cV_NP
\]
on the excitation space. For every fixed real $r$,
\begin{align}
	\norm{M^rTz}&\le C_{r,T}N^{3/2}\norm{M^{r+1}z},
	&&T=S,\ \nabla_tS,\ i[M,S],\label{eq:S-refined}\\
	\norm{M^rBz}&\le C_{r,T}N\norm{M^{r+5/2}z},
	&&B=[V_P,C],\ \nabla_t[V_P,C],\
	i[M,[V_P,C]].\label{eq:VC-refined}
\end{align}
In particular the estimates hold for negative balanced weights.
\end{lemma}
\begin{proof}
Put $W_N(x)=N^2v(Nx)$ and $\kappa_N(x)=k_N(x)$, with the
notation of Lemma~\ref{lem:op-inputs}. Scaling gives
\begin{equation}\label{eq:bare-scales}
	\norm{W_N}_2\le CN^{1/2},\quad \norm{\kappa_N}_2\le C\epsilon,\quad \norm{\kappa_N}_\infty\le CN^{1/2},\qquad \epsilon=N^{-3/4}.
\end{equation}
The bounds for $\eta,p,\sigma$ in
\cite[Lemma A.3]{COS} give bounded row and column $L^2$
norms, and
\[
	|\eta(x,y)|+|\sigma(x,y)|+|p(x,y)| \le C_TN|\phi(x)\phi(y)|.
\]
One time derivative replaces the right side by a finite sum of
products of $|\phi|$ and $|\dot\phi|$, with the same factor $N$.
The row and column bounds remain uniform.
These functions belong to $L^2\cap L^\infty$.

We first describe how quadratic conjugation and its derivative act on the
operator monomials used below. They consist of
generalized pairs, one generalized field times an ordinary
density, two generalized fields times an ordinary density,
and number-conserving one-body terms. View each kernel as a map from its annihilation variables to its creation
variables, and denote its norm by $k$. If the monomial
contains $d\le4$ fields, its weighted operator bound is
$C_r k M^{d/2}$.
For these families quadratic conjugation preserves this
power, and its derivative is bounded by
\begin{equation}\label{eq:few-field-dot}
	C_{r,T}\big(\dot k+k\norm{\nabla_t\eta}_2\big)M^{d/2}.
\end{equation}
Here $\dot k$ denotes a bound for the norm of the
differentiated kernel.

To justify this assertion use the field expansion in the proof
of Lemma~\ref{lem:R0-dot}. An ordinary density is conjugated
by integrating its commutator with $B(\eta)$, which replaces
it by two generalized fields. Thus there are at most $d$
terminal fields. Every additional pair in the generalized-field expansion retains its
factor $N^{-1}$. The associated number weight is bounded
because $M/N\le2$ on the physical space.
Contractions of a few-particle map with a pair kernel obey
the same bound with an extra factor $\norm\eta_2$.
When a contraction changes an output variable into an input variable, use
a Schmidt expansion of $\eta$. Cauchy--Schwarz in the
Schmidt index bounds the resulting map by $k\norm\eta_2$.
Contraction of two variables is composition with the pair vector. A closed loop contains at least two pair factors,
whose product is trace class. These observations also apply
to a density kernel retaining an input coordinate.

After the factorial coefficients in
\cite[Lemmas 3.2--3.3]{BS} are included, the total
kernel-degree-$J$ majorant, with weights $M^r$ and
$M^{-r-d/2}$, is
$C_0^J(1+J)^{b_{r,d}} k\norm\eta_2^J$.
There are at most four series indices; their partitions and
number shifts contribute only the polynomial factor.
The normalized internal factors have exactly the bounds used
in Lemma~\ref{lem:R0-dot}. Differentiation replaces the last
factor by
$\dot k\norm\eta_2^J+ J k\norm{\nabla_t\eta}_2\norm\eta_2^{J-1}$.
The two series converge on the same sufficiently small
Hilbert--Schmidt ball. This proves \eqref{eq:few-field-dot}
in the weighted operator norm, first for finite-rank
approximations and then by their uniform bounds.
The argument also applies to the number-gauge derivative.
Normalized number polynomials in the excitation generator
are treated by writing them as products of ordinary
densities and retaining their $N^{-1}$ factors.
The normalized factors compensate the number weights introduced
by differentiation.

Consider now $S$. Before quadratic conjugation the bare
pair terms have size $C_TN^{3/2}M$; the cubic terms have
size $C_TNM^{3/2}$. These are the $L^2$ kernel scalings
in Lemma~\ref{lem:op-inputs}. The scalar, linear,
one-body and normalized number-polynomial terms are bounded
by their sum, both before and after a time or number-gauge
derivative. Convolution coefficients use
$\norm{N^3v(N\cdot)}_1=\norm v_1$.
The kinetic conjugation correction is the integral of
$e^{-aB}[K_P,B]e^{aB}$; its pair kernel and its derivative
have norm at most $C_TN^{3/2}$. Formula
\eqref{eq:few-field-dot} applies to it as well.

For the quartic conjugation correction use
\[
	e^{-B}V_Pe^B-V_P=\int_0^1e^{-aB}[V_P,B]e^{aB}\d a.
\]
The unprojected commutator has the pair kernel
$W_N(u-v)\eta(u,v)$ and the density-pair kernels
$(W_N(x-u)+W_N(x-v))\eta(u,v)$ displayed in
Lemma~\ref{lem:R0-dot}. Their map norms satisfy, respectively,
\begin{align*}
	\norm{W_N(u-v)\eta(u,v)}_{L^2_{u,v}}&\le C_TN^{3/2},\\
	\sup_x\norm{W_N(x-u)\eta(u,v)}_{L^2_{u,v}}
	&\le C_T\norm{W_N}_2\le C_TN^{1/2}.
\end{align*}
The first line uses the pointwise kernel bound and
$\phi\in L^2\cap L^\infty$; the second uses the uniform
row bound. Thus
$[V_P,B]$ has size
$C_T(N^{3/2}M+N^{1/2}M^2)$.
The same estimates hold when $\eta$ is differentiated.

The projection and transport terms satisfy the following bounds. For
$f,g\in L^2\cap L^\infty$, multiplication on two coordinates
obeys
\begin{equation}\label{eq:rankone-potential}
	\norm{W_N(x-y)(|f\rangle\langle g|\otimes1)} \le \norm{W_N}_2\norm f_\infty\norm g_2 .
\end{equation}
This follows by first applying Cauchy--Schwarz in the input
$x$ variable and then integrating the output $x$ variable.
Its adjoint has the same bound. The commutators of the potential with the finitely many active projections
therefore have two-particle operator norm at most $C_TN^{1/2}$.
Moreover
$\nabla_tV_P=P[\cV_N,\mathscr Z]P$ has a two-body kernel
of norm at most $C_TN^{1/2}$, since $Z$ is a sum of rank-one
operators with vectors $\phi,q\dot\phi$.
Consequently $[\nabla_tV_P,B]$ has size at most
$C_TN^{1/2}M^2$ by the two-body contraction argument above.
Together with \eqref{eq:few-field-dot}, this controls the
derivative of the quartic conjugation correction.

The quadratic logarithmic derivative is an integral of a
conjugated generalized pair. Its derivative uses
$\nabla_t^2\eta$, uniformly Hilbert--Schmidt, and
\eqref{eq:few-field-dot}; both have size $C_TM$.
The scalar subtraction in \cite[Eq.~(4.25)]{BS} and
its first derivative are $O_T(N)$: its convolution terms
have this scaling, and its gradient-square and
potential-square terms have the same bound by the scattering
kernel estimates. Combining the preceding bounds and
$M\le N+1$ gives \eqref{eq:S-refined}, since
$NM^{3/2}+N^{1/2}M^2\le C N^{3/2}M$.
For $i[M,S]$, use $\eta\mapsto e^{2i\theta}\eta$
and the finitely many particle-number shifts of the bare terms; all
the preceding estimates are unchanged.

It remains to verify \eqref{eq:VC-refined}. In the
identity-field part of $C$, the one-to-two map is
$f(x)\longmapsto \kappa_N(x-y)\phi(y)f(x)$.
In its commutator with the bare potential all interactions
between retained old coordinates cancel. The new internal
interaction and each new spectator interaction have map norms
bounded by
\[
	\sup_x\norm{W_N(x-y)\kappa_N(x-y)\phi(y)}_{L^2_y}, \quad \sup_{x,a}\norm{W_N(a-y)\kappa_N(x-y)\phi(y)}_{L^2_y} \le C_TN .
\]
Here \eqref{eq:bare-scales} is sufficient.
The internal term has three fields and the spectator term has five. Their
number weights are therefore $M^{3/2}$ and
$M^{5/2}$.

For the dressed terms put
$F(x,y,z)=\kappa_N(x-y)\phi(y)h(x,z)$, where
$h=p$ or $\sigma$. An insertion $W_N(x-y)$ has
$L^2$ norm at most $C_TN\norm h_2$.
An insertion $W_N(x-z)$ or $W_N(y-z)$ has norm at most
$C_TN^{3/2}\epsilon\le C_TN$, using
$|h(x,z)|\le C_TN|\phi(x)\phi(z)|$.
For a fixed spectator coordinate $a$, an insertion
$W_N(a-x)$, $W_N(a-y)$ or $W_N(a-z)$ has norm at most
$C_TN^{1/2}\epsilon$, using the row and column bounds.
This list includes every internal and spectator interaction of the
three-variable kernel, as well as the negative input interaction of its
two-to-one adjoint.
Their Wick maps again have at most five fields.
Equation \eqref{eq:rankone-potential} controls the additional commutators
with projections on the active coordinates. These are composed with the
original cubic map, whose norm is $O_T(\epsilon)$.

The potential commutes with both the number cutoff and
the scalar number coefficients in products of generalized
$b$ fields. Thus no difference of cutoff values and no
extra field pair occurs in this first commutator.
Finite bandwidth proves the estimate for every fixed
weight $r$ and for adjoint terms.
Replacing one coefficient by its time derivative gives the
same bound for $[V_P,X]$, by the kernel bounds stated above.
For the other differentiated term, use
$\nabla_t[V_P,C]=[V_P,X]+[\nabla_tV_P,C]$.
The second summand has the coarse bound
$C_TN^{1/2}\epsilon M^{7/2}$ and two-sided support in
$E_N$. On that support $M\le C\sqrt N$, so it is
bounded by $C_TN M^{5/2}$.
Finally $V_P$ preserves number; differentiating its
commutator in the number gauge only multiplies the cubic
blocks by their particle-number shifts of absolute value at most three.
This completes the proof.
\end{proof}

\begin{proposition}
\label{prop:refined-remainder}
Let $v\in C_c^\infty(\mathbb R^3)$ be nonnegative and radial.
Let $\phi_t$ solve the modified GP equation with a fixed normalized
initial datum in $H^4(\mathbb R^3)$. Fix a sufficiently small
quadratic scattering radius $\ell\in(0,1/2)$, independent of $N$.
Use the cubic phase \eqref{spec:eq:actual-phase} at this background,
with $m=N^{-1/2}$ and the original COS cutoff $0\le\Theta_N\le1$.
It equals one for $n\le N^{1/2}/2+10$ and zero for
$n\ge N^{1/2}-10$, for sufficiently large $N$.
Assume that $\phi\in C^1([-T,T];H^4)\cap C^2([-T,T];H^2)$,
with both norms bounded independently of $N$.
For every fixed integer $s\ge2$ and real $r$,
\begin{equation}\label{eq:refined-general}
	\norm{M^rR_{s,N}z} \le C_{s,r,T}\left( N^{7/4-3s/4}\norm{M^{r+3s/2+1}z} +N^{-3(s-1)/4}\norm{M^{r+3s/2}A_P^{1/2}z}\right).
\end{equation}
For the original difference
$R=\mathfrak J-\mathfrak J^{[3]}$, each of
$T=R,\nabla_tR,i[M,R]$ satisfies
\begin{align}
	\norm{M^rTz}
	&\le C_{r,T}\left(N^{-5/4}\norm{M^{r+7}z}
	+N^{-9/4}\norm{M^{r+6}A_P^{1/2}z}\right),
	\label{eq:refined-operator}\\
	|T[z,z]|&\le\delta A_P[z]+
	C_{\delta,T}N^{-5/4}\norm{M^{7/2}z}^2 .
	\label{eq:refined-form}
\end{align}
With $D=e^{-C}Me^C-M$ and $L=-ie^{-C}\nabla_t(e^C)$,
the sum of $|i[R,D][z,z]|$ and $|i[R,L][z,z]|$
has the same bound as \eqref{eq:refined-form}.
\end{proposition}
\begin{proof}
Use the decomposition in Lemma~\ref{lem:bare-split}.
The finite-word bound for the $S$ remainder is
$C_{s,r,T}N^{3/2}\epsilon^sM^{r+3s/2+1}$.
For the potential first commute once, and then write
every further commutator as a sum of
$C^j[V_P,C]C^k$, $j+k=s-1$.
Equation \eqref{eq:VC-refined} gives
$C_{s,r,T}N\epsilon^{s-1}M^{r+3s/2+1}$.
The first coefficient is smaller than the second by
$N^{-1/4}$. The kinetic part is unchanged from
Theorem~\ref{thm:operator}. Its coefficients
$N\epsilon^{s-1}+\epsilon^{s-2}$ are bounded by
$2N\epsilon^{s-1}$.
In the integral remainder the additional graph loss
from the inner phase is
$\epsilon^{s-1}M^{r+3s/2+3/2}$; it is absorbed by
$N\epsilon^{s-1}M^{r+3s/2+1}$, since $M\le N+1$.
The weighted phase bounds give
\eqref{eq:refined-general}.

At $s=4$ the potential coefficient is $N\epsilon^3=N^{-5/4}$.
The time remainder is $O_T(\epsilon^2M^3)$ and is smaller.
Differentiation replaces a cubic block by $X$, or the
first commutator or $S$ by its derivative from
Lemma~\ref{lem:bare-split}. Derivatives of the exterior
phases use Lemma~\ref{lem:A-phase}. The kinetic and
time contributions have already been controlled in
Proposition~\ref{prop:partial-derivative}.
This proves \eqref{eq:refined-operator}, also for the
number-gauge derivative.

For the nonkinetic remainder use the balanced weight
$r=-7/2$ and pair with $M^{7/2}z$.
For the kinetic remainder retain its separate bound and
two-sided support on $E_N$ from Lemma~\ref{lem:inf-form}.
The Young inequality there gives
$\delta A_P+C_{\delta,T}N^{-3/2}M^6$, in addition to
$C_TN^{-5/4}M^6$. Both are bounded by the right side of
\eqref{eq:refined-form}; the time term is smaller as well.
The same decomposition applies to the two derivatives.
For commutators with $D,L$, use their boundedness on
$M^{7/2}$ for the nonkinetic pairing, and retain the
kinetic mixed pairing of
Proposition~\ref{prop:bounded-corrections}.
The additional term from the graph estimate is bounded by
$C_TN^{-3/2}M^6$.
Hermiticity and smaller intermediate Young constants
prove the last assertion.
\end{proof}

\begin{corollary}
\label{cor:refined-energy}
There are fixed $b,d>0$ for which
\[
	\widehat{\mathcal E}_{N,t} =\mathfrak J_t^{[3]}-\mathcal L_{C,t} +b e^{-C_t}Me^{C_t}+dN^{-5/4}M^7
\]
is a closed positive form uniformly comparable to
$Q_{N,P_t}+N^{-5/4}M^7$. Along the same Taylor evolution,
\begin{equation}\label{eq:refined-balance}
	\left|\frac{\d}{\d t}\widehat{\mathcal E}_{N,t}[y,y] -dN^{-5/4}i[\mathfrak J_t^{[3]},M^7][y,y]\right| \le C_T\widehat{\mathcal E}_{N,t}[y,y].
\end{equation}
Furthermore the forcing estimate of
Corollary~\ref{cor:quadratic-test} improves to
\[
	\sup_{|t|\le T}\norm{R_tU_{2,N}(t;0)\xi} \le C_TN^{-5/4}\norm{(\cK^2+M^{24})^{1/2}\xi}.
\]
This last assertion concerns the quadratic reference flow.
\end{corollary}
\begin{proof}
Repeat the positivity and exact connection cancellation
in Theorem~\ref{thm:adapted-energy}, now using
Proposition~\ref{prop:refined-remainder}.
Every term other than the displayed penalty commutator
is bounded by $C_T(Q+N^{-5/4}M^7)$.
For the forcing assertion, $M^{14}\le M^{24}$ and
$M^{12}A_P\le C(K_P^2+M^{24})$; the proof of
Corollary~\ref{cor:quadratic-test} therefore applies with
\eqref{eq:refined-operator}.
\end{proof}

\begin{lemma}
\label{lem:hardy-VC}
Let $v\in C_c^\infty(\mathbb R^3)$ be nonnegative and radial.
Let $\phi_t$ solve the modified GP equation with a fixed normalized
initial datum in $H^4(\mathbb R^3)$. Fix a sufficiently small
quadratic scattering radius $\ell\in(0,1/2)$, independent of $N$.
Use the cubic phase \eqref{spec:eq:actual-phase} at this background,
with $m=N^{-1/2}$ and the original COS cutoff $0\le\Theta_N\le1$.
It equals one for $n\le N^{1/2}/2+10$ and zero for
$n\ge N^{1/2}-10$, for sufficiently large $N$.
For every fixed real $r$ and
$z\in D(M^{r+3/2})\cap D(M^{r+2}A_P^{1/2})$,
\begin{equation}\label{eq:hardy-VC}
	\norm{M^r[V_P,C]z} \le C_{r,T}\left( N\norm{M^{r+3/2}z} +\norm{M^{r+2}A_P^{1/2}z}\right).
\end{equation}
The estimate also holds with $[V_P,C]$ replaced by
$i[M,[V_P,C]]$. If $\phi\in C^1([-T,T];H^4)\cap C^2([-T,T];H^2)$
with norms bounded independently of $N$, it also holds for
$\nabla_t[V_P,C]$.
\end{lemma}
\begin{proof}
The pointwise bound in \cite[Lemma A.5]{COS} gives
\[
	|\kappa_N(r)|\le\frac{C}{\sqrt N(|r|+N^{-1})}.
\]
If $\operatorname{supp}v\subset B_R$, then
\begin{equation}\label{eq:hardy-convolution}
	F_N(r)^2:=\int |W_N(s)|^2|\kappa_N(r-s)|^2\d s \le C|r|^{-2}\quad(r\ne0).
\end{equation}
Indeed, for $|r|\ge2R/N$ use $|r-s|\ge|r|/2$ and
$\norm{W_N}_2^2=N\norm v_2^2$. For $|r|<2R/N$ use
$\norm{\kappa_N}_\infty^2\le CN$, obtaining
$F_N(r)^2\le CN^2\le C_R|r|^{-2}$.

First omit the projections on the active coordinates and write the
identity family as $T=\int\kappa_N(x-y)\phi(y)a_x^*a_y^*a_x\d x\d y$.
Direct normal ordering gives
\[
	[\cV_N,T]=I+\mathbb B ,
\]
where the internal term $I$ has kernel
$W_N(x-y)\kappa_N(x-y)\phi(y)$ and three fields, whereas
\[
	\mathbb B=\int W_N(y-a)\kappa_N(x-y)\phi(y) a_x^*a_y^*a_a^*a_a a_x\d x\d y\d a .
\]
The internal map has norm at most $C_TN$.
The two-to-three map underlying $\mathbb B$ is
\[
	(Bf)(x,y,a)=W_N(y-a)\kappa_N(x-y)\phi(y)f(x,a).
\]
Equation \eqref{eq:hardy-convolution} and Hardy's inequality
\cite{LiebLoss}
in the $x$ coordinate, with $a$ fixed, imply
\[
	\norm{Bf}_2^2\le C_T\int\frac{|f(x,a)|^2}{|x-a|^2} \d x\d a \le C_T\norm{\nabla_x f}_2^2 .
\]
For the adjoint, estimate the derivative of the input vector.
Cauchy--Schwarz in $y$ gives
\[
	\norm{B^*g}_2^2 \le C_T\int\frac{|g(x,y,a)|^2}{|x-a|^2}\d x\d y\d a \le C_T\norm{\nabla_x g}_2^2 .
\]
The Wick lifting factor for a two-to-three map on the
$n$-particle sector and the derivative of a symmetric input obey
\[
	\frac{\sqrt{n!(n+1)!}}{(n-2)!}\le C(n+1)^{5/2}, \qquad \norm{\nabla_{x_1}z_n}=n^{-1/2}\norm{K_n^{1/2}z_n}.
\]
Output symmetrization is a contraction. Applying these facts
also to the three-to-two adjoint proves
\[
	\norm{\mathbb Bz}+\norm{\mathbb B^*z} \le C_T\norm{M^2K^{1/2}z}.
\]
The internal term and its adjoint are bounded by $C_TN\norm{M^{3/2}z}$.

For a dressed three-variable kernel
$\kappa_N(x-y)\phi(y)h(x,z)$, with $h=p,\sigma$, we use
the internal/spectator distinction from the proof of
Lemma~\ref{lem:bare-split}. Internal insertions produce
three-field terms with kernel norm at most $C_TN$, whereas
spectator insertions produce five-field terms with norm at
most $C_TN^{1/2}\epsilon$. To see the latter estimate, fix
the retained spectator $a$ and consider $W_N(a-z)$:
\[
	\int |W_N(a-z)\kappa_N(x-y)\phi(y)h(x,z)|^2 \d x\d y\d z \le C_T\epsilon^2\norm{W_N}_2^2 ,
\]
where we used the uniform column bound for $h$; its row
bound controls the other two spectator insertions. Thus,
with the retained spectator coordinate fixed, the kernels
are uniformly Hilbert--Schmidt in the remaining three
variables. The same estimates consequently hold for their
direct-integral maps and adjoints.

We now include the projections on the active coordinates.
Let $q=1-|\phi\rangle\langle\phi|$.
For an unprojected two-body kernel and a cubic Wick map,
the commutator has only one- and two-contraction terms:
the zero-contraction products cancel.
On $qL^2$, each contraction contains $q$,
and exterior $q$ projections act on the remaining
variables. Subtracting the unprojected formula therefore gives
$q-1$ in at least one contraction.
For two contractions use
$q\otimes q-1\otimes1=(q-1)\otimes q+1\otimes(q-1)$.
Each resulting composition contains the potential next to a rank-one
projection on one of its two coordinates. Its
two-particle operator norm is at most $C_TN^{1/2}$ by
\eqref{eq:rankone-potential}, or its adjoint.
The original cubic map has norm at most $C_T\epsilon$.
Tensoring these maps with identities and composing them
bounds each correction by $C_TN^{1/2}\epsilon$.
There are only finitely many such terms, with at most
five fields. This argument also applies to the
zero-to-three family and all adjoints; it uses operator
composition.
The exterior projections are contractions. Inputs already
lie in the excitation space, so the preceding Hardy bound
uses precisely their restricted kinetic form.

The projection corrections and dressed spectator terms are bounded by
$C_TN^{1/2}\epsilon M^{5/2}$. This is at most $C_TNM^{3/2}$ on the
physical space $M\le N+1$.
In each cubic monomial all generalized $b$ factors can be
collected into a sector scalar of absolute value at most
one. These factors and the original number cutoff commute
with $V_P$. The first commutator therefore keeps these scalar factors
and the field degrees of the canonical terms.
Finite bandwidth supplies every fixed positive or negative
weight in \eqref{eq:hardy-VC}; the number commutator multiplies each block by its particle-number shift,
whose absolute value is at most three. Closure proves the assertion on the stated domain.

For the covariant derivative, differentiate these actual
few-particle maps before lifting them to Fock space.
The profile $\kappa_N$ and the number coefficients are
time independent. Replacing $\phi$ by $\dot\phi$ in the
identity family preserves its internal and Hardy bounds.
The row, column and pointwise estimates for $\dot p,\dot\sigma$
in \cite[Lemma A.3]{COS} give exactly the same orders for
the differentiated dressed maps. Each differentiated
projection correction still has a rank-one factor adjacent
to the potential: $\dot q$ is a sum of rank-one maps with
vectors $\phi,\dot\phi$. Equation \eqref{eq:rankone-potential}
applies with those vectors, and differentiating the original
cubic map preserves its $O_T(\epsilon)$ norm.

We next estimate transport in the Hardy maps. Its covariant
derivative is controlled in the input-gradient norm.
Put $Z^{(j)}=\sum_{\ell=1}^jZ_\ell$. A map $B$ from $p$
to $q$ particles has covariant derivative
$\dot B-Z^{(q)}B+BZ^{(p)}$.
This identity follows by commuting $\dGamma(Z)$ with its Wick
expression. Only the active field coordinates contribute.
The outer excitation projections have zero covariant
derivative. Moreover $Z$ is a finite sum of rank-one
maps and is uniformly bounded from $L^2$ to $H^1$.
Thus, in the coordinate used for Hardy's inequality,
\[
	\norm{\nabla_x Z_x f}_2\le C_T\norm f_2, \qquad \norm{\nabla_x Z_a f}_2\le C_T\norm{\nabla_x f}_2 \quad(a\ne x).
\]
Output transport is bounded in $L^2$. Applying these
facts separately to $B$ and $B^*$ gives the same
input-gradient bound, with an additional $C_T\norm f_2$.
Lifting this additional term to Fock space gives a number weight
$M^{5/2}$, bounded by $M^2A_P^{1/2}$.
Transport of all the other few-particle maps is bounded
in $L^2$ and leaves their orders unchanged.
This proves the claimed bound for $\nabla_t[V_P,C]$,
including the derivative of the compressed potential.
\end{proof}

\begin{proposition}
\label{prop:hardy-remainder}
For every fixed integer $s\ge2$, the conjugation remainder
satisfies, for each fixed real $r$,
\begin{align}
	\norm{M^rR_{s,N}z}
	\le C_{s,r,T}\big(&N^{7/4-3s/4}\norm{M^{r+3s/2}z}
	+N^{3/2-3s/4}\norm{M^{r+3s/2+1}z}\notag\\
	&+N^{-3(s-1)/4}
	\norm{M^{r+3s/2+1/2}A_P^{1/2}z}\big).
	\label{eq:hardy-general}
\end{align}
Consequently the complete difference $R=\mathfrak J-\mathfrak
J^{[3]}$, and also $i[M,R]$ and $\nabla_tR$, satisfy
\begin{align}
	\norm{M^rRz}
	\le C_{r,T}\big(&N^{-5/4}\norm{M^{r+6}z}
	+N^{-3/2}\norm{M^{r+7}z}\notag\\
	&+N^{-9/4}\norm{M^{r+13/2}A_P^{1/2}z}\big).
	\label{eq:hardy-operator}
\end{align}
In particular, put
$\Phi_N=N^{-5/4}M^6+N^{-3/2}M^7$. Then
\begin{equation}\label{eq:hardy-form}
	|R[z,z]|+|i[M,R][z,z]|+|\nabla_tR[z,z]| \le\delta A_P[z]+C_{\delta,T}\Phi_N[z].
\end{equation}
This gives a positive closed comparison form
$\mathfrak J^{[3]}-\mathcal L_C+b e^{-C}Me^C+d\Phi_N$
uniformly comparable to $Q+\Phi_N$ for fixed suitable $b,d$.
Estimate \eqref{eq:refined-operator} remains useful:
it requires a smaller kinetic number weight than
\eqref{eq:hardy-operator}.
\end{proposition}
\begin{proof}
In $C^j[V_P,C]C^k$, $j+k=s-1$, the internal terms of Lemma~\ref{lem:hardy-VC} are bounded by
$N\epsilon^{s-1}M^{3s/2}$.
For the Hardy term the first-energy block bounds give
\[
	\epsilon^{s-1}M^{3s/2+1/2}A_P^{1/2} +C_s\epsilon^{s-2}M^{3s/2+1/2}.
\]
These are weighted operator bounds. The two exterior phases
in the integral remainder preserve the first term up to
an additional
$\epsilon^{s-1}M^{3s/2+2}$.
The potential remainder is supported on both sides in $E_N$,
since $V_P$ preserves number. The strong cutoff condition
therefore bounds this additional term by
$C\epsilon^{s-2}M^{3s/2+1/2}$.
The $S$ remainder has the bound
$N^{3/2}\epsilon^s M^{3s/2+1}$; because
$\epsilon^{s-2}=N^{3/2}\epsilon^s$, it also bounds the preceding zero-order contribution to the Hardy
estimate.
The kinetic remainder has the smaller bound already proved.
This proves \eqref{eq:hardy-general}. For $s=4$ add the
time remainder $C_T\epsilon^2M^3$, which is smaller.
The number-gauge argument uses the second assertion of
Lemma~\ref{lem:hardy-VC} and the already proved gauge
bound for $S$.
For the time derivative use the covariant assertion of that
lemma and the bound for $\nabla_tS$ in
Lemma~\ref{lem:bare-split}. Each differentiated cubic factor
is $X=\nabla_tC$, with the same $L^2$ and first-energy
blocks. Differentiating either exterior phase uses
Lemma~\ref{lem:A-phase} and has the same graph loss already
absorbed above. The support projection $E_N$ is covariantly
constant. The differentiated kinetic and time remainders
are covered by Proposition~\ref{prop:partial-derivative}.

For the diagonal form estimate keep the parts separate.
Balance the internal potential term with $M^3$ on each
argument. Balance the $S$ and Hardy terms with $M^{7/2}$.
For the derivative part of the Hardy term this leaves
\[
	C_TN^{-9/4}\norm{M^{7/2}E_Nz} \norm{M^3 A_P^{1/2}E_Nz}.
\]
Since $M^3E_N\le C N^{3/2}E_N$, Young's inequality
bounds it by
$\delta A_P[z]+C_{\delta,T}N^{-3/2}M^7[z]$.
The zero-order Hardy term has power $M^{13/2}$ and is
bounded by the same $N^{-3/2}M^7$ form.
The separate kinetic and time remainders are smaller.
The same estimates hold for the number-gauge and time derivatives.
Finally use the positivity of $e^{-C}\mathfrak G e^C+b
e^{-C}Me^C$ and absorb the small multiple of $A_P$.
The remaining commutator for the improved comparison
energy is identified next.
\end{proof}

\begin{corollary}
\label{cor:hardy-balance}
Let $\Phi_N=N^{-5/4}M^6+N^{-3/2}M^7$.
For $D=e^{-C}Me^C-M$ and $L=\mathcal L_C$,
\begin{equation}\label{eq:hardy-mixed}
	|i[R,D][z,z]|+|i[R,L][z,z]| \le\delta A_P[z]+C_{\delta,T}\Phi_N[z].
\end{equation}
There are fixed $b,d>0$ such that
\[
	\mathcal E^{\rm H}_{N,t} =\mathfrak J_t^{[3]}-\mathcal L_{C,t} +b e^{-C_t}Me^{C_t}+d\Phi_N \asymp_T Q_{N,P_t}+\Phi_N
\]
is a closed positive form. Along a common-domain solution
of the same Taylor evolution,
\begin{equation}\label{eq:hardy-balance}
	\left|\frac{\d}{\d t}\mathcal E^{\rm H}_{N,t}[y,y] -d\,i[\mathfrak J_t^{[3]},\Phi_N][y,y]\right| \le C_T\mathcal E^{\rm H}_{N,t}[y,y].
\end{equation}
Section~\ref{sec:taylor-first-energy} proves first-energy
propagation using a comparison energy with a linear number correction.
\end{corollary}
\begin{proof}
For $T=D,L$, use both bounds in
\eqref{eq:bounded-correction-graph}.
The internal potential remainder paired with $T$ is
bounded by $C_TN^{-5/4}M^6[z]$; the $S$ remainder by
$C_TN^{-3/2}M^7[z]$.
Keep the Hardy part $R_B$ separately. Its balanced
operator bound with weight $r=-7/2$ gives, for $z_E=E_Nz$,
\begin{align*}
	|R_B[z,Tz]|\le C_T\norm{M^{7/2}z_E}\big(
	&N^{-3/2}\norm{M^3z_E}\\
	+&N^{-9/4}\norm{M^3A_P^{1/2}z_E}
	+N^{-9/4}\norm{M^{9/2}z_E}\big).
\end{align*}
The first product is bounded by $C_TN^{-3/2}M^7[z]$.
Using $M^3E_N\le C N^{3/2}E_N$ in the second product
and then Young's inequality gives
$\delta A_P[z]+C_{\delta,T}N^{-3/2}M^7[z]$.
For the last product use
$M^{9/2}E_N\le C N^{1/2}M^{7/2}E_N$.
Its coefficient is $N^{-7/4}\le N^{-3/2}$.
Proposition~\ref{prop:bounded-corrections} bounds the mixed kinetic term
by $N^{-5/4}M^6+N^{-3/2}M^6$. The time remainder is bounded by
$N^{-3/2}M^3$. These are bounded by $C_T\Phi_N$.
Hermiticity and smaller intermediate Young constants
give \eqref{eq:hardy-mixed}.

Positivity follows from Proposition~\ref{prop:hardy-remainder}.
Write $U=e^C$ and $\mathcal M=U^*MU$.
The exact connection cancellation in
\eqref{eq:adapted-source} now leaves
\[
	U^*(\nabla_t\mathfrak G+i b[\mathfrak G,M])U -\nabla_tR+i[R,L]-i b[R,\mathcal M] +d\,i[\mathfrak J^{[3]},\Phi_N].
\]
Here $\nabla_t\Phi_N=0$. The first term is $Q$-bounded
by Lemma~\ref{lem:moving-G} and the phase comparison.
The other terms before the last one are bounded by
$C_T(Q+\Phi_N)$, using
\eqref{eq:hardy-form}, \eqref{eq:hardy-mixed} and
$\mathcal M=M+D$. This proves \eqref{eq:hardy-balance}.
\end{proof}

\subsection{Coercivity and weighted energy}
\label{sec:coercivity-limits}

The following graded model separates conservation of the first
energy from propagation of number-weighted energy.

\begin{proposition}\label{prop:weighted-obstruction}
Uniform coercivity, compatibility with every fixed power of $M$,
conservation of the first energy, and a form bound on $[M,G]$
do not imply uniform propagation of $M^pQ$ for $p>0$,
even from normalized data with every
fixed weighted first-energy moment uniformly bounded initially.
This remains false for time-independent generators of number
bandwidth one.
\end{proposition}
\begin{proof}
For integers $N\ge16$ put $L=\lfloor\sqrt N\rfloor$.
Consider a graded Hilbert space with orthonormal basis
$\Omega,e_0,\ldots,e_L$, and set
\[
	M\Omega=Q\Omega=\Omega,\qquad Me_j=(j+2)e_j,\qquad Qe_j=N^2e_j.
\]
Thus $Q\ge M$ and $K=Q-M\ge0$. Define the self-adjoint
tridiagonal matrix $S_L$ by $S_L\Omega=0$ and
\[
	S_Le_j=\sqrt{(j+1)(L-j)}\,e_{j+1} +\sqrt{j(L-j+1)}\,e_{j-1},
\]
omitting endpoint terms. Let $G_N=Q+NS_L$.
The matrix $S_L$ is the restriction of
$\sum_{\ell=1}^L\sigma_x^{(\ell)}$ to the symmetric tensor
power of $\mathbb C^2$, in its normalized occupation basis.
Hence $\norm{S_L}=L$, and
\[
	\tfrac12 Q\le G_N\le\tfrac32 Q,\qquad [Q,G_N]=0.
\]
Furthermore,
$\norm{Q^{-1/2}[M,G_N]Q^{-1/2}}\le 2L/N$.
For every fixed real $r$, each shifted weight ratio is at most
$(3/2)^{|r|}$, and the two shifts give
\[
	\norm{M^rQ^{-1/2}G_NQ^{-1/2}M^{-r}} \le1+2(3/2)^{|r|}L/N.
\]
These constants are uniform; the generator has zero time
derivative. This example also satisfies the coarse structural
inputs used above: take $A=Q$, $V=0$, $C=X=0$, and
$R_0=M+NS_L$. The phase is the identity, so its bounds hold and the Taylor remainder
vanishes. Moreover,
$\norm{M^rR_0z}\le C_rN^2\norm{M^{r+2}z}$.

Choose the normalized initial vector
\[
	\xi_N=(1-N^{-2})^{1/2}\Omega+N^{-1}e_0.
\]
For every fixed $p>0$,
$\langle\xi_N,M^pQ\xi_N\rangle=1-N^{-2}+2^p$.
At the time $\tau_N=\pi/(2N)$, however,
\[
	e^{-i\tau_NNS_L}e_0=(-i)^Le_L.
\]
To verify this identity, apply
$e^{-i(\pi/2)\sigma_x}=-i\sigma_x$ to each of the $L$
tensor factors. The scalar $Q=N^2$ on the chain only supplies
a phase. Consequently
\[
	\langle e^{-i\tau_NG_N}\xi_N,\, M^pQ\,e^{-i\tau_NG_N}\xi_N\rangle =1-N^{-2}+(L+2)^p\longrightarrow\infty.
\]
The first energy remains exactly $2-N^{-2}$.
Since $\tau_N\to0$, uniform propagation fails on every
nontrivial compact time interval. The small initial component
has large energy and moves through the number sectors;
first-energy conservation does not control its weighted energy.
\end{proof}

For the GP generator, the required propagation estimates use
its number-changing kernels and the initial correlated-energy bound.

\subsection{Correlated energy domains}
\label{sec:static-domain}

The operator remainder can be controlled on a domain defined by
powers of a positive energy associated with the full generator.
The weighted inverse estimate below provides the connection.
Its hypotheses include both polynomial number compatibility and
a bound on the first number commutator.
Related preservation of localization under inversion is studied
in the theory of inverse-closed matrix algebras; see
\cite[Sections 3--6]{KohldorferBalazs} for operator-valued matrices.
We use a direct proof with shifted number weights, which keeps
the constants uniform in the particle cutoff. The assumptions
below are stated as operator bounds on weighted spaces, rather
than as entrywise decay or weighted Schur bounds. The lemma is
an auxiliary domain estimate for the Taylor comparison.

\begin{lemma}\label{lem:shifted-inverse}
Consider a family of truncated graded Hilbert spaces with number weight
$M\ge1$. Let $Q\ge M$ strongly commute with
$M$. Let $F$ be a closed positive form
satisfying $cQ\le F\le CQ$, with constants independent of the
truncation.
Put $T=Q^{-1/2}FQ^{-1/2}$. Assume
\begin{equation}\label{eq:inverse-inputs}
	\sup_N\norm{M^aTM^{-a}}<\infty\quad(a\in\mathbb R), \qquad \sup_N\norm{[M,T]}<\infty.
\end{equation}
The commutator is understood as its bounded extension.
For every fixed real $r$ there exists a fixed $\rho\ge1$ such that,
with $W=M+\rho$,
\begin{equation}\label{eq:weighted-inverse}
	\norm{Q^{1/2}W^rF^{-1}W^{-r}Q^{1/2}}\le 2/c.
\end{equation}
In particular, for every fixed integer $k\ge1$,
\begin{equation}\label{eq:power-domain}
	\norm{M^kF^{-k}}+ \norm{M^{k-1/2}Q^{1/2}F^{-k}}\le C_k .
\end{equation}
All constants are uniform in the truncation.
\end{lemma}
\begin{proof}
We first show the uniform smallness
\begin{equation}\label{eq:shift-small}
	\norm{W^rTW^{-r}-T}\le C_r\rho^{-1/5},\qquad \rho\ge2.
\end{equation}
Polynomial compatibility gives
$\norm{W^aTW^{-a}}\le C_a$
uniformly also in $\rho$.
For a nonnegative integer $a$, expand $(M+\rho)^a$ on
the left of $T$; its $j$th summand is bounded using
$M^jTM^{-j}$ and
$\rho^{a-j}M^j(M+\rho)^{-a}\le1$.
Since $T=T^*$, the adjoint of $W^aTW^{-a}$ is
$W^{-a}TW^a$. This gives the negative integer bounds
with the same constants. For integers $a_0<a_1$, apply
the three-lines theorem to
$W^{a_0+(a_1-a_0)z}TW^{-a_0-(a_1-a_0)z}$.
On $\operatorname{Re}z=0,1$, the imaginary powers
are unitary and the bounds are $C_{a_0},C_{a_1}$.
The interior bound is $C_{a_0}^{1-\theta}C_{a_1}^\theta$
at $a=(1-\theta)a_0+\theta a_1$, uniformly in $N,\rho$.
This proves the real-exponent assertion
\cite{BerghLofstrom}.

Choose a nonnegative Schwartz function $h$ on $\mathbb R$
with integral one and Fourier transform supported in $[-1,1]$.
We use $\widehat h(\omega)=\int h(s)e^{-is\omega}\d s$.
Such a function is obtained by squaring the absolute value
of the inverse Fourier transform of a smooth function
supported in $[-1/2,1/2]$, and normalizing.
For $D\ge1$ define the norm-convergent integral
\[
	T_D=\int_{\mathbb R}D h(Ds)e^{isM}Te^{-isM}\d s .
\]
Only number shifts $|d|\le D$ occur in $T_D$.
Indeed its $(m,n)$ block is
$\widehat h(-(m-n)/D)\Pi_mT\Pi_n$, where $M\Pi_n=(n+1)\Pi_n$.
The bounded commutator and the unitary fundamental theorem
of calculus give
\[
	\norm{T-T_D}\le D^{-1}\norm{[M,T]} \int_{\mathbb R}|s|h(s)\d s .
\]
Gauge conjugation commutes with $W$, so $T_D$ has all the
same uniform weighted bounds as $T$.
In particular, for $E_D=T-T_D$,
\[
	\|W^{2r}E_DW^{-2r}\|
	\le\|W^{2r}TW^{-2r}\|
	+\int_{\mathbb R}Dh(Ds)\|W^{2r}TW^{-2r}\|\d s
	\le2C_{2r} .
\]
Together with $\|E_D\|\le C D^{-1}$, this supplies
both boundary estimates for the three-lines theorem
\cite{BerghLofstrom}. Applying it to
$W^{2rz}E_DW^{-2rz}$ on $0\le\operatorname{Re}z\le1$
and evaluating at $z=1/2$ gives
\begin{equation}\label{eq:weighted-tail}
	\norm{W^rE_DW^{-r}}\le C_rD^{-1/2}.
\end{equation}
The weights in these analytic families are bounded at each
fixed truncation. The estimates on both boundary lines are
uniform, which gives uniform interpolation constants.
Explicitly $W^{2rz}=\exp(2rz\log W)$ is a bounded-operator
analytic function at fixed truncation, defined on the whole
Hilbert space. Apply the scalar three-lines theorem to
$\langle u,W^{2rz}E_DW^{-2rz}v\rangle$ for unit vectors $u,v$.
The imaginary powers are unitary, so the two boundary bounds
are $C D^{-1}$ and $C_r$, with no truncation-dependent factor.
Their geometric mean gives \eqref{eq:weighted-tail}.

For $D\le\rho$, each single-shift piece of $T_D$ has norm
at most $\norm{T_D}$, since it is an orthogonal sum of
number-sector blocks. On a block $m=n+d$,
\[
	\left|\left(\frac{m+1+\rho}{n+1+\rho}\right)^r-1\right| \le C_r |d|/\rho .
\]
In detail the shift-$d$ piece is
$\sum_{n:\,0\le n,n+d\le N}\Pi_{n+d}T_D\Pi_n$; its input
spaces and output spaces are separately orthogonal, so its
norm is $\sup_n\|\Pi_{n+d}T_D\Pi_n\|\le\|T_D\|$.
There are at most $2\lfloor D\rfloor+1$ shifts. Consequently
$\norm{W^rT_DW^{-r}-T_D}\le C_rD^2/\rho$.
Combine this with \eqref{eq:weighted-tail} and the unweighted
tail bound, and choose $D=\rho^{2/5}$.
This proves \eqref{eq:shift-small}.

The form inequalities give $cI\le T\le CI$, so we choose
$\rho$ to make the right-hand side of \eqref{eq:shift-small}
smaller than $c/2$. A Neumann series then shows that
$W^rTW^{-r}$ is invertible, with inverse norm at most $2/c$,
by perturbing $T$. The representation theorem gives
$F^{-1}=Q^{-1/2}T^{-1}Q^{-1/2}$ for the closed form.
Commuting $Q$ with $W$ yields \eqref{eq:weighted-inverse};
the products involving $Q^{1/2}$ are understood through
these bounded extensions.
More precisely, put $V=D(Q^{1/2})$ with norm
$\|Q^{1/2}u\|$. The form operator $F:V\to V^*$ is
an isomorphism, and its inverse agrees on $\mathcal H$
with the self-adjoint inverse. The dual extension of
$Q^{-1/2}$ maps $V^*$ onto $\mathcal H$ isometrically.
At each fixed number cutoff, $W^{\pm r}$ preserve
$V$ and $V^*$. Thus \eqref{eq:weighted-inverse} also gives
\begin{equation}\label{eq:weighted-form-inverse}
	\|W^rQ^{1/2}F^{-1}f\|
	\le(2/c)\|W^rQ^{-1/2}f\|,\qquad f\in V^* .
\end{equation}
Indeed, set $g=W^rQ^{-1/2}f$. The vector on the left
is $\mathcal I_rg$, where $\mathcal I_r$ denotes the
bounded operator in \eqref{eq:weighted-inverse}.
This identifies the ordered products on the full dual
space as well as on Hilbert-space sources.

For a fixed $k$, choose a common $\rho$ for the finitely
many exponents $r=j+1/2$, $0\le j\le k-1$.
Because $W\le(1+\rho)Q$, \eqref{eq:weighted-inverse}
implies boundedness of
\[
	B_j=W^{j+1}F^{-1}W^{-j},\qquad D_j=W^{j+1/2}Q^{1/2}F^{-1}W^{-j}.
\]
For example $B_j$ is the operator in
\eqref{eq:weighted-inverse} at $r=j+1/2$,
multiplied on both sides by $W^{1/2}Q^{-1/2}$.
For $D_j$ only the right factor is needed.
Explicitly, if
$\mathcal I_r=Q^{1/2}W^rF^{-1}W^{-r}Q^{1/2}$ and
$S=W^{1/2}Q^{-1/2}$, then
\[
	B_j=S\mathcal I_{j+1/2}S,\qquad D_j=\mathcal I_{j+1/2}S,\qquad \|S\|\le\sqrt{1+\rho}.
\]
These identities first hold on the spectral core of $Q$ and
then give the bounded extensions. At fixed number cutoff $W$
is bounded and preserves the form domain, so the inverse maps
used here are defined on that core. Closure of $Q^{1/2}$
gives the corresponding range statements.
The telescoping products give
\[
	W^kF^{-k}=B_{k-1}\cdots B_0,\qquad W^{k-1/2}Q^{1/2}F^{-k} =D_{k-1}B_{k-2}\cdots B_0 .
\]
The empty product for $k=1$ is the identity.
Finally $M^aW^{-a}$ is a contraction for $a\ge0$.
These bounds prove \eqref{eq:power-domain}, including
the asserted domain inclusions by closure.
\end{proof}

\begin{proposition}
\label{prop:correlated-domain}
Fix a compact time interval. For a sufficiently large constant
$b$, independent of $N,t$, define the positive form
\[
	F_{N,t}=\mathfrak G_{N,t}+bM .
\]
It satisfies, for each fixed integer $k\ge1$,
\begin{equation}\label{eq:GP-power-domain}
	\norm{M^kz}+ \norm{M^{k-1/2}Q_{N,P_t}^{1/2}z} \le C_{k,T}\norm{F_{N,t}^kz}, \qquad z\in D(F_{N,t}^k).
\end{equation}
In particular the original cubic Taylor remainder obeys
\begin{equation}\label{eq:correlated-remainder}
	\norm{R_tz}\le C_TN^{-5/4}\norm{F_{N,t}^7z}, \qquad z\in D(F_{N,t}^7).
\end{equation}
The normalized initial vectors
\begin{equation}\label{eq:prepared-data}
	\xi_N=\frac{F_{N,0}^{-7}\Omega} {\norm{F_{N,0}^{-7}\Omega}}
\end{equation}
are well defined and satisfy
$\sup_N\norm{F_{N,0}^7\xi_N}<\infty$.
These are initial-time estimates.
\end{proposition}
\begin{proof}
The lower and upper form bounds in \cite[Eq.~(4.26)]{BS}
give $c_TQ\le F\le C_TQ$ after increasing $b$.
Proposition~\ref{prop:GP-weight} gives the first condition
in \eqref{eq:inverse-inputs}. The additional term $bM$
preserves number and is bounded by $bQ$.
Since $M$ commutes with $Q$,
\[
	[M,Q^{-1/2}FQ^{-1/2}] =Q^{-1/2}[M,\mathfrak G]Q^{-1/2}.
\]
Its norm is uniformly bounded by the number-commutator
estimate in the same external theorem, as also used in
Lemma~\ref{lem:moving-G}.
Thus Lemma~\ref{lem:shifted-inverse} applies directly to
the actual generator; the directly $M$-bounded families
in Proposition~\ref{prop:GP-weight} need no additional
termwise decomposition.
This proves \eqref{eq:GP-power-domain}.
At $k=7$ it controls both $M^7$ and
$M^6A_P^{1/2}$, because $A_P\le Q$ and $M\ge1$.
Equation \eqref{eq:refined-operator} now proves
\eqref{eq:correlated-remainder}.

Finally $\Omega$ lies in the form domain and
$F_{N,0}[\Omega]\le C_TQ[\Omega]=C_T$.
For its spectral probability measure, convexity of
$x\mapsto x^{-7}$ on $(0,\infty)$ gives
\[
	\norm{F_{N,0}^{-7}\Omega} \ge\ip{\Omega}{F_{N,0}^{-7}\Omega} \ge \big(F_{N,0}[\Omega]\big)^{-7}\ge C_T^{-7}.
\]
The spectral lower bound for $F$ also makes its inverse
powers bounded. Hence \eqref{eq:prepared-data} belongs
to $D(F_{N,0}^7)$ and
$\norm{F_{N,0}^7\xi_N}=\norm{F_{N,0}^{-7}\Omega}^{-1}
\le C_T^7$, as claimed.
\end{proof}

\begin{remark}
\label{rem:power-propagation}
Equation \eqref{eq:correlated-remainder} gives a sufficient condition
based on propagation of $F^7$. This condition is not
necessary for the comparison.
Even for time-independent matrices satisfying
Lemma~\ref{lem:shifted-inverse}, high powers of $F$ need not
propagate under $G=F-M$, although every number-weighted
first energy does propagate.

Here is an explicit example, distinct from
Proposition~\ref{prop:weighted-obstruction}. On $\mathbb C^2$ put
\[
	M=\begin{pmatrix}1&0\\0&2\end{pmatrix},\quad
	Q_N=\begin{pmatrix}1&0\\0&N^2\end{pmatrix},\quad
	F_N=\begin{pmatrix}2&N\\N&N^2\end{pmatrix},\quad G_N=F_N-M.
\]
For $N\ge2$, $Q_N\ge M$ and
$Q_N^{-1/2}F_NQ_N^{-1/2}=\left(\begin{smallmatrix}2&1\\1&1\end{smallmatrix}\right)$.
All the lemma's constants are uniform.
Let $v_-,v_+$ be normalized eigenvectors of $F_N$,
whose eigenvalues are
$\lambda_\pm=(N^2+2\pm\delta_N)/2$ with
$\delta_N=\sqrt{N^4+4}$.
Thus $\lambda_-\to1$ and $\lambda_+\sim N^2$.
In this eigenbasis the off-diagonal entry of $G_N$
has absolute value $N/\delta_N$.
Its eigenvalue gap is $\Delta_N=\sqrt{N^4-2N^2+9}$.
The two-by-two exponential formula at $t_N=\pi/\Delta_N$
therefore gives
\[
	|\ip{v_+}{e^{-it_NG_N}v_-}|=\frac{2N}{\delta_N\Delta_N}, \qquad \norm{F_N^ke^{-it_NG_N}v_-}\ge c_kN^{2k-3}.
\]
Initially $\norm{F_N^kv_-}=\lambda_-^k$ is bounded for
every fixed $k$, but the evolved bound fails for $k\ge2$.
On the other hand, conservation of $G_N$ and $1\le M\le2$
give
\[
	F_N[e^{-itG_N}v_-]\le \lambda_-+1 .
\]
Hence every fixed $M^pQ_N$ expectation stays uniformly bounded
for all times. In this model, the weighted kinetic norm in
Corollary~\ref{cor:single-kinetic-norm} remains bounded even
when higher spectral powers of $F_N$ grow.
\end{remark}

\subsection{Conditional comparison estimates}
\label{sec:conditional-comparisons}

We write the integrals below for $0\le t\le T$; for negative
times reverse the integration direction.
Let $x_N(t)$ solve the exact covariant equation and $y_N(t)$
the equation generated by $\mathfrak J_t^{[3]}$, with the same
initial vector. At fixed $N$, the preceding construction gives
the weak pairings on a common first-energy space after transport.
The weak product rule for a Hilbert space and its form-domain
dual \cite[Chapter III]{LionsMagenes} yields
\[
	\frac{\d}{\d t}\ip{x_N(t)}{y_N(t)} =i(\mathfrak J_t-\mathfrak J_t^{[3]})[x_N(t),y_N(t)].
\]
For form-space initial data it follows by density or the
$L^2(V)\cap H^1(V^*)$ product rule. Consequently,
\begin{align}
	\norm{x_N(t)-y_N(t)}^2
	\le C_TN^{-3/2}\int_0^t&
	\norm{Q_{N,P_r}^{1/2}x_N(r)}\notag\\[-1mm]
	&{}\times\norm{M^3Q_{N,P_r}^{1/2}y_N(r)}\d r.
	\label{eq:dynamic}
\end{align}
Here we use the improved remainder proved in
Proposition~\ref{prop:smaller-asymmetric}; the earlier
Theorem~\ref{thm:full} would require $M^6$ instead of $M^3$.
The energy form bound for the exact evolution follows from
\cite[Proposition 6.1]{BS} and the phase-energy comparison in
\cite{COS}. If the second factor in \eqref{eq:dynamic} is uniformly
bounded along the evolution generated by $\mathfrak J^{[3]}$,
this estimate gives an $O(N^{-3/4})$ vector error on compact
intervals. Theorems~\ref{thm:energy-class-comparison} and
\ref{thm:full-model-comparison} prove the $O(N^{-3/16})$ and
$O(N^{-5/4})$ bounds on their respective initial domains.

There is also a direct Duhamel alternative. If
$y_N(t)$ belongs to the operator-remainder domain almost
everywhere, and the norm below is time-integrable,
Corollary~\ref{cor:single-kinetic-norm} gives
\begin{equation}
	\norm{x_N(t)-y_N(t)} \le C_TN^{-5/4}\int_0^t \norm{M^6A_{P_r}^{1/2}y_N(r)}\d r . \label{eq:operator-dynamic}
\end{equation}
For operator-domain data this is the ordinary variation-of-constants
identity with the exact propagator on the left of the forcing.
For form solutions with the stated integrability, the same identity
follows by testing against exact backward form solutions:
the form difference has the $L^2$ representative supplied by
Corollary~\ref{cor:single-kinetic-norm}. Density gives the displayed bound.
An $N$-uniform bound on this integral gives $O(N^{-5/4})$
directly. Uniform propagation of the $M^{12}A_P$ form
is sufficient. The energy norm here is the kinetic norm
$A_P^{1/2}$.
Proposition~\ref{prop:correlated-domain} supplies another
precise sufficient condition: if
$\int_0^T\norm{F_{N,r}^7y_N(r)}\d r$ is uniformly bounded,
then \eqref{eq:operator-dynamic} is $O_T(N^{-5/4})$.
The prepared data in \eqref{eq:prepared-data} satisfy
this energy condition initially, but its propagation
along $\mathfrak J^{[3]}$ remains an additional requirement for this direct Duhamel estimate. Theorem~\ref{thm:full-model-comparison} gives a comparison
on $D(F_3(s)^{3/2})$ by inverse-energy integration by parts.

\subsection{Kinetic graph bounds for the cubic phase}
\label{sec:kinetic-graph}

The spatial structure of the cubic kernels improves the
zero-order term in the general block bound \eqref{eq:blocks}.
We establish this refinement for the cubic phase and its
exponential. Propagation under the Taylor generator will be
treated separately.

\begin{proposition}\label{prop:sharp-kinetic-block}
Let $v\in C_c^\infty(\mathbb R^3)$ be nonnegative and radial.
Let $\phi_t$ solve the modified GP equation with a fixed normalized
initial datum in $H^4(\mathbb R^3)$. Fix a sufficiently small
quadratic scattering radius $\ell\in(0,1/2)$, independent of $N$.
Use the cubic phase \eqref{spec:eq:actual-phase} at this background,
with $m=N^{-1/2}$ and the original COS cutoff $0\le\Theta_N\le1$.
It equals one for $n\le N^{1/2}/2+10$ and zero for
$n\ge N^{1/2}-10$, for sufficiently large $N$.
The blocks of $C$ satisfy
\begin{equation}\label{eq:sharp-kinetic-block}
	\norm{A_{P,j}^{1/2}C_{jn}z_n} \le C_T\left( \epsilon(n+1)^{3/2}\norm{A_{P,n}^{1/2}z_n} +(n+1)^{1/2}\norm{z_n}\right), \qquad \epsilon=N^{-3/4}.
\end{equation}
The same bound holds for $[M,C]$. Consequently, for every
fixed real $r$ and $|s|\le1$,
\begin{align}
	\norm{M^rA_P^{1/2}e^{sC}z}
	&\le C_{r,T}\norm{M^rA_P^{1/2}z},\label{eq:sharp-A-phase}\\
	\norm{M^rA_P^{1/2}(e^{-C}Me^C-M)z}
	&\le C_{r,T}\norm{M^rA_P^{1/2}z}.\notag
\end{align}
If $\phi,\dot\phi$ are continuous in $H^4$ and uniformly bounded
there on compact time intervals, the time derivative satisfies the
estimates in Proposition~\ref{prop:sharp-time-graph}.
\end{proposition}
\begin{proof}
We first work with canonical, unprojected fields.
Write $\kappa=\kappa_N$ and $g=\nabla\kappa$.
The pointwise gradient bound in \cite[Lemma A.1]{COS},
scaled to the ball of radius $m$, implies
\[
	|g(r)|\le\frac{C}{\sqrt N\,|r|^2}\,\mathbf1_{|r|\le m}.
\]
Together with the already established $L^2$ bounds this gives
\begin{equation}\label{eq:gradient-small-norms}
	\norm\kappa_2\le C\epsilon,\quad \norm g_2\le C,\quad \norm g_1\le CN^{-1},\quad \norm g_{6/5}\le C\epsilon .
\end{equation}
All gradients below are first derivatives. The matched
Neumann extension introduces no boundary measure.

Let $\psi_n$ be a smooth symmetric $n$-particle vector.
For a canonical cubic monomial $T$, the identity
\[
	\nabla_z a_zT\psi_n =T\nabla_z a_z\psi_n+\nabla_z[a_z,T]\psi_n
\]
and $\norm{K^{1/2}\psi}^2=\int\norm{\nabla_z a_z\psi}^2\d z$
reduce the proof to the differentiated commutator.
The first term is bounded by $C_T\epsilon(n+1)^{3/2}\norm{K^{1/2}\psi_n}$, using the cubic
$L^2$ block estimate.

For the identity family
$T_0=\int\kappa(x-y)\phi(y)a_x^*a_y^*a_x\d x\d y$,
put $f_z(y)=\kappa(z-y)\phi(y)$. Then
\[
	[a_z,T_0]=a^*(f_z)a_z+ \phi(z)\dGamma(\kappa(\,\cdot-z)).
\]
The contribution from $a^*(f_z)\nabla_z a_z$ is bounded by
$C_T\epsilon\sqrt{n+1}\norm{K^{1/2}\psi_n}$.
For the derivative of $f_z$, the creation identity gives
\begin{align*}
	\int\norm{a^*(\nabla f_z)a_z\psi_n}^2\d z
	=\int \norm{\nabla f_z}_2^2\norm{a_z\psi_n}^2\d z
	+\int\norm{a(\nabla f_z)a_z\psi_n}^2\d z .
\end{align*}
The first integral is at most $C_Tn\norm{\psi_n}^2$.
In the second, integrate the derivative of $\kappa(z-y)$
by parts in $y$. Cauchy--Schwarz with $\norm\kappa_2$
then bounds it by
\[
	C_T\epsilon^2\bigl( n^2\norm{\psi_n}^2+n\norm{K^{1/2}\psi_n}^2\bigr).
\]
Here $\phi,\nabla\phi$ are bounded; the derivative on
$a_y a_z\psi_n$ is an input derivative.

In the density family, the term containing $\nabla\phi(z)$
is bounded by $C_T\epsilon n\norm{\psi_n}$. Normal ordering
the remaining term separates a diagonal contribution, at most
$C_Tn\norm{\psi_n}^2$, from a two-particle contribution bounded by
\[
	C_T\int J(x-y)\norm{a_xa_y\psi_n}^2\d x\d y, \qquad J=|g|*|g| .
\]
Young's inequality and \eqref{eq:gradient-small-norms}
give $\norm J_{3/2}\le C\epsilon^2$.
The Sobolev inequality in $x$, with $y$ fixed, bounds
the last display by
$C_T\epsilon^2(n-1)\norm{K^{1/2}\psi_n}^2$.
For $T_0^*$, the corresponding commutator is the
two-annihilator expression $a(f_z)a_z$, up to conjugations.
The same integration by parts bounds its derivatives on the input vector.

We give the dressed-field estimates as well.
Let $h=p$ or $\sigma$. Its Hilbert--Schmidt norm and its
row and column $L^2$ norms are bounded uniformly, while
$\norm{\nabla_jh}_2\le C_T\sqrt N$.
The families are
\[
	T_h^-=\int\kappa(x-y)\phi(y)a_x^*a_y^*a(h_x)\d x\d y, \quad T_h^+=\int\kappa(x-y)\phi(y)a_x^*a_y^*a^*(h_x)\d x\d y .
\]
Their annihilator commutators contain
$a^*(f_z)a^\sharp(h_z)$ and
$\phi(z)\int\kappa(x-z)a_x^*a^\sharp(h_x)\d x$.
The all-creation family has in addition the pair-creation
kernel $\kappa(x-y)\phi(y)h(x,z)$.
The estimates below are invariant under complex conjugation.

We record explicitly the bounds for a derivative on
$\kappa$. If $u_z(y)=g(z-y)\phi(y)$ and
$Q_z^\sharp=\int g(x-z)a_x^*a^\sharp(h_x)\d x$, then
\begin{align}
	\int\norm{a^*(u_z)a^\sharp(h_z)\psi_n}^2\d z
	&\le C_T\bigl((n+1)\norm g_2^2
	+(n+1)^2\norm g_1^2\bigr)\norm{\psi_n}^2,
	\notag\\
	\int\norm{Q_z^\sharp\psi_n}^2\d z
	&\le C_T\bigl((n+1)\norm g_2^2
	+(n+1)^2\norm g_1^2\bigr)\norm{\psi_n}^2.
	\label{eq:dressed-quadratic-sources}
\end{align}
We verify these bounds for one component of $g$ and then sum
over the three components. Write $R_h=\sup_x\|h_x\|_2$
and $H_h=\|h\|_{\mathrm{HS}}$. The Hilbert-space-valued
Young inequality gives
\begin{align*}
	\int\|a(u_z)\psi_n\|^2\d z
	&\le n\|\phi\|_\infty^2\|g\|_1^2\|\psi_n\|^2,\\
	\int\|a(u_z)a(h_z)\psi_n\|^2\d z
	&\le n(n-1)R_h^2\|\phi\|_\infty^2\|g\|_1^2\|\psi_n\|^2.
\end{align*}
The second line uses commutativity of the annihilators and
the row bound after applying $a(u_z)$. For the mixed product,
the canonical commutation relations give exactly
\[
	\|a^*(u_z)a(h_z)\psi_n\|^2
	=\|a(u_z)a(h_z)\psi_n\|^2
	+\|u_z\|_2^2\|a(h_z)\psi_n\|^2.
\]
For the product of two creators, the same relations give
\begin{align*}
	\|a^*(u_z)a^*(h_z)\psi_n\|^2
	\le{}&\|a(u_z)a(h_z)\psi_n\|^2
	+2\|u_z\|_2^2\|a(h_z)\psi_n\|^2\\
	&+2\|h_z\|_2^2\|a(u_z)\psi_n\|^2
	+2\|u_z\|_2^2\|h_z\|_2^2\|\psi_n\|^2.
\end{align*}
Here Cauchy--Schwarz bounds the cross contraction.
Since $\sup_z\|u_z\|_2^2\le\|\phi\|_\infty^2\|g\|_2^2$
and $\int\|h_z\|_2^2\d z=H_h^2$, the first contraction
and the vacuum term are bounded by
$C(n+1)\|g\|_2^2\|\psi_n\|^2$ after integration.
The other contraction is bounded by the first displayed
Young estimate and $R_h^2$. This proves the first line of
\eqref{eq:dressed-quadratic-sources}, including $n=0,1$.

For the number-preserving $Q_z^-$, set
$c_g(x,y)=\int\bar g(x-z)g(y-z)\d z$. Normal ordering gives
\begin{align*}
	\int\|Q_z^-\psi_n\|^2\d z
	={}&\|g\|_2^2\int\|a(h_x)\psi_n\|^2\d x\\
	&+\int c_g(x,y)
	\langle a(h_x)a_y\psi_n,a(h_y)a_x\psi_n\rangle\d x\d y.
\end{align*}
The first term is at most $nH_h^2\|g\|_2^2\|\psi_n\|^2$.
Since $|c_g(x,y)|\le J(x-y)$, the absolute value of the second
term is at most
\[
	C_T\int J(x-y) \norm{a(h_x)a_y\psi_n}\, \norm{a(h_y)a_x\psi_n}\d x\d y .
\]
Cauchy--Schwarz under interchange of $x,y$ reduces this to
\[
	C\int J(x-y)\|a(h_x)a_y\psi_n\|^2\d x\d y
	\le Cn(n-1)R_h^2\|g\|_1^2\|\psi_n\|^2.
\]
We used $\|a(h_x)a_y\psi_n\|^2
\le(n-1)R_h^2\|a_y\psi_n\|^2$ before integrating in $x$.

For $Q_z^+$, use $a(f)=\int\bar f(w)a_w\d w$ and
$h_x(w)=h(x,w)$. Its pair-creation kernel is
$L_z(x,w)=g(x-z)h(x,w)$. Put $S_z=(L_z+L_z^{\mathsf T})/2$
and define $\mathcal B^*(S)=\int S(x,w)a_x^*a_w^*\d x\d w$,
with $\mathcal B(S)=\mathcal B^*(S)^*$.
Commutativity of the creators gives
$Q_z^+=\mathcal B^*(L_z)=\mathcal B^*(S_z)$.
Normal ordering gives
\begin{equation}\label{eq:quadratic-source-normal-order}
	\|\mathcal B^*(S_z)\psi_n\|^2
	=\|\mathcal B(S_z)\psi_n\|^2
	+4\langle\psi_n,\dGamma(S_zS_z^*)\psi_n\rangle
	+2\|S_z\|_{\mathrm{HS}}^2\|\psi_n\|^2.
\end{equation}
Here $(S_zS_z^*)(x,y)=\int S_z(x,w)\bar S_z(y,w)\d w$
acts on the uncontracted creation coordinate.
Also $\mathcal B(S_z)=\int\bar g(x-z)a(h_x)a_x\d x$, so
\[
	\int\|\mathcal B(S_z)\psi_n\|^2\d z
	\le\|g\|_1^2\int\|a(h_x)a_x\psi_n\|^2\d x
	\le n(n-1)R_h^2\|g\|_1^2\|\psi_n\|^2.
\]
This is Young's inequality for the pair-annihilation field.
For the one-particle contraction, use
$S_zS_z^*\le(L_zL_z^*+L_z^{\mathsf T}\bar L_z)/2$.
The kernel of $\int L_zL_z^*\d z$ is bounded in absolute
value by $R_h^2J(x-y)$. The other orientation has kernel
\[
	\int(L_z^{\mathsf T}\bar L_z)(x,y)\d z
	=\int\!\int|g(w-z)|^2h(w,x)\bar h(w,y)\d w\d z
	=\|g\|_2^2(h^{\mathsf T}\bar h)(x,y).
\]
Thus
\[
	\left\|\int S_zS_z^*\d z\right\|_{\mathrm{op}}
	\le\tfrac12\bigl(R_h^2\|g\|_1^2+H_h^2\|g\|_2^2\bigr),
	\qquad \int\|S_z\|_{\mathrm{HS}}^2\d z\le H_h^2\|g\|_2^2.
\]
The second line of \eqref{eq:dressed-quadratic-sources}
now follows from \eqref{eq:quadratic-source-normal-order}.
Both right sides in \eqref{eq:dressed-quadratic-sources}
are bounded by $C_T(n+1)\norm{\psi_n}^2$ on the active sectors.

Terms with a derivative on $\phi$ are estimated using
$\norm\kappa_2$ and the elementary quadratic bound. Their
contribution is at most $C_T\epsilon(n+1)\norm{\psi_n}$.
For terms with a derivative on $h$, the integrated
Hilbert--Schmidt norm of the remaining quadratic kernel
is at most $C_T\epsilon\norm{\nabla h}_2$.
The quadratic number factor is at most $C(n+1)$.
These terms are therefore bounded by
\[
	C_T\epsilon\sqrt N(n+1)\norm{\psi_n} \le C_T\sqrt{n+1}\norm{\psi_n},
\]
where the last inequality uses $n\le C\sqrt N$.
The same argument treats the derivative of the additional
pair-creation kernel in $[a_z,T_h^+]$.
For $(T_h^-)^*$ the commutator is just that type of pair
annihilator; for $(T_h^+)^*$ it is zero.
This covers all adjoints separately.

Combining these bounds proves the canonical kinetic estimate
in \eqref{eq:sharp-kinetic-block}. Its number-energy part
follows from the original $L^2$ block bound.
Now restore the generalized fields and the cutoff separately
on each number block. Their coefficients are sector scalars
of absolute value at most one and have no spatial derivatives.
For an excitation-space input, the projected block is the
canonical block followed by its output projection.
The estimate
$\norm{A^{1/2}P\psi}\le C_T\norm{A^{1/2}\psi}$
has a constant independent of the sector: apply
$\norm{\nabla q f}\le\norm{\nabla f}+\norm{\nabla\phi}_2\norm f$
in each coordinate and sum the squared bounds.
This proves \eqref{eq:sharp-kinetic-block} for the actual phase.
Multiplication of a block by its particle-number shift proves the
same assertion for $[M,C]$.

Since $\epsilon(n+1)^{3/2}$ is uniformly bounded on the
support of $C$, finite bandwidth and
$A_{P,n}\ge n+1$ show that $C$ and $[M,C]$ are uniformly
bounded on every space with norm $\norm{M^rA_P^{1/2}z}$.
Their exponential series and the integral formula for
$e^{-C}Me^C-M$ prove \eqref{eq:sharp-A-phase}.
All identities first hold on smooth finite-sector cores
and extend by the displayed graph bounds.
\end{proof}

\begin{corollary}\label{cor:sharp-kinetic-words}
For every fixed integer $j\ge1$ and real $r$,
\[
	\norm{M^rA_P^{1/2}C^jz} \le C_{j,r,T}\left( \epsilon^j\norm{M^{r+3j/2}A_P^{1/2}z} +\epsilon^{j-1}\norm{M^{r+3j/2-1}z}\right).
\]
\end{corollary}
\begin{proof}
Iterate \eqref{eq:sharp-kinetic-block} along the finitely
many number paths of length $j$. The zero-order source
has one factor $M^{1/2}$ and $j-1$ factors
$\epsilon M^{3/2}$, giving the displayed second term.
Shifted number weights change only the fixed constant.
\end{proof}

The uniform kinetic graph bound follows from the normal-ordered
contractions and the small $L^1$ and $L^{6/5}$ gradient norms.
Section~\ref{sec:taylor-first-energy} uses this bound to construct
a comparison energy with a linear number correction.

\begin{proposition}
\label{prop:sharp-time-graph}
Let $v\in C_c^\infty(\mathbb R^3)$ be nonnegative and radial.
Let $\phi_t$ solve the modified GP equation with a fixed normalized
initial datum in $H^4(\mathbb R^3)$. Fix a sufficiently small
quadratic scattering radius $\ell\in(0,1/2)$, independent of $N$.
Use the cubic phase \eqref{spec:eq:actual-phase} at this background,
with $m=N^{-1/2}$ and the original COS cutoff $0\le\Theta_N\le1$.
It equals one for $n\le N^{1/2}/2+10$ and zero for
$n\ge N^{1/2}-10$, for sufficiently large $N$.
Assume $\phi\in C^1([-T,T];H^4)$, with this norm bounded
independently of $N$.
Then \eqref{eq:sharp-kinetic-block} holds also with $C$
replaced by $X=\nabla_tC$ or by $[M,X]$.
For every fixed real $r$ and $|s|\le1$,
\begin{equation}\label{eq:sharp-time-graph}
	\norm{M^rA_P^{1/2}\nabla_t(e^{sC})z} +\norm{M^rA_P^{1/2}\mathcal L_Cz} \le C_{r,T}\norm{M^rA_P^{1/2}z}.
\end{equation}
The finite-word bound in Corollary~\ref{cor:sharp-kinetic-words}
holds for any word of length $j$ in $C$ and $X$.

\end{proposition}
\begin{proof}
Ordinary time differentiation preserves the kernel classes used
in Proposition~\ref{prop:sharp-kinetic-block}: it replaces
$\phi$ by $\dot\phi$, or a dressed kernel $h$ by $\dot h$.
The condensate and its gradient remain bounded, while the
differentiated kernels retain bounded Hilbert--Schmidt, row and
column norms and a first-derivative norm at most $C_T\sqrt N$.
For $\dot p$, these estimates follow from
\cite[Lemma A.3(iv),(vi)]{COS}; the bounds for $\dot\sigma$
follow from $\dot\sigma=\dot\eta+\dot r$ and the same lemma.

We check the transport terms separately. On each number block
write the actual cubic expression as $P_jT_{jn}P_n$, with
the generalized-field and cutoff coefficients collected into
sector scalars. Since $\nabla_tP=0$, its covariant derivative is
\[
	P_j\bigl(\dot T_{jn}-Z^{(j)}T_{jn}+T_{jn}Z^{(n)}\bigr)P_n.
\]
In the Wick expression, the last two terms form the negative
commutator with $\dGamma(Z)$. The spectator contributions cancel, leaving only the three active field
operators. Moreover, $Z=[\dot q,q]$
is a finite sum of rank-one maps $|u\rangle\langle v|$, with
$u,v$ bounded in $H^4$. We can therefore estimate the transport terms by applying one such map to
each active field operator.
In addition to \eqref{eq:gradient-small-norms}, we use
\[
	\norm\kappa_1\le C N^{-3/2},\qquad \norm{\nabla\kappa}_1\le C N^{-1}.
\]
The first follows by integrating the pointwise bound for
$\kappa$ over its support of radius $N^{-1/2}$.

We begin with $T_0$. Acting on its annihilation operator gives
a term in the family $T_h^-$ already considered, with a smooth
rank-one $h$. It therefore remains to examine the two creation
operators. The contribution at $y$ is $a^*(u)\dGamma(k)$,
where $k$ is multiplication by
\[
	k(x)=\int\kappa(x-y)\phi(y)\overline{v(y)}\d y.
\]
The convolution bounds give $\norm k_\infty\le C_TN^{-3/2}$
and $\norm{\nabla k}_\infty\le C_TN^{-1}$.
For the $x$ contribution, the multiplication operator is replaced
by an integral operator: the expression is $a^*(u)\dGamma(B)$, with
$B(y,x)=\kappa(x-y)\phi(y)\overline{v(x)}$.
Young's inequality, also after differentiating either
output variable, gives
\[
	\norm B_{\mathrm{op}}\le C_TN^{-3/2},\qquad \norm{\nabla B}_{\mathrm{op}}+\norm{\nabla B^*}_{\mathrm{op}} \le C_TN^{-1}.
\]
Here the bounded first derivatives of $\phi,u,v$ suffice.
For either of these two types, the annihilator commutator
with $\dGamma(B)$, followed by the one with $a^*(u)$, gives
\begin{align*}
	\norm{K^{1/2}a^*(u)\dGamma(B)\psi_n}
	\le C_T\bigl(&N^{-3/2}(n+1)^{3/2}
	\norm{K^{1/2}\psi_n}\\
	&+N^{-1}(n+1)\norm{\psi_n}\bigr).
\end{align*}
For the multiplication operator, the two terms are estimated
with $\nabla k$ and with $k$ acting on the input derivative,
respectively. The same bound holds for the adjoint
$\dGamma(B^*)a(u)$ because $a(u)$ commutes with $\nabla_z a_z$
and the output derivative of $B^*$ obeys the displayed estimate.
The $L^2$ block bounds then supply the number-energy terms.
These estimates imply \eqref{eq:sharp-kinetic-block}, since
$N^{-3/2}\le\epsilon$ and
$N^{-1}(n+1)\le C\sqrt{n+1}$ on the active sectors.

For $T_h^\pm$, we first consider the field containing the kernel
$h$. Applying $|u\rangle\langle v|$ or its adjoint to this
field composes the two kernels and preserves the row, column,
Hilbert--Schmidt and first-derivative bounds.
We next consider the two creation operators. The contribution at
$y$ is $a^*(u)$ times a normally ordered
quadratic operator with kernel $k(x)h(x,z)$. The contribution at
$x$ involves
\[
	B_h(y,z)=\phi(y)\int\kappa(x-y)\overline{v(x)}h(x,z)\d x .
\]
These two kernels have Hilbert--Schmidt norms at most
$C_TN^{-3/2}$, with the sum of their first-derivative norms
bounded by $C_TN^{-1}$. When a derivative falls on $\kappa$,
we use its $L^1$ gradient norm; when it falls on $h$, the contribution is
$\norm\kappa_1\norm{\nabla h}_2\le C_TN^{-1}$.
Applying the usual cubic block estimate to each operator and
its adjoint therefore bounds their kinetic norm by
\[
	C_T(n+1)^{3/2}\bigl( N^{-3/2}\norm{A_n^{1/2}\psi_n} +N^{-1}\norm{\psi_n}\bigr).
\]
Its second term is at most $C_T\sqrt{n+1}\norm{\psi_n}$
because $n\le C\sqrt N$. This covers the transport terms for all active field operators, including
the all-creation family.

The sector scalars are time independent and commute with
$\dGamma(Z)$, so they introduce no further transport terms.
The outer projections are uniformly bounded in the kinetic form
norm. Restoring them proves the block estimate for $X$. Multiplication
by the particle-number shift gives the same estimate for $[M,X]$.
Thus $X$ is uniformly bounded on the space with norm
$\norm{M^rA_P^{1/2}z}$, and we may use the exact formula
\[
	\nabla_t(e^{sC})=\int_0^s e^{(s-a)C}Xe^{aC}\d a
\]
and Proposition~\ref{prop:sharp-kinetic-block} to get
\eqref{eq:sharp-time-graph}, including
$\mathcal L_C=-ie^{-C}\nabla_t(e^C)$.
Iterating the two identical block bounds, with the common
$L^2$ bound $C_T\epsilon(n+1)^{3/2}$, proves the mixed-word
assertion. Extension from the same smooth cores gives the
stated domains.
\end{proof}

\subsection{First-energy propagation}
\label{sec:taylor-first-energy}

The full-energy graph bound for the cubic phase yields first-energy
propagation with a linear number correction.
We prove this bound using the spatial structure of the kernels,
with the cutoffs and generators fixed above.

\begin{proposition}
\label{prop:full-graph}
Let $v\in C_c^\infty(\mathbb R^3)$ be nonnegative and radial.
Let $\phi_t$ solve the modified GP equation with a fixed normalized
initial datum in $H^4(\mathbb R^3)$. Fix a sufficiently small
quadratic scattering radius $\ell\in(0,1/2)$, independent of $N$.
Use the cubic phase \eqref{spec:eq:actual-phase} at this background,
with $m=N^{-1/2}$ and the original COS cutoff $0\le\Theta_N\le1$.
It equals one for $n\le N^{1/2}/2+10$ and zero for
$n\ge N^{1/2}-10$, for sufficiently large $N$.
The block estimate
\eqref{eq:sharp-kinetic-block} holds with $A_P$ replaced by
$Q=K_P+V_P+M$. In particular $C$, $[M,C]$, $e^{sC}$ and
$e^{-C}Me^C-M$ are uniformly bounded on every graph space
with norm $\norm{M^rQ^{1/2}z}$, for fixed $r$ and $|s|\le1$.
If $\phi\in C^1([-T,T];H^4)$ with a bound independent of $N$,
$X=\nabla_tC$ is also uniformly bounded on these spaces.
The same is then true of $\nabla_t(e^{sC})$ and $\mathcal L_C$.

\end{proposition}
\begin{proof}
Only the potential part of the block estimate remains.
We first use canonical fields and put $W(r)=N^2v(Nr)$.
For a cubic monomial $T$ set
$D_x=[a_x,T]$ and $E_{xy}=[a_x,[a_y,T]]$.
Normal ordering gives the exact identity
\begin{equation}\label{eq:pair-comm-identity}
	a_xa_yT=T a_xa_y+D_xa_y+D_ya_x+E_{xy}.
\end{equation}
Use this in
$\norm{V^{1/2}\psi}^2=\frac12\int W(x-y)\norm{a_xa_y\psi}^2\d x\d y$.
The first term in \eqref{eq:pair-comm-identity} is bounded by
$C_T\epsilon(n+1)^{3/2}\norm{V^{1/2}\psi_n}$.

For $T_0$ in Proposition~\ref{prop:sharp-kinetic-block},
\[
	D_x=a^*(f_x)a_x+\phi(x)\dGamma(\kappa(\,\cdot-x)),\qquad E_{xy}=\kappa(x-y)(\phi(x)a_y+\phi(y)a_x).
\]
The creation term in $D_xa_y$ is bounded by
$C_T\epsilon\sqrt{n+1}\norm{V^{1/2}\psi_n}$.
For its density term the sector Cauchy--Schwarz inequality
and Hardy's inequality give
\begin{align}
	&\int W(x-y)|\phi(x)|^2
	\norm{\dGamma(\kappa(\,\cdot-x))a_y\psi_n}^2\d x\d y
	\notag\\
	&\quad\le C_T(n+1)\int W(x-y)|\kappa(z-x)|^2
	\norm{a_za_y\psi_n}^2\d x\d y\d z
	\notag\\
	&\quad\le C_T\frac{(n+1)^2}{N^2}\norm{K^{1/2}\psi_n}^2.
	\label{eq:potential-density-source}
\end{align}
Indeed
\begin{equation}\label{eq:single-W-hardy}
	\int W(s)|\kappa(r-s)|^2\d s\le\frac{C}{N^2|r|^2}.
\end{equation}
Choose a fixed $a>0$ such that $\operatorname{supp}v\subset B_a$.
For $|r|>2a/N$ and $s\in\operatorname{supp}W$, we have
$|r-s|\ge |r|/2$. Thus $\norm W_1=O(N^{-1})$ and
$|\kappa(r-s)|\le C N^{-1/2}/|r-s|$ give the required
$C N^{-2}|r|^{-2}$ bound.
The actual Neumann kernel satisfies
$\|\kappa\|_\infty\le C\sqrt N$, by
\cite[Lemma A.1(ii)]{COS} and
$\kappa(r)=\sqrt Nw^{\rm Neu}_{Nm}(Nr)$.
For $0<|r|\le2a/N$, use
$\norm W_1\norm\kappa_\infty^2\le C$ and
$N^{-2}|r|^{-2}\ge(4a^2)^{-1}$.
This proves \eqref{eq:single-W-hardy} in both regions.
The two-annihilator norm in \eqref{eq:potential-density-source}
then contributes $(n-1)\norm{K^{1/2}\psi_n}^2$ after Hardy.
Since $\int W|\kappa|^2\le C$, the $E_{xy}$ term is bounded by
$C_T\sqrt{n+1}\norm{\psi_n}$.
For $T_0^*$, $D_x$ is the two-annihilator expression
$a(f_x)a_x$, up to conjugation, and $E_{xy}=0$.
The same creation-term estimate, now with an annihilator,
applies directly to this adjoint.

For the dressed families, all terms in $D_x$ have operator
norm at most $C_T\epsilon(n+1)$ on the $n$ sector,
uniformly in $x$. Besides the two quadratic expressions
already used in Section~\ref{sec:kinetic-graph}, the all-creation family has
the pair kernel
$F_x(z,w)=\kappa(z-w)\phi(w)h(z,x)$.
Its Hilbert--Schmidt norm is at most $C_T\epsilon$ by
the column bound for $h$; the other kernels use its row bound.
Consequently
\begin{equation}\label{eq:dressed-potential-D}
	\int W(x-y)\norm{D_xa_y\psi_n}^2\d x\d y \le C_T\frac{\epsilon^2(n+1)^3}{N}\norm{\psi_n}^2.
\end{equation}
Here integration of $W$ in $x$
contributes a factor at most $C/N$.

The double contractions of $T_h^+$ give the creation
kernel $e_{xy}(z)$, where
\begin{align*}
	e_{xy}(z)={}&\kappa(x-y)\bigl(\phi(y)h(x,z)+\phi(x)h(y,z)\bigr)\\
	&+\kappa(y-z)\phi(z)h(y,x)+\kappa(x-z)\phi(z)h(x,y)\\
	&+\kappa(z-y)\phi(y)h(z,x)+\kappa(z-x)\phi(x)h(z,y).
\end{align*}
The first line has squared norm integrated against $W(x-y)$
bounded uniformly by $\int W|\kappa|^2$ and the row bound.
For the other lines use
$|h(x,y)|\le C_TN|\phi(x)\phi(y)|$, a consequence of
\cite[Lemma A.3(ii)]{COS}. Their corresponding squared
norms are at most $C_TN\epsilon^2\le C_T$.
For example, integrate $W(x-y)|\phi(x)|^2$ first in the
last line, then $|\kappa(z-y)|^2$; both integrations use
bounded condensate factors.
Thus $E_{xy}=a^*(e_{xy})$ contributes at most $C_T\sqrt{n+1}\norm{\psi_n}$ to
the potential norm.
The annihilation choices simplify this calculation. For $T_h^-$,
only the first line survives, with annihilation on the remaining
dressed field. Passing to $(T_h^-)^*$ makes $D_x$ an $F_x$-type
pair annihilator and gives $E_{xy}=0$; for $(T_h^+)^*$,
$D_x$ also vanishes. These observations cover all adjoints.

The same estimates hold if the amplitude $\phi$ is replaced
by a function bounded in $L^2\cap L^\infty$, and the pointwise
bound on $h$ is replaced by a finite sum
\begin{equation}\label{eq:profile-envelope}
	|h(x,y)|\le C_TN\sum_{\ell=1}^J |u_\ell(x)v_\ell(y)|, \qquad \norm{u_\ell}_{2\cap\infty} +\norm{v_\ell}_{2\cap\infty}\le C_T.
\end{equation}
The row and column bounds are still assumed. The profiles
may depend on $N$, subject to the displayed uniform bounds.
This observation will also cover the time derivatives.
To verify the double contractions for this extension,
replace $\phi$ by $a\in L^2\cap L^\infty$ and put
$R(h)=\sup_x\|h(x,\cdot)\|_2$ and
$\mathcal P_h=\sum_{\ell=1}^J\|u_\ell\|_2^2\|v_\ell\|_\infty^2$.
The six terms reduce, by interchanging $x,y$, to three
integrals. Their bounds are
\begin{align}
	&\int W(x-y)|\kappa(x-y)a(y)|^2\|h(x,\cdot)\|_2^2\d x\d y
	\le C\|a\|_2^2R(h)^2,\notag\\
	&\int W(x-y)|\kappa(y-z)a(z)h(y,x)|^2\d x\d y\d z
	\le CN\epsilon^2\|a\|_\infty^2\mathcal P_h,\notag\\
	&\int W(x-y)|\kappa(z-y)a(y)h(z,x)|^2\d x\d y\d z
	\le CN\epsilon^2\|a\|_\infty^2\mathcal P_h .
	\label{eq:profile-double-contractions}
\end{align}
The first line uses $\int W|\kappa|^2\le C$ and
then integrates $|a(y)|^2$. For the other two lines,
square \eqref{eq:profile-envelope} and sum its $J$
terms. Integration in $x$ gives
$\int W(x-y)|v_\ell(x)|^2\d x\le CN^{-1}\|v_\ell\|_\infty^2$.
The remaining relative integration contributes
$\|\kappa\|_2^2\le C\epsilon^2$, and the last one
gives $\|u_\ell\|_2^2$. The fixed profile count is
absorbed in $C$. Since $N\epsilon^2=N^{-1/2}\le1$,
all six kernels have a uniformly bounded integrated
squared norm. The associated creation or annihilation operators
therefore contribute at most $C\sqrt{n+1}$.

On $n\le C\sqrt N$, both $(n+1)/N$ in
\eqref{eq:potential-density-source} and
$\epsilon\sqrt{n+1}$ are bounded by
$C\epsilon(n+1)^{3/2}$. The square root of
\eqref{eq:dressed-potential-D} is at most $C\sqrt{n+1}$.
Combining all terms proves the claimed canonical $Q$ block.

We restore the projections explicitly. The pair-annihilation
map obeys $a_xa_yP=P_{\rm rest}(q\otimes q)a_xa_y$.
Expand just these two active $q$'s. Each rank-one term is
controlled by
\[
	\sup_y\int W(x-y)|\phi(x)|^2\d x\le C_T/N.
\]
The usual annihilation bounds therefore give
\begin{equation}\label{eq:V-projection}
	\norm{V^{1/2}P\psi_n} \le\norm{V^{1/2}\psi_n} +C_T\frac{n+1}{\sqrt N}\norm{\psi_n}.
\end{equation}
On the physical space $n\le N$, this and the kinetic
projection estimate show $\norm{Q^{1/2}P\psi}\le C_T\norm{Q^{1/2}\psi}$.
All generalized-field and cutoff factors are collected
separately on each original block before this calculation;
they are scalar contractions. This proves the full-energy
block estimate for the actual $C$.
The strong cutoff condition and finite bandwidth prove the
claimed graph bounds for $C$, $[M,C]$ and their exponentials.

To treat $X$, we separate ordinary coefficient derivatives from
the rank-one transport terms. The former satisfy the same
estimates: $\dot\phi$ is bounded in $H^4$, and $\dot h$ has
the required row, column and profile bounds by
\cite[Lemma A.3(v),(vi)]{COS}.
For the transport terms listed in
Proposition~\ref{prop:sharp-time-graph}, acting on the annihilation or
dressed field preserves \eqref{eq:profile-envelope}.
For example, $h|u\rangle\langle v|$ has profile $(hu)(x)v(y)$,
and the Hilbert--Schmidt and row bounds place $hu$ uniformly
in $L^2\cap L^\infty$. The remaining field operators are controlled by the more precise kinetic
estimates proved there:
\[
	C_TN^{-3/2}(n+1)^{3/2}A_n^{1/2} +C_TN^{-1}(n+1)^{3/2}
\]
as block estimates; the identity family's zero-order term
even has only $n+1$. Use $Q\le CMA$ to multiply this
bound by $C\sqrt{n+1}$.
Relative to $Q_n^{1/2}\ge A_n^{1/2},\sqrt{n+1}$ the
two coefficients are at most
$C_TN^{-3/2}(n+1)^2$ and $C_TN^{-1}(n+1)^{3/2}$,
both uniformly bounded on the active sectors.
Restore the outer projections with \eqref{eq:V-projection}.
Thus $X$ is uniformly $Q$-graph bounded. Finite bandwidth
gives every fixed weight $M^r$. The exact time-integral
formula for the phase gives its last two asserted bounds.
\end{proof}

\begin{lemma}
\label{lem:inf-HC}
Let $H=K_P+V_P=Q-M$. For every $\delta>0$,
\begin{equation}\label{eq:inf-HC}
	|[H,C][z,z]|\le\delta Q[z]+C_{\delta,T}M[z].
\end{equation}
The constants are independent of $N$ on compact time intervals.
\end{lemma}
\begin{proof}
We first omit the excitation projections, but retain the
actual number coefficients. Write the anti-self-adjoint
canonical expression as a sum of a cubic monomial with
its sector coefficient and the negative adjoint.
On the active sectors those coefficients $b_N(n)$ satisfy
\begin{equation}\label{eq:first-cutoff-difference}
	|b_N(n)|\le C,\qquad |b_N(n+d)-b_N(n)|\le C_d N^{-1/2}\quad (|d|\le3).
\end{equation}
Indeed the original piecewise linear $\Theta$ has slope
$O(N^{-1/2})$, including across its corners. Products
of at most three generalized-field square roots have
increments $O(N^{-1})$ on $n\le C\sqrt N$.
The finitely shifted coefficients obey the same bounds.
The argument below uses these first differences.

Denote this unprojected anti-self-adjoint operator by $\widehat C$.
For a canonical monomial with no sector coefficient, the
proof of Proposition~\ref{prop:sharp-kinetic-block} gives
the following stronger estimate for its differentiated
annihilator commutator alone:
\begin{equation}\label{eq:kinetic-source-only}
	\left(\int\norm{\nabla_x[a_x,T]\psi_n}^2\d x\right)^{1/2} \le C_T\bigl(\epsilon\sqrt{n+1}\norm{K^{1/2}\psi_n} +\sqrt{n+1}\norm{\psi_n}\bigr).
\end{equation}
The term $T\nabla_xa_x\psi_n$ is excluded here.
For the identity family, the input-derivative terms and
the normal-ordered convolution $|g|*|g|$ give the first
summand. The diagonal contraction gives the second.
For the dressed families, \eqref{eq:dressed-quadratic-sources}
and $\epsilon\sqrt N(n+1)\le C\sqrt{n+1}$ give the second;
all adjoints were checked separately there.

In $\nabla_xa_x\widehat C-\widehat C\nabla_xa_x$,
commuting a sector coefficient adds its first difference
times $T\nabla_xa_x$. By \eqref{eq:first-cutoff-difference}
its norm is at most
$C_TN^{-1/2}\epsilon(n+1)^{3/2}\norm{K^{1/2}\psi_n}$.
Consequently, after summing the finite number shifts,
\begin{equation}\label{eq:full-kinetic-source}
	\norm{\nabla a\widehat C z-\widehat C\nabla a z}_{L^2} \le C_TN^{-1/2}\norm{K^{1/2}z} +C_T\norm{M^{1/2}z}.
\end{equation}
The left side uses the natural direct integral over the
removed coordinate and remaining Fock space.

Likewise, \eqref{eq:pair-comm-identity} and its estimates
give, without the old-particle term $T a_xa_y$,
\begin{equation}\label{eq:full-potential-source}
	\left(\int W(x-y)\norm{a_xa_y\widehat C z -\widehat C a_xa_yz}^2\d x\d y\right)^{1/2} \le C_TN^{-1/2}\norm{Q^{1/2}z} +C_T\norm{M^{1/2}z}.
\end{equation}
The identity creation term is bounded by $\epsilon\sqrt{n+1}\norm{V^{1/2}\psi_n}$, and the
density term by $(n+1)\norm{K^{1/2}\psi_n}/N$. The dressed $D$
terms and all $E$ terms are bounded by
$C_T\sqrt{n+1}\norm{\psi_n}$.
Shifting a number coefficient across two annihilators
adds $C_TN^{-1/2}\epsilon(n+1)^{3/2}\norm{V^{1/2}\psi_n}$.
These are exactly the two terms on the right of
\eqref{eq:full-potential-source} on the standard cutoff.

When evaluating the real part of $H[z,\widehat C z]$,
the old-particle terms in these two identities vanish:
$\widehat C$ is anti-self-adjoint on the remaining Fock
space for each fixed removed coordinate or pair.
Cauchy--Schwarz therefore proves the canonical bound
\begin{equation}\label{eq:canonical-HC}
	|[K+V,\widehat C][z,z]| \le C_TN^{-1/2}Q[z] +C_T\norm{Q^{1/2}z}\norm{M^{1/2}z}.
\end{equation}

We must still check the compressed form. For $Pz=z$ put
$w=\widehat C z$, so $Cz=Pw$.
The remaining-particle projections can be moved to the
left in $K[z,Pw]$; the removed coordinate gives the error
\[
	-\int\ip{\nabla_xa_xz}{\nabla\phi(x)a(\phi)w}\d x.
\]
Its absolute value is at most
$C_T\norm{K^{1/2}z}\norm{M^{1/2}w}$.
For the potential pairing, move the remaining projections
in the same way and expand $q\otimes q$ only on the two removed coordinates. The argument for \eqref{eq:V-projection}
bounds this pairing error by
$C_T\norm{V^{1/2}z}\norm{N^{-1/2}Mw}$.
The number bounds for $\widehat C$ under the strong cutoff condition, together with
$M\le N+1$ on its support, bound both errors by
$C_T\norm{Q^{1/2}z}\norm{M^{1/2}z}$.
This proves \eqref{eq:canonical-HC} also for $[H,C]$,
with $Q$ restricted to the excitation space.

For $N$ sufficiently large in terms of $\delta$, absorb
the first term of \eqref{eq:canonical-HC}; Young's inequality
bounds the second. The finitely many remaining $N$ have
the same infinitesimal form bound with a larger constant:
use \eqref{eq:KC-op} on their finite number spaces and the
bounded potential commutator. This proves \eqref{eq:inf-HC}.
The argument first applies to smooth finite-sector cores;
the displayed form bounds give its closed-form extension.
\end{proof}

\begin{theorem}
\label{thm:Taylor-first-energy}
Let $v\in C_c^\infty(\mathbb R^3)$ be nonnegative and radial.
Let $\phi_t$ solve the modified GP equation with a fixed normalized
initial datum in $H^4(\mathbb R^3)$. Fix a sufficiently small
quadratic scattering radius $\ell\in(0,1/2)$, independent of $N$.
Use the cubic phase \eqref{spec:eq:actual-phase} at this background,
with $m=N^{-1/2}$ and the original COS cutoff $0\le\Theta_N\le1$.
It equals one for $n\le N^{1/2}/2+10$ and zero for
$n\ge N^{1/2}-10$, for sufficiently large $N$.
Assume that $\phi\in C^1([-T,T];H^4)\cap C^2([-T,T];H^2)$,
with both norms bounded independently of $N$.
There is $b_T>0$, independent of $N$, such that
\begin{equation}\label{eq:penalty-free-comparison}
	F_t^{[3]}=\mathfrak J_t^{[3]}+b_TM, \qquad c_TQ_{N,P_t}\le F_t^{[3]}\le C_TQ_{N,P_t}.
\end{equation}
The propagator constructed in Theorem~\ref{thm:selfadjoint}
therefore satisfies
\begin{equation}\label{eq:Taylor-first-energy}
	Q_{N,P_t}[U_N^{[3]}(t,s)z]\le C_TQ_{N,P_s}[z], \qquad |s|,|t|\le T.
\end{equation}

\end{theorem}
\begin{proof}
We use a further estimate from the literature, with the scalar convention
fixed in Section~\ref{sec:number-compatibility}. The proof of
\cite[Theorem 4.4, Eqs.~(5.130)--(5.131), pp.~80--81]{BS}
gives, after compressing to $P\cF^{\le N}$,
\begin{equation}\label{eq:BS-inf-S}
	\mathfrak G=H+S,\qquad |S[z,z]|\le C_T\norm{Q^{1/2}z}\norm{M^{1/2}z}.
\end{equation}
Its remaining Neumann-eigenvalue pair term is bounded by
$C_TM$ and is included in this inequality because $Q\ge M$.

Infinitesimal bounds extend to mixed form pairings as follows.
If a Hermitian form $B$ satisfies
$|B[z,z]|\le\delta Q[z]+C_\delta M[z]$ for every $\delta>0$,
then it is a contraction form relative to
$D_\delta=\delta Q+C_\delta M$:
\[
	|B[u,v]|\le\norm{D_\delta^{1/2}u}\norm{D_\delta^{1/2}v}.
\]
For fixed words $L,R$ in $C$, Proposition~\ref{prop:full-graph}
and the number graph bounds imply
\[
	|B[Lz,Rz]|\le \delta C_{L,R,T}Q[z] +C'_{\delta,L,R,T}M[z].
\]
Rescale $\delta$. Thus every fixed finite sum of such
pairings has the same infinitesimal relative form property.

Expand the Taylor polynomial as
\[
	\mathfrak J^{[3]} =H+\sum_{j=1}^3\frac{\ad_C^{j-1}[H,C]}{j!} +\sum_{j=0}^3\frac{\ad_C^jS}{j!}-iX.
\]
Lemma~\ref{lem:inf-HC} and \eqref{eq:BS-inf-S} show that
all displayed finite-word forms other than $H$ are
infinitesimally $Q$-bounded with an $M$ remainder.
Also $X$ is uniformly bounded in $L^2$ by its block bound
and the strong cutoff condition. It follows that, for every
$\delta>0$,
\[
	|(\mathfrak J^{[3]}-H)[z,z]| \le\delta Q[z]+C_{\delta,T}M[z].
\]
Choose $\delta<1/2$ and then $b_T$ large enough. This proves
\eqref{eq:penalty-free-comparison} on the common form domain.
At fixed $N$ it is the closed form of the self-adjoint
Taylor generator plus the bounded operator $b_TM$.

For the derivative, Lemma~\ref{lem:moving-G} supplies
the $Q$-bounded forms $\nabla_t\mathfrak G$ and
$i[\mathfrak G,M]$. Differentiate the finite Taylor
polynomial using the derivation rule. Each resulting form contains one of these derivatives or
$\mathfrak G$, with a finite product of $C$ and
$X$ on either side.
For the number commutator, replace one occurrence of $C$
by $[M,C]$ instead. All these words are uniformly bounded
in the $Q$ graph norm by Proposition~\ref{prop:full-graph}.
The derivative of $-iX$ is bounded in $L^2$: the block
bound for $\nabla_tX$ was proved in
Proposition~\ref{prop:partial-derivative}. Its number
commutator is bounded in $L^2$ as well, by finite bandwidth.
Consequently
\begin{equation}\label{eq:Taylor-first-source}
	|(\nabla_t\mathfrak J^{[3]})[z,z]| +|i[\mathfrak J^{[3]},M][z,z]| \le C_TQ[z].
\end{equation}

Along the same covariant Taylor equation,
\[
	\frac{\d}{\d t}F_t^{[3]}[y(t),y(t)] =\bigl(\nabla_t\mathfrak J_t^{[3]} +i b_T[\mathfrak J_t^{[3]},M]\bigr)[y(t),y(t)].
\]
Equations \eqref{eq:penalty-free-comparison} and
\eqref{eq:Taylor-first-source} now give Gronwall's inequality.
First use common operator-domain data for the fixed-$N$
propagator, or bounded form regularizations, to justify
the product rule. The estimate is independent of those
regularizations and of $N$. Density in the first-energy
space and lower semicontinuity extend it to all form data.
Reverse time for $t<s$. This proves \eqref{eq:Taylor-first-energy}.
\end{proof}

Theorem~\ref{thm:Taylor-first-energy} proves the first-energy
bound using a linear number correction.

\subsection{Comparison on initial data in the energy form domain}
\label{sec:first-energy-comparison}

The full-energy block estimate reduces the number weight in
the asymmetric remainder. Together with first-energy propagation
for both evolutions, it gives a comparison by localizing their
overlap in particle number.

\begin{proposition}
\label{prop:smaller-asymmetric}
Let $v\in C_c^\infty(\mathbb R^3)$ be nonnegative and radial.
Let $\phi_t$ solve the modified GP equation with a fixed normalized
initial datum in $H^4(\mathbb R^3)$. Fix a sufficiently small
quadratic scattering radius $\ell\in(0,1/2)$, independent of $N$.
Use the cubic phase \eqref{spec:eq:actual-phase} at this background,
with $m=N^{-1/2}$ and the original COS cutoff $0\le\Theta_N\le1$.
It equals one for $n\le N^{1/2}/2+10$ and zero for
$n\ge N^{1/2}-10$, for sufficiently large $N$.
The order-$s$ conjugation remainder of $\mathfrak G$, for each
fixed integer $s\ge2$, satisfies
\begin{equation}\label{eq:smaller-asymmetric}
	|R_{s,N}[u,v]|\le C_{s,T}\epsilon^{s-2} \norm{Q^{1/2}u} \norm{M^{3(s-2)/2}Q^{1/2}v},\qquad \epsilon=N^{-3/4}.
\end{equation}
In particular the complete generator difference obeys
\begin{equation}\label{eq:smaller-full}
	|(\mathfrak J-\mathfrak J^{[3]})[u,v]| \le C_TN^{-3/2}\norm{Q^{1/2}u}\norm{M^3Q^{1/2}v}.
\end{equation}
All forms are understood on the displayed domains and
then by their bounded extensions. Here $\epsilon=N^{-3/4}$ and
the active particle-number range $n\le\Lambda_N=O(\sqrt N)$
satisfies $\epsilon(\Lambda_N+1)^{3/2}\le D$ uniformly in $N$.
\end{proposition}
\begin{proof}
Iterating the full-energy block in Proposition~\ref{prop:full-graph}
as in Corollary~\ref{cor:sharp-kinetic-words} gives, for
$j\ge1$ and fixed real $r$,
\begin{align*}
	\norm{M^rQ^{1/2}C^jz}
	\le C_{j,r,T}\bigl(&\epsilon^j\norm{M^{r+3j/2}Q^{1/2}E_Nz}\\
	&+\epsilon^{j-1}\norm{M^{r+3j/2-1}E_Nz}\bigr).
\end{align*}
Here $E_N$ contains all active input and output sectors.
Since $Q\ge M$ and $\epsilon M^{3/2}E_N$ is uniformly
bounded, both terms imply the sharper word bound
\begin{equation}\label{eq:smaller-Q-word}
	\norm{M^rQ^{1/2}C^jz} \le C_{j,r,T}\epsilon^{j-1} \norm{M^{r+3(j-1)/2}Q^{1/2}E_Nz}.
\end{equation}

Expand $\ad_C^s\mathfrak G$ as the finite sum of forms
$\mathfrak G[C^ju,C^{s-j}v]$ with binomial coefficients,
using $C^*=-C$. When $1\le j\le s-1$, put
$\alpha=3(j-1)/2$. Insert $M^{-\alpha}$ on the left
energy factor and $M^\alpha$ on the right, using
the number compatibility of Proposition~\ref{prop:GP-weight}.
Applying \eqref{eq:smaller-Q-word} to both factors gives
$\epsilon^{s-2}$ and total right weight
$3(j-1)/2+3(s-j-1)/2=3(s-2)/2$.

For the two endpoint terms apply
\eqref{eq:smaller-Q-word} with $j=s$ and absorb
one additional $\epsilon M^{3/2}E_N$ on the input of
the word $C^s$. This gives
\[
	\norm{M^rQ^{1/2}C^sz} \le C_{s,r,T}\epsilon^{s-2} \norm{M^{r+3(s-2)/2}Q^{1/2}z}.
\]
For the left endpoint use $r=-3(s-2)/2$ and number
compatibility with that weight; for the right endpoint
use $r=0$. These choices place each number weight on its corresponding
argument of $\mathfrak G$.
The weighted full-energy bounds for the two exterior
phases in the integral remainder prove
\eqref{eq:smaller-asymmetric}.

For the logarithmic time remainder, Lemma~\ref{lem:log} gives
\[
	|R_{\rm time}[u,v]| \le C_T\epsilon^2\norm u\norm{M^3v} \le C_T\epsilon^2\norm{Q^{1/2}u}\norm{M^3Q^{1/2}v}.
\]
It is controlled by the same weights as the $s=4$ bound. This proves
\eqref{eq:smaller-full} for the complete covariant difference.
\end{proof}

\begin{theorem}
\label{thm:energy-class-comparison}
Let $v\in C_c^\infty(\mathbb R^3)$ be nonnegative and radial.
Let $\phi_t$ solve the modified GP equation with a fixed normalized
initial datum in $H^4(\mathbb R^3)$. Fix a sufficiently small
quadratic scattering radius $\ell\in(0,1/2)$, independent of $N$.
Use the cubic phase \eqref{spec:eq:actual-phase} at this background,
with $m=N^{-1/2}$ and the original COS cutoff $0\le\Theta_N\le1$.
It equals one for $n\le N^{1/2}/2+10$ and zero for
$n\ge N^{1/2}-10$, for sufficiently large $N$.
Assume that $\phi\in C^1([-T,T];H^4)\cap C^2([-T,T];H^2)$,
with both norms bounded independently of $N$.
Let $U_N(t,s)$ be the exact covariant propagator generated
by $\mathfrak J_t$, with the scalar convention of Section~\ref{sec:covariant-generator}.
For every $z\in D(Q_{N,P_s}^{1/2})$ and $|s|,|t|\le T$,
\begin{equation}\label{eq:energy-class-comparison}
	\norm{(U_N(t,s)-U_N^{[3]}(t,s))z} \le C_TN^{-3/16}\norm{Q_{N,P_s}^{1/2}z}.
\end{equation}
Thus the estimate is uniform also for $N$-dependent
initial states with bounded first energy.
\end{theorem}
\begin{proof}
Choose a real $\chi\in C_c^\infty(\mathbb R)$ with
$0\le\chi\le1$, $\chi=1$ on $[-1,1]$ and
$\chi=0$ outside $[-2,2]$, and set $\chi_L=\chi(M/L)$,
where $L\ge1$. This is an observable used in the proof;
the original cutoff in $C$ is unchanged.

We first estimate its boundary commutator as a form.
Put $B_t=Q_t^{-1/2}\mathfrak J_t^{[3]}Q_t^{-1/2}$.
Theorem~\ref{thm:Taylor-first-energy} gives uniform bounds
for $B_t$ and for
$[M,B_t]=Q_t^{-1/2}[M,\mathfrak J_t^{[3]}]Q_t^{-1/2}$.
Here $M$ strongly commutes with $Q_t$.
Fourier inversion and
\[
	[e^{iaM},B_t] =i\int_0^a e^{i(a-b)M}[M,B_t]e^{ibM}\d b
\]
imply
\begin{equation}\label{eq:localized-boundary}
	|[\mathfrak J_t^{[3]},\chi_L][u,v]| \le \frac{C_T}{L}\norm{Q_t^{1/2}u}\norm{Q_t^{1/2}v}.
\end{equation}
Indeed $\int |a|\,|\widehat\chi(a)|\d a<\infty$.
The estimate uses the bounded form commutator.

Let $x(r)=U_N(r,s)z$ and $y(r)=U_N^{[3]}(r,s)z$.
The exact first-energy bound from \cite[Proposition 6.1]{BS}
and the phase comparison in \cite{COS}, together with
Theorem~\ref{thm:Taylor-first-energy}, give
$Q_r[x(r)]+Q_r[y(r)]\le C_T Q_s[z]$.
Since $\nabla_r\chi_L=0$, the weak product rule yields
\begin{equation}\label{eq:localized-overlap}
	\frac{\d}{\d r}\ip{x(r)}{\chi_Ly(r)} =iR_r[x(r),\chi_Ly(r)] +i[\mathfrak J_r^{[3]},\chi_L][x(r),y(r)].
\end{equation}
The two terms have bounds
\begin{align*}
	|R_r[x(r),\chi_Ly(r)]|&\le C_TN^{-3/2}L^3Q_s[z],\\
	|[\mathfrak J_r^{[3]},\chi_L][x(r),y(r)]|
	&\le C_TL^{-1}Q_s[z],
\end{align*}
by \eqref{eq:smaller-full} and \eqref{eq:localized-boundary}.
The product rule is valid for form solutions after unitary
transport to a fixed space. Indeed $\chi_L$ preserves the
common form domain, both generators map that domain to its
dual, and both pairings above are integrable.

For the initial and final tails, functional calculus gives
$0\le1-\chi_L\le M/L$. Hence
\begin{align*}
	\norm z^2-\ip z{\chi_Lz}&\le L^{-1}M[z],\\
	|\ip{x(r)}{(1-\chi_L)y(r)}|
	&\le L^{-1}\norm{M^{1/2}x(r)}\norm{M^{1/2}y(r)}.
\end{align*}
Both tail factors are used in the second estimate;
estimating just one would lose a square root of $L$.
Integrate \eqref{eq:localized-overlap} and use unitarity
to conclude
\[
	\norm{x(t)-y(t)}^2 \le C_T\bigl(L^{-1}+N^{-3/2}L^3\bigr)Q_s[z].
\]
Taking $L=N^{3/8}$ and then the square root proves
\eqref{eq:energy-class-comparison}. Reverse the time
integration for $t<s$.
\end{proof}

The localization argument retains the boundary term from the
localized Duhamel equation. First-energy bounds for both evolutions
control this term in the overlap estimate.

\begin{corollary}
\label{cor:single-kinetic-norm}
Let $v\in C_c^\infty(\mathbb R^3)$ be nonnegative and radial.
Let $\phi_t$ solve the modified GP equation with a fixed normalized
initial datum in $H^4(\mathbb R^3)$. Fix a sufficiently small
quadratic scattering radius $\ell\in(0,1/2)$, independent of $N$.
Use the cubic phase \eqref{spec:eq:actual-phase} at this background,
with $m=N^{-1/2}$ and the original COS cutoff $0\le\Theta_N\le1$.
It equals one for $n\le N^{1/2}/2+10$ and zero for
$n\ge N^{1/2}-10$, for sufficiently large $N$.
Assume that $\phi\in C^1([-T,T];H^4)\cap C^2([-T,T];H^2)$,
with both norms bounded independently of $N$.
Each of
$T=R,\nabla_tR,i[M,R]$ satisfies, for fixed real $r$,
\begin{equation}\label{eq:single-kinetic-norm}
	\norm{M^rTz}\le C_{r,T}N^{-5/4} \norm{M^{r+6}A_P^{1/2}z}.
\end{equation}
Consequently, for the same two evolutions with a common
initial vector at time zero,
\begin{equation}\label{eq:single-kinetic-dynamic}
	\norm{x_N(t)-y_N(t)} \le C_TN^{-5/4}\int_0^t \norm{M^6A_{P_r}^{1/2}y_N(r)}\d r, \qquad 0\le t\le T,
\end{equation}
whenever the displayed norm is integrable. Reverse the
integration direction for negative times. Uniform propagation
of $M^{12}A_P$ is a sufficient condition for this rate.
\end{corollary}
\begin{proof}
We retain the separate parts of the proof of
Proposition~\ref{prop:hardy-remainder} and localize them individually.
Write $\mathfrak G=K_P+V_P+S$, and choose
$E_N=\mathbf1_{M\le\Lambda_N}$ containing every active
sector of $C$, with $\Lambda_N\le C\sqrt N$.
The kinetic and potential remainders are supported on
both sides in $E_N$, since $K_P,V_P$ preserve number.
Their separate bounds in that proof are
\begin{align*}
	\norm{M^rR_{K+V}z}\le C_{r,T}\bigl(
	&N^{-5/4}\norm{M^{r+6}E_Nz}
	+N^{-3/2}\norm{M^{r+13/2}E_Nz}\\
	&+N^{-9/4}\norm{M^{r+13/2}A_P^{1/2}E_Nz}\bigr).
\end{align*}
The middle term includes the graph loss from the inner
phase: its original coefficient and weight are
$\epsilon^3M^{r+8}$, which are bounded by
$C\epsilon^2M^{r+13/2}$ on $E_N$.
In the last term use $M^{1/2}E_N\le\Lambda_N^{1/2}$.
It becomes at most
$C_{r,T}N^{-2}\norm{M^{r+6}A_P^{1/2}z}$.
The other two terms obey \eqref{eq:single-kinetic-norm}
because $A_P\ge M\ge1$.

For the endpoint terms of the $S$ remainder, retain
$E_N$ on the adjacent cubic factor.
For $j\ge1$, the cubic $L^2$ block bound gives
\[
	\norm{M^rC^jw} \le C_{j,r,T}\epsilon^j\Lambda_N^{1/2} \norm{M^{r+3j/2-1/2}w}.
\]
Indeed the usual word estimate contains $E_Nw$;
absorb just half a number power there, then drop $E_N$.
For a term $C^jSC^k$ with $j+k=4$ and $j\ge1$, use
this estimate on the left word, followed by
\eqref{eq:S-refined} at weight $r+3j/2-1/2$ and the
ordinary $L^2$ word estimate on the right. The result is
\[
	\norm{M^rC^jSC^kz} \le C_{r,T}N^{3/2}\epsilon^4\Lambda_N^{1/2} \norm{M^{r+13/2}z}.
\]
The case $j=0$ gives the same bound by applying
$\norm{M^rSw}\le C_{r,T}N^{3/2}\norm{M^{r+1}w}$
first and then absorbing half a number power in $C^4z$.
The projections thus remain adjacent to the cubic factors on either side of $S$.
The coefficient is at most $C_TN^{-5/4}$, and the exterior
phases preserve every fixed number norm. Integration therefore
gives the estimate for $R_S$, with the final energy weight
controlled by $M^{r+13/2}\le M^{r+6}A_P^{1/2}$.

The time remainder is bounded by
$C_{r,T}\epsilon^2\norm{M^{r+3}z}$ and is smaller.
For the covariant or number derivative, use the derivative
estimates of Lemmas~\ref{lem:bare-split},
\ref{lem:hardy-VC} and
Proposition~\ref{prop:partial-derivative}.
Each differentiated cubic word has the same $L^2$ bound;
the derivatives of the two exterior phases have the graph
bounds used in Proposition~\ref{prop:hardy-remainder}.
Their potential and kinetic terms retain two-sided support
in $E_N$, while the $S$ terms use the half-power argument
just given, with $S$ or a cubic factor differentiated.
This proves \eqref{eq:single-kinetic-norm} for all three
operators on the displayed domain by closure.

Finally use the weak Duhamel justification from
Section~\ref{sec:conditional-comparisons} with \eqref{eq:single-kinetic-norm} at $r=0$.
This proves \eqref{eq:single-kinetic-dynamic} under the stated
time-integrability assumption.
\end{proof}

\section{Time regularity and inverse-energy comparison}
\label{sec:propagation-comparison}

We turn to propagation on a stronger energy domain. The required
time regularity is verified on a common form space, with constants
uniform in particle number. This gives the domain control used both
in the inverse-energy comparison below and in the exact moment
bound of Section~\ref{sec:exact-moment-stability}.
The model input is Proposition~\ref{prop:full-second-form}; the
propagation estimate is Lemma~\ref{lem:three-halves-propagation}.

\subsection{Time regularity and propagation in the form space}
\label{sec:second-regularity}

We establish the domain propagation needed for integration by
parts in time and for the exact higher-moment bound. The common
normalized initial background belongs to $H^6(\mathbb R^3)$.
The radius $\ell$ in $B(\eta)$ is fixed as in Section~\ref{sec:number-compatibility}.
Only the radius in the cubic phase is $m=N^{-1/2}$.
The scalar subtracted from $\mathfrak G$ remains exactly
\cite[Eq.~(4.25)]{BS}. Write $\nabla=\nabla_t$ and
$Q=Q_{N,P_t}$. All constants below are uniform in $N$ on a
fixed compact time interval $I$.

\subsubsection{The common form space}

Let $\mathcal W(t,s)$ be the second quantization of the
one-particle parallel transport generated by $Z=[\dot q,q]$.
It intertwines $P_s$ and $P_t$ and commutes with $M$.
On the $n$-particle sector, differentiation of the compressed
energy gives
\begin{align}
	\nabla Q_{n,P}&=P_n[Q_n,\mathscr Z]P_n,\notag\\
	\nabla^2 Q_{n,P}
	&=P_n\bigl([Q_n,\dot{\mathscr Z}]
	+[[Q_n,\mathscr Z],\mathscr Z]\bigr)P_n .
	\label{eq:Q-two-derivatives}
\end{align}
Both forms are bounded by $C Q_{n,P}$, uniformly for $n\le N$.
To verify this bound on the moving excitation space, write
the kinetic commutators as sums
of bounded finite-rank one-particle operators of norm at most
$C n$: the vectors in $Z,\dot Z$ belong to $H^2$.
For a pair potential $W_{ij}=N^2v(N(x_i-x_j))$, only
$B_{ij}=Z_i+Z_j$ acts on its active coordinates. Young's
inequality and $\|W_N\|_1=O(N^{-1})$ give
\[
	\norm{W_{ij}^{1/2}B_{ij}u} +\norm{W_{ij}^{1/2}B_{ij}^2u} +\norm{W_{ij}^{1/2}\dot B_{ij}u} \le C N^{-1/2}\norm{u}.
\]
For example a factor $|f\rangle\langle g|$ on coordinate $i$
uses $\sup_{x_j}\int W_N(x_i-x_j)|f(x_i)|^2\d x_i
\le C N^{-1}$ and $\|g\|_2\le C$. Products of two such
factors have the same bound. Expanding the two commutators
and applying Cauchy--Schwarz to the pair sum therefore yields
\[
	C nN^{-1/2}\norm{\cV_N^{1/2}u}\norm{u} +C n^2N^{-1}\norm{u}^2\le C Q_n[u].
\]
Indeed, Young's inequality first bounds the mixed term by
$C\cV_N[u]+Cn^2N^{-1}\|u\|^2$, and $n^2/N\le n$
on the full physical space $0\le n\le N$.
The same inequality holds for excitation vectors by
evaluating the ambient form on $u=P_nu$.
The coefficients are continuous in these form estimates.

In parallel coordinates, the first bound in
\eqref{eq:Q-two-derivatives} and Gronwall's inequality give
uniform equivalence with $Q_{N,P_s}$. Hence
\[
	V_N=D(Q_{N,P_s}^{1/2})\subset\mathcal H_N\subset V_N^*
\]
is a fixed Hilbert scale for the rest of the argument.
Form operators and their inverses below act between these
spaces. The same statements hold with any fixed number weight.

\subsubsection{The second graph derivative of the cubic phase}

\begin{lemma}\label{lem:phase-second-regularity}
For each fixed real $r$,
\begin{align}
	\norm{M^rQ^{1/2}(\nabla^j C)z}
	&\le C_{r,I}\norm{M^rQ^{1/2}z},
	&&0\le j\le2,\label{eq:C-two-graph}\\
	|(\nabla^3 C)[u,v]|
	&\le C_I\norm{Q^{1/2}u}\norm{Q^{1/2}v}.
	\label{eq:C-three-form}
\end{align}
The corresponding graph and form families are continuous
after parallel transport. The second line also holds with
fixed weights on the two arguments.
\end{lemma}
\begin{proof}
The modified GP equation is \cite[Eq.~(4.8)]{BS}.
It has the time-independent convolution
potential $v_{\mathrm{eff},N}(x)=N^3v(Nx)f_\ell(Nx)$, with
uniformly bounded $L^1$ norm:
$\|v_{\mathrm{eff},N}\|_1\le\|v\|_1$ because
$0\le f_\ell\le1$. The uniform spatial Sobolev bounds for
the background are given in \cite[Proposition A.2(ii)]{COS}.
Convolution is bounded on each
$H^j$ by this same $L^1$ norm, with derivatives placed
on the other factor. Its
differentiated equation and the algebra property of
$H^2(\mathbb R^3)$ \cite{Cazenave2003,LiebLoss} give, on $I$,
\[
	\phi\in C(H^6)\cap C^1(H^4)\cap C^2(H^2)\cap C^3(L^2).
\]
For the last inclusion, with $\rho=|\phi|^2$ the identity is
\[
	i\phi^{(3)} =-\Delta\ddot\phi+(v_{\mathrm{eff},N}*\ddot\rho)\phi +2(v_{\mathrm{eff},N}*\dot\rho)\dot\phi +(v_{\mathrm{eff},N}*\rho)\ddot\phi .
\]
Every term is continuous in $L^2$, uniformly bounded there.
The potential enters this bound through its $L^1$ norm.
The Hartree nonlinearity is locally Lipschitz on $H^6$,
with constants controlled by $\|v_{\mathrm{eff},N}\|_1$
and the $H^6$ radius. The Duhamel equation gives
continuity in $H^6$, and the equation then bounds
$\dot\phi$ in $H^4$.
Then $\dot\rho=2\operatorname{Re}(\bar\phi\dot\phi)$
belongs to $H^4$, and one differentiated equation
bounds $\ddot\phi$ in $H^2$. The identity
$\ddot\rho=2|\dot\phi|^2+2\operatorname{Re}(\bar\phi\ddot\phi)$
places $\ddot\rho$ in $H^2$.
In the displayed third derivative, $-\Delta\ddot\phi$
is in $L^2$ and all three nonlinear terms are in $H^2$.
The multiplication and convolution bounds are uniform
on $I$, so the same computation gives continuity in
the indicated spaces.

We first lower the background regularity required by the
static graph calculation of Propositions~\ref{prop:sharp-kinetic-block}
and~\ref{prop:full-graph} to $H^2$.
Let $a\in H^2$ be an amplitude and let $\kappa=\kappa_N$
be the relative kernel of the actual cubic phase.
Sobolev embedding and interpolation give
$\|a\|_\infty+\|\nabla a\|_3\le C\|a\|_{H^2}$.
Replace the estimates using $\nabla a\in L^\infty$
by the following bounds:
\begin{align*}
	\int\d z\left\|\int\d y\,\kappa(z-y)\nabla a(y)
	a_ya_z\psi_n\right\|^2
	&\le C\epsilon^2(n-1)\|\nabla a\|_3^2
	\|\cK^{1/2}\psi_n\|^2,\\
	\int\d z\,|\nabla a(z)|^2
	\|\dGamma(\kappa(\cdot-z))\psi_n\|^2
	&\le C\epsilon^2 n\|\nabla a\|_3^2
	\|\cK^{1/2}\psi_n\|^2 .
\end{align*}
For the first line apply Cauchy--Schwarz to $\kappa$ and
then the Hilbert-space-valued Sobolev inequality to
$y\mapsto a_y\psi_n$. For the second line use
$|\sum_{j=1}^n\kappa(x_j-z)|^2\le n\sum_j|\kappa(x_j-z)|^2$.
The resulting multiplier
$|\kappa|^2*|\nabla a|^2$ has $L^{3/2}$ norm at most
$\epsilon^2\|\nabla a\|_3^2$.
For a dressed field $h=p,\sigma$, the uniform row bound gives
\[
	\int|\nabla a(z)|^2 \|\kappa(x-z)h(x,w)\|_{L^2_{x,w}}^2\d z \le C\epsilon^2\|\nabla a\|_2^2 .
\]
We list the remaining contributions and their bounds explicitly. In the following
table $d=n+1$, $k_n=\|\cK^{1/2}z_n\|$, $v_n=\|\cV_N^{1/2}z_n\|$
and $z_n^\circ=\|z_n\|$. Each entry bounds the square root of
the integrated squared source norm. Constants depend on the
$H^2$ norm of $a$, the row and column bounds of $h$, its
Hilbert--Schmidt norm, and its profile constants, but not on $N$.
\begin{equation}\label{eq:H2-source-table}
	\begin{array}{c|l}
	\text{kinetic source}&\text{bound}\\ \hline
	T\nabla a_z z_n& C\epsilon d^{3/2}k_n\\
	a^*(f_z)\nabla a_z z_n&C\epsilon d^{1/2}k_n\\
	\nabla\kappa,\ \text{identity family}
	&C(d^{1/2}z_n^\circ+\epsilon d^{1/2}k_n)\\
	\nabla a,\ \text{identity family}
	&C\epsilon d^{1/2}k_n\\
	\nabla\kappa,\ \text{dressed families}
	&C(d^{1/2}+dN^{-1})z_n^\circ\\
	\nabla a,\ \text{dressed families}&C\epsilon d z_n^\circ\\
	\nabla h,\ \text{dressed families}&C\epsilon\sqrt N\,d z_n^\circ
	\end{array}
\end{equation}
Here $f_z(y)=\kappa(z-y)a(y)$; the first row denotes the
input derivative in the annihilation-field identity. To verify that the list is complete, write the identity commutator as
$a^*(f_z)a_z+a(z)\dGamma(\kappa(\cdot-z))$.
Its derivative has precisely an input, a relative-kernel,
or an amplitude derivative. The dressed commutators consist
of $a^*(f_z)a^\sharp(h_z)$,
$a(z)\int\kappa(x-z)a_x^*a^\sharp(h_x)\d x$, and, for
the all-creation cubic term, the extra pair kernel
$\kappa(x-y)a(y)h(x,z)$. Differentiating these gives exactly
the last three rows and the input row.

The third row separates into a diagonal contraction,
$d\|\nabla\kappa\|_2^2$, and an off-diagonal convolution
controlled by $\||\nabla\kappa|*|\nabla\kappa|\|_{3/2}\le C\epsilon^2$.
For the other relative-kernel term, integration by parts leaves
the two amplitude estimates above and a derivative on the input.
The remaining rows follow from the dressed-kernel bounds. Equation
\eqref{eq:dressed-quadratic-sources}, together with $\|a\|_\infty$
and the row and column estimates, gives the fifth row. The sixth uses the
integrated $\nabla a$ bound and the quadratic field factor
$d$. The last is controlled by $\epsilon\|\nabla h\|_{\mathrm{HS}}d$.
All three field arrangements obey these bounds. Indeed, for
a two-variable kernel $F$ with symmetric part $F_{\rm s}$,
\begin{align*}
	\|\dGamma(F)\psi_n\|&\le n\|F\|_{\mathrm{HS}}\|\psi_n\|,\\
	\|\mathcal B(F)\psi_n\|&\le\sqrt{n(n-1)}\|F_{\rm s}\|_{\mathrm{HS}}\|\psi_n\|,\\
	\|\mathcal B^*(F)\psi_n\|&\le\sqrt{(n+1)(n+2)}\|F_{\rm s}\|_{\mathrm{HS}}\|\psi_n\|.
\end{align*}
Each factor is at most $C d$; the annihilation term is zero
for $n<2$. These inequalities can be integrated in the
removed coordinate before taking the square root.

For the adjoint identity term, integration by parts gives
the following three sources on the smooth core:
\begin{align*}
	\nabla_z[a_z,T_0^*]\psi_n
	={}&\int\bar\kappa(z-y)\nabla\bar a(y)a_ya_z\psi_n\d y\\
	&+\int\bar\kappa(z-y)\bar a(y)(\nabla_y a_y)a_z\psi_n\d y\\
	&+\int\bar\kappa(z-y)\bar a(y)a_y\nabla_z a_z\psi_n\d y.
\end{align*}
The first term uses the $L^3$ amplitude estimate above.
For the second, use
$\int\|a_z\chi_{n-1}\|^2\d z=(n-1)\|\chi_{n-1}\|^2$
with $\chi_{n-1}=\nabla_y a_y\psi_n$.
Cauchy--Schwarz in $y$ and
$\int\!\int\|a_z\nabla_y a_y\psi_n\|^2\d y\d z
=(n-1)k_n^2$ give $C\epsilon\sqrt{n-1}\,k_n$.
The last term uses $\sup_z\|f_z\|_2\le C\epsilon$ on
the $(n-1)$-particle input and has the same bound.
All three terms vanish for $n<2$.
For the adjoint dressed term the commutator is instead
\[
	[a_z,(T_h^-)^*]
	=\int\bar\kappa(x-y)\bar a(y)h_x(z)a_ya_x\d x\d y.
\]
Only $h_x(z)$ is differentiated. The integrated squared
Hilbert--Schmidt norm of this pair kernel is at most
$C\epsilon^2\|\nabla_2h\|_{\mathrm{HS}}^2$.
Its pair-annihilation factor is $\sqrt{n(n-1)}$, which
gives the last row of \eqref{eq:H2-source-table}.
The commutator for $(T_h^+)^*$ vanishes.

For the potential, apply \eqref{eq:pair-comm-identity}.
All nonzero source types are
\begin{equation}\label{eq:H2-potential-source-table}
	\begin{array}{c|l}
	T a_xa_y z_n&C\epsilon d^{3/2}v_n\\
	D_xa_y,\ \text{identity creation part}&C\epsilon d^{1/2}v_n\\
	D_xa_y,\ \text{identity density part}&CdN^{-1}k_n\\
	D_xa_y,\ \text{dressed parts}&C\epsilon d^{3/2}N^{-1/2}z_n^\circ\\
	E_{xy},\ \text{identity or dressed parts}&Cd^{1/2}z_n^\circ
	\end{array}
\end{equation}
The norms in this table include the integration against
$W(x-y)=N^2v(N(x-y))$. The third row is
\eqref{eq:single-W-hardy}; the fourth follows by integrating
$W$ in its free variable. The six double contractions listed
in Proposition~\ref{prop:full-graph} give the last row:
their squared kernel norms are bounded by
$C\int W|\kappa|^2$ or $CN\epsilon^2$.
They use the row and column bounds and
\eqref{eq:profile-envelope}. The two-annihilator adjoints
have $E_{xy}=0$, and the all-annihilator adjoint has
$D_x=E_{xy}=0$. These observations cover every term of
the pair identity.

On $d\le C\sqrt N$,
$\epsilon\sqrt N\,d\le C\sqrt d$,
$d/N\le C\epsilon d^{3/2}$ and
$\epsilon d^{3/2}/\sqrt N\le C\sqrt d$.
The two tables therefore give the canonical block bound.
The number part uses the $L^2$ block bound.
The output projection is bounded in the $Q$ form norm by
\eqref{eq:V-projection} and its kinetic counterpart,
using only $\|a\|_{H^2}$ for the actual condensate projection.
Generalized-field factors and the cutoff are scalar
contractions on each original number block; no spatial
derivative acts on them. Hence the block estimate is
\begin{equation}\label{eq:H2-phase-block}
	\|Q_{j,P}^{1/2}C_{jn}z_n\| \le C\{\epsilon(n+1)^{3/2}\|Q_{n,P}^{1/2}z_n\| +(n+1)^{1/2}\|z_n\|\}.
\end{equation}
The estimate holds for each differentiated coefficient term
whose amplitude and dressed kernel obey the stated bounds.

For $k_\ell(x,y)=-Nw_\ell(N(x-y))\phi(x)\phi(y)$ and
$\eta=(q\otimes q)k_\ell$, two ordinary time derivatives
replace the amplitudes by $H^2$ vectors. They preserve the
Hilbert--Schmidt, row and column bounds and the gradient
bound $C\sqrt N$. Indeed $Nw_\ell(Nr)\le C/|r|$ gives
the row bounds by Hardy's inequality, and its gradient has
squared integral $O(N)$. The finite-rank projection terms
are estimated by the same contractions. The alternating
$\eta,\bar\eta$ series defining $p$ and $\sigma$ retain
these bounds. A spatial derivative acts on an end factor.
Two time derivatives give a number of terms that grows at
most quadratically with the series index; the factorial
coefficients make the sum convergent. The profile bounds,
including mixed derivatives, follow from the finite-rank
formula below. Put $a_N(x)=Nw_\ell(Nx)$,
$c_\psi=a_N*|\psi|^2$ and
$c_{\psi,0}=\int|\psi|^2c_\psi$. Projection gives the exact
identity
\begin{equation}\label{eq:eta-finite-rank-correction}
	\eta_\psi(x,y)=-a_N(x-y)\psi(x)\psi(y) +\psi(x)\psi(y)\{c_\psi(x)+c_\psi(y)-c_{\psi,0}\}.
\end{equation}
Since $\|a_N\|_1\le C$, convolution and the $H^2$ algebra
estimate bound $c_\psi$ in $H^2$ on bounded $H^2$ sets.
Explicitly, for $|\alpha|\le2$,
$\partial^\alpha c_\psi=a_N*\partial^\alpha|\psi|^2$
in distributions, with both derivatives on the amplitude.
Likewise
$\|\partial_t^j c_\phi\|_{H^2}
\le C\|\partial_t^j|\phi|^2\|_{H^2}\le C_I$ for $j\le2$,
since $\phi,\dot\phi,\ddot\phi$ are bounded in $H^2$.
For the first two time derivatives, every amplitude in this
formula is in $H^2$ with a uniform bound. Thus there is a
finite sum $U$ of absolute values of such amplitudes, with
$\|U\|_2+\|U\|_\infty\le C_I$, such that, for $j\le2$,
\[
	|\partial_t^j\eta(x,y)|\le CN U(x)U(y),\quad \|\partial_t^j\eta(x,\cdot)\|_2\le C U(x),\quad \|\partial_t^j\eta(\cdot,y)\|_2\le C U(y).
\]
The row estimates for the singular part use Hardy's
inequality on its other amplitude. For a product of at
least two kernels, apply row/column Cauchy--Schwarz at
the ends and operator norms to the interior factors.
Its pointwise bound is $C^lU(x)U(y)$; a one-factor term
retains the displayed factor $N$. After factorial summation,
$p,\sigma$ and their first two derivatives satisfy
\eqref{eq:profile-envelope}, with the same type of finite
profile sum. This also verifies the potential entries in
\eqref{eq:H2-potential-source-table} for every derivative.
Set $R(A)=\sup_x\|A(x,\cdot)\|_2$ and
$S(A)=\sup_y\|A(\cdot,y)\|_2$. Then
\begin{align*}
	\|A_1\cdots A_l\|_{\mathrm{HS}}
	&\le\|A_1\|_{\mathrm{HS}}\prod_{j=2}^l\|A_j\|_{\mathrm{op}},\\
	R(A_1\cdots A_l)&\le R(A_1)\prod_{j=2}^l\|A_j\|_{\mathrm{op}},\\
	S(A_1\cdots A_l)&\le S(A_l)\prod_{j=1}^{l-1}\|A_j\|_{\mathrm{op}}.
\end{align*}
Spatial differentiation affects only the ends of the product:
an output derivative replaces $A_1$ by $\nabla A_1$, while an
input derivative acts on the last factor. Since every factor is
either one of $\eta,\bar\eta$ or one of its first two time
derivatives, its operator norm is controlled by its Hilbert--Schmidt
norm. A derivative of total time order $j\le2$ is therefore
bounded by $C^l(l+1)^j$. The bound becomes $C^l(l+1)^j\sqrt N$
when a spatial gradient acts on an end factor. The original factorials absorb these factors,
also when the time derivatives fall on different kernels.

It remains to check connection insertions. For
$R=|u\rangle\langle v|$ with $u,v\in H^2$, acting on a creation
operator in the identity term gives
$a^*(u)\dGamma(k)$ or $a^*(u)\dGamma(B)$, where
\[
	k=\kappa*(a\bar v),\qquad B(y,x)=a(y)\kappa(x-y)\bar v(x).
\]
The relative kernel obeys
\[
	\|\kappa\|_1\le C N^{-3/2},\qquad \|\kappa\|_{3/2}+\|\nabla\kappa\|_1\le C N^{-1}.
\]
Consequently
\begin{align*}
	\|k\|_\infty+\|B\|_{\mathrm{op}}&\le C N^{-3/2},\\
	\|\nabla k\|_\infty+\|\nabla B\|_{\mathrm{op}}
	+\|\nabla B^*\|_{\mathrm{op}}&\le C N^{-1}.
\end{align*}
For the only less immediate term use
$\|(\nabla a)(\kappa*(\bar v f))\|_2
\le\|\nabla a\|_3\|\kappa\|_{3/2}\|v\|_\infty\|f\|_2$.

For a dressed field, the same contractions produce
$k(x)h(x,z)$ and $a(y)\int\kappa(x-y)\bar v(x)h(x,z)\d x$.
Their Hilbert--Schmidt norms are $O(N^{-3/2})$, as follows
by bounding them with $\|k\|_\infty\|h\|_{\mathrm{HS}}$ and
$\|a\|_\infty\|\kappa\|_1\|v\|_\infty\|h\|_{\mathrm{HS}}$,
respectively. Differentiating at either end gives $O(N^{-1})$:
the contributions are controlled by $\|\nabla\kappa\|_1$ or
$\|\kappa\|_1\|\nabla h\|_{\mathrm{HS}}$, with the preceding
$L^3$ estimate covering a derivative on $a$.
The contraction is essential to these bounds, since the
uncontracted three-variable kernel has only the $O(\epsilon)$
estimate. Acting directly on the dressed field instead amounts
to left or right composition with $R$.

The second insertion preserves these kernel classes. For a
second rank-one operator $R_1$,
\[
	[\dGamma(R_1),a^*(u)\dGamma(B)]
	=a^*(R_1u)\dGamma(B)+a^*(u)\dGamma([R_1,B]).
\]
The same identity holds when $B$ is multiplication by $k$.
For $R_1=|u_1\rangle\langle v_1|$, an output gradient gives
$\nabla R_1=|\nabla u_1\rangle\langle v_1|$ and
$\nabla R_1^*=|\nabla v_1\rangle\langle u_1|$.
Both map $L^2$ boundedly into $L^2(\mathbb R^3;\mathbb C^3)$.
For an integral kernel $B$, use
$\nabla[R_1,B]=(\nabla R_1)B-(\nabla B)R_1$ and
$\nabla[R_1,B]^*=(\nabla B^*)R_1^*-(\nabla R_1^*)B^*$.
Thus
\[
	\|[R_1,B]\|_{\mathrm{op}}\le C N^{-3/2},\qquad
	\|\nabla[R_1,B]\|_{\mathrm{op}}
	+\|\nabla[R_1,B]^*\|_{\mathrm{op}}\le C N^{-1}.
\]
For multiplication by $k$, apply the product rule to
$kR_1$ and use $\|\nabla k\|_\infty$; $R_1k$ uses
$\nabla R_1$ instead. For a contracted two-variable kernel,
the second insertion composes either end with $R_1$ or
$R_1^*$ and preserves the Hilbert--Schmidt and gradient
bounds. In each case the remaining created vector stays
in a bounded set of $H^2$.
The annihilation operator of the original identity term produces
a dressed kernel whose $h$ is rank one and of class $H^2$,
so the two tables apply. On a dressed field, $h$ becomes $Rh$
or $hR$, preserving its row/column, profile and gradient bounds.
Indeed, a rank-one factor $R=|u\rangle\langle v|$ maps $L^2$
into $H^2$ with norm $\|u\|_{H^2}\|v\|_2$.
These cases cover all three active field operators and both choices for a
second insertion. The spectator terms cancel in the commutator with
$\dGamma(R)$.

We now combine the kernel estimates. For these contracted terms
and their adjoints, the kinetic
graph bounds are
$C N^{-3/2}(n+1)^{3/2}\|A^{1/2}z_n\|
+C N^{-1}(n+1)^{3/2}\|z_n\|$.
The adjoint estimate uses the output derivative of $B^*$
or the corresponding end derivative of the pair kernel.
Let $S_{jn}$ denote any such block and put $d=n+1$.
Since $|j-n|\le3$, $Q_j\le C(j+1)A_j$ and $Q_n\ge A_n\ge d$,
\begin{equation}\label{eq:contracted-insertion-Q-bound}
	\|Q_j^{1/2}S_{jn}z_n\|
	\le C\bigl(N^{-3/2}d^2+N^{-1}d^{3/2}\bigr)
	\|Q_n^{1/2}z_n\|.
\end{equation}
Both coefficients are bounded for $d\le C\sqrt N$.
Insertions that only compose a dressed field use
\eqref{eq:H2-phase-block} directly. Fixed number weights
change these bounds by a constant because the number shifts
are bounded. Compression is applied after the canonical
estimates, using the uniform $Q$-form bound for $P$ on
every output sector from Proposition~\ref{prop:full-graph}.
Together with \eqref{eq:H2-phase-block}, this proves
\eqref{eq:C-two-graph}, since
\[
	\nabla^2 C =P\{\ddot C_{\rm can}-[\dot{\mathscr Z},C_{\rm can}] -2[\mathscr Z,\dot C_{\rm can}] +[\mathscr Z,[\mathscr Z,C_{\rm can}]]\}P .
\]

\textit{Third covariant derivative.}
The full third covariant derivative has the ordered expansion
\begin{align}
	\nabla^3 C=P\bigl\{&C_{\rm can}^{(3)}
	-[\ddot{\mathscr Z},C_{\rm can}]
	-3[\dot{\mathscr Z},\dot C_{\rm can}]\notag\\
	&-3[\mathscr Z,\ddot C_{\rm can}]
	+[\dot{\mathscr Z},[\mathscr Z,C_{\rm can}]]\notag\\
	&+2[\mathscr Z,[\dot{\mathscr Z},C_{\rm can}]]
	+3[\mathscr Z,[\mathscr Z,\dot C_{\rm can}]]\notag\\
	&-[\mathscr Z,[\mathscr Z,[\mathscr Z,C_{\rm can}]]]\bigr\}P .
	\label{eq:phase-third-covariant-expansion}
\end{align}
This follows by applying $\partial_t-[\mathscr Z,\cdot]$
to the second-derivative formula and using $\nabla P=0$.
At the one-particle level,
\[
	\dot Z=[\ddot q,q],\qquad
	\ddot Z=[q^{(3)},q]+[\ddot q,\dot q] .
\]
In particular,
\[
	q^{(3)}=-|\phi^{(3)}\rangle\langle\phi|
	-3|\ddot\phi\rangle\langle\dot\phi|
	-3|\dot\phi\rangle\langle\ddot\phi|
	-|\phi\rangle\langle\phi^{(3)}| .
\]
Thus $Z,\dot Z$ have $H^2$ end vectors, and the new
terms in $\ddot Z$ have one end vector in $L^2$.
Only $C_{\rm can}^{(3)}$ and
$[\ddot{\mathscr Z},C_{\rm can}]$ contain the highest
derivative $\zeta=\phi^{(3)}$. All other terms use the
$H^2$ coefficient and insertion classes already estimated.
The Neumann estimate \cite[Lemma A.1]{COS}, at radius $m$, gives
\[
	|\kappa(r)|=\sqrt N\,|w^{\rm Neu}_{Nm}(Nr)| \le\frac{C}{\sqrt N|r|}\mathbf1_{|r|\le m},\qquad \|\kappa\|_2^2\le C m/N=C\epsilon^2.
\]
In the identity term, $\kappa(r)^2\le C/(N|r|^2)$
and Hardy's inequality in the retained input coordinate
give
\[
	\|T_\zeta z_n\|
	\le C N^{-1/2}\sqrt{n(n+1)}\|\zeta\|_2\|\cK^{1/2}z_n\| .
\]
Here $T_\zeta=\int\kappa(x-y)\zeta(y)a_x^*a_y^*a_x\d x\d y$
before the bounded number coefficients and excitation projections.
The scalar coefficient is $\kappa(x-y)\zeta(y)$.
The field product has two creation operators and one
annihilation operator, so $[M,T_\zeta]=T_\zeta$ and
$T_\zeta$ maps the $n$-particle sector to the $(n+1)$-particle sector.
With the standard sector normalization
$(a_xz_n)^{(n-1)}=\sqrt n\,z_n(x,\cdot)$, its exact
coordinate expression is
\[
	(T_\zeta z_n)(X_{n+1})= \frac1{\sqrt{n+1}}\sum_{j=1}^{n+1}\zeta(x_j) \left(\sum_{i\ne j}\kappa(x_i-x_j)\right) z_n(X_{n+1}\setminus x_j).
\]
The two creation factors contribute $1/\sqrt{n(n+1)}$,
which cancels the $\sqrt n$ from annihilation. The
annihilated coordinate is restored by the $x$ creation,
so the remaining wave function is precisely the one
displayed, by symmetry. Cauchy--Schwarz in the outer
sum cancels $1/(n+1)$. All $n+1$ resulting integrals
are equal after relabeling the remaining coordinates
as $X$ and the omitted coordinate as $y$. Hence
\[
	\|T_\zeta z_n\|^2 \le(n+1)\int |\zeta(y)|^2 \left|\sum_{j=1}^n\kappa(x_j-y)\right|^2 |z_n(X)|^2\d X\d y.
\]
Bound the square of the sum by $n\sum_j|\kappa(x_j-y)|^2$.
For every $y$, Hardy in each $x_j$ gives
$\sum_j\int |x_j-y|^{-2}|z_n(X)|^2\d X
\le C\|\cK^{1/2}z_n\|^2$. This proves the displayed bound,
including the vanishing $n=0$ sector.
Output projections and generalized-field coefficients are
contractions on each block. Hardy's inequality on the ambient form domain also holds
on the excitation subspace.
Pair with the output using
$\|u_{n+1}\|\le(n+2)^{-1/2}\|Q^{1/2}u_{n+1}\|$.
The resulting form estimate is
\[
	|\langle u_{n+1},T_\zeta z_n\rangle|
	\le C\sqrt{n/N}\,\|\zeta\|_2
	\|Q_{n+1,P}^{1/2}u_{n+1}\|\|Q_{n,P}^{1/2}z_n\| .
\]
Its coefficient is uniformly bounded. The adjoint
is handled by reversing this form pairing.
For a dressed cubic term the row bound gives the smaller
kernel norm $C\epsilon\|\zeta\|_2$.
Indeed the squared norm of
$\kappa(x-y)\zeta(y)h(x,z)$ is at most
$R(h)^2\|\kappa\|_2^2\|\zeta\|_2^2$.
If the highest derivative acts on $\eta$, the kernel
$Nw_\ell(N(x-y))\zeta(x)\phi(y)$ has Hilbert--Schmidt
norm at most $C\|\zeta\|_2$ by Hardy's inequality in $y$.
Explicitly, the squared norm is bounded by
\[
	C\int |\zeta(x)|^2 \left(\int\frac{|\phi(y)|^2}{|x-y|^2}\d y\right)\d x \le C\|\zeta\|_2^2\|\nabla\phi\|_2^2.
\]
Interchanging $x,y$ treats the other end. The third derivatives
of the projection have rank-one factors with bounded $L^2$
norms, so composition with the bounded $k_\ell$ gives the same
Hilbert--Schmidt estimate. The product estimate above applies
to the factorial series with at most $C(l+1)^3$ insertion
choices; at the highest derivative only its Hilbert--Schmidt
norm is used.
The third projection derivative is finite rank with one
$L^2$ vector. We estimate the third derivatives of $p,\sigma$
in Hilbert--Schmidt norm.
For example, a third derivative $h_3$ on the dressed field gives
$\|\kappa(x-y)\phi(y)h_3(x,z)\|_2
\le\epsilon\|\phi\|_\infty\|h_3\|_{\mathrm{HS}}$.
For any such normally ordered three-field kernel $F$,
the sector estimate is
$\|T_Fz_n\|\le C(n+1)^{3/2}\|F\|_2\|z_n\|$,
for each creation/annihilation choice and its adjoint.
Here $\|F\|_2\le C_I\epsilon$, and
$\epsilon(n+1)^{3/2}\le C$ on the phase cutoff.
Finite bandwidth therefore gives a uniform Hilbert-space
operator bound for these terms. Their form bound follows
from $Q\ge M\ge1$ on both arguments. More explicitly,
for an allowed block $n\to j$ with $|j-n|\le3$,
\[
	|\langle u_j,T_Fz_n\rangle|
	\le C\epsilon\sqrt{n+1}\,
	\|Q_{j,P}^{1/2}u_j\|\|Q_{n,P}^{1/2}z_n\| .
\]
The ratio $(n+1)/(j+1)$ is at most four on a nonzero
block. This gives the stated form estimate for every
creation and annihilation choice, including its adjoint.
Finally a rank-one insertion with one $L^2$ vector has
either the scalar contraction
$\|\kappa*(\phi\bar v)\|_\infty
\le\epsilon\|\phi\|_\infty\|v\|_2$ or a two-variable
Hilbert--Schmidt contraction of that size. These give
$C\epsilon(n+1)^{3/2}$ field bounds, which are uniform.
For the identity term, we check the action on each of the three field
operators by writing $R=|u\rangle\langle v|$.
The $y$ creation operator contributes
$a^*(u)\dGamma(\kappa*(\phi\bar v))$, whereas either field operator at $x$ produces the
three-variable kernel
$u(z)\bar v(x)\kappa(x-y)\phi(y)$ or its adjoint pairing.
The latter has Hilbert--Schmidt norm bounded by
$\epsilon\|u\|_2\|v\|_2\|\phi\|_\infty$, so an $L^2$
end vector introduces no additional number power. For a dressed field, composition with $R$ preserves the
same Hilbert--Schmidt bound. The Leibniz partitions of total order three are $(3)$,
$(2,1)$ and $(1,1,1)$. In the first case all other
factors are undifferentiated. The last two cases use
only the $H^2$ coefficient and insertion classes verified
above. In particular no term contains both a third
derivative and a further positive-order derivative.
For $R=|u\rangle\langle v|$, the commutator with
$\dGamma(R)$ replaces one of the three active field
operators. Repeated insertion therefore produces at
most $3^k$ terms for $k\le3$. The kernel classes
described above are closed under each $H^2$ insertion,
so they also cover the three nested commutators in
\eqref{eq:phase-third-covariant-expansion}. Number
coefficients commute with $\dGamma(R)$, and spectator
coordinates give no extra particle-number factor.
This proves \eqref{eq:C-three-form}. Continuity follows
in the same norms by taking differences of the displayed
multilinear expressions. In particular the identity
cubic term satisfies
\[
	\|Q^{-1/2}(T_{\zeta_1}-T_{\zeta_2})Q^{-1/2}\| \le C_I\|\zeta_1-\zeta_2\|_2 ,
\]
by the same Hardy estimate and cutoff. For a dressed
highest-derivative term use
$\|\kappa(x-y)(\zeta_1-\zeta_2)(y)h(x,z)\|_2
\le\epsilon R(h)\|\zeta_1-\zeta_2\|_2$;
differences of $h$ use the row norm instead.
Terms with $h_3$ use its HS-norm difference.
The coefficient maps through order two are continuous
in $H^2$, while $\phi^{(3)}$ and the highest rank-one
vectors are continuous in $L^2$. The same difference
argument applies to the two graph tables. Factorial
majorants permit passage to the infinite kernel series.
First proving the identities on the finite-particle
form core and then closing them gives the asserted
extensions. Parallel transport is strongly continuous
on the common form space; the covariant derivative
identities and these continuous coefficient bounds give
the stated graph and form continuity there.
Fixed number weights introduce only constants determined by the bandwidth
and the chosen weight.
\end{proof}

\subsubsection{Differentiating the cancelled quadratic generator}

The phase estimates above supply the derivatives of the correlation
transformation. We now establish the corresponding form bounds for
the full generator. The proof treats the quadratic time term,
the spatial kinetic kernels and the differentiated scattering
cancellations. The estimates on a common parameter neighborhood
justify differentiating the convergent kernel series.

\begin{proposition}\label{prop:full-second-form}
For $T=\mathfrak G,\mathfrak J,\mathfrak J^{[3]}$, the forms
\[
	\nabla^2T,\qquad [M,\nabla T],\qquad [M,[M,T]]
\]
are continuous and uniformly $Q$-bounded. These assertions
hold after conjugation by each fixed power of $M$ as well.
\end{proposition}
\begin{proof}
We prove the second-derivative estimate for $\mathfrak G$.
\textit{Quadratic time term.}
We work on the fixed excitation space supplied by parallel transport.
Section~\ref{sec:fixed-kernel-estimates} gives the maps
$\mathcal T_J$ and their number-weighted bounds.
At each fixed $N$, the quadratic generators are bounded and
the exact logarithmic derivative is
\[
	L_B=\sum_{J\ge0}\mathcal T_J(\eta,\nabla\eta).
\]
We now prove convergence together with two time derivatives.

The original strict small-kernel condition gives
$\rho_*:=\sup_{N,t\in I}\|\eta_t\|_{\mathrm{HS}}<c_{\rm time}^{-1}$,
with the constant in \eqref{eq:time-homogeneous-weight}.
Choose once $\rho_*<\rho_1<c_{\rm time}^{-1}$ and
$0<\delta<\rho_1-\rho_*$, and put $\theta=c_{\rm time}\rho_1<1$.
These choices are independent of $N$ and $t$.
On the real HS ball of radius $\delta$ about any $\eta_t$,
\eqref{eq:time-homogeneous-weight} is bounded by
$C_r(1+J)^{b_r}\theta^J\|\beta\|_{\mathrm{HS}}$.
Restrict a scalar matrix element to the real span of the
chosen directions $h_1,h_2$ through $\eta$.
Apply the one-variable polynomial Markov inequality along
each line in this ball, then use real polarization
\cite[Theorem 1]{Harris}. Only subspaces of dimension at most
two occur, and the constants are independent of the chosen
directions and of $N$. Taking the supremum over unit matrix
elements gives, for $j\le2$,
\begin{equation}\label{eq:time-HS-parameter-derivatives}
	\|D_\zeta^j\mathcal T_J(\eta,\beta)
	[h_1,\ldots,h_j]\|_{M,r}
	\le C_{r,j}\delta^{-j}(1+J)^{b_r+2j}\theta^J
	\|\beta\|_{\mathrm{HS}}\prod_{a=1}^j\|h_a\|_{\mathrm{HS}} .
\end{equation}
The polynomial derivative estimate holds for arbitrary HS
directions $\beta,h_1,h_2$.

In parallel coordinates write $\eta_j=\nabla^j\eta$.
The exact second derivative of each summand is
\begin{align}
	\nabla^2\mathcal T_J(\eta,\eta_1)
	={}&D_\zeta^2\mathcal T_J(\eta,\eta_1)[\eta_1,\eta_1]
	+D_\zeta\mathcal T_J(\eta,\eta_1)[\eta_2]\notag\\
	&+2D_\zeta\mathcal T_J(\eta,\eta_2)[\eta_1]
	+\mathcal T_J(\eta,\eta_3).
	\label{eq:time-second-chain-rule}
\end{align}
The HS bounds and continuity of $\eta_1,\eta_2,\eta_3$
were proved in Lemma~\ref{lem:phase-second-regularity}.
Equation \eqref{eq:time-HS-parameter-derivatives} makes
the series and its first two derivatives uniformly summable
in every $\|\cdot\|_{M,r}$ norm. It also gives continuity
and the required number commutators. To pass to the
$Q$ form norm, set $E=M^{1/2}Q^{-1/2}$.
The relations $M\le Q$ and $[M,Q]=0$ give $\|E\|\le1$ and
\[
	Q^{-1/2}M^rTM^{-r}Q^{-1/2}
	=E\bigl(M^{r-1/2}TM^{-r-1/2}\bigr)E .
\]
First on the form core and then by continuity, this gives
exactly the $Q$ form bounds with fixed $M$ weights used here.
The compression uses the connection already defined in
Section~\ref{sec:covariant-generator}:
$P=\Gamma(q)$, $Z=[\dot q,q]$ and
$\mathscr Z=\dGamma(Z)$. The identity $qZq=0$
implies $P\mathscr ZP=0$ on every excitation sector.
Since $\eta=(q\otimes q)\eta$, all creation and annihilation
operators in $B$ act in $qL^2$. Their number factors commute
with $P$, so $B$ and the bounded unitary $U=e^B$ preserve
$P\cF^{\le N}$ and its orthogonal complement.
Consequently
\[
	PU^{-1}[\mathscr Z,U]P
	=U^{-1}P\mathscr ZPU-P\mathscr ZP=0 .
\]
Thus $PU^{-1}\partial_tUP=PU^{-1}\nabla_tUP$, which
identifies the compressed logarithmic derivatives.

\textit{Spatial groups and scattering cancellations.}
For the other terms, let $\mathcal A_N(\psi)$ be the
five bare conjugated groups in \cite[Eq.~(5.3)]{BS},
with its original scalar subtracted and compressed to
$\Gamma(q_\psi)$. The modified GP equation has already
removed the explicit $\dot\psi$ in these groups.
Hence $\mathfrak G(t)=\mathcal A_N(\phi(t))+L_B(t)$.
In this identity the condensate-extraction connection
is included in the bare groups. More explicitly,
\cite[Eqs.~(5.2)--(5.3)]{BS} is the compressed identity
\[
	P_t\{(i\partial_t\mathcal U_{N,t})\mathcal U_{N,t}^* +\mathcal U_{N,t}H_N^{\rm mb}\mathcal U_{N,t}^*\}P_t =P_t\sum_{j=0}^4L_{N,t}^{(j)}P_t .
\]
The modified GP equation is substituted in this identity
before defining $\mathcal A_N$.
Both $e^{B_t}$ and its inverse preserve $P_t$.
For a section in the moving excitation space the
off-diagonal derivative is $\mathscr Z_t$ times that
section, and $P_t\mathscr Z_tP_t=0$. Thus replacing
the ordinary derivative by the common covariant one
gives exactly the displayed compressed generator.
The off-diagonal transport is implemented by the
parallel-coordinate unitary.
To specify the cancellations away from the trajectory,
write $f_N(r)=f_\ell(Nr)$ and
$h_\psi=(N^3v(N\cdot)w_\ell(N\cdot)*|\psi|^2)\psi$.
The linear and cubic extractions in
\cite[Eqs.~(5.45),(5.98)]{BS} are the same generalized
field expression with coefficients $+\sqrt N h_\psi$
and $-\sqrt N h_\psi$. They cancel for each $\psi$.
The three singular pair coefficients from the kinetic,
pair and quartic groups combine to
\[
	N^3\bigl\{(\Delta w_\ell)(Nr) +\tfrac12v(Nr)(1-w_\ell(Nr))\bigr\} \psi(x)\psi(y) =N^3\lambda_\ell f_N(r)\mathbf1_{|r|\le\ell} \psi(x)\psi(y),\quad r=x-y .
\]
The remaining pair kernel is uniformly Hilbert--Schmidt
since $N^3\lambda_\ell\le C$ and $0\le f_N\le1$.
The scalar subtracted in $\mathcal A_N$ is, explicitly,
\begin{align*}
	c_N(\psi)={}&\tfrac12\langle\psi,
	[N^3v(N\cdot)(N-1-2Nf_N)*|\psi|^2]\psi\rangle
	+\|\nabla_x k_\psi\|_{\mathrm{HS}}^2\\
	&+\tfrac12\int N^2v(N(x-y))|k_\psi(x,y)|^2\d x\d y\\
	&+\operatorname{Re}\int N^3v(N(x-y))
	\overline{\psi(x)\psi(y)}\,k_\psi(x,y)\d x\d y ,
\end{align*}
where $k_\psi(x,y)=-Nw_\ell(N(x-y))\psi(x)\psi(y)$.
These are precisely the scalar extractions in
\cite[Eqs.~(5.3),(5.63),(5.89),(5.115)]{BS}.
We estimate the combined scalar and pair terms. For smooth normalized $\psi$
the identities also follow by taking $\psi$ as the initial
datum of the modified GP equation and evaluating the
algebraic expansion at time zero.

\textit{Bounds on a common parameter neighborhood.}
We first verify that the degree majorant
\eqref{eq:degree-majorant} is uniform on a small
normalized $H^2$ neighborhood of any background on $I$.
The convolution and field estimates for the linear,
pair, cubic and quartic groups require the $L^2$,
$L^\infty$, Hardy and row/column bounds, all controlled
by $H^2$. The four-product check following
\eqref{eq:degree-majorant} is retained for the cubic
group, preserving the decay in each expansion index.

For the linear group, write
$\widetilde h_\psi=(N^3v(N\cdot)*|\psi|^2)\psi$.
The bound $0\le w_\ell\le1$ gives
\[
	\|h_\psi\|_2+\|\widetilde h_\psi\|_2
	\le2\|v\|_1\|\psi\|_\infty^2\|\psi\|_2\le C_R .
\]
The canonical terms in \cite[Eqs.~(5.45),(5.98)]{BS}
cancel before differentiation.
We use the coefficients $\mathcal L_j(g)$ defined in
Section~\ref{sec:fixed-kernel-estimates}.
Denote by $\mathcal R_j(g)$ the remainder after subtracting
the canonical one-field term at order $j$.
Equations (5.56)--(5.57) of \cite{BS} give
\begin{equation}\label{eq:linear-canonical-remainder-degree}
	\|M^{-1/2}\mathcal R_j(g)M^{-1/2}\|
	\le C N^{-1/2}(1+j)(C_0\rho)^j\|g\|_2,
	\qquad j\ge1,\quad \rho=\|\eta_\psi\|_{\mathrm{HS}} .
\end{equation}
For a noncanonical term $W_j(g)$, the improved bound in
\cite[Lemma 5.2(i), Eq.~(5.14)]{BS} is
\[
	\|M^{-1/2}W_j(g)z\|
	\le C_0^jN^{-1/2}\rho^j\|g\|_2\|M^{1/2}z\|.
\]
Taking $z=M^{-1/2}u$ gives its balanced operator norm.
The canonical coefficient in \cite[Eq.~(5.54)]{BS} is a
product of $j$ factors chosen from
$(N-\cN)/N$ and $(N+1-\cN)/N$.
On an output sector with particle number $m$, each factor
has modulus at most two and differs from one by at most
$(m+1)/N$. A telescoping product therefore gives
$|c_{j,N}(m)-1|\le j2^{j-1}(m+1)/N$.
The remaining one-field kernel has norm at most
$\rho^j\|g\|_2$, and its number shift is $\pm1$.
Balancing this field by $M^{-1/2}$ on both sides bounds
the block by
\[
	C_0^jj\frac{\sqrt{m+1}}{N}\rho^j\|g\|_2
	\le C_0^jj\sqrt{\frac{N+1}{N}}\,
	N^{-1/2}\rho^j\|g\|_2 .
\]
Only $0\le m\le N$ contributes.
The $2^jj!$ terms and the coefficient $1/j!$ leave a
fixed exponential in $j$. The term of order zero vanishes.
The factor $\sqrt N$ in the linear generator cancels
$N^{-1/2}$ in this estimate.

The remaining linear term uses
\eqref{eq:linear-number-product-degree} with
$g=\widetilde h_\psi$; adjoints follow by duality.
The number polynomials use \eqref{eq:number-square-degree}.
The one-body terms use \eqref{eq:one-body-degree} and
the exact number identity \eqref{eq:normal-one-body-number}.
These estimates were established in
Section~\ref{sec:fixed-kernel-estimates}.

Their coefficients are uniformly bounded on the $H^2$
neighborhood. Indeed, set $V_N^{(3)}=N^3v(N\cdot)$ and
$H_\psi(x,y)=V_N^{(3)}(x-y)\psi(x)\overline{\psi(y)}$.
Young's inequality gives
\[
	\|V_N^{(3)}*|\psi|^2\|_\infty
	+\|H_\psi\|_{\mathrm{op}}
	\le2\|v\|_1\|\psi\|_\infty^2\le C_R .
\]
The same bound holds with $v$ replaced by $vw_\ell$.
Compression by $q_\psi$ preserves the operator norm bound.
The scalar coefficients in the number polynomial obey
the same estimate after integration against $|\psi|^2$.
These coefficient maps and projections have fixed polynomial
degree in $\psi,\overline\psi$. Each term above has degree
$O(J+1)$, as in the linear group. Its number bandwidth is
at most $2J$, so two number commutators add at most
$(2J)^2$. The products and their sums converge in the
displayed weighted form norms, uniformly on the common
neighborhood. They therefore enter
\eqref{eq:neighborhood-majorant} and the parameter derivative
estimate below with the same geometric decay.

\textit{Pair terms.}
For the pair group, the low orders in
\cite[Eq.~(5.93)]{BS} contain a scalar subtraction and
three field terms. Put $V_N^{(3)}(r)=N^3v(Nr)$,
$\rho=\|\eta_\psi\|_{\mathrm{HS}}$ and
$r_\psi(x)=\|\eta_\psi(x,\cdot)\|_2$.
The scalar is multiplied by
$\alpha_N=(N-1-\cN)(N-\cN)/N^2$.
On $\cF^{\le N}$ one has
$|\alpha_N-1|\le CM/N$ and $|\alpha_N|\le C$.
The bounds $|k_\psi(x,y)|\le CN|\psi(x)\psi(y)|$
and $|\mu_\psi(x,y)|\le C_R|\psi(x)\psi(y)|$ give
\[
	\left|\int V_N^{(3)}(x-y)
	\overline{\psi(x)\psi(y)}k_\psi(x,y)\d x\d y\right|
	\le C_RN,\qquad
	\left|\int V_N^{(3)}(x-y)
	\overline{\psi(x)\psi(y)}\mu_\psi(x,y)\d x\d y\right|
	\le C_R .
\]
Thus subtracting the displayed scalar leaves an
$M$-bounded form, uniformly on the neighborhood.
The next two field terms use the row norm $r_\psi$.
The last uses the interaction form and
$N^{-1/2}\|Mz\|\le\sqrt2\|M^{1/2}z\|$.
These are the three estimates in
\cite[Eqs.~(5.93)--(5.94)]{BS}.

For the remaining words, retain the total degree $J=n+k$
in \cite[Eqs.~(5.95)--(5.97)]{BS}.
Writing $Z_x=\|a_xz\|$ and $Z_{xy}=\|a_xa_yz\|$,
their four spatial integrals obey
\begin{align}
	\int V_N^{(3)}(x-y)|\psi(x)\psi(y)|
	r_\psi(x)r_\psi(y)\d x\d y
	&\le C_R\rho^2,\notag\\
	\int V_N^{(3)}(x-y)|\psi(x)\psi(y)|
	r_\psi(y)Z_x\d x\d y
	&\le C_R\rho\|\cN^{1/2}z\|,\notag\\
	\int V_N^{(3)}(x-y)|\psi(x)\psi(y)|\d x\d y
	&\le C_R,\notag\\
	N^{-1/2}\int V_N^{(3)}(x-y)|\psi(x)\psi(y)|
	Z_{xy}\d x\d y
	&\le C_R\|\cV_N^{1/2}z\|.
	\label{eq:pair-neighborhood-integrals}
\end{align}
The first three lines follow from Young's inequality,
$\|V_N^{(3)}\|_1=\|v\|_1$,
$\|r_\psi\|_2=\rho$ and $\|\psi\|_\infty\le C_R$.
For the last line, Cauchy--Schwarz uses
\[
	\int V_N^{(3)}(x-y)Z_{xy}^2\d x\d y
	=2N\|\cV_N^{1/2}z\|^2,\qquad
	\int V_N^{(3)}(x-y)|\psi(x)\psi(y)|^2\d x\d y
	\le C_R .
\]
In the estimate following \cite[Eq.~(5.97)]{BS}, these
four integrals carry the coefficients
$n$, $\rho$, $C_R\rho$ and $\rho^2$, respectively,
with the common factor
$C_0^J\rho^{J-2}\|M^{1/2}z\|$.
The first and third terms also contain $\|M^{1/2}z\|$.
Consequently each word, for $J\ge1$, satisfies
\[
	|E_\iota[z,z]|
	\le C_RC_0^J\bigl((J+1)\rho^J+\rho^{J-1}\bigr)Q_\psi[z].
\]
The estimate at $\rho=0$ follows by continuity after
collecting these nonnegative powers.
For fixed $(n,k)$ there are $2^Jn!k!$ words and the
series coefficient is $1/(n!k!)$.
Summing over the $J+1$ index pairs leaves at most a
quadratic factor in $J+1$ and a fixed exponential in $J$.
The small-kernel condition therefore gives the pair contribution
to \eqref{eq:neighborhood-majorant}.
The scalar, adjoint and field estimates hold on the form core
and extend by their $Q_\psi$ bounds.

We next consider the three cubic families in
\cite[Eqs.~(5.104)--(5.110)]{BS}. Put
$a(x)=\|\eta_\psi(x,\cdot)\|_2$ and
$F_N=N^{5/2}v(N\cdot)$. For nonnegative $g$, define
\[
	\mathcal I_N(g)=\int F_N(x-y)|\psi(y)|g(x,y)\d x\d y .
\]
Use $L,L_1,Z,Y$ as in \eqref{eq:cubic-four-products},
and write $Z_{xy}=\|a_xa_yz\|$ and
$D=\|\cV_N^{1/2}z\|$. The kernel formula
\eqref{eq:eta-finite-rank-correction} gives
\[
	\|a\|_2=\rho,\qquad \|a\|_\infty\le C_R,\qquad
	N^{-1}|\eta_\psi(x,y)|\le C_R|\psi(x)\psi(y)| .
\]
All constants below are uniform on the same $H^2$ neighborhood.
The spatial integrals needed for (5.104) and (5.106) are
\begin{align}
	\mathcal I_N(a(x)^2a(y))
	&\le C_RN^{-1/2}\rho^2,\notag\\
	\mathcal I_N(a(x)N^{-1}|\eta_\psi(x,y)|)
	&\le C_RN^{-1/2}\rho,\notag\\
	\mathcal I_N(a(x)a(y)Y(x))
	+\mathcal I_N(a(x)^2Y(y))
	&\le C_RN^{-1/2}\rho L_1,\notag\\
	\mathcal I_N(a(x)Z_{xy})
	&\le C_R\rho D .
	\label{eq:cubic-dressed-source-integrals}
\end{align}
To prove the first and third lines, let
$H_N=F_N*(|\psi|a)$. Then
$\|H_N\|_2\le C_RN^{-1/2}\rho$ and
$\|H_N\|_\infty\le C_RN^{-1/2}$.
The first integral is at most $\|H_N\|_\infty\rho^2$.
For the third line use, respectively,
$\|a\|_\infty\|H_N\|_2\|Y\|_2$ and
$\|\psi\|_\infty\|Y\|_2\|F_N*a^2\|_2$.
Here $\|a^2\|_2\le\|a\|_\infty\rho$.
The second line follows from
\[
	\mathcal I_N(a(x)N^{-1}|\eta_\psi(x,y)|)
	\le C_R\|a\psi\|_2\|F_N*|\psi|^2\|_2
	\le C_RN^{-1/2}\rho .
\]
For the last line, write $W_N=N^2v(N\cdot)$ and use
\[
	\int W_N(x-y)Z_{xy}^2\d x\d y=2D^2,\qquad
	\int W_N(x-y)|\psi(y)|^2a(x)^2\d x\d y
	\le C_RN^{-1}\rho^2 .
\]
Cauchy--Schwarz and $F_N=\sqrt N W_N$ give the claim.

For a word in (5.104), let $J=n+k+r$, with $k\ge1$.
The two factors called $Q_1,Q_2$ in \cite{BS} satisfy
\begin{align*}
	Q_1&\le C_0^r\rho^ra(x)L,\\
	Q_2&\le C_0^{n+k}\rho^{n+k-2}
	\bigl\{[(n+1)a(x)a(y)+\rho N^{-1}|\eta_\psi(x,y)|]L_1\\
	&\hspace{35mm}+\rho a(y)Y(x)+\rho^2Z_{xy}\bigr\} .
\end{align*}
These are the individual-word bounds preceding (5.105).
After multiplication, the first three terms in
\eqref{eq:cubic-dressed-source-integrals} carry the common
factor $C_0^J\rho^{J-2}L$.
Their coefficients are $(n+1)L_1$, $\rho L_1$ and $\rho$.
The interaction term has the additional factor $\rho^2$.
Thus the expectation of each word is at most
\[
	C_RC_0^J\bigl((n+1)N^{-1/2}\rho^JLL_1
	+\rho^{J+1}LD\bigr)
	\le C_RC_0^J(n+1)\rho^J Q_\psi[z] .
\]
We used $\rho\le1$, $L_1\le\sqrt{N+1}L$ and
$L^2+D^2\le Q_\psi[z]$.
The apparent negative power when $J=1$ disappears after
the spatial integration. The resulting bound extends to
$\rho=0$ by continuity.

For (5.106), put $J=n+r$ and first take $n\ge2$.
The bound on $R_1$ is the displayed bound on $Q_1$.
The estimate on $R_2$ following (5.108) is
\begin{align*}
	R_2\le{}&C_0^n\rho^{n-2}
	\bigl\{[(1+n/N)a(x)a(y)+\rho N^{-1}|\eta_\psi(x,y)|]L_1\\
	&\hspace{18mm}+\rho a(x)Y(y)
	+(n/N)\rho a(y)Y(x)+\rho^2Z_{xy}\bigr\} .
\end{align*}
All five integrals are in
\eqref{eq:cubic-dressed-source-integrals}.
Their sum is bounded by
$C_RC_0^J(n+1)\rho^J Q_\psi[z]$.
At $n=0$, the left product is $b_yb_x$ and has norm
at most $Z_{xy}$. The last line of
\eqref{eq:cubic-dressed-source-integrals} then gives
$C_RC_0^r\rho^{r+1}LD$, which satisfies the same bound.

For the single sum in (5.110), the right factor is $Z(x)$.
For $n\ge2$, use the bound on $R_2$ with $a_y$ in place
of $b_y$, as in \cite[Eq.~(5.110)]{BS}.
The additional spatial integrals obey
\begin{align}
	\mathcal I_N(a(x)a(y)Z(x))
	&\le C_RN^{-1/2}\rho L,\notag\\
	\mathcal I_N(N^{-1}|\eta_\psi(x,y)|Z(x))
	&\le C_RN^{-1/2}L,\notag\\
	\mathcal I_N(a(x)Y(y)Z(x))
	+\mathcal I_N(a(y)Y(x)Z(x))
	&\le C_RN^{-1/2}LL_1,\notag\\
	\mathcal I_N(Z_{xy}Z(x))
	&\le C_RDL .
	\label{eq:cubic-bare-source-integrals}
\end{align}
The first line uses $\|a\|_\infty\|H_N\|_2\|Z\|_2$.
For the second, bound
$|\psi|(F_N*|\psi|^2)$ in $L^2$ by $C_RN^{-1/2}$.
In the third line, Young's inequality bounds the first
integral by
$\|a\|_\infty\|\psi\|_\infty\|F_N\|_1\|Z\|_2\|Y\|_2$;
the second is at most $\|H_N\|_\infty\|Y\|_2\|Z\|_2$.
The last line follows by Cauchy--Schwarz with $W_N$,
since
\[
	\int W_N(x-y)|\psi(y)|^2Z(x)^2\d x\d y
	\le C_RN^{-1}L^2 .
\]
Inserting these bounds in $R_2Z(x)$ gives
$C_RC_0^n(n+1)\rho^{n-1}Q_\psi[z]$ for $n\ge2$.
At $n=0$, the form is bounded directly by
$\mathcal I_N(Z_{xy}Z(x))\le C_RDL$.
The value $n=1$ belongs to the linear extraction in (5.100).

For (5.104), the word count $2^Jn!k!r!$ is paired with
$1/[n!k!r!(k+r+1)]$.
For (5.106), the count is $2^Jn!r!$ and the coefficient
is $1/[n!(r+1)!]$.
For (5.110), they are $2^nn!$ and $1/n!$.
There are at most $(J+1)(J+2)/2$ index tuples in each case.
The remaining factors are polynomial in $J+1$.
These three families therefore satisfy
\eqref{eq:neighborhood-majorant}, with a loss of at most
one power of $\rho$ and the same small-kernel choice.
For each elementary sesquilinear form $e$, write its
diagonal bound as $|e[z,z]|\le A_JQ_\psi[z]$.
Complex polarization on the form core gives
\[
	e[u,v]=\frac14\sum_{j=0}^3i^{-j}
	e[u+i^jv,u+i^jv],\qquad
	|e[u,v]|\le A_J\bigl(Q_\psi[u]+Q_\psi[v]\bigr) .
\]
Replace $(u,v)$ by $(tu,t^{-1}v)$ and minimize over $t>0$.
The result is
$|e[u,v]|\le2A_JQ_\psi[u]^{1/2}Q_\psi[v]^{1/2}$.
Density now gives a unique continuous extension to
$D(Q_\psi^{1/2})\times D(Q_\psi^{1/2})$.
The additional factor two preserves the geometric majorant.

The remaining cubic terms are the two $n=1$ groups in
\cite[Eq.~(5.100)]{BS}. We keep the normalization
\[
	T_r(x)=\frac{\mathcal D_B^r b^*(\eta_x)}{r!},\qquad
	\|T_r(x)z\|\le C_0^r\rho^ra(x)L .
\]
Their series coefficient is $1/(r+1)$, since
$\mathcal D_B=[\,\cdot\,,B]$ absorbs the sign in the
commutator convention of \cite{BS}.
The norm estimate is \cite[Eq.~(5.107)]{BS} after division by $r!$.
Write $S_2,S_{12},S_{112},S_{111}$ for the four contributions
in the decomposition following \cite[Eq.~(5.111)]{BS},
with this same division by $r!$.

We first give the field estimate used in $S_2$.
Define $\beta_N(m)=\sqrt{(1-m/N)_+}$ and
\[
	\mathcal A_\eta^*=\int\eta(z,w)a_z^*a_w^*\d z\d w,
	\qquad D_x=N^{-1}\mathcal A_\eta^*\beta_N(\cN+1)a_x .
\]
The operator $D_x$ is the $N^{-1}\Pi^{(1)}$ term in
\cite[Eq.~(5.111)]{BS}. Here the canonical fields act on
full Fock space; the displayed $\beta_N$ factors restrict
their product to $\cF^{\le N}$. On the finite-particle core,
the canonical commutation relations give
\[
	a_yD_x=N^{-1}\mathcal A_\eta^*\beta_N(\cN+2)a_ya_x
	+2N^{-1}a^*(\eta_y)\beta_N(\cN+1)a_x .
\]
The pair creation estimate
$\|\mathcal A_\eta^*M^{-1}\|\le C\rho$ and the
one-field estimate therefore imply
\begin{equation}\label{eq:cubic-extraction-field}
	\|a_yD_xz\|+\|b_yD_xz\|
	\le C\rho Z_{xy}+CN^{-1}a(y)Y(x) .
\end{equation}
For the first term, $a_ya_xz$ has at most $N-2$ particles,
so $M/N\le1$ on its nonzero sectors.
For the second, $\|M^{1/2}a_xz\|\le Y(x)$.
The bound for $b_y$ follows from $b_y=\beta_N(\cN)a_y$.
The positive part in $\beta_N$ also specifies all boundary
sectors, including $N=1$.

Two integrals in \eqref{eq:cubic-dressed-source-integrals} give
\begin{align*}
	|\langle z,S_2z\rangle|
	&\le C C_0^r\rho^r L
	\bigl\{N^{-1}\mathcal I_N(a(x)a(y)Y(x))
	+\rho\mathcal I_N(a(x)Z_{xy})\bigr\}\\
	&\le C_R C_0^r
	\bigl(N^{-3/2}\rho^{r+1}LL_1+\rho^{r+2}LD\bigr) .
\end{align*}
For $S_{12}$, its left factor is the adjoint of
$\nu_N\{b_y^*b(\eta_x)-N^{-1}a_y^*a(\eta_x)\}$,
where $\nu_N=(N-1-\cN)/N$.
It has norm at most $Ca(x)Y(y)$ on $z$.
Indeed $|\nu_N|\le1$ on the physical sectors, and each
creation field has norm at most its kernel norm times $M^{1/2}$.
Consequently
\[
	|\langle z,S_{12}z\rangle|
	\le C C_0^r\rho^r L\mathcal I_N(a(x)^2Y(y))
	\le C_R C_0^rN^{-1/2}\rho^{r+1}LL_1 .
\]
The finite-rank bound
$|\mu_\psi(x,y)|\le C_R|\psi(x)\psi(y)|$ gives
\[
	\mathcal I_N(|\mu_\psi(x,y)|a(x))
	\le C_R\|a\psi\|_2\|F_N*|\psi|^2\|_2
	\le C_RN^{-1/2}\rho .
\]
Since $|\alpha_N|\le1$, it follows that
$|\langle z,S_{112}z\rangle|
\le C_R C_0^rN^{-1/2}\rho^{r+1}L\|z\|$.
All three bounds are at most
$C_R C_0^r\rho^{r+1}Q_\psi[z]$.

For $S_{111}$, let $C_r(x)$ be the canonical one-field
term of $T_r(x)$. Its kernel has norm at most
$\rho^ra(x)/r!$ and its number shift is $\pm1$.
The adjoint of \eqref{eq:linear-canonical-remainder-degree}
gives
\[
	\|M^{-1/2}(T_r(x)-C_r(x))M^{-1/2}\|
	\le C N^{-1/2}(1+r)(C_1\rho)^ra(x) .
\]
At $r=0$ this difference is zero.
Multiplication on the left by $\alpha_N$ preserves the bound.
The remaining difference $(\alpha_N-1)C_r(x)$ is controlled
by $|\alpha_N-1|\le CM/N$ and its single number shift.
On an output sector $m$, its balanced norm is at most
$C\rho^ra(x)\sqrt{m+1}/(Nr!)$.
Thus
\begin{equation}\label{eq:cubic-extraction-canonical}
	\|M^{-1/2}(\alpha_NT_r(x)-C_r(x))M^{-1/2}\|
	\le C N^{-1/2}(1+r)(C_1\rho)^ra(x),\qquad r\ge0 .
\end{equation}
The absolute value of its scalar kernel is
$F_N(x-y)|\psi(y)k_\psi(x,y)|$. Since
$|k_\psi(x,y)|\le N|\psi(x)\psi(y)|$,
\[
	\mathcal I_N(|k_\psi(x,y)|a(x))
	\le N\|a\psi\|_2\|F_N*|\psi|^2\|_2
	\le C_R\sqrt N\rho .
\]
The factor $\sqrt N$ cancels the $N^{-1/2}$ in
\eqref{eq:cubic-extraction-canonical}.
After the canonical term is subtracted, the $M$-balanced
norm of $S_{111}$ is therefore at most
$C_R(1+r)(C_1\rho)^r\rho$.
The canonical terms are precisely those in
\cite[Eqs.~(5.112)--(5.113)]{BS}.
Multiplication by $1/(r+1)$ changes their factorial
from $r!$ to $(r+1)!$.
Their even and odd sums give the two field terms in
\cite[Eq.~(5.114)]{BS}, which cancel the corresponding
linear extraction already displayed above.
The remainders have a geometric majorant uniform on
the $H^2$ neighborhood.

It remains to check the first, finite term of (5.100).
The exact commutator is
\[
	[B,b_x^*]=-b(\eta_x)\theta_N+D_x^*,
	\qquad \theta_N=(N-\cN)/N .
\]
In $b(\eta_x)\theta_Na_y^*a_x$, first move $\theta_N$
past $a_y^*$ and then use $b(\eta_x)=\beta_N(\cN)a(\eta_x)$.
The two number shifts satisfy
$\theta_Na_y^*=a_y^*\nu_N$ and $\nu_Na_x=a_x\theta_N$.
Hence the full contraction identity is
\begin{align*}
	b(\eta_x)\theta_Na_y^*a_x
	&=\beta_N(\cN)a(\eta_x)a_y^*\nu_Na_x\\
	&=\beta_N(\cN)a_y^*a(\eta_x)\nu_Na_x
	+\overline{\eta(x,y)}\,\beta_N(\cN)\nu_Na_x\\
	&=\beta_N(\cN)a_y^*a(\eta_x)\nu_Na_x
	+\overline{\eta(x,y)}\,b_x\theta_N .
\end{align*}
On an input sector $n\ge1$, the scalar multiplying $a_x$
in the last contraction is
$\beta_N(n-1)(N-n)/N$; it vanishes at $n=N$.
On the vacuum both sides vanish.
Its $k_\psi$ part is $-\sqrt N b(h_\psi)\theta_N$.
After subtraction of $-\sqrt N b(h_\psi)$, the remainder is
$N^{-1/2}b(h_\psi)\cN=N^{-1/2}Mb(h_\psi)$,
bounded by \eqref{eq:linear-number-product-degree} with $l=j=0$.
The $\mu_\psi$ contraction has a one-field kernel of norm
at most $C_RN^{-1/2}$, since its absolute value is bounded by
$C_R|\psi|(F_N*|\psi|^2)$.
The uncontracted part is bounded by
\[
	C\mathcal I_N(a(x)Y(x)Z(y))
	\le C\|a\|_\infty\|\psi\|_\infty
	\|F_N\|_1\|Y\|_2\|Z\|_2
	\le C_RN^{-1/2}LL_1 .
\]
Finally, \eqref{eq:cubic-extraction-field} bounds the term
with $D_x^*a_y^*a_x$ by
\[
	C N^{-1}\mathcal I_N(a(y)Y(x)Z(x))
	+C\rho\mathcal I_N(Z_{xy}Z(x))
	\le C_RN^{-3/2}LL_1+C_R\rho DL .
\]
These are $Q_\psi$ form bounds. The finite term and the
series above therefore complete the cubic contribution to
\eqref{eq:neighborhood-majorant}. Their adjoints obey the
same bounds, and the preceding polarization argument
extends them to the common form domain.

\textit{Kinetic terms.}
For the kinetic terms, set $a_N(r)=Nw_\ell(Nr)$.
The radial Neumann ground state in
\cite[Eq.~(4.1) and Lemma 4.1]{BS} is defined on $B_{N\ell}$,
normalized by $f_\ell(N\ell)=1$, and satisfies
$\partial_\nu f_\ell=0$ on $\partial B_{N\ell}$.
It is extended by $1$ outside that ball. Accordingly,
$w_\ell=1-f_\ell$ is extended by zero and has both
zero value and zero normal derivative at the boundary.
After scaling, $a_N$ is supported in $B_\ell$ and
$a_N=\partial_\nu a_N=0$ on $\partial B_\ell$.
For a smooth compactly supported test function, Green's
identity on $B_\ell$ therefore has zero boundary terms.
There is no surface distribution in its Laplacian:
\begin{equation}\label{eq:scaled-Neumann-Laplacian}
	\Delta a_N(r)=N^3 \{-\tfrac12v(Nr)f_\ell(Nr)+\lambda_\ell f_\ell(Nr)\} \mathbf1_{|r|\le\ell}.
\end{equation}
Since $0\le f_\ell\le1$ and
$N^3\lambda_\ell\le C_\ell$ by
\cite[Lemma 4.1(i)]{BS}, this gives
\[
	\|\Delta a_N\|_1 \le\tfrac12\|v\|_1+N^3\lambda_\ell|B_\ell| \le C_\ell .
\]
In particular the matching sphere is $|r|=\ell$ in
these scaled coordinates. The other Neumann estimates give
\[
	\|a_N\|_1+\|a_N\|_{3/2}+\|\nabla a_N\|_1 +\|\Delta a_N\|_1\le C,\qquad |a_N(r)|\le C/|r|,\quad |\nabla a_N(r)|\le C/|r|^2.
\]
Write $\nabla k=U+V$ with kernels
$U(x,y)=\nabla\psi(x)a_N(x-y)\psi(y)$ and
$V(x,y)=\psi(x)\nabla a_N(x-y)\psi(y)$.
Hardy's inequality bounds $\|U\|_{\mathrm{HS}}$ uniformly;
Young's inequality bounds $\|V\|_{\mathrm{op}}$ uniformly.
The kernel of $V^*V$ is bounded in absolute value by
$C|\psi(z)\psi(y)|/|z-y|$, since the convolution of
two $|x|^{-2}$ kernels in three dimensions is
$C|x|^{-1}$. Hardy's inequality bounds its
Hilbert--Schmidt norm. The three products involving
$U$ are Hilbert--Schmidt by the preceding bounds.
Thus the gradient contractions in
\cite[Eq.~(5.71)]{BS} are uniform on this neighborhood.
Formula \eqref{eq:eta-finite-rank-correction} writes
$\mu=\eta-k$ as a sum of three rank-one kernels with
uniformly bounded $H^2$ factors, so
$\|\nabla\mu\|_{\mathrm{HS}}\le C$. For each term $f(x)g(y)$,
integration by parts gives
\[
	|\langle\nabla_x k,\nabla f\otimes g\rangle| =|\langle k,-\Delta f\otimes g\rangle| \le\|k\|_{\mathrm{HS}}\|\Delta f\|_2\|g\|_2\le C.
\]
Expanding the square therefore controls
$\|\nabla\eta\|_{\mathrm{HS}}^2-\|\nabla k\|_{\mathrm{HS}}^2$
uniformly. The same formula shows
$\|\nabla\eta\|_{\mathrm{op}}\le C$, by the bounds on $U,V$
and $\nabla\mu$.
End Laplacians of kernel products require a further
operator estimate. For $f\in L^2$, differentiating the
output coordinate of $k$ gives the following three bounds:
\begin{align*}
	\|\psi((\Delta a_N)*(\psi f))\|_2
	&\le\|\psi\|_\infty^2\|\Delta a_N\|_1\|f\|_2,\\
	\|(\nabla\psi)((\nabla a_N)*(\psi f))\|_2
	&\le C\|\nabla\psi\|_3\|\psi\|_\infty\|f\|_2,\\
	\|(\Delta\psi)(a_N*(\psi f))\|_2
	&\le\|\Delta\psi\|_2\|a_N\|_2\|\psi\|_\infty\|f\|_2.
\end{align*}
The middle line uses the Hardy--Littlewood--Sobolev inequality
\cite{LiebLoss}, in the form of the map $|x|^{-2}*:L^2\to L^6$ and
$|\nabla a_N(x)|\le C|x|^{-2}$. The last line uses
$a_N*:L^2\to L^\infty$. The scattering equation gives
the uniform $L^1$ bound on $\Delta a_N$, including its
Neumann-eigenvalue part. The matched extension has no
boundary measure. The finite-rank formula for $\mu$
also gives $\|\Delta_x\mu\|_{\mathrm{HS}}\le C$. Consequently
\begin{equation}\label{eq:eta-end-operator}
	\|\nabla_x\eta\|_{\mathrm{op}}+\|\Delta_x\eta\|_{\mathrm{op}} \le C_R .
\end{equation}
The input-coordinate estimates follow by adjunction
and complex conjugation. For every alternating product
$\eta^{(l)}$ with $l\ge2$ this proves
\begin{equation}\label{eq:eta-product-endpoint}
	\|\nabla_j\eta^{(l)}\|_{\mathrm{HS}} +\|\Delta_j\eta^{(l)}\|_{\mathrm{HS}} \le C_R\|\eta\|_{\mathrm{HS}}^{\,l-2},\qquad j=1,2,
\end{equation}
by composing the differentiated end factor with its
HS neighbor, and bounding all further factors in
operator norm. These are the estimates for derivatives at the ends of kernel products in
\cite[Lemma 4.3(i)]{BS}, now with constants controlled on the
$H^2$ ball.
The uncontracted gradient has the uniform scaled bound
\begin{equation}\label{eq:kinetic-gradient-scaled}
	N^{-1}\|\nabla_x\eta_\psi\|_{\mathrm{HS}}^2\le C_R .
\end{equation}
Indeed \cite[Lemma 4.1(iii)]{BS} gives
$|\nabla a_N(r)|\le CN^2/(1+N^2|r|^2)$.
Integration in three dimensions yields
$\|\nabla a_N\|_2^2\le CN$.
The term with $\nabla\psi$ is bounded by Hardy's inequality,
while the other term obeys
$\|\psi(x)\nabla a_N(x-y)\psi(y)\|_2^2
\le\|\psi\|_\infty^2\|\psi\|_2^2\|\nabla a_N\|_2^2$.
Together with the bound on $\nabla\mu$, this proves
\eqref{eq:kinetic-gradient-scaled}.
We estimate the two amplitude-derivative terms in
\cite[Eq.~(5.75)]{BS} by Hardy's inequality.
On the finite-particle form core, Hardy's inequality gives
\[
	\int\frac{\|a_xa_y z\|^2}{|x-y|^2}\d x\d y \le C N\|\cK^{1/2}z\|^2.
\]
Consequently, Cauchy--Schwarz yields
\begin{align*}
	\int\frac{|\Delta\psi(x)||\psi(y)|}{|x-y|}
	\|a_xa_yz\|\d x\d y
	&\le C\sqrt N\|\Delta\psi\|_2\|\psi\|_2
	\|\cK^{1/2}z\|,\\
	\int\frac{|\nabla\psi(x)||\psi(y)|}{|x-y|^2}
	\|a_xa_yz\|\d x\d y
	&\le C\sqrt N\|\nabla\psi\|_2\|\psi\|_{H^1}
	\|\cK^{1/2}z\|.
\end{align*}
For the second line use Hardy's inequality in $y$ in the
coefficient factor. The retained $N^{-1/2}\|M^{1/2}z\|$
in the estimate following \cite[Eq.~(5.87)]{BS} absorbs
$\sqrt N$ and leaves a $Q$ form bound. The interaction
piece of $\Delta a_N$ is paired with $\cV_N^{1/2}$ exactly
as in \cite[Eq.~(5.77)]{BS}. These estimates also hold for
the adjoint terms by reversing the form pairing.
More explicitly, put $Z_{xy}=\|a_xa_yz\|$ and
$W_N(r)=N^2v(Nr)$ as above.
For the interaction part of $\Delta a_N$,
\begin{align*}
	\int N^3v(N(x-y))|\psi(x)\psi(y)|Z_{xy}\d x\d y
	&\le N\left(\int W_N(x-y)|\psi(x)\psi(y)|^2\d x\d y\right)^{1/2}
	\left(\int W_N(x-y)Z_{xy}^2\d x\d y\right)^{1/2}\\
	&\le C_R\sqrt N\|\cV_N^{1/2}z\| .
\end{align*}
Here $\|W_N\|_1=N^{-1}\|v\|_1$.
The Neumann-eigenvalue part has coefficient $N^3\lambda_\ell\le C$.
Its integral is at most
$C_R\|\psi\otimes\psi\|_2\|Z_{xy}\|_2
\le C_R\|Mz\|\le C_R\sqrt{N+1}\|M^{1/2}z\|$.
Combining these terms with the two amplitude derivatives gives
\begin{equation}\label{eq:kinetic-bare-Laplacian}
	\int|\Delta_x k_\psi(x,y)|\|a_xa_yz\|\d x\d y
	\le C_R\sqrt N\|Q_\psi^{1/2}z\| .
\end{equation}
We give the form estimates for the two orders extracted before
the kinetic series in \cite[Eq.~(5.66)]{BS}.
Write $L=\|M^{1/2}z\|$ and $K_0=\|\cK^{1/2}z\|$.
First, the projection correction satisfies
\[
	\left|\int\langle z,b(\nabla_x\mu_x)\nabla_xb_xz\rangle\d x\right|
	\le\|\nabla_x\mu\|_{\mathrm{HS}}LK_0\le C_RLK_0.
\]
For the $k$ part, integration by parts gives the singular
scattering pair already extracted in \eqref{eq:scaled-Neumann-Laplacian}
and the two amplitude terms of \cite[Eq.~(5.68)]{BS}.
Put $s_x(y)=a_N(x-y)\psi(y)$. Hardy's inequality gives
\[
	\sup_x\|s_x\|_2\le C\|\nabla\psi\|_2.
\]
Thus the term with $\Delta\psi$ has absolute form value at most
$C\sup_x\|s_x\|_2\|\Delta\psi\|_2L^2$.
For the term with $\nabla a_N\cdot\nabla\psi$, integrate by
parts once more in $x$. The derivative acts on $\nabla\psi$
or on $b_x$. The two resulting form values are bounded by
\[
	C\sup_x\|s_x\|_2
	\bigl(\|\Delta\psi\|_2L^2+\|\nabla\psi\|_2LK_0\bigr).
\]
These estimates include the adjoints and are uniform on the
$H^2$ ball. They use the form pairing of two fields, as in
\cite[Eq.~(5.69)]{BS}.

The two products involving $b(\nabla\eta_x)b^*(\nabla\eta_x)$
can be evaluated on each number sector.
Set $g_{x,j}=\partial_{x_j}\eta(x,\cdot)$ and
\[
	G_N=\sum_{j=1}^3\int\|g_{x,j}\|_2^2\d x,
	\qquad D_\eta=\sum_{j=1}^3\int
	|g_{x,j}\rangle\langle g_{x,j}|\d x.
\]
The latter is a positive operator. The preceding gradient bounds give
$G_N\le C_RN$, $\|D_\eta\|\le C_R$ and
$|G_N-\|\nabla_xk\|_{\mathrm{HS}}^2|\le C_R$.
With $\theta_n=1-n/N$, the convention
$b(f)=\sqrt{1-\cN/N}\,a(f)$ gives the exact sector identities
\begin{align}\label{eq:kinetic-isolated-number-identities}
	\left.\sum_j\int b(g_{x,j})b^*(g_{x,j})\d x\right|_{\cF^{(n)}}
	&=\theta_n\bigl(G_N+\dGamma(D_\eta)\bigr),\notag\\
	\left.N^{-1}\sum_j\int b(g_{x,j})\cN b^*(g_{x,j})\d x\right|_{\cF^{(n)}}
	&=\frac{n+1}{N}\theta_n
	\bigl(G_N+\dGamma(D_\eta)\bigr).
\end{align}
The first identity follows from
$a(f)a^*(f)=a^*(f)a(f)+\|f\|^2$; the intermediate number
in the second identity is $n+1$.
For $0\le n\le N$, subtracting
$\|\nabla_xk\|_{\mathrm{HS}}^2$ from the first line leaves
an operator of norm at most $C_R(n+1)$.
The second line has the same bound, including $n=0$ and $n=N$.
These are the scalar and number terms in
\cite[Eqs.~(5.70)--(5.72)]{BS}.

It remains to bound the four-field term
\[
	\mathcal T=\frac1{2N}\sum_j\int\eta(z,y)
	b(g_{x,j})b_y^*a_z^*\partial_{x_j}a_x\d x\d y\d z.
\]
Its number shift is zero. On sector $n$, the two generalized
fields contribute the common factor $\theta_n$.
Ordinary normal ordering and symmetry of $\eta$ therefore give
\begin{equation}\label{eq:kinetic-isolated-four-field}
	\mathcal T\big|_{\cF^{(n)}}
	=\frac{\theta_n}{2N}\mathcal P^*(\eta)\mathcal A_\eta
	+\frac{\theta_n}{N}\mathcal R_\eta,
\end{equation}
where
\begin{align*}
	\mathcal P^*(\eta)&=\int\eta(z,y)a_y^*a_z^*\d y\d z,\\
	\mathcal A_\eta&=\sum_j\int a(g_{x,j})\partial_{x_j}a_x\d x,\\
	\mathcal R_\eta&=\sum_j\int\eta(z,y)\overline{\partial_{x_j}\eta(x,y)}
	a_z^*\partial_{x_j}a_x\d x\d y\d z.
\end{align*}
We use $a(f)=\int\overline{f(x)}a_x\d x$ throughout.
Both contractions in the normal ordering are equal after
interchanging $y,z$, which accounts for the factor two.
The identity also holds at $n=N$, where $b_y^*$ vanishes
after $a_z^*\partial_{x_j}a_x$.

Put $\rho=\|\eta\|_{\mathrm{HS}}$ and $z=z_n$ for these estimates.
The pair-annihilation bound gives
$\|\mathcal P(\eta)z_n\|\le\rho\sqrt{n(n-1)}\|z_n\|$.
Split $\mathcal A_\eta=\mathcal A_k+\mathcal A_\mu$.
Cauchy--Schwarz in $x$ and the annihilation bound imply
\[
	\|\mathcal A_\mu z_n\|\le
	C\sqrt N\|\nabla_x\mu\|_{\mathrm{HS}}\|\cK^{1/2}z_n\|.
\]
Integration by parts and \eqref{eq:kinetic-bare-Laplacian} give
\[
	\mathcal A_k=\int\overline{-\Delta_xk(x,w)}a_wa_x\d x\d w,
	\qquad \|\mathcal A_kz_n\|\le C_R\sqrt N\|Q_\psi^{1/2}z_n\|.
\]
Since $n/\sqrt N\le\sqrt n$ on the physical sectors,
the first term in \eqref{eq:kinetic-isolated-four-field}
has absolute form value at most
$C_R\rho\|M^{1/2}z_n\|\|Q_\psi^{1/2}z_n\|$.
For the contraction term, Cauchy--Schwarz in $(x,y,z)$ gives
\[
	N^{-1}|\langle z_n,\mathcal R_\eta z_n\rangle|
	\le N^{-1}\rho G_N^{1/2}
	\|\cN^{1/2}z_n\|\|\cK^{1/2}z_n\|
	\le C_RN^{-1/2}\rho
	\|M^{1/2}z_n\|\|\cK^{1/2}z_n\|.
\]
Summing sectors and adding the adjoint yields
\[
	|\langle z,(\mathcal T+\mathcal T^*)z\rangle|
	\le C_R\|M^{1/2}z\|\|Q_\psi^{1/2}z\|.
\]
Thus all isolated kinetic terms satisfy the required form bound.
The calculation starts on the finite-particle smooth core;
polarization and closure extend the forms to $D(Q_\psi^{1/2})$.
Their number shifts are $0,2,-2$, so fixed number weights
only change the constants. After the scalar and scattering
subtractions, their coefficient expressions have fixed finite
degree in $\psi,\overline\psi$ and their spatial derivatives.
The common-parameter argument below therefore applies to these
terms with the same $H^2$ neighborhood bounds.

We summarize the estimates used from the static proof of \cite{BS},
grouped by operator type. The last column
lists all spatial norms used after the stated extractions;
the field and number combinatorics are those of
\cite[Lemmas 3.2 and 5.1]{BS}.
\begin{center}
\small
\begin{tabular}{p{.16\linewidth}p{.23\linewidth}p{.49\linewidth}}
\hline
Family & Location in \cite{BS} & Spatial norms on an $H^2$
ball\\
\hline
Linear & (5.52)--(5.58) &
$\|h_\psi\|_2$, $\|\psi\|_\infty$,
the HS and row norms of $\eta$.\\
Kinetic & (5.66)--(5.87) &
$\|\nabla\mu\|_{\mathrm{HS}}$, the scalar difference above,
$N^{-1/2}\|\nabla\eta\|_{\mathrm{HS}}$,
$\|\nabla\eta\|_{\mathrm{op}}$,
$\|(\nabla\eta)^*(\nabla\eta)\|_{\mathrm{HS}}$,
\eqref{eq:eta-product-endpoint}, and the uncontracted
end-Laplacian bounds.\\
Pair & (5.93)--(5.97) &
$\|N^3v(N\cdot)\|_1$, $\|\psi\|_{2\cap\infty}$,
the HS and row norms, and the scattering pair extraction.\\
Cubic & (5.102)--(5.114) &
The same norms, $\|N^{5/2}v(N\cdot)\|_1=O(N^{-1/2})$,
and \eqref{eq:cubic-four-products},
\eqref{eq:cubic-dressed-source-integrals} and
\eqref{eq:cubic-bare-source-integrals}.\\
Quartic & (5.119)--(5.126) &
$\|N^2v(N\cdot)\|_1=O(N^{-1})$,
$\|\psi\|_\infty$, HS and row norms; the remaining
interaction pair is kept in $\cV_N^{1/2}$.\\
\hline
\end{tabular}
\end{center}
Here $\|h_\psi\|_2\le
\|N^3v(N\cdot)w_\ell(N\cdot)\|_1\|\psi\|_\infty^2
\|\psi\|_2\le C$.
For the kinetic line, the isolated orders are controlled by
\eqref{eq:kinetic-isolated-number-identities} and
\eqref{eq:kinetic-isolated-four-field}, together with the
amplitude pair estimates above.
We now retain the kernel degree in each of the six
cases following \cite[Eq.~(5.78)]{BS}.
Set $J=n+k$, $\rho=\|\eta_\psi\|_{\mathrm{HS}}\le1$,
$L=\|M^{1/2}z\|$, $K_0=\|\cK^{1/2}z\|$ and
$E_0=\|Q_\psi^{1/2}z\|$.
Each word contains $J+1$ kernels.
Its two differentiated terminal chains have lengths
$l_1+1$ and $l_2$; the remaining chains contain
$J-l_1-l_2$ kernels.
These counts follow from \cite[Lemma 3.2]{BS}.
The field estimates of \cite[Lemmas 2.1 and 2.4]{BS}
leave the factor $C_0^J\rho^{J-l_1-l_2}$ outside
the two terminal estimates.
The diagonal terms in these field estimates have chain
length $m\ge2$ by \cite[Lemma 3.2(vi)]{BS}.
Hilbert--Schmidt Cauchy--Schwarz gives
$\int|\eta^{(m)}(x,x)|\d x\le\rho^m$,
so they retain the same degree count.

For the first four cases, the resulting bounds for one word
are as follows; the common factor $C_RC_0^J$ is suppressed:
\[
	\begin{array}{c|c|c}
	(l_1,l_2)&\text{terminal factors}&\text{bound}\\
	\hline
	l_1\ge1,\ l_2\ge2&
	\|\nabla\eta^{(l_1+1)}\|_{\mathrm{HS}}
	\|\nabla\eta^{(l_2)}\|_{\mathrm{HS}}&\rho^{J-3}L^2\\
	l_1\ge1,\ l_2=1&
	\|\Delta\eta^{(l_1+1)}\|_{\mathrm{HS}}\|\eta\|_{\mathrm{HS}}&
	\rho^{J-1}L^2\\
	l_1\ge1,\ l_2=0&
	\|\nabla\eta^{(l_1+1)}\|_{\mathrm{HS}}&\rho^{J-1}LK_0\\
	l_1=0,\ l_2\ge2&
	\|\eta\|_{\mathrm{HS}}\|\Delta\eta^{(l_2)}\|_{\mathrm{HS}}&
	\rho^{J-1}L^2
	\end{array}
\]
In the first row, \eqref{eq:eta-product-endpoint}
contributes $\rho^{l_1-1}\rho^{l_2-2}$, so $J\ge3$.
In the second and fourth rows, move the derivative off
the single $\eta$ kernel. The two remaining norms contribute
$\rho^{l_1-1}\rho$ and $\rho\rho^{l_2-2}$, respectively.
In the third row, keep $\nabla_xa_x$ as the last field and use
$\int\|\nabla_xa_xz\|^2\d x=K_0^2$.
Thus every displayed exponent is nonnegative on its index set.

For the fifth case, $l_1=0,l_2=1$, use the decomposition
$E=F_2+F_{11}+F_{12}$ in
\cite[Eqs.~(5.82)--(5.85)]{BS}.
Here $J\ge2$, since the pair $(n,k)=(0,1)$ has been extracted.
The contraction in $F_{11}$ consumes two kernels and gives
$\|\nabla\eta\|_{\mathrm{HS}}^2$.
There remain $J-1\ge1$ kernels in the preceding field factors.
Let $\alpha$ be the sum of the orders of their $\Pi^{(1)}$
and $\Pi^{(2)}$ factors, as in (5.85).
The kernel count gives $1\le\alpha\le J-1$.
Retaining the physical-sector endpoint yields
\[
	N^{-\alpha}\|M^{\alpha/2}z\|^2
	\le \frac{2^{\alpha-1}}{N}L^2 .
\]
Together with \eqref{eq:kinetic-gradient-scaled}, this bounds
$F_{11}$ by $C_RC_1^J\rho^{J-1}L^2$.
The uncontracted pair in $F_{12}$ is
\[
	\int a^*(\nabla\eta_x)a(\nabla\eta_x)\d x
	=\dGamma(D_\eta),\qquad
	0\le\dGamma(D_\eta)\le C_R\cN .
\]
Here $D_\eta$ is defined by
$\langle f,D_\eta f\rangle
=\int|\langle\nabla\eta_x,f\rangle|^2\d x$.
Its operator norm is bounded by \eqref{eq:eta-end-operator};
complex conjugation preserves that norm.
The preceding factors contain $2\alpha$ fields.
Write their ordered product as $P=P_s\cdots P_1$.
Include the number coefficients among these factors.
Let $p_i$ be the order of $P_i$ and $j_i$ its kernel degree;
for a number coefficient both are zero.
Then $\sum_i p_i=\alpha$, $\sum_i j_i=J-1$ and
$s\le C(J+1)$.
Each factor has a definite number shift $d_i$ with
$|d_i|\le2p_i$.

Keep the two marked terminal fields when locating the
sector cutoffs in (5.80)--(5.84).
The uncontracted term preserves the relative order of all
unmarked fields.
Removing either field changes the intermediate number by one.
Moving a marked field past a number coefficient shifts
that coefficient's argument by the same one unit.
Thus every nonzero intermediate sector at a factor boundary
has at most $N+2$ particles.
Inside a chain, the interior field pairs preserve number,
so a partial chain changes this number by at most two more.
The pair estimates of \cite[Lemmas 2.1 and 2.4]{BS}
therefore use $M\le N+5$ on all these intermediate sectors.
Since $(N+5)/N\le6$, they give the block bound
\[
	\|\Pi_{m_i}N^{-p_i}P_i\Pi_{m_{i-1}}\|
	\le C^{p_i+1}\rho^{j_i},\qquad m_i=m_{i-1}+d_i .
\]
Here $\Pi_m$ is the $m$-particle projection.
Negative sectors vanish. The shifted generalized-field
coefficients impose the upper cutoffs just described.
For a fixed input number $m$, the intermediate sequence
$m_i=m+\sum_{h\le i}d_h$ is unique.
Induction through this sequence gives
\[
	\|\Pi_{m+d}N^{-\alpha}P\Pi_m\|
	\le\prod_{i=1}^s C^{p_i+1}\rho^{j_i}
	=C^{\alpha+s}\rho^{J-1}\le C_1^J\rho^{J-1},
	\qquad d=\sum_i d_i .
\]
For this single word, the total shift $d$ is fixed.
Distinct inputs $m$ have distinct outputs $m+d$.
Orthogonality therefore gives the same bound for the norm
of $N^{-\alpha}P$ on the physical space.
On each nonzero $m\to m+d$ block, conjugation by
$M^{-1/2}$ multiplies the norm by
$\sqrt{(m+1)/(m+d+1)}\le(1+|d|)^{1/2}$,
with $|d|\le2J+2$.
Since $M^{-1}\dGamma(D_\eta)$ is bounded,
factor the balanced form as
\[
	M^{-1/2}N^{-\alpha}P\,\dGamma(D_\eta)M^{-1/2}
	=\bigl(M^{-1/2}N^{-\alpha}PM^{1/2}\bigr)
	\bigl(M^{-1}\dGamma(D_\eta)\bigr) .
\]
This bounds $F_{12}$ by
$C_R(J+1)^{1/2}C_1^J\rho^{J-1}L^2$.
If terms with several total shifts are grouped, apply the
triangle inequality to their at most $4J+5$ shift components.
This adds only a polynomial factor in $J+1$.

The commutator terms in $F_2$ have the form (5.82).
Commuting the differentiated field through a chain of
$m\ge1$ kernels produces $\nabla\eta^{(m+1)}$.
Move the other derivative onto this chain.
Its norm is then at most $C_R\rho^{m-1}$ by
\eqref{eq:eta-product-endpoint}.
The other terminal kernel contributes $\rho$, and the
remaining chains contribute $\rho^{J-m-1}$.
Their product is $\rho^{J-1}$.
There are at most $C(J+1)$ commutation positions.
Thus this part has the bound
$C_R(J+1)C_1^J\rho^{J-1}L^2$.

For the sixth case, $l_1=l_2=0$, use
$E=G_2+G_{12}+G_{11}$ from
\cite[Eqs.~(5.86)--(5.87)]{BS}.
The terms in $G_2$ contain the same differentiated chain.
There is now a bare terminal field; the other chains have
total degree $J-m$.
The bound $\rho^{m-1}\rho^{J-m}=\rho^{J-1}$ and the
same commutation count give
$C_R(J+1)C_1^J\rho^{J-1}L^2$.
The term $G_{12}$ uses $\nabla\mu$ and is at most
$C_RC_1^J\rho^{J-1}LK_0$.
For $G_{11}$, the bare Laplacian consumes one kernel.
There is at least one remaining kernel because $J\ge1$.
The order count in (5.87) therefore gives $1\le\alpha\le J$.
On $\cF^{\le N}$,
$N^{-\alpha}\|M^\alpha z\| \le 2^{\alpha-1/2}N^{-1/2}L$.
Equation (5.87), with \eqref{eq:kinetic-bare-Laplacian}, now yields
\begin{align*}
	|\langle z,G_{11}z\rangle|
	&\le C_RC_1^J\rho^{J-1}N^{-1/2}L
	\int|\Delta_xk_\psi(x,y)|Z_{xy}\d x\d y\\
	&\le C_RC_1^J\rho^{J-1}LE_0 .
\end{align*}

All six cases are bounded by
$C_R(J+1)^bC_1^J\rho^{\max(J-3,0)}Q_\psi[z]$.
At $\rho=0$, take the continuous extension of these
nonnegative powers.
For each $(n,k)$, the count $2^Jn!k!$ is paired with the
coefficient $1/[n!k!(J+1)]$ in (5.66).
There are $J+1$ pairs of total degree $J$.
The commutation positions and number shifts add polynomial
factors, so the sum satisfies \eqref{eq:neighborhood-majorant}
with $b_0=3$ for the kinetic family.
Spatial integrations by parts are first made with smooth
amplitudes and form-core vectors.
The uniform bounds above and density in the normalized
$H^2$ neighborhood give the same forms for $H^2$ amplitudes.
Polarization then extends them to the common form domain.

\textit{Summation of the kernel degrees.}
To track the uniform geometric factor, retain
$\|\eta_\psi\|_{\mathrm{HS}}^{\max(J-b_0,0)}$ in each
elementary word, as in Section~\ref{sec:number-compatibility}. The finitely many
row or gradient endpoints remove at most $b_0$ factors.
The $2^j j!$ word count is cancelled by the series
factorials up to a fixed exponential, and the number
of index tuples and endpoint choices is polynomial
in $J$. The common constants in the last column are
uniform on the chosen ball. This gives, for the
cancelled groups $E_\iota(\psi)$ used in Section~\ref{sec:number-compatibility},
\begin{equation}\label{eq:neighborhood-majorant}
	\sum_{J(\iota)=J}\sup_{\psi\in\mathcal U} \|Q_\psi^{-1/2}E_\iota(\psi)Q_\psi^{-1/2}\| \le C(1+J)^b\vartheta^{\max(J-b_0,0)},\qquad \vartheta<1 .
\end{equation}
The unchanged scalar and scattering cancellations take
place inside the groups, for every $\psi\in\mathcal U$.
The neighborhood can be chosen uniformly in $N$.
Indeed $\|a_N\|_2\le C_\ell$, by
$|a_N(r)|\le C|r|^{-1}\mathbf1_{|r|\le\ell}$.
On a bounded $H^2$ ball, splitting the two amplitude
differences gives
\[
	\|k_\psi-k_\chi\|_{\mathrm{HS}} \le \|a_N\|_2(\|\psi\|_\infty+\|\chi\|_\infty) \|\psi-\chi\|_2.
\]
For normalized vectors
$\|q_\psi-q_\chi\|_{\mathrm{op}}\le2\|\psi-\chi\|_2$.
Applying the tensor projections to the preceding
difference therefore yields
\begin{equation}\label{eq:eta-local-lipschitz}
	\|\eta_\psi-\eta_\chi\|_{\mathrm{HS}} \le C_R\|\psi-\chi\|_{H^2},
\end{equation}
uniformly in $N$. The strict small-kernel margin at
the background hence persists on a common ball.
The cancellations hold algebraically for each normalized
$\psi$. The scattering profile is independent of
$\psi$. The large linear terms contain the modified GP
vector field $-\Delta\psi+(v_{\mathrm{eff},N}*|\psi|^2)\psi$, and the scalar is the polynomial in
$\psi,\bar\psi$ specified in \cite[Eq.~(4.25)]{BS}. The original strict smallness and the
uniform Lipschitz bound give a radius $\delta_0>0$ and a common
$\vartheta<1$ in \eqref{eq:neighborhood-majorant}, independent
of $N$. The finitely many spatial constants are bounded on the
same $H^2$ ball.

\textit{Parameter derivatives on the fixed form space.}
We estimate the derivatives using the polynomial structure
of the groups. Fix a normalized $\psi_0$. For
$\xi\in H^2$ with $\langle\psi_0,\xi\rangle=0$, put
\[
	R_\xi=|\xi\rangle\langle\psi_0| -|\psi_0\rangle\langle\xi|,\quad W_\xi=(1+R_\xi)(1-R_\xi)^{-1},\quad d(\xi)=1+\|\xi\|_2^2.
\]
Then $W_\xi$ is unitary, is the identity off the span of
$\psi_0,\xi$, and
\[
	\psi_\xi=W_\xi\psi_0 =\frac{(1-\|\xi\|_2^2)\psi_0+2\xi}{d(\xi)} .
\]
The identity $R_\xi^3=-\|\xi\|_2^2R_\xi$ gives
\[
	W_\xi=1+\frac{2(R_\xi+R_\xi^2)}{d(\xi)} .
\]
Thus its finite-rank part has a real quadratic numerator.
We regard
$X_0=\{\xi\in H^2:\langle\psi_0,\xi\rangle=0\}$
as a real Banach space. Complex conjugation is then
real linear, and $d(\xi)\ge1$ throughout the chart.
Along $\tau\mapsto W_{\tau\xi}$ its generator is
$2R_\xi/(1+\tau^2\|\xi\|_2^2)$.
Let $z_\tau=\Gamma(W_{\tau\xi})z$ and evaluate the
ambient form $\cK+\cV_N+M$ on $z_\tau$. The rank-one
estimate in \eqref{eq:Q-two-derivatives}, applied to this
radial generator, gives
$|\partial_\tau Q[z_\tau]|\le C_R Q[z_\tau]$.
Gronwall's inequality on $0\le\tau\le1$ proves
\[
	C_R^{-1}Q_{\psi_0}[z]\le Q_{\psi_\xi}[\Gamma(W_\xi)z]\le C_R Q_{\psi_0}[z],
\]
uniformly in $N$ and in the chart ball.
For fixed $N$, the potential is a bounded multiplier and
the form domain on each sector is the symmetric $H^1$
space. Finite-rank operators with $H^2$ vectors preserve
this space. We first differentiate on finite sums of
symmetrized $H^2$ tensors, projected with $q_{\psi_0}$.
They form a core, and $\Gamma(W_{\tau\xi})$ maps them to
the corresponding moving core. The form estimate above
extends the radial identity to the whole form domain.
Thus all subsequent operator norms refer to the fixed
space $V_N=D(Q_{\psi_0}^{1/2})$.
Only the active field operators of a monomial are transformed; the
spectator transformations cancel. Their number is $O(J+1)$.
Consequently, for some integers $a_\iota\le C(J+1)$,
\[
	d(\xi)^{a_\iota} \Gamma(W_\xi)^*E_\iota(\psi_\xi)\Gamma(W_\xi)
\]
is a real polynomial of degree $O(J+1)$ in $\xi$,
valued in $\mathcal B(V_N,V_N^*)$ on the fixed
$\psi_0$ space. This identity follows on the form
core and extends by interpolation at finitely many
parameter values. A cancellation group uses one common
denominator before this step.
To see the degree count, each $k_\psi$ has two
amplitudes, each $q_\psi$ has degree two, and
$\eta_\psi=(q_\psi\otimes q_\psi)k_\psi$ has degree
at most six in $(\psi,\bar\psi)$. A word of total
kernel degree $J$ has $O(J+1)$ such factors and active field operators. Scattering profiles and number
coefficients are independent of the parameter. Spatial
derivatives preserve the parameter degree.
Substitution of $\psi_\xi$ and transformation of an active
field operator each supply a numerator of degree at most two and a power
of $d^{-1}$. Thus both the denominator
power and the numerator degree are bounded by a
fixed constant times $J+1$. This argument includes
the scalar subtractions and the compressed forms.

To extend each word to the form space, write its numerator as $\sum_{k=0}^m p_k(\xi)$,
where $p_k$ is homogeneous of degree $k$ on the core.
For each unit direction $h$, interpolation at $m+1$
fixed points of $t\mapsto p(th)$ on $[-\delta,\delta]$
expresses $p_k(h)$ as a finite linear combination of
bounded form values. The resulting bound is uniform
in $h$: if $B_\iota$ is the numerator supremum on the
chart ball, then
\[
	\sup_{\|h\|_{H^2}\le1}\|p_k(h)\|_{\mathcal B(V_N,V_N^*)}
	\le C_m\delta^{-k}B_\iota .
\]
Real polarization bounds the associated $k$-linear map
by $(k^k/k!)C_m\delta^{-k}B_\iota$ for $k\ge1$.
It therefore extends continuously from $X_0^k$ into
$\mathcal B(V_N,V_N^*)$. Density identifies
its values with the original form on the chart ball.
This constructs a continuous polynomial for each word.
The estimates below follow directly from its supremum
norm and are uniform in the degree.

Put $b_\iota=\sup_{\psi\in\mathcal U}
\|Q_\psi^{-1/2}E_\iota(\psi)Q_\psi^{-1/2}\|$.
The chart comparison gives the numerator bound
$C_R(1+\delta^2)^{C(J+1)}b_\iota$ on
$\|\xi\|_{H^2}\le\delta$. Harris's Markov inequality
\cite[Theorem 1]{Harris} applies to real normed spaces
and vector-valued polynomials. For degree $m$, it
bounds diagonal derivatives by $T_m^{(j)}(1)$ times
the supremum norm on the unit ball. In particular,
$T_m'(1)=m^2$ and $T_m''(1)=m^2(m^2-1)/3$.
Real polarization of the second derivative
introduces a factor at most two. After scaling the ball,
\[
	\|D^j p(\xi)\|\le C_j\delta^{-j}m^{2j}
	\sup_{\|\zeta\|_{H^2}\le\delta}\|p(\zeta)\|,
	\qquad j=1,2,\quad \|\xi\|_{H^2}\le\delta .
\]
We regard the complex form space as a real normed
space for this application. Constants are independent
of its dimension and hence of $N$.
Choose $\delta$ small enough that
$\theta=\vartheta(1+\delta^2)^C<1$.
For example, with $C$ enlarged if necessary, take
$0<\delta\le1$ such that $C_R\delta\le\delta_0$ and
$(1+\delta^2)^C<(1+\vartheta^{-1})/2$.
The first condition places the chart ball inside
$\mathcal U$, since $\|\psi_\xi-\psi_0\|_{H^2}\le C_R\delta$.
For $g(\xi)=d(\xi)^{-a}$, direct differentiation gives
\begin{align*}
	Dg[h]&=-a d^{-a-1}Dd[h],\\
	D^2g[h,k]&=a(a+1)d^{-a-2}Dd[h]Dd[k]
	-a d^{-a-1}D^2d[h,k] .
\end{align*}
On this ball, $d\ge1$, $\|Dd\|\le2\delta$ and
$\|D^2d\|\le2$. Thus $\|Dg\|\le2a\delta$ and
$\|D^2g\|\le4a(a+1)\delta^2+2a$.
The product rule restores the denominator with
a factor polynomial in $J$. Summing the resulting
$b_\iota$ bounds and using
\eqref{eq:neighborhood-majorant} gives
\begin{equation}\label{eq:parameter-derivatives}
	\sum_{J(\iota)=J} \left\|D^j\left[ \Gamma(W_\xi)^*E_\iota(\psi_\xi)\Gamma(W_\xi) \right]_{\xi=0}\right\| \le C_j(1+J)^{b_j}\theta^{\max(J-b_0,0)},\quad j\le2 .
\end{equation}
The norm is the fixed balanced form norm.
We transport excitation vectors with $W_\xi$ and
retain the condensate phase in the coefficients:
\[
	\mathcal E_\iota(\alpha,\xi)
	=\Gamma(W_\xi)^*E_\iota(e^{i\alpha}\psi_\xi)\Gamma(W_\xi) .
\]
For a constant phase, $q_{e^{i\alpha}\psi}=q_\psi$,
$k_{e^{i\alpha}\psi}=e^{2i\alpha}k_\psi$ and
$\eta_{e^{i\alpha}\psi}=e^{2i\alpha}\eta_\psi$.
The linear coefficient $h_\psi$ gains $e^{i\alpha}$;
normal one-body kernels and the subtracted scalar
are phase invariant. The cubic coefficient gains
$e^{i\alpha}$, in agreement with its two creation
fields and one annihilation field. These identities
give, term by term in the commutator expansions,
\begin{equation}\label{eq:chart-phase-gauge}
	\mathcal E_\iota(\alpha,\xi)
	=e^{i\alpha\mathcal N}\mathcal E_\iota(0,\xi)e^{-i\alpha\mathcal N} .
\end{equation}
For a term of number shift $d$, the right side is
$e^{id\alpha}\mathcal E_\iota(0,\xi)$.
One or two phase derivatives therefore introduce factors bounded by
$2J+4$ or $(2J+4)^2$. Mixed phase and chart derivatives
satisfy the same geometric majorant. The number gauge
is unitary in $V_N$ since $[M,Q_{\psi_0}]=0$.
This is a full local chart of the normalized sphere.
If $\psi$ is sufficiently close to $\psi_0$, choose
$\alpha=\arg\langle\psi_0,\psi\rangle$, set
$a=|\langle\psi_0,\psi\rangle|$, and take
$\xi=q_0e^{-i\alpha}\psi/(1+a)$.
Then $\psi=e^{i\alpha}W_\xi\psi_0$.
The chart and its inverse are twice continuously
differentiable on their real Banach spaces, with
uniform bounds on a smaller ball. Along the actual
background, their first two velocities satisfy
\[
	|\dot\alpha|+\|\dot\xi\|_{H^2} \le C_R\|\dot\phi\|_{H^2},\qquad |\ddot\alpha|+\|\ddot\xi\|_{H^2} \le C_R(\|\ddot\phi\|_{H^2}+\|\dot\phi\|_{H^2}^2).
\]
The second chain derivative of the transported form is
$D^2\mathcal E_\iota[\dot\gamma,\dot\gamma]
 +D\mathcal E_\iota[\ddot\gamma]$, with
$\gamma=(\alpha,\xi)$. Thus the degree majorants
control both the acceleration term and the quadratic
velocity term.

We identify these derivatives with the prescribed
covariant ones. Along the chart let $A=W_\xi^*\dot W_\xi$
and $a=q_0Aq_0$. Since
$q=W_\xi q_0W_\xi^*$, direct differentiation gives
\[
	W_\xi^*ZW_\xi=[[A,q_0],q_0],\qquad Z=[\dot q,q] .
\]
The double commutator is the off-diagonal part of $A$
relative to $q_0$. The excitation block of the
remaining part is $a$. For a compressed form $E=PEP$,
put $\widehat E=\Gamma(W_\xi)^*E\Gamma(W_\xi)$,
extended by zero off $P_0\cF^{\le N}$.
Before compression, the product rule on the common
core reads
\[
	\partial_t\widehat E
	=-\dGamma(A)\widehat E
	+\Gamma(W_\xi)^*(\partial_t E)\Gamma(W_\xi)
	+\widehat E\,\dGamma(A) .
\]
Here $(\Gamma(W_\xi)^*)'=-\dGamma(A)\Gamma(W_\xi)^*$.
Substituting $\partial_t E=\nabla_tE+[\mathscr Z,E]$
leaves $[\widehat E,\dGamma(A-W_\xi^*ZW_\xi)]$.
Compression therefore gives
\begin{equation}\label{eq:chart-covariant-identity}
	\partial_t\widehat E
	=\Gamma(W_\xi)^*(\nabla_tE)\Gamma(W_\xi)
	+[\widehat E,\dGamma(a)] .
\end{equation}
The condensate block vanishes on the excitation space.

Differentiating the Cayley transform gives the exact formula
$A=2(1+R_\xi)^{-1}\dot R_\xi(1-R_\xi)^{-1}$.
At a reference time $t_0$ with $\xi(t_0)=0$, this gives
$A(t_0)=2\dot R_\xi(t_0)$ and
$\dot A(t_0)=2\ddot R_\xi(t_0)$; the two products
of $\dot R_\xi(t_0)$ cancel.
Both $R_\xi$ and its time derivatives are off-diagonal.
Consequently $a(t_0)=0$, and
$\dot a(t_0)=2q_0\ddot R_\xi(t_0)q_0=0$.
Applying \eqref{eq:chart-covariant-identity} twice
therefore gives
\[
	\partial_t^j\widehat E(t_0)
	=\Gamma(W_{\xi(t_0)})^*(\nabla_t^jE)(t_0)
	\Gamma(W_{\xi(t_0)}),\qquad j=1,2 .
\]
The phase derivatives are already included in $E(t)$
through \eqref{eq:chart-phase-gauge}; the transport
$W_\xi$ depends only on the moving projection.
At fixed $N$, all finite-rank connection operators
preserve the form domain, so the core identities
extend by continuity. Their uniform bounds follow
from the chart majorants and the bounded $H^2$
velocities. Recentering the chart at any $t_0\in I$
gives the asserted estimates throughout $I$.
Equations \eqref{eq:parameter-derivatives} and
\eqref{eq:Q-two-derivatives}, together with the $L_B$
estimate, prove the assertions for $\mathfrak G$.
The series and both derivatives converge locally
uniformly. Fixed number weights and one or two
number commutators add only polynomial degree factors.
If a cancellation group contains several number shifts,
we first take its Fourier components under the number gauge:
\[
	E_{\iota,d}=\frac{1}{2\pi}\int_0^{2\pi}
	e^{-id\beta}e^{i\beta\mathcal N}E_\iota e^{-i\beta\mathcal N}\d\beta .
\]
This projection is contractive in the balanced form norm
because $Q$ commutes with $M$. It commutes with the
parameter derivatives and retains the cancellations.
There are at most $4J+9$ shifts, so summing their bounds
adds only a polynomial factor.
Indeed, for an elementary term of definite shift $d$,
$[M,E_\iota]=dE_\iota$ and
$[M,[M,E_\iota]]=d^2E_\iota$: these commutators preserve
the shift. The field count gives $|d|\le2J+4$.
For a diagonal form bound
$|E_\iota[z,z]|\le b_\iota Q[z]$, the numerical-radius
inequality gives
$\|Q^{-1/2}E_\iota Q^{-1/2}\|\le2b_\iota$, including
non-Hermitian terms. On a nonzero $n\to n+d$ block,
conjugation by $M^r$ multiplies it by
$(n+d+1)^r/(n+1)^r$, whose absolute value is at most
$(1+|d|)^{|r|}$. Thus, for arbitrary form-domain
vectors $u,v$,
\[
	|E_\iota[u,v]| \le2b_\iota(1+|d|)^{|r|} \|M^{-r}Q^{1/2}u\|\,\|M^rQ^{1/2}v\|.
\]
The derivative of a definite-shift term retains that
shift, because the connection and chart transport
preserve number. The same argument applies to the
derivative majorants; the powers of $1+|d|$ and
$d^2$ are absorbed by the existing polynomial in $J$.
More explicitly, for each of the five cancelled families
$\mathscr I$ in Section~\ref{sec:number-compatibility} and each fixed number weight $r$,
the preceding argument gives constants $b_r,C_{r,I}$ with
\[
	\sum_{\substack{\iota\in\mathscr I\\J(\iota)=J}}
	\bigl\|M^r Q^{-1/2}(\nabla^jE_\iota)Q^{-1/2}M^{-r}\bigr\|
	\le C_{r,I}(1+J)^{b_r}\theta^{\max(J-b_0,0)},
	\qquad j=0,1,2.
\]
The quadratic time group was estimated separately above.
This summable derivative bound follows from the polynomial
identities and their neighborhood bounds.
The same fixed space $V_N$ is used for the spatial groups,
the quadratic time term and their derivatives.
Their common majorants permit termwise differentiation in
$\mathcal B(V_N,V_N^*)$. Number commutators and fixed number
weights preserve summability by the shift bounds just proved.
This completes the passage from the static group estimates
to the asserted $C^2$ form bounds for $\mathfrak G$.

Twice differentiating the finite polynomial defining
$\mathfrak J^{[3]}$ now puts $\mathfrak G$ or one
of its first two derivatives between words in
$C,\nabla C,\nabla^2 C$. Lemma~\ref{lem:phase-second-regularity}
controls these words on $V_N$; its form estimate
controls the last term $-i\nabla^3 C$.
For the exact $\mathfrak J$ use the integral formula
$-i\int_0^1e^{-aC}(\nabla C)e^{aC}\d a$ for its
time term. Its second derivative contains at most
one $\nabla^3 C$, paired between graph-bounded
words and exponentials. Number derivatives use the
same Duhamel formula with $[M,C]$. These arguments
also prove continuity in the asserted topologies.
\end{proof}

\subsubsection{A positive representation of the evolution}

Choose one sufficiently large constant $b_I$ so that
\[
	F_3=\mathfrak J^{[3]}+b_IM\asymp Q,\qquad F_{\rm ex}=\mathfrak J+b_IM\asymp Q,\qquad F_3,F_{\rm ex}\ge c_I Q .
\]
The first assertion is Theorem~\ref{thm:Taylor-first-energy}.
For the second, conjugate the corresponding bound on
$\mathfrak G$ by $e^C$, and use the full-energy phase
bound, number-weight bounds and the bounded logarithmic
derivative. In parallel coordinates set
\[
	L_t=e^{-ib_I(t-s)M},\qquad H_j(t)=L_tF_j(t)L_t^*,\quad \widehat U_j(t,r)=L_tU_j(t,r)L_r^*, \qquad j=3,{\rm ex}.
\]
Here $U_3=U_N^{[3]}$ and $U_{\rm ex}$ denotes the exact
cubic-conjugated evolution. These are unitary changes
of representation of the same two evolutions.
Since $L_t$ commutes with $Q$ and $M$, all their
energy and number norms are unchanged. In particular
\[
	H_j''=L_t\{\nabla^2F_j-2ib_I[M,\nabla F_j] -b_I^2[M,[M,F_j]]\}L_t^* .
\]
Proposition~\ref{prop:full-second-form} shows that
$H_j$ is a positive $C^2$ form family on the fixed
space $V_N$, with uniform form and derivative bounds.

\begin{lemma}
\label{lem:three-halves-propagation}
Let $H(t):V\to V^*$ be a positive $C^2$ form family,
uniformly equivalent to the squared norm of $V$.
Its unitary form propagator satisfies, for fixed $s$,
\begin{equation}\label{eq:three-halves-propagation}
	\sup_{t\in I}\{\|u(t)\|_V+\|H(t)u(t)\|_V\} \le C_{I,s}\|H(s)^{3/2}u_s\|,\qquad u(t)=U(t,s)u_s ,
\end{equation}
for $u_s\in D(H(s)^{3/2})$. In particular $u\in C^1(I;V)$.
The constant is uniform in a family with uniform form
bounds, first two derivative bounds and form-propagator
bounds.
\end{lemma}
\begin{proof}
This is the form-space operator-domain conclusion of
Kisy\'nski's theorem \cite{Kisynski}; see the precise formulation
in \cite[Proposition 3.9(iii), Theorem 3.10(iv)]{BLP}.
In that notation $\Lambda_0(t)$ is the self-adjoint
operator $H(t)$ on $\mathcal H$, and
\[
	D(\Lambda_1(t))
	=\{u\in D(H(t)):H(t)u\in V\}
	=D(H(t)^{3/2}) .
\]
The last equality follows from $V=D(H(t)^{1/2})$
and the spectral calculus. Equivalently, if $u\in V$
and the form value $H(t)u\in V^*$ belongs to $V$,
the representation theorem first gives $u\in D(H(t))$.
Thus the domain in the cited result is exactly the one
used here. Theorem 3.10(i)--(ii) of the same reference
gives strong continuity on $V^*$ and $V$.
We derive the uniform estimate from the form-space
propagator in parts (i)--(iii).
The first-energy inequality gives a uniform bound for
$U(t,r)$ on $V$, and duality gives the bound on $V^*$.
In the notation of this lemma, that inequality is
$|H'(t)[v,v]|\le C H(t)[v,v]$ and hence
$H(t)[U(t,r)v,U(t,r)v]\le e^{C|t-r|}H(r)[v,v]$.
It holds for the form propagator by the usual form
regularization in the cited theorem. Uniform coercivity
therefore gives the asserted $V$ bound; the dual estimate
uses $U(r,t)=U(t,r)^*$.
Set $B=H^{-1}H'$ as a bounded operator on $V$.
Here $H^{-1}$ is the inverse of the coercive form operator
$H:V\to V^*$, so
\[
	HB=H(H^{-1}H')=H'\quad\text{in }\mathcal B(V,V^*).
\]
The products are compositions of form operators.
Both $B$ and
$B'=H^{-1}H''-H^{-1}H'H^{-1}H'$ are uniformly
bounded on $V$. Put
$v_s=-iH(s)u_s\in V$, $w_s=v_s+B(s)u_s$.
The coupled Volterra equations on $V\times V$ are
\begin{align*}
	u(t)&=u_s+\int_s^t(w-Bu)(r)\d r,\\
	w(t)&=U(t,s)w_s+
	\int_s^tU(t,r)\{(B'-B^2)u+Bw\}(r)\d r .
\end{align*}
Let $C_U$ bound $U(t,r)$ on $V$,
and let $B_0,B_1$ bound $B,B'$ there. On an interval
$I_0$ of length $h$, the right sides define a map on
$C(I_0;V)^2$ with Lipschitz constant at most
$h\max\{1+B_0,\ C_U(B_1+B_0^2+B_0)\}$
in the maximum norm. Choose $h$ so this constant is
less than one, solve by contraction, and concatenate
the solutions. All choices are uniform under the
stated bounds. Gronwall's inequality gives
\[
	\sup_{t\in I}\{\|u(t)\|_V+\|w(t)\|_V\}
	\le C\{\|u_s\|_V+\|H(s)u_s\|_V\}.
\]
The inverse $H^{-1}:V^*\to V$ is $C^2$ by coercivity and
the inverse differentiation rule, so $B\in C^1(I;\mathcal B(V))$.
The first integral equation gives $u\in C^1(I;V)$.
The forcing $g=(B'-B^2)u+Bw$ is continuous in $V$.
The forward form equation gives
\[
	\frac{\d}{\d t}\int_s^tU(t,r)g(r)\d r
	=g(t)-iH(t)\int_s^tU(t,r)g(r)\d r
	\quad\text{in }V^* .
\]
This follows by difference quotients, since
$H(t):V\to V^*$ is bounded and the propagator is
strongly continuous on both spaces. Thus the second
Volterra equation has derivative
$w'=-iHw+(B'-B^2)u+Bw$.
Set $v=w-Bu\in C(I;V)$. Then in $V^*$,
\[
	u'=v,\qquad v'=-iHw=-iHv-iH'u,\qquad (Hu)'=H'u+Hv.
\]
In particular $(iv-Hu)'=0$ in $V^*$, with zero initial value,
so $iu'=Hu$. Uniqueness identifies $u$ with the
given propagator, and $Hu=iv\in C(I;V)$.
The spectral lower bound and equivalence of form norms give
$\|u_s\|_V+\|H(s)u_s\|_V\le C\|H(s)^{3/2}u_s\|$.
This proves the estimate.
Every coefficient used in this argument has the stated
uniform bound. Reverse the integrals for $t<s$.
\end{proof}

\begin{corollary}\label{cor:Taylor-three-halves}
For $z\in D(F_3(s)^{3/2})$, the actual Taylor evolution obeys
\begin{align}
	\sup_{t\in I}\|Q_t^{1/2}F_3(t)U_3(t,s)z\|
	&\le C_{I,s}\|F_3(s)^{3/2}z\|,\label{eq:Taylor-three-halves}\\
	\sup_{t\in I}\bigl\{\|M Q_t^{1/2}U_3(t,s)z\|
	+\|M^{3/2}U_3(t,s)z\|\bigr\}
	&\le C_{I,s}\|F_3(s)^{3/2}z\| .
	\label{eq:Taylor-one-number}
\end{align}
\end{corollary}
\begin{proof}
Apply Lemma~\ref{lem:three-halves-propagation} to $H_3$
and undo the number phase and parallel transport.
For the second line apply Lemma~\ref{lem:shifted-inverse}
to $F_3$ with $r=1$. Its hypotheses follow from
Proposition~\ref{prop:GP-weight}, the weighted full-energy
phase bounds, and \eqref{eq:Taylor-first-source}.
With $W=M+\rho$ and $u=U_3(t,s)z$ this yields
\[
	\|WQ^{1/2}u\| \le C\|Q^{-1/2}W F_3u\| \le C_\rho\|Q^{1/2}F_3u\| .
\]
Here $W\le(1+\rho)Q$ and $Q$ commutes with $M$.
Finally $M^{3/2}\le M Q^{1/2}$ in their joint
functional calculus.
\end{proof}

\subsection{The inverse-energy comparison principle}
\label{sec:inverse-comparison}

We now integrate the remainder against an inverse energy.
The preceding propagation theorem controls the right trajectory;
the static estimate below supplies a remainder bound with one
number weight.

\begin{theorem}\label{thm:inverse-comparison}
Let $v\in C_c^\infty(\mathbb R^3)$ be nonnegative and radial.
Let $\phi_t$ solve the modified GP equation with an $N$-independent
normalized initial datum in $H^6(\mathbb R^3)$. Fix a sufficiently
small quadratic scattering radius $\ell\in(0,1/2)$, independent of $N$.
Use the cubic phase \eqref{spec:eq:actual-phase} at this background,
with $m=N^{-1/2}$ and the original COS cutoff $0\le\Theta_N\le1$.
It equals one for $n\le N^{1/2}/2+10$ and zero for
$n\ge N^{1/2}-10$, for sufficiently large $N$.
Fix a compact interval $I$ and $s\in I$. Choose a fixed $b_I$
such that $F_{\rm ex}=\mathfrak J+b_IM$ and
$F_3=\mathfrak J^{[3]}+b_IM$ satisfy
$c_IQ_t\le F_{\rm ex},F_3\le C_IQ_t$ on $I$, uniformly in $N$.
Suppose the exact
difference $R=\mathfrak J-\mathfrak J^{[3]}$ admits a
form-convergent decomposition
$R=\sum_{J\le J_*}R_J+R_>$, where $J_*\le c_0\sqrt N$.
Let $\Lambda_N\le C\sqrt N$ and
$E_J=\mathbf1_{M\le\Lambda_N+c(J+1)}$.
Assume, for each $T_J=R_J,\nabla R_J,i[M,R_J]$,
\begin{align}
	T_J&=E_JT_JE_J,\notag\\
	|T_J[u,v]|&\le a_J\epsilon^2
	\|M Q_t^{1/2}E_Ju\|\,\|Q_t^{1/2}E_Jv\|,
	\qquad \epsilon=N^{-3/4},\label{eq:one-weight-static}\\
	\sum_{J\ge0}(1+J)^b a_J&<\infty
	\quad\hbox{for each fixed }b .
	\notag
\end{align}
The forms are $C^1$ with the same summable majorants.
For $T_>=R_>,\nabla R_>,i[M,R_>]$, assume
\begin{equation}\label{eq:geometric-tail-static}
	|T_>[u,v]|\le C_KN^{-K} \|Q_t^{1/2}u\|\|Q_t^{1/2}v\| \quad\hbox{for every fixed }K>0 .
\end{equation}
Then
\begin{equation}\label{eq:inverse-comparison-result}
	\|(U_{\rm ex}(t,s)-U_3(t,s))z\| \le C_{I,s}N^{-5/4}\|F_3(s)^{3/2}z\|,\qquad t\in I,
\end{equation}
for $z\in D(F_3(s)^{3/2})$.
\end{theorem}
\begin{proof}
All calculations first take place in parallel coordinates.
Put $\widehat R=L_tRL_t^*$ and $H=H_{\rm ex}(t)$.
The polynomial compatibility and first number commutator
needed for Lemma~\ref{lem:shifted-inverse} hold for $H$,
by the weighted phase bounds and the exact integral
formula for the number derivative of $e^C$.
The same weighted form compatibility holds for $H'$ by
Proposition~\ref{prop:full-second-form} and its proof.
That inverse lemma at exponent $r=1/2$ therefore gives
\begin{align}
	\|M^{1/2}Q_t^{1/2}H^{-1}\psi\|&\le C\|\psi\|,
	\notag\\
	\|M^{1/2}Q_t^{1/2}H^{-1}H'H^{-1}\psi\|&\le C\|\psi\|.
	\label{eq:half-number-inverse}
\end{align}
To specify both ordered products, fix $t$ and write
$Q=Q_t$, $W=M+\rho$. Set
\begin{align*}
	\mathcal I&=Q^{1/2}W^{1/2}H^{-1}W^{-1/2}Q^{1/2},\\
	\mathcal K&=W^{1/2}Q^{-1/2}H'Q^{-1/2}W^{-1/2},
	\qquad S=W^{1/2}Q^{-1/2} .
\end{align*}
The shifted inverse estimate and the weighted form
bound for $H'$ give uniform bounds on $\mathcal I$
and $\mathcal K$. Also $\|S\|\le\sqrt{1+\rho}$,
since $Q\ge M\ge1$. Using
\eqref{eq:weighted-form-inverse}, the exact factorizations are
\begin{align*}
	W^{1/2}Q^{1/2}H^{-1}&=\mathcal I S,\\
	W^{1/2}Q^{1/2}H^{-1}H'H^{-1}
	&=\mathcal I\mathcal K\mathcal I S .
\end{align*}
Here the rightmost inverse maps $\mathcal H_N$ into
$V_N$, then $H'$ maps $V_N$ into $V_N^*$, and the
left inverse maps $V_N^*$ back into $V_N$.
The identities hold first on the form core and then
on $\mathcal H_N$ by their bounded extensions.
In detail the adjacent commuting factors cancel in
\[
	\mathcal I\mathcal K
	=Q^{1/2}W^{1/2}H^{-1}H'Q^{-1/2}W^{-1/2} .
\]
Multiplying this by $\mathcal I S=W^{1/2}Q^{1/2}H^{-1}$
uses $Q^{-1/2}W^{-1/2}W^{1/2}Q^{1/2}=1$ and gives
the second factorization. All inverses adjacent to
$H'$ are the form inverses just described.
Replacing $W^{1/2}$ by $M^{1/2}$ proves
\eqref{eq:half-number-inverse}.

Define $D(t)=H(t)^{-1}\widehat R(t):V_N\to V_N$.
For each fixed $N$ this is a $C^1$ family of bounded
operators $V_N\to V_N$.
Indeed $H:V_N\to V_N^*$ is $C^2$ in parallel
coordinates, so its form inverse is $C^2$ from
$V_N^*$ to $V_N$, with derivative $-H^{-1}H'H^{-1}$.
The number weight is bounded at fixed $N$; the summable
source bounds therefore give
$\widehat R\in C^1(I;\mathcal B(V_N,V_N^*))$.
For $\psi\in\mathcal H_N$ and $v\in V_N$,
test the $J$th term against $H^{-1}\psi$.
Since $E_J$ commutes with $M,Q_t$,
\[
	\|M Q_t^{1/2}E_JH^{-1}\psi\| \le(\Lambda_N+c(J+1))^{1/2} \|M^{1/2}Q_t^{1/2}H^{-1}\psi\| .
\]
Sum \eqref{eq:one-weight-static} over $J$ and use
\eqref{eq:geometric-tail-static} for the tail.
The derivative
$D'=H^{-1}\widehat R'-H^{-1}H'H^{-1}\widehat R$
is handled by the same pairing: its second term tests
the remainder against $H^{-1}H'H^{-1}\psi$ and uses
the second line of \eqref{eq:half-number-inverse}.
Indeed $H'$ is Hermitian as a form, and duality gives
\[
	\langle\psi,H^{-1}H'H^{-1}\widehat Rv\rangle
	=\widehat R[H^{-1}H'H^{-1}\psi,v] .
\]
Moreover
$\widehat R'=L_t(\nabla R-ib_I[M,R])L_t^*$,
so the stated number derivative is part of the estimate.
We obtain
\begin{equation}\label{eq:inverse-remainder-small}
	\|D\|_{V_N\to\mathcal H_N} +\|D'\|_{V_N\to\mathcal H_N} \le C\{\epsilon^2(1+\Lambda_N)^{1/2}+N^{-2}\} \le C N^{-5/4}.
\end{equation}

Let $y(r)=\widehat U_3(r,s)z$.
Lemma~\ref{lem:three-halves-propagation} gives
$y\in C^1(I;V_N)$ and
$\|y(r)\|_{V_N}+\|H_3(r)y(r)\|_{V_N}
\le C\|F_3(s)^{3/2}z\|$.
The derivative in the second propagator variable follows
from its composition law. For $f\in V_N$,
\[
	\frac{\widehat U_{\rm ex}(t,r+h)-\widehat U_{\rm ex}(t,r)}{h}f
	=\widehat U_{\rm ex}(t,r+h)
	\frac{1-\widehat U_{\rm ex}(r+h,r)}{h}f
	\longrightarrow i\widehat U_{\rm ex}(t,r)H(r)f
	\quad\text{in }V_N^* .
\]
The limit uses strong continuity on $V_N^*$ and the
forward form equation. Both propagator families are
locally bounded on this space.
The weak product rule in $V_N^*$ is
\begin{align}
	\partial_r\{\widehat U_{\rm ex}(t,r)D(r)y(r)\}
	={}&i\widehat U_{\rm ex}(t,r)\widehat R(r)y(r)\notag\\
	&+\widehat U_{\rm ex}(t,r)D'(r)y(r)
	-i\widehat U_{\rm ex}(t,r)D(r)H_3(r)y(r).
	\label{eq:inverse-product-rule}
\end{align}
We have $D y\in C^1(I;V_N)$ for each fixed $N$. The
form propagator is differentiable from $V_N$ to
$V_N^*$ and extends continuously to $V_N^*$.
These properties justify the product rule. It can equivalently be obtained by testing
against a vector in $V_N$ and taking difference quotients.
The same rule applied to
$\widehat U_{\rm ex}(t,r)y(r)$ gives the weak Duhamel
identity. Combining the two integrated identities yields
\begin{align}
	(\widehat U_{\rm ex}(t,s)-\widehat U_3(t,s))z
	={}&\widehat U_{\rm ex}(t,s)D(s)z-D(t)y(t)\notag\\
	&+\int_s^t\widehat U_{\rm ex}(t,r)D'(r)y(r)\d r\notag\\
	&-i\int_s^t\widehat U_{\rm ex}(t,r)D(r)H_3(r)y(r)\d r .
	\label{eq:inverse-comparison-Hilbert-identity}
\end{align}
After integration the boundary vectors and the last
two integrals are in $\mathcal H_N$, by
\eqref{eq:inverse-remainder-small}. Their norms are
bounded by $C_{I,s}N^{-5/4}\|F_3(s)^{3/2}z\|$.
Thus \eqref{eq:inverse-product-rule} identifies the
weak Duhamel integral of $\widehat R y$ as an
$L^2$ vector through its boundary and integrated terms.
The weak Duhamel identity proves
\eqref{eq:inverse-comparison-result}. Unitarity of the left flow gives the final norm estimate. Undo both unitary changes of coordinates;
reverse the integrals if $t<s$.
\end{proof}

\begin{remark}
\label{rem:enlarged-support}
For a source group $S_J$ of bandwidth at most $c(J+1)$,
each term $C^jS_JC^k$, $j+k=4$, is supported on both
sides in $E_J$. At least one of $j,k$ is positive;
that side is in the original particle-number cutoff range of $C$,
and the bandwidth controls the other side.
The exterior $e^{aC}$ is the identity off the original
particle-number cutoff range, so it does not enlarge $E_J$.
The same reasoning applies when a derivative replaces
a $C$ by $\nabla C$ or $[M,C]$.
This proves the support part of
\eqref{eq:one-weight-static} for the exact integral
remainders. The spatial source estimates give its one-weight
coefficient in Theorem~\ref{thm:full-model-comparison}.
\end{remark}

\begin{remark}
\label{rem:inverse-preparation}
The initial norm in \eqref{eq:inverse-comparison-result}
defines a nonempty uniform class. For example put
\[
	z_N=\frac{F_3(s)^{-2}\Omega} {\|F_3(s)^{-2}\Omega\|}.
\]
Since $F_3(s)\ge c_I$ and
$F_3(s)[\Omega]\le C_IQ_s[\Omega]=C_I$, Jensen's
inequality for its spectral probability measure gives
$\|F_3(s)^{-2}\Omega\|^2
=\int\lambda^{-4}\d\mu(\lambda)
\ge(\int\lambda\d\mu(\lambda))^{-4}\ge C_I^{-4}$.
The numerator after applying $F_3(s)^{3/2}$ is
$F_3(s)^{-1/2}\Omega$, uniformly bounded.
The preparation uses the initial operator $F_3(s)$.
\end{remark}

\begin{proposition}
\label{prop:Taylor-resolvent-preparation}
At the common initial time $s$, write
$G=\mathfrak J^{[3]}(s)$, $F=F_3(s)=G+b_IM$.
There is a fixed $\kappa_0>0$, independent of $N$,
such that the normalized vectors
\begin{equation}\label{eq:Taylor-resolvent-data}
	z_{7,N} =\frac{(G-i\kappa_0)^{-7}\Omega} {\|(G-i\kappa_0)^{-7}\Omega\|}
\end{equation}
satisfy $\sup_N\|F^{3/2}z_{7,N}\|<\infty$.
The parameter $\kappa_0$ is a spectral parameter,
unrelated to the relative kernel $\kappa_N$.
\end{proposition}
\begin{proof}
We prove the needed number-weighted resolvent estimate.
For any fixed integer $k$, one can choose
$W=M+\rho$ and $\kappa_0$, uniformly in $N$, such that
\begin{equation}\label{eq:weighted-Taylor-resolvent}
	\|W^r(G-i\kappa_0)^{-1}W^{-r}\|\le C_k, \qquad r=0,\tfrac12,1,\ldots,k .
\end{equation}
The weighted form bounds for $G$ and $[G,M]$ follow
from the same inputs as in
Corollary~\ref{cor:Taylor-three-halves}.
For $u=(G-i\kappa_0)^{-1}f$, set
$U_r=\|W^ru\|$, $B_r=\|W^rf\|$ and
$D_r=\|W^{r-1/2}Q^{1/2}u\|$.
Lemma~\ref{lem:shifted-inverse} at exponent $r-1/2$,
applied to $Fu=f+i\kappa_0u+b_IMu$, gives
\[
	D_r\le C\{\rho^{-1/2}B_r+ (\kappa_0\rho^{-1/2}+b_I)U_r\}.
\]
Here $Q^{-1/2}W^{-1/2}\le\rho^{-1/2}$, while
$Q^{-1/2}M W^{-1/2}\le1$; the last term introduces no factor $\sqrt\rho$.
The telescoping identity for $[G,W^{2r}]$ and the
weighted form bounds for $[G,M]$ imply
$|[G,W^{2r}][u,u]|\le C_rD_r^2$.
Take the imaginary part of the resolvent equation
paired with $W^{2r}u$. It follows that
\[
	\kappa_0 U_r^2 \le U_rB_r+C'_r\rho^{-1}B_r^2 +C'_r(\kappa_0^2/\rho+b_I^2)U_r^2 .
\]
First choose $\kappa_0$ so that
$C'_r b_I^2<\kappa_0/4$ for every $r\le k$,
then $\rho$ so that
$C'_r\kappa_0^2/\rho<\kappa_0/4$ and all inverse
estimates apply. Absorption proves
\eqref{eq:weighted-Taylor-resolvent}. At fixed $N$
the weights are bounded and these form pairings
are legitimate on $D(G)$.

Put $R_+=(G-i\kappa_0)^{-1}$ and $u=R_+\Omega$.
The weighted estimate gives $\|Mu\|\le C$.
Coercivity and the real part of the equation give
$cQ[u]\le\operatorname{Re}\langle u,\Omega\rangle
+b_I\langle u,Mu\rangle\le C\|u\|$.
Pair the same equation with the vacuum and use
$|G[\Omega,u]|\le C Q[\Omega]^{1/2}Q[u]^{1/2}$.
Then $1\le C\|u\|^{1/2}+\kappa_0\|u\|$, so
$\|R_+\Omega\|\ge c_0>0$.
For the spectral probability measure of $G$ on
$\Omega$, Jensen's inequality now yields
\[
	\|R_+^k\Omega\|^2 =\int(\lambda^2+\kappa_0^2)^{-k}\d\mu(\lambda) \ge\left(\int(\lambda^2+\kappa_0^2)^{-1} \d\mu(\lambda)\right)^k \ge c_0^{2k}.
\]
Using $G R_+=1+i\kappa_0R_+$ and
\eqref{eq:weighted-Taylor-resolvent} repeatedly gives
\[
	\sup_N\|W^{k-j}G^jz_{k,N}\|<\infty,\qquad 0\le j\le k,\qquad z_{k,N}=R_+^k\Omega/\|R_+^k\Omega\| .
\]
In particular $G^2z_{7,N}$, $WGz_{7,N}$ and
$W^2z_{7,N}$ are uniformly bounded.
The unweighted inverse estimate, applied to $Gz_{7,N}$,
then bounds $Q^{1/2}Gz_{7,N}$.
Its weight-one version, applied to $z_{7,N}$, bounds
$M Q^{1/2}z_{7,N}$ by
$C(\|WGz_{7,N}\|+\|W^2z_{7,N}\|)$.
Consequently $Fz_{7,N}=Gz_{7,N}+b_IMz_{7,N}$
belongs to $D(Q^{1/2})$ uniformly there.
Since $F\asymp Q$, this is precisely the asserted
$F^{3/2}$ domain and bound.
The vectors are prepared at time $s$ only.
\end{proof}

\section{Higher moments under the exact evolution}
\label{sec:exact-moment-stability}

We now prove Theorem~\ref{thm:D}. The argument uses the exact-generator
form bounds of Proposition~\ref{prop:full-second-form}, the
propagation estimate of Lemma~\ref{lem:three-halves-propagation},
and a first number-commutator bound. The next lemma gives the required
weighted coercivity directly from the inverse-commutator identity.
We then propagate its third moment and derive the physical
fluctuation bounds. The model assumptions are those of
Theorem~\ref{thm:D}, and all constants on a fixed compact interval
$I$ are independent of $N$. The propagator $U_{\rm ex}$ is the
exact conjugation of the full $N$-body evolution on all
physical particle sectors. The cutoff in $C_t$ enters the
correlation transformation.

\subsection{Correlated energy and weighted coercivity}

Fix $b=b_I$ as in Section~\ref{sec:second-regularity}, and write
\begin{equation}\label{eq:exact-higher-energy}
	F_t=\mathfrak J_t+bM,\qquad \mathscr E_{3,N}(t;z)=\|F_t^{3/2}z\|^2, \qquad z\in D(F_t^{3/2}).
\end{equation}
Here and below $F_t$ denotes its positive self-adjoint realization.
The moment in \eqref{eq:exact-higher-energy} is the quadratic form
of $F_t^3$, defined on $D(F_t^{3/2})$. The scalar convention for $\mathfrak J$ is the fixed one
in Sections~\ref{sec:number-compatibility}--\ref{sec:covariant-generator}. This energy is determined by the instantaneous
generator and its number correction.

\begin{lemma}
\label{lem:exact-moment-coercivity}
Let $v\in C_c^\infty(\mathbb R^3)$ be nonnegative and radial.
Let $\phi_t$ solve the modified GP equation with an $N$-independent
normalized initial datum in $H^6(\mathbb R^3)$. Fix a sufficiently
small quadratic scattering radius $\ell\in(0,1/2)$, independent of $N$.
Use the cubic phase \eqref{spec:eq:actual-phase} at this background,
with $m=N^{-1/2}$ and the original COS cutoff $0\le\Theta_N\le1$.
It equals one for $n\le N^{1/2}/2+10$ and zero for
$n\ge N^{1/2}-10$, for sufficiently large $N$.
Fix a compact time interval $I$. Choose a fixed $b_I$ such that
$F_t=\mathfrak J_t+b_IM$ satisfies $c_IQ_t\le F_t\le C_IQ_t$
for $t\in I$, uniformly in $N$.
The exact $F_t$ satisfies the hypotheses of
Lemma~\ref{lem:shifted-inverse}, uniformly for $t\in I$.
Moreover,
\begin{equation}\label{eq:exact-moment-coercivity}
	\|M Q_t^{1/2}z\|+\|M^{3/2}z\| \le C_I\|F_t^{3/2}z\|, \qquad z\in D(F_t^{3/2}).
\end{equation}
\end{lemma}
\begin{proof}
The positive form comparison $c_IQ_t\le F_t\le C_IQ_t$ was
proved in Section~\ref{sec:second-regularity}. The form is
Hermitian: $\mathfrak G_t$ is Hermitian, $C_t$ and its covariant
derivative $X_t$ are skew-adjoint, and hence
$-ie^{-aC_t}X_te^{aC_t}$ is Hermitian. At fixed $N$ these
phase derivatives are bounded on the Hilbert space. Coercivity
and continuity make the form norm equivalent to the complete
$Q_t$ form norm. The representation theorem
\cite[Chapter VI, Section 2]{KatoBook} therefore defines a unique
positive self-adjoint $F_t$. At fixed $N$, $bM$ is
a bounded self-adjoint operator. The bounded perturbation theorem
therefore makes $\mathfrak J_t=F_t-bM$ self-adjoint on $D(F_t)$.
We use the spectral identity
\[
	D(F_t^{3/2})= \{z\in D(F_t):F_tz\in D(F_t^{1/2})\}.
\]
Since $M$ commutes
with $Q_t$, polynomial number compatibility for
$Q_t^{-1/2}e^{-C_t}\mathfrak G_te^{C_t}Q_t^{-1/2}$ follows by
factoring it into the balanced form of $\mathfrak G_t$ and the two
graph-bounded exterior exponentials. Use
Proposition~\ref{prop:GP-weight} and
Proposition~\ref{prop:full-graph}, with positive and negative
number weights; taking adjoints supplies the bounds on the
left exterior factor. The exact time term
\[
	\mathcal L_{C,t}=-i\int_0^1e^{-aC_t}X_te^{aC_t}\d a
\]
has the same weighted form bounds. The term $bM$ also has them.
More explicitly, put
\[
	R_r(a)=M^rQ_t^{1/2}e^{aC_t}Q_t^{-1/2}M^{-r},\qquad T_{G,r}=M^rQ_t^{-1/2}\mathfrak G_tQ_t^{-1/2}M^{-r}.
\]
The weighted graph estimates give $\|R_{\pm r}(a)\|\le C_{r,I}$
for $|a|\le1$. The balanced conjugated form factors exactly as
\[
	M^rQ_t^{-1/2}e^{-C_t}\mathfrak G_te^{C_t}Q_t^{-1/2}M^{-r} =R_{-r}(1)^*T_{G,r}R_r(1).
\]
For the time term replace $T_{G,r}$ by
$T_{X,r}=M^rQ_t^{-1/2}X_tQ_t^{-1/2}M^{-r}$, use $R_{\pm r}(a)$,
and integrate over $a$. This proves the two-sided estimates
with the corresponding opposite number weights.
To make the sector convention explicit, let $\Pi_n$ denote the
$n$-particle projection and $P_{t,n}=P_t\Pi_n$. A term of shift
$d$ has compressed blocks
$P_{t,n+d}E_\iota P_{t,n}$, set to zero when either index lies
outside $[0,N]$. Its restricted form bound is the original
bound evaluated on excitation vectors. As $P_t$, $Q_t$, the
generalized-field coefficients and the cutoff all preserve
number, the weight ratio is still
$(n+d+1)^r/(n+1)^r$. Its bound depends only on $r,d$.
For the non-Hermitian terms the numerical-radius inequality
in Section~\ref{sec:number-compatibility} gives the balanced operator bound before summation.
Thus compression and endpoint sectors introduce no new factor
depending on $N$.

For the first number commutator, expand $[M,e^{aC}]$ as
\[
	[M,e^{aC}]=\int_0^a e^{(a-r)C}[M,C]e^{rC}\d r.
\]
Each factor is graph bounded, including $[M,C]$; the middle
$[M,\mathfrak G]$ term is controlled by
Lemma~\ref{lem:moving-G}. The commutator of the time integral
is treated by the same identity and $[M,X]$. The latter has the
same graph bounds as $X$, since its number shifts have absolute
value at most three. Thus, with $T=Q_t^{-1/2}F_tQ_t^{-1/2}$,
$[M,T]$ has a uniform bounded extension. This verifies both
conditions of \eqref{eq:inverse-inputs} for the exact $F_t$.

To prove \eqref{eq:exact-moment-coercivity}, recall that $M=\cN+1\ge1$.
Write $T\ge c_I$ and $\|[M,T]\|\le D_I$. At fixed $N$, $M$ is
bounded, while $\|M^{-1}\|\le1$ uniformly in $N$. The standard
inverse-commutator identity, obtained by commuting $M$ with
$TT^{-1}=1$ (see \cite[Eq.~(5.3)]{GoleniaJecko2013}), gives
\[
	[M,T^{-1}]=-T^{-1}[M,T]T^{-1},\qquad \|MT^{-1}M^{-1}\|\le c_I^{-1}+D_Ic_I^{-2}=:A_I.
\]
The representation theorem for closed forms
\cite[Chapter VI, Section 2]{KatoBook} gives
$F_t^{-1}=Q_t^{-1/2}T^{-1}Q_t^{-1/2}$.
For $z\in D(F_t^{3/2})$, form comparison implies
$F_tz\in D(F_t^{1/2})=D(Q_t^{1/2})$. Hence
\[
	MQ_t^{1/2}z=(MT^{-1}M^{-1})(MQ_t^{-1/2}F_tz).
\]
Strong commutation and $M\le Q_t$ imply
$\|MQ_t^{-1/2}y\|\le\|Q_t^{1/2}y\|$ on $D(Q_t^{1/2})$. Thus
\[
	\|MQ_t^{1/2}z\|\le A_I\|Q_t^{1/2}F_tz\|\le A_Ic_I^{-1/2}\|F_t^{3/2}z\|.
\]
Finally, $\|M^{3/2}z\|\le\|MQ_t^{1/2}z\|$ by the joint spectral
calculus. All constants depend only on the uniform form and
commutator bounds. The boundedness of $M$ at fixed $N$ justifies
the products above. The displayed estimates are uniform in $N$.
\end{proof}

In sector notation, the quantity controlled on the left is
\begin{equation}\label{eq:weighted-energy-content}
	\|M Q_t^{1/2}z\|^2 =\sum_{n=0}^N(n+1)^2 \left\{\sum_{j=1}^n\|\nabla_jz^{(n)}\|^2 +\mathcal V_N[z^{(n)}]+(n+1)\|z^{(n)}\|^2\right\}.
\end{equation}
The restricted forms have these values on the excitation vectors.
Thus the chosen higher energy controls a third number moment
and number-weighted kinetic and interaction energy.

\subsection{Propagation by the exact generator}

The time regularity established above allows the form-space
propagation theorem to be applied to the exact generator. Combined
with the weighted coercivity estimate, it gives the following bound.

\begin{theorem}
\label{thm:exact-third-moment}
Let $v\in C_c^\infty(\mathbb R^3)$ be nonnegative and radial.
Let $\phi_t$ solve the modified GP equation with an $N$-independent
normalized initial datum in $H^6(\mathbb R^3)$. Fix a sufficiently
small quadratic scattering radius $\ell\in(0,1/2)$, independent of $N$.
Use the cubic phase \eqref{spec:eq:actual-phase} at this background,
with $m=N^{-1/2}$ and the original COS cutoff $0\le\Theta_N\le1$.
It equals one for $n\le N^{1/2}/2+10$ and zero for
$n\ge N^{1/2}-10$, for sufficiently large $N$.
Fix a compact time interval $I$. Choose a fixed $b_I$ such that
$F_t=\mathfrak J_t+b_IM$ satisfies $c_IQ_t\le F_t\le C_IQ_t$
for $t\in I$, uniformly in $N$.
Let $s\in I$, $z\in D(F_s^{3/2})$, and
$\xi(t)=U_{\rm ex}(t,s)z$. Then $\xi(t)\in D(F_t^{3/2})$ and
\begin{equation}\label{eq:exact-third-moment}
	\sup_{t\in I} \left\{\mathscr E_{3,N}(t;\xi(t)) +\|M Q_t^{1/2}\xi(t)\|^2\right\} \le C_{I,s}\mathscr E_{3,N}(s;z).
\end{equation}
\end{theorem}
\begin{proof}
First pass to the parallel coordinates of
Section~\ref{sec:second-regularity}. Their common form space is
$V_N=D(Q_s^{1/2})$. Set
$L_t=e^{-ib(t-s)M}$, $H(t)=L_tF_tL_t^*$ and
$u(t)=L_t\xi(t)$. The equation for this same trajectory is
$iu'=H(t)u$. The additional $bM$ is exactly the derivative
of the number phase.

The exact-generator case of Proposition~\ref{prop:full-second-form}
gives the $C^2$ form regularity of $H$. In particular its second
derivative is the combination
\[
	L_t\{\nabla_t^2F_t-2ib[M,\nabla_tF_t] -b^2[M,[M,F_t]]\}L_t^*.
\]
Its norm in $\mathcal B(V_N,V_N^*)$ is uniform. The same is true
of the form comparison and first derivative bounds.
Lemma~\ref{lem:three-halves-propagation}, whose Volterra estimate
keeps these constants uniform, therefore yields
\[
	\sup_{t\in I}\|H(t)^{3/2}u(t)\| \le C_{I,s}\|H(s)^{3/2}u(s)\|.
\]
Here $D(H^{3/2})$ is exactly the domain where $Hu\in V_N$;
the norm identification uses the first-energy form equivalence.
Undoing the two unitary changes of coordinates gives the first
term of \eqref{eq:exact-third-moment}. Apply
Lemma~\ref{lem:exact-moment-coercivity} at time $t$ to obtain the
second term.
\end{proof}

\begin{remark}
\label{rem:exact-number-shift}
At a fixed time let $F_a=\mathfrak J+aM$ and
$F_b=\mathfrak J+bM$ be positive self-adjoint energies on
$P\cF^{\le N}$. Assume, for $\gamma=a,b$, the uniform bounds
\[
	c_\gamma Q\le F_\gamma\le C_\gamma Q,\qquad
	\|MQ^{1/2}z\|\le C_\gamma\|F_\gamma^{3/2}z\|,
	\quad z\in D(F_\gamma^{3/2}),
\]
where $Q\ge M\ge1$ strongly commutes with $M$.
Their $3/2$ domains agree,
with graph norms equivalent uniformly in $N$ (constants may
depend on $a,b$). Indeed, $M$ is bounded at each fixed $N$,
so $D(F_a)=D(F_b)$, and
\[
	\|Q^{1/2}F_bz\| \le \|Q^{1/2}F_az\|+|b-a|\|M Q^{1/2}z\| \le C_{a,b}\|F_a^{3/2}z\|.
\]
The domain characterization and the reverse estimate prove
the assertion. Consequently, an initial family prepared once
with a fixed positive $F_a(s)$ can be used on every compact
time interval on which the background bounds hold. The same
initial family works with interval-dependent $b_I$ and propagation
constants.
\end{remark}

The initial third correlated-energy moment is a condition on the state in the
correlation representation. It is imposed only at the initial time;
Theorem~\ref{thm:exact-third-moment} proves its subsequent propagation.
The following spectral preparation gives an explicit class of such
states from normalized reference vectors with uniformly bounded
$Q_s$ form energy.

\begin{proposition}
\label{prop:exact-moment-preparation}
Let $\chi_N$ be normalized and satisfy
$\sup_N Q_s[\chi_N]<\infty$. Then
\begin{equation}\label{eq:exact-moment-preparation}
	z_N=\frac{F_s^{-3/2}\chi_N}{\|F_s^{-3/2}\chi_N\|} \quad\hbox{satisfies}\quad \sup_N\|F_s^{3/2}z_N\|<\infty.
\end{equation}
In particular $\chi_N=\Omega$ is allowed.
\end{proposition}
\begin{proof}
By the spectral functional calculus \cite{ReedSimonI},
the lower spectral bound on $F_s$ makes $F_s^{-3/2}$ bounded,
and its range is contained in $D(F_s^{3/2})$. For the spectral
probability measure of $F_s$ in $\chi_N$, Jensen's inequality
for $x\mapsto x^{-3/2}$ gives
\[
	\|F_s^{-3/2}\chi_N\| \ge\langle\chi_N,F_s^{-3/2}\chi_N\rangle \ge\bigl(F_s[\chi_N]\bigr)^{-3/2}\ge c>0.
\]
Only the first form moment of $\chi_N$ is used. Therefore
$\|F_s^{3/2}z_N\|=\|F_s^{-3/2}\chi_N\|^{-1}$ is uniformly
bounded. Since $Q_s[\Omega]=1$, the vacuum reference vector is admissible.
The spectral preparation is performed at the initial time.
\end{proof}

A single inverse already suffices when the reference vector has
finite first form energy. The following construction defines
the preparation through a positive form equation.

\begin{proposition}\label{prop:variational-preparation}
Let $\chi_N\in D(Q_s^{1/2})$ be normalized, and let
$F_s=F_{\rm ex}(s)$. There is a unique $y_N\in D(Q_s^{1/2})$
satisfying
\begin{equation}\label{eq:variational-preparation}
	F_s[v,y_N]=\langle v,\chi_N\rangle,\qquad v\in D(Q_s^{1/2}).
\end{equation}
It is nonzero. The normalized vector $z_N=y_N/\|y_N\|$ belongs
to $D(F_s^{3/2})$ and obeys
\begin{equation}\label{eq:variational-preparation-bound}
	\|F_s^{3/2}z_N\|^2\le \bigl(F_s[\chi_N]\bigr)^3
	\le C_I^3\bigl(Q_s[\chi_N]\bigr)^3.
\end{equation}
In particular, fix an integer $m\ge0$ and a normalized
$f\in H^2(\mathbb R^3)\cap\{\phi_0\}^{\perp}$. For $N\ge m$,
the reference vector
\[
	\chi_N=\bigoplus_{n=0}^{m}c_n f^{\otimes n},\qquad \sum_{n=0}^{m}|c_n|^2=1,
\]
has uniformly bounded $Q_0$ form energy. Its preparation by
\eqref{eq:variational-preparation}, with $s=0$, satisfies the
uniform initial-moment condition in Theorem~\ref{thm:D}.
\end{proposition}
\begin{proof}
The closed-form representation theorem
\cite[Chapter VI, Section 2]{KatoBook} identifies the solution as
$y_N=F_s^{-1}\chi_N$. The bound $F_s\ge cQ_s\ge cI$
gives existence and uniqueness, with $c>0$ independent of $N$.
Since $D(Q_s^{1/2})=D(F_s^{1/2})$, spectral calculus gives
\[
	F_s^{3/2}y_N=F_s^{1/2}\chi_N,\qquad
	\|F_s^{3/2}y_N\|^2=F_s[\chi_N].
\]
For the spectral probability measure of $F_s$ in $\chi_N$,
Jensen's inequality yields
\[
	\|y_N\|\ge\langle\chi_N,F_s^{-1}\chi_N\rangle
	\ge\bigl(F_s[\chi_N]\bigr)^{-1}>0.
\]
These identities prove \eqref{eq:variational-preparation-bound}.

For the stated reference vectors, number conservation of $Q_0$
eliminates cross terms between sectors. In the $n$-particle sector,
\[
	\langle f^{\otimes n},\cV_N f^{\otimes n}\rangle
	\le\binom n2 N^{-1}\|v\|_1\|f\|_\infty^2,
\]
because $\|N^2v(N\cdot)\|_1=N^{-1}\|v\|_1$ and
$\|f\|_2=1$. Also,
$\langle f^{\otimes n},\cK f^{\otimes n}\rangle
=n\|\nabla f\|_2^2$. The embedding $H^2(\mathbb R^3)\subset
L^\infty(\mathbb R^3)$ therefore gives
\[
	Q_0[\chi_N]\le m+1+m\|\nabla f\|_2^2
	+\binom m2 N^{-1}\|v\|_1\|f\|_\infty^2.
\]
For $m=0$ the last two terms vanish. The orbitals are orthogonal
to the common initial condensate, so the same expectation is
obtained for the restricted form on the excitation space.
\end{proof}

The vacuum reference in
Proposition~\ref{prop:exact-moment-preparation} produces a
spectrally prepared state. For the centered inverse-energy
construction, Theorem~\ref{thm:common-preparation} also determines
its limiting initial microscopic spectrum and proves positivity
of the contact coefficient. Proposition~\ref{prop:variational-preparation}
constructs admissible states through one form equation for the
initial correlated generator.

\subsection{Physical observables}

To interpret the correlated-energy estimate, we undo the two
correlation transformations and reconstruct the many-body state.
Number-moment bounds then give control of depletion and collective
one-particle fluctuations. The relation between excitation fields
and collective observables is also used in the central limit
arguments of \cite{BAKS,BSSCLT,RSCLT,COS}. Here the same relation gives physical moment bounds
from the correlated-energy estimate.

Let $\mathcal U_{N,t}$ be the condensate-extraction unitary from
the symmetric $N$-particle space to
$\mathcal F_{\perp\phi_t}^{\le N}$, as in \cite{LNS2015};
see the definition preceding \cite[Eq.~(2.4)]{COS} and
the explicit unitary in \cite[Eq.~(4.13)]{BS}.
It maps the orthogonal decomposition $\psi=\sum_{n=0}^N\phi_t^{\otimes(N-n)}\otimes_s\zeta^{(n)}$ to
$(\zeta^{(0)},\ldots,\zeta^{(N)})$. Here $\otimes_s$ denotes the isometric
symmetrized embedding, and each $\zeta^{(n)}$ belongs to
$(q_tL^2)^{\otimes_s n}$. This also specifies its inverse
and shows that it is onto.
Define the instantaneous reconstruction
\begin{equation}\label{eq:physical-reconstruction}
	\mathcal W_{N,t}=\mathcal U_{N,t}^*e^{B_t}e^{C_t},\qquad \Psi_N(t)=\mathcal W_{N,t}\xi_N(t).
\end{equation}
Here the quadratic phase is
\begin{align*}
	B_t&=\frac12\int\left\{\eta_t(x,y)b_x^*b_y^*
	-\overline{\eta_t(x,y)}b_xb_y\right\}\d x\d y,\\
	\eta_t&=(q_t\otimes q_t)
	[-Nw_\ell(N(x-y))\phi_t(x)\phi_t(y)],
\end{align*}
with the fixed Neumann profile of Section~\ref{sec:number-compatibility}.
This is the generalized quadratic transformation of
\cite[Eqs.~(3.1), (4.15)--(4.16)]{BS}.
Set
$\mathcal T_t=e^{-B_t}\mathcal U_{N,t}$ and
$\nabla_t\mathcal T_t=\partial_t\mathcal T_t-\mathscr Z_t\mathcal T_t$.
At fixed $N$, these are differentiable unitary maps between the
physical space and the moving excitation space. The derivatives
of $\mathcal U_{N,t}$ are finite sums of insertions of
$\dot\phi_t$, and $\dot B_t$ is bounded by its Hilbert--Schmidt
kernel and the finite number cutoff. They preserve the first
form domains: for fixed $N$ the smooth interaction is bounded,
extraction uses $\phi_t\in H^1$, and the quadratic phase uses
$\eta_t,\nabla\eta_t\in L^2$. The cubic graph bound was proved
above. Constants in these domain identifications may depend
on $N$.
Since $P_t\mathcal T_t=\mathcal T_t$ and
$P_t\mathscr Z_tP_t=0$, differentiation yields
$(1-P_t)\partial_t\mathcal T_t=\mathscr Z_t\mathcal T_t$.
Thus $\nabla_t\mathcal T_t=P_t\partial_t\mathcal T_t$,
with the same connection as in Section~\ref{sec:covariant-generator}.
The definition of the exact generator in \cite[Eq.~(4.24)]{BS},
after compression and scalar subtraction, is the form identity
\[
	\mathfrak G_t=\mathcal T_tH_N^{\rm mb}\mathcal T_t^* +i(\nabla_t\mathcal T_t)\mathcal T_t^*-c_{N,t}.
\]
For a physical solution $\psi(t)$, put
$\zeta(t)=e^{i\int_s^t c_{N,r}\d r}\mathcal T_t\psi(t)$.
The product rule gives $i\nabla_t\zeta=\mathfrak G_t\zeta$.
Putting $\xi=e^{-C_t}\zeta$ then gives
\[
	i\nabla_t\xi =\left\{e^{-C_t}\mathfrak G_te^{C_t} -ie^{-C_t}\nabla_t(e^{C_t})\right\}\xi =\mathfrak J_t\xi.
\]
These identities hold first on the form core and extend to form
solutions by the graph bounds and uniqueness. Consequently
$\psi(t)=e^{-i\int_s^t c_{N,r}\d r}\Psi_N(t)$ with the same
initial state in \eqref{eq:physical-reconstruction}.
All expectations below are insensitive to this scalar phase.

\begin{lemma}
\label{lem:reconstruction-number-moment}
Uniformly on $I$,
\begin{equation}\label{eq:reconstruction-number-moment}
	\|M^{3/2}e^{B_t}e^{C_t}z\| \le C_I\|M^{3/2}z\|.
\end{equation}
\end{lemma}
\begin{proof}
The quadratic phase has number bandwidth two and blocks bounded
by $C_I(n+1)$, because $\|\eta_t\|_{\mathrm{HS}}$ is uniform and
the generalized-field factors are contractions. On a nonzero
block the difference $|(n+3)^3-(n+1)^3|$ is bounded by
$C(n+1)^2$. Differentiating
$\|M^{3/2}e^{aB_t}z\|^2$ for $|a|\le1$, summing the finitely
many neighboring blocks, and applying $2xy\le x^2+y^2$ gives
an absolute derivative bound by $C_I$ times that same norm
squared. Gronwall proves the quadratic bound; it is also
\cite[Lemma 2.1]{COS} with exponent three.

For the cubic phase, the corresponding difference of number
cubes is again $O((n+1)^2)$ and the block norm is
$C_I\epsilon(n+1)^{3/2}$. The relative coefficient is
$C_I\epsilon\sqrt{n+1}$, uniformly bounded on its particle-number cutoff support by \eqref{eq:weak-budget}. The same argument gives
$\|M^{3/2}e^{C_t}z\|\le C_I\|M^{3/2}z\|$.
The estimates include the boundary sectors since absent blocks
contribute zero. Composition proves the assertion.
\end{proof}

\begin{theorem}
\label{thm:physical-sixth-moment}
Let $v\in C_c^\infty(\mathbb R^3)$ be nonnegative and radial.
Let $\phi_t$ solve the modified GP equation with an $N$-independent
normalized initial datum in $H^6(\mathbb R^3)$. Fix a sufficiently
small quadratic scattering radius $\ell\in(0,1/2)$, independent of $N$.
Use the cubic phase \eqref{spec:eq:actual-phase} at this background,
with $m=N^{-1/2}$ and the original COS cutoff $0\le\Theta_N\le1$.
It equals one for $n\le N^{1/2}/2+10$ and zero for
$n\ge N^{1/2}-10$, for sufficiently large $N$.
Fix a compact interval $I$ and $s\in I$. Choose a fixed $b_I$
so that $F_t=\mathfrak J_t+b_IM$ satisfies
$c_IQ_t\le F_t\le C_IQ_t$ on $I$, uniformly in $N$.
Let $z_N\in P_s\cF^{\le N}$ be normalized, with
$E_{*,N}=\|F_s^{3/2}z_N\|^2<\infty$.
Set $\xi_N(t)=U_{\rm ex}(t,s)z_N$ and
$\Psi_N(t)=\mathcal U_{N,t}^*e^{B_t}e^{C_t}\xi_N(t)$,
up to the scalar reconstruction phase. Set
\[
	\mathcal N_{\perp,N}(t)=\sum_{j=1}^N(q_t)_j.
\]
Then
\begin{equation}\label{eq:physical-depletion-third}
	\sup_{t\in I}\langle\Psi_N(t), (\mathcal N_{\perp,N}(t)+1)^3\Psi_N(t)\rangle \le C_{I,s}E_{*,N}.
\end{equation}
For bounded self-adjoint $O$ on the one-particle space, define
\[
	\mu_t=\langle\phi_t,O\phi_t\rangle,\qquad \mathcal X_{O,N}(t)=N^{-1/2} \left(\sum_{j=1}^N O_j-N\mu_t\right).
\]
One has
\begin{equation}\label{eq:physical-sixth-moment}
	\sup_{t\in I}\langle\Psi_N(t), \mathcal X_{O,N}(t)^6\Psi_N(t)\rangle \le C_{I,s}\|O\|^6E_{*,N}.
\end{equation}
The same estimate, with a different numerical constant, holds
for the sixth moment centered at the actual expectation.
\end{theorem}
\begin{proof}
Write $\zeta_N(t)=e^{B_t}e^{C_t}\xi_N(t)$. The extraction
identity is
$\mathcal U_{N,t}\mathcal N_{\perp,N}(t)\mathcal U_{N,t}^*=\mathcal N$.
Thus \eqref{eq:physical-depletion-third} follows directly from
\eqref{eq:reconstruction-number-moment},
\eqref{eq:exact-moment-coercivity} and
\eqref{eq:exact-third-moment}.

For the second assertion put $f_t=q_tO\phi_t$,
$h_t=q_tOq_t$ on the excitation one-particle space, and
$g_N(\mathcal N)=\sqrt{1-\mathcal N/N}$. The exact extraction identities of \cite[Eq.~(2.4)]{COS},
applied to the one-particle observable, give
\begin{equation}\label{eq:collective-field-exact}
	\begin{aligned}
	Y_{O,N}(t)
	&:=\mathcal U_{N,t}\mathcal X_{O,N}(t)\mathcal U_{N,t}^*\\
	&=a^*(f_t)g_N(\mathcal N)+g_N(\mathcal N)a(f_t)\\
	&\quad+N^{-1/2}\{\dGamma(h_t)-\mu_t\mathcal N\}.
	\end{aligned}
\end{equation}
The order of the factors in the first two terms is retained.
Since $\|f_t\|,\|h_t\|,|\mu_t|\le\|O\|$, all blocks satisfy
\[
	(Y_{O,N})_{mn}=0\quad\hbox{if }|m-n|>1,\qquad \|(Y_{O,N})_{mn}\|\le C\|O\|\sqrt{n+1}.
\]
For the diagonal block use $n/\sqrt N\le\sqrt n$ for
$0\le n\le N$; for the other blocks use the creation and
annihilation bounds and $0\le g_N\le1$.
Along each of the at most $3^3$ paths of length three,
all intermediate numbers differ from the starting one by at
most three. Therefore finite-band Cauchy--Schwarz gives
\begin{equation}\label{eq:collective-cube-bound}
	\|Y_{O,N}(t)^3\zeta\| \le C\|O\|^3\|M^{3/2}\zeta\|.
\end{equation}
At fixed $N$ all operators in this calculation are bounded,
so the identity and its powers apply on the whole truncated
space. Self-adjointness of $\mathcal X_{O,N}$ shows that its
sixth moment equals $\|Y_{O,N}^3\zeta_N\|^2$. Apply
\eqref{eq:collective-cube-bound} and the already proved bound
on $\|M^{3/2}\zeta_N\|$.

Finally, for the spectral probability measure of
$\mathcal X_{O,N}$, Jensen and
$|x-a|^6\le32(|x|^6+|a|^6)$ give
\[
	\langle|\mathcal X_{O,N}-\langle\mathcal X_{O,N}\rangle|^6\rangle \le64\langle\mathcal X_{O,N}^6\rangle.
\]
This proves the centered assertion.
\end{proof}

For a bounded region $D$, taking $O$ to be multiplication by
$\mathbf1_D$ gives a concrete density statement. If
$\mathcal N_D=\sum_{j=1}^N\mathbf1_D(x_j)$, then
\begin{equation}\label{eq:local-density-sixth}
	\sup_{t\in I} \left\langle\left| \frac{\mathcal N_D-N\int_D|\phi_t(x)|^2\d x}{\sqrt N} \right|^6\right\rangle_{\Psi_N(t)} \le C_{I,s}E_{*,N}.
\end{equation}
The same constant works for every measurable $D$, since
$\|\mathbf1_D\|_\infty\le1$. The particle-number fluctuation
is normalized by $\sqrt N$.

\subsection{Tail estimates}

The moment bounds imply polynomial decay of the measurement tails.
The estimates below are uniform in particle number and in time
on the prescribed compact interval.

\begin{corollary}
\label{cor:exact-moment-tails}
Suppose $E_*:=\sup_N E_{*,N}<\infty$. For $\lambda,R,L>0$,
the spectral measurement probabilities satisfy
\begin{align}
	\sup_{N,\,t\in I}
	\|\mathbf1_{[\lambda,\infty)}(F_t)\xi_N(t)\|^2
	&\le C_{I,s}E_*\lambda^{-3},\label{eq:exact-energy-tail}\\
	\sup_{N,\,t\in I}
	\mathbb P_{\Psi_N(t)}(\mathcal N_{\perp,N}(t)\ge L)
	&\le C_{I,s}E_*L^{-3},\label{eq:depletion-tail}\\
	\sup_{N,\,t\in I}
	\mathbb P_{\Psi_N(t)}(|\mathcal X_{O,N}(t)|\ge R)
	&\le C_{I,s}\|O\|^6E_*R^{-6}.
	\label{eq:collective-tail}
\end{align}
In particular, at fixed depletion fraction $\delta>0$,
$\sup_{t\in I}\mathbb P(\mathcal N_{\perp,N}(t)\ge\delta N)
\le C_{I,s}E_*\delta^{-3}N^{-3}$, and
\[
	\sup_{t\in I}N^{-1} \langle\mathcal N_{\perp,N}(t)\rangle_{\Psi_N(t)} \le C_{I,s}E_*^{1/3}N^{-1}.
\]
\end{corollary}
\begin{proof}
The spectral theorem \cite{ReedSimonI} reduces the assertion to
the scalar Markov inequality. For a positive operator $A$, the
resulting spectral inequality
$\mathbf1_{[r,\infty)}(A)\le r^{-p}A^p$ gives the first two
bounds with $p=3$. Apply it to $|\mathcal X_{O,N}|$ with $p=6$
for the third. The last expectation bound follows by H\"older
from \eqref{eq:physical-depletion-third}.
The probability bounds are uniform for $t\in I$.
\end{proof}

\section{Microscopic energy carried by cubic correlations}
\label{sec:microscopic-spectrum}

We now examine the energy distribution produced by the same cubic
correlation transformation, beginning at the initial time. Unless
an evolving background is specified, we fix $\phi=\phi_0$ and
impose $\phi\in C_c^\infty(\mathbb R^3)$.
Section~\ref{sec:evolving-spectrum} removes compact support
at positive times and allows the condensate to depend on $N$.
The potential is nonzero. The observable is the excitation
energy $Q_{N,P}$, before conjugation by the cubic phase. It differs
from the positive correlated generator $F_{\rm ex}$ in
Theorem~\ref{thm:D}.

\subsection{Scattering data and the initial phase}

Let $v\in C_c^\infty(\mathbb R^3)$ be nonzero, nonnegative and radial.
Let $f$ be the zero-energy scattering solution
\[
	(-\Delta+v/2)f=0,\qquad f(x)\longrightarrow1,
	\qquad w=1-f.
\]
Write $a>0$ for its scattering length. Outside the support of $v$,
$w(x)=a/|x|$. In particular, $w\notin L^2(\mathbb R^3)$, while
$\nabla w\in L^2$. Set
\begin{equation}\label{spec:eq:D}
	D_v=\int vw=2\int|\nabla w|^2+\int vw^2
	=\int v-8\pi a>0.
\end{equation}
These identities follow by integrating the scattering equation and
then multiplying $-2\Delta w=vf$ by $w$; the boundary term in the
second computation tends to zero.

On bosonic Fock space $\cF$ over $L^2(\mathbb R^3)$, put
\[
	M=\cN+1,\qquad A=\cK+M,\qquad
	Q_N=A+\cV_N,
	\qquad \cV_N=\frac12\int N^2v(N(x-y))a_x^*a_y^*a_ya_x\d x\d y.
\]
Here $N$ is a positive integer, distinct from the particle number
$\cN$. The operators are the positive operators associated with the
displayed forms. Equivalently, they are direct sums of their
sector restrictions. For fixed $N$ and particle number $n$, the
potential is a bounded multiplication operator and $D(Q_{N,n})=H^2_s$.

Fix normalized $\phi\in C_c^\infty(\mathbb R^3)$ and a real radial
$0\le\rho\in C_c^\infty(\mathbb R^3)$ equal to one near zero. For
$0<\alpha<1$, define
\[
	m=N^{-\alpha},\qquad \epsilon=(m/N)^{1/2},\qquad
	k_N(r)=\sqrt N\,w(Nr)\rho(r/m).
\]
Let $0\le\chi_N\le1$ be a real function of particle number, supported
in $n\le L_N<\infty$, and equal to one at $n=1,2$. On the smooth
finite-particle core set
\begin{equation}\label{spec:eq:phase}
	G_N=\int k_N(x-y)\phi(y)a_x^*a_y^*a_x\d x\d y,
	\qquad C_N=\chi_N(\cN)(G_N-G_N^*)\chi_N(\cN).
\end{equation}
The estimates below show that $C_N$ extends to a bounded
anti-self-adjoint operator for each $N$. Write
\[
	U_N=e^{C_N},\qquad \mathscr H_N=U_N^*Q_NU_N.
\]
We first use the undressed, unprojected phase \eqref{spec:eq:phase};
Section~\ref{spec:sec:projected-profile} treats its condensate-orthogonal
compression.
Except for the inverse-operator estimates, we impose the cutoff condition
\begin{equation}\label{spec:eq:budget}
	\epsilon(L_N+2)^{3/2}\le D
\end{equation}
with fixed $D$. A fixed cutoff $L_N\ge2$ is allowed, as is a growing
cutoff satisfying \eqref{spec:eq:budget}. The phase acts on every sector
within that cutoff.

For a fixed $h\in C_c^\infty(\mathbb R^3)$ in the one-particle sector, define
the finite energy-weighted measure on $[0,\infty)$ by
\begin{equation}\label{spec:eq:energy-measure}
	\nu_{N,h}(B)
	=\ip{\mathscr H_N^{1/2}h}
	{\mathbf1_B(\mathscr H_N/N^2)\mathscr H_N^{1/2}h}.
\end{equation}
The domain assertion and a uniform bound for its total mass are
proved below. This measure resolves the microscopic energy contribution.

Section~\ref{spec:sec:finite-sectors} extends the result to fixed smooth
finite-sector vectors $\psi$, assuming that the cutoff is one on
their occupied sectors and on the immediately higher sector.
The contact coefficient then becomes
$2\ip\psi{\dGamma(|\phi|^2)\psi}$.
The extension to a fixed weighted energy domain requires a cutoff
plateau eventually containing every fixed finite set of particle
numbers. These additional assumptions are stated with the results.
Section~\ref{spec:sec:dressed-profile} proves the corresponding statement
for $N$-dependent dressed sources and applies it to the actual
cubic phase of \cite{COS} at its initial time.

Nonzero energy effects of cubic GP phases are already present in
\cite[Section 3 and Eq.~(8.24)]{COS}. The relative Hamiltonian
$-2\Delta+v$ and short-scale correlations also occur in the
analysis of correlation formation in \cite{EMS}. Here we compute a spectral measure
for the explicitly prepared auxiliary state $U_Nh$.
Section~\ref{sec:evolving-spectrum} treats the corresponding
question for the common initial family under exact many-body
evolution. Throughout, $Q_N$ denotes the excitation energy,
which differs from the complete many-body Hamiltonian.

\subsubsection{Main result for the dressed excitation phase}

Let $q=1-|\phi\rangle\langle\phi|$ and $P=\Gamma(q)$.
Write $A_P,Q_{N,P}$ for the operators of the corresponding
restricted forms on $P\cF$, and put
\[
	\mathscr X_P=P\cF\cap D(M^2A^{1/2}).
\]
Let $f_R^{\mathrm{Neu}}$ be the ground state of
$-\Delta+v/2$ on the ball of radius $R$, with Neumann boundary
condition and boundary value one, extended by one outside.
Set $w_R^{\mathrm{Neu}}=1-f_R^{\mathrm{Neu}}$.
Fix $\ell\in(0,1/2)$ and define
\[
	\eta_N=(q\otimes q)
	\big[-Nw_{N\ell}^{\mathrm{Neu}}(N(x-y))\phi(x)\phi(y)\big],
	\quad \gamma_N=\cosh\eta_N,\quad \sigma_N=\sinh\eta_N,
\]
where, for a complex symmetric kernel, the series conventions are
\[
	\cosh\eta_N=\sum_{j\ge0}\frac{(\eta_N\overline{\eta_N})^j}{(2j)!},
	\qquad
	\sinh\eta_N=\sum_{j\ge0}
	\frac{(\eta_N\overline{\eta_N})^j\eta_N}{(2j+1)!}.
\]
The limiting kernel is
\begin{align}
	\eta_\infty&=(q\otimes q)
	[-w_{\infty,\ell}(x-y)\phi(x)\phi(y)],\notag\\
	w_{\infty,\ell}(r)
	&=a\left(\frac1{|r|}-\frac3{2\ell}
	+\frac{|r|^2}{2\ell^3}\right)
	\mathbf1_{\{|r|\le\ell\}}.
	\label{spec:eq:actual-eta-limit}
\end{align}
Put $\gamma_\infty=\cosh\eta_\infty$ and
$\sigma_\infty=\sinh\eta_\infty$, with the same series.
The phase at the initial background in \cite[Eq.~(3.2)]{COS} is
\begin{align}
	\mathcal C_N^{\mathrm{COS}}
	&=-\Theta_N(\cN)\int \sqrt N\,
	w_{Nm}^{\mathrm{Neu}}(N(x-y))\phi(y)
	\widetilde b_x^*\widetilde b_y^*
	[\widetilde b(\gamma_{N,x})
	+\widetilde b^*(\sigma_{N,x})]\d x\d y
	-\mathrm{h.c.},\label{spec:eq:actual-phase}
\end{align}
where $m=N^{-1/2}$, $\widetilde a(f)=a(qf)$ and
$\widetilde b(f)=\sqrt{1-\cN/N}\,\widetilde a(f)$ on
$P\cF^{\le N}$. The adjoints have the reversed order.
Here $a(f)=\int\overline{f(z)}a_z\d z$; the subscript $x$
denotes the column of a kernel, including the identity
distribution in $\gamma_{N,x}$.
For large $N$, the real cutoff $0\le\Theta_N\le1$ equals one
below $N^{1/2}/2+10$ and vanishes above $N^{1/2}-10$.
Extend the phase by zero above the physical particle number $N$.
The definitions and bounds below also justify its exponential.

Let $L=-2\Delta+v$ on $L^2(\mathbb R^3)$ and
\[
	\xi_v=L^{-1/2}v,\qquad
	\mu_v(B)=\ip{\xi_v}{\mathbf1_B(L)\xi_v}.
\]
Lemma~\ref{spec:lem:scattering-vector} proves that this inverse power
is defined on $v$, that $\mu_v([0,\infty))=D_v$, and that
$\mu_v(\{0\})=0$. Define
\[
	c_B[\psi]=\int|\phi(x)|^2
	\norm{[a(\gamma_{\infty,x})+
	a^*(\sigma_{\infty,x})]\psi}^2\d x.
\]
This form is continuous on $D(M^{1/2})$.

\begin{theorem}
\label{spec:thm:actual-COS-profile}
Let $v\in C_c^\infty(\mathbb R^3)$ be nonzero, nonnegative and
radial. Fix an $N$-independent normalized
$\phi\in C_c^\infty(\mathbb R^3)$ and $\ell\in(0,1/2)$.
Let $q=1-|\phi\rangle\langle\phi|$, $P=\Gamma(q)$, and use
the Neumann pair kernel $\eta_N$ and its limit
$\eta_\infty$ in \eqref{spec:eq:actual-eta-limit}.
Let $C_N=\mathcal C_N^{\mathrm{COS}}$ be the phase
\eqref{spec:eq:actual-phase}, with $m=N^{-1/2}$.
Its real cutoff $0\le\Theta_N\le1$ equals one for
$n\le N^{1/2}/2+10$ and zero for $n\ge N^{1/2}-10$.
Extend $C_N$ by zero above sector $N$.
Let
\[
	\mathscr H_N^{\mathrm{COS}}
	=e^{-\mathcal C_N^{\mathrm{COS}}}Q_{N,P}
	e^{\mathcal C_N^{\mathrm{COS}}}.
\]
For every fixed $\psi\in\mathscr X_P$, define
\[
	\nu_{N,\psi}^{\mathrm{COS}}(B)
	=\ip{(\mathscr H_N^{\mathrm{COS}})^{1/2}\psi}
	{\mathbf1_B(\mathscr H_N^{\mathrm{COS}}/N^2)
	(\mathscr H_N^{\mathrm{COS}})^{1/2}\psi}.
\]
These finite measures converge against every bounded continuous
test on $[0,\infty)$ to
\begin{equation}\label{spec:eq:actual-COS-profile}
	A_P[\psi]\delta_0+
	2\left(\int|\phi(x)|^2
	\norm{[a(\gamma_{\infty,x})+
	a^*(\sigma_{\infty,x})]\psi}^2\d x\right)\mu_v.
\end{equation}
For each fixed test this convergence is uniform on compact
subsets of $\mathscr X_P$. On the fixed Hilbert space $P\cF$,
the extended operators also satisfy
\[
	\norm{(\mathscr H_N^{\mathrm{COS}})^{-1}-A_P^{-1}}
	\le C N^{-1/2}.
\]
\end{theorem}

The coefficient in \eqref{spec:eq:actual-COS-profile} is $2c_B[\psi]$.
It retains interference between creation and annihilation sources.
The theorem concerns the excitation energy $Q_{N,P}$ after
the stated preparation.

The proof is given in Section~\ref{spec:sec:dressed-profile}, by verifying
the hypotheses of Theorem~\ref{spec:thm:dressed-profile}; the inverse
estimate follows from Corollary~\ref{spec:cor:dressed-inverse}.
We first derive the spectral limit for the undressed phase.
The growing-vector perturbation lemma and pair orthogonality in
Section~\ref{spec:sec:finite-sectors} extend it to multiple excitations.
Section~\ref{spec:sec:projected-profile} treats compressed spectral
operators, before the approximation of the dressed source.

\subsection{Excitation number and microscopic energy in the same state}
\label{sec:spec:joint-number-energy}

The vacuum gives a direct comparison between number fluctuations and
microscopic energy. The creation part of the dressed field produces
three excitations at its first nonzero order. Their probability tends
to zero, but their contribution to the first energy moment does not.
We first show that the coefficient of this contribution is positive.

\begin{lemma}
\label{lem:spec:vacuum-contact-positive}
Let $v\in C_c^\infty(\mathbb R^3)$ be nonzero, nonnegative and
radial. Fix an $N$-independent normalized
$\phi\in C_c^\infty(\mathbb R^3)$ and $\ell\in(0,1/2)$.
Let $q=1-|\phi\rangle\langle\phi|$, $P=\Gamma(q)$, and use
the Neumann pair kernel $\eta_N$ and its limit
$\eta_\infty$ in \eqref{spec:eq:actual-eta-limit}.
With $\sigma_\infty=\sinh\eta_\infty$,
\[
	c_0:=c_B[\Omega]=\int|\phi(x)|^2\|\sigma_\infty(x,\cdot)\|_2^2\d x>0.
\]
\end{lemma}
\begin{proof}
Let $S=\{x:\phi(x)\ne0\}$. The kernel $\eta_\infty$ vanishes
almost everywhere outside $S\times S$. To check this and
nonvanishing, write $w_\infty=w_{\infty,\ell}$,
$a_\phi=w_\infty*|\phi|^2$ and
$b_\phi=\int|\phi|^2a_\phi$. Expanding the two projections gives
\[
	\eta_\infty(x,y)=\phi(x)\phi(y)\{-w_\infty(x-y)+a_\phi(x)+a_\phi(y)-b_\phi\}.
\]
The last three terms are bounded because $w_\infty\in L^1$
and $\phi\in L^\infty$. Choose a ball on which $|\phi|$ is
bounded below. For two points in that ball with sufficiently
small positive separation, $a/|x-y|$ exceeds the bounded terms
and the regular part of $w_\infty$. The kernel is therefore
nonzero on a set of positive measure. Thus $\eta_\infty\ne0$.

Regard $\eta_\infty$ as a compact operator. Its kernel is symmetric,
so $\overline{\eta_\infty}=\eta_\infty^*$ and
$T=\eta_\infty\overline{\eta_\infty}\ge0$. The defining series gives
\[
	\sigma_\infty=f(T)\eta_\infty,\qquad f(t)=\frac{\sinh\sqrt t}{\sqrt t},\quad f(0)=1.
\]
Since $f(T)\ge1$, this operator is nonzero. The same series shows
that its kernel vanishes almost everywhere outside $S\times S$.
If the displayed integral were zero, positivity of $|\phi|^2$
on $S$ would force the kernel to vanish there as well, contradicting
$\sigma_\infty\ne0$. Its finiteness follows from the
Hilbert--Schmidt bound and $\phi\in L^\infty$.
Finally, $a(\gamma_{\infty,x})\Omega=0$ and
$\|a^*(\sigma_{\infty,x})\Omega\|=\|\sigma_{\infty,x}\|_2$.
These identities give the stated normalization of $c_B[\Omega]$.
\end{proof}

\begin{proposition}
\label{prop:spec:joint-vacuum-profile}
Let $v\in C_c^\infty(\mathbb R^3)$ be nonzero, nonnegative and
radial. Fix an $N$-independent normalized
$\phi\in C_c^\infty(\mathbb R^3)$ and $\ell\in(0,1/2)$.
Let $q=1-|\phi\rangle\langle\phi|$, $P=\Gamma(q)$, and use
the Neumann pair kernel $\eta_N$ and its limit
$\eta_\infty$ in \eqref{spec:eq:actual-eta-limit}.
Let $C_N=\mathcal C_N^{\mathrm{COS}}$ be the phase
\eqref{spec:eq:actual-phase}, with $m=N^{-1/2}$.
Its real cutoff $0\le\Theta_N\le1$ equals one for
$n\le N^{1/2}/2+10$ and zero for $n\ge N^{1/2}-10$.
Extend $C_N$ by zero above sector $N$.
Set
$u_N=e^{\mathcal C_N^{\mathrm{COS}}}\Omega$.
For every $r\ge0$,
\begin{equation}\label{eq:spec:vacuum-number-bound}
	\|M^{r/2}(u_N-\Omega)\|\le C_rN^{-3/4},\qquad 0\le\langle u_N,M^ru_N\rangle-1\le C_rN^{-3/2}.
\end{equation}
For $h:\mathbb N_0\to\mathbb C$ bounded and
$g\in C_b([0,\infty))$,
\begin{align}
	&\left\langle Q_{N,P}^{1/2}u_N,
	h(\cN)g(Q_{N,P}/N^2)Q_{N,P}^{1/2}u_N\right\rangle\notag\\
	&\qquad\longrightarrow h(0)g(0)+2c_0h(3)\int g(\lambda)\d\mu_v(\lambda).
	\label{eq:spec:joint-vacuum-profile}
\end{align}
In particular, for every real $p>1$,
\begin{equation}\label{eq:spec:vacuum-energy-growth}
	\mathfrak m_{p,N}(u_N):=\int_{[0,\infty)}E^p\,\d\langle u_N,\mathbf1_{(-\infty,E]}(Q_{N,P})u_N\rangle\ge c_pN^{2p-2}
\end{equation}
for all sufficiently large $N$, with $c_p>0$. The integral is an
extended spectral moment. It equals $\|Q_{N,P}^{p/2}u_N\|^2$
when $u_N\in D(Q_{N,P}^{p/2})$ and is $+\infty$ otherwise.
\end{proposition}
\begin{proof}
Write $C_N=\mathcal C_N^{\mathrm{COS}}$, $\epsilon=N^{-3/4}$,
and let $\mathbf1_n$ be the projection onto sector $n$.
The sector estimates in Theorem~\ref{spec:thm:dressed-profile} give
\[
	\|\mathbf1_jC_N\mathbf1_n\|\le K\epsilon(n+1)^{3/2},\qquad |j-n|\le3,
\]
with zero blocks outside the cutoff support. The support satisfies
$\epsilon(L_N+6)^{3/2}\le D$. Adjacent number weights are
comparable, and there are at most seven blocks in each row and
column. The Schur bound therefore gives
\[
	\|M^{r/2}C_NM^{-r/2}\|\le K_rD,\qquad \|M^{r/2}C_N\Omega\|\le K_r\epsilon.
\]
The exponential series on the weighted number space yields
\[
	\|M^{r/2}e^{sC_N}M^{-r/2}\|\le e^{K_rD},\qquad 0\le s\le1.
\]
Integrating $e^{sC_N}C_N\Omega$ proves the first bound in
\eqref{eq:spec:vacuum-number-bound}. Since $u_N$ is normalized,
\[
	\langle u_N,M^ru_N\rangle-1=\sum_{n\ge1}\bigl((n+1)^r-1\bigr)\|\mathbf1_nu_N\|^2\le\|M^{r/2}(u_N-\Omega)\|^2.
\]
This proves the second bound.

The first-insertion estimate \eqref{spec:eq:dressed-reduction},
including its sign and generalized-field variants, gives
\[
	u_N=\Omega+z_N+r_N,\qquad \mathbf1_3z_N=z_N,\qquad \|Q_{N,P}^{1/2}r_N\|\longrightarrow0.
\]
Only the three-creation term acts on vacuum; the annihilation
terms and the adjoint vanish there. The energies of $u_N$ and
$\Omega+z_N$ are uniformly bounded. The operator $Q_{N,P}$
preserves particle number and satisfies $Q_{N,P}\Omega=\Omega$.
Its bounded joint functional calculus with $\cN$ thus removes
the mixed vacuum--three-particle term exactly. The contribution
of $r_N$ tends to zero by Cauchy--Schwarz in the energy norm.
Theorem~\ref{spec:thm:actual-COS-profile}, applied with spectral test $g$,
then identifies the remaining three-particle contribution as
$2c_0\int g\d\mu_v$. Multiplication by $h(3)$ gives
\eqref{eq:spec:joint-vacuum-profile}.

Choose $0<b<c<\infty$ and $0\le g\le1$ supported in $[b,c]$
with $\int g\d\mu_v>0$. Such a test exists because
$\mu_v([0,\infty))=D_v>0$ and $\mu_v(\{0\})=0$.
The energy in this window has a positive limit by
\eqref{eq:spec:joint-vacuum-profile} and
Lemma~\ref{lem:spec:vacuum-contact-positive}.
The pointwise inequality
\[
	E^p\ge(bN^2)^{p-1}E g(E/N^2),\qquad E\ge0,
\]
proves \eqref{eq:spec:vacuum-energy-growth} by the spectral theorem.
\end{proof}

\begin{corollary}
\label{cor:spec:vacuum-physical-moments}
Let $v\in C_c^\infty(\mathbb R^3)$ be nonzero, nonnegative and
radial. Fix an $N$-independent normalized
$\phi\in C_c^\infty(\mathbb R^3)$ and $\ell\in(0,1/2)$.
Let $q=1-|\phi\rangle\langle\phi|$, $P=\Gamma(q)$, and use
the Neumann pair kernel $\eta_N$ and its limit
$\eta_\infty$ in \eqref{spec:eq:actual-eta-limit}.
Let $C_N=\mathcal C_N^{\mathrm{COS}}$ be the phase
\eqref{spec:eq:actual-phase}, with $m=N^{-1/2}$.
Its real cutoff $0\le\Theta_N\le1$ equals one for
$n\le N^{1/2}/2+10$ and zero for $n\ge N^{1/2}-10$.
Extend $C_N$ by zero above sector $N$.

For the many-body state
\[
	\Phi_N=\mathcal U_{N,0}^*e^{B_0}e^{C_0}\Omega,
\]
one has, uniformly in $N$,
\[
	\langle\Phi_N,(\mathcal N_{\perp,N}(0)+1)^3\Phi_N\rangle\le C,\qquad \langle\Phi_N,\mathcal X_{O,N}(0)^6\Phi_N\rangle\le C\|O\|^6
\]
for every bounded self-adjoint one-particle observable $O$.
At the same initial background, the cubic phase is not uniformly
bounded from $D(A_P^{p/2})$ to $D(Q_{N,P}^{p/2})$ for any $p>1$.
\end{corollary}
\begin{proof}
The initial phase $C_0$ is precisely
$\mathcal C_N^{\mathrm{COS}}$. Apply
Lemma~\ref{lem:reconstruction-number-moment} to $\Omega$, and then
the extraction identity and \eqref{eq:collective-cube-bound}.
The graph assertion follows from $A_P\Omega=\Omega$ and
\eqref{eq:spec:vacuum-energy-growth}.
\end{proof}

The energy in \eqref{eq:spec:vacuum-energy-growth} is measured
in the excitation representation after the cubic transformation.
The corollary concerns the reconstructed state at the initial time.
Theorem~\ref{thm:D} propagates its physical moments for initial
families satisfying the third correlated-energy bound.

\subsection{The second microscopic energy moment}
\label{sec:spec:second-moment}

For the vacuum preparation, the second-derivative estimates
used in the cubic comparison determine the second energy moment.

\begin{theorem}
\label{thm:spec:vacuum-second-moment}
Let $v\in C_c^\infty(\mathbb R^3)$ be nonzero, nonnegative and
radial. Fix an $N$-independent normalized
$\phi\in C_c^\infty(\mathbb R^3)$ and $\ell\in(0,1/2)$.
Let $q=1-|\phi\rangle\langle\phi|$, $P=\Gamma(q)$, and use
the Neumann pair kernel $\eta_N$ and its limit
$\eta_\infty$ in \eqref{spec:eq:actual-eta-limit}.
Let $C_N=\mathcal C_N^{\mathrm{COS}}$ be the phase
\eqref{spec:eq:actual-phase}, with $m=N^{-1/2}$.
Its real cutoff $0\le\Theta_N\le1$ equals one for
$n\le N^{1/2}/2+10$ and zero for $n\ge N^{1/2}-10$.
Extend $C_N$ by zero above sector $N$.
Let $u_N=e^{\mathcal C_N^{\mathrm{COS}}}\Omega$. Then
$u_N\in D(Q_{N,P})$ for every sufficiently large $N$, and
\begin{equation}\label{eq:spec:vacuum-second-moment}
	N^{-2}\|Q_{N,P}u_N\|^2\longrightarrow2c_0\|v\|_2^2.
\end{equation}
The same limit holds for $N^{-2}\operatorname{Var}_{u_N}(Q_{N,P})$.
\end{theorem}
\begin{proof}
We first control the full exponential in the operator domain.
Write $C=C_0=\mathcal C_N^{\mathrm{COS}}$, $A=A_P$ and
$\epsilon=N^{-3/4}$. In this proof all number-weight exponents
are fixed, and all constants are independent of $N$.
The sector estimates and \eqref{eq:KC-op} imply
\[
	\|M^r C\Omega\|\le C_r\epsilon,\qquad \|M^rA^{1/2}C\Omega\|\le C_r,\qquad \|M^rAC\Omega\|\le C_rN.
\]
Here $A\Omega=\Omega$. For the last bound, use
$AC=CA+[K_P,C]+[M,C]$; the number commutator has the same
sector bound as $C$. Applying these estimates once more gives
\[
	\|M^rC^2\Omega\|\le C_r\epsilon^2,\quad \|M^rA^{1/2}C^2\Omega\|\le C_r\epsilon,\quad \|M^rAC^2\Omega\|\le C_r(N\epsilon+1).
\]
For example, the last commutator term uses the first-energy
bound on $C\Omega$. The weighted norm of $CA C\Omega$ is
$O_r(N\epsilon)$. All inputs occupy finitely many sectors.

The same commutator identity, integrated along $e^{sC}$, gives
for $0\le s\le1$
\begin{align}
	\|M^rAe^{sC}w\|\le C_r\bigl(&\|M^rAw\|+N\|M^{r+3}w\|\notag\\
	&+\|M^{r+3/2}A^{1/2}w\|\bigr).
	\label{eq:spec:operator-phase-bound}
\end{align}
Indeed, write $Ae^{sC}w=e^{sC}Aw+
\int_0^se^{(s-\tau)C}[A,C]e^{\tau C}w\d\tau$.
The number-weight bound and \eqref{eq:A-phase} control the
integrand in \eqref{eq:KC-op}. The extra number term is absorbed
by $NM^{r+3}$, since $M\ge1$ and $N\ge1$. In particular,
\[
	\|M^{r+3/2}A^{1/2}e^{\tau C}w\|\le C_r\bigl(\|M^{r+3/2}A^{1/2}w\|+\|M^{r+3}w\|\bigr).
\]
At fixed $N$, the number cutoff is finite. The commutator bound
and $AC=CA+[A,C]$ show that $C$ is bounded on $D(A)$ with its
graph norm. Its exponential series therefore converges in that
norm. This justifies the integral identity on $D(A)$ and proves
invariance of that domain at each fixed $N$.

Set $r_N=(e^C-1-C)\Omega$. The integral Taylor formula applied
to $C^2\Omega$, together with \eqref{eq:spec:operator-phase-bound},
yields
\[
	\|M^rr_N\|\le C_r\epsilon^2,\qquad \|Ar_N\|\le C(N\epsilon+1).
\]
On sector $n$, the potential has operator norm at most
$CN^2(n+1)^2$. The same bound holds after compression.
Consequently
\begin{equation}\label{eq:spec:operator-vacuum-remainder}
	\|Q_{N,P}r_N\|\le C(N\epsilon+1+N^2\epsilon^2)=O(N^{1/2})=o(N).
\end{equation}
This also proves $u_N\in D(Q_{N,P})$: for each fixed $N$ the
phase preserves finitely many sectors and their $H^2$ domains,
on which the interaction is bounded.

It remains to identify $N^{-1}Q_{N,P}C\Omega$.
In \eqref{spec:eq:actual-phase}, the term containing
$\widetilde b(\gamma_{N,x})$ and the adjoint annihilate vacuum.
The remaining term has three creation fields. Its dressing kernel
is $s_N=\sigma_N$. The estimates proved with \eqref{eq:KC-op} give
\[
	\|s_N\|_2\le C,\qquad \|\nabla s_N\|_2\le CN^{1/2},\qquad \|\nabla^2s_N\|_2\le CN^{3/2}.
\]
Choose fixed smooth finite-rank kernels $s^{(j)}$ tending to
$s_\infty$ in Hilbert--Schmidt norm, with their second variable
in $qL^2$. They are obtained by applying $1\otimes q$ to smooth
finite sums of tensor products approximating $s_\infty$.
This operation preserves smooth compact support and converges
in $L^2$, because $(1\otimes q)s_\infty=s_\infty$.
To justify this approximation in the scaled operator
norm, put $J_{N,h}(x,y,z)=k_N^{\mathrm{Neu}}(x-y)\phi(y)h(x,z)$.
The product rule and the kernel bounds give
\[
	\|J_{N,h}\|_{H^1}\le C(\|h\|_2+\epsilon\|\nabla h\|_2),\qquad \|J_{N,h}\|_{H^2}\le C(N\|h\|_2+\|\nabla h\|_2+\epsilon\|\nabla^2h\|_2).
\]
The constants are independent of $h,N$. The Sobolev norms take
derivatives in all spatial variables.
Terms with two derivatives on $k_N$ are bounded by $CN\|h\|_2$.
With one derivative on each factor, the bound is $C\|\nabla h\|_2$.
With both derivatives on $h$, it is $C\epsilon\|\nabla^2h\|_2$.
Integrating in $y$ first proves these estimates. Derivatives on
$\phi$ use its bounded derivatives and give smaller terms.
On a fixed particle sector,
Sobolev's inequality applied to the square of each pair potential
gives
$\|\mathcal V_{N,n}f\|_2\le C_nN\|f\|_{H^1}$.
Indeed, $\|N^4v(N\cdot)^2\|_{3/2}=N^2\|v^2\|_{3/2}$.
For fixed spectator coordinates, apply H\"older and the
three-dimensional Sobolev inequality in $x_i$ to obtain
$\|N^2v(N(x_i-x_j))f\|_2^2\le CN^2\|\nabla_{x_i}f\|_2^2$.
Integration over the spectators and summation over the finitely
many pairs prove the claim.
Projection is bounded on $H^1$ and $H^2$ for the fixed smooth
condensate. Thus the compressed three-particle insertion
$T_{N,h}=P_3\int k_N^{\mathrm{Neu}}(x-y)\phi(y)h(x,z)
a_x^*a_y^*a_z^*\Omega\d x\d y\d z$ obeys
\begin{equation}\label{eq:spec:scaled-insertion-approximation}
	N^{-1}\|Q_{N,P}T_{N,h}\|\le C\left(\|h\|_2+(\epsilon+N^{-1})\|\nabla h\|_2+\frac\epsilon N\|\nabla^2h\|_2\right).
\end{equation}
Here $P_3=q^{\otimes3}$. Symmetrization has a fixed normalization
factor $\sqrt{3!}$ and is bounded on both Sobolev spaces.
The kinetic term divided by $N$ is bounded by
$CN^{-1}\|J_{N,h}\|_{H^2}$; the potential term divided by
$N$ is bounded by $C\|J_{N,h}\|_{H^1}$.
Their sum gives \eqref{eq:spec:scaled-insertion-approximation}.
For $h=s_N-s^{(j)}$, the derivative terms tend to zero for
each fixed $j$. Hilbert--Schmidt convergence therefore implies
\[
	\limsup_{N\to\infty}N^{-1}\|Q_{N,P}(T_{N,s_N}-T_{N,s^{(j)}})\|\le C\|s_\infty-s^{(j)}\|_2.
\]
The cutoff equals one on the vacuum insertion. Applying the
three generalized creation fields from right to left gives the
exact identity
\[
	C\Omega=-\beta_{3,0,N}(0)T_{N,s_N},\qquad \beta_{3,0,N}(0)=\sqrt{(1-N^{-1})(1-2N^{-1})}=1+O(N^{-1}).
\]
Equation~\eqref{eq:spec:scaled-insertion-approximation} shows
that replacing this coefficient by one gives a vanishing error
after applying $Q_{N,P}/N$. The common minus sign is irrelevant
for the squared norm.

For fixed $j$, resolve $T_{N,s^{(j)}}$ into its three unordered
pairs. Before applying $P_3$, its three-particle wave function is
\[
	\frac1{\sqrt6}\sum_{\pi\in S_3}k_N^{\mathrm{Neu}}(x_{\pi(1)}-x_{\pi(2)})\phi(x_{\pi(2)})s^{(j)}(x_{\pi(1)},x_{\pi(3)}).
\]
Each unordered pair occurs twice. At zero pair separation,
these two terms have the same amplitude. For pair $a$ it is
\[
	a_a^{(j)}(R,z)=\sqrt{\frac23}\,\phi(R)s^{(j)}(R,z).
\]
The operator-vector calculation
\eqref{spec:eq:pair-operator-vector}, with the Neumann equation
\eqref{spec:eq:neumann-energy-vector}, gives the limit
$a_a^{(j)}\otimes v$ under that pair's dilation.
For the compressed operator, the removed part is negligible
by \eqref{spec:eq:projected-removed}. The remaining projection
converges to the spectator projection and fixes this limiting
amplitude, since $s^{(j)}(R,\cdot)\in qL^2$.
Different pair dilations are asymptotically orthogonal by
Lemma~\ref{spec:lem:pair-orthogonality}. Hence
\[
	N^{-2}\|Q_{N,P}T_{N,s^{(j)}}\|^2\longrightarrow 2\left(\int|\phi(R)|^2\|s^{(j)}(R,\cdot)\|_2^2\d R\right)\|v\|_2^2.
\]
Let $j\to\infty$ using the scaled operator estimate above.
Finally $Q_{N,P}\Omega=\Omega$ and
\eqref{eq:spec:operator-vacuum-remainder} allow replacement of
$C\Omega$ by $u_N$. This proves
\eqref{eq:spec:vacuum-second-moment}.
The first moments converge to $1+2c_0D_v$ by
Theorem~\ref{spec:thm:actual-COS-profile}. Their squared contribution
to the variance vanishes after division by $N^2$.
\end{proof}

The vacuum calculation extends to regular reference states. The
additional input is an operator estimate for the pair potential,
with explicit dependence on excitation number. The weighted
first-energy bound for this phase is
\cite[Lemma 3.1]{COS}. The estimate below retains the leading
operator vector at the scale needed for the second moment.

\begin{theorem}
\label{thm:spec:reference-second-moment}
Let $v\in C_c^\infty(\mathbb R^3)$ be nonzero, nonnegative and
radial. Fix an $N$-independent normalized
$\phi\in C_c^\infty(\mathbb R^3)$ and $\ell\in(0,1/2)$.
Let $q=1-|\phi\rangle\langle\phi|$, $P=\Gamma(q)$, and use
the Neumann pair kernel $\eta_N$ and its limit
$\eta_\infty$ in \eqref{spec:eq:actual-eta-limit}.
Let $C_N=\mathcal C_N^{\mathrm{COS}}$ be the phase
\eqref{spec:eq:actual-phase}, with $m=N^{-1/2}$.
Its real cutoff $0\le\Theta_N\le1$ equals one for
$n\le N^{1/2}/2+10$ and zero for $n\ge N^{1/2}-10$.
Extend $C_N$ by zero above sector $N$.
Put
\[
	\mathscr X_{2,P}=P\cF\cap D(M^2A_P),\qquad u_{N,\psi}=e^{\mathcal C_N^{\mathrm{COS}}}\psi .
\]
Give $\mathscr X_{2,P}$ the norm $\|M^2A_P\psi\|$.
For every $\psi\in\mathscr X_{2,P}$ and sufficiently large $N$,
one has $u_{N,\psi}\in D(Q_{N,P})$ and
\begin{equation}\label{eq:spec:reference-second-moment}
	N^{-2}\|Q_{N,P}u_{N,\psi}\|^2\longrightarrow 2c_B[\psi]\|v\|_2^2 .
\end{equation}
The convergence is uniform on compact subsets of
$\mathscr X_{2,P}$. For normalized $\psi$, the same limit holds
for $N^{-2}\operatorname{Var}_{u_{N,\psi}}(Q_{N,P})$.
Both limits also hold for the normalized physical preparations
$e^{\mathcal C_N^{\mathrm{COS}}}\mathbf1_{\{\cN\le N\}}\psi/
\|\mathbf1_{\{\cN\le N\}}\psi\|$.
\end{theorem}
\begin{proof}
Write $A=A_P$, $Q=Q_{N,P}$, $C=\mathcal C_N^{\mathrm{COS}}$
and $\epsilon=N^{-3/4}$. Write
$\kappa_N=k_N^{\mathrm{Neu}}$ for its relative kernel.
We first prove a bound on the full reference domain. On sector
$n$, let $W_{ij}=N^2v(N(x_i-x_j))$. The Sobolev estimate in the
preceding proof gives
$\|W_{ij}f\|^2\le CN^2\|\nabla_{x_i}f\|^2$.
Cauchy--Schwarz over the pairs, followed by summation over their
coordinates, yields
\[
	\left\|\sum_{i<j}W_{ij}f\right\|^2\le CN^2(n+1)^3\|\mathcal K_n^{1/2}f\|^2 .
\]
Here $\mathcal K_n=\sum_{i=1}^n-\Delta_{x_i}$.
For an excitation vector the restricted kinetic form equals
the unrestricted one, by definition. In particular,
$\|K_{P,n}^{1/2}f\|=\|\mathcal K_n^{1/2}f\|$ on the
restricted form domain. The projection preserves $H^1$ by
\eqref{spec:eq:projection-H1}. On each fixed sector it also
preserves $H^2$, since $\phi$ is smooth.
The potential is the sectorwise compression
$\mathcal V_{N,P,n}=P_n\mathcal V_{N,n}P_n$, whose output
projection is a contraction.
Thus, on the direct sum of all excitation sectors,
\begin{equation}\label{eq:spec:potential-operator-number}
	N^{-1}\|\mathcal V_{N,P}w\|\le C\|M^{3/2}K_P^{1/2}w\| .
\end{equation}
This operator estimate follows from Sobolev's inequality.
Combining it with \eqref{eq:A-phase} and
\eqref{eq:spec:operator-phase-bound} gives
\begin{align}
	N^{-1}\|Qe^C\psi\|
	&\le C\bigl(N^{-1}\|A\psi\|+\|M^3\psi\|+\|M^{3/2}A^{1/2}\psi\|\bigr)\notag\\
	&\le C\|M^2A\psi\| .
	\label{eq:spec:reference-operator-bound}
\end{align}
The operators $A$ and $M$ strongly commute. Their joint
spectral variables obey $a\ge m\ge1$, so $a,m^3$ and
$m^{3/2}a^{1/2}$ are bounded by $m^2a$.
For the domain assertion, first truncate excitation number.
On each retained sector, $D(A_P)=P_nH^2$; this follows by
removing the bounded off-diagonal part of $\mathcal K_n$
relative to $P_n$, as in the proof of
\eqref{spec:eq:microscopic-block}.
Smooth compactly supported approximation followed by $P_n$
is therefore dense in this operator domain. It gives core
vectors $\psi_j\to\psi$ in $\|M^2A\cdot\|$.
At fixed $N$, unitarity gives $e^C\psi_j\to e^C\psi$,
and \eqref{eq:spec:reference-operator-bound} makes
$Qe^C\psi_j$ Cauchy. Closedness of $Q$ proves the assertion.
The constant in that estimate was already independent of $N$
on the core and remains so after taking the limit.

We now work on this projected smooth core. If $\psi$ vanishes
above $n_0$, then $C\psi$ and $C^2\psi$ vanish above
$n_0+3$ and $n_0+6$, respectively. For large $N$ these
sectors lie in the cutoff plateau. The block estimates for $C$ and
\eqref{eq:KC-op}, for every fixed $r\ge0$, give
\[
	\|M^rC\psi\|=O_\psi(\epsilon),\qquad \|M^rA^{1/2}C\psi\|=O_\psi(1),\qquad \|M^rAC\psi\|=O_\psi(N).
\]
Apply the same estimates to $C\psi$. In the operator bound use
$AC^2=CAC+[A,C]C$, and in $CAC$ use the $L^2$ block bound for
$C$. This gives
\[
	\|M^rC^2\psi\|=O_\psi(\epsilon^2),\qquad \|M^rA^{1/2}C^2\psi\|=O_\psi(\epsilon),\qquad \|M^rAC^2\psi\|=O_\psi(N\epsilon+1).
\]
The integral Taylor remainder and
\eqref{eq:spec:operator-phase-bound} therefore imply, for
$r_N=(e^C-1-C)\psi$,
\[
	\|M^rr_N\|=O_\psi(\epsilon^2),\qquad \|Ar_N\|=O_\psi(N\epsilon+1).
\]
The elementary operator bound
$\|\mathcal V_{N,P}w\|\le CN^2\|M^2w\|$ now yields
\begin{equation}\label{eq:spec:reference-operator-remainder}
	\|Qr_N\|=O_\psi(N\epsilon+1+N^2\epsilon^2)=o(N).
\end{equation}
Also $\|Q\psi\|=O_\psi(1+\sqrt N)$: on each fixed smooth
sector the integral of a squared pair potential is
$N\|v\|_2^2$, and the remaining coordinates have bounded
compactly supported amplitudes. It remains to evaluate
$QC\psi/N$.

Use the canonical raising operator $G_N$ of
\eqref{spec:eq:dressed-phase}, with the coefficient convention
of \eqref{spec:eq:actual-phase}. Its annihilation field is
$a(\gamma_{N,x})$ and its creation field is
$a^*(\sigma_{N,x})$. With $a(f)=\int\overline{f(z)}a_z\d z$,
the coefficient kernels in \eqref{spec:eq:dressed-phase} are
\[
	p_N(x,z)=\overline{(\gamma_N-1)(z,x)},\qquad s_N(x,z)=\sigma_N(z,x).
\]
Their limits define exactly the field in $c_B$.
The proof of Lemma~\ref{lem:op-inputs}, from the Neumann
equation and the hyperbolic series, gives for either kernel $h_N$
\[
	\|h_N\|_2\le C,\qquad \|\nabla h_N\|_2\le CN^{1/2},\qquad \|\nabla^2h_N\|_2\le CN^{3/2}.
\]
We use these bounds for the actual COS kernels in the
operator estimate below.
On the sectors of $\psi$, the cutoff
equals one for large $N$. The generalized-field coefficients
in each cubic monomial are $1+O_\psi(N^{-1})$.
We next show that the raising part has bounded norm after
applying $Q/N$, whereas the adjoint tends to zero in that norm.
These estimates also justify replacing those coefficients by one.

For the identity term in $G_N^*\psi$, integrate derivatives
of $\kappa_N(x-y)$ by parts in the annihilated $y$ coordinate.
They then fall on $\phi$ and on the fixed input.
Since $\|\kappa_N\|_1=O(N^{-3/2})$, the resulting vector has
$H^2$ norm $O_\psi(N^{-3/2})$. In the adjoint of the
$p_N$ term, the only new output coordinate is $z$. Its
derivatives act on the conjugate kernel. More explicitly,
before symmetrization its contraction on $\psi_n$ is
\[
	I_{N,p}(z,Y)=\int\kappa_N(x-y)\overline{\phi(y)p_N(x,z)}\psi_n(x,y,Y)\d x\d y .
\]
The adjoint of the $s_N$ term instead has contraction
\[
	I_{N,s}(Y)=\int\kappa_N(x-y)\overline{\phi(y)s_N(x,z)}\psi_n(x,y,z,Y)\d x\d y\d z .
\]
They have $n-1$ and $n-3$ output coordinates.
The sector factors from contraction and symmetrization are
bounded by $C(n_0+3)^{3/2}$. Cauchy--Schwarz in the
annihilated variables, followed by integration in the
outputs, bounds these contractions by the $L^2$ norm of
the kernel times fixed input norms. Thus
\begin{align}
	\|(G_{N,p_N})^*\psi\|_{H^1}
	&\le C_\psi\epsilon(\|p_N\|_2+\|\nabla p_N\|_2),\notag\\
	\|(G_{N,p_N})^*\psi\|_{H^2}
	&\le C_\psi\epsilon(\|p_N\|_2+\|\nabla p_N\|_2+\|\nabla^2p_N\|_2).
	\label{eq:spec:adjoint-operator-estimates}
\end{align}
Here the Sobolev norms include all output coordinates;
derivatives on spectators act on $\psi$.
The adjoint of the $s_N$ term creates no particles, so its
$H^2$ norm is at most $C_\psi\epsilon\|s_N\|_2$.
The same bounds hold after projection, since there are only
finitely many sectors and the condensate is fixed and smooth.
The kernel estimates
$\|\nabla p_N\|_2=O(N^{1/2})$ and
$\|\nabla^2p_N\|_2=O(N^{3/2})$, together with
\eqref{eq:spec:potential-operator-number}, show that
\begin{align}
	N^{-1}\|QPG_N^*\psi\|&\le C_\psi\bigl(N^{-1}\|PG_N^*\psi\|_{H^2}+\|PG_N^*\psi\|_{H^1}\bigr)\notag\\
	&\longrightarrow0 .
	\label{eq:spec:adjoint-scaled-vanishing}
\end{align}

For the raising part, choose fixed smooth finite-rank
$p^{(j)},s^{(j)}$ tending in Hilbert--Schmidt norm to
$p_\infty,s_\infty$. Project the second variable of
$s^{(j)}$ into $qL^2$. For a dressing kernel $h$, the
product-rule argument for
\eqref{eq:spec:scaled-insertion-approximation} gives
\begin{equation}\label{eq:spec:reference-insertion-approximation}
	N^{-1}\|QP G_{N,h}\psi\|\le C_\psi\left(\|h\|_2+(\epsilon+N^{-1})\|\nabla h\|_2+\frac{\epsilon}{N}\|\nabla^2h\|_2\right).
\end{equation}
This holds for both the annihilation and the creation term.
For a creation term, the kernel is
$\kappa_N(x-y)\phi(y)h(x,z)$ times a fixed spectator wave
function. Its $H^1$ and $H^2$ bounds are the ones in the
vacuum proof, with derivatives on spectators contributing
only $C_\psi$.
For an annihilation term, contract the $z$ variable against
the smooth input before symmetrizing. Cauchy--Schwarz in $z$
gives the same bounds; derivatives on the input again
contribute only $C_\psi$.
All contraction factors are bounded by $C(n_0+3)^{3/2}$,
independently of $N,h$.
Projection and symmetrization are bounded in $H^1,H^2$ on
these finitely many sectors. The kinetic term is estimated
in $H^2/N$ and the interaction in $H^1$, using
\eqref{eq:spec:potential-operator-number}.
The identity part obeys the same product estimates without
$h$ and is bounded in the scaled operator norm.
Consequently $N^{-1}\|QPG_N\psi\|\le C_\psi$.
For $h=p_N-p^{(j)}$ or $s_N-s^{(j)}$, the derivative terms
in \eqref{eq:spec:reference-insertion-approximation} vanish
as $N\to\infty$ at each fixed $j$: their largest orders are
$N^{-1/4}$ and $N^{-1/4}$, respectively. Hence
\begin{align}
	\limsup_{N\to\infty}N^{-1}\|QP(G_N-G_N^{(j)})\psi\|
	\le C_\psi\bigl(\|p_\infty-p^{(j)}\|_2+\|s_\infty-s^{(j)}\|_2\bigr).
	\label{eq:spec:reference-insertion-limit}
\end{align}

For fixed $j$, use the sector formula
\eqref{spec:eq:dressed-sector}. In output sector $\ell$,
the diagonal amplitude for each unordered pair is
\[
	a_\ell^{(j)}(R,Y)=\frac{2\phi(R)(B_R^{(j)}\psi)_{\ell-2}(Y)}{\sqrt{\ell(\ell-1)}} .
\]
The operator-vector limit
\eqref{spec:eq:pair-operator-vector}, with the Neumann
identity \eqref{spec:eq:neumann-energy-vector}, sends
each such contribution under its pair dilation to
$a_\ell^{(j)}\otimes v$ after applying $Q/N$.
These are smooth compactly supported amplitudes with
spectators in $qL^2$, as required in that single-pair proof.
For each pair contribution $g_{N,a}$,
\eqref{spec:eq:projected-removed} gives
$\|Q_{N,\ell}^{\rm full}(1-P_\ell)g_{N,a}\|/N\to0$.
Use $Q_{N,P,\ell}P_\ell g=P_\ell Q_{N,\ell}^{\rm full}P_\ell g$.
Under the pair dilation the output projection converges
strongly to the spectator projection, which fixes the limit.
This proves the compressed operator-vector limit.
Different pair dilations are asymptotically orthogonal by
Lemma~\ref{spec:lem:pair-orthogonality}, and different output
sectors are orthogonal. Summing their squared norms gives
\begin{align*}
	N^{-2}\|QPG_N^{(j)}\psi\|^2&\longrightarrow 2\|v\|_2^2\sum_{\ell\ge2}\int|\phi(R)|^2\|(B_R^{(j)}\psi)_{\ell-2}\|^2\d R\\
	&=2c_{B^{(j)}}[\psi]\|v\|_2^2 .
\end{align*}
The sum is finite. Creation and annihilation contributions
to the same $(B_R^{(j)}\psi)_{\ell-2}$ have been added
before taking the squared norm.
Let $j\to\infty$ in
\eqref{eq:spec:reference-insertion-limit}. The contact
converges by the creation and annihilation bounds.
Equation~\eqref{eq:spec:adjoint-scaled-vanishing}, the
generalized-field coefficients and
\eqref{eq:spec:reference-operator-remainder} then prove
\eqref{eq:spec:reference-second-moment} on the core.

To pass to $\mathscr X_{2,P}$, approximate in
$\|M^2A\cdot\|$ and use
\eqref{eq:spec:reference-operator-bound}. The limiting
quadratic form is continuous in this norm, since $c_B$
is continuous on $D(M^{1/2})$. These bounds also give
uniform convergence on compact sets by finite approximation.
Indeed, the maps $\psi\mapsto N^{-1}\|Qe^C\psi\|$
are uniformly Lipschitz in $\|M^2A\cdot\|$.
The limit is $\sqrt2\|v\|_2\|T_B\psi\|$, where
$(T_B\psi)(x)=\phi(x)B_x\psi$ and
$\|T_B\psi\|\le C\|M^{1/2}\psi\|$.
Finite core approximations of a compact set therefore give
uniform convergence of these norms and of their squares.
For normalized inputs, Theorem~\ref{spec:thm:actual-COS-profile}
applied to the constant test $1$ keeps the first moments
bounded, so their squares divided by $N^2$ vanish.
The number truncations commute with $M^2A$. Their tails
thus tend to zero in $\mathscr X_{2,P}$, and their norms
tend to one. The normalized truncations and their limit
form a compact set in $\mathscr X_{2,P}$ and in
$\mathscr X_P$. The compact-uniform second- and first-energy
limits prove both assertions for the physical preparations.
\end{proof}

The theorem determines the second excitation-energy moment
of the cubic transform of a reference in $D(M^2A_P)$.

\subsection{Initial states for the spectral and propagation results}
\label{sec:common-preparation}

The spectral theorem and the propagation theorem can be applied to
the same prepared family. The required step is convergence of the
initial preparation in the energy form norm. We prove this at the
common initial background, using the quadratic-generator comparison
of \cite{COS}.

In this subsection write $Q_N=Q_{N,P}$, $A=A_P$ and $C_N=C_0$.
The operators $Q_N,A$ also denote their extensions to the full
excitation Fock space by the same sector forms. Let
$\Pi_N=\mathbf1_{\{\cN\le N\}}$ and
$\mathfrak J_N=\mathfrak J(0)$ on $\Pi_NP\cF$.
As in the spectral theorem, extend $C_N$ by zero above sector $N$.
Set
\begin{equation}\label{eq:common-centered-energy}
	\alpha_N=\mathfrak J_N[\Omega],\qquad \widehat F_{N,b}=\mathfrak J_N-\alpha_N I+bM.
\end{equation}
The subtraction fixes the scalar energy reference at vacuum.
The form bounds already proved imply $|\alpha_N|\le C$.
For every sufficiently large fixed $b$, the form
$\widehat F_{N,b}$ is positive and uniformly equivalent to $Q_N$.

Let $\mathcal J_{2,N}$ and $\mathcal J_{2,\infty}$ denote the
quadratic generators at time zero in
\cite[Eqs.~(2.16), (2.28)]{COS}, compressed to the common
excitation space. Their vacuum expectations vanish. Define
\[
	F_{\infty,b}=\mathcal J_{2,\infty}+bM=K_P+bM+S_\infty.
\]
Comparisons on the truncated space use the compressed form
$\Pi_N\mathcal J_{2,\infty}\Pi_N$; the definition of
$F_{\infty,b}$ itself is on the full excitation space.
The normal-ordered expression in \cite[Eq.~(2.28)]{COS} has
a bounded one-particle coefficient and a Hilbert--Schmidt pair
coefficient; see also \cite[Lemma 4.2 and Proposition 5.1]{COS}.
Thus $S_\infty$ changes the excitation number by at most two and, for
every fixed real $r$,
\begin{equation}\label{eq:common-quadratic-weight}
	\|M^rS_\infty M^{-r}M^{-1}\|\le C_r.
\end{equation}
In particular $F_{\infty,b}$ is positive and self-adjoint on
$D(A)$ for sufficiently large $b$.

\begin{lemma}
\label{lem:common-form-comparison}
Uniformly for $u,v\in\Pi_NP\cF$ in the indicated domains,
\begin{align}
	| (\widehat F_{N,b}-F_{\infty,b})[u,v] |
	\le C\|Q_N^{1/2}u\|\bigl\{&N^{-1/4}
	(\|MQ_N^{1/2}v\|+\|M^{5/2}v\|)\notag\\
	&+N^{-1/2}(\|Av\|+\|M^{3/2}v\|+\|K_P^{1/2}v\|)\bigr\}.
	\label{eq:common-form-comparison}
\end{align}
\end{lemma}
\begin{proof}
The exact reconstruction identity identifies $\mathfrak J_N$
with the generator in \cite[Eq.~(3.7)]{COS}, up to the scalar
subtracted in our definition. Both use the same modified GP
background and the same quadratic and cubic transformations.
Compression removes the off-diagonal connection from the form,
since $P\mathscr ZP=0$.
By \cite[Proposition 3.3]{COS}, for a real scalar $d_N$,
\[
	\mathfrak J_N=d_N I+\mathcal J_{2,N}+\mathcal V_{N,P}+E_N,
\]
where
\[
	|E_N[u,v]|\le CN^{-1/4}\|Q_N^{1/2}u\|\|(Q_N+M^3)^{1/2}Mv\|.
\]
Indeed, put $\mathcal H_N=\cK+\mathcal V_N$ as in COS.
The first squared norm in their bound equals $Q_N[u]$ after
restriction to the excitation space. The second squared norm obeys
\[
	\mathcal H_N[Mv]+\|\cN^{3/2}Mv\|^2+\|Mv\|^2\le Q_N[Mv]+\|M^{5/2}v\|^2.
\]
Since $\mathcal J_{2,N}[\Omega]=\mathcal V_{N,P}[\Omega]=0$,
one has $\alpha_N=d_N+E_N[\Omega]$ and
$|E_N[\Omega]|\le CN^{-1/4}$. Therefore
\[
	\widehat F_{N,b}-F_{\infty,b}=\mathcal V_{N,P}+(\mathcal J_{2,N}-\mathcal J_{2,\infty})+E_N-E_N[\Omega]I
\]
as forms on the truncated space.

The potential estimate of Proposition~\ref{spec:prop:projected}
gives $\|\mathcal V_{N,P}^{1/2}v\|\le CN^{-1/2}\|Av\|$.
Together with $\mathcal V_{N,P}\le Q_N$, this controls its form.
The error $E_N$ is bounded by the first line of
\eqref{eq:common-form-comparison}, because $Q_N$ and $M$ strongly
commute. Its scalar expectation is absorbed by $M\ge1$.

For $R_N=\mathcal J_{2,N}-\Pi_N\mathcal J_{2,\infty}\Pi_N$,
\cite[Proposition 5.2]{COS} at $t=0$ states
\[
	|R_N[u,v]|\le CN^{-1/2}\bigl(\|Mu\|\|Mv\|+\|M^{1/2}u\|\|K_P^{1/2}v\|\bigr).
\]
The same bound holds after compression. Let $\mathbf1_n$
project onto sector $n$. Both generators have number shifts
$0,2,-2$. For $m,n\le N$ with $m-n\in\{0,2,-2\}$,
duality in sector $m$ and comparability of neighboring weights give
\[
	\|(m+1)^{-1/2}\mathbf1_mR_N\mathbf1_nv_n\|\le CN^{-1/2}\bigl((n+1)^{3/2}\|v_n\|+\|K_P^{1/2}v_n\|\bigr).
\]
This defines the block as a Hilbert-space vector for each
$v_n\in D(K_P^{1/2})$. All other blocks vanish. Compression
only removes blocks whose output sector exceeds $N$.
Squaring and summing the three diagonals therefore gives
\[
	\|M^{-1/2}R_Nv\|\le CN^{-1/2}\bigl(\|M^{3/2}v\|+\|K_P^{1/2}v\|\bigr).
\]
Here $R_N$ is compressed to the truncated space. Pairing this
vector with $M^{1/2}u$ and using $M\le Q_N$ proves the remaining
terms in \eqref{eq:common-form-comparison}.
\end{proof}

\begin{theorem}
\label{thm:common-preparation}
Let $v\in C_c^\infty(\mathbb R^3)$ be nonzero, nonnegative and
radial. Fix an $N$-independent normalized
$\phi\in C_c^\infty(\mathbb R^3)$ and a sufficiently small fixed $\ell\in(0,1/2)$.
Let $q=1-|\phi\rangle\langle\phi|$, $P=\Gamma(q)$, and use
the Neumann pair kernel $\eta_N$ and its limit
$\eta_\infty$ in \eqref{spec:eq:actual-eta-limit}.
Let $C_N=\mathcal C_N^{\mathrm{COS}}$ be the phase
\eqref{spec:eq:actual-phase}, with $m=N^{-1/2}$.
Its real cutoff $0\le\Theta_N\le1$ equals one for
$n\le N^{1/2}/2+10$ and zero for $n\ge N^{1/2}-10$.
Extend $C_N$ by zero above sector $N$.
Use the centered initial energies
$\widehat F_{N,b}=\mathfrak J_N-\mathfrak J_N[\Omega]I+bM$
on $\Pi_NP\cF$ and $F_{\infty,b}=\mathcal J_{2,\infty}+bM$
on $P\cF$, where $\Pi_N=\mathbf1_{\{\cN\le N\}}$. Let $\chi\in P\cF$
be normalized, belong to $D(A)$, and have finitely many nonzero
number sectors. There exists $b_\chi>0$ such that, for every
fixed $b\ge b_\chi$, the normalized vectors
\begin{equation}\label{eq:common-prepared-vectors}
	z_N=\frac{\widehat F_{N,b}^{-1}\chi}{\|\widehat F_{N,b}^{-1}\chi\|},\qquad z_\infty=\frac{F_{\infty,b}^{-1}\chi}{\|F_{\infty,b}^{-1}\chi\|}
\end{equation}
satisfy
\begin{equation}\label{eq:common-energy-convergence}
	\begin{gathered}
	\|Q_N^{1/2}(z_N-z_\infty)\|\le C_{b,\chi}N^{-1/4},\\
	z_\infty\in D(M^2A^{1/2}),\qquad c_B[z_\infty]>0.
	\end{gathered}
\end{equation}
Here $N$ is at least the largest occupied sector of $\chi$, and
$z_N$ is extended by zero above sector $N$. Moreover,
\begin{equation}\label{eq:common-third-initial}
	\sup_N\|F_{\rm ex}(0)^{3/2}z_N\|<\infty,
\end{equation}
where $F_{\rm ex}(0)=\mathfrak J(0)+bM$. The same prepared
family is admissible on every compact time interval in
Theorem~\ref{thm:D}.

Let $u_N=e^{C_N}z_N$. For every $g\in C_b([0,\infty))$,
\begin{equation}\label{eq:common-microscopic-profile}
	\langle Q_N^{1/2}u_N,g(Q_N/N^2)Q_N^{1/2}u_N\rangle\longrightarrow A[z_\infty]g(0)+2c_B[z_\infty]\int g\d\mu_v.
\end{equation}
Thus these same initial vectors satisfy both the microscopic
spectral conclusion and the hypotheses of Theorem~\ref{thm:D}.
\end{theorem}
\begin{proof}
First construct the limiting vector with enough number weights.
Put $H_b=K_P+bM$, which commutes with $M$. Equation
\eqref{eq:common-quadratic-weight} implies
$\|M^rS_\infty H_b^{-1}M^{-r}\|\le C_r/b$.
Choose $b>2\max(C_0,C_2)$ and also large enough for positivity.
On $\mathcal H_2=D(M^2)$ with norm $\|M^2\cdot\|$, the
operator $S_\infty H_b^{-1}$ has a bounded extension of norm
at most $C_2/b$. Its extension agrees with the unweighted
operator, by density of finite-sector vectors and closedness
of $M^2$. The Neumann series for
$(1+S_\infty H_b^{-1})^{-1}$ thus converges on both spaces.
Since $H_b^{-1}$ commutes with $M$, its product with this
inverse maps $\mathcal H_2$ into $D(M^2H_b)$.
For $y=F_{\infty,b}^{-1}\chi$ this gives
\[
	\|M^2H_by\|\le2\|M^2\chi\|,\qquad \|M^2Ay\|\le C_{b,\chi}.
\]
All weights required below are therefore finite. Positivity
follows from the form estimate $|S_\infty[u,u]|\le C\|M^{1/2}u\|^2$;
self-adjointness on $D(A)$ follows from the relative operator
bound $\|S_\infty H_b^{-1}\|<1$.

Let $y_N=\widehat F_{N,b}^{-1}\chi$ and $v_N=\Pi_Ny$.
The weak equations for $y_N,y$ imply, for $u\in\Pi_NP\cF$,
\[
	\widehat F_{N,b}[u,y_N-v_N]=(F_{\infty,b}-\widehat F_{N,b})[u,v_N]+F_{\infty,b}[u,(1-\Pi_N)y].
\]
All the right-hand norms of $v_N$ in
\eqref{eq:common-form-comparison} are uniformly bounded.
For the $N$-dependent form norm use $Q_N\le CMA$, which gives
$\|MQ_N^{1/2}v_N\|\le C\|M^{3/2}A^{1/2}y\|$.
The first term is consequently bounded by
$C_{b,\chi}N^{-1/4}\|Q_N^{1/2}u\|$.

Since $H_b$ preserves number, only $S_\infty$ contributes to
the last term. Its number-form bound gives
\[
	|F_{\infty,b}[u,(1-\Pi_N)y]|\le C\|M^{1/2}u\|\|M^{1/2}(1-\Pi_N)y\|\le C_{b,\chi}N^{-1}\|Q_N^{1/2}u\|.
\]
Testing with $u=y_N-v_N$ and using uniform coercivity of
$\widehat F_{N,b}$ yields
$\|Q_N^{1/2}(y_N-v_N)\|\le C_{b,\chi}N^{-1/4}$.
The full-space tail is also small:
\[
	\|Q_N^{1/2}(1-\Pi_N)y\|\le C\|M^{1/2}A^{1/2}(1-\Pi_N)y\|\le C_{b,\chi}N^{-3/2}.
\]
Thus $\|Q_N^{1/2}(y_N-y)\|\le C_{b,\chi}N^{-1/4}$.
The reference state has uniformly bounded first energy, since
$Q_N[\chi]\le A[\chi]+CN^{-1}\|A\chi\|^2$.
The same holds for $\widehat F_{N,b}[\chi]$, by form comparison.
Jensen's inequality gives
$\|y_N\|\ge\widehat F_{N,b}[\chi]^{-1}\ge c_{b,\chi}>0$.
Also $0<F_{\infty,b}[\chi]\le C_bA[\chi]$, so
$\|y\|\ge F_{\infty,b}[\chi]^{-1}>0$.
Their first energies are uniformly bounded. Normalization
therefore preserves the convergence rate and proves the first
two assertions in \eqref{eq:common-energy-convergence}.

For strict positivity of the limiting contact, set $\eta=M^{-1}\chi$.
This vector belongs to $D(A)$ and has finitely many number sectors.
The resolvent equation gives
\[
	bF_{\infty,b}^{-1}\chi-\eta=-F_{\infty,b}^{-1}(K_P+S_\infty)\eta.
\]
Since $F_{\infty,b}\ge(b-C)M$, one has
$\|M^{1/2}F_{\infty,b}^{-1}\|\le(b-C)^{-1}$.
Consequently $z_\infty\to\eta/\|\eta\|$ in $D(M^{1/2})$
as $b\to\infty$. Let $m$ be the largest occupied sector of
$\eta$. In the $(m+1)$-particle component of
$B_x\eta$, where $B_x=a(\gamma_{\infty,x})+a^*(\sigma_{\infty,x})$,
only $a^*(\sigma_{\infty,x})\eta_m$ occurs. By definition,
$c_B[\eta]=\int|\phi(x)|^2\|B_x\eta\|^2\d x$.
For a finite-sector vector, the canonical commutation relation gives
$\|a^*(f)\eta_m\|^2=\|a(f)\eta_m\|^2+\|f\|_2^2\|\eta_m\|^2$.
Use this identity with $f=\sigma_{\infty,x}\in qL^2$ and
integrate to obtain
\[
	c_B[\eta]\ge\int|\phi(x)|^2\|a^*(\sigma_{\infty,x})\eta_m\|^2\d x\ge c_B[\Omega]\|\eta_m\|^2>0,
\]
where the last step is Lemma~\ref{lem:spec:vacuum-contact-positive}.
Continuity of the contact form on $D(M^{1/2})$ therefore gives
$c_B[z_\infty]>0$ for all sufficiently large $b$.

The preparation argument of
Proposition~\ref{prop:variational-preparation}, applied to
$\widehat F_{N,b}$ and the reference vector $\chi$, gives
\[
	\|\widehat F_{N,b}^{3/2}z_N\|^2\le\widehat F_{N,b}[\chi]^3\le C_{b,\chi}.
\]
At time zero, $F_{\rm ex}(0)=\widehat F_{N,b}+\alpha_N I$.
This is a uniformly bounded scalar shift of a positive operator.
Spectral calculus, with $|\alpha_N|\le C$, proves
\eqref{eq:common-third-initial}. Remark~\ref{rem:exact-number-shift}
then allows the number correction to be enlarged for any fixed
compact interval without changing the prepared initial family.

Finally, Proposition~\ref{prop:full-graph} gives
\[
	\|Q_N^{1/2}e^{C_N}(z_N-z_\infty)\|\le C\|Q_N^{1/2}(z_N-z_\infty)\|\le C_{b,\chi}N^{-1/4}.
\]
Above sector $N$ the cubic phase is the identity, so this
estimate also holds on the extended space. For a bounded test
$g$, the difference of the two energy-weighted spectral forms
is bounded by this norm times the sum of the two bounded
energy form norms and $\|g\|_\infty$.
Theorem~\ref{spec:thm:actual-COS-profile} applies to the fixed
vector $z_\infty\in D(M^2A^{1/2})$ and proves
\eqref{eq:common-microscopic-profile}.
\end{proof}

\begin{corollary}
\label{cor:common-physical-propagation}
For the initial vectors in Theorem~\ref{thm:common-preparation},
let
\[
	\Psi_N(t)=\mathcal U_{N,t}^*e^{B_t}e^{C_t}U_{\rm ex}(t,0)z_N.
\]
For every bounded self-adjoint one-particle observable $O$ and
every fixed compact interval $I$ containing zero,
\begin{gather*}
	\sup_{N,t\in I}\langle\Psi_N(t),(\mathcal N_{\perp,N}(t)+1)^3\Psi_N(t)\rangle<\infty,\\
	\sup_{N,t\in I}\langle\Psi_N(t),\mathcal X_{O,N}(t)^6\Psi_N(t)\rangle\le C_{I,b,\chi}\|O\|^6.
\end{gather*}
At the initial time, $u_N=e^{C_0}z_N$ has the nonzero microscopic
profile \eqref{eq:common-microscopic-profile}. Every excitation-energy
moment of order $p>1$ has an extended spectral lower bound
$c_{p,b,\chi}N^{2p-2}$ for sufficiently large $N$, with
$c_{p,b,\chi}>0$.
\end{corollary}
\begin{proof}
Apply Theorem~\ref{thm:D} using
\eqref{eq:common-third-initial}. The spectral lower bound follows
from a positive compact window in
\eqref{eq:common-microscopic-profile}, as in the proof of
Proposition~\ref{prop:spec:joint-vacuum-profile}.
\end{proof}

This result places the initial microscopic spectrum and the exact
moment propagation on one explicitly specified family.
Sections~\ref{sec:evolving-spectrum} and
\ref{sec:uniform-evolved-spectrum} determine the microscopic
measure at later times, uniformly on compact intervals.
The reference vector can be the vacuum or a finite superposition
of smooth excitations. The initial state is its normalized
inverse-energy preparation.

\subsection{Quasi-free reference states}
\label{sec:quasifree-preparation}

The construction extends to quasi-free states with a regular pair
kernel. These are standard reference states in Bogoliubov theory
\cite{Shale,NNS2016}; the discussion following
\cite[Theorem 2.2]{COS} also considers initial states obtained
by a second quadratic transformation.
The additional inverse-energy preparation below gives the initial
third moment required for the exact evolution.
We first treat the kinetic operator domain. The weaker form-domain
condition is established in Section~\ref{sec:form-energy-preparation}.

\begin{proposition}
\label{prop:weighted-reference-preparation}
Let $v\in C_c^\infty(\mathbb R^3)$ be nonzero, nonnegative and
radial. Fix an $N$-independent normalized
$\phi\in C_c^\infty(\mathbb R^3)$ and a sufficiently small fixed $\ell\in(0,1/2)$.
Let $q=1-|\phi\rangle\langle\phi|$, $P=\Gamma(q)$, and use
the Neumann pair kernel $\eta_N$ and its limit
$\eta_\infty$ in \eqref{spec:eq:actual-eta-limit}.
Let $C_N=\mathcal C_N^{\mathrm{COS}}$ be the phase
\eqref{spec:eq:actual-phase}, with $m=N^{-1/2}$.
Its real cutoff $0\le\Theta_N\le1$ equals one for
$n\le N^{1/2}/2+10$ and zero for $n\ge N^{1/2}-10$.
Extend $C_N$ by zero above sector $N$.
Use the centered initial energies
$\widehat F_{N,b}=\mathfrak J_N-\mathfrak J_N[\Omega]I+bM$
on $\Pi_NP\cF$ and $F_{\infty,b}=\mathcal J_{2,\infty}+bM$
on $P\cF$, where $\Pi_N=\mathbf1_{\{\cN\le N\}}$.
Let $\chi\in P\cF\cap D(A)\cap D(M^2)$ be normalized and assume
$c_B[M^{-1}\chi]>0$. There exists $b_0(\chi)$ such that the
following holds for every fixed $b\ge b_0(\chi)$. Define
\begin{equation}\label{eq:weighted-reference-preparation}
	z_N=\frac{\widehat F_{N,b}^{-1}\Pi_N\chi}{\|\widehat F_{N,b}^{-1}\Pi_N\chi\|},\qquad z_\infty=\frac{F_{\infty,b}^{-1}\chi}{\|F_{\infty,b}^{-1}\chi\|}.
\end{equation}
These vectors are defined for all $N\ge N_0(b,\chi)$.
They satisfy \eqref{eq:common-energy-convergence},
\eqref{eq:common-third-initial} and
\eqref{eq:common-microscopic-profile}. Their exact many-body
reconstruction satisfies Corollary~\ref{cor:common-physical-propagation}.
The uniform bounds are understood for sufficiently large $N$.
\end{proposition}
\begin{proof}
Put $H_b=K_P+bM$. The weighted inverse construction gives
\[
	y=H_b^{-1}(1+S_\infty H_b^{-1})^{-1}\chi,\qquad \|M^2H_by\|\le2\|M^2\chi\|.
\]
Here \eqref{eq:common-quadratic-weight} bounds
$S_\infty H_b^{-1}$ by $C_2/b<1/2$ on $D(M^2)$.
The inverse is its convergent Neumann series on that space.
Since $H_b^{-1}$ commutes with $M$ and $AH_b^{-1}$ is bounded,
one has $y\in D(M^2A)$. Put
$y_N=\widehat F_{N,b}^{-1}\Pi_N\chi$ and $v_N=\Pi_Ny$.
For $u\in\Pi_NP\cF$, one has
$\langle u,\Pi_N\chi\rangle=\langle u,\chi\rangle$.
Hence the weak comparison equation and both tail estimates in
Theorem~\ref{thm:common-preparation} apply without an additional
source term. They give
$\|Q_N^{1/2}(y_N-y)\|\le C_{b,\chi}N^{-1/4}$.
More explicitly, the solution tail
$F_{\infty,b}[u,(1-\Pi_N)y]$ is bounded by
$C_{b,\chi}N^{-1}\|Q_N^{1/2}u\|$, and
$\|Q_N^{1/2}(1-\Pi_N)y\|\le C_{b,\chi}N^{-3/2}$.
All norms of $v_N$ in the form comparison are controlled by
$\|M^2Ay\|$.

To justify normalization and the third moment, write
$t_N=\|\Pi_N\chi\|\to1$. Number conservation of $A$ gives
$Q_N[\Pi_N\chi]\le A[\chi]+CN^{-1}\|A\chi\|^2$.
Thus the normalized reference $t_N^{-1}\Pi_N\chi$ has a
uniform first-energy bound. For $x=\Pi_N\chi$ and
$F=\widehat F_{N,b}$, Cauchy--Schwarz gives
\[
	\|x\|^4\le F[x]\langle x,F^{-1}x\rangle\le F[x]\|x\|\|F^{-1}x\|.
\]
These are form identities; only $x\in D(F^{1/2})$ is needed.
It follows that
\[
	\|y_N\|\ge\frac{t_N^3}{\widehat F_{N,b}[\Pi_N\chi]}\ge c_{b,\chi}>0
\]
for large $N$. Normalization preserves the convergence rate.
Proposition~\ref{prop:variational-preparation}, followed by the
bounded scalar shift from $\widehat F_{N,b}$ to $F_{\rm ex}(0)$,
gives \eqref{eq:common-third-initial}.

For positivity, $\eta=M^{-1}\chi$ belongs to $D(A)$ and
$(K_P+S_\infty)\eta\in\cF$. The same resolvent identity gives
\[
	\|M^{1/2}(bF_{\infty,b}^{-1}\chi-\eta)\|\le\frac{\|(K_P+S_\infty)\eta\|}{b-C}.
\]
After normalization, $z_\infty(b)\to\eta/\|\eta\|$ in
$D(M^{1/2})$ as $b\to\infty$. The hypothesis on $c_B[\eta]$
and its continuity imply
$c_B[z_\infty]>0$ for all sufficiently large $b$.
The spectral transfer and the exact moment bounds now follow
from the same arguments as in
Theorem~\ref{thm:common-preparation} and
Corollary~\ref{cor:common-physical-propagation}.
\end{proof}

We next verify both assumptions on $\chi$ for a class of pure
quasi-free states. Write $\mathfrak h=qL^2(\mathbb R^3)$ and
let $h$ be the nonnegative one-particle operator obtained by
restricting the form of $-\Delta$ to $\mathfrak h$. Thus
$K_P=\dGamma(h)$. A symmetric pair kernel
$\kappa\in\mathfrak h\otimes_s\mathfrak h$ also defines a
Hilbert--Schmidt map $\overline{\mathfrak h}\to\mathfrak h$.
Assume
\begin{equation}\label{eq:quasifree-kernel-domain}
	\|\kappa\|_{\mathrm{op}}<1,\qquad \kappa\in D(h\otimes1+1\otimes h).
\end{equation}
The second condition is an operator-domain assumption, so
$(h\otimes1+1\otimes h)\kappa$ is a two-particle Hilbert-space vector.
Define
\begin{equation}\label{eq:quasifree-reference}
	\begin{gathered}
	\chi_\kappa=Z_\kappa\exp\!\left(\frac12\mathcal P^*(\kappa)\right)\Omega,\qquad Z_\kappa=\det(1-\kappa^*\kappa)^{1/4},\\
	\mathcal P^*(\kappa)=\int\kappa(x,y)a_x^*a_y^*\d x\d y.
	\end{gathered}
\end{equation}
The exponential is defined by its Fock-space series. The determinant
is a Fredholm determinant and is strictly positive. This is the
usual pair representation of a centered pure quasi-free state;
see \cite[Eqs.~(3.28), (3.38)]{Derezinski2017}.

\begin{lemma}
\label{lem:quasifree-reference-contact}
Let $v\in C_c^\infty(\mathbb R^3)$ be nonzero, nonnegative and
radial. Fix an $N$-independent normalized
$\phi\in C_c^\infty(\mathbb R^3)$ and $\ell\in(0,1/2)$.
Let $q=1-|\phi\rangle\langle\phi|$, $P=\Gamma(q)$, and use
the Neumann pair kernel $\eta_N$ and its limit
$\eta_\infty$ in \eqref{spec:eq:actual-eta-limit}.
Let $h$ be the operator associated with the restriction of the
kinetic form to $\mathfrak h=qL^2$. Let
$\kappa\in\mathfrak h\otimes_s\mathfrak h$ be Hilbert--Schmidt,
with $\|\kappa\|_{\mathrm{op}}<1$ and
$\kappa\in D(h\otimes1+1\otimes h)$.
The vector $\chi_\kappa$ in \eqref{eq:quasifree-reference} is normalized, belongs to $D(A)$, and has every
fixed number moment finite. Moreover,
\begin{equation}\label{eq:quasifree-positive-contact}
	c_B[M^{-1}\chi_\kappa]>0.
\end{equation}
More precisely, put
$\rho_0=|\langle\Omega,\chi_\kappa\rangle|^2$ and
$\rho_2=\|\mathbf1_2\chi_\kappa\|^2$.
If $\kappa\ne0$, then
\begin{equation}\label{eq:quasifree-contact-lower}
	c_B[M^{-1}\chi_\kappa]\ge\frac{16\rho_0\rho_2}{9(25\rho_0+9\rho_2)}c_B[\Omega]>0.
\end{equation}
For $\kappa=0$, the left-hand side is $c_B[\Omega]$.
\end{lemma}
\begin{proof}
The mode decomposition for a compact symmetric pair kernel
\cite[Lemma 5 and its proof in the appendix]{NNS2016} gives
$\kappa=\sum_j\lambda_j f_j\otimes f_j$, with orthonormal
$f_j\in\mathfrak h$, $0\le\lambda_j<1$, and
$\sum_j\lambda_j^2<\infty$.
In one mode,
\[
	\left\|\exp\!\left(\frac\lambda2a^*(f)^2\right)\Omega\right\|^2=\sum_{j\ge0}\frac{(2j)!}{2^{2j}(j!)^2}\lambda^{2j}=(1-\lambda^2)^{-1/2}.
\]
Taking products over the modes proves normalization and, for
$\tau>0$ with $e^{2\tau}\|\kappa\|_{\mathrm{op}}^2<1$, gives
\begin{equation}\label{eq:quasifree-number-generating}
	\langle\chi_\kappa,e^{\tau\cN}\chi_\kappa\rangle=\frac{\det(1-\kappa^*\kappa)^{1/2}}{\det(1-e^{2\tau}\kappa^*\kappa)^{1/2}}<\infty.
\end{equation}
The infinite products converge because $\kappa$ is
Hilbert--Schmidt. In particular $\chi_\kappa\in D(M^r)$
for every fixed $r$.

Set $\kappa_h=(h\otimes1+1\otimes h)\kappa$.
The pair-creation action in \cite[Eq.~(3.18)]{Derezinski2017}
gives the bound on sector $n$,
\[
	\|\mathcal P^*(\kappa_h)\psi_n\|\le\sqrt{(n+1)(n+2)}\|\kappa_h\|_2\|\psi_n\|.
\]
Since creation operators commute, the kinetic commutator gives
\begin{equation}\label{eq:quasifree-kinetic-vector}
	K_P\chi_\kappa=\frac12\mathcal P^*(\kappa_h)\chi_\kappa.
\end{equation}
To justify this identity on the operator domain, truncate the mode decomposition
to kernels $\kappa_j$. Positivity of $h$ and joint spectral
calculus give
$(h\otimes1)\kappa,(1\otimes h)\kappa\in\mathfrak h\otimes\mathfrak h$.
Partial contraction with $f_j$ in the second variable then gives
$\lambda_j f_j\in D(h)$, so every nonzero mode lies in $D(h)$.
Let $E_j$ project onto the first
$j$ nonzero modes. The identity
\[
	(h\otimes1)\kappa_j=(1\otimes E_j)(h\otimes1)\kappa
\]
proves convergence after applying $h$ in the first variable.
Symmetry gives the corresponding convergence
in the second variable. Thus $(\kappa_j)_h\to\kappa_h$.
The vacuum overlap of the omitted modes gives
$\langle\chi_{\kappa_j},\chi_\kappa\rangle=Z_\kappa/Z_{\kappa_j}\to1$.
Together with the uniform exponential number bound
\eqref{eq:quasifree-number-generating}, this proves convergence
of the normalized states in $\|M\cdot\|$.
For a fixed finite-mode kernel, put
$\chi_{\kappa_j}^{(L)}=\mathbf1_{\{\cN\le L\}}\chi_{\kappa_j}$.
Each of its finitely many sectors is a tensor of vectors in
$D(h)$, so $\chi_{\kappa_j}^{(L)}\in D(K_P)$. The sectorwise
commutator identity gives
\[
	K_P\chi_{\kappa_j}^{(L)}=\frac12\mathbf1_{\{\cN\le L\}}\mathcal P^*((\kappa_j)_h)\chi_{\kappa_j}.
\]
The right-hand side converges in Hilbert norm as $L\to\infty$,
by the pair bound and $M\chi_{\kappa_j}\in\cF$.
Closedness gives $\chi_{\kappa_j}\in D(K_P)$ and
\eqref{eq:quasifree-kinetic-vector} for this kernel. Next, the pair bound yields
\begin{align*}
	&\|\mathcal P^*((\kappa_j)_h)\chi_{\kappa_j}-\mathcal P^*(\kappa_h)\chi_\kappa\|\\
	&\quad\le C\|(\kappa_j)_h-\kappa_h\|_2\|M\chi_{\kappa_j}\|\\
	&\qquad+C\|\kappa_h\|_2\|M(\chi_{\kappa_j}-\chi_\kappa)\|\longrightarrow0.
\end{align*}
Thus both $\chi_{\kappa_j}$ and $K_P\chi_{\kappa_j}$ converge
in Hilbert norm. Closedness of $K_P$ proves
\eqref{eq:quasifree-kinetic-vector} for $\chi_\kappa$.
Hence $\chi_\kappa\in D(A)$.

It remains to establish strict contact positivity. The pair series
and the canonical commutation relations give
\begin{equation}\label{eq:quasifree-annihilation-relation}
	a(f)\chi_\kappa=a^*(\kappa\overline f)\chi_\kappa,\qquad f\in\mathfrak h.
\end{equation}
To treat its distributional coefficient, define the bounded map
$G_\gamma:\mathfrak h\to L^2(\mathbb R^3)$ by
$(G_\gamma f)(x)=\langle q\gamma_{\infty,x},f\rangle$, with
the inner product linear in its second variable.
This convention fixes the conjugations. Define the kernel
$r=(G_\gamma\otimes1)\kappa$ in
$L^2(\mathbb R^3)\otimes\mathfrak h$, and let $r_x=r(x,\cdot)$.
It satisfies
\[
	\int|\phi(x)|^2\|r_x\|^2\d x\le\|\phi\|_\infty^2\|G_\gamma\|^2\|\kappa\|_2^2.
\]
For a test function $g$, the definition and symmetry of $\kappa$ give
\[
	\int\overline{g(x)}r_x\d x=\kappa\overline{G_\gamma^*g}.
\]
Apply \eqref{eq:quasifree-annihilation-relation} with
$f=G_\gamma^*g$. It identifies the smeared annihilation field
with the smeared creation field $a^*(r_x)$ on $\chi_\kappa$.
Both are Hilbert-space-valued $L^2$ kernels: for
$\psi\in D(M^{1/2})$,
\begin{align*}
	\int\|a(q\gamma_{\infty,x})\psi\|^2\d x&\le\|G_\gamma\|^2\|\cN^{1/2}\psi\|^2,\\
	\int\|a^*(r_x)\psi\|^2\d x&\le\|r\|_2^2\|M^{1/2}\psi\|^2.
\end{align*}
Consequently the smeared identities hold almost everywhere and
in $L^2(|\phi(x)|^2\d x;\cF)$. The ordinary-vector identity
\eqref{eq:quasifree-annihilation-relation} itself follows by
the same mode approximations and the number bounds above.

Put $\eta=M^{-1}\chi_\kappa$ and
$\chi_{\kappa,n}=\mathbf1_n\chi_\kappa$.
Its first and third output sectors under $B_x$ are
\begin{align*}
	\mathbf1_1B_x\eta&=Z_\kappa\bigl(\sigma_{\infty,x}+r_x/3\bigr),\\
	\mathbf1_3B_x\eta&=a^*(\sigma_{\infty,x}/3+r_x/5)\chi_{\kappa,2}.
\end{align*}
Indeed, \eqref{eq:quasifree-annihilation-relation} pairs the
annihilation of sector $n+2$ with creation from sector $n$,
whereas $M^{-1}$ gives the different factors $(n+3)^{-1}$
and $(n+1)^{-1}$. Orthogonality of output sectors and the
creation norm identity therefore imply
\[
	\|B_x\eta\|^2\ge\rho_0\|\sigma_{\infty,x}+r_x/3\|^2+\rho_2\|\sigma_{\infty,x}/3+r_x/5\|^2.
\]
For $\kappa\ne0$ one has $\rho_0=Z_\kappa^2>0$ and
$\rho_2=Z_\kappa^2\|\kappa\|_2^2/2>0$.
Completing the square in $r_x$ bounds the right-hand side below by
\[
	\frac{16\rho_0\rho_2}{9(25\rho_0+9\rho_2)}\|\sigma_{\infty,x}\|_2^2.
\]
Integration against $|\phi(x)|^2\d x$ and
Lemma~\ref{lem:spec:vacuum-contact-positive} prove
\eqref{eq:quasifree-contact-lower}. The case $\kappa=0$ is the vacuum.
\end{proof}

\begin{corollary}
\label{cor:quasifree-common-preparation}
Let $\kappa$ satisfy \eqref{eq:quasifree-kernel-domain} and take
$\chi=\chi_\kappa$ in \eqref{eq:weighted-reference-preparation}.
For every sufficiently large fixed $b$, these prepared vectors
have the nonzero initial microscopic spectral limit
\eqref{eq:common-microscopic-profile} and satisfy the exact
physical-moment bounds of
Corollary~\ref{cor:common-physical-propagation} on every compact
time interval, uniformly for sufficiently large $N$.
\end{corollary}
\begin{proof}
Lemma~\ref{lem:quasifree-reference-contact} verifies the assumptions
of Proposition~\ref{prop:weighted-reference-preparation}.
The vacuum component gives $\|\Pi_N\chi_\kappa\|\ge Z_\kappa>0$.
Together with the uniform coercivity of $\widehat F_{N,b}$,
this ensures normalization throughout the stated range of $N$.
The same fixed initial family
is used on every compact interval.
\end{proof}

For example, let $f\in\mathfrak h\cap H^2(\mathbb R^3)$ be
normalized and take $\kappa=\lambda f\otimes f$ with
$0\le\lambda<1$. Then
\[
	\chi_\kappa=(1-\lambda^2)^{1/4}\sum_{j\ge0}\frac{\sqrt{(2j)!}}{2^j j!}\lambda^j f^{\otimes2j}.
\]
This is a single-mode squeezed vacuum. The corollary applies
for every fixed $\lambda<1$, with constants allowed to depend
on $\lambda$ and $f$. The positive form equation prepares
an admissible initial state from this squeezed reference.

\subsection{The initial energy domain}
\label{sec:vacuum-domain}

The inverse-energy preparation has a further role. For the fixed
Neumann radius used here, the unprepared vacuum need not satisfy
the initial third correlated-energy condition. The obstruction can already
be seen in the limiting quadratic generator. We first give the
resolvent argument that transfers this information to the exact
energies.

\begin{proposition}
\label{prop:initial-energy-resolvent}
Let $v\in C_c^\infty(\mathbb R^3)$ be nonzero, nonnegative and
radial. Fix an $N$-independent normalized
$\phi\in C_c^\infty(\mathbb R^3)$ and a sufficiently small fixed $\ell\in(0,1/2)$.
Let $q=1-|\phi\rangle\langle\phi|$, $P=\Gamma(q)$, and use
the Neumann pair kernel $\eta_N$ and its limit
$\eta_\infty$ in \eqref{spec:eq:actual-eta-limit}.
Let $C_N=\mathcal C_N^{\mathrm{COS}}$ be the phase
\eqref{spec:eq:actual-phase}, with $m=N^{-1/2}$.
Its real cutoff $0\le\Theta_N\le1$ equals one for
$n\le N^{1/2}/2+10$ and zero for $n\ge N^{1/2}-10$.
Extend $C_N$ by zero above sector $N$.
Use the centered initial energies
$\widehat F_{N,b}=\mathfrak J_N-\mathfrak J_N[\Omega]I+bM$
on $\Pi_NP\cF$ and $F_{\infty,b}=\mathcal J_{2,\infty}+bM$
on $P\cF$, where $\Pi_N=\mathbf1_{\{\cN\le N\}}$.
Choose a fixed $b$ so that
$c_bQ_N\le\widehat F_{N,b}\le C_bQ_N$ on $\Pi_NP\cF$
and $c_bA\le F_{\infty,b}\le C_bA$, uniformly in $N$.
Extend $\widehat F_{N,b}$ to $P\cF$ by
\[
	\widetilde F_{N,b}=\widehat F_{N,b}\oplus (K_P+bM)|_{(1-\Pi_N)P\cF}.
\]
Then $\widetilde F_{N,b}$ converges to $F_{\infty,b}$ in the
strong resolvent sense. If $\xi_N\in\Pi_NP\cF$ and
$\xi_N\to\xi$ in norm, then, for every $s>0$,
\begin{equation}\label{eq:initial-energy-liminf}
	\|F_{\infty,b}^{s}\xi\|^2\le\liminf_{N\to\infty}\|\widehat F_{N,b}^{s}\xi_N\|^2.
\end{equation}
Both sides are extended spectral moments. A finite right-hand
side therefore implies $\xi\in D(F_{\infty,b}^{s})$.
\end{proposition}
\begin{proof}
For every finite-sector $\chi\in D(A)$, the unnormalized
resolvent comparison in the proof of
Theorem~\ref{thm:common-preparation} gives
\[
	\|Q_N^{1/2}(\widehat F_{N,b}^{-1}\chi-F_{\infty,b}^{-1}\chi)\|\le C_{b,\chi}N^{-1/4}.
\]
This step requires $b>2\max(C_0,C_2)$ in
\eqref{eq:common-quadratic-weight}, together with uniform
positivity. On this dense set, $\Pi_N\chi=\chi$ for
large $N$. The inverses are uniformly bounded by coercivity.
Density thus gives
$\widetilde F_{N,b}^{-1}\to F_{\infty,b}^{-1}$ strongly on
$P\cF$. Both energies are bounded below by $c_bI$, with
$c_b>0$ independent of $N$. Continuous functional calculus for their bounded
inverses implies strong resolvent convergence
\cite{ReedSimonI}.

For $R>0$ apply bounded continuous functional calculus to
$f_R(\lambda)=\min(\lambda^{2s},R^{2s})$ on the positive
spectra. Strong resolvent convergence and $\xi_N\to\xi$ give
\begin{align*}
	\langle\xi,f_R(F_{\infty,b})\xi\rangle&=\lim_{N\to\infty}\langle\xi_N,f_R(\widetilde F_{N,b})\xi_N\rangle\\
	&\le\liminf_{N\to\infty}\|\widehat F_{N,b}^{s}\xi_N\|^2.
\end{align*}
Let $R\to\infty$ and use monotone convergence for the spectral
measure of $\xi$. This proves \eqref{eq:initial-energy-liminf}.
\end{proof}

Retain the scattering length $a>0$ of $v$ and the
fixed quadratic Neumann radius $\ell$. This radius is distinct
from the shrinking radius $N^{-1/2}$ in the cubic phase.

\begin{theorem}
\label{thm:vacuum-third-obstruction}
Let $v\in C_c^\infty(\mathbb R^3)$ be nonzero, nonnegative and
radial. Fix an $N$-independent normalized
$\phi\in C_c^\infty(\mathbb R^3)$ and a sufficiently small fixed $\ell\in(0,1/2)$.
Let $q=1-|\phi\rangle\langle\phi|$, $P=\Gamma(q)$, and use
the Neumann pair kernel $\eta_N$ and its limit
$\eta_\infty$ in \eqref{spec:eq:actual-eta-limit}.
Let $C_N=\mathcal C_N^{\mathrm{COS}}$ be the phase
\eqref{spec:eq:actual-phase}, with $m=N^{-1/2}$.
Its real cutoff $0\le\Theta_N\le1$ equals one for
$n\le N^{1/2}/2+10$ and zero for $n\ge N^{1/2}-10$.
Extend $C_N$ by zero above sector $N$.
Use the centered initial energies
$\widehat F_{N,b}=\mathfrak J_N-\mathfrak J_N[\Omega]I+bM$
on $\Pi_NP\cF$ and $F_{\infty,b}=\mathcal J_{2,\infty}+bM$
on $P\cF$, where $\Pi_N=\mathbf1_{\{\cN\le N\}}$.
Suppose in addition that $\phi$ is real and
\begin{equation}\label{eq:vacuum-radius-overlap}
	\phi(x_0)\phi(y_0)\ne0\quad\text{for some }x_0,y_0\text{ with }|x_0-y_0|=\ell.
\end{equation}
For every sufficiently large fixed $b$,
\[
	\Omega\in D(F_{\infty,b}),\qquad \Omega\notin D(F_{\infty,b}^{3/2}).
\]
Consequently, every sequence $\xi_N\in\Pi_NP\cF$ with
$\xi_N\to\Omega$ in norm satisfies
\begin{equation}\label{eq:vacuum-third-divergence}
	\lim_{N\to\infty}\|\widehat F_{N,b}^{3/2}\xi_N\|^2=+\infty.
\end{equation}
The same conclusion holds with $\widehat F_{N,b}$ replaced by
$F_{\rm ex}(0)=\mathfrak J_N+bM$. The moments in
\eqref{eq:vacuum-third-divergence} are understood in the extended
sense; membership in the finite-$N$ power domain is not assumed.
\end{theorem}
\begin{proof}
Let $H_\infty$ be the symmetric pair-creation kernel of the
compressed generator $\mathcal J_{2,\infty}$ at time zero, with
the convention
\[
	\mathcal J_{2,\infty}=\dGamma(h_\infty)+\int H_\infty(x,y)a_x^*a_y^*\d x\d y+\text{adjoint}.
\]
The pair kernel is Hilbert--Schmidt. In this convention,
\begin{equation}\label{eq:vacuum-generator-image}
	F_{\infty,b}\Omega=b\Omega+\sqrt2\,H_\infty,
\end{equation}
where $H_\infty$ on the right denotes its two-particle component.
We will show that this component is not in $H^1(\mathbb R^6)$.

The limiting scattering profile in \cite[Eq.~(2.25)]{COS} is
\[
	w(x)=a\left(\frac1{|x|}-\frac3{2\ell}+\frac{|x|^2}{2\ell^3}\right)\mathbf1_{\{|x|\le\ell\}}.
\]
Its value and first radial derivative vanish at $|x|=\ell$.
In distributions,
\begin{equation}\label{eq:vacuum-profile-laplacian}
	-\Delta w=4\pi a\delta_0-\frac{3a}{\ell^3}\mathbf1_{\{|x|<\ell\}}.
\end{equation}
In particular, convolution by $w$ maps $H^j$ boundedly into
$H^{j+2}$ for every integer $j\ge0$. Indeed, $w\in L^1$ controls
its Fourier multiplier near zero, and
\eqref{eq:vacuum-profile-laplacian} bounds
$|p|^2\widehat w(p)$ at large $p$.

We check the other pair terms before identifying the jump.
Put $k(x,y)=-w(x-y)\phi(x)\phi(y)$ and
$\eta=(q\otimes q)k$. Multiplication by $\phi$ and the
rank-one correction $1-q$ preserve every integer Sobolev space.
Thus $\eta:H^j\to H^{j+2}$ is bounded. It is also
Hilbert--Schmidt. At this real initial background,
$p=\cosh\eta-1$ and $r=\sinh\eta-\eta$ are convergent sums of
powers of the same real symmetric operator. They satisfy
\begin{equation}\label{eq:vacuum-smooth-compositions}
	p\in H^2(\mathbb R^6),\qquad r\in H^4(\mathbb R^6).
\end{equation}
Indeed, the Hilbert--Schmidt ideal property \cite{SimonTrace}
gives
\begin{align*}
	\|\eta^2\|_{\mathfrak S_2(L^2,H^2)}&\le\|\eta\|_{L^2\to H^2}\|\eta\|_{\mathfrak S_2},\\
	\|\eta^3\|_{\mathfrak S_2(L^2,H^4)}&\le\|\eta\|_{H^2\to H^4}\|\eta\|_{L^2\to H^2}\|\eta\|_{\mathfrak S_2}.
\end{align*}
Right multiplication by another $\eta$ introduces a factor
at most $\|\eta\|_{\mathrm{op}}$ in either estimate.
Transpose symmetry gives the same estimates in the second
variable. The pure derivative bounds control all mixed
derivatives of these orders by Fourier interpolation. The
factorial coefficients make the remaining powers summable in
these norms.

The projection correction $\mu=\eta-k$ is smooth. Its explicit
formula is \eqref{eq:BS-projection-profile}, with $a_N$ replaced
by $w$; all derivatives can be transferred from $w$ to
$|\phi|^2$. The kinetic pair terms in
\cite[Eqs.~(2.28)--(2.29)]{COS} are, up to transpose and
symmetrization, the kernels of
\[
	-\Delta\mu,\qquad p(-\Delta)\eta,\qquad p(-\Delta)r,\qquad -\Delta r.
\]
Here the Laplacian in $-\Delta_xp_x$ differentiates the
shared integration variable. For example its pair kernel is
\[
	\int[-\Delta_zp(x,z)]\eta(y,z)\d z=(p(-\Delta)\eta)(x,y).
\]
All other kinetic terms in that formula contain one creation
and one annihilation operator. Their vacuum contributions to $H_\infty$, including
the adjoint terms, vanish.
Each of the four displayed pair kernels belongs to
$H^1(\mathbb R^6)$. For example,
$p(-\Delta)\eta$ has an $L^2$ derivative in its first variable
because $\nabla p$ is Hilbert--Schmidt and
$(-\Delta)\eta$ is bounded. Its transpose is
$\eta(-\Delta)p$; its first derivative is Hilbert--Schmidt
because $\nabla\eta$ is bounded and $(-\Delta)p$ is
Hilbert--Schmidt. Explicitly, with $D=-\Delta$,
\begin{align*}
	\|\nabla(pD\eta)\|_{\mathfrak S_2}&\le\|\nabla p\|_{\mathfrak S_2}\|D\eta\|_{\mathrm{op}},\\
	\|\nabla(\eta Dp)\|_{\mathfrak S_2}&\le\|\nabla\eta\|_{\mathrm{op}}\|Dp\|_{\mathfrak S_2}.
\end{align*}
Replace
$\eta$ by $r$ for $pDr$, and use $r\in H^4$ for the direct
kernel $Dr$. These estimates first hold on smooth test
functions and then extend by the displayed bounds.
The output-derivative products are controlled as well:
\[
	(Dp)\eta=D(p\eta),\qquad (D\eta)p=D(\eta p),\qquad (Dp)r=D(pr).
\]
The kernels $p\eta=\eta p$ and $pr$ belong to $H^4$ by the
same power-series estimates, so these three kernels belong to
$H^2$ by taking products of functions of $\eta$.

Write $V_f$ for multiplication by a smooth compactly supported
function $f$. Up to numerical coefficients, transpose and
symmetrization, the interaction pair terms in the same formula
are of the types
\[
	pV_f,\qquad \gamma V_fp,\qquad \sigma V_f\sigma,\qquad \gamma V_f\sigma,\qquad \sigma V_f\gamma,
\]
where $f$ is $\phi^2$ or $|\phi|^2$.
The product $\gamma V_{\phi^2}p$ is also determined directly
by the second pair term in \cite[Eq.~(2.18)]{COS}.
Indeed, let $T_N$ be convolution by $N^3v(N\cdot)f_{N\ell}(N\cdot)$,
where $f_{N\ell}$ is the Neumann scattering solution
on the ball of radius $N\ell$. Its mass tends to $8\pi a$,
and $\sup_N\|T_N\|_{\mathrm{op}}<\infty$.
The approximate identity gives $T_N\to8\pi aI$ strongly.
The Hilbert--Schmidt convergence of $p_N$ and the operator
convergence of $\gamma_N$ therefore give
\[
	\gamma_NV_\phi T_NV_\phi p_N\longrightarrow8\pi a\,\gamma V_{\phi^2}p\quad\text{in Hilbert--Schmidt norm}.
\]
Strong convergence is sufficient here: approximate the fixed
Hilbert--Schmidt factor $V_\phi p$ by finite-rank operators and
use the uniform bound on $T_N$. The composition fixes the
common integration variable.
Expand $\gamma=1+p$ and $\sigma=\eta+r$.
All terms other than $V_f\eta$ and $\eta V_f$ have $H^1$
kernels by the composition estimates above. For these two
terms, the only nonsmooth factor is $w(x-y)$. It is $H^1$
on every annulus separated from $x=y$, including across
$|x-y|=\ell$.

The time-derivative pair terms have kernels of the types
$\gamma_s\dot\eta\gamma_s^{\mathrm T}$ and
$\sigma_s\overline{\dot\eta}\sigma_s^{\mathrm T}$.
Their leading kernel is $\dot\eta$; every other term contains
it together with at least two factors of $\eta$.
At time zero the GP equation gives
$\dot\phi=i\Delta\phi-i8\pi a\phi^3\in C_c^\infty$.
Hence $\dot\eta$ has the same Sobolev mapping bounds and the
same regularity away from $x=y$ as $\eta$. The power series
are uniformly summable for the auxiliary parameter in $[0,1]$.
In detail,
\begin{gather*}
	\dot\eta=(\dot q\otimes q+q\otimes\dot q)k+(q\otimes q)\dot k,\\
	\dot k(x,y)=-w(x-y)\{\dot\phi(x)\phi(y)+\phi(x)\dot\phi(y)\}.
\end{gather*}
The vectors in $\dot q=-|\dot\phi\rangle\langle\phi|-|\phi\rangle\langle\dot\phi|$
are smooth and compactly supported. Contracting one of them
against $k$ gives a smooth vector by convolution. Thus the
$\dot q$ terms are smooth finite-rank kernels; $\dot k$ has
the stated mapping bounds. If $T$ is any of these smoothing
Hilbert--Schmidt operators, then $pT$ and $Tp^{\mathrm T}$ have
$H^1$ kernels by the ideal estimates already used, and so does
$\sigma T\sigma^{\mathrm T}$. This proves the assertion for
all time-derivative pair terms, including their adjoints.

The remaining explicit kernels, apart from the Neumann
eigenvalue term, are
\begin{gather*}
	\frac12w(x-y)\{\Delta\phi(x)\phi(y)+\phi(x)\Delta\phi(y)\},\\
	\nabla w(x-y)\cdot\{\nabla\phi(x)\phi(y)-\phi(x)\nabla\phi(y)\}.
\end{gather*}
They are $H^1$ on an annulus about $|x-y|=\ell$, since $w$
and $\nabla w$ have matching traces there. Projection of either
coordinate adds smooth kernels: integration against the smooth
condensate transfers derivatives to smooth compactly supported
factors. This also holds for the projection of the eigenvalue
term, whose correction involves
$\mathbf1_{\{|\cdot|<\ell\}}*|\phi|^2$.
For example, if $j$ is $w$ or $\mathbf1_{\{|\cdot|<\ell\}}$
and $g_j(x,y)=j(x-y)\phi(x)\phi(y)$, then, with
$\rho=|\phi|^2$,
\begin{align*}
	((q\otimes q-I)g_j)(x,y)=-\phi(x)\phi(y)\bigl\{&(j*\rho)(x)+(j*\rho)(y)\\
	&-\langle\rho,j*\rho\rangle\bigr\}.
\end{align*}
For every multi-index $\alpha$,
$\partial^\alpha(j*\rho)=j*\partial^\alpha\rho$ is bounded
and continuous since $j\in L^1$. The kernels containing
$\nabla w$ have the same property, with smooth differentiated
amplitudes in place of $\rho$; here $\nabla w\in L^1$.
Projection preserves the global $H^1$ bounds for all composed
kernels.

It follows from the coefficient of the eigenvalue term in
\cite[Eq.~(2.28)]{COS} that, on
$\mathcal A_\ell=\{(x,y):\ell/2<|x-y|<3\ell/2\}$,
\begin{equation}\label{eq:vacuum-pair-jump}
	H_\infty(x,y)=\frac{3a}{\ell^3}\mathbf1_{\{|x-y|<\ell\}}\phi(x)\phi(y)+R_\ell(x,y),\qquad R_\ell\in H^1(\mathcal A_\ell).
\end{equation}
This verification retains the interaction and time-derivative
terms of the limiting generator. By
\eqref{eq:vacuum-radius-overlap}, the first term has a nonzero
jump on an open part of the hypersurface $|x-y|=\ell$.
Its distributional normal derivative contains a nonzero
surface measure. Such a measure is not an $L^2$ function,
whereas the derivatives of $R_\ell$ are in $L^2$.
Consequently $H_\infty\notin H^1(\mathbb R^6)$.

For large $b$, $F_{\infty,b}\asymp A$ as positive forms and
$D(F_{\infty,b})=D(A)$. Thus $\Omega\in D(F_{\infty,b})$,
while spectral calculus and the form-domain identity give
\[
	\Omega\in D(F_{\infty,b}^{3/2})\quad\Longleftrightarrow\quad F_{\infty,b}\Omega\in D(A^{1/2}).
\]
Equation~\eqref{eq:vacuum-generator-image} and the jump just
proved rule out the right-hand side. Apply
Proposition~\ref{prop:initial-energy-resolvent} with $s=3/2$
to obtain \eqref{eq:vacuum-third-divergence}.
Finally $F_{\rm ex}(0)=\widehat F_{N,b}+\alpha_N I$, with
$|\alpha_N|\le C$. The scalar inequality
$\lambda^3\le C'(1+(\lambda+\alpha_N)^3)$ on their positive
spectra transfers the divergence to $F_{\rm ex}(0)$.
\end{proof}

Condition~\eqref{eq:vacuum-radius-overlap} holds whenever
$\phi$ is nonzero on a ball of radius greater than $\ell/2$.
For any fixed nonzero real smooth condensate, it therefore holds
after choosing the allowed fixed radius sufficiently small.
The conclusion concerns an initial energy domain for this
Neumann correlation representation. The prepared families in
Theorem~\ref{thm:common-preparation} and
Corollary~\ref{cor:quasifree-common-preparation} retain their
uniform third-energy bounds. No family with a uniform bound in
the fixed third-energy norm can converge in norm to the
unprepared vacuum under the assumptions of the theorem.

The same obstruction applies to every nonzero finite-excitation
vector in the kinetic-energy operator domain.

\begin{corollary}
\label{cor:finite-excitation-obstruction}
Let $v\in C_c^\infty(\mathbb R^3)$ be nonzero, nonnegative and
radial. Fix an $N$-independent normalized
$\phi\in C_c^\infty(\mathbb R^3)$ and a sufficiently small fixed $\ell\in(0,1/2)$.
Let $q=1-|\phi\rangle\langle\phi|$, $P=\Gamma(q)$, and use
the Neumann pair kernel $\eta_N$ and its limit
$\eta_\infty$ in \eqref{spec:eq:actual-eta-limit}.
Let $C_N=\mathcal C_N^{\mathrm{COS}}$ be the phase
\eqref{spec:eq:actual-phase}, with $m=N^{-1/2}$.
Its real cutoff $0\le\Theta_N\le1$ equals one for
$n\le N^{1/2}/2+10$ and zero for $n\ge N^{1/2}-10$.
Extend $C_N$ by zero above sector $N$.
Use the centered initial energies
$\widehat F_{N,b}=\mathfrak J_N-\mathfrak J_N[\Omega]I+bM$
on $\Pi_NP\cF$ and $F_{\infty,b}=\mathcal J_{2,\infty}+bM$
on $P\cF$, where $\Pi_N=\mathbf1_{\{\cN\le N\}}$.
Assume that $\phi$ is real and that
$\phi(x_0)\phi(y_0)\ne0$ for some $|x_0-y_0|=\ell$.
Fix $b$ sufficiently large that the two initial energies are
positive with uniform form comparisons to $Q_N$ and $A$, respectively.
Let $0\ne\chi\in P\cF\cap D(A)$ have finitely many occupied
number sectors. Then $\chi\notin D(F_{\infty,b}^{3/2})$.
Every sequence $\xi_N\in\Pi_NP\cF$ with $\xi_N\to\chi$ in
norm has divergent extended third moments for both
$\widehat F_{N,b}$ and $F_{\rm ex}(0)$.
\end{corollary}
\begin{proof}
Let $m$ be the highest occupied sector, and put $n=m+2$.
Only pair creation contributes to sector $n$ of
$F_{\infty,b}\chi$. Its wave function is
\begin{align*}
	&(F_{\infty,b}\chi)_n(x_1,\ldots,x_n)\\
	&\qquad=\frac{2}{\sqrt{n(n-1)}}\sum_{i<j}H_\infty(x_i,x_j)\chi_m(x_{\widehat{i,j}}),
\end{align*}
where $x_{\widehat{i,j}}$ denotes the remaining $m$ coordinates.
This is the usual normalized symmetrization of
$\sqrt{n(n-1)}H_\infty\otimes\chi_m$.
The kernel analysis above also shows that $H_\infty$ is locally
$H^1$ away from the diagonal and the surface $|x-y|=\ell$.
Consider the hypersurface $\Sigma_{12}=\{|x_1-x_2|=\ell\}$.
Choose a bounded open patch that meets no other pair surface
$\{|x_i-x_j|=\ell\}$ and no pair diagonal. It can be chosen
so that
$\phi(x_1)\phi(x_2)\chi_m(x_3,\ldots,x_n)$
is nonzero on a subset of positive surface measure. Indeed,
the first two factors are nonzero on an open subset of
$\Sigma_{12}$ by \eqref{eq:vacuum-radius-overlap}, and
$\chi_m$ is nonzero on a set of positive measure in the
spectator coordinates. The finitely many other pair surfaces
and diagonals have zero measure in this product surface.

On this patch only the summand $(i,j)=(1,2)$ has a jump.
Its coefficient is
$6a\phi(x_1)\phi(x_2)\chi_m(x_3,\ldots,x_n)/
(\ell^3\sqrt{n(n-1)})$, which is nonzero on the chosen set.
The remainder is locally $H^1$: each kernel remainder and
each other pair kernel is $H^1$ there, and
$\chi_m\in H^2$ in its own coordinates. Their products use
disjoint coordinate groups, so the local $H^1$ bounds follow
by Fubini. The normal derivative therefore retains a nonzero
surface measure. Thus $(F_{\infty,b}\chi)_n\notin H^1$.
Since $\chi\in D(F_{\infty,b})=D(A)$, the same domain
criterion as in \eqref{eq:vacuum-generator-image} excludes
$\chi\in D(F_{\infty,b}^{3/2})$.
Equation~\eqref{eq:initial-energy-liminf} and the bounded
scalar shift prove the two divergence statements.
\end{proof}

\subsection{Energy convergence under the exact evolution}
\label{sec:evolved-energy}

The prepared initial family also admits an energy comparison at
later times. The Hilbert-norm estimate follows from the quadratic
approximation of \cite{COS} and unitarity. To obtain an energy
norm, we additionally use the propagated third moment and a
comparison of the instantaneous positive forms.

In this subsection distinguish the modified GP condensate
$\phi_{N,t}$ from its limit $\phi_{\infty,t}$. Put
$q_{j,t}=1-|\phi_{j,t}\rangle\langle\phi_{j,t}|$ and
$P_{j,t}=\Gamma(q_{j,t})$ for $j=N,\infty$.
Write $A_{j,t}=K_{P_{j,t}}+M$ and $Q_N(t)=Q_{N,P_{N,t}}$.
The limiting quadratic propagator is
$\mathcal U_{2,\infty}(t,0)$ from \cite[Eq.~(2.30)]{COS},
viewed on the ambient Fock space. Let $z_N,z_\infty$ be any of
the prepared pairs in Theorem~\ref{thm:common-preparation} or
Proposition~\ref{prop:weighted-reference-preparation}. Their
proofs give the stronger regularity
$z_\infty\in D(M^2A_{\infty,0})$.

We use a one-particle unitary to compare the two excitation
spaces at a fixed time. Set $p_{j,t}=1-q_{j,t}$ and, for large
$N$, define
\begin{equation}\label{eq:evolved-direct-rotation}
	R_N(t)=(q_{N,t}q_{\infty,t}+p_{N,t}p_{\infty,t})\{1-(p_{N,t}-p_{\infty,t})^2\}^{-1/2},\qquad \mathcal R_N(t)=\Gamma(R_N(t)).
\end{equation}
This is the direct rotation of the two orthogonal projections
\cite{KatoBook}. It obeys
$R_Nq_{\infty,t}R_N^*=q_{N,t}$ and commutes with number after
Fock lifting. The $H^4$ comparison in
\cite[Proposition A.2(iv)]{COS} implies, on each compact $I$,
\begin{equation}\label{eq:evolved-rotation-bounds}
	\|R_N-1\|_{L^2\to H^2}+\|R_N^*-1\|_{L^2\to H^2}\le C_I N^{-1}.
\end{equation}
Indeed, $p_{N,t}-p_{\infty,t}$ has this bound by its rank-two
formula. Write $\delta=p_{N,t}-p_{\infty,t}$ and
$W=q_{N,t}q_{\infty,t}+p_{N,t}p_{\infty,t}$.
Then $W-1=\delta(2p_{\infty,t}-1)$, and
$\|\delta^2\|_{L^2\to H^2}\le C_IN^{-2}$.
The power series for $(1-\delta^2)^{-1/2}-1$ therefore
converges in $\mathcal B(L^2,H^2)$, with norm $O_I(N^{-2})$.
The same calculation for the adjoint uses the uniform
$H^2\to H^2$ bound of $2p_{\infty,t}-1$ and proves
\eqref{eq:evolved-rotation-bounds}. In particular,
\[
	\|R_N^*(-\Delta)R_N+\Delta\|_{\mathrm{op}}\le C_I N^{-1},\qquad \|\mathcal R_N^*A_{N,t}\mathcal R_N-A_{\infty,t}\|_{D(M)\to\cF}\le C_I N^{-1},
\]
where $D(M)$ has norm $\|M\cdot\|$ and the second assertion
is on $P_{\infty,t}\cF$. To verify the operator assertion,
put $D=-\Delta$ and $B_N=D(R_N-1)\in\mathcal B(L^2)$.
On $D(D)$ the difference is $R_N^*B_N+B_N^*$, since
$(R_N^*-1)D=B_N^*$ there. It therefore extends to a bounded
operator of norm at most $2\|B_N\|\le C_IN^{-1}$.
The direct rotation preserves $D(D)$ by
\eqref{eq:evolved-rotation-bounds}, so this is an operator
identity on that domain. Second quantization and compression
give the second assertion. On sector $n$ the difference has
operator norm at most $C_IN^{-1}(n+1)$, hence also satisfies
the form bound $\pm(\mathcal R_N^*A_{N,t}\mathcal R_N-A_{\infty,t})\le C_IN^{-1}M$.
Both compressed kinetic energies
preserve particle number; thus $A_{j,t}$ strongly commutes
with $M$ at each fixed time.

\begin{lemma}
\label{lem:quadratic-weighted-operator-domain}
Put $A_{\rm amb}=\cK+M$. For $v\in D(M^2A_{\rm amb})$,
\[
	\sup_{t\in I}\|M^2A_{\rm amb}\mathcal U_{2,\infty}(t,0)v\|\le C_I\|M^2A_{\rm amb}v\|.
\]
The evolving vector is continuous in this graph norm.
\end{lemma}
\begin{proof}
Write $J(t)=\mathcal J_{2,\infty}(t)=\cK+S(t)$ in ambient
coordinates. The bounds in \cite[Proposition 5.1 and its proof]{COS}
give, for every fixed $r\ge0$,
\begin{equation}\label{eq:quadratic-weighted-coefficients}
	\|M^rS(t)v\|+\|M^r\dot J(t)v\|\le C_{I,r}\|M^{r+1}v\|.
\end{equation}
The cited form bounds give this operator estimate because both $S$ and $\dot J$ have only number shifts $0,2,-2$.
The inequalities analogous to (4.1) and (4.3) in that reference
bound each block $m,n$ by
$C_I\sqrt{(m+1)(n+1)}$. Neighboring weights are comparable;
squaring and summing the three diagonals proves
\eqref{eq:quadratic-weighted-coefficients}. The time derivative
includes the projected fields and the bounded one-particle
multiplication term. The $H^6$ condensate regularity is the
one used for these estimates in that reference.

The number estimate in \cite[Proposition 5.1]{COS} makes
$M^r\mathcal U_{2,\infty}(t,s)M^{-r}$ bounded uniformly for
$s,t\in I$. The same estimate holds on ambient Fock space:
the three number diagonals bound the commutator with $M^{2r}$
by $C_{I,r}M^{2r}$ as a form, and Gronwall's inequality applies.
Set $v(t)=\mathcal U_{2,\infty}(t,0)v$. The
operator identity
\[
	J(t)v(t)=\mathcal U_{2,\infty}(t,0)J(0)v+\int_0^t\mathcal U_{2,\infty}(t,s)\dot J(s)v(s)\d s
\]
therefore gives
\[
	\|M^2J(t)v(t)\|\le C_I\bigl(\|M^2J(0)v\|+\|M^3v\|\bigr).
\]
For the domain argument use
$J_L(t)=\cK+\Pi_LS(t)\Pi_L$ and initial vector $\Pi_Lv$.
At fixed $L$, the perturbation is a strongly $C^1$ bounded
operator. This follows from the differentiated quadratic
coefficients in the proof of \cite[Proposition 5.1]{COS}.
Thus the displayed identity holds on $D(\cK)$ \cite{Kato}.
Number truncation commutes with $\cK$ and preserves its domain;
$\Pi_L\cF$ is invariant under $J_L$. The number commutator
estimate and the displayed graph estimate are uniform in $L$.
In particular, for $v_L(t)=\mathcal U_L(t,0)\Pi_Lv$,
\[
	\sup_{L,t\in I}\bigl(\|M^3v_L(t)\|+\|M^2\cK v_L(t)\|\bigr)\le C_I\|M^2A_{\rm amb}v\|.
\]
The finite bandwidth gives
$\|(S-\Pi_LS\Pi_L)u\|\le C_I L^{-2}\|M^3u\|$.
Duhamel's formula makes $v_L$ converge uniformly in Hilbert
norm to the quadratic solution. Interpolation gives convergence
in $D(M)$ as well. The graph of the closed operator $M^2\cK$
is weakly closed; the uniform graph bound therefore places the
limit in its domain with the same bound. Since
$\|M^3v(t)\|$ is also bounded, this proves the claimed estimate.

The unweighted integral identity passes to the limit in norm.
Indeed, $S(t)$ and $\dot J(t)$ are strongly continuous maps
$D(M)\to\cF$, with the bounds above, and the truncated
integrands converge in that norm. It follows that both
$J(t)v(t)$ and $S(t)v(t)$ are continuous, hence so is
$\cK v(t)$. For finite-sector initial vectors, repeat the
estimate with $M^3$ in place of $M^2$. It gives a uniform
bound on $M^3\cK v(t)$, so the tails of $M^2\cK v(t)$ tend
to zero uniformly. The finitely many remaining sectors are
continuous by the unweighted assertion. Finite-sector
approximation in $D(M^2A_{\rm amb})$ and the proved bound
give graph continuity for all stated data.

On $P_{\infty,t}\cF$, the omitted one-particle block of
$-\Delta$ is bounded by $C_I$, since $\phi_{\infty,t}\in H^2$.
Second quantization bounds its action by $C_IM$. For every
fixed $r\ge0$, compression and this bound give
\[
	\|M^rA_{\infty,t}u\|\le\|M^rA_{\rm amb}u\|\le C_I\|M^rA_{\infty,t}u\|.
\]
The first inequality uses the orthogonal compression; the
second also uses $A_{\infty,t}\ge M$ and strong commutation
with $M$. This proves the required domain identification and
allows the lemma to be applied to $z_\infty$.
\end{proof}

\begin{proposition}
\label{prop:prepared-quadratic-comparison}
Let
\[
	\xi_N(t)=U_{\rm ex}(t,0)z_N,\qquad \xi_\infty(t)=\mathcal U_{2,\infty}(t,0)z_\infty.
\]
There is a real scalar phase $\theta_N(t)$, with
$\theta_N(0)=0$, such that
\begin{equation}\label{eq:prepared-quadratic-norm}
	\sup_{t\in I}\|e^{i\theta_N(t)}\xi_N(t)-\xi_\infty(t)\|\le C_I N^{-1/8}.
\end{equation}
More generally, for $0\le r<3/2$,
\begin{equation}\label{eq:prepared-quadratic-number}
	\sup_{t\in I}\|M^r(e^{i\theta_N(t)}\xi_N(t)-\xi_\infty(t))\|\le C_{I,r}N^{-(3-2r)/24}.
\end{equation}
All vectors in these comparisons are embedded in the ambient
Fock space, even though their condensate projections differ.
\end{proposition}
\begin{proof}
Put $\zeta_N=\Pi_Nz_\infty/\|\Pi_Nz_\infty\|$.
Since number truncation commutes with $\cK$ and $M$,
\[
	\|(1-\Pi_N)z_\infty\|\le(N+1)^{-3}\|M^3z_\infty\|,\qquad \|\Pi_Nz_\infty\|\longrightarrow1.
\]
The normalization is therefore bounded below for large $N$.
The stated regularity gives uniform bounds on
$\|\cK\zeta_N\|$ and $\|M^3\zeta_N\|$.
The cubic block estimate gives
$\|(e^{\pm C_t}-1)v\|\le C_I N^{-3/4}\|M^{3/2}v\|$,
on the physical excitation space, by integrating
$e^{aC_t}C_tv$. At time zero, the preparation estimate and
the number tail therefore give
\[
	\|e^{C_0}z_N-\zeta_N\|\le\|z_N-\zeta_N\|+\|(e^{C_0}-1)\zeta_N\|\le C_I N^{-1/4}.
\]
Here unitarity is used on $e^{C_0}(z_N-\zeta_N)$; the cubic
bound is applied only to the regular vector $\zeta_N$.
Reconstruct both initial vectors using the same
$\mathcal U_{N,0}^*e^{B_0}$. Their physical exact evolutions
remain this close by unitarity. The initial vector $\zeta_N$
satisfies precisely the $\cK^2+\cN^6$ hypothesis of
\cite[Theorem 2.2]{COS}. Applying that theorem gives its
quadratic approximation with error $C_IN^{-1/8}$.
Theorem 2.3 of the same reference then replaces
$\mathcal U_{2,N}(t,0)\zeta_N$ by
$\mathcal U_{2,\infty}(t,0)z_\infty$, with error
$C_IN^{-1/4}$. The number estimate in its Proposition 4.3
also bounds the removal of $e^{C_t}$ from the finite-$N$
quadratic vector by $C_IN^{-3/4}$.

The scalar convention is fixed by
$\theta_N(t)=\int_0^t(\kappa_N(s)-c_{N,s})\d s$, where
$\kappa_N$ is \cite[Eq.~(2.20)]{COS} and $c_{N,s}$ is the
scalar in the physical reconstruction above. Reversing that
reconstruction, and multiplying the physical comparison by
$e^{i\int_0^t\kappa_N(s)\d s}$, gives
\[
	\|e^{i\theta_N(t)}e^{C_t}\xi_N(t)-\mathcal U_{2,N}(t,0)\zeta_N\|\le C_IN^{-1/8}.
\]
Removal of the cubic phase and the preceding quadratic
comparison prove \eqref{eq:prepared-quadratic-norm}.
The regular comparison vector is $\zeta_N$.

Theorem~\ref{thm:D} bounds $\|M^{3/2}\xi_N(t)\|$ uniformly,
and the quadratic number estimate gives the same bound for
$\xi_\infty(t)$. Spectral interpolation for $M$ yields
\[
	\|M^rv\|\le\|v\|^{1-2r/3}\|M^{3/2}v\|^{2r/3},
\]
which proves \eqref{eq:prepared-quadratic-number}.
\end{proof}

We next compare the positive energies at each fixed time.
Center the exact generator at its vacuum expectation and set
\begin{gather*}
	\alpha_N(t)=\mathfrak J_t[\Omega],\qquad \widehat F_N(t)=\mathfrak J_t-\alpha_N(t)I+b_IM,\\
	F_\infty(t)=P_{\infty,t}\mathcal J_{2,\infty}(t)P_{\infty,t}+b_IM.
\end{gather*}
Choose $b_I$ sufficiently large for positivity and the weighted
inverse estimates on $I$. The preparation itself is kept
fixed, as in Remark~\ref{rem:exact-number-shift}. On
$P_{\infty,t}\cF$, extend
$\mathcal R_N^*\widehat F_N\mathcal R_N$ above sector $N$
by $K_{P_{\infty,t}}+b_IM$, and denote this positive operator
by $\mathcal F_N(t)$.
This is a direct sum with no coupling across sector $N$.
Its form domain is
\[
	\mathcal R_N^*D(\widehat F_N^{1/2})\oplus\bigl((1-\Pi_N)D(A_{\infty,t}^{1/2})\bigr).
\]
Both summands are closed positive forms and have a common
lower bound $c_I I$, independent of $N$.

\begin{lemma}
\label{lem:evolved-form-comparison}
For each fixed $t\in I$, $\mathcal F_N(t)$ converges to
$F_\infty(t)$ in the strong resolvent sense. If
$v\in D(M^2A_{\infty,t})$, then
\begin{equation}\label{eq:evolved-form-consistency}
	\mathcal F_N(t)[\Pi_Nv]\longrightarrow F_\infty(t)[v].
\end{equation}
\end{lemma}
\begin{proof}
We check that the comparison in
Lemma~\ref{lem:common-form-comparison} survives the moving
condensate. Put $\overline Q_N=\mathcal R_N^*Q_N(t)\mathcal R_N$.
Uniformly on $I$, for $u,v\in\Pi_NP_{\infty,t}\cF$ in
the indicated form and operator domains,
\begin{align}
	| (\mathcal F_N-F_\infty)[u,v] |
	\le C_I\|\overline Q_N^{1/2}u\|\bigl\{&N^{-1/4}
	(\|M\overline Q_N^{1/2}v\|+\|M^{5/2}v\|)\notag\\
	&+N^{-1/2}(\|A_{\infty,t}v\|+\|M^{3/2}v\|)\bigr\}.
	\label{eq:evolved-form-difference}
\end{align}
To prove this, apply \cite[Proposition 3.3]{COS} to
$\mathcal R_Nu,\mathcal R_Nv$ in the actual excitation
space. Vacuum centering removes the scalar in that proposition;
the residual vacuum error is $O(N^{-1/4})$.
The potential contribution obeys
\[
	\|\mathcal V_{N,P_{N,t}}^{1/2}\mathcal R_Nv\|\le C_IN^{-1/2}\|A_{\infty,t}v\|,
\]
by Proposition~\ref{spec:prop:projected} and
\eqref{eq:evolved-rotation-bounds}. Its constants depend only
on the uniformly bounded background norms, not on compact
support at time $t$.

Proposition 5.2 of \cite{COS} is an ambient-space comparison.
Its three number diagonals give, as in the proof of
Lemma~\ref{lem:common-form-comparison},
\[
	\|M^{-1/2}(\mathcal J_{2,N}-\Pi_N\mathcal J_{2,\infty}\Pi_N)w\|\le C_IN^{-1/2}(\|M^{3/2}w\|+\|\cK^{1/2}w\|).
\]
Use $w=\mathcal R_Nv$. It remains in the same number cutoff,
and the compressed kinetic comparison gives
\[
	\|\cK^{1/2}\mathcal R_Nv\|^2=K_{P_{N,t}}[\mathcal R_Nv]\le K_{P_{\infty,t}}[v]+C_IN^{-1}\|M^{1/2}v\|^2.
\]
This also proves that the rotation maps the relevant kinetic
form domains into one another.

Write the ambient quadratic remainder as
$S(t)=\dGamma(h_t)+\mathcal P^*(k_t)+\mathcal P(k_t)$,
with the pair normalization absorbed into $k_t$.
The bounds for $S$ on its one-particle block and its
vacuum-to-two-particle block give
$\|h_t\|_{\mathrm{op}}+\|k_t\|_2\le C_I$.
Unitary conjugation obeys
\begin{gather*}
	\|R_N^*h_tR_N-h_t\|_{\mathrm{op}}\le2\|h_t\|_{\mathrm{op}}\|R_N-1\|_{\mathrm{op}},\\
	\|(R_N^*\otimes R_N^*)k_t-k_t\|_2\le2\|k_t\|_2\|R_N-1\|_{\mathrm{op}}.
\end{gather*}
Both are $O_I(N^{-1})$. The kinetic difference was proved
after \eqref{eq:evolved-rotation-bounds}. The usual quadratic
creation bounds therefore control the total form difference by
$C_IN^{-1}\|M^{1/2}u\|\|M^{1/2}v\|$.
Compression to $P_{\infty,t}$ changes none of these bounds.
This proves \eqref{eq:evolved-form-difference}.

The lifted rotation also gives
$\overline Q_N\le C_IMA_{\infty,t}$ and
$\mathcal R_N^*\widehat F_N\mathcal R_N\asymp\overline Q_N$
on the physical sectors. For the first inequality, use
$Q_N\le C_IMA_{N,t}$, the fact that $\mathcal R_N$ commutes
with $M$, and the operator difference just proved:
\[
	\overline Q_N\le C_IM\mathcal R_N^*A_{N,t}\mathcal R_N\le C_I(MA_{\infty,t}+N^{-1}M^2)\le C_I'MA_{\infty,t}.
\]
This first inequality holds also on the extended sector forms.
The equivalence with $\mathcal F_N$ is used only below sector
$N$. The weighted bounds for $S(t)$ are uniform on $I$.
Consequently one choice of $b_I>2\sup_{t\in I}\max(C_0(t),C_2(t))$
makes the weighted Neumann inverse argument of
Theorem~\ref{thm:common-preparation} valid throughout $I$.
For a finite-sector $\chi\in D(A_{\infty,t})$, put
$y=F_\infty(t)^{-1}\chi$ and $v_N=\Pi_Ny$; then
$y\in D(M^2A_{\infty,t})$.
For $N$ containing the support of $\chi$,
$x_N=\mathcal F_N^{-1}\chi$ lies in the physical block.
The error $e_N=x_N-v_N$ obeys, for physical form vectors $u$,
\[
	\mathcal F_N[u,e_N]=(F_\infty-\mathcal F_N)[u,v_N]+F_\infty[u,(1-\Pi_N)y].
\]
The first term is bounded by
$C_{I,\chi}N^{-1/4}\|\overline Q_N^{1/2}u\|$.
In the last term, the number-preserving part vanishes.
The remaining quadratic part satisfies
\begin{align*}
	|\langle u,S(t)(1-\Pi_N)y\rangle|
	&\le C_I\|M^{1/2}u\|\|M^{1/2}(1-\Pi_N)y\|\\
	&\le C_I(N+1)^{-1}\|\overline Q_N^{1/2}u\|\|M^{3/2}y\|.
\end{align*}
Here the negative-weight block bound for $S$ follows by the
same three-diagonal calculation. Testing with $u=e_N$ gives
$\|\overline Q_N^{1/2}e_N\|\le C_{I,\chi}N^{-1/4}$.
Thus $\mathcal F_N^{-1}\chi\to F_\infty^{-1}\chi$.
Finite-sector $D(A_{\infty,t})$ vectors are dense. The
common positive lower bound extends inverse convergence to
all vectors and implies strong resolvent convergence
\cite{ReedSimonI}. This explicitly controls the coupling at
the number cutoff; the complementary block contributes zero
for these source vectors.

For a general $v\in D(M^2A_{\infty,t})$, all right-hand
norms of $\Pi_Nv$ in \eqref{eq:evolved-form-difference} are
uniformly bounded. Its diagonal form difference tends to
zero. Number truncation converges in the $F_\infty^{1/2}$
norm, which proves \eqref{eq:evolved-form-consistency}.
\end{proof}

We first state the fixed-time limit. Its uniform extension is
proved in Theorem~\ref{thm:uniform-evolved-energy-spectrum}.

\begin{theorem}
\label{thm:evolved-first-energy-convergence}
Let $v\in C_c^\infty(\mathbb R^3)$ be nonzero, nonnegative and
radial. Fix an $N$-independent normalized
$\phi\in C_c^\infty(\mathbb R^3)$ and a sufficiently small fixed $\ell\in(0,1/2)$.
Let $q=1-|\phi\rangle\langle\phi|$, $P=\Gamma(q)$, and use
the Neumann pair kernel $\eta_N$ and its limit
$\eta_\infty$ in \eqref{spec:eq:actual-eta-limit}.
Let $C_N=\mathcal C_N^{\mathrm{COS}}$ be the phase
\eqref{spec:eq:actual-phase}, with $m=N^{-1/2}$.
Its real cutoff $0\le\Theta_N\le1$ equals one for
$n\le N^{1/2}/2+10$ and zero for $n\ge N^{1/2}-10$.
Extend $C_N$ by zero above sector $N$.
Use the centered initial energies
$\widehat F_{N,b}=\mathfrak J_N-\mathfrak J_N[\Omega]I+bM$
on $\Pi_NP\cF$ and $F_{\infty,b}=\mathcal J_{2,\infty}+bM$
on $P\cF$, where $\Pi_N=\mathbf1_{\{\cN\le N\}}$.
Let $\chi\in P\cF\cap D(A)$ be normalized. Assume either that
$\chi$ has finitely many occupied number sectors, or that
$\chi\in D(M^2)$ and $c_B[M^{-1}\chi]>0$.
Fix $b$ sufficiently large that $c_B[F_{\infty,b}^{-1}\chi]>0$,
$c_bQ_N\le\widehat F_{N,b},F_{\rm ex}(0)\le C_bQ_N$ on the
physical space, and $c_bA\le F_{\infty,b}\le C_bA$.
With $H_b=K_P+bM$ and $S_\infty=F_{\infty,b}-H_b$, require
$\|M^rS_\infty H_b^{-1}M^{-r}\|<1/2$ for $r=0,2$.
Set
\[
	z_N=\frac{\widehat F_{N,b}^{-1}\Pi_N\chi}
	{\|\widehat F_{N,b}^{-1}\Pi_N\chi\|},\qquad
	z_\infty=\frac{F_{\infty,b}^{-1}\chi}{\|F_{\infty,b}^{-1}\chi\|}.
\]
Let $\xi_N(t)=U_{\rm ex}(t,0)z_N$ and
$\xi_\infty(t)=\mathcal U_{2,\infty}(t,0)z_\infty$.
Use the scalar phase $\theta_N$ of
Proposition~\ref{prop:prepared-quadratic-comparison}.
For every fixed $t\in I$,
\begin{equation}\label{eq:evolved-first-energy-convergence}
	\|Q_N(t)^{1/2}(e^{i\theta_N(t)}\xi_N(t)-\mathcal R_N(t)\xi_\infty(t))\|\longrightarrow0.
\end{equation}
Here $Q_N(t)$ is extended by its sector forms above particle
number $N$. Corollary~\ref{cor:prepared-evolved-energy-rate}
gives a quantitative bound uniform on compact time intervals.
\end{theorem}
\begin{proof}
Fix $t$ and suppress it from the notation. Put
$w_N=\mathcal R_N^*e^{i\theta_N}\xi_N$ and
$v_N=\Pi_N\xi_\infty$ on $P_{\infty,t}\cF$.
The lifted one-particle rotation satisfies
$\| (\mathcal R_N-1)v\|\le C_IN^{-1}\|Mv\|$,
by a sectorwise tensor-product telescoping sum.
Proposition~\ref{prop:prepared-quadratic-comparison} therefore
gives $w_N\to\xi_\infty$ in norm. Theorem~\ref{thm:D}, the
bounded scalar $\alpha_N(t)$ and
Remark~\ref{rem:exact-number-shift} give
\[
	\sup_N\|\mathcal F_N^{3/2}w_N\|<\infty.
\]
This controls the high-energy tail:
\[
	\|\mathcal F_N^{1/2}\mathbf1_{\{\mathcal F_N>L\}}w_N\|\le L^{-1}\|\mathcal F_N^{3/2}w_N\|.
\]
Strong resolvent convergence from
Lemma~\ref{lem:evolved-form-comparison}, applied first to
bounded continuous spectral cutoffs, now implies
\begin{equation}\label{eq:evolved-square-root-vector}
	\mathcal F_N^{1/2}w_N\longrightarrow F_\infty^{1/2}\xi_\infty.
\end{equation}
Indeed, the cutoff terms converge strongly, and the displayed
tail tends to zero uniformly as $L\to\infty$. To control the
limit independently of its kinetic regularity, use the bounded
continuous functions $g_R(\lambda)=\min(\lambda^3,R)$:
\[
	\langle\xi_\infty,g_R(F_\infty)\xi_\infty\rangle=\lim_N\langle w_N,g_R(\mathcal F_N)w_N\rangle\le\sup_N\|\mathcal F_N^{3/2}w_N\|^2.
\]
Here $g_R-R$ vanishes at infinity on the common positive
spectral half-line, so the resolvent limit applies to $g_R$
as well. Monotone convergence as $R\to\infty$ gives
$\xi_\infty\in D(F_\infty^{3/2})$ and the same third-moment
bound. Thus its first-energy tail also tends to zero. For
the low-energy comparison take $\lambda^{1/2}\chi(\lambda/L)$,
where $\chi$ is continuous, equals one on $[0,1]$ and vanishes
on $[2,\infty)$. The remaining terms are bounded by the tails
above. This proves \eqref{eq:evolved-square-root-vector}.

Lemma~\ref{lem:quadratic-weighted-operator-domain} gives
$\xi_\infty\in D(M^2A_{\infty,t})$. By
\eqref{eq:evolved-form-consistency},
$\|\mathcal F_N^{1/2}v_N\|^2\to F_\infty[\xi_\infty]$.
The uniform positive lower bound makes $\lambda^{-1/2}$
a bounded continuous function on all the spectra; its strong
convergence follows from the resolvent limit.
These square-root vectors also converge weakly to
$F_\infty^{1/2}\xi_\infty$. To identify a weak limit, test
against $\mathcal F_N^{-1/2}f\to F_\infty^{-1/2}f$ and use
\[
	\langle\mathcal F_N^{1/2}v_N,\mathcal F_N^{-1/2}f\rangle=\langle v_N,f\rangle.
\]
The range of $F_\infty^{-1/2}$ is dense. Convergence of norms
then makes this weak convergence strong. Subtracting it from
\eqref{eq:evolved-square-root-vector} and using coercivity on
the physical sectors yields
\[
	\|Q_N^{1/2}\mathcal R_N(w_N-v_N)\|\longrightarrow0.
\]
The omitted tail satisfies
\[
	\|Q_N^{1/2}\mathcal R_N(1-\Pi_N)\xi_\infty\|\le C_I(N+1)^{-3/2}\|M^2A_{\infty,t}^{1/2}\xi_\infty\|\longrightarrow0.
\]
Here $M$ and $A_{\infty,t}$ strongly commute, and the last
norm is finite by Lemma~\ref{lem:quadratic-weighted-operator-domain}.
This proves \eqref{eq:evolved-first-energy-convergence}.
\end{proof}

Theorem~\ref{thm:evolved-first-energy-convergence} supplies the
energy comparison needed to transfer an energy-weighted
spectral statement to the exact evolution at a fixed time.
The fixed-background spectral theorem also requires an extension
to the evolving condensate. Section~\ref{sec:evolving-spectrum}
proves that extension before using the present energy comparison.

\subsection{The microscopic measure}

Let $L=-2\Delta+v$ on $L^2(\mathbb R^3)$, with domain $H^2(\mathbb R^3)$.
It is nonnegative and has trivial kernel: its zero-energy form
vanishes only on an $L^2$ function with zero gradient, hence only
on zero. Its spectrum may approach zero, so $L^{-1/2}$ is understood
as an unbounded spectral-calculus operator on its proper domain.

\begin{lemma}\label{spec:lem:scattering-vector}
There is a vector $\xi_v\in L^2(\mathbb R^3)$ such that, for $R\to\infty$,
\begin{equation}\label{spec:eq:xi-limit}
	\norm{L^{1/2}(w\rho(\cdot/R))-\xi_v}_2\le C R^{-1/2}.
\end{equation}
It is independent of the admissible cutoff $\rho$. Moreover,
\[
	\xi_v=L^{-1/2}v,\qquad L^{1/2}\xi_v=v,
	\qquad \norm{\xi_v}_2^2=D_v.
\]
Define
\begin{equation}\label{spec:eq:mu}
	\mu_v(B)=\ip{\xi_v}{\mathbf1_B(L)\xi_v}.
\end{equation}
Then $\mu_v$ is positive, has mass $D_v$, has no atom at zero, and
\begin{equation}\label{spec:eq:mu-first}
	\int_0^\infty\lambda\d\mu_v(\lambda)=\norm{v}_2^2.
\end{equation}
\end{lemma}
\begin{proof}
Put $w_R=w\rho(\cdot/R)$. If $R,S$ are large, $w_R-w_S$ vanishes
on the support of $v$. On the remaining region use $w=a/|x|$.
The gradient of its tail and the cutoff derivative each have
squared $L^2$ norm at most $C/R$ for a cutoff at radius $R$.
Consequently
$\norm{L^{1/2}(w_R-w_S)}_2 \le C(R^{-1/2}+S^{-1/2})$.
This proves existence of $\xi_v$ and \eqref{spec:eq:xi-limit}. Comparing
two different cutoffs by the same argument proves independence.

For $u\in C_c^\infty$, $w_R=w$ on its support for large $R$.
Since $Lw=v$ distributionally, integration by parts gives
\[
	\ip{\xi_v}{L^{1/2}u}
	=\lim_R\ip{L^{1/2}w_R}{L^{1/2}u}
	=\ip v u.
\]
The space $C_c^\infty$ is a core for $L^{1/2}$: its graph norm is
equivalent to the $H^1$ norm because $v$ is bounded. Thus the last
identity extends to $D(L^{1/2})$. Self-adjointness implies
$\xi_v\in D(L^{1/2})$ and $L^{1/2}\xi_v=v$.
Let $\sigma_v(B)=\langle v,\mathbf1_B(L)v\rangle$.
The spectral theorem gives
\begin{equation}\label{spec:eq:source-measure-identity}
	\d\sigma_v(\lambda)=\lambda\d\mu_v(\lambda),\qquad
	\int_{(0,\infty)}\lambda^{-1}\d\sigma_v(\lambda)
	=\|\xi_v\|_2^2<\infty .
\end{equation}
Here the zero spectral projection vanishes because $\ker L=0$.
Thus $v\in D(L^{-1/2})$, and spectral calculus gives
$L^{-1/2}v=\xi_v$.

The limit of the squared energy norm of $w_R$ equals the energy
of $w$ in \eqref{spec:eq:D}. Alternatively, integration by parts yields
the exact identity, for large $R$,
\[
	\norm{L^{1/2}w_R}^2
	=D_v+\frac{2a^2}{R}\int\frac{|\nabla\rho(z)|^2}{|z|^2}\d z.
\]
To obtain this identity, put $\rho_R=\rho(\cdot/R)$ and
test $Lw=v$ against $\rho_R^2w$. The product rule gives
\[
	2\int|\nabla(\rho_Rw)|^2+\int v\rho_R^2w^2
	=\int v\rho_R^2w+2\int w^2|\nabla\rho_R|^2.
\]
For large $R$, the first term on the right is $D_v$.
On the support of $\nabla\rho_R$ one has $w(x)=a/|x|$;
the change of variables $x=Rz$ gives the stated correction.
This proves the mass assertion. The absence of a zero eigenvector
gives $\mu_v(\{0\})=0$. Finally, the spectral theorem and
$L^{1/2}\xi_v=v$ prove \eqref{spec:eq:mu-first}.
\end{proof}

\begin{proposition}
\label{prop:radial-microscopic-density}
Let $R>0$ contain the support of the radial potential $v$.
For $k\ge0$, let $s(k,r)$ solve
\begin{equation}\label{eq:radial-regular-solution}
	-s''(k,r)+\tfrac12v(r)s(k,r)=k^2s(k,r),\qquad
	s(k,0)=0,\qquad s'(k,0)=1.
\end{equation}
Define
\begin{equation}\label{eq:radial-jost-data}
	\mathcal J_v(k)=e^{ikR}\bigl(s'(k,R)-ik s(k,R)\bigr),
	\qquad I_v(k)=\int_0^R r v(r)s(k,r)\d r.
\end{equation}
The function $\mathcal J_v$ is independent of the choice of
$R$ outside the support of $v$, and $\mathcal J_v(k)\ne0$
for $k>0$. The microscopic measure has the density
\begin{equation}\label{eq:radial-microscopic-density}
	\d\mu_v(\lambda)=\rho_v(\lambda)\d\lambda,\qquad
	\rho_v(2k^2)=\frac{|I_v(k)|^2}{k|\mathcal J_v(k)|^2},
	\qquad k>0.
\end{equation}
It is continuous on $(0,\infty)$ and positive except at
a discrete set. In particular, every nonempty open interval
in $(0,\infty)$ has positive $\mu_v$ mass.
At the threshold,
\begin{equation}\label{eq:radial-threshold-density}
	\rho_v(\lambda)=4\sqrt2\,a^2\lambda^{-1/2}
	+O_v(\lambda^{1/2}),\qquad \lambda\downarrow0.
\end{equation}
Consequently, for $b\downarrow0$,
\begin{equation}\label{eq:radial-threshold-intensity}
	m_v(b):=\int_b^\infty\frac{\rho_v(\lambda)}{\lambda}\d\lambda
	=8\sqrt2\,a^2 b^{-1/2}+O_v(1).
\end{equation}
\end{proposition}
\begin{proof}
The unitary radial map $u(x)\mapsto\sqrt{4\pi}\,r u(r)$
sends the radial restriction of $L$ to
\[
	L_{\rm rad}=-2\frac{\d^2}{\d r^2}+v(r)
	=2H_{\rm rad},\qquad
	D(H_{\rm rad})=H^2(0,\infty)\cap H_0^1(0,\infty).
\]
It sends the source $v$ to $g_v(r)=\sqrt{4\pi}\,r v(r)$.
All spectral measures of this source are radial.
We use the half-line Green function and Stone's formula
in the conventions of \cite[Lemma 9.7 and Theorem 4.3]{Teschl2009}.
The calculation below fixes the normalization for $L=2H_{\rm rad}$.

Let $j(k,r)$ be the outgoing solution of
$H_{\rm rad}j=k^2j$, with $j(k,r)=e^{ikr}$ for $r\ge R$.
Use the Wronskian convention $W(u,v)=uv'-u'v$.
Then $W(j,s)=\mathcal J_v(k)$, proving its independence
of $R$. For real $k>0$, $s(k,\cdot)$ is real and
$W(\overline j,j)=2ik$. If $\mathcal J_v(k)=0$, the
nonzero real solution $s$ would be proportional to both
$j$ and $\overline j$, contradicting this Wronskian.
Thus the denominator in \eqref{eq:radial-microscopic-density}
does not vanish.

For $\operatorname{Im}k>0$, the Dirichlet Green kernel of
$(H_{\rm rad}-k^2)^{-1}$ is
\[
	G(k^2;r,r')=
	\frac{s(k,\min(r,r'))j(k,\max(r,r'))}{\mathcal J_v(k)}.
\]
The derivative jump at $r=r'$ is
\[
	\partial_rG(k^2;r'+,r')-\partial_rG(k^2;r'-,r')
	=\frac{s(k,r')j'(k,r')-s'(k,r')j(k,r')}
	{\mathcal J_v(k)}=-1 .
\]
This is the normalization for the leading operator $-\d^2/\d r^2$.
On the positive real axis the Wronskian identity gives
\[
	s(k,r)=\frac{\overline{\mathcal J_v(k)}j(k,r)
	-\mathcal J_v(k)\overline{j(k,r)}}{2ik},
	\qquad
	\operatorname{Im}G(k^2+i0;r,r')
	=\frac{k\,s(k,r)s(k,r')}{|\mathcal J_v(k)|^2}.
\]
The notation $i0$ denotes the limit from the upper spectral
half-plane. On compact positive energy intervals this limit
is uniform for $r,r'\in[0,R]$. This follows from the
ordinary differential equation and the nonvanishing denominator.
Since $g_v$ has this compact support, Stone's formula gives
an absolutely continuous source measure
$\sigma_v(B)=\langle v,\mathbf1_B(L)v\rangle$ on $(0,\infty)$,
with
\begin{equation}\label{eq:radial-source-density}
	\frac{\d\sigma_v}{\d\lambda}(2k^2)
	=\frac{k}{2\pi|\mathcal J_v(k)|^2}
	\left|\int_0^R s(k,r)g_v(r)\d r\right|^2
	=\frac{2k|I_v(k)|^2}{|\mathcal J_v(k)|^2}.
\end{equation}
The factor $1/2$ comes from
$(L_{\rm rad}-\lambda)^{-1}
=\tfrac12(H_{\rm rad}-\lambda/2)^{-1}$.
To identify the whole source measure on a compact positive
interval, put $\mathfrak m_v(z)=\langle v,(L-z)^{-1}v\rangle$.
The Green kernel gives bounded, uniform boundary values of
$\operatorname{Im}\mathfrak m_v$ there. For each $\lambda_0>0$,
\[
	\sigma_v(\{\lambda_0\})
	\le\delta\operatorname{Im}\mathfrak m_v(\lambda_0+i\delta)
	\longrightarrow0\qquad(\delta\downarrow0).
\]
For $0<\alpha<\beta$, Stone's formula and uniform convergence yield
\[
	\sigma_v([\alpha,\beta])
	=\frac1\pi\lim_{\delta\downarrow0}
	\int_\alpha^\beta\operatorname{Im}\mathfrak m_v(\lambda+i\delta)\d\lambda
	=\int_\alpha^\beta
	\frac{\d\sigma_v}{\d\lambda}(\lambda)\d\lambda.
\]
These interval identities exclude a singular component on
$(0,\infty)$. Equation~\eqref{spec:eq:source-measure-identity} gives
$\d\mu_v=\lambda^{-1}\d\sigma_v$ on this half-line.
The zero spectral projection vanishes, and $L\ge0$. Thus
\eqref{eq:radial-microscopic-density} describes the measure
$\mu_v$ on $[0,\infty)$.

For the threshold, put $s_0(r)=s(0,r)$.
The integral equation for \eqref{eq:radial-regular-solution}
shows that $s_0(r)>0$ for $r>0$ and $s_0'(R)>0$.
Write $\mathfrak a_v=s_0'(R)$.
The zero-energy scattering solution then gives
$s_0(r)=\mathfrak a_v r f(r)$ and, for $r\ge R$,
$s_0(r)=\mathfrak a_v(r-a)$.
Since $v s_0=2s_0''$,
\[
	I_v(0)=2\int_0^R r s_0''(r)\d r
	=2\bigl(Rs_0'(R)-s_0(R)\bigr)
	=2\mathfrak a_v a,\qquad
	\mathcal J_v(0)=\mathfrak a_v.
\]
The solution $s(k,r)$ and its derivative are analytic in
$k^2$ on bounded $r$ intervals. For real $k$,
$|\mathcal J_v(k)|^2=s'(k,R)^2+k^2s(k,R)^2$.
Hence $I_v(k)^2/|\mathcal J_v(k)|^2=4a^2+O_v(k^2)$.
Substitution of $k=\sqrt{\lambda/2}$ proves
\eqref{eq:radial-threshold-density}.
The same analytic dependence shows that $I_v$ is not
identically zero, since $I_v(0)=2\mathfrak a_v a>0$.
Its positive zeros are therefore isolated. This proves
the asserted positivity on intervals.

Integrate \eqref{eq:radial-threshold-density} divided by
$\lambda$ between $b$ and a fixed small positive number.
The remainder is integrable at zero. The rest is finite
because $\mu_v$ is finite. This proves
\eqref{eq:radial-threshold-intensity}.
\end{proof}

This is the standard radial scattering representation of the
source measure, applied to the particular vector $L^{-1/2}v$.
The full density depends on the finite-range potential.
Only its leading threshold coefficient reduces to the
scattering length.

\begin{theorem}\label{spec:thm:profile}
Let $v\in C_c^\infty(\mathbb R^3)$ be nonzero, nonnegative and
radial, and let $\phi\in C_c^\infty(\mathbb R^3)$ be normalized.
Fix $0<\alpha<1$ and a real radial $0\le\rho\in C_c^\infty$
equal to one near zero. Set $m=N^{-\alpha}$,
$\epsilon=(m/N)^{1/2}$ and $k_N(r)=\sqrt N\,w(Nr)\rho(r/m)$.
Let $0\le\chi_N\le1$ be real, supported in $n\le L_N<\infty$,
and use the phase $C_N$ in \eqref{spec:eq:phase}.
Assume $\chi_N(1)=\chi_N(2)=1$ and
$\epsilon(L_N+2)^{3/2}\le D$, with $D$ independent of $N$.
For fixed
$h\in C_c^\infty(\mathbb R^3)$ in the one-particle sector,
\begin{equation}\label{spec:eq:profile-limit}
	\nu_{N,h}\ \Longrightarrow\
	q_1[h]\delta_0+2\norm{\phi h}_2^2\mu_v,
	\qquad q_1[h]=\ip h{(-\Delta+2)h}.
\end{equation}
Here convergence holds against every bounded continuous function
on $[0,\infty)$. In particular the microscopic measure is independent
of $\rho$ and $\alpha\in(0,1)$.

For every bounded Lipschitz function $F$ on $[0,\infty)$, one has
the quantitative estimate
\begin{align}
	&\left|\int F\d\nu_{N,h}
	-F(0)q_1[h]-2\norm{\phi h}_2^2\int F\d\mu_v\right|\notag\\
	&\qquad\le C_{h,\phi,v,\rho,D}\bigg[\norm F_\infty
	\left(m+\epsilon+(Nm)^{-1/2}\right)
	+\frac{\operatorname{Lip}(F)}{N^2}\bigg].
	\label{spec:eq:profile-rate}
\end{align}
\end{theorem}

The proof is given below. For $m=N^{-1/2}$,
\eqref{spec:eq:profile-rate} gives $O(N^{-1/4})$ on a bounded Lipschitz
unit ball.
The choice $\alpha=1/3$ instead gives $O(N^{-1/3})$ with this proof.

\begin{corollary}\label{spec:cor:probability}
For $g\in C_c((0,\infty))$,
\begin{equation}\label{spec:eq:probability-profile}
	N^2\ip h{g(\mathscr H_N/N^2)h}
	\longrightarrow
	2\norm{\phi h}_2^2\int\frac{g(\lambda)}{\lambda}\d\mu_v(\lambda).
\end{equation}
The vector may be normalized if this is to be read as a measurement
probability. For an interval $[b,c]\subset(0,\infty)$ whose endpoints
are not atoms of $\mu_v$, the analogous assertion holds for its
indicator function. If $\phi h\ne0$, some such interval has positive
limiting coefficient.
\end{corollary}
\begin{proof}
Extend $g(\lambda)/\lambda$ by zero near zero and apply
Theorem~\ref{spec:thm:profile}. The same conclusion for continuity
intervals follows from weak convergence of the finite measures,
with the bounded weight $1/\lambda$ on that interval. Nonzero total
mass and no atom at zero ensure that an interval of this kind with
positive mass exists.
\end{proof}

\begin{corollary}
\label{spec:cor:high-moments}
If $\phi h\ne0$, then, for every real $p>1$, there are
$c_{p,h}>0$ and $N_{p,h}$ such that
\begin{equation}\label{spec:eq:high-moment}
	\ip h{\mathscr H_N^p h}
	=\norm{Q_N^{p/2}U_Nh}^2
	\ge c_{p,h}N^{2p-2}\qquad(N\ge N_{p,h}).
\end{equation}
The first energies converge to $q_1[h]+2D_v\norm{\phi h}^2$.
Thus $U_N$ is not uniformly bounded from any fixed $A^{p/2}$
graph domain to the $Q_N^{p/2}$ graph domain for $p>1$.
\end{corollary}
\begin{proof}
Choose $0<b<c<\infty$ and $0\le F\le1$, supported in $[b,c]$,
with $\int F\d\mu_v>0$. By Theorem~\ref{spec:thm:profile},
$\int F\d\nu_{N,h}$ has a positive limit. Pointwise for $E\ge0$,
\[
	E^p\ge (bN^2)^{p-1} E F(E/N^2).
\]
The spectral theorem gives \eqref{spec:eq:high-moment}. For the
smooth-cutoff phase \eqref{spec:eq:phase} in this corollary, all
these moments are finite for each fixed $N$. In fact the
cutoff leaves finitely many sectors; the smooth kernel of $C_N$
maps each integer Sobolev space into itself by the differentiated
sector formulas below. Its exponential has the same property.
On these sectors $Q_N$ has a bounded smooth potential, so its
integer-power domains contain these vectors. Fractional powers
follow by spectral interpolation at fixed $N$.
For the Neumann phase in
Theorem~\ref{spec:thm:actual-COS-profile}, higher moments are
understood in the extended spectral sense on their stated domains.
\end{proof}

\begin{corollary}
\label{spec:cor:micro-time}
For each fixed real $\tau$,
\begin{equation}\label{spec:eq:micro-time}
	N^2\norm{(e^{-i\tau\mathscr H_N/N^2}-1)h}^2
	\longrightarrow
	4\norm{\phi h}^2\int\frac{1-\cos(\tau\lambda)}{\lambda}
	\d\mu_v(\lambda).
\end{equation}
As $\tau\to0$, the right side equals
$2\tau^2\norm{\phi h}^2\norm v_2^2+o(\tau^2)$.
Here $\mathscr H_N$ generates the evolution.
\end{corollary}
\begin{proof}
Apply Theorem~\ref{spec:thm:profile} with
$F_\tau(\lambda)=2(1-\cos(\tau\lambda))/\lambda$ for $\lambda>0$
and $F_\tau(0)=0$. This is bounded and continuous; the spectral
theorem identifies its integral against $\nu_{N,h}$ with the left
side of \eqref{spec:eq:micro-time}. Finally use
$0\le1-\cos(\tau\lambda)\le\tau^2\lambda^2/2$,
\eqref{spec:eq:mu-first}, and dominated convergence, with
$N\to\infty$ followed by $\tau\to0$.
\end{proof}

\subsection{Proof of the spectral profile}

\subsubsection{An estimate for the full cutoff phase}

For $r\ge0$ write
\[
	\norm\psi_{r,0}=\norm{M^{r/2}\psi},\qquad
	\norm\psi_{r,1}=\norm{M^{r/2}A^{1/2}\psi}.
\]
The following proof also verifies that the estimates refer to the
full phase in \eqref{spec:eq:phase}.

\begin{lemma}\label{spec:lem:phase}
Let $v\in C_c^\infty(\mathbb R^3)$ be nonzero, nonnegative and
radial, and let $\phi\in C_c^\infty(\mathbb R^3)$ be normalized.
Fix $0<\alpha<1$ and a real radial $0\le\rho\in C_c^\infty$
equal to one near zero. Set $m=N^{-\alpha}$,
$\epsilon=(m/N)^{1/2}$ and $k_N(r)=\sqrt N\,w(Nr)\rho(r/m)$.
Let $0\le\chi_N\le1$ be real, supported in $n\le L_N<\infty$,
and use the phase $C_N$ in \eqref{spec:eq:phase}.
Assume $\chi_N(1)=\chi_N(2)=1$.
Uniformly in $N,L_N$,
\begin{align}
	\norm{C_N\psi}_{r,0}&\le B_r\epsilon\norm\psi_{r+3,0},
	\label{spec:eq:phase-L2}\\
	\norm{C_N\psi}_{r,1}&\le B_r\big(
	\epsilon\norm\psi_{r+3,1}+\norm\psi_{r+3,0}\big).
	\label{spec:eq:phase-H1}
\end{align}
For fixed one-particle $h\in C_c^\infty$, let
\begin{equation}\label{spec:eq:T}
	(T_Nh)(x,y)=\frac{k_N(x-y)}{\sqrt2}
	(\phi(y)h(x)+\phi(x)h(y)).
\end{equation}
If $\epsilon(L_N+2)^{3/2}\le D$ uniformly in $N$,
\begin{equation}\label{spec:eq:phase-reduction}
	U_Nh=h\oplus T_Nh+r_N,\qquad
	\norm{Q_N^{1/2}r_N}\le C_{h,D}\epsilon,
	\qquad \norm{Q_N^{1/2}U_Nh}\le C_{h,D}.
\end{equation}
\end{lemma}
\begin{proof}
The scattering tail gives
\[
	\norm{k_N}_2\le C\epsilon,\qquad
	\norm{\nabla k_N}_2+\norm{(N^2v(N\cdot))^{1/2}k_N}_2\le C.
\]
The cutoff derivative has squared norm $O((Nm)^{-1})$; the other
derivative term is controlled by $\norm{\nabla w}_2$.
On sector $n$, the exact raising formula is
\[
	(G_N\psi_n)(X_{n+1})=\frac1{\sqrt{n+1}}
	\sum_{i\ne j}k_N(x_i-x_j)\phi(x_j)
	\psi_n(X_{n+1}\setminus x_j).
\]
Each summand has $L^2$ norm at most $C\epsilon\norm{\psi_n}$.
For the full gradient, the derivative of $k_N$ has norm bounded
by $C$. Derivatives on the remaining factors give the bound
$C\epsilon\norm{\psi_n}_{H^1}$. Summing the $n(n+1)$ terms gives
the number factor $(n+1)^{3/2}$.

On symmetric inputs the adjoint is
\[
	(G_N^*\psi_{n+1})(X_n)=\sqrt{n+1}\sum_{i=1}^n
	\int k_N(x_i-y)\overline{\phi(y)}\psi_{n+1}(X_n,y)\d y.
\]
In an $x_i$ derivative, transfer a derivative on $k_N$ to $y$ and
integrate by parts. All derivatives then fall on $\phi$ or the input,
and this block satisfies the stronger estimate with a factor
$\epsilon$ in its energy form norm. Other coordinate derivatives
fall directly on the input. The $L^2$ estimates follow by
Cauchy--Schwarz in $y$. Multiplication by the number cutoff is a
contraction; adjacent number weights are comparable. Squaring and
summing sectors proves \eqref{spec:eq:phase-L2}--\eqref{spec:eq:phase-H1}.

The scaled Sobolev inequality gives $\cV_N\le C\cN\cK$, hence
\begin{equation}\label{spec:eq:energy-comparison}
	A\le Q_N\le CMA.
\end{equation}
Apply the one-coordinate homogeneous Sobolev
inequality to each pair using
$\norm{N^2v(N\cdot)}_{3/2}=\norm v_{3/2}$, average over the two
coordinates, and sum pairs. This proves the comparison with a
constant independent of $N,n$.

For fixed $N,L_N$, the sector bounds imply that $C_N$ is bounded
on both Fock space and the $Q_N$ form domain. Thus its exponential
and its Taylor series are valid there. On a vector $\zeta$ supported
in $n\le L_N$, put $\delta=\epsilon(L_N+2)^{3/2}$. The estimates
with $r=1$ imply
\[
	\norm{C_N\zeta}_{1,0}\le B\delta\norm\zeta_{1,0},\qquad
	\norm{C_N\zeta}_{1,1}\le B\delta
	(\norm\zeta_{1,1}+\epsilon^{-1}\norm\zeta_{1,0}).
\]
On the fixed one-particle vector, two applications of the sector
estimates give $\norm{C_N^2h}_{1,0}\le C_h\epsilon^2$ and
$\norm{C_N^2h}_{1,1}\le C_h\epsilon$. Triangular iteration gives,
for $j\ge2$,
\[
	\norm{C_N^jh}_{1,1}
	\le C_h(j-1)\epsilon(B\delta)^{j-2}.
\]
The series with coefficients $1/j!$ is uniformly summable when
$\delta\le D$. Since $C_Nh=T_Nh$ and $G_N^*h=0$, this proves
\eqref{spec:eq:phase-reduction}, using \eqref{spec:eq:energy-comparison}.
\end{proof}

If $F$ is bounded, the energy-weighted functional calculus is
continuous in the energy norm:
\begin{align}\label{spec:eq:form-continuity}
	&\left|\ip{Q^{1/2}u}{F(Q/N^2)Q^{1/2}u}
	-\ip{Q^{1/2}z}{F(Q/N^2)Q^{1/2}z}\right|\notag\\
	&\qquad\le\norm F_\infty\norm{Q^{1/2}(u-z)}
	(\norm{Q^{1/2}u}+\norm{Q^{1/2}z}).
\end{align}
Consequently, \eqref{spec:eq:phase-reduction} permits replacing $U_Nh$
by $h\oplus T_Nh$ with error at most $C_h\norm F_\infty\epsilon$.
The two remaining sectors are orthogonal also under functional
calculus, because $Q_N$ preserves particle number.

\subsubsection{Dilation in the relative variable}

Set $R=(x+y)/2$, $r=x-y$, and $a_h(R)=\phi(R)h(R)$. The Jacobian
has absolute value one, and
\[
	Q_{2,N}=-\tfrac12\Delta_R-2\Delta_r+N^2v(Nr)+3.
\]
Replace $T_Nh$ by
\begin{equation}\label{spec:eq:separated}
	z_N(R,r)=\sqrt2\,a_h(R)k_N(r).
\end{equation}
The smooth amplitude in \eqref{spec:eq:T} equals $\sqrt2a_h(R)$ at
$r=0$. Its difference has $L^2_R$ norm at most $C_h|r|$, and its
first derivatives have uniformly bounded $L^2_R$ norms on
$|r|\le Cm$. The kernel estimates from Lemma~\ref{spec:lem:phase} yield
\begin{equation}\label{spec:eq:separated-error}
	\norm{Q_{2,N}^{1/2}(T_Nh-z_N)}
	\le C_h(m+\epsilon).
\end{equation}
When the derivative acts on the kernel, the amplitude difference
gives a factor $O(m)$. When it acts on the amplitude, the term is
$O(\epsilon)$. On the support of the potential, the amplitude
difference gives a factor $O(N^{-1})$. Since $N^{-1}\le m$, this is absorbed in the bound.
All center-coordinate supports remain in a fixed compact set.

The unitary dilation from $(R,r)$ to $(R,z)$ is
$(S_Nu)(R,z)=N^{-3/2}u(R,z/N)$.
Writing $H_R=-\frac12\Delta_R+3$, we have
\begin{equation}\label{spec:eq:dilation}
	N^{-2}S_NQ_{2,N}S_N^*=L_N:=L+N^{-2}H_R,
	\qquad S_Nz_N=N^{-1}\sqrt2\,a_h\otimes w_{Nm}.
\end{equation}
The two summands of $L_N$ strongly commute, as they act in different
coordinates. Define
\[
	\eta_N=\sqrt2\,L_N^{1/2}(a_h\otimes w_{Nm}),\qquad
	\eta=\sqrt2\,a_h\otimes\xi_v.
\]
The scalar inequality $(\sqrt{s+t}-\sqrt s)^2\le t$ and joint
spectral calculus give
\[
	\norm{(L_N^{1/2}-L^{1/2})(a_h\otimes w_{Nm})}
	\le N^{-1}\norm{H_R^{1/2}a_h}\norm{w_{Nm}}_2
	\le C_h\epsilon.
\]
Here $\norm{w_R}_2\le C\sqrt R$ for $R\ge1$. Combined with
Lemma~\ref{spec:lem:scattering-vector}, this proves
\begin{equation}\label{spec:eq:eta-convergence}
	\norm{\eta_N-\eta}\le C_h(\epsilon+(Nm)^{-1/2}).
\end{equation}
The left side of the two-particle spectral expression for $z_N$ is
exactly $\ip{\eta_N}{F(L_N)\eta_N}$, by \eqref{spec:eq:dilation}.

For bounded continuous $F$, joint spectral calculus gives
$F(L_N)\to I\otimes F(L)$ strongly, by dominated convergence in
the joint spectral measure of $H_R$ and $L$. Thus
\[
	\ip{\eta_N}{F(L_N)\eta_N}
	\longrightarrow 2\norm{a_h}^2\ip{\xi_v}{F(L)\xi_v}.
\]
For bounded Lipschitz $F$, the same joint calculus gives the more
explicit estimate
\[
	\norm{(F(L_N)-F(L))(a_h\otimes\xi_v)}
	\le N^{-2}\operatorname{Lip}(F)\norm{H_Ra_h}\norm{\xi_v}.
\]
Together with \eqref{spec:eq:eta-convergence}, this controls the error
by $C_h[\norm F_\infty(\epsilon+(Nm)^{-1/2})+
N^{-2}\operatorname{Lip}(F)]$.

On the one-particle sector $Q_1=-\Delta+2$ is independent of $N$.
Its contribution converges to $F(0)q_1[h]$. For Lipschitz $F$,
the error is at most
$N^{-2}\operatorname{Lip}(F)\norm{Q_1h}^2$.
Combining this bound, \eqref{spec:eq:form-continuity},
\eqref{spec:eq:separated-error} and the preceding two-particle estimates
proves both assertions of Theorem~\ref{spec:thm:profile}.

\subsection{General finite-sector states}\label{spec:sec:finite-sectors}

We evaluate individual pair contributions using the sector
Hamiltonian on the full tensor space $L^2((\mathbb R^3)^p)$ by the
same differential expression. It commutes with particle
permutations. Its restriction to the symmetric subspace is the
bosonic sector Hamiltonian used above. The projection
$P_p=q^{\otimes p}$ also commutes with permutations, so the
same convention applies to the restricted forms and their spectral
functions. After summing the pair contributions, all vectors
belong to the symmetric subspace.

We first establish the spectral perturbation and pair
orthogonality estimates for the particle-pair contributions.

\begin{lemma}
\label{spec:lem:growing-vector}
On a fixed Hilbert space, let $B_N,A_N\ge0$ be self-adjoint and
$A_N=B_N+W_N$, where $W_N$ is bounded and self-adjoint. Suppose
\[
	\sup_N\norm{W_N}<\infty,\qquad W_N\longrightarrow0
	\ \hbox{strongly},\qquad B_N\longrightarrow B
	\ \hbox{in strong resolvent}.
\]
Let $u_N\in D(B_N^{1/2})=D(A_N^{1/2})$ satisfy
\begin{equation}\label{spec:eq:growing-input}
	B_N^{1/2}u_N\longrightarrow\eta,\qquad
	\sup_N\norm{A_N^{1/2}u_N}<\infty,\qquad
	\norm{W_Nu_N}\longrightarrow0.
\end{equation}
Then, for every $g\in C_c((0,\infty))$,
\begin{equation}\label{spec:eq:growing-conclusion}
	\norm{g(A_N)u_N-g(B_N)u_N}\longrightarrow0.
\end{equation}
\end{lemma}
\begin{proof}
Write $R_{B,N}=(B_N+1)^{-1}$, $R_{A,N}=(A_N+1)^{-1}$ and
$\eta_N=B_N^{1/2}u_N$. For each integer $j\ge1$ define
\[
	k_j(\lambda)=\frac{1-(1+\lambda)^{-j}}{\sqrt\lambda}
	\quad(\lambda>0),\qquad k_j(0)=0.
\]
This is a continuous function vanishing at infinity. Functional
calculus gives
$R_{B,N}^j u_N=u_N-k_j(B_N)\eta_N$.
Strong resolvent convergence and $\eta_N\to\eta$ imply convergence
in norm of $k_j(B_N)\eta_N$. Uniform boundedness and strong
convergence of $W_N$, together with \eqref{spec:eq:growing-input}, give
$\norm{W_NR_{B,N}^j u_N}\longrightarrow0$.
The resolvent identity and a telescoping sum now imply
\[
	(R_{A,N}^j-R_{B,N}^j)u_N
	=-\sum_{\ell=0}^{j-1}
	R_{A,N}^{j-\ell}W_NR_{B,N}^{\ell+1}u_N\longrightarrow0.
\]
All inverse powers on the left of $W_N$ are contractions. It
follows that \eqref{spec:eq:growing-conclusion} holds for any polynomial
in $(1+\lambda)^{-1}$; its constant term cancels.

To extend the class of functions, set
$r=(1+\lambda)^{-1}$ and $b(\lambda)=\lambda/(1+\lambda)=1-r$.
For the given $g$, the function $g(\lambda)/b(\lambda)$ extends
continuously by zero to the endpoints of $r\in[0,1]$.
Choose a polynomial $P$ approximating this function uniformly
within $\delta$. Then
\[
	|g(\lambda)-b(\lambda)P((1+\lambda)^{-1})|
	\le\delta\,\frac{\lambda}{1+\lambda}
	\le\frac{\delta}{2}\sqrt\lambda.
\]
The approximating function is itself a polynomial in the inverse
resolvent. The error on $u_N$ is at most $\delta/2$ times its
$A_N^{1/2}$ or $B_N^{1/2}$ norm, both uniformly bounded.
First let $N\to\infty$ and then $\delta\downarrow0$.
\end{proof}

\begin{lemma}
\label{spec:lem:pair-orthogonality}
Fix a particle number $p\ge3$. For an unordered pair $a=\{i,j\}$,
write $R_a=(x_i+x_j)/2$ and $r_a=x_i-x_j$. Denote the remaining
coordinates by $Y_a$. Define the unitary map
\[
	(S_{N,a}u)(R_a,z,Y_a)
	=N^{-3/2}u(R_a,z/N,Y_a).
\]
If $a\ne b$ and $\zeta_a,\zeta_b$ are fixed $L^2$ vectors in the
respective coordinate spaces, then
\begin{equation}\label{spec:eq:pair-orthogonality}
	\ip{S_{N,a}^*\zeta_a}{S_{N,b}^*\zeta_b}\longrightarrow0.
\end{equation}
\end{lemma}
\begin{proof}
First take bounded compactly supported profiles. The two physical
vectors have amplitudes bounded by $C N^{3/2}$, and their joint
support lies in a fixed bounded set with
$|r_a|,|r_b|\le C/N$. Two distinct relative coordinates are
independent linear coordinates, also when the pairs share a vertex.
For pairs $\{1,2\}$ and $\{1,3\}$, use
\[
	t=x_1,\qquad r=x_1-x_2,\qquad s=x_1-x_3,
	\qquad (x_1,x_2,x_3)=(t,t-r,t-s).
\]
The absolute Jacobian is one. For disjoint pairs $\{1,2\}$
and $\{3,4\}$, use $(R_a,r_a,R_b,r_b)$ and the unchanged
spectators; its absolute Jacobian is also one.
In either case the remaining coordinates range over a fixed
bounded set. The overlap volume is $O(N^{-6})$.
Multiplication by the two amplitude bounds gives $O(N^{-3})$.

For general profiles choose bounded compactly supported
$\zeta_a^{(j)},\zeta_b^{(j)}$ converging in $L^2$.
Unitarity bounds the change in the inner product by
\[
	\|\zeta_a-\zeta_a^{(j)}\|_2\|\zeta_b\|_2
	+\|\zeta_a^{(j)}\|_2\|\zeta_b-\zeta_b^{(j)}\|_2 .
\]
First let $N\to\infty$ at fixed $j$, then $j\to\infty$.
The same argument is uniform when the two profiles range over
fixed compact subsets of their respective $L^2$ spaces:
choose finite nets and apply this bound at each net point.
\end{proof}

Put $T_\phi=\dGamma(|\phi|^2)$. For a vector in the
$\mathscr H_N$ form domain use the notation
\[
	\nu_{N,\psi}(B)
	=\ip{\mathscr H_N^{1/2}\psi}
	{\mathbf1_B(\mathscr H_N/N^2)\mathscr H_N^{1/2}\psi}.
\]

\begin{theorem}
\label{spec:thm:finite-profile}
Let $v\in C_c^\infty(\mathbb R^3)$ be nonzero, nonnegative and
radial, and let $\phi\in C_c^\infty(\mathbb R^3)$ be normalized.
Fix $0<\alpha<1$ and a real radial $0\le\rho\in C_c^\infty$
equal to one near zero. Set $m=N^{-\alpha}$,
$\epsilon=(m/N)^{1/2}$ and $k_N(r)=\sqrt N\,w(Nr)\rho(r/m)$.
Let $0\le\chi_N\le1$ be real, supported in $n\le L_N<\infty$,
and use the phase $C_N$ in \eqref{spec:eq:phase}.
Let $\psi$ be fixed, with finitely many nonzero particle sectors,
each smooth and compactly supported. Suppose $\psi_n=0$ for $n>n_0$,
that $\chi_N(n)=1$ for $0\le n\le n_0+1$ for all sufficiently
large $N$, and that $\epsilon(L_N+2)^{3/2}\le D$ with $D$
independent of $N$. Then
\begin{equation}\label{spec:eq:finite-profile}
	\nu_{N,\psi}\ \Longrightarrow\
	A[\psi]\delta_0+2\ip\psi{T_\phi\psi}\mu_v
\end{equation}
against every bounded continuous function on $[0,\infty)$.

\end{theorem}
\begin{proof}
\textit{Step 1: the full exponential reduces to its raising part.}
The proof of Lemma~\ref{spec:lem:phase}, with the fixed finite-sector
vector in place of $h$, gives
\[
	\norm{Q_N^{1/2}(U_N\psi-\psi-C_N\psi)}
	\le C_{\psi,D}\epsilon.
\]
Indeed, two insertions have weighted $L^2$ norm $O_\psi(\epsilon^2)$
and weighted energy form norm $O_\psi(\epsilon)$, and the same
triangular iteration sums the tail. The cutoff plateau implies
$C_N\psi=G_N\psi-G_N^*\psi$. The stronger adjoint estimate in the
proof of Lemma~\ref{spec:lem:phase}, and \eqref{spec:eq:energy-comparison}
on its finitely many output sectors, give
$\norm{Q_N^{1/2}G_N^*\psi}\le C_\psi\epsilon$. Consequently,
\begin{equation}\label{spec:eq:finite-reduction}
	U_N\psi=\psi+G_N\psi+r_{N,\psi},\qquad
	\norm{Q_N^{1/2}r_{N,\psi}}\le C_{\psi,D}\epsilon.
\end{equation}
The first energies of $\psi$ and $G_N\psi$ are uniformly bounded.
Equation~\eqref{spec:eq:form-continuity} applies to the replacement.

\textit{Step 2: one particle-pair contribution.}
Fix an input sector $n\ge1$ and put $p=n+1$.
The Fock-space normalization gives
\begin{equation}\label{spec:eq:raising-sector-normalization}
	(G_N\psi_n)(X_p)=\frac1{\sqrt p}
	\sum_{j=1}^p\sum_{i\ne j}
	k_N(x_i-x_j)\phi(x_j)\psi_n(X_p\setminus x_j).
\end{equation}
Indeed, annihilation contributes $\sqrt n$ and the two
creation operators contribute $1/\sqrt{p(p-1)}$ in the
ordered sum. Since $p-1=n$, their product is $1/\sqrt p$.
The radial kernel is even. Combining the terms $(i,j)$ and
$(j,i)$ therefore gives, on the full tensor space,
\begin{align}
	G_N\psi_n&=\sum_a g_{N,a},\qquad
	g_{N,a}=\frac{k_N(r_a)}{\sqrt p}F_a(X_p),\notag\\
	F_a(X_p)&=\phi(x_j)\psi_n(X_p\setminus x_j)
	+\phi(x_i)\psi_n(X_p\setminus x_i),
	\qquad a=\{i,j\}. \label{spec:eq:pair-profile-input}
\end{align}
Evaluate the individual summands with the permutation-invariant
Hamiltonian on the full tensor space. Its restriction to the
symmetric space is the original operator.
Let
\[
	a_a(R_a,Y_a)=\frac{2}{\sqrt p}\phi(R_a)\psi_n(R_a,Y_a),
	\qquad u_{N,a}=N S_{N,a}g_{N,a}.
\]
The choice and order of spectator coordinates in $\psi_n$ is
irrelevant by symmetry. With $r_a=z/N$, the dilated vector is
\[
	u_{N,a}(R_a,z,Y_a)=w(z)\rho(z/(Nm))b_{N,a}(R_a,z,Y_a),
	\qquad b_{N,a}=p^{-1/2}F_a\big|_{r_a=z/N}.
\]
On the support of the relative kernel, smoothness gives
\[
	|b_{N,a}-a_a|\le C_\psi m,\qquad
	|\nabla_z b_{N,a}|\le C_\psi/N .
\]
The remaining coordinate derivatives are bounded on a fixed
compact set. In particular $\|u_{N,a}\|_2\le C_\psi\sqrt{Nm}$.
After dilation the total energy is
\begin{align}
	\mathcal A_{N,a}
	&=N^{-2}S_{N,a}Q_{N,p}S_{N,a}^*
	=\mathcal B_{N,a}+W_{N,a},\notag\\
	\mathcal B_{N,a}
	&=L_z+N^{-2}\left(-\tfrac12\Delta_{R_a}
	-\sum_{k\notin a}\Delta_{x_k}+p+1\right).
	\label{spec:eq:pair-scaled-energy}
\end{align}
Here $W_{N,a}$ is the sum of the unmatched potentials, each
multiplication by $v(Nr_b)$ in the dilated coordinates, $b\ne a$.
It is nonnegative and uniformly bounded for fixed $p$. At almost
every fixed $(R_a,z,Y_a)$ each such factor tends to zero. For a
pair sharing a vertex with $a$, its argument has the form
$N(R_a-x_k)\pm z/2$; for a disjoint pair it is
$N(x_k-x_\ell)$. The exceptional equalities are null sets.
Thus $W_{N,a}\to0$ strongly by dominated convergence.

The smooth-amplitude replacement used in
\eqref{spec:eq:separated-error}, now with spectator derivatives included,
and Lemma~\ref{spec:lem:scattering-vector} give
\begin{equation}\label{spec:eq:pair-energy-vector}
	\mathcal B_{N,a}^{1/2}u_{N,a}
	\longrightarrow a_a\otimes\xi_v
	\quad\hbox{in }L^2.
\end{equation}
Write $w_{Nm}(z)=w(z)\rho(z/(Nm))$. The amplitude bounds give
\[
	\|\mathcal B_{N,a}^{1/2}
	(u_{N,a}-a_a\otimes w_{Nm})\|_2
	\le C_\psi(m+\epsilon).
\]
The relative derivative of $w_{Nm}$ is uniformly bounded,
and its potential energy is bounded, so the factor
$b_{N,a}-a_a$ contributes $O_\psi(m)$.
A derivative of $b_{N,a}$ contributes at most
$C_\psi N^{-1}\|w_{Nm}\|_2\le C_\psi\epsilon$.
The center, spectator and number terms have the same bound
after taking the square root of their factor $N^{-2}$.
The separated vector is $a_a\otimes w_{Nm}$. The residual center and spectator kinetic
term changes its square-root energy vector by $O_\psi(\epsilon)$,
by the commuting square-root inequality used after
\eqref{spec:eq:dilation}. The relative energy vector differs from
$a_a\otimes\xi_v$ by $O_\psi((Nm)^{-1/2})$.

Two distinct relative coordinates can be integrated independently.
Boundedness and compact support of the smooth amplitudes imply,
for $b\ne a$,
\begin{equation}\label{spec:eq:unmatched-spectral}
	\norm{V_{N,b}^{1/2}g_{N,a}}^2
	\le C_\psi\norm{N^2v(N\cdot)}_1\norm{k_N}_2^2
	\le C_\psi\epsilon^2/N.
\end{equation}
Here $V_{N,b}=N^2v(Nr_b)$. Scaling preserves the relevant energy
pairing:
\[
	\ip{u_{N,a}}{W_{N,a}u_{N,a}}
	=\sum_{b\ne a}\norm{V_{N,b}^{1/2}g_{N,a}}^2\longrightarrow0.
\]
It follows that $\norm{W_{N,a}u_{N,a}}\to0$, since
$0\le W_{N,a}\le C_p$, and the total scaled energy is bounded.
Also $\mathcal B_{N,a}\to L_z$ in strong resolvent, by joint
spectral calculus in the relative and remaining variables.

We may therefore apply Lemma~\ref{spec:lem:growing-vector}. For
$0\le F\in C_c((0,\infty))$ take $g(\lambda)=\sqrt{\lambda F(\lambda)}$.
Together with \eqref{spec:eq:pair-energy-vector}, the lemma gives
\begin{equation}\label{spec:eq:pair-spectral-vector}
	\left\|Q_{N,p}^{1/2}\sqrt{F(Q_{N,p}/N^2)}\,g_{N,a}
	-S_{N,a}^*\big(a_a\otimes\sqrt{F(L)}\,\xi_v\big)\right\|
	\longrightarrow0.
\end{equation}
Indeed, after dilation the first vector is
$g(\mathcal A_{N,a})u_{N,a}$; the corresponding expression for
$\mathcal B_{N,a}$ is
$\sqrt{F(\mathcal B_{N,a})}\mathcal B_{N,a}^{1/2}u_{N,a}$.
The latter converges strongly to the displayed fixed profile.

\textit{Step 3: different pairs decouple under the spectral function.}
Apply Lemma~\ref{spec:lem:pair-orthogonality} to the profiles in
\eqref{spec:eq:pair-spectral-vector}. The norm squared of their sum
converges to the sum of their squared norms. There are $n(n+1)/2$
unordered pairs. Each diagonal amplitude has the factor
$2/\sqrt p$, so their squared sum has coefficient
\[
	\frac{p(p-1)}2\frac4p=2(p-1)=2n .
\]
Consequently,
\begin{equation}\label{spec:eq:pair-count-spectral}
	\sum_a\norm{a_a}^2
	=2n\int|\phi(x)|^2|\psi_n(x,Y)|^2\d x\d Y
	=2\ip{\psi_n}{T_\phi\psi_n}.
\end{equation}
Thus the energy-weighted measure of $G_N\psi_n$ converges against
$F$ to $2\ip{\psi_n}{T_\phi\psi_n}\int F\d\mu_v$.
Splitting real and imaginary parts and positive and negative
parts extends this assertion to every $F\in C_c((0,\infty))$.
Different input sectors of $G_N$ have different output particle
numbers, so their spectral pairings vanish exactly. Summing over
the finitely many input sectors proves the same assertion for
$G_N\psi$ with coefficient $2\ip\psi{T_\phi\psi}$.

\textit{Step 4: no energy is lost at zero or infinity in the raising part.}
We verify convergence of its total first energy. For each diagonal
pair term in \eqref{spec:eq:pair-profile-input}, the derivatives falling
on the smooth amplitude contribute $O_\psi(\epsilon)$ to the
energy. The remaining matching-pair density is
\[
	J_N(r)=2|\nabla k_N(r)|^2+N^2v(Nr)|k_N(r)|^2.
\]
It is supported in $|r|\le Cm$ and has mass tending to $D_v$,
by Lemma~\ref{spec:lem:scattering-vector} and scaling. Replacing the
smooth amplitude by its value at $r=0$ therefore gives
$Q_{N,p}[g_{N,a}]\longrightarrow D_v\norm{a_a}^2$.
The unmatched potentials are negligible by
\eqref{spec:eq:unmatched-spectral}. In a kinetic cross term between
different pairs, the leading derivatives satisfy
$\norm{\nabla k_N}_1\le Cm/\sqrt N$. Independent relative
coordinates bound that term by $C_\psi m^2/N$; terms with a
derivative on a smooth amplitude are $O_\psi(\epsilon)$.
For a potential cross term between $g_{N,a}$ and $g_{N,c}$,
$a\ne c$, at least one of them is unmatched with its potential
pair. Equation~\eqref{spec:eq:unmatched-spectral} and Cauchy--Schwarz
bound the term by $O_\psi(\epsilon/\sqrt N)$.
The number term is $O_\psi(\epsilon^2)$.
Consequently
\begin{equation}\label{spec:eq:raising-total}
	Q_N[G_N\psi]\longrightarrow2D_v\ip\psi{T_\phi\psi}.
\end{equation}

The limiting measure found in Step 3 has precisely this mass and
no atom at zero. To see explicitly that the convergence extends
to bounded continuous tests on $[0,\infty)$, choose
$0\le\theta\le1$ in $C_c((0,\infty))$ whose integral against that
limit is within $\delta$ of its total mass. Convergence of the
$\theta$ integral and \eqref{spec:eq:raising-total} show that the
limsup of the mass weighted by $1-\theta$ is at most $\delta$.
For bounded continuous $F$, apply Step 3 to $F\theta$ and bound
the two remaining integrals by $\norm F_\infty$ times these
small masses. Let $\delta\downarrow0$. If the limiting mass is
zero, \eqref{spec:eq:raising-total} gives the result directly.

\textit{Step 5: the initial smooth part and its mixed spectral form.}
For a fixed smooth finite-sector vector,
\begin{equation}\label{spec:eq:smooth-input-spectrum}
	Q_N[\psi]=A[\psi]+O_\psi(N^{-1}),
	\qquad \norm{Q_N\psi}^2\le C_\psi N.
\end{equation}
The first assertion follows from the $L^1$ norm of each potential.
For the second, $\norm{N^4v(N\cdot)^2}_1=N\norm v_2^2$ bounds
the squared norm of each potential acting on the smooth vector
by $C_\psi N$. There are finitely many pairs, so the triangle
inequality and the fixed kinetic term suffice.
The first-energy mass at scaled energies at least $\delta>0$ is
at most $\norm{Q_N\psi}^2/(\delta N^2)$ and tends to zero.
Thus the energy-weighted measure of $\psi$ converges to
$A[\psi]\delta_0$, and
\[
	\norm{(F(Q_N/N^2)-F(0))Q_N^{1/2}\psi}\longrightarrow0
\]
for every bounded continuous $F$.
The ordinary mixed energy obeys
$|Q_N[\psi,G_N\psi]|\le C_\psi(\epsilon+N^{-1/2})$.
For its kinetic and number parts, move $A$ onto the fixed smooth
vector and use $\norm{G_N\psi}=O_\psi(\epsilon)$. For the potential
part, use $\norm{\cV_N^{1/2}\psi}=O_\psi(N^{-1/2})$ and the bounded
energy of $G_N\psi$. For real $F$, its mixed spectral form is
$F(0)Q_N[\psi,G_N\psi]+o(1)$ by the preceding vanishing norm.
It tends to zero. Complex $F$ follows by linearity.

Combining Steps 3--5 gives the claimed limit for
$\psi+G_N\psi$. The energy-norm remainder
\eqref{spec:eq:finite-reduction} and \eqref{spec:eq:form-continuity}
then prove \eqref{spec:eq:finite-profile}.
\end{proof}

\begin{proposition}
\label{spec:prop:domain-profile}
Let $v\in C_c^\infty(\mathbb R^3)$ be nonzero, nonnegative and
radial, and let $\phi\in C_c^\infty(\mathbb R^3)$ be normalized.
Fix $0<\alpha<1$ and a real radial $0\le\rho\in C_c^\infty$
equal to one near zero. Set $m=N^{-\alpha}$,
$\epsilon=(m/N)^{1/2}$ and $k_N(r)=\sqrt N\,w(Nr)\rho(r/m)$.
Let $0\le\chi_N\le1$ be real, supported in $n\le L_N<\infty$,
and use the phase $C_N$ in \eqref{spec:eq:phase}.
Assume $\epsilon(L_N+2)^{3/2}\le D$ uniformly in $N$.
Require $\chi_N=1$ on every fixed finite set of particle numbers
for all sufficiently large $N$. Set
\[
	\mathscr X=D(M^2A^{1/2}),\qquad
	\norm\psi_{\mathscr X}=\norm{M^2A^{1/2}\psi}.
\]
Then \eqref{spec:eq:finite-profile} holds for every fixed
$\psi\in\mathscr X$. For each fixed bounded continuous test
function, the convergence is uniform on compact subsets of
$\mathscr X$.
\end{proposition}
\begin{proof}
First we show the uniform mapping bound
\begin{equation}\label{spec:eq:domain-phase-map}
	\norm{Q_N^{1/2}U_N\psi}\le C_D\norm\psi_{\mathscr X}.
\end{equation}
Use Lemma~\ref{spec:lem:phase} with weight $r=1$. Its estimates on
the first insertion give
\[
	\norm{C_N\psi}_{1,0}\le C\epsilon\norm\psi_{\mathscr X},
	\qquad \norm{C_N\psi}_{1,1}\le C\norm\psi_{\mathscr X}.
\]
This inserted vector is cutoff-supported. The triangular iteration
in that lemma yields, for $j\ge1$,
\[
	\norm{C_N^j\psi}_{1,1}
	\le Cj(B\delta_N)^{j-1}\norm\psi_{\mathscr X},
	\qquad \delta_N=\epsilon(L_N+2)^{3/2}\le D.
\]
Summing with $1/j!$ and using $Q_N\le CMA$ proves
\eqref{spec:eq:domain-phase-map} first on smooth finite-sector vectors
and then by density and closedness of the energy form.

The quadratic forms $\nu_{N,\psi}(F)$ are uniformly continuous
on bounded $\mathscr X$ sets:
\[
	|\nu_{N,\psi}(F)-\nu_{N,\zeta}(F)|
	\le C_D\norm F_\infty\norm{\psi-\zeta}_{\mathscr X}
	(\norm\psi_{\mathscr X}+\norm\zeta_{\mathscr X}).
\]
The limiting form in \eqref{spec:eq:finite-profile} is also continuous
there, since $T_\phi\le\norm\phi_\infty^2\cN\le C A$ and
$|\int F\d\mu_v|\le D_v\norm F_\infty$.
Smooth finite-sector vectors are dense in $\mathscr X$ by particle
number truncation and sectorwise Sobolev density. Applying
Theorem~\ref{spec:thm:finite-profile} to such approximants proves the
fixed-vector result. A finite net in each compact set, followed by
the displayed equicontinuity bound, proves compact-uniform
convergence.
\end{proof}

\begin{corollary}
\label{spec:cor:general-consequences}
Let $v\in C_c^\infty(\mathbb R^3)$ be nonzero, nonnegative and
radial, and let $\phi\in C_c^\infty(\mathbb R^3)$ be normalized.
Fix $0<\alpha<1$ and a real radial $0\le\rho\in C_c^\infty$
equal to one near zero. Set $m=N^{-\alpha}$,
$\epsilon=(m/N)^{1/2}$ and $k_N(r)=\sqrt N\,w(Nr)\rho(r/m)$.
Let $0\le\chi_N\le1$ be real, supported in $n\le L_N<\infty$,
and use the phase $C_N$ in \eqref{spec:eq:phase}.

Assume $\epsilon(L_N+2)^{3/2}\le D$ uniformly in $N$.
Let $\psi$ be fixed and satisfy either of the following conditions:
\begin{enumerate}
\item Its sectors are smooth and compactly supported, $\psi_n=0$
for $n>n_0$, and $\chi_N(n)=1$ for $0\le n\le n_0+1$ eventually.
\item $\psi\in\mathscr X=D(M^2A^{1/2})$, and $\chi_N=1$ on each
fixed finite set of particle numbers for all sufficiently large $N$.
\end{enumerate}
Write $c_\psi=\ip\psi{T_\phi\psi}$. Then
\begin{align}
	N^2\ip\psi{g(\mathscr H_N/N^2)\psi}
	&\longrightarrow2c_\psi
	\int\frac{g(\lambda)}{\lambda}\d\mu_v(\lambda),
	\qquad g\in C_c((0,\infty)),\label{spec:eq:general-window}\\
	N^2\norm{(e^{-i\tau\mathscr H_N/N^2}-1)\psi}^2
	&\longrightarrow4c_\psi
	\int\frac{1-\cos(\tau\lambda)}{\lambda}\d\mu_v(\lambda),
	\qquad \tau\in\mathbb R.\label{spec:eq:general-time}
\end{align}
If $c_\psi>0$, every $p>1$ satisfies
$\ip\psi{\mathscr H_N^p\psi}\ge c_{p,\psi}N^{2p-2}$ for large $N$.
For $\psi\in\mathscr X$ outside the higher-power domain, this is
understood as an inequality for the possibly infinite nonnegative
spectral moment. It is finite for the smooth finite-sector inputs.
\end{corollary}
\begin{proof}
Apply the proved measure convergence to $g(\lambda)/\lambda$
extended by zero near zero, and to
$2(1-\cos(\tau\lambda))/\lambda$ extended by zero at zero.
These are bounded continuous functions. For the moment lower
bound use a positive compact spectral window as in
Corollary~\ref{spec:cor:high-moments}. The limiting window mass is
positive because $c_\psi>0$, $\mu_v$ has mass $D_v>0$, and its
zero atom vanishes.
\end{proof}

\subsection{The condensate-orthogonal microscopic profile}
\label{spec:sec:projected-profile}

Let $q=1-|\phi\rangle\langle\phi|$ and $P=\Gamma(q)$.
On $P\cF$, let $A_P,Q_{N,P}$ be the operators of the restricted
energy forms, and set
\[
	C_{N,P}=PC_NP,\qquad U_{N,P}=e^{C_{N,P}},\qquad
	\mathscr H_{N,P}=U_{N,P}^*Q_{N,P}U_{N,P}.
\]
In particular, $Q_{N,P}$ is defined by form restriction, rather
than by assuming that $P$ commutes with the kinetic energy.
During a pair decomposition, the same notation on a fixed sector
denotes the extension to the full tensor space. These operators
commute with permutations, so the resulting identities restrict
to the bosonic space.

\begin{lemma}
\label{spec:lem:dilated-projection}
Fix $p\ge2$ and a pair $a=\{i,j\}$. With the dilation of
Lemma~\ref{spec:lem:pair-orthogonality}, put
\[
	P_{N,a}=S_{N,a}q^{\otimes p}S_{N,a}^*,\qquad
	P_{\mathrm{sp},a}=\prod_{k\notin a}q_k.
\]
The empty product is the identity. Then
$P_{N,a}\to P_{\mathrm{sp},a}$ strongly.
Moreover, uniformly in the particle number,
\begin{equation}\label{spec:eq:projection-H1}
	\norm{A^{1/2}P\zeta}\le C_\phi\norm{A^{1/2}\zeta}.
\end{equation}
\end{lemma}
\begin{proof}
For a bounded compactly supported profile $\zeta(R,z,Y)$,
partial contraction of $S_{N,a}^*\zeta$ against $\phi(x_i)$ is
\[
	N^{-3/2}\int
	\overline{\phi(x_j+z/N)}
	\zeta(x_j+z/(2N),z,Y)\d z.
\]
It has $L^2_{x_j,Y}$ norm at most $C_\zeta N^{-3/2}$.
Thus the dilated rank-one projection in coordinate $i$ tends
strongly to zero on a dense set, and hence everywhere by
contractivity. The same holds for $j$. Spectator projections are
unchanged by $S_{N,a}$ and commute with these two projections.
This proves the strong limit, retaining the spectator projections.

For a Cartesian derivative $\partial_\alpha$, integration by parts gives
\begin{align*}
	[\partial_\alpha,q]
	&=-|\partial_\alpha\phi\rangle\langle\phi|
	-|\phi\rangle\langle\partial_\alpha\phi|,\\
	[-\Delta,q]
	&=|\phi\rangle\langle-\Delta\phi|
	-|-\Delta\phi\rangle\langle\phi|.
\end{align*}
Their norms are at most $2\|\partial_\alpha\phi\|_2$ and
$2\|\Delta\phi\|_2$, respectively. In $[\nabla_i,P_p]$,
all factors except $q_i$ remain contractions. Consequently,
\[
	\sum_{i=1}^p\norm{\nabla_iP_p\zeta}^2
	\le 2\sum_{i=1}^p\norm{\nabla_i\zeta}^2+C_\phi p\norm\zeta^2.
\]
Adding the number term and summing sectors proves
\eqref{spec:eq:projection-H1}. For $\zeta\in H^2$, the rank-one
formula gives, distributionally,
\[
	-\Delta_iP_p\zeta
	=P_p(-\Delta_i\zeta)+\Bigl(\prod_{k\ne i}q_k\Bigr)[-\Delta_i,q_i]\zeta.
\]
The right-hand side is in $L^2$. Summing over $i$ proves
$P_pH^2\subset H^2$ and $\|[A_p,P_p]\|\le C_\phi p$.
\end{proof}

\begin{lemma}
\label{spec:lem:projected-pair}
Let $\psi_n$ be smooth, compactly supported and in
$q^{\otimes n}L_s^2$, with $n\ge1$. Use $p=n+1$, $g_{N,a}$ and
$a_a$ from \eqref{spec:eq:pair-profile-input}. Then for every
$0\le F\in C_c((0,\infty))$,
\begin{equation}\label{spec:eq:projected-pair-spectrum}
	\left\|Q_{N,P,p}^{1/2}\sqrt{F(Q_{N,P,p}/N^2)}\,P_pg_{N,a}
	-S_{N,a}^*\big(a_a\otimes\sqrt{F(L)}\,\xi_v\big)\right\|
	\longrightarrow0.
\end{equation}
The vectors in $P_pL^2$ are embedded in the full tensor space
in this formula.
\end{lemma}
\begin{proof}
We first record estimates for the removed part
$r_{N,a}=(1-P_p)g_{N,a}$. All spectator coordinates of $g_{N,a}$
are already in the range of their $q_k$. Thus
$r_{N,a}=(1-q_iq_j)g_{N,a}$.
Write $d_N=m^2/\sqrt N$. The scattering tail gives
$\norm{k_N}_1\le C d_N$. For a smooth pair amplitude $b$, set
\[
	h_N(y,Y)=\int k_N(x-y)\overline{\phi(x)}b(x,y,Y)\d x.
\]
Distributional integration by parts gives, for every multi-index $\alpha$,
\[
	\partial_y^\alpha h_N
	=\int k_N(x-y)(\partial_x+\partial_y)^\alpha
	[\overline\phi(x)b(x,y,Y)]\d x.
\]
Derivatives in $Y$ act on $b$. Thus every derivative through
a fixed order $s$ is bounded by $C_{\psi,s}d_N$.
The support in $(y,Y)$ is fixed and compact.
Reinserting $\phi(x)$ gives the removed projection in coordinate $i$.
The contraction in $j$ has the same bound. Use
$1-q_iq_j=(1-q_i)+(1-q_j)-(1-q_i)(1-q_j)$
and the boundedness of the second projection on $H^2$ to obtain
\begin{equation}\label{spec:eq:projected-removed}
	\norm{r_{N,a}}_{H^2}\le C_\psi d_N,\qquad
	\norm{Q_{N,p}^{1/2}r_{N,a}}\le C_\psi d_N,\qquad
	\norm{Q_{N,p}r_{N,a}}\le C_\psi d_N\sqrt N.
\end{equation}
The middle bound follows from \eqref{spec:eq:energy-comparison} at
fixed $p$. For the last one, the kinetic part is controlled in
$H^2$; each squared potential has integral
$N\norm v_2^2$, and the pointwise bound and compact support of
$r_{N,a}$ give an $L^2$ norm at most $C_\psi d_N\sqrt N$.

Use the scaled operators $\mathcal A_{N,a},\mathcal B_{N,a}$
and $W_{N,a}$ from \eqref{spec:eq:pair-scaled-energy}, and abbreviate
$E_N=P_{N,a}$, $u_N=N S_{N,a}g_{N,a}$. In addition to
\eqref{spec:eq:pair-energy-vector}, we have the operator-level limit
\begin{equation}\label{spec:eq:pair-operator-vector}
	\mathcal A_{N,a}u_N\longrightarrow a_a\otimes v
	\quad\hbox{in }L^2.
\end{equation}
To verify this limit, write
$u_N=w_{Nm}(z)b_N(R,z,Y)$, where $b_N$ is the smooth amplitude
in \eqref{spec:eq:pair-profile-input} at relative coordinate $z/N$.
It tends locally to $a_a$; its relative derivatives of orders
one and two are bounded by $C_\psi/N$ and $C_\psi/N^2$.
On its support in the relative variable, all center and
spectator supports and derivative bounds are uniform.
For $R$ large the scattering tail gives
\[
	Lw_R=v+e_R,\qquad \norm{e_R}_2\le CR^{-3/2},
	\qquad \norm{\nabla w_R}_2\le C,\qquad \norm{w_R}_2\le C\sqrt R.
\]
Indeed $e_R=-4\nabla w\cdot\nabla\rho(\cdot/R)
-2w\Delta\rho(\cdot/R)$, with the derivatives here including
their scaling factors, is supported in an annulus of size $R$.
The product rule therefore shows that
$\mathcal B_{N,a}u_N-a_a\otimes v$ tends to zero: its errors
are bounded by
$C_\psi((Nm)^{-3/2}+N^{-1}+N^{-2}\sqrt{Nm})$.
The unmatched term $W_{N,a}u_N$ tends to zero by
\eqref{spec:eq:unmatched-spectral}. This proves
\eqref{spec:eq:pair-operator-vector}.

Define the block-diagonal operator
\begin{align}
	\mathcal D_N
	&=E_N\mathcal A_{N,a}E_N+
	(1-E_N)\mathcal A_{N,a}(1-E_N)\notag\\
	&=\mathcal A_{N,a}-\mathcal O_N,\qquad
	\mathcal O_N=E_N\mathcal A_{N,a}(1-E_N)
	+(1-E_N)\mathcal A_{N,a}E_N.
	\label{spec:eq:microscopic-block}
\end{align}
The off-diagonal kinetic part is bounded in norm by
$C_\phi p/N^2$, by Lemma~\ref{spec:lem:dilated-projection}.
The scaled potentials are bounded uniformly for fixed $p$.
Thus $\mathcal O_N$ is a uniformly bounded self-adjoint
operator. Subtracting it gives a self-adjoint, nonnegative
$\mathcal D_N$ on $D(\mathcal A_{N,a})$, reducing $E_N$.
Its $E_N$ block is unitarily equivalent to $N^{-2}Q_{N,P,p}$.
The operator domains in this step can be checked before scaling.
For fixed $N,p$, the interaction is bounded and
$D(Q_{N,p})=H^2$. Since $\phi\in H^2$, each rank-one condensate
projection preserves $H^2$, and $[-\Delta,q]$ is bounded.
Thus $P_p$ preserves $D(Q_{N,p})$. The off-diagonal kinetic
operator is bounded, as is the off-diagonal interaction for fixed
$N,p$. Removing their sum gives a self-adjoint block-diagonal
operator on this same domain. Its quadratic form is the sum of
the two restricted forms, which identifies its $P_p$ block with
the operator associated with the restricted form. The scaled
argument above applies this construction to the restricted operator.

Moreover $\mathcal O_N\to0$ strongly. The unmatched potentials
tend strongly to zero. For the matching multiplier $v(z)$,
use $E_N\to P_{\mathrm{sp},a}$ and
$[P_{\mathrm{sp},a},v(z)]=0$; its limiting off-diagonal part
is zero. The kinetic part tends to zero in norm.
Consequently
\[
	Z_N:=\mathcal D_N-\mathcal B_{N,a}
	=W_{N,a}-\mathcal O_N
	\longrightarrow0\quad\hbox{strongly},\qquad
	\sup_N\norm{Z_N}<\infty.
\]
We next estimate the action on the growing vector $u_N$.
The profile $a_a\otimes v$ is in the range of
$P_{\mathrm{sp},a}$. Equation~\eqref{spec:eq:pair-operator-vector}
therefore gives
$(1-E_N)\mathcal A_{N,a}u_N\to0$. Also
\[
	\norm{\mathcal A_{N,a}(1-E_N)u_N}
	=N^{-1}\norm{Q_{N,p}r_{N,a}}
	\le C_\psi d_N/\sqrt N\longrightarrow0.
\]
The identity
$\mathcal O_Nu_N=(1-E_N)\mathcal A_{N,a}u_N+
(2E_N-1)\mathcal A_{N,a}(1-E_N)u_N$
implies $\mathcal O_Nu_N\to0$, and hence $Z_Nu_N\to0$.
Finally,
\[
	\norm{\mathcal D_N^{1/2}u_N}^2
	=Q_{N,p}[P_pg_{N,a}]+Q_{N,p}[r_{N,a}]
	\le C_\psi
\]
by \eqref{spec:eq:projected-removed} and the bounded energy of
$g_{N,a}$.

All hypotheses of Lemma~\ref{spec:lem:growing-vector} now hold for
$\mathcal D_N=\mathcal B_{N,a}+Z_N$ and $u_N$.
With $g(\lambda)=\sqrt{\lambda F(\lambda)}$, it follows that
\[
	g(\mathcal D_N)u_N\longrightarrow
	a_a\otimes\sqrt{F(L)}\,\xi_v.
\]
Removing $(1-E_N)u_N$ changes this vector by at most
\[
	\norm F_\infty^{1/2}
	\norm{\mathcal D_N^{1/2}(1-E_N)u_N}
	\le C_\psi\norm F_\infty^{1/2}d_N.
\]
The remaining $E_N$ block, transformed back by $S_{N,a}^*$,
is precisely the first vector in
\eqref{spec:eq:projected-pair-spectrum}. This proves the lemma.
\end{proof}

\begin{theorem}
\label{spec:thm:projected-profile}
Let $v\in C_c^\infty(\mathbb R^3)$ be nonzero, nonnegative and
radial, and let $\phi\in C_c^\infty(\mathbb R^3)$ be normalized.
Fix $0<\alpha<1$ and a real radial $0\le\rho\in C_c^\infty$
equal to one near zero. Set $m=N^{-\alpha}$,
$\epsilon=(m/N)^{1/2}$ and $k_N(r)=\sqrt N\,w(Nr)\rho(r/m)$.
Let $0\le\chi_N\le1$ be real, supported in $n\le L_N<\infty$,
and use the phase $C_N$ in \eqref{spec:eq:phase}.
Let $P=\Gamma(1-|\phi\rangle\langle\phi|)$ and let $\psi\in P\cF$
have smooth compactly supported sectors, with $\psi_n=0$ for $n>n_0$.
Assume $\chi_N(n)=1$ for $0\le n\le n_0+1$ eventually and
$\epsilon(L_N+2)^{3/2}\le D$ uniformly in $N$.
Use the compressed phase $PC_NP$ and define
\[
	\nu^P_{N,\psi}(B)
	=\ip{\mathscr H_{N,P}^{1/2}\psi}
	{\mathbf1_B(\mathscr H_{N,P}/N^2)
	\mathscr H_{N,P}^{1/2}\psi}.
\]
Then, against every bounded continuous function,
\begin{equation}\label{spec:eq:projected-profile}
	\nu^P_{N,\psi}\ \Longrightarrow\
	A_P[\psi]\delta_0+2\ip\psi{T_\phi\psi}\mu_v.
\end{equation}
The last coefficient denotes the expectation of the original
multiplication form on $\psi\in P\cF$; it need not preserve that
subspace as an operator.

If the plateau eventually contains every fixed finite set of
particle numbers, the same limit holds on
$\mathscr X_P=P\cF\cap\mathscr X$, uniformly on its compact
subsets for each fixed bounded continuous test.
The conclusions of Corollary~\ref{spec:cor:general-consequences}
hold with $\mathscr H_{N,P}$ in place of $\mathscr H_N$.
\end{theorem}
\begin{proof}
By \eqref{spec:eq:projection-H1}, $P$ is uniformly bounded on the
weighted first-energy spaces in Lemma~\ref{spec:lem:phase}, and it
is a contraction on their weighted $L^2$ counterparts.
The estimates of that lemma hold for $PC_NP$ with constants
independent of the cutoff. The same two-insertion and
triangular-series argument gives, on the fixed smooth input,
\[
	U_{N,P}\psi=\psi+PG_N\psi+\widetilde r_N,\qquad
	\norm{Q_{N,P}^{1/2}\widetilde r_N}\le C_{\psi,D}\epsilon.
\]
Here the projected lowering term is $O_\psi(\epsilon)$ in
first energy, by its differentiated sector formula and
\eqref{spec:eq:projection-H1}.

For the raising term, Lemma~\ref{spec:lem:projected-pair} and the
orthogonality Lemma~\ref{spec:lem:pair-orthogonality} give the
positive-energy spectral profile exactly as in Step 3 of
Theorem~\ref{spec:thm:finite-profile}; the pair count
\eqref{spec:eq:pair-count-spectral} is unchanged.
Equation~\eqref{spec:eq:projected-removed}, summed over the finitely
many pairs and sectors, also gives
\[
	\norm{Q_N^{1/2}(1-P)G_N\psi}\le C_\psi d_N.
\]
Thus the total energy of $PG_N\psi$ has the same limit
$2D_v\ip\psi{T_\phi\psi}$. The positive measures have the
same limiting mass as their profile on $(0,\infty)$, so the
argument in Step 4 extends convergence to every bounded
continuous test and excludes loss at zero or infinity.

On the fixed smooth input, $Q_{N,P}\psi=PQ_N\psi$ on each
sector: $P$ preserves $H^2$, and the block-diagonal construction
\eqref{spec:eq:microscopic-block} identifies the restricted operator.
Consequently $\norm{Q_{N,P}\psi}^2\le C_\psi N$, and its form
on $\psi$ equals $Q_N[\psi]$.
Its energy-weighted measure therefore concentrates at zero
with mass $A_P[\psi]$.
The mixed form with $PG_N\psi$ tends to zero: its kinetic part
is $O_\psi(\epsilon)$ by moving $A$ onto the smooth input,
and its potential part is $O_\psi(N^{-1/2})$ by
Cauchy--Schwarz. Step 5 and energy-norm continuity now prove
\eqref{spec:eq:projected-profile}.

For the domain extension, the projected sector bounds give
\[
	\norm{Q_{N,P}^{1/2}U_{N,P}\psi}
	\le C_{\phi,D}\norm\psi_{\mathscr X}
	\qquad(\psi\in\mathscr X_P),
\]
by the same summation as \eqref{spec:eq:domain-phase-map}.
The projection is bounded on $\mathscr X$ by
\eqref{spec:eq:projection-H1} and commutation with $M$.
Applying it to smooth finite-sector approximants proves
density of the projected core in $\mathscr X_P$; its images
are still smooth and compactly supported since $\phi$ is.
Uniform form continuity, density, and finite nets prove the
remaining convergence statements. The spectral and time
consequences use the same bounded test functions and positive
spectral window as Corollary~\ref{spec:cor:general-consequences}.
\end{proof}

\subsection{Neumann kernels and the dressed GP phase}
\label{spec:sec:dressed-profile}

The following comparison uses the first-energy limit of the
constant Neumann extension and its equation in $L^2$.

\begin{lemma}
\label{spec:lem:neumann-profile}
Let $f_R^{\mathrm{Neu}}$ be the positive radial ground state of
$-\Delta+v/2$ on the ball of radius $R$, with Neumann boundary
condition and boundary value one. Extend it by one and put
$w_R^{\mathrm{Neu}}=1-f_R^{\mathrm{Neu}}$. Then
\begin{equation}\label{spec:eq:neumann-energy-vector}
	L^{1/2}w_R^{\mathrm{Neu}}\longrightarrow\xi_v,\qquad
	Lw_R^{\mathrm{Neu}}\longrightarrow v
	\quad\hbox{in }L^2,\qquad R\longrightarrow\infty.
\end{equation}
The second error is $O(R^{-3/2})$.
For $R_N=Nm$ and
$k_N^{\mathrm{Neu}}(r)=\sqrt N\,w_{R_N}^{\mathrm{Neu}}(Nr)$,
all qualitative single-pair arguments in
Sections~\ref{spec:sec:finite-sectors}--\ref{spec:sec:projected-profile}
remain valid with the same measure $\mu_v$.
\end{lemma}
\begin{proof}
The radial Neumann ground state has nonzero boundary value.
Otherwise its value and derivative would both vanish there,
and uniqueness of the radial equation would force it to be zero.
Normalize this value to one. The constant trial function in the
Neumann Rayleigh quotient gives,
once the ball contains the support of $v$,
\[
	0\le\lambda_R\le\frac{1}{2|B_R|}\int v=O(R^{-3}).
\]
The bounds of \cite[Lemma A.1(ii)--(iii)]{COS}, expressed in the
ball radius $R$, give $0\le f_R^{\mathrm{Neu}}\le1$ and
\[
	|w_R^{\mathrm{Neu}}(z)|\le\frac C{1+|z|},\qquad
	|\nabla w_R^{\mathrm{Neu}}(z)|\le\frac C{1+|z|^2}.
\]
These bounds concern the ball radius $R$; hence they apply also
when $R=Nm$ tends to infinity. Local elliptic estimates and
$\lambda_R\to0$ show that every sequence has a subsequence
converging locally in $H^1$ to a solution of the zero-energy
scattering equation. The bound by $C/(1+|z|)$ identifies its
value at infinity as one. Uniqueness follows by multiplying
the difference of two such solutions by itself and integrating
by parts: outside the potential its radial decaying difference
is a multiple of $1/|z|$, so the boundary term vanishes.
Thus the whole family converges locally to $f$.

The uniform gradient tail has squared integral at most $C/T$
outside radius $T$. Local convergence and then $T\to\infty$
give $\nabla w_R^{\mathrm{Neu}}\to\nabla w$ in $L^2$.
The potential part converges as well, since $v$ is compactly
supported. Comparing with $w\rho(\cdot/S)$ in the energy
seminorm, and then taking $R,S\to\infty$, identifies the
first limit in \eqref{spec:eq:neumann-energy-vector} with the
vector of Lemma~\ref{spec:lem:scattering-vector}.

Both the function and its normal derivative agree across the
Neumann boundary. Its constant extension is in $H^2_{\rm loc}$,
and its equation has no surface delta term. Distributionally,
\[
	Lw_R^{\mathrm{Neu}}
	=v-2\lambda_R f_R^{\mathrm{Neu}}\mathbf1_{\{|z|\le R\}}.
\]
The error has $L^2$ norm at most $C R^{-3/2}$.

Scaling the displayed bounds gives
\begin{equation}\label{spec:eq:neumann-kernel-bounds}
	\norm{k_N^{\mathrm{Neu}}}_2\le C\epsilon,\quad
	\norm{\nabla k_N^{\mathrm{Neu}}}_2\le C,\quad
	\norm{k_N^{\mathrm{Neu}}}_1\le C m^2/\sqrt N,\quad
	\norm{\nabla k_N^{\mathrm{Neu}}}_1\le C m/\sqrt N.
\end{equation}
Its support is $|r|\le m$. The single-pair energy vector
therefore converges by the same separated-amplitude argument,
using \eqref{spec:eq:neumann-energy-vector}. Its matching energy
density has mass tending to $D_v$. The unmatched-potential
estimate and pair orthogonality are unchanged. The operator
vector needed in the projected proof follows from the second
limit in \eqref{spec:eq:neumann-energy-vector} and the same product
rule. Finally, contraction against a smooth condensate transfers
all derivatives to the smooth factors, so the estimates
\eqref{spec:eq:projected-removed} hold using the $L^1$ bound in
\eqref{spec:eq:neumann-kernel-bounds}. A removed pair term contains an integral of the form
\[
	R_N(y,Y)=\int\overline{\phi(x)}k_N^{\mathrm{Neu}}(x-y)b(x,y,Y)\d x
\]
with fixed smooth amplitude $b$. In a $y$ derivative on the
relative kernel, use $\partial_y k(x-y)=-\partial_x k(x-y)$
distributionally and integrate by parts in $x$. Repeating this
identity transfers every such derivative to $\overline\phi b$.
Derivatives in the other coordinates already fall on smooth
factors. Each derivative through any fixed order is therefore
bounded by $C_b\|k_N^{\mathrm{Neu}}\|_1$, with fixed compact
support. In particular this proves the $H^2$ bound for the removed
term. The first-energy comparison and the squared-potential
integral $N\|v\|_2^2$ give the other two estimates in
\eqref{spec:eq:projected-removed}.
For varying dressed kernels this argument is applied only after
the fixed smooth finite-rank approximation below; its error is
controlled separately in the energy form norm.
\end{proof}

For the next theorem let $\kappa_N$ denote either $k_N$ or
$k_N^{\mathrm{Neu}}$. Introduce $p_N,s_N\in H^1(\mathbb R^6)$ such that
\begin{equation}\label{spec:eq:dressed-assumptions}
	p_N\to p,\quad s_N\to s\quad\hbox{in }L^2(\mathbb R^6),\qquad
	t_N:=\epsilon(\norm{\nabla p_N}_2+\norm{\nabla s_N}_2)\to0.
\end{equation}
The gradients include both variables. The limiting kernels
belong to the stated Hilbert--Schmidt class. Our coefficient
convention is
\begin{align}
	B_{N,x}
	&=a_x+\int p_N(x,z)a_z\d z+\int s_N(x,z)a_z^*\d z,\notag\\
	G_N^{\mathrm d}
	&=\int\kappa_N(x-y)\phi(y)a_x^*a_y^*B_{N,x}\d x\d y,\qquad
	C_N^{\mathrm d}=\chi_N(G_N^{\mathrm d}-(G_N^{\mathrm d})^*)\chi_N.
	\label{spec:eq:dressed-phase}
\end{align}
Define $B_x$ by the same formula with $p,s$. For finite-number
vectors, and by closure on $D(M^{1/2})$, set
\begin{equation}\label{spec:eq:dressed-contact}
	c_B[\psi]=\int|\phi(x)|^2\norm{B_x\psi}_{\cF}^2\d x.
\end{equation}
This is a continuous nonnegative quadratic form on that domain:
the annihilation and creation bounds give
$c_B[\psi]\le C_{\phi,p,s}\norm{M^{1/2}\psi}^2$.

\begin{theorem}
\label{spec:thm:dressed-profile}
Let $v\in C_c^\infty(\mathbb R^3)$ be nonzero, nonnegative and
radial, and let $\phi\in C_c^\infty(\mathbb R^3)$ be normalized.
Fix $0<\alpha<1$, set $m=N^{-\alpha}$ and
$\epsilon=(m/N)^{1/2}$, and let $\kappa_N$ be either $k_N$
or $k_N^{\mathrm{Neu}}$. In the first case the cutoff
$0\le\rho\in C_c^\infty$ is real, radial and equal to one near zero.
Let $p_N,s_N\in H^1(\mathbb R^6)$ satisfy
\[
	p_N\to p,\qquad s_N\to s\quad\text{in }L^2(\mathbb R^6),
	\qquad \epsilon(\|\nabla p_N\|_2+\|\nabla s_N\|_2)\to0.
\]
The gradients act in both variables. Let $0\le\chi_N\le1$ be real
and supported in $n\le L_N<\infty$.
Let $\Pi=I$ or $P=\Gamma(1-|\phi\rangle\langle\phi|)$.
In the case $\Pi=P$, require in addition that
$s_N(x,\cdot)\in qL^2$ for almost every $x$; the limit $s$
then has the same property. Use the restricted operators
$A_\Pi,Q_{N,\Pi}$ and let
\[
	U_N^{\mathrm d}=\exp(\Pi C_N^{\mathrm d}\Pi),\qquad
	\mathscr H_N^{\mathrm d}
	=(U_N^{\mathrm d})^*Q_{N,\Pi}U_N^{\mathrm d}
	\quad\hbox{on }\Pi\cF.
\]
Suppose $\psi\in\Pi\cF$ has smooth compactly supported sectors
and vanishes above $n_0$. Require $\chi_N=1$ on $0\le n\le n_0+3$
eventually and
\[
	\epsilon(L_N+6)^{3/2}\le D.
\]
Then the energy-weighted spectral measure of
$\mathscr H_N^{\mathrm d}$ on $\psi$ converges, against all
bounded continuous tests, to
\begin{equation}\label{spec:eq:dressed-profile}
	A_\Pi[\psi]\delta_0+2c_B[\psi]\mu_v.
\end{equation}
If the plateau eventually contains every fixed finite set, the
same conclusion holds on $\mathscr X_\Pi=\Pi\cF\cap\mathscr X$
and is uniform on its compact subsets for each fixed test.

The conclusion is unchanged if the phase uses the one-sided
cutoff $\Theta_NG_N^{\mathrm d}-(G_N^{\mathrm d})^*\Theta_N$,
with the same plateau and support conditions, or if all fields
in its normally ordered monomials are replaced by
\[
	b_x=\sqrt{1-\cN/N}\,a_x,\qquad
	b_x^*=a_x^*\sqrt{1-\cN/N}
\]
on $\cF^{\le N}$, provided $L_N\le N-3$ eventually.
A common change of sign of the phase also leaves the limit
unchanged. For this finite-space variant the fixed-input claim
is read for sufficiently large $N$; the domain extension may be
read either after extending the phase by zero, or on
$\mathbf1_{\{\cN\le N\}}\psi$.
\end{theorem}
\begin{proof}
\textit{Uniform phase bounds and its first insertion.}
For a kernel $h(x,z)$ let
$J_{N,h}(x,y,z)=\kappa_N(x-y)\phi(y)h(x,z)$.
Fubini and the kernel bounds give
\begin{equation}\label{spec:eq:dressed-kernel-estimates}
	\norm{J_{N,h}}_2\le C\epsilon\norm h_2,\qquad
	\norm{\nabla J_{N,h}}_2
	\le C(\norm h_2+\epsilon\norm{\nabla h}_2).
\end{equation}
The norms are over all three variables. For example, in a
derivative on $\kappa_N$, integrate $y$ first and use
$\norm\phi_\infty\norm{\nabla\kappa_N}_2$.
A derivative on $h$ instead has the factor
$\norm\phi_\infty\norm{\kappa_N}_2$.

A normally ordered monomial with $r$ creations, $s$ annihilations
and an $L^2$ kernel has sector norm at most
\[
	\left(\frac{n!}{(n-s)!}
	\frac{(n-s+r)!}{(n-s)!}\right)^{1/2}\norm J_2
	\qquad(n\ge s).
\]
This follows by partial contraction and symmetrization.
For $r+s=3$, it is bounded by $C(n+1)^{3/2}\norm J_2$.
For a full gradient, differentiating an output kernel uses
$\norm{\nabla J}_2$; differentiating the remaining coordinates
uses $\norm J_2\norm{\cK^{1/2}\psi_n}$.
Using \eqref{spec:eq:dressed-kernel-estimates} for $p_N,s_N$ and their
adjoints, together with Lemma~\ref{spec:lem:phase} for the identity
term, proves \eqref{spec:eq:phase-L2}--\eqref{spec:eq:phase-H1} for
$C_N^{\mathrm d}$ with uniform constants. The changes of
particle number are now $\pm1,\pm3$; bounded shifts of the
weights change only the constants. Projection is allowed by
\eqref{spec:eq:projection-H1}.

On a fixed smooth finite-sector input, the adjoint has first
energy $O_\psi(\epsilon+t_N)$. The identity term uses the
integration by parts in Lemma~\ref{spec:lem:phase}. In the adjoint
of the $p_N$ term, the only created coordinate is $z$.
The kernel derivative gives the bound $\epsilon\norm{\nabla_z p_N}_2$.
The adjoint of the $s_N$ term only annihilates, so output
derivatives act on the original vector. Spectator derivatives
are bounded by $O_\psi(\epsilon)$ in both cases.
The two-insertion and triangular-series proof, with bandwidth
three, consequently gives
\begin{equation}\label{spec:eq:dressed-reduction}
	U_N^{\mathrm d}\psi=\psi+\Pi G_N^{\mathrm d}\psi+r_N,\qquad
	\norm{Q_{N,\Pi}^{1/2}r_N}\le C_\psi(\epsilon+t_N)\to0.
\end{equation}
Indeed, two insertions have weighted $L^2$ norm $O_\psi(\epsilon^2)$
and weighted first energy $O_\psi(\epsilon)$; the cutoff condition controls
all further insertions as before.
To bound the remaining series uniformly, put
$\delta_N=\epsilon(L_N+6)^{3/2}$ and use the norms
$\|\cdot\|_{1,0}$ and $\|\cdot\|_{1,1}$ of
Lemma~\ref{spec:lem:phase}. On the cutoff support the dressed
sector bounds imply
\[
	\|C_N^{\mathrm d}\zeta\|_{1,0}\le K\delta_N\|\zeta\|_{1,0},\qquad \|C_N^{\mathrm d}\zeta\|_{1,1}\le K\delta_N\bigl(\|\zeta\|_{1,1}+\epsilon^{-1}\|\zeta\|_{1,0}\bigr).
\]
The constant $K$ is uniform because the Hilbert--Schmidt norms
of $p_N,s_N$ and the quantities $t_N$ are bounded.
Starting with the two-insertion bounds gives, for $j\ge2$,
\[
	\|(C_N^{\mathrm d})^j\psi\|_{1,1}\le C_\psi(j-1)\epsilon(K\delta_N)^{j-2}.
\]
Thus the terms of order at least two have total first energy
$O_\psi(\epsilon)$ after division by $j!$ and summation.
The omitted adjoint in the first insertion contributes
$O_\psi(\epsilon+t_N)$. Projection changes only the fixed
constants in these estimates. This proves the stated remainder
for the full exponential.

\textit{Fixed smooth coefficients and coherent pair counting.}
First replace $p_N,s_N$ by fixed finite sums $p^{(j)},s^{(j)}$
of products of smooth compactly supported functions. In the
projected case take $s^{(j)}(x,\cdot)\in qL^2$.
Put $b_{\ell-2}^{(j)}(x,Y)=(B_x^{(j)}\psi)_{\ell-2}(Y)$.
The exact output-sector formula for the raising operator is
\begin{equation}\label{spec:eq:dressed-sector}
	(G_N^{(j)}\psi)_\ell(X_\ell)
	=\frac1{\sqrt{\ell(\ell-1)}}\sum_{i\ne k}
	\kappa_N(x_i-x_k)\phi(x_k)
	b_{\ell-2}^{(j)}(x_i,X_\ell\setminus\{x_i,x_k\}),
	\qquad\ell\ge2.
\end{equation}
It follows directly from the formula for two creation fields.
All amplitudes here are fixed smooth compactly supported
functions. In the projected case their spectator variables
already lie in $q$, by the condition on $s^{(j)}$.

For an unordered pair in sector $\ell$, its diagonal amplitude
is
\[
	\frac{2\phi(R)b_{\ell-2}^{(j)}(R,Y)}
	{\sqrt{\ell(\ell-1)}}.
\]
The proved single-pair spectral argument, or its projected
version, applies to these amplitudes. Lemma~\ref{spec:lem:neumann-profile}
allows either choice of $\kappa_N$.
The $\ell(\ell-1)/2$ pairs have total squared amplitude
$2\int|\phi(R)|^2\norm{b_{\ell-2}^{(j)}(R)}^2\d R$.
Different output sectors are orthogonal. To evaluate each sector,
write $b_{\ell-2}^{(j)}=b_\ell^-+b_\ell^+$, where
\begin{align*}
	b_\ell^-(x)&=\left(a_x+\int p^{(j)}(x,z)a_z\d z\right)\psi_{\ell-1},\\
	b_\ell^+(x)&=\int s^{(j)}(x,z)a_z^*\psi_{\ell-3}\d z.
\end{align*}
Here sectors with negative indices are zero. Both vectors lie
in the $(\ell-2)$-particle sector, so
\begin{equation}\label{spec:eq:coherent-contact-sectors}
	c_{B^{(j)}}[\psi]=\sum_{\ell\ge2}\int|\phi(x)|^2
	\left(\|b_\ell^-(x)\|^2+\|b_\ell^+(x)\|^2
	+2\operatorname{Re}\langle b_\ell^-(x),b_\ell^+(x)\rangle\right)\d x.
\end{equation}
The pair count therefore gives $2c_{B^{(j)}}[\psi]$, including
interference between inputs whose particle numbers differ by two.

The energy mass argument in Step 4 of
Theorem~\ref{spec:thm:finite-profile} applies to the same smooth
amplitudes. The initial smooth vector and its mixed spectral
form are handled by Step 5, since the raising vector has
bounded first energy and ordinary norm $O_\psi(\epsilon)$.
This proves the limit for $\psi+\Pi G_N^{(j)}\psi$.

\textit{Energy-norm approximation of the varying kernels.}
Choose these smooth finite-rank kernels with
$p^{(j)}\to p$, $s^{(j)}\to s$ in $L^2$.
For the projected assertion, applying $q$ to the second
variable of the approximants preserves smooth compact support
and gives the required approximation to $s$.
Equation~\eqref{spec:eq:dressed-kernel-estimates}, the sector estimates
and \eqref{spec:eq:energy-comparison} on the finitely many output
sectors give
\begin{align}
	&\norm{Q_{N,\Pi}^{1/2}\Pi
	(G_N^{\mathrm d}-G_N^{(j)})\psi}\notag\\
	&\quad\le C_\psi\big[
	\norm{p_N-p^{(j)}}_2+\norm{s_N-s^{(j)}}_2
	+t_N+\epsilon(\norm{\nabla p^{(j)}}_2+
	\norm{\nabla s^{(j)}}_2)\big].
	\label{spec:eq:dressed-approximation}
\end{align}
The constants in this bound do not depend on $j$.
The first energies of both sides' states have bounded limsup
for each fixed $j$, with a bound independent of $j$ when the
approximants are bounded in $L^2$.
Thus \eqref{spec:eq:form-continuity} permits $N\to\infty$ first,
then $j\to\infty$, for any bounded continuous spectral test.
The contact coefficients converge, since
\[
	\left(\int|\phi(x)|^2
	\norm{(B_x^{(j)}-B_x)\psi}^2\d x\right)^{1/2}
	\le C_\phi(\norm{p^{(j)}-p}_2+\norm{s^{(j)}-s}_2)
	\norm{M^{1/2}\psi}.
\]
Together with \eqref{spec:eq:dressed-reduction}, this proves
\eqref{spec:eq:dressed-profile}.

\textit{Domains, cutoffs and generalized fields.}
The uniform sector estimates also prove
\[
	\norm{Q_{N,\Pi}^{1/2}U_N^{\mathrm d}\psi}
	\le C_D\norm\psi_{\mathscr X}.
\]
The limit form is continuous on $\mathscr X_\Pi$, by the bound
following \eqref{spec:eq:dressed-contact}. Density and finite nets
give the stated domain and compact-uniform extensions.
For a one-sided cutoff the same estimates hold because number
multipliers commute with spatial derivatives; the first
insertion on the specified input has multiplier one.
Each normally ordered product of generalized fields is the
corresponding canonical monomial times a scalar product of
number square roots. More precisely, a word with $r$ creation
and $s$ annihilation fields, $r+s=3$, acting on sector $n\ge s$
has scalar coefficient
\[
	\beta_{r,s,N}(n)=\prod_{j=1}^{s}\sqrt{1-\frac{n-j}{N}}\prod_{j=0}^{r-1}\sqrt{1-\frac{n-s+j}{N}}.
\]
This follows by applying the fields from right to left, using
$b=f_N(\cN)a$, $b^*=a^*f_N(\cN)$ and
$f_N(k)=\sqrt{1-k/N}$. Each factor is evaluated at the number
present immediately before a creation or after an annihilation.
Retain the blocks satisfying $0\le n-s+r\le N$.
The one-sided cutoff multiplies a raising block by
$\Theta_N(n+r-s)$ and its adjoint block by the corresponding
input factor. Its support keeps every active intermediate sector
below $N$ when $L_N+3\le N$.
Consequently $0\le\beta_{r,s,N}(n)\le1$, so the full phase
and the entire higher-order remainder satisfy the same bounds.
On the finitely many sectors in the first insertion,
$\beta_{r,s,N}(n)=1+O_\psi(N^{-1})$ and the cutoff equals one.
That insertion changes by $O_\psi(N^{-1})$ in energy norm.
This proves the generalized-field assertion. Particle-number
truncation converges in $\mathscr X$, so the uniform mapping
bound also justifies its stated domain interpretation.
Changing the overall sign reverses the raising contribution;
its squared coefficient is unchanged and its mixed form with
the initial smooth part still tends to zero.
\end{proof}

\begin{proof}[Proof of Theorem~\ref{spec:thm:actual-COS-profile}]
We use the following specific external inputs from
\cite[Lemmas A.3--A.4]{COS} at $t=0$:
\[
	\norm{\gamma_N-I}_2+\norm{\sigma_N}_2\le C,\qquad
	\norm{\nabla(\gamma_N-I)}_2+\norm{\nabla\sigma_N}_2
	\le C\sqrt N,
\]
and Hilbert--Schmidt convergence of $\gamma_N-I,\sigma_N$ to
$\gamma_\infty-I,\sigma_\infty$.
The latter limits are given by \eqref{spec:eq:actual-eta-limit};
the series converge in Hilbert--Schmidt norm for their
nonidentity terms. Since $\epsilon=N^{-3/4}$,
the derivative quantity $t_N$ is $O(N^{-1/4})$ and tends to
zero. This verifies \eqref{spec:eq:dressed-assumptions}.

The kernel of $\gamma_N$ is Hermitian and that of $\sigma_N$
is symmetric. Consequently their column-field expression in
\eqref{spec:eq:actual-COS-profile} is exactly the coordinate
convention \eqref{spec:eq:dressed-phase} with
$p_N=\gamma_N-I$, $s_N=\sigma_N$.
The range condition on $s_N(x,\cdot)$ follows from
$\eta_N\in(q\otimes q)L^2$ and its defining series.

In \cite[Eq.~(3.1)]{COS}, the relative coefficient of the cubic
phase after division by $\sqrt N$ is
$-\sqrt N\,w_{Nm}^{\mathrm{Neu}}(N(x-y))\phi(y)$.
The minus sign is common to the raising terms.
Lemma~\ref{spec:lem:neumann-profile} supplies its microscopic limit.
The number cutoff in that paper is one below
$N^{1/2}/2+10$ and zero above $N^{1/2}-10$, for large $N$.
Thus its plateau includes every fixed finite sector, its
support is below $N-3$, and
$\epsilon(L_N+6)^{3/2}$ is uniformly bounded.
The projected fields are the compression of the normally
ordered expressions; the generalized number factors are
exactly those treated in Theorem~\ref{spec:thm:dressed-profile}.
All hypotheses of that theorem are now verified. Its contact
form is the coefficient in \eqref{spec:eq:actual-COS-profile}.
The fixed-domain and compact-set assertions follow from the
corresponding assertions of Theorem~\ref{spec:thm:dressed-profile}.
Finally, Corollary~\ref{spec:cor:dressed-inverse}, with
$\epsilon=N^{-3/4}$, gives the stated $O(N^{-1/2})$ inverse bound.
\end{proof}

For all the dressed results, the rare-window and microscopic-time
formulae of Corollary~\ref{spec:cor:general-consequences} hold with
$c_\psi$ replaced by $c_B[\psi]$ and with the appropriate
conjugated operator. If $c_B[\psi]>0$, all moments of order
$p>1$ have the stated $N^{2p-2}$ lower bound in the extended
spectral sense. Moreover,
\[
	c_B[\Omega]=\int|\phi(x)|^2\norm{s(x,\cdot)}_2^2\d x:
\]
the creation component can give a nonzero microscopic energy
even on vacuum.

\subsection{Microscopic spectrum along the exact evolution}
\label{sec:evolving-spectrum}

We now allow the condensate in the spectral calculation to
depend on $N$. This removes the compact-support restriction
at positive time. The spectral operator retains the actual
condensate projection throughout the comparison.

First suppose that normalized condensates satisfy
\begin{equation}\label{eq:spectral-moving-background}
	\phi_N\longrightarrow\phi\quad\hbox{in }H^4(\mathbb R^3).
\end{equation}
Put $q_N=1-|\phi_N\rangle\langle\phi_N|$, $q=1-|\phi\rangle\langle\phi|$,
$P_N=\Gamma(q_N)$ and $P=\Gamma(q)$.
The direct rotation $R_NqR_N^*=q_N$ is defined by
\eqref{eq:evolved-direct-rotation}, and $\mathcal R_N=\Gamma(R_N)$.
We use $A_N=K_{P_N}+M$, $A=K_P+M$ and $Q_N=Q_{N,P_N}$
in this subsection. For fixed particle number, a subscript
$p$ denotes the corresponding sector on the full tensor space.
Write $Q_{N,p}^{\rm full}$ for the uncompressed sector
operator; $Q_{N,p}$ denotes its restriction to $P_{N,p}$.
As before, $m=N^{-1/2}$, $\epsilon=N^{-3/4}$ and
$\kappa_N=k_N^{\mathrm{Neu}}$.

\begin{lemma}
\label{lem:moving-projected-pair}
Fix $p\ge2$ and a pair $a=\{i,j\}$. Let
\[
	b_N(X_p)=\sum_{\ell=1}^J\beta_\ell(x_i,x_j)\prod_{k\notin a}q_Nf_{\ell,k}(x_k),\qquad
	g_{N,a}=\kappa_N(x_i-x_j)b_N(X_p),
\]
where $\beta_\ell\in C_c^\infty(\mathbb R^6)$ and
$f_{\ell,k}\in C_c^\infty(\mathbb R^3)$ are fixed.
Define $b$ by replacing $q_N$ with $q$, and put
$a_a(R,Y)=b(R,R,Y)$. For $0\le F\in C_c((0,\infty))$,
\begin{equation}\label{eq:moving-pair-spectrum}
	\left\|Q_{N,p}^{1/2}\sqrt{F(Q_{N,p}/N^2)}\,P_{N,p}g_{N,a}-S_{N,a}^*(a_a\otimes\sqrt{F(L)}\xi_v)\right\|\longrightarrow0.
\end{equation}
Moreover,
\[
	Q_{N,p}[P_{N,p}g_{N,a}]\longrightarrow D_v\|a_a\|^2.
\]
For finite sums over distinct pairs, the mixed spectral and
total-energy terms tend to zero.
\end{lemma}
\begin{proof}
The identity $q_Nf=f-\phi_N\langle\phi_N,f\rangle$ gives
$\|q_Nf-qf\|_{H^4}\le C_{\phi,f}\|\phi_N-\phi\|_{H^4}\longrightarrow0$.
Thus the spectator factors are uniformly bounded in $H^4$,
and hence in $L^2$, $L^\infty$ and $H^2$.
The relative-coordinate estimates use these bounds.

Let $E_N=S_{N,a}P_{N,p}S_{N,a}^*$ and let $P_{{\rm sp},a}$
be the product of the limiting $q$ projections on spectators.
Here $P_{N,p}=q_N^{\otimes p}$ on the full tensor space;
all dilated operators act on the fixed space $L^2(R,z,Y)$.
Then $E_N\to P_{{\rm sp},a}$ strongly. Indeed, the contraction
in a pair coordinate used in Lemma~\ref{spec:lem:dilated-projection}
is bounded by $C_\zeta N^{-3/2}\|\phi_N\|_\infty$ on bounded
compactly supported profiles. This bound is uniform in $N$.
Spectator projections converge in operator norm. Both $E_N$
and $P_{{\rm sp},a}$ are orthogonal projections, of norm at
most one. For any $h\in L^2$ and bounded compactly supported
$h_j\to h$,
\[
	\|(E_N-P_{{\rm sp},a})h\|\le2\|h-h_j\|+\|(E_N-P_{{\rm sp},a})h_j\|.
\]
Taking $N\to\infty$ and then $j\to\infty$ proves the
assertion for every $h$.

Write $\mathcal A_{N,a}$ and $\mathcal B_{N,a}$ for the
uncompressed scaled operators in \eqref{spec:eq:pair-scaled-energy},
and put $u_N=NS_{N,a}g_{N,a}$. The Neumann equation and the
product rule give
\begin{equation}\label{eq:moving-pair-vectors}
	\mathcal B_{N,a}^{1/2}u_N\longrightarrow a_a\otimes\xi_v,\qquad
	\mathcal A_{N,a}u_N\longrightarrow a_a\otimes v.
\end{equation}
Here the pair amplitudes are fixed smooth functions of two
coordinates. Their relative derivatives after scaling introduce factors
$N^{-1}$ and $N^{-2}$. The spectator factors converge in
$H^2$. Thus the product estimates proving
\eqref{spec:eq:pair-energy-vector} and
\eqref{spec:eq:pair-operator-vector} apply with uniform
constants and limiting amplitude $a_a$.
For the unmatched potential, integration in its relative
coordinate gives
\[
	\|V_{N,c}^{1/2}g_{N,a}\|^2\le C N^{-1}\|\kappa_N\|_2^2\longrightarrow0\qquad(c\ne a).
\]
There are two possible pair configurations. Write a spectator
factor as $f_N=q_Nf$. If the potential joins a pair coordinate
$x$ to this spectator, then
\[
	\int N^2v(N(x-y))|f_N(y)|^2\d y\le N^{-1}\|v\|_1\|f_N\|_\infty^2.
\]
If it joins two spectators, the corresponding integral is
bounded by
$N^{-1}\|v\|_1\|f_N\|_\infty^2\|h_N\|_2^2$.
All unused spectator coordinates contribute their $L^2$
norms; the pair amplitude is bounded with fixed support.
Integrating its relative coordinate contributes
$\|\kappa_N\|_2^2$. Finite sums use
$|\sum_{\ell=1}^J z_\ell|^2\le J\sum_\ell|z_\ell|^2$,
so the constants depend only on fixed $p,J$ and the stated
uniform bounds. For the scaled unmatched operator $W_N$,
$\|W_Nu_N\|^2\le\|W_N\|\langle u_N,W_Nu_N\rangle\longrightarrow0$.

Only the two pair projections can remove part of $g_{N,a}$.
Set $r_{N,a}=(1-P_{N,p})g_{N,a}$ and
$d_N=\|\kappa_N\|_1=O(N^{-3/2})$.
In a contraction against $\phi_N$, transfer derivatives on
$\kappa_N(x_i-x_j)$ to $\overline{\phi_N}\beta_\ell$.
Explicitly, for one pair amplitude put
\[
	h_N(y)=\int\kappa_N(x-y)\overline{\phi_N(x)}\beta_\ell(x,y)\d x.
\]
For every multi-index $|\alpha|\le2$, distributional
integration by parts gives
\[
	\partial_y^\alpha h_N(y)=\int\kappa_N(x-y)(\partial_x+\partial_y)^\alpha
	[\overline{\phi_N(x)}\beta_\ell(x,y)]\d x.
\]
Thus $\|\partial^\alpha h_N\|_\infty\le C_\ell d_N$.
The $y$ support is contained in the fixed projection of
$\operatorname{supp}\beta_\ell$, so $\|h_N\|_{H^2}\le C_\ell d_N$.
The $H^4$ bound of $\phi_N$ controls its derivatives through
order two in $L^\infty$; reinsertion of $\phi_N$ is bounded
in $H^2$. Tensor products with the spectator factors give
\[
	\|r_{N,a}\|_{H^2}+\|(Q_{N,p}^{\rm full})^{1/2}r_{N,a}\|\le C d_N.
\]
The uncompressed potential is bounded by $C_pN^2$, so
\[
	N^{-1}\|Q_{N,p}^{\rm full}r_{N,a}\|\le C_p(N^{-1}+N)d_N\longrightarrow0.
\]
For the two removed coordinates use
$1-q_{N,i}q_{N,j}=p_{N,i}+p_{N,j}-p_{N,i}p_{N,j}$,
where $p_{N,k}=1-q_{N,k}$.
Each single contraction was just estimated; the second
projection is uniformly bounded on $H^2$.
For the operator estimate, the kinetic part of $Q_{N,p}^{\rm full}r_{N,a}$ is bounded
by $C\|r_{N,a}\|_{H^2}$, and the potential part by
$C_pN^2\|r_{N,a}\|$. In particular $r_{N,a}$ lies in its
operator domain.

Construct the block operator $\mathcal D_N$ as in
\eqref{spec:eq:microscopic-block}, now using $E_N$.
The off-diagonal kinetic part has norm at most $C_p/N^2$,
because $\sup_N\|\phi_N\|_{H^2}<\infty$.
The scaled potentials are bounded. Their off-diagonal part
converges strongly to zero: the unmatched terms do so before
compression, and the matching $v(z)$ commutes with
$P_{{\rm sp},a}$. Thus
$Z_N=\mathcal D_N-\mathcal B_{N,a}$ is a uniformly bounded
self-adjoint operator and tends strongly to zero.
Removing the bounded off-diagonal part of $\mathcal A_{N,a}$
gives a self-adjoint operator on the same domain. Its form
is the sum of the two restricted nonnegative forms, so
$\mathcal D_N\ge0$.
On the full coordinate space the comparison operator is
\[
	\mathcal B_{N,a}=L_z+N^{-2}T,\qquad
	T=-\tfrac12\Delta_R-\sum_{k\notin a}\Delta_{x_k}+p+1.
\]
$L_z$ and $T$ act on different variables. Joint spectral
calculus and dominated convergence give
$(\mathcal B_{N,a}+1)^{-1}\to(L_z+1)^{-1}$ strongly.
Neither $T$ nor this convergence depends on the condensate.
The resolvent identity then gives the same strong resolvent
limit for $\mathcal D_N$, since $Z_N\to0$ strongly and
both positive resolvents are uniformly bounded.
The limit $a_a\otimes v$ in \eqref{eq:moving-pair-vectors}
lies in $P_{{\rm sp},a}$. The removed-vector estimate gives
\begin{align*}
	\|(1-E_N)\mathcal A_{N,a}u_N\|
	&\le\|\mathcal A_{N,a}u_N-a_a\otimes v\|+\|(1-E_N)(a_a\otimes v)\|\longrightarrow0,\\
	\|\mathcal A_{N,a}(1-E_N)u_N\|
	&=N^{-1}\|Q_{N,p}^{\rm full}r_{N,a}\|\longrightarrow0.
\end{align*}
The equality uses $(1-E_N)u_N=NS_{N,a}r_{N,a}$.
For the off-diagonal operator $\mathcal O_N$,
\[
	\mathcal O_Nu_N=(1-E_N)\mathcal A_{N,a}u_N+(2E_N-1)\mathcal A_{N,a}(1-E_N)u_N.
\]
Together with $W_Nu_N\to0$, this proves $Z_Nu_N\to0$.
The form bound follows from
$\mathcal D_N[u_N]=Q_{N,p}^{\rm full}[P_{N,p}g_{N,a}]+Q_{N,p}^{\rm full}[r_{N,a}]$,
the uniform $H^1$ bound for projection and
$Q_{N,p}^{\rm full}\le C_p(\cK_p+p+1)$ as forms.
All domains in this construction remain $H^2$ at fixed
$N,p$, since $q_N$ preserves $H^2$.
Lemma~\ref{spec:lem:growing-vector} proves
\eqref{eq:moving-pair-spectrum}.

For the total energy, the matching density
$2|\nabla\kappa_N|^2+N^2v(N\cdot)|\kappa_N|^2$
has support of radius $m$ and mass tending to $D_v$.
Its amplitude converges on the diagonal, in the spectator
$L^2$ norm. More explicitly, with $r=x_i-x_j$,
\[
	\|b_N(R+r/2,R-r/2,Y)-a_a(R,Y)\|_{L^2_{R,Y}}\le C|r|+o(1).
\]
The $o(1)$ is independent of $|r|\le m$ and follows from
the $H^2$ convergence of the finitely many spectator factors.
All these norms are bounded. Writing the matching density
as $J_N(r)$, the difference between
$\int J_N(r)\|b_N(R+r/2,R-r/2,Y)\|_{L^2_{R,Y}}^2\d r$
and $\|a_a\|^2\int J_N$ is therefore $O(m)+o(1)$.
Amplitude derivatives contribute $O(\epsilon)$,
and unmatched potentials tend to zero by the estimate above.
The removed part has vanishing first energy. This proves the
total-energy assertion. Distinct-pair spectral terms vanish
by Lemma~\ref{spec:lem:pair-orthogonality}.
For disjoint pairs, a derivative in one coordinate cannot
hit both relative kernels. At least one factor then has an
amplitude derivative, and Cauchy--Schwarz gives $O(\epsilon)$.
For pairs sharing one coordinate, the only remaining term
has a derivative on both kernels. Their two relative
coordinates are independent; the common center lies in a
fixed bounded set because of the two pair amplitudes.
All factors in these coordinates are bounded in $L^\infty$;
the unused coordinates contribute products of $L^2$ norms.
The bound is
$C\|\nabla\kappa_N\|_1^2=O(m^2/N)$.
Terms with the same pair have already been combined in
$b_N$. For any potential pair $c$
and distinct $a,b$, at least one of $c\ne a,c\ne b$ holds.
Thus at least one factor is unmatched, including when the
pairs share a coordinate. Cauchy--Schwarz and the unmatched
bound give $O(\epsilon/\sqrt N)$ for that cross term.
Removing $r_{N,a}$ changes all these energy pairings by
$o(1)$. This completes the proof.
\end{proof}

\begin{theorem}
\label{thm:moving-dressed-spectrum}
Let $v\in C_c^\infty(\mathbb R^3)$ be nonzero, nonnegative and radial.
Let $\phi_N,\phi$ be normalized, with $\phi_N\to\phi$ in
$H^4(\mathbb R^3)$. Put $q_N=1-|\phi_N\rangle\langle\phi_N|$,
$q=1-|\phi\rangle\langle\phi|$, $P_N=\Gamma(q_N)$ and $P=\Gamma(q)$.
Set $m=N^{-1/2}$, $\epsilon=N^{-3/4}$ and
$\kappa_N=k_N^{\mathrm{Neu}}$.
Let $p_N,s_N\in H^1(\mathbb R^6)$ satisfy
\[
	p_N\to p,\qquad s_N\to s\quad\text{in }L^2(\mathbb R^6),
	\qquad \epsilon(\|\nabla p_N\|_2+\|\nabla s_N\|_2)\to0.
\]
Assume $s_N(x,\cdot)\in q_NL^2$ for almost every $x$. Use $\phi_N$ in the dressed
phase \eqref{spec:eq:dressed-phase}, compress it with $P_N$,
and write its exponential as $V_N^{\rm d}$.
Let $0\le\chi_N\le1$ be real and supported in $n\le L_N$,
with $\epsilon(L_N+6)^{3/2}\le D$ uniformly in $N$.
Require $\chi_N=1$ on every fixed finite set of particle numbers
for all sufficiently large $N$.
For every fixed $\psi\in P\cF\cap D(M^2A_{\rm amb}^{1/2})$
and $g\in C_b([0,\infty))$,
\begin{align}\label{eq:moving-dressed-spectrum}
	&\langle Q_N^{1/2}V_N^{\rm d}\mathcal R_N\psi,
	g(Q_N/N^2)Q_N^{1/2}V_N^{\rm d}\mathcal R_N\psi\rangle\notag\\
	&\qquad\longrightarrow A[\psi]g(0)+2c_{\phi,B}[\psi]\int g\d\mu_v,
\end{align}
where $A_{\rm amb}=\cK+M$ and
\[
	c_{\phi,B}[\psi]=\int|\phi(x)|^2\|B_x\psi\|^2\d x.
\]
The field $B_x$ uses the limiting kernels $p,s$ in
\eqref{spec:eq:dressed-phase}. The one-sided cutoff,
generalized-field and common-sign variants in
Theorem~\ref{spec:thm:dressed-profile} have the same limit.
\end{theorem}
\begin{proof}
The background norms used in
\eqref{spec:eq:dressed-kernel-estimates} are
$\|\phi_N\|_\infty$ and $\|\nabla\phi_N\|_\infty$.
Projection uses a uniform $H^1$ bound. These follow from
\eqref{eq:spectral-moving-background}. The proof of the phase mapping
bound therefore gives
\begin{equation}\label{eq:moving-phase-map}
	\|Q_N^{1/2}V_N^{\rm d}\zeta\|\le C\|M^2A_{\rm amb}^{1/2}\zeta\|\qquad(\zeta\in P_N\cF).
\end{equation}
The direct-rotation argument after
\eqref{eq:evolved-rotation-bounds} gives
$\mathcal R_N^*A_N\mathcal R_N-A=o(1)M$ as forms.
Let $X=D(M^2A_{\rm amb}^{1/2})$, with norm
$\|\zeta\|_X=\|M^2A_{\rm amb}^{1/2}\zeta\|$.
Particle number commutes with each energy and with $\mathcal R_N$.
For $\zeta\in P_N\cF\cap X$, restriction of the energy form gives
\[
	\|M^2A_N^{1/2}\zeta\|^2
	=A_N[M^2\zeta]=A_{\rm amb}[M^2\zeta]=\|\zeta\|_X^2.
\]
Consequently, for $\psi\in P\cF\cap X$ and some $\eta_N\to0$,
\begin{equation}\label{eq:moving-weighted-rotation}
	\left|\|\mathcal R_N\psi\|_X^2-\|\psi\|_X^2\right|
	\le\eta_N\|M^{5/2}\psi\|^2\le\eta_N\|\psi\|_X^2.
\end{equation}
The last inequality uses $A_{\rm amb}\ge M$.
These identities hold first on finite-sector form vectors and
extend by closure. Thus $\mathcal R_N$ is uniformly bounded
from $P\cF\cap X$ to $P_N\cF\cap X$.
Its convergence to the identity in $X$ follows on finite
sectors from the one-particle $H^1$ convergence, and then by density.

It suffices first to take $\psi=P\chi$, where $\chi$ is a
finite linear combination of symmetrized products of
$C_c^\infty$ one-particle functions. Put
$\psi_N=\mathcal R_N\psi$.
These vectors occupy a fixed finite set of number sectors
and converge to $\psi$ in $H^2$ on those sectors.
Indeed the rank-two formula also gives
$\|R_N-1\|_{L^2\to H^4}=o(1)$ under
\eqref{eq:spectral-moving-background}.
The two-insertion and adjoint estimates in the proof of
Theorem~\ref{spec:thm:dressed-profile} then imply
\begin{equation}\label{eq:moving-phase-reduction}
	V_N^{\rm d}\psi_N=\psi_N+P_NG_N^{\rm d}\psi_N+r_N,\qquad
	\|Q_N^{1/2}r_N\|\le C_\psi(\epsilon+t_N)\longrightarrow0.
\end{equation}
Only uniform first derivatives of the input are used in
this reduction. The coefficient and input derivatives have
the same bounds as in the fixed-background proof.

We next approximate the raising part while keeping $P_N$
and $Q_N$ fixed. Choose compactly supported smooth
approximants to $\phi$ in $H^4$, to $p,s$ in Hilbert--Schmidt
norm, and to $\psi$ in the finite-sector $H^1$ norm.
Take the kernel and vector approximants to be finite sums
of products of one-particle functions. The raising terms
then have fixed smooth pair amplitudes and smooth spectator
factors. Apply $q_N$ to each spectator factor. This preserves their outer $P_N$ projection and places them
in the class of Lemma~\ref{lem:moving-projected-pair}.

To justify the successive approximations, let $H_N^{(j)}$
be the resulting raising vector after spectator projection,
with approximation error $\delta_j\to0$ in the norms just
listed. Fubini, the cubic sector bounds and
\eqref{spec:eq:dressed-kernel-estimates} give
\begin{equation}\label{eq:moving-source-approximation}
	\limsup_{N\to\infty}\|Q_N^{1/2}(P_NG_N^{\rm d}\psi_N-P_NH_N^{(j)})\|\le C_\psi\delta_j.
\end{equation}
Here the background error is bounded in $W^{1,\infty}$;
the input error is bounded in $H^1$.
The errors for the two dressed kernels are bounded by their
Hilbert--Schmidt differences, by $t_N$, and by $\epsilon$
times the first derivatives of the fixed approximants.
The last two contributions vanish as $N\to\infty$ at fixed
$j$. Spectator projection is uniformly bounded in $H^1$.
All approximants can be chosen bounded in the norms without
derivatives of $p,s$, so $C_\psi$ is independent of $j$.
In detail, before taking the limsup the right-hand side is
bounded by a constant times
\begin{gather*}
	\|\psi_N-\chi^{(j)}\|_{H^1}+\|\phi_N-\phi^{(j)}\|_{W^{1,\infty}}+
	\|p_N-p^{(j)}\|_2+\|s_N-s^{(j)}\|_2+t_N+C_j\epsilon.
\end{gather*}
Only $C_j$ contains the derivatives of the smooth kernel
approximants. The other constant depends on their bounded
Hilbert--Schmidt norms, the bounded input $H^1$ norms and
the bounded background $W^{1,\infty}$ norms. This estimate
uses the kernel norms before their finite tensor decomposition.
We first fix $j$, then let $N\to\infty$, and finally let
$j\to\infty$. This proves \eqref{eq:moving-source-approximation}
for the same spectral operator.

Apply Lemma~\ref{lem:moving-projected-pair} at fixed $j$.
In output sector $\ell$, the two ordered contributions for
each pair have diagonal amplitude
\[
	\frac{2\phi(R)}{\sqrt{\ell(\ell-1)}}
	\bigl(P_{{\rm sp},\ell-2}(B_R\psi)_{\ell-2}\bigr)(Y)
\]
in the limit $j\to\infty$.
Since $(1\otimes q_N)s_N=s_N$, the limiting creation kernel satisfies
\[
	\|(1\otimes(1-q))s\|_2
	\le\|s-s_N\|_2+\|q_N-q\|\,\|s_N\|_2\longrightarrow0.
\]
Here $\|q_N-q\|\le2\|\phi_N-\phi\|_2$.
Annihilation preserves the $q$ condition in every remaining coordinate
of $\psi\in P\cF$. The creation term inserts $s(R,\cdot)\in qL^2$.
Thus the displayed spectator projection fixes $(B_R\psi)_{\ell-2}$.
Its two source terms combine as in
\eqref{spec:eq:coherent-contact-sectors}.
The annihilation and creation estimates also give
\[
	\|\phi^{(j)}P B^{(j)}\chi^{(j)}-\phi P B\psi\|_{L^2_x(\cF)}\le C_\psi\delta_j.
\]
Thus the sum of the two source amplitudes converges before
its squared norm is taken, including its interference term.
Coherent pair counting, including interference between the
two source sectors, gives total coefficient $2c_{\phi,B}[\psi]$.
The same lemma gives convergence of the total raising energy.
For each fixed $j$, let $\nu_N^{(j)}$ be its positive
energy-weighted measure and $\nu^{(j)}$ its limiting measure
on $(0,\infty)$. Their total masses converge and
$\nu^{(j)}$ has no atom at zero. Given $\delta>0$, choose
$0\le\vartheta\le1$ in $C_c((0,\infty))$ with
$\nu^{(j)}(1-\vartheta)<\delta$. Then
$\limsup_N\nu_N^{(j)}(1-\vartheta)<\delta$.
For $g\in C_b([0,\infty))$, apply positive-window convergence
to $g\vartheta$ and bound the two complementary integrals
by $\|g\|_\infty\delta$. Let $\delta\downarrow0$.
This uses the total first energy. Equation
\eqref{eq:moving-source-approximation} and first-energy
continuity then allow $j\to\infty$.

The regular part has spectral mass only at zero in the limit.
Its uniform $H^2$ bound and the projected potential estimate give
\[
	\mathcal V_{N,P_N}[\psi_N]=O_\psi(N^{-1}),\qquad
	Q_N[\psi_N]\longrightarrow A[\psi].
\]
On these finitely many sectors the uncompressed multiplication
potential is bounded between zero and $C_\psi N^2$.
Its compression $\mathcal V_{N,P_N}$ has the same operator
bound. Consequently
$\|\mathcal V_{N,P_N}\psi_N\|^2\le C_\psi N^2\mathcal V_{N,P_N}[\psi_N]=O_\psi(N)$,
and $\|Q_N\psi_N\|^2=O_\psi(N)$.
This makes its first-energy mass above $\delta N^2$ tend
to zero for every $\delta>0$.
The kinetic mixed form with the raising vector tends to
zero by moving $A_N$ onto $\psi_N$ and using its
$O_\psi(\epsilon)$ ordinary norm. Its potential part is
$O_\psi(N^{-1/2})$ by Cauchy--Schwarz.
The spectral mixed form then vanishes as in Step 5 of
Theorem~\ref{spec:thm:finite-profile}.
Together with \eqref{eq:moving-phase-reduction}, this proves
\eqref{eq:moving-dressed-spectrum} on the stated core.

Projected finite sums of smooth product vectors are dense
in $P\cF\cap D(M^2A_{\rm amb}^{1/2})$.
The uniform bound \eqref{eq:moving-phase-map}, the bound
for $\mathcal R_N$ and continuity of the contact form
extend the conclusion by density.
Number cutoffs and the generalized fields introduce exactly
the sector multipliers already checked in
Theorem~\ref{spec:thm:dressed-profile}. On the fixed first
insertion they differ from one by $O_\psi(N^{-1})$;
all higher terms retain the same bounds. The limiting squared amplitudes are invariant under
a common change of sign.
\end{proof}

Return to the exact evolution and notation of
Section~\ref{sec:evolved-energy}. Define the limiting dressed
contact along its quadratic reference state by
\begin{equation}\label{eq:evolved-contact}
	c(t)=\int|\phi_{\infty,t}(x)|^2
	\|[a(\gamma_{\infty,t,x})+a^*(\sigma_{\infty,t,x})]\xi_\infty(t)\|^2\d x.
\end{equation}
The excitation vector after the cubic transformation is
$u_N(t)=e^{C_t}\xi_N(t)$.
Its excitation energy is $Q_N(t)$.

\begin{theorem}
\label{thm:exact-evolved-microscopic-spectrum}
For the prepared family of
Theorem~\ref{thm:evolved-first-energy-convergence}, every fixed
$t$ and $g\in C_b([0,\infty))$ satisfy
\begin{align}\label{eq:exact-evolved-microscopic-spectrum}
	&\langle Q_N(t)^{1/2}u_N(t),g(Q_N(t)/N^2)Q_N(t)^{1/2}u_N(t)\rangle\notag\\
	&\qquad\longrightarrow A_{\infty,t}[\xi_\infty(t)]g(0)+2c(t)\int g\d\mu_v.
\end{align}
Thus the positive microscopic measure is the same relative
scattering measure $\mu_v$, with a time-dependent coefficient.
The limit holds at each fixed macroscopic time.
\end{theorem}
\begin{proof}
Fix a compact interval containing zero and $t$.
Proposition A.2(iv) of \cite{COS} gives
$\phi_{N,t}\to\phi_{\infty,t}$ in $H^4$.
Lemmas A.3--A.4 of that reference give, for
$p_N=\gamma_{N,t}-1$ and $s_N=\sigma_{N,t}$,
\begin{gather*}
	\|p_N-p_{\infty,t}\|_2+\|s_N-s_{\infty,t}\|_2\le C_I N^{-1/2},\\
	\epsilon(\|\nabla p_N\|_2+\|\nabla s_N\|_2)\le C_I N^{-1/4}.
\end{gather*}
Both gradients include the two kernel variables, and the
creation kernel takes values in the actual $q_{N,t}$ space.
The actual phase has the sign, one-sided cutoff and
generalized fields allowed in
Theorem~\ref{thm:moving-dressed-spectrum}.
Lemma~\ref{lem:quadratic-weighted-operator-domain} places
$\xi_\infty(t)$ in its weighted energy domain.
That theorem therefore proves the claimed limit for
$e^{C_t}\mathcal R_N(t)\xi_\infty(t)$.

By Proposition~\ref{prop:full-graph} and
Theorem~\ref{thm:evolved-first-energy-convergence},
\[
	\|Q_N(t)^{1/2}(e^{i\theta_N(t)}u_N(t)-e^{C_t}\mathcal R_N(t)\xi_\infty(t))\|\longrightarrow0.
\]
The two first energies are bounded.
Equation~\eqref{spec:eq:form-continuity} transfers every bounded
spectral test on the same $Q_N(t)$, and the scalar phase
drops out of the quadratic form. This proves
\eqref{eq:exact-evolved-microscopic-spectrum}.
\end{proof}

\begin{corollary}
\label{cor:evolved-positive-microscopic-energy}
The function $c(t)$ is continuous and $c(0)>0$ for the
common prepared families. Hence there is a fixed $\tau_0>0$
such that $c(t)>0$ for $|t|<\tau_0$.
At each fixed time in this interval, and for every $\rho>1$,
\[
	\langle u_N(t),Q_N(t)^\rho u_N(t)\rangle\ge c_{\rho,t}N^{2\rho-2}
\]
for all sufficiently large $N$, with $c_{\rho,t}>0$.
The expectation is interpreted as an extended nonnegative
spectral moment. On every compact time interval the same
exact many-body states retain the uniform third depletion
and sixth collective-moment bounds of Theorem~\ref{thm:D}.
\end{corollary}
\begin{proof}
The map
$\psi\mapsto\phi_{\infty,t}(x)[a(\gamma_{\infty,t,x})+a^*(\sigma_{\infty,t,x})]\psi$
is bounded from $D(M^{1/2})$ to $L^2_x(\cF)$.
Its norm is controlled by the condensate supremum norm and
the two Hilbert--Schmidt norms. These coefficients vary
continuously in the same norms by \cite[Proposition A.2 and
Lemma A.4]{COS}.
The identity part of $\gamma$ uses
$\int\|a_x\psi\|^2\d x=\|\cN^{1/2}\psi\|^2$.
Lemma~\ref{lem:quadratic-weighted-operator-domain} gives
continuity of $\xi_\infty(t)$ in $D(M^{1/2})$.
Thus $c(t)$ is continuous. Its initial positivity is
Theorem~\ref{thm:common-preparation} or
Proposition~\ref{prop:weighted-reference-preparation}.

Choose a nonnegative nonzero test supported in a compact
positive-energy window with positive $\mu_v$ mass.
Equation~\eqref{eq:exact-evolved-microscopic-spectrum} gives
a positive limiting first-energy mass there. Write $\nu_{N,t}$
for that energy-weighted measure, and take the test $0\le h\le1$
supported in $[a,b]\subset(0,\infty)$. Spectral calculus gives
\[
	\langle u_N(t),Q_N(t)^\rho u_N(t)\rangle
	=N^{2\rho-2}\int\lambda^{\rho-1}\d\nu_{N,t}(\lambda)
	\ge N^{2\rho-2}a^{\rho-1}\nu_{N,t}(h).
\]
The last factor has a positive limit, which proves the bound.
The physical-moment bounds follow from Theorem~\ref{thm:D}
for the same unchanged initial family.
\end{proof}

\subsection{Strict positivity at finite kinetic energy}
\label{sec:contact-positive-kinetic}

Continuity alone gives the local conclusion above. The short-distance
structure of the limiting pair kernel gives a stronger result.
We use the kernel convention of \eqref{spec:eq:dressed-contact}:
\[
	B_x=a_x+\int p(x,z)a_z\d z+\int\sigma(x,z)a_z^*\d z,
	\qquad p=\gamma-1.
\]
All kernels in this subsection are the actual limiting Neumann
kernels.

\begin{proposition}
\label{prop:contact-positive-kinetic}
Let $\phi\in H^6(\mathbb R^3)$ be normalized, let $q=1-|\phi\rangle\langle\phi|$
and $P=\Gamma(q)$, and assume that the scattering length $a$ is positive.
For every nonzero $\psi\in P\cF\cap D((\cK+M)^{1/2})$,
\begin{equation}\label{eq:contact-positive-kinetic}
	c_{\phi,B}[\psi]=\int|\phi(x)|^2\|B_x\psi\|^2\d x>0.
\end{equation}
\end{proposition}
\begin{proof}
Write $w=w_{\infty,\ell}$ and
$\eta=(q\otimes q)[-w(x-y)\phi(x)\phi(y)]$.
The explicit function $w$ in \cite[(2.25)]{COS} satisfies
\[
	w(r)=a\left(\frac1{|r|}-\frac3{2\ell}+\frac{|r|^2}{2\ell^3}\right)
	\mathbf1_{\{|r|\le\ell\}}.
\]
Put $f=w*|\phi|^2$ and $d_0=\int|\phi|^2f$.
Expanding the two projections, with complex conjugates in their
inner products, gives
\[
	\eta(x,y)=-\phi(x)\phi(y)\{w(x-y)-f(x)-f(y)+d_0\}.
\]
Since $w\in L^1$ and $|\phi|^2\in H^6$, Young's inequality
gives $f\in H^6$. The projection correction is therefore in
$H^2(\mathbb R^6)$. By \cite[Lemma A.4(i)]{COS}, both
$p=\gamma-1$ and $r=\sigma-\eta$ belong to $H^2(\mathbb R^6)$.
Indeed, their Hilbert--Schmidt norms and the two partial
Laplacians are finite, which is the Fourier characterization of
this space. Consequently
\begin{equation}\label{eq:contact-collision-kernel}
	\sigma(x,y)=-\frac{a\phi(x)\phi(y)}{|x-y|}+R(x,y),
	\qquad R\in H^1_{\rm loc}(\{|x-y|<\ell\}).
\end{equation}
Away from $x=y$, the full kernel $\sigma$ is locally $H^1$.
This also holds across $|x-y|=\ell$: $w$ vanishes continuously
there and has a locally bounded first weak derivative away
from the origin.

Suppose that the contact vanishes. The number form bound for
$B_x$ makes this a well-defined equality in $L^2_x(\cF)$
after multiplication by $\phi$. Choose a ball $U$ on which
$|\phi|$ is bounded below by a positive constant. Then
$B_x\psi=0$ for almost every $x\in U$.
Choose any nonzero sector $\psi_m$, where $m\ge0$. Its contribution to output
sector $m+1$, with $Y=(y_1,\ldots,y_{m+1})$, gives
\begin{align}\label{eq:contact-sector-cancellation}
	0={}&\sqrt{m+2}\,\psi_{m+2}(x,Y)
	+\sqrt{m+2}\int p(x,z)\psi_{m+2}(z,Y)\d z\notag\\
	&+\frac1{\sqrt{m+1}}\sum_{i=1}^{m+1}
	\sigma(x,y_i)\psi_m(Y\setminus y_i).
\end{align}
This identity holds in $L^2(U\times\mathbb R^{3(m+1)})$.
Every sector of $\psi$ has finite kinetic energy, hence lies
in $H^1$. The first line is thus $H^1$ in $(x,Y)$.
For the integral term, an $x$ derivative acts on $p$ and a
$Y$ derivative acts on $\psi_{m+2}$. The Hilbert--Schmidt
bounds for $p$ and $\nabla_xp$, followed by Cauchy--Schwarz
in $z$, bound both derivatives in $L^2$.

We localize the second line near just one collision.
Fix $x_0\in U$. A sufficiently small ball $V$ about $x_0$
has closure in $U$ and diameter less than $\ell$.
For $m\ge1$ it can also be chosen so that
\[
	\int_{(\mathbb R^3\setminus\overline V)^m}|\psi_m(Y')|^2\d Y'>0.
\]
As $V$ shrinks to $x_0$, this integral tends to
$\|\psi_m\|^2$ by dominated convergence.
Choose a smooth compactly supported product cutoff $\zeta(Y')$
in this open set such that $\zeta\psi_m\ne0$.
Its support has positive distance from $\overline V$ in
each coordinate. For $m=0$, set $\zeta=1$ and regard
$\psi_0\ne0$ as a scalar. Finally take a nonzero real
$\chi\in C_c^\infty(V)$ and multiply
\eqref{eq:contact-sector-cancellation} by
$\chi(x)\chi(y_1)\zeta(y_2,\ldots,y_{m+1})$.

The $i\ge2$ summands are now $H^1$. Their kernel variables
$(x,y_i)$ stay away from their collision, and their state
variables are disjoint from those two variables. A derivative
therefore acts on one factor of an $H^1$ tensor product.
The same argument applies to the $R(x,y_1)$ part of the
$i=1$ summand, using \eqref{eq:contact-collision-kernel}.
All cutoffs are smooth and compactly supported.
The only remaining term, apart from a nonzero constant, is
\[
	\frac{\chi(x)\chi(y_1)\phi(x)\phi(y_1)}{|x-y_1|}
	(\zeta\psi_m)(y_2,\ldots,y_{m+1}).
\]
It is not in $H^1$. To see this, use relative coordinate
$r=x-y_1$ and center $R_0=(x+y_1)/2$. On a smaller set of
centers the numerator at $r=0$ is bounded away from zero.
Its derivatives are bounded, whereas the relative derivative
of $|r|^{-1}$ has size $|r|^{-2}$. The derivative of the
numerator contributes only $O(|r|^{-1})$. The squared
relative derivative consequently has divergent radial integral
$\int_0^\delta r^{-2}\d r$. Fubini's theorem preserves this
divergence after multiplication by the nonzero $L^2$ factor
$\zeta\psi_m$. This contradicts the $H^1$ regularity of
all the other localized terms in
\eqref{eq:contact-sector-cancellation}, and proves the claim.
\end{proof}

The energy dependence of this positivity follows from the high
spatial frequencies of the creation kernel.
Fourier analysis of a short-distance $1/r$ singularity also
underlies the contact relations for momentum distributions
\cite[Section III and Table I]{WernerCastin2012},
\cite[Eq.~(2)]{BKP2011}.
Here the Fourier cutoff acts in the $x$ coordinate of
$\phi(x)\sigma(x,z)$. The canonical commutation relation
then bounds the contribution of the creation field, while
the reference kinetic energy controls the annihilation field.

\begin{proposition}\label{prop:contact-energy-coercivity}
Let $\phi\in H^6(\mathbb R^3)$ be normalized, with
$q=1-|\phi\rangle\langle\phi|$ and $P=\Gamma(q)$.
Let the scattering length satisfy $a>0$, fix $\ell\in(0,1/2)$,
and define $B_x$ from the limiting Neumann kernel
$\eta_\infty$ in \eqref{spec:eq:actual-eta-limit}.
There is $\kappa_\phi>0$ such that
\begin{equation}\label{eq:contact-energy-lower}
	c_{\phi,B}[\psi]\,\|(\cK+M)^{1/2}\psi\|^2
	\ge\kappa_\phi\|\psi\|^4,
	\qquad \psi\in P\cF\cap D((\cK+M)^{1/2}).
\end{equation}
For the actual limiting kernels along a fixed condensate evolution,
one constant $\kappa_I>0$ works at every $t$ in a compact interval $I$.
In particular, normalized reference vectors with
$\|(\cK+M)^{1/2}\psi\|^2\le E$ satisfy
$c_{\phi_t,B_t}[\psi]\ge\kappa_I/E$.
\end{proposition}
\begin{proof}
We use the unitary Fourier transform
\[
	\widehat f(k)=(2\pi)^{-3/2}\int e^{-ik\cdot x}f(x)\d x.
\]
Let $P_\Lambda$ be the Fourier projection onto $|k|\ge\Lambda$
in the source variable $x$.
For $\mathfrak s(x,z)=\phi(x)\sigma(x,z)$, we first prove
\begin{equation}\label{eq:contact-source-tail}
	\lim_{\Lambda\to\infty}
	\Lambda\|P_\Lambda\mathfrak s\|_{L^2_{x,z}}^2
	=8a^2\|\phi^3\|_2^2>0.
\end{equation}
Choose a real radial $\rho\in C_c^\infty(\{|r|<\ell\})$
equal to one near zero, and put $h(r)=\rho(r)/|r|$.
The decomposition in the preceding proof gives
\[
	\mathfrak s(x,z)=-a\phi(z)^3h(x-z)+\mathcal R(x,z),
	\qquad \mathcal R,\nabla_x\mathcal R\in L^2(\mathbb R^6).
\]
Indeed, $w-ah$ is regular near zero and has compact support.
It is continuous across $|r|=\ell$ and has a bounded piecewise
first derivative, so it belongs to $H^1(\mathbb R^3)$.
Multiplication by $\phi(x)^2\phi(z)$ gives an $L^2_{x,z}$ kernel
with an $L^2$ first $x$ derivative: use $\phi\in W^{1,\infty}$
in $x$ and $\phi\in L^2$ in $z$.
The projection correction $\eta+w(x-z)\phi(x)\phi(z)$ and
$\sigma-\eta$ belong to $H^2(\mathbb R^6)$.
Multiplication by $\phi(x)\in W^{1,\infty}$ preserves their
first derivatives. The remaining difference is
$-a\phi(z)\bigl(\phi(x)^2-\phi(z)^2\bigr)h(x-z)$.
Its first $x$ derivative is bounded near the diagonal by
$C|\phi(z)|/|x-z|$, and its relative support is compact.
This proves the claimed $L^2$ bounds.

Distributionally, $-\Delta h=4\pi\delta_0+e$, with
$e\in C_c^\infty(\mathbb R^3)$ supported away from zero.
Consequently
\[
	\widehat h(k)=\frac{4\pi(2\pi)^{-3/2}}{|k|^2}
	+\frac{\widehat e(k)}{|k|^2},\qquad
	\lim_{\Lambda\to\infty}\Lambda\|P_\Lambda h\|_2^2=8.
\]
The Fourier transform in $x$ of $\phi(z)^3h(x-z)$ is
$\phi(z)^3e^{-ik\cdot z}\widehat h(k)$.
Also $\|P_\Lambda\mathcal R\|_2\le
\Lambda^{-1}\|\nabla_x\mathcal R\|_2$.
These facts prove \eqref{eq:contact-source-tail}.

Write $\phi(x)B_x\psi=\mathcal D_\psi(x)+\mathcal C_\psi(x)$, where
\[
	\mathcal D_\psi(x)=\phi(x)\left(a_x+\int p(x,z)a_z\d z\right)\psi,
	\qquad \mathcal C_\psi(x)=a^*(\mathfrak s(x,\cdot))\psi.
\]
The annihilation identities and the Hilbert--Schmidt bounds for
$p,\nabla_xp$ give
\[
	\|\nabla_x\mathcal D_\psi\|_{L^2_x(\cF)}
	\le C_\phi\|(\cK+M)^{1/2}\psi\|.
\]
Here $\int\|\nabla_xa_x\psi\|^2\d x=\|\cK^{1/2}\psi\|^2$
is an identity of closed quadratic forms.
For the integral term, Cauchy--Schwarz gives
\[
	\left\|\int g(x,z)a_z\psi\d z\right\|_{L^2_x(\cF)}
	\le\|g\|_2\|(M-1)^{1/2}\psi\|,
	\qquad g=p,\ \nabla_xp.
\]
Derivatives on $\phi$ use the same number form norm.
We use $a^*(f)=\int f(z)a_z^*\d z$, so $a^*(f)$ is linear
and $a(f)$ is antilinear in $f$.
For fixed $\psi\in D(M^{1/2})$, the map
$f\mapsto a^*(f)\psi$ is bounded from $L^2$ to $\cF$.
The projection in $x$ therefore gives
\[
	(P_\Lambda\mathcal C_\psi)(x)
	=a^*((P_\Lambda\mathfrak s)(x,\cdot))\psi.
\]
The identity $\|a^*(f)\psi\|^2=\|a(f)\psi\|^2+
\|f\|^2\|\psi\|^2$ now yields
\begin{align*}
	\|P_\Lambda\mathcal C_\psi\|_{L^2_x(\cF)}^2
	&=\int\|a((P_\Lambda\mathfrak s)(x,\cdot))\psi\|^2\d x
	+\|P_\Lambda\mathfrak s\|_2^2\|\psi\|^2\\
	&\ge\|P_\Lambda\mathfrak s\|_2^2\|\psi\|^2.
\end{align*}
Take first finite sums of sectors with smooth tensor factors
projected by $q$. They are dense in $P\cF$ for the form norm of
$\cK+M$, since $q$ preserves $H^1$.
The estimates above make $\mathcal D_\psi$ continuous in
$H^1_x(\cF)$ and $\mathcal C_\psi$ continuous in $L^2_x(\cF)$
under this approximation. They thus give the displayed identities
and bounds on the stated form domain.

Set $\mathcal E_\psi=\|(\cK+M)^{1/2}\psi\|^2$.
The triangle inequality, \eqref{eq:contact-source-tail} and the
bound for $\mathcal D_\psi$ give constants $A_\phi>0$, $D_\phi<\infty$
and $\Lambda_0\ge1$ such that, for $\Lambda\ge\Lambda_0$,
\[
	c_{\phi,B}[\psi]^{1/2}
	\ge A_\phi\Lambda^{-1/2}\|\psi\|
	-D_\phi\Lambda^{-1}\mathcal E_\psi^{1/2}.
\]
For $\psi\ne0$, choose
\[
	L_\phi\ge\max\{\Lambda_0,(2D_\phi/A_\phi)^2\},\qquad
	\Lambda=L_\phi\mathcal E_\psi/\|\psi\|^2.
\]
Since $\mathcal E_\psi\ge\|\psi\|^2$, this choice is allowed.
It gives \eqref{eq:contact-energy-lower} with
$\kappa_\phi=A_\phi^2/(4L_\phi)$; the zero vector is immediate.

For the condensate solution, \cite[Proposition A.2(i)--(ii)]{COS}
gives $\phi_t\in C(\mathbb R;H^1)$, $\|\phi_t\|_2=1$ and
$\sup_{t\in I}\|\phi_t\|_{H^6}<\infty$.
The $H^2$ norms of $p,\sigma-\eta$ are uniformly bounded on $I$
by \cite[Lemma A.4(i)]{COS}.
The remainder and annihilation bounds above are therefore uniform.
The embedding $H^1(\mathbb R^3)\subset L^6(\mathbb R^3)$ makes
$t\mapsto\|\phi_t^3\|_2=\|\phi_t\|_6^3$ continuous and positive,
so its minimum on $I$ is positive. The explicit Fourier estimate
for $h$ then permits common $A_I>0$, $D_I$ and $\Lambda_0$.
The same choice of $L_I$ proves the uniform assertion.
\end{proof}

\begin{corollary}
\label{cor:contact-all-finite-times}
For the common prepared families, the coefficient in
\eqref{eq:evolved-contact} satisfies
\[
	c(t)>0\quad(t\in\mathbb R),\qquad
	\min_{t\in I}c(t)>0
\]
on every compact interval $I$. Moreover, in
Proposition~\ref{prop:weighted-reference-preparation}, the condition
$c_B[M^{-1}\chi]>0$ is automatic for every nonzero reference
$\chi\in P\cF\cap D(A)\cap D(M^2)$.
\end{corollary}
\begin{proof}
The limiting condensate remains normalized and belongs to $H^6$
at each finite time \cite[Proposition A.2]{COS}.
Lemma~\ref{lem:quadratic-weighted-operator-domain} places
$\xi_\infty(t)$ in the kinetic form domain, and unitarity
keeps its norm equal to one. On a compact interval $I$, that lemma
also bounds
$E_I=\sup_{t\in I}\|(\cK+M)^{1/2}\xi_\infty(t)\|^2$.
Proposition~\ref{prop:contact-energy-coercivity} gives
$c(t)\ge\kappa_I/E_I>0$ throughout $I$.
For the reference statement, $M^{-1}$ commutes with $A$ and
is injective, so $M^{-1}\chi$ is a nonzero vector in the same
kinetic form domain. Apply
Proposition~\ref{prop:contact-positive-kinetic} at time zero.
\end{proof}

The lower bound $\kappa_I/E$ holds for every normalized reference
vector with $\|(\cK+M)^{1/2}\psi\|^2\le E$.

\subsection{Reference states with finite form energy}
\label{sec:form-energy-preparation}

The regularity needed for the reference can now be reduced.
The inverse-energy equation provides the operator regularity
of the limiting prepared vector. The reference itself only
needs a weighted kinetic form bound.
Use the common initial background of
Theorem~\ref{thm:common-preparation} and define
\begin{equation}\label{eq:form-reference-class}
	\mathcal Y=P\cF\cap D(M^2)\cap D(M^{1/2}A^{1/2}),
	\qquad A=K_P+M.
\end{equation}
The product is defined by joint spectral calculus, since
$A$ preserves particle number. For finite-excitation vectors
the last condition is precisely finite kinetic form energy.
Throughout this subsection, $Q_N$ includes its actual sector-form
extension above particle number $N$, as in
Section~\ref{sec:common-preparation}. In particular, the
form inequality $Q_N\le CMA$ also controls the full-space tails.

\begin{theorem}
\label{thm:form-reference-preparation}
Let $v\in C_c^\infty(\mathbb R^3)$ be nonzero, nonnegative and
radial. Fix an $N$-independent normalized
$\phi\in C_c^\infty(\mathbb R^3)$ and a sufficiently small fixed $\ell\in(0,1/2)$.
Let $q=1-|\phi\rangle\langle\phi|$, $P=\Gamma(q)$, and use
the Neumann pair kernel $\eta_N$ and its limit
$\eta_\infty$ in \eqref{spec:eq:actual-eta-limit}.
Let $C_N=\mathcal C_N^{\mathrm{COS}}$ be the phase
\eqref{spec:eq:actual-phase}, with $m=N^{-1/2}$.
Its real cutoff $0\le\Theta_N\le1$ equals one for
$n\le N^{1/2}/2+10$ and zero for $n\ge N^{1/2}-10$.
Extend $C_N$ by zero above sector $N$.
Use the centered initial energies
$\widehat F_{N,b}=\mathfrak J_N-\mathfrak J_N[\Omega]I+bM$
on $\Pi_NP\cF$ and $F_{\infty,b}=\mathcal J_{2,\infty}+bM$
on $P\cF$, where $\Pi_N=\mathbf1_{\{\cN\le N\}}$.
Choose a fixed $b$ so that, uniformly in $N$,
$c_bQ_N\le\widehat F_{N,b},F_{\rm ex}(0)\le C_bQ_N$
on the physical space and $c_bA\le F_{\infty,b}\le C_bA$.
Writing $H_b=K_P+bM$ and $S_\infty=F_{\infty,b}-H_b$, require
\[
	\|M^rS_\infty H_b^{-1}M^{-r}\|<\tfrac12,
	\qquad r=0,2.
\]
All these conditions hold for sufficiently large $b$, independently of $N$.
Let $\chi\in\mathcal Y=P\cF\cap D(M^2)\cap D(M^{1/2}A^{1/2})$
be normalized. The choice of $b$ is independent of $\chi$.
The estimates are uniform in $N$; their constants may depend
on $b$ and $\chi$.
The preparations in \eqref{eq:weighted-reference-preparation}
are defined for all sufficiently large $N$ and satisfy
\begin{gather}
	\|Q_N^{1/2}(z_N-z_\infty)\|\le C_{b,\chi}N^{-1/4},
	\qquad z_\infty\in D(M^2A),\label{eq:form-reference-convergence}\\
	c_B[z_\infty]>0,\qquad
	\sup_{N\ge N_0}\|F_{\rm ex}(0)^{3/2}z_N\|<\infty.
	\label{eq:form-reference-third-moment}
\end{gather}
The first-energy and microscopic spectral limits in
Theorem~\ref{thm:uniform-evolved-energy-spectrum}, the excitation-energy
lower bounds in Corollary~\ref{cor:uniform-positive-microscopic-moments},
and the exact physical-moment bounds of Theorem~\ref{thm:D}
hold for this same family. The reference and its preparation
are independent of the observation interval.
\end{theorem}
\begin{proof}
Set $H_b=K_P+bM$ and take
$b>2\max(C_0,C_2)$ in \eqref{eq:common-quadratic-weight},
also large enough for the uniform positive form comparisons.
More explicitly, the two bounded extensions satisfy
\[
	\|S_\infty H_b^{-1}\|_{\cF\to\cF}<\frac12,\qquad
	\|M^2S_\infty H_b^{-1}M^{-2}\|_{\cF\to\cF}<\frac12.
\]
The second norm is the operator norm on $D(M^2)$ equipped
with $\|M^2\cdot\|$. The extensions agree on their common
domain by finite-sector approximation and closedness of $M^2$,
as in the proof of Theorem~\ref{thm:common-preparation}.
The factorization
\[
	y=F_{\infty,b}^{-1}\chi
	=H_b^{-1}(1+S_\infty H_b^{-1})^{-1}\chi,
	\qquad \|M^2H_by\|\le2\|M^2\chi\|
\]
therefore gives $y\in D(M^2A)$. Indeed, $H_b$ and $A$
are functions of the strongly commuting positive operators
$K_P,M$, and $A\le C_bH_b$. Their joint spectral calculus
gives $\|M^2Ay\|\le C_b\|M^2H_by\|$, with the domain
inclusion obtained by monotone spectral truncation.
This step uses
$\chi\in D(M^2)$, with no kinetic derivative of $\chi$.
Since $F_{\infty,b}$ is positive and invertible, $y\ne0$.

Let $x_N=\Pi_N\chi$, $y_N=\widehat F_{N,b}^{-1}x_N$
and $v_N=\Pi_Ny$. For physical form vectors $u$ the source
terms agree, since $\langle u,x_N\rangle=\langle u,\chi\rangle$.
All following pairings are closed-form pairings.
On $\Pi_NP\cF$, the bounds $A\le Q_N\le CMA$ and
$M\le N+1$ identify their common form domain with
$D(A^{1/2})\cap\Pi_NP\cF$.
Thus
\[
	\widehat F_{N,b}[u,y_N-v_N]
	=(F_{\infty,b}-\widehat F_{N,b})[u,v_N]
	+F_{\infty,b}[u,(1-\Pi_N)y].
\]
The right-hand variable in the asymmetric comparison is $v_N$.
$\Pi_N$ commutes with $A,M$ and preserves $D(M^2A)$.
Lemma~\ref{lem:common-form-comparison} and $Q_N\le CMA$ give
\[
	\|MQ_N^{1/2}v_N\|\le C\|M^{3/2}A^{1/2}y\|,
	\qquad \|M^{5/2}v_N\|+\|Av_N\|\le C\|M^2Ay\|.
\]
For the first estimate, apply the form comparison to $Mv_N$:
\[
	\|MQ_N^{1/2}v_N\|^2=Q_N[Mv_N]
	\le C\|M^{3/2}A^{1/2}v_N\|^2.
\]
Both $v_N$ and $Mv_N$ belong to the $Q_N$ form domain
by that comparison. The equality uses the joint spectral
calculus of $Q_N,M$.
The remaining estimates follow from $A\ge M\ge1$ by joint
spectral calculus. The operator $H_b$ preserves number, so
$H_b[u,(1-\Pi_N)y]=0$.
Equation~\eqref{eq:common-quadratic-weight} with $r=-1/2$
bounds $M^{-1/2}S_\infty M^{-1/2}$. Consequently
\[
	|F_{\infty,b}[u,(1-\Pi_N)y]|
	\le C\|M^{1/2}u\|\|M^{1/2}(1-\Pi_N)y\|
	\le C(N+1)^{-1}\|Q_N^{1/2}u\|\|M^{3/2}y\|.
\]
Coercivity, with $u=y_N-v_N$, now yields an
$O_{b,\chi}(N^{-1/4})$ bound in the $Q_N^{1/2}$ norm.
The omitted tail is bounded by
\[
	\|Q_N^{1/2}(1-\Pi_N)y\|
	\le C\|M^{1/2}A^{1/2}(1-\Pi_N)y\|
	\le C(N+1)^{-3/2}\|M^2A^{1/2}y\|.
\]
The right-hand norm is finite since $A\ge1$ and
$y\in D(M^2A)$.
Consequently $\|Q_N^{1/2}(y_N-y)\|=O_{b,\chi}(N^{-1/4})$.

The new reference condition supplies the uniform source energy:
\begin{equation}\label{eq:form-reference-source-bound}
	Q_N[x_N]\le C\|M^{1/2}A^{1/2}x_N\|^2
	\le C\|M^{1/2}A^{1/2}\chi\|^2.
\end{equation}
Thus $x_N\in D(\widehat F_{N,b}^{1/2})$ and
$\widehat F_{N,b}[x_N]\le C_{b,\chi}$.
Writing $t_N=\|x_N\|\to1$, the form identity and
Cauchy--Schwarz give
\[
	t_N^4\le\widehat F_{N,b}[x_N]
	\langle x_N,\widehat F_{N,b}^{-1}x_N\rangle
	\le\widehat F_{N,b}[x_N]t_N\|y_N\|.
\]
Hence $\|y_N\|$ is bounded below for large $N$.
Since $Q_N\ge1$ and $Q_N[y]\le C\|M^{1/2}A^{1/2}y\|^2$,
normalization preserves the energy-norm convergence.
Spectral calculus also gives the exact identity
\[
	\|\widehat F_{N,b}^{3/2}z_N\|^2
	=\frac{\widehat F_{N,b}[x_N]}{\|y_N\|^2}\le C_{b,\chi}.
\]
At time zero $F_{\rm ex}(0)=\widehat F_{N,b}+\alpha_N I$,
with $|\alpha_N|\le C_\alpha$ uniformly.
Our choice of $b$ gives uniform positive lower bounds for
both energies. For their scalar spectral variable,
$0\le(\lambda+\alpha_N)^3\le4(\lambda^3+C_\alpha^3)$.
Functional calculus therefore gives
\[
	\|F_{\rm ex}(0)^{3/2}z_N\|^2
	\le4\|\widehat F_{N,b}^{3/2}z_N\|^2+4C_\alpha^3.
\]
Finally, $z_\infty=y/\|y\|$ lies in the kinetic form
domain and is nonzero. On $P\cF$ the ambient kinetic form
equals the compressed form $K_P$, so this also gives the
ambient form regularity required by
Proposition~\ref{prop:contact-positive-kinetic}. That proposition
therefore gives strict contact positivity for this fixed $b$.

The following estimates control the subsequent evolution.
On the excitation space, the kinetic off-diagonal block satisfies
$\|\cK u\|\le\|K_Pu\|+C\|Mu\|\le C\|Au\|$.
Indeed, $\phi\in H^2$ makes the one-particle off-diagonal
block bounded; second quantization bounds its norm by $CM$.
It preserves total particle number, so the same estimate
applies after multiplication by $M^2$.
As $A$ strongly commutes with $M$ and $A\ge M$,
$z_\infty\in D(M^2A)$ also gives $\|M^3z_\infty\|<\infty$.
Thus, for $\zeta_N=\Pi_Nz_\infty/\|\Pi_Nz_\infty\|$,
\[
	\sup_{N\ge N_0}
	\bigl(\|\cK\zeta_N\|^2+\|\cN^3\zeta_N\|^2\bigr)<\infty.
\]
This is the $\cK^2+\cN^6$ expectation bound in
\cite[Theorem 2.2]{COS}, evaluated on $\zeta_N\in D(\cK)$.
The weaker initial condition of its Theorem 2.3 follows as well.
The common smooth radial nonnegative compactly supported
potential lies in $L^3$, and the common smooth condensate
lies in $H^6$. These are the model assumptions of both
theorems; the Neumann radius is the same fixed small radius.
The physical comparison state
$\mathcal U_{N,0}^*e^{B_0}\zeta_N$ has excitation vector
$\zeta_N$ in the convention of \cite[Eq.~(2.21)]{COS}.
The number tail and the cubic block bound give
\[
	\|e^{C_0}z_N-\zeta_N\|
	\le\|z_N-\zeta_N\|+\|(e^{C_0}-1)\zeta_N\|
	=O_{b,\chi}(N^{-1/4}).
\]
Unitarity of the physical evolution therefore transfers the COS
comparison from $\zeta_N$ to the new prepared family, exactly
as in Proposition~\ref{prop:prepared-quadratic-comparison}.
Only the regular comparison vector enters the COS hypothesis.

We verify the remaining domain requirements explicitly.
\begin{enumerate}
\item Theorem~\ref{thm:D} requires the initial
$F_{\rm ex}(0)^{3/2}$ bound just proved. It propagates the
third correlated energy and the stated physical moments.
\item Lemma~\ref{lem:quadratic-weighted-operator-domain}
requires $z_\infty\in D(M^2A_{\rm amb})$, where
$A_{\rm amb}=\cK+M$. The weighted kinetic comparison above
verifies this domain. The lemma gives
$\xi_\infty(t)\in D(M^2A_{\rm amb})$ and continuity in that
graph norm on each compact interval.
\item Lemma~\ref{lem:evolved-form-comparison} requires
$D(M^2A_{\infty,t})$ regularity of its right-hand test vector.
Compression bounds this norm by the ambient norm in the
previous item. With the Hilbert comparison and the propagated
third-energy tail, the proof of
Theorem~\ref{thm:evolved-first-energy-convergence} applies
to the present $\xi_N,\xi_\infty$.
\item Theorem~\ref{thm:moving-dressed-spectrum} requires
$D(M^2A_{\rm amb}^{1/2})$ for its fixed reference vector.
The stronger graph bound in the second item supplies this
condition. Graph continuity also supplies it at the limiting
time of every convergent time sequence. These are the state
inputs in the proof of
Theorem~\ref{thm:uniform-evolved-energy-spectrum};
its remaining estimates concern only the model coefficients.
\item Proposition~\ref{prop:contact-positive-kinetic}
requires a nonzero vector in $D(A_{\rm amb}^{1/2})$.
Unitarity and the second item verify this at every finite time.
The proof of Corollary~\ref{cor:evolved-positive-microscopic-energy}
gives continuity of the contact from $D(M^{1/2})$ continuity
of the state and continuity of the condensate and pair kernels.
The same hypotheses hold here. Its positive minimum on a
compact interval and the uniform spectral limit give
Corollary~\ref{cor:uniform-positive-microscopic-moments}
for the new family.
\end{enumerate}
Thus each evolution and spectral estimate has its required
initial or reference domain under \eqref{eq:form-reference-class}.
Remark~\ref{rem:exact-number-shift} keeps the preparation fixed
when the observation interval is enlarged.
\end{proof}

The lower bound can be expressed directly in terms of the
reference before preparation.

\begin{proposition}\label{prop:prepared-contact-energy}
Let $v\in C_c^\infty(\mathbb R^3)$ be nonzero, nonnegative and
radial. Fix an $N$-independent normalized
$\phi\in C_c^\infty(\mathbb R^3)$ and a sufficiently small fixed $\ell\in(0,1/2)$.
Let $q=1-|\phi\rangle\langle\phi|$, $P=\Gamma(q)$, and use
the Neumann pair kernel $\eta_N$ and its limit
$\eta_\infty$ in \eqref{spec:eq:actual-eta-limit}.
Let $C_N=\mathcal C_N^{\mathrm{COS}}$ be the phase
\eqref{spec:eq:actual-phase}, with $m=N^{-1/2}$.
Its real cutoff $0\le\Theta_N\le1$ equals one for
$n\le N^{1/2}/2+10$ and zero for $n\ge N^{1/2}-10$.
Extend $C_N$ by zero above sector $N$.
Use the centered initial energies
$\widehat F_{N,b}=\mathfrak J_N-\mathfrak J_N[\Omega]I+bM$
on $\Pi_NP\cF$ and $F_{\infty,b}=\mathcal J_{2,\infty}+bM$
on $P\cF$, where $\Pi_N=\mathbf1_{\{\cN\le N\}}$.
Choose a fixed $b$ so that, uniformly in $N$,
$c_bQ_N\le\widehat F_{N,b},F_{\rm ex}(0)\le C_bQ_N$
on the physical space and $c_bA\le F_{\infty,b}\le C_bA$.
Writing $H_b=K_P+bM$ and $S_\infty=F_{\infty,b}-H_b$, require
\[
	\|M^rS_\infty H_b^{-1}M^{-r}\|<\tfrac12,
	\qquad r=0,2.
\]
All these conditions hold for sufficiently large $b$, independently of $N$.
For a normalized
$\chi\in\mathcal Y=P\cF\cap D(M^2)\cap D(M^{1/2}A^{1/2})$, set
\[
	e_\chi=\|A^{1/2}\chi\|^2,\qquad
	z_\infty=\frac{F_{\infty,b}^{-1}\chi}
	{\|F_{\infty,b}^{-1}\chi\|},\qquad
	\xi_\infty(t)=\mathcal U_{2,\infty}(t,0)z_\infty.
\]
On every compact time interval $I$, there is a constant
$C_{I,b}$ independent of $\chi$ such that
\begin{equation}\label{eq:prepared-contact-energy}
	\sup_{t\in I}\|(\cK+M)^{1/2}\xi_\infty(t)\|^2
	\le C_{I,b}e_\chi,\qquad
	\inf_{t\in I}c(t)\ge\frac{\kappa_I}{C_{I,b}e_\chi}.
\end{equation}
Here $\kappa_I$ is the constant in
Proposition~\ref{prop:contact-energy-coercivity}.
\end{proposition}
\begin{proof}
Write $F=F_{\infty,b}$ and $y=F^{-1}\chi$.
We use $F[u]=\|F^{1/2}u\|^2$ on the form domain.
The inverse gives $y\in D(F)$ and $Fy=\chi$.
The positive form comparison gives
$c_bA\le F\le C_bA$, with $c_b>0$.
Cauchy--Schwarz in the spectral representation of $F$ yields
$1\le F[\chi]\langle\chi,F^{-1}\chi\rangle \le F[\chi]\|y\|$.
Since $F[y]=\langle\chi,y\rangle$, we obtain
\begin{equation}\label{eq:prepared-first-form-bound}
	A[z_\infty]\le c_b^{-1}\frac{F[y]}{\|y\|^2}
	\le\frac{c_b^{-1}}{\|y\|}
	\le c_b^{-1}F[\chi]\le c_b^{-1}C_be_\chi.
\end{equation}
Only the form value $F[\chi]$ enters this normalization bound.

We next propagate this first energy under the quadratic evolution.
Put $A_{\rm amb}=\cK+M$ and write
$J(t)=\mathcal J_{2,\infty}(t)=\cK+S(t)$ as in
Lemma~\ref{lem:quadratic-weighted-operator-domain}.
Enlarge $I$ to contain zero if necessary.
The three number diagonals and
\eqref{eq:quadratic-weighted-coefficients} give, as forms,
\[
	\pm S(t)\le C_IM,\qquad
	\pm\dot S(t)\le C_IM,\qquad
	\pm i[S(t),M]\le C_IM.
\]
Indeed, a block from sector $n$ to sector $m$ has norm at most
$C_I(n+1)$ and vanishes unless $m-n=0,2,-2$.
Its commutator with $M$ multiplies it by $n-m$.
Cauchy--Schwarz on these three diagonals gives the displayed
form bounds. These are the quadratic coefficient estimates
used in \cite[Proposition 5.1 and its proof]{COS}.

Choose $\beta>C_I+1$ and set $D(t)=J(t)+\beta M$.
Then $c_IA_{\rm amb}\le D(t)\le C_I'A_{\rm amb}$.
For $\Pi_L=\mathbf1_{\{\cN\le L\}}$, define
\[
	J_L(t)=\cK+\Pi_LS(t)\Pi_L,\qquad
	D_L(t)=J_L(t)+\beta M.
\]
Let $v_L$ solve $i\partial_tv_L=J_Lv_L$ with
$v_L(0)=\Pi_Lz_\infty$. The preparation belongs to
$D(M^2A_{\rm amb})$ by the preceding theorem.
At fixed $L$ the perturbation and its time derivative are bounded.
Thus the energy identity is valid on the kinetic operator domain:
\[
	\frac{\d}{\d t}D_L(t)[v_L(t)]
	=\langle v_L(t),(\Pi_L\dot S(t)\Pi_L
	+i\beta[\Pi_LS(t)\Pi_L,M])v_L(t)\rangle.
\]
The same form bounds and equivalence with $A_{\rm amb}$ hold
uniformly in $L$. Gronwall's inequality in either time direction
therefore gives
\[
	\sup_{t\in I}A_{\rm amb}[v_L(t)]
	\le C_I A_{\rm amb}[z_\infty].
\]
The number truncations converge in Hilbert norm to
$\xi_\infty(t)$, as proved in
Lemma~\ref{lem:quadratic-weighted-operator-domain}.
Lower semicontinuity of the closed kinetic form gives the same
bound for $\xi_\infty(t)$ at every $t\in I$.
The constant is independent of $t$, so it also bounds the supremum.
At time zero $A_{\rm amb}[z_\infty]=A[z_\infty]$.
Combining this identity with
\eqref{eq:prepared-first-form-bound} proves the first estimate.
Unitarity gives $\|\xi_\infty(t)\|=1$; now apply
Proposition~\ref{prop:contact-energy-coercivity} for the second.
\end{proof}

The inclusion of the earlier reference class is continuous.
Indeed, joint spectral calculus gives
\[
	\|M^{1/2}A^{1/2}\chi\|^2
	\le\frac12\bigl(\|A\chi\|^2+\|M\chi\|^2\bigr)
	\qquad(\chi\in D(A)).
\]
The extension is strict. A normalized one-particle vector
$f\in qH^1\setminus H^2$ belongs to $\mathcal Y$ but not
to $D(A)$. Such vectors exist: apply $q$ to an $H^1\setminus H^2$
function and normalize; subtracting its smooth condensate
component cannot restore $H^2$ regularity.
Its prepared family has all the conclusions of the theorem.

\begin{lemma}
\label{lem:quasifree-form-energy}
Let $\kappa\in\mathfrak h\otimes_s\mathfrak h$ be Hilbert--Schmidt,
with $\|\kappa\|_{\mathrm{op}}<1$, where $\mathfrak h=qL^2$ and
$K_P=\dGamma(h)$ as in Section~\ref{sec:quasifree-preparation}.
Assume only
\begin{equation}\label{eq:quasifree-form-kernel}
	\kappa\in D((h\otimes1+1\otimes h)^{1/2}).
\end{equation}
The normalized pair state $\chi_\kappa$ in
\eqref{eq:quasifree-reference} then satisfies
\[
	\chi_\kappa\in D(M^r)\cap D(M^rA^{1/2})
	\qquad\text{for every fixed }r\ge0.
\]
In particular $\chi_\kappa\in\mathcal Y$.
\end{lemma}
\begin{proof}
The normalization and exponential number formula
\eqref{eq:quasifree-number-generating} only use the
Hilbert--Schmidt and contraction assumptions. They remain
valid under \eqref{eq:quasifree-form-kernel}.
Use the mode decomposition
$\kappa=\sum_j\lambda_j f_j\otimes f_j$ of
\cite[Lemma 5]{NNS2016}, with $\lambda_j>0$ on its nonzero modes.
Set $h_L=\min(h,L)$ by spectral calculus.
For each finite $L$, orthogonality in the second variable gives
\[
	\langle\kappa,(h_L\otimes1)\kappa\rangle
	=\sum_j\lambda_j^2\langle f_j,h_Lf_j\rangle.
\]
Monotone convergence of the positive series and of the
spectral forms shows that each nonzero mode lies in
$D(h^{1/2})$ and gives
\begin{equation}\label{eq:quasifree-pair-form-trace}
	E_\kappa:=\|(h^{1/2}\otimes1)\kappa\|^2
	=\sum_j\lambda_j^2\|h^{1/2}f_j\|^2<\infty.
\end{equation}
Symmetry makes the full pair form equal to $2E_\kappa$.

For a bounded nonnegative one-particle operator $d$, the
one-mode pair series has mean occupation
$\lambda_j^2/(1-\lambda_j^2)$. Distinct modes have zero
off-diagonal one-particle expectation. Hence the usual
one-particle density identity \cite[Eqs.~(4)--(5)]{NNS2016}
takes the form
\[
	\langle\chi_\kappa,\dGamma(d)\chi_\kappa\rangle
	=\sum_j\frac{\lambda_j^2}{1-\lambda_j^2}\langle f_j,df_j\rangle.
\]
For infinitely many modes, truncate the pair kernel first.
The states converge in $D(M^{1/2})$ by their vacuum overlaps
and uniform exponential number bounds. Since
$0\le\dGamma(d)\le\|d\|\cN$, the expectations converge.
Take $d=\min(h,L)$ and let $L\to\infty$. Monotone
convergence of the sector forms and of the positive series gives
\begin{equation}\label{eq:quasifree-kinetic-form}
	\|K_P^{1/2}\chi_\kappa\|^2
	=\sum_j\frac{\lambda_j^2}{1-\lambda_j^2}\|h^{1/2}f_j\|^2
	\le\frac{E_\kappa}{1-\|\kappa\|_{\mathrm{op}}^2}.
\end{equation}
This identity only uses the form domain of $h$.

To include number weights, choose $\tau>0$ with
$e^{2\tau}\|\kappa\|_{\mathrm{op}}^2<1$. The Fock series
gives an identity of Hilbert-space vectors,
\begin{equation}\label{eq:quasifree-number-tilt}
	e^{\tau\cN/2}\chi_\kappa
	=\frac{Z_\kappa}{Z_{e^\tau\kappa}}\chi_{e^\tau\kappa}.
\end{equation}
Indeed, the $2m$-particle term is multiplied by $e^{\tau m}$.
The scaled kernel still satisfies
\eqref{eq:quasifree-form-kernel}, with form energy
$e^{2\tau}E_\kappa$.
Applying \eqref{eq:quasifree-kinetic-form} to it yields
\[
	\|K_P^{1/2}e^{\tau\cN/2}\chi_\kappa\|^2
	\le\frac{Z_\kappa^2}{Z_{e^\tau\kappa}^2}
	\frac{e^{2\tau}E_\kappa}{1-e^{2\tau}\|\kappa\|_{\mathrm{op}}^2}<\infty.
\]
The operators $K_P$ and $\cN$ strongly commute.
On sector $n$, $(n+1)^{2r}\le C_{r,\tau}e^{\tau n}$.
Summing the kinetic forms therefore proves
$\|M^rK_P^{1/2}\chi_\kappa\|<\infty$ for every fixed $r$.
The exponential number bound controls the additional $M$
term in $A=K_P+M$. More explicitly, joint spectral calculus gives
\[
	\|M^rA^{1/2}\chi_\kappa\|^2
	=\|M^rK_P^{1/2}\chi_\kappa\|^2
	+\|M^{r+1/2}\chi_\kappa\|^2<\infty.
\]
This proves the assertion.
\end{proof}

\begin{corollary}
\label{cor:quasifree-form-preparation}
Let $v\in C_c^\infty(\mathbb R^3)$ be nonzero, nonnegative and
radial. Fix an $N$-independent normalized
$\phi\in C_c^\infty(\mathbb R^3)$ and a sufficiently small fixed $\ell\in(0,1/2)$.
Let $q=1-|\phi\rangle\langle\phi|$, $P=\Gamma(q)$, and use
the Neumann pair kernel $\eta_N$ and its limit
$\eta_\infty$ in \eqref{spec:eq:actual-eta-limit}.
Let $C_N=\mathcal C_N^{\mathrm{COS}}$ be the phase
\eqref{spec:eq:actual-phase}, with $m=N^{-1/2}$.
Its real cutoff $0\le\Theta_N\le1$ equals one for
$n\le N^{1/2}/2+10$ and zero for $n\ge N^{1/2}-10$.
Extend $C_N$ by zero above sector $N$.
Use the centered initial energies
$\widehat F_{N,b}=\mathfrak J_N-\mathfrak J_N[\Omega]I+bM$
on $\Pi_NP\cF$ and $F_{\infty,b}=\mathcal J_{2,\infty}+bM$
on $P\cF$, where $\Pi_N=\mathbf1_{\{\cN\le N\}}$.
Choose a fixed $b$ so that, uniformly in $N$,
$c_bQ_N\le\widehat F_{N,b},F_{\rm ex}(0)\le C_bQ_N$
on the physical space and $c_bA\le F_{\infty,b}\le C_bA$.
Writing $H_b=K_P+bM$ and $S_\infty=F_{\infty,b}-H_b$, require
\[
	\|M^rS_\infty H_b^{-1}M^{-r}\|<\tfrac12,
	\qquad r=0,2.
\]
All these conditions hold for sufficiently large $b$, independently of $N$.
Let $h$ be the operator of the kinetic form on $\mathfrak h=qL^2$.
Suppose $\kappa\in\mathfrak h\otimes_s\mathfrak h$ is Hilbert--Schmidt,
$\|\kappa\|_{\mathrm{op}}<1$, and
$\kappa\in D((h\otimes1+1\otimes h)^{1/2})$.
Then the normalized pair state $\chi_\kappa$ in
\eqref{eq:quasifree-reference} admits the preparations and all
conclusions of Theorem~\ref{thm:form-reference-preparation}.
\end{corollary}
\begin{proof}
That lemma places $\chi_\kappa$ in $\mathcal Y$.
Apply Theorem~\ref{thm:form-reference-preparation}.
\end{proof}

For these references, the contact bound has a direct expression
in terms of the pair kernel. The occupation and kinetic formulas
in Lemma~\ref{lem:quasifree-form-energy} give
\[
	e_{\chi_\kappa}=1+
	\sum_j\frac{\lambda_j^2}{1-\lambda_j^2}
	\bigl(1+\|h^{1/2}f_j\|^2\bigr)
	\le1+\frac{\|\kappa\|_{\mathrm{HS}}^2+E_\kappa}
	{1-\|\kappa\|_{\mathrm{op}}^2}.
\]
Consequently, for the evolved limiting preparation of
$\chi_\kappa$, Proposition~\ref{prop:prepared-contact-energy} yields
\begin{equation}\label{eq:quasifree-prepared-contact-bound}
	\inf_{t\in I}c(t)\ge
	\frac{\kappa_I}{C_{I,b}}
	\left(1+\frac{\|\kappa\|_{\mathrm{HS}}^2+E_\kappa}
	{1-\|\kappa\|_{\mathrm{op}}^2}\right)^{-1}.
\end{equation}
This gives a common positive lower bound when the pair form
energy and Hilbert--Schmidt norm are bounded and the operator
norm stays below a fixed number less than one.

For the compressed Laplacian, the pair form domain is
$(\mathfrak h\otimes_s\mathfrak h)\cap H^1(\mathbb R^6)$.
Thus the new condition requires one weak spatial derivative
of the pair kernel; the earlier operator condition required
two. A concrete strict example is
$\kappa=\lambda f\otimes f$, with $0<\lambda<1$ and a
normalized $f\in qH^1\setminus H^2$.
This kernel satisfies the form condition but cannot lie in
$D(h\otimes1+1\otimes h)$.
Indeed, positivity and joint spectral calculus give
$D(h\otimes1+1\otimes h)\subset D(h\otimes1)$.
On the product vector the squared norm of $h_L\otimes1$
is $\lambda^2\|h_Lf\|^2$.
A bound uniform in $L$ would imply $f\in D(h)=qH^2$,
contrary to the choice of $f$.

The pair-state normalization, density formula and kinetic
expectation used here are standard facts about quasi-free
states \cite{NNS2016,Derezinski2017}. Their role is to verify
the weaker input condition for the same inverse-energy
construction.

\subsection{Preparation from arbitrary reference vectors}
\label{sec:arbitrary-reference-preparation}

The reference regularity can be removed by repeating the positive
form equation. Three inverse powers suffice for the domains used
in the energy and spectral comparisons. This is a further
application of spectral preparation, as in
Proposition~\ref{prop:exact-moment-preparation}.
Write, on the common initial excitation space,
\[
	T_N=\widehat F_{N,b}^{-1}\Pi_N,\qquad
	T=F_{\infty,b}^{-1}.
\]
The inverse in $T_N$ acts on the physical sectors and is zero on
their complement. The operator $Q_N$ retains its actual
sector-form extension on the full space.

\begin{theorem}
\label{thm:arbitrary-reference-preparation}
Let $v\in C_c^\infty(\mathbb R^3)$ be nonzero, nonnegative and
radial. Fix an $N$-independent normalized
$\phi\in C_c^\infty(\mathbb R^3)$ and a sufficiently small fixed $\ell\in(0,1/2)$.
Let $q=1-|\phi\rangle\langle\phi|$, $P=\Gamma(q)$, and use
the Neumann pair kernel $\eta_N$ and its limit
$\eta_\infty$ in \eqref{spec:eq:actual-eta-limit}.
Let $C_N=\mathcal C_N^{\mathrm{COS}}$ be the phase
\eqref{spec:eq:actual-phase}, with $m=N^{-1/2}$.
Its real cutoff $0\le\Theta_N\le1$ equals one for
$n\le N^{1/2}/2+10$ and zero for $n\ge N^{1/2}-10$.
Extend $C_N$ by zero above sector $N$.
Use the centered initial energies
$\widehat F_{N,b}=\mathfrak J_N-\mathfrak J_N[\Omega]I+bM$
on $\Pi_NP\cF$ and $F_{\infty,b}=\mathcal J_{2,\infty}+bM$
on $P\cF$, where $\Pi_N=\mathbf1_{\{\cN\le N\}}$.
Choose a fixed $b$ so that, uniformly in $N$,
$c_bQ_N\le\widehat F_{N,b},F_{\rm ex}(0)\le C_bQ_N$
on the physical space and $c_bA\le F_{\infty,b}\le C_bA$.
Writing $H_b=K_P+bM$ and $S_\infty=F_{\infty,b}-H_b$, require
\[
	\|M^rS_\infty H_b^{-1}M^{-r}\|<\tfrac12,
	\qquad r=0,1,2.
\]
All these conditions hold for sufficiently large $b$, independently of $N$.
Put $T_N=\widehat F_{N,b}^{-1}\Pi_N$ and $T=F_{\infty,b}^{-1}$;
$T_N$ is zero on $(1-\Pi_N)P\cF$.
Let $\chi_N,\chi\in P\cF$ satisfy $\chi_N\to\chi$ in norm
and $\|\chi\|=1$. For all sufficiently large $N$, define
\begin{equation}\label{eq:arbitrary-prepared-vectors}
	z_N=\frac{T_N^3\chi_N}{\|T_N^3\chi_N\|},\qquad
	z_\infty=\frac{T^3\chi}{\|T^3\chi\|}.
\end{equation}
Then
\begin{gather}
	\|Q_N^{1/2}(z_N-z_\infty)\|
	\le C_{b,\chi}\bigl(N^{-3/28}+\|\chi_N-\chi\|\bigr),\qquad
	z_\infty\in D(M^2A),\label{eq:arbitrary-energy-convergence}\\
	\sup_{N\ge N_0}\|F_{\rm ex}(0)^{3/2}z_N\|<\infty,
	\qquad c_B[z_\infty]>0.\label{eq:arbitrary-third-initial}
\end{gather}
The exact physical-moment bounds of Theorem~\ref{thm:D}
and the uniform first-energy and microscopic spectral limits
of Theorem~\ref{thm:uniform-evolved-energy-spectrum} hold
for this family. The positive excitation-energy lower bounds of
Corollary~\ref{cor:uniform-positive-microscopic-moments}
hold as well. The preparation is fixed independently of the
compact observation interval.

If additionally $\chi\in D(M^2)$, then
\begin{equation}\label{eq:arbitrary-reference-rate}
	\|Q_N^{1/2}(z_N-z_\infty)\|
	\le C_{b,\chi}\bigl(N^{-1/4}+\|\chi_N-\chi\|\bigr).
\end{equation}
Particle-number truncation yields the exponent $3/28$.
\end{theorem}
\begin{proof}
The choice of $b$ is uniform in $N$:
$\widehat F_{N,b}\ge c_bQ_N$ and
$F_{\rm ex}(0)\ge c_bQ_N\ge c_bI$.
The common positive form bounds give these inequalities after
one fixed increase of $b$; the bounded $\alpha_N$ is absorbed
using $M\ge1$.
We first record a bound for each inverse step.
Put $H_b=K_P+bM$ and $S=S_\infty$.
The weighted Neumann argument in
Theorem~\ref{thm:form-reference-preparation} gives, for $r=0,1,2$,
\begin{equation}\label{eq:inverse-number-gain}
	\|M^rATf\|\le C_b\|M^rf\|,
	\qquad f\in D(M^r).
\end{equation}
Indeed, $(1+SH_b^{-1})^{-1}$ is bounded on each of these
weighted spaces, and
$T=H_b^{-1}(1+SH_b^{-1})^{-1}$.
Joint spectral calculus for $K_P,M$ gives
$A\le C_bH_b$ and $A\ge M$.
Thus the inverse maps $D(M^r)$ into both $D(M^rA)$
and $D(M^{r+1})$. Applying this successively with $r=0,1,2$,
\begin{equation}\label{eq:triple-inverse-regularity}
	\|M^2AT^3f\|\le C_b\|f\|,\qquad f\in P\cF.
\end{equation}
In particular $T^3\chi\ne0$, since $T$ is injective.

Positivity and form equivalence also imply
\[
	\|Q_N^{1/2}T_Nf\|
	\le C_b\|\widehat F_{N,b}^{-1/2}\Pi_Nf\|
	\le C_b\|f\|.
\]
For the limiting inverse, $Q_N\le CMA$ and $M\le A$ give
\[
	\|Q_N^{1/2}Tf\|
	\le C\|M^{1/2}A^{1/2}Tf\|
	\le C\|ATf\|\le C_b\|f\|.
\]
The middle inequality only uses the strongly commuting
operators $A,M$.
We first compare the inverses for $f\in D(M^2)$.
Set $y=Tf$ and $v_N=\Pi_Ny$.
For a physical form vector $u$, the inverse equations give
\[
	\widehat F_{N,b}[u,T_Nf-v_N]
	=(F_{\infty,b}-\widehat F_{N,b})[u,v_N]
	+F_{\infty,b}[u,(1-\Pi_N)y].
\]
The source terms cancel because
$\langle u,\Pi_Nf\rangle=\langle u,f\rangle$.
Estimate \eqref{eq:inverse-number-gain} with $r=2$ supplies
$y\in D(M^2A)$. Lemma~\ref{lem:common-form-comparison},
the number-form bound on $S$, and the full-space tail
estimate in Theorem~\ref{thm:form-reference-preparation}
then give
\[
	\|Q_N^{1/2}(T_N-T)f\|\le C_bN^{-1/4}\|M^2f\|.
\]
This argument uses only the inverse equations and the
regularity of $y$.

All $T_N,T$ are uniformly bounded on Hilbert space, and
$T$ preserves $D(M^2)$ by the weighted inverse bound.
The noncommutative telescoping identity
\[
	T_N^3-T^3
	=T_N^2(T_N-T)+T_N(T_N-T)T+(T_N-T)T^2
\]
and the preceding bounds yield, for $f\in D(M^2)$,
\begin{equation}\label{eq:weighted-triple-inverse-comparison}
	\|Q_N^{1/2}(T_N^3-T^3)f\|
	\le C_bN^{-1/4}\|M^2f\|.
\end{equation}
For the first two terms, use the uniform energy bound for
the leftmost $T_N$ and the inverse comparison at $f,Tf$.
The last term uses that comparison at $T^2f$.
The bounds for $M^2T$ and $M^2T^2$ on $D(M^2)$ keep
the right-hand side linear in $\|M^2f\|$.

We now control references with large particle number.
Lemma~\ref{lem:exact-moment-coercivity} at time zero,
together with the bounded scalar shift from $F_{\rm ex}(0)$
to $\widehat F_{N,b}$, gives
$\|M^{3/2}\widehat F_{N,b}^{-3/2}\|\le C_b$.
The shift uses the bounded scalar function
$(\lambda+\alpha_N)^{3/2}\lambda^{-3/2}$ on the common
positive spectra: $\lambda\ge c_b>0$ and
$|\alpha_N|\le C_\alpha$. More precisely,
\[
	M^{3/2}\widehat F_{N,b}^{-3/2}
	=(M^{3/2}F_{\rm ex}(0)^{-3/2})
	(F_{\rm ex}(0)^{3/2}\widehat F_{N,b}^{-3/2}).
\]
The first factor is bounded by the coercivity lemma, and
the second by the scalar spectral ratio just given.
Taking adjoints bounds
$\widehat F_{N,b}^{-3/2}M^{3/2}$.
Form comparison separately gives
$\|Q_N^{1/2}\widehat F_{N,b}^{-1/2}\|\le C_b$:
its range is in the common form domain, and the bound
follows by testing the two forms on
$\widehat F_{N,b}^{-1/2}f$.
We can therefore use the ordered product
\begin{equation}\label{eq:finite-triple-right-weight}
	Q_N^{1/2}T_N^3M^{3/2}
	=(Q_N^{1/2}\widehat F_{N,b}^{-1/2})
	\widehat F_{N,b}^{-1}
	(\widehat F_{N,b}^{-3/2}M^{3/2})\Pi_N.
\end{equation}
Each factor on the right is bounded uniformly in $N$.
The inverse powers act on $\Pi_NP\cF$; at fixed $N$,
$M^{3/2}\Pi_N$ is bounded. Their product agrees there
with $\widehat F_{N,b}^{-3}M^{3/2}$ by functional calculus
of this one positive operator. It is zero on the complement.
This ordered factorization proves a bounded extension
of the left-hand side, with $Q_N^{1/2}$ on the left.

For the limiting inverse, \eqref{eq:inverse-number-gain}
gives bounded operators $MT$ and $M^2TM^{-1}$, since
\[
	\|MTf\|\le\|ATf\|\le C_b\|f\|,\qquad
	\|M^2TM^{-1}f\|\le\|MATM^{-1}f\|\le C_b\|f\|.
\]
In the second bound $M^{-1}f\in D(M)$, as required
for the $r=1$ inverse estimate.
Self-adjointness of $T$ gives bounded extensions of their
adjoints $TM$ and $M^{-1}TM^2$.
On $D(M^2)$,
$AT^3M^2=(AT)(TM)(M^{-1}TM^2)$.
Each factor on the right is bounded.
For $f\in D(M^2)$ the last factor produces
$M^{-1}TM^2f\in D(M)$, so the intermediate $TM$ acts
on its original domain. Also $T$ maps Hilbert space into
$D(A)$. These facts justify the equality on $D(M^2)$.
By closure and $Q_N\le CMA\le CA^2$ as forms,
\[
	\sup_N\|Q_N^{1/2}T^3M^2\|\le C_b.
\]
This calculation fixes the order of the factors.

Let $\Pi_L=\mathbf1_{\{\cN\le L\}}$.
Apply \eqref{eq:weighted-triple-inverse-comparison} to
$\Pi_Lf$ and the last two bounds to $(1-\Pi_L)f$.
These are cutoffs on the input: factor
$M^{-3/2}(1-\Pi_L)$ and $M^{-2}(1-\Pi_L)$ on the right,
respectively.
For all $f\in P\cF$ this gives
\[
	\|Q_N^{1/2}(T_N^3-T^3)f\|
	\le C_b\bigl(N^{-1/4}(L+1)^2+(L+1)^{-3/2}
	+(L+1)^{-2}\bigr)\|f\|.
\]
Taking $L=\lfloor N^{1/14}\rfloor$ proves the operator estimate
\begin{equation}\label{eq:triple-inverse-operator-rate}
	\|Q_N^{1/2}(T_N^3-T^3)\|_{P\cF\to\cF}
	\le C_bN^{-3/28}.
\end{equation}
All energy norms use the same actual $Q_N$.
Since $\sup_N\|Q_N^{1/2}T_N^3\|<\infty$, the additional
error from replacing $\chi$ by $\chi_N$ is bounded by
$C_b\|\chi_N-\chi\|$.

Since $Q_N\ge1$, these energy bounds imply
$T_N^3\chi_N\to T^3\chi$ in norm. The denominator in
\eqref{eq:arbitrary-prepared-vectors} is therefore bounded
below for large $N$. Moreover,
$\sup_N\|Q_N^{1/2}T^3\chi\|<\infty$ by
\eqref{eq:triple-inverse-regularity}.
Normalization preserves the convergence and its stated rate.

The third initial energy follows directly from functional
calculus for the positive inverse \cite{ReedSimonI}:
\[
	\|\widehat F_{N,b}^{3/2}z_N\|^2
	=\frac{\|\widehat F_{N,b}^{-3/2}\Pi_N\chi_N\|^2}
	{\|T_N^3\chi_N\|^2}\le C_{b,\chi}.
\]
Here $\chi_N$ is bounded in norm and the lower spectral
bound for $\widehat F_{N,b}$ is uniform.
The bounded scalar shift
$F_{\rm ex}(0)=\widehat F_{N,b}+\alpha_N I$ transfers
this estimate as in the proof of
Theorem~\ref{thm:form-reference-preparation}.
Equation~\eqref{eq:triple-inverse-regularity} and
Proposition~\ref{prop:contact-positive-kinetic} give
$c_B[z_\infty]>0$.

The later-time comparison needs no initial convergence rate.
To see this, use the regular comparison vector
$\zeta_N=\Pi_Nz_\infty/\|\Pi_Nz_\infty\|$.
Its kinetic and sixth number moments are uniformly bounded
by $z_\infty\in D(M^2A)$ and the ambient kinetic comparison
in Theorem~\ref{thm:form-reference-preparation}.
The proof of Proposition~\ref{prop:prepared-quadratic-comparison}
then gives, with $\xi_N(t)=U_{\rm ex}(t,0)z_N$ and
$\xi_\infty(t)=\mathcal U_{2,\infty}(t,0)z_\infty$,
\[
	\sup_{t\in I}\|e^{i\theta_N(t)}\xi_N(t)-\xi_\infty(t)\|
	\le C_{I,b,\chi}\bigl(N^{-1/8}+\|z_N-z_\infty\|\bigr)
	\longrightarrow0.
\]
The scalar convention is the same as in that proposition.
Only $\zeta_N$ enters the regularity assumptions of
\cite[Theorems 2.2--2.3]{COS}.

The remaining domain checks are precisely those listed in the
proof of Theorem~\ref{thm:form-reference-preparation}:
the exact third-energy propagation, the quadratic
$D(M^2A_{\rm amb})$ graph bound, the instantaneous form
comparison, and the moving-background spectral theorem.
The proof of Theorem~\ref{thm:uniform-evolved-energy-spectrum}
uses only Hilbert convergence uniformly on $I$, graph
continuity of $\xi_\infty$, and the uniform third-energy tail.
The displayed estimate and \eqref{eq:arbitrary-third-initial}
provide these state inputs.
Contact positivity and its continuity give the same positive
spectral window and excitation-energy lower bounds.
Finally, Remark~\ref{rem:exact-number-shift} allows the
observation interval to be enlarged without changing
\eqref{eq:arbitrary-prepared-vectors}.
\end{proof}

The three powers have a direct variational meaning: solve the
same positive form equation three times, beginning with
$\Pi_N\chi_N$, and normalize the last solution.
For example, the reference may be a pure quasi-free vector with
an arbitrary symmetric Hilbert--Schmidt pair kernel on $qL^2$,
of operator norm less than one. Its normalization follows from
\eqref{eq:quasifree-number-generating}, independently of any
spatial derivative of the kernel. The state to which the
moment theorem applies is the resulting prepared vector.
The single-inverse construction requires only one form
equation under \eqref{eq:form-reference-class}.

The vacuum of the limiting dressed fields has finite number
moments, zero contact and infinite kinetic form energy.

\begin{proposition}
\label{prop:dressed-vacuum-reference}
Let $v\in C_c^\infty(\mathbb R^3)$ be nonzero, nonnegative and
radial. Fix an $N$-independent normalized
$\phi\in C_c^\infty(\mathbb R^3)$ and a sufficiently small fixed $\ell\in(0,1/2)$.
Let $q=1-|\phi\rangle\langle\phi|$, $P=\Gamma(q)$, and use
the Neumann pair kernel $\eta_N$ and its limit
$\eta_\infty$ in \eqref{spec:eq:actual-eta-limit}.
Let $C_N=\mathcal C_N^{\mathrm{COS}}$ be the phase
\eqref{spec:eq:actual-phase}, with $m=N^{-1/2}$.
Its real cutoff $0\le\Theta_N\le1$ equals one for
$n\le N^{1/2}/2+10$ and zero for $n\ge N^{1/2}-10$.
Extend $C_N$ by zero above sector $N$.
Suppose in addition that $\phi$ is real.
Let $\eta=\eta_\infty$ be its real symmetric limiting Neumann
pair kernel on $qL^2$, and set
\[
	\kappa_{\rm d}=-\tanh\eta,\qquad
	\chi_{\rm d}=\chi_{\kappa_{\rm d}}.
\]
Then $\chi_{\rm d}$ is normalized and has an exponential
number moment. It satisfies
\begin{equation}\label{eq:dressed-vacuum-zero-contact}
	B_x\chi_{\rm d}=0\ \text{for almost every }x,\qquad
	c_B[\chi_{\rm d}]=0,\qquad
	\chi_{\rm d}\notin D(\cK^{1/2}).
\end{equation}
The three-step preparations of
Theorem~\ref{thm:arbitrary-reference-preparation} with
$\chi_N=\chi=\chi_{\rm d}$ have initial first-energy
error $O(N^{-1/4})$ and all its moment and spectral conclusions.
\end{proposition}
\begin{proof}
The operator $\eta$ is real, self-adjoint and Hilbert--Schmidt.
Spectral calculus gives
\[
	\|\kappa_{\rm d}\|_{\mathfrak S_2}\le
	\|\eta\|_{\mathfrak S_2},\qquad
	\|\kappa_{\rm d}\|_{\mathrm{op}}\le
	\tanh\|\eta\|_{\mathrm{op}}<1.
\]
The pair-state representation and number-generating function
\eqref{eq:quasifree-reference}--\eqref{eq:quasifree-number-generating}
therefore define $\chi_{\rm d}$ and give
$\|e^{\tau\cN/2}\chi_{\rm d}\|<\infty$ for some $\tau>0$.

The annihilation identity for a normalized pair state is
\[
	a_x\chi_\kappa=\int\kappa(x,z)a_z^*\chi_\kappa\,\d z.
\]
It follows first for finite-mode pair series and then in
$L^2_x(\cF)$ by their convergence in $D(M^{1/2})$.
Indeed, the vacuum overlaps give Hilbert convergence and the
common exponential number bound controls the number tails
uniformly. The creation terms converge by the
Hilbert--Schmidt kernel bound and the same weighted norm.
This is the usual quasi-free vacuum relation
\cite{Shale,Derezinski2017}. At this real background,
$\gamma=\cosh\eta$ and $\sigma=\sinh\eta$ commute with
$\kappa_{\rm d}$. Substitution into the definition of $B_x$
gives
\[
	B_x\chi_{\rm d}
	=\int(\gamma\kappa_{\rm d}+\sigma)(x,z)
	a_z^*\chi_{\rm d}\,\d z=0.
\]
The kernel in this expression is Hilbert--Schmidt, and
$\gamma\kappa_{\rm d}+\sigma=0$ by scalar functional calculus.
This proves the zero-contact assertion.
If $\chi_{\rm d}$ had finite kinetic form energy,
Proposition~\ref{prop:contact-positive-kinetic} would give
$c_B[\chi_{\rm d}]>0$, since its norm is one and its mean
number is finite. Hence its kinetic form energy is infinite.
Finally, the exponential number moment implies
$\chi_{\rm d}\in D(M^2)$, so
\eqref{eq:arbitrary-reference-rate} applies.
\end{proof}

\subsection{Rates in the energy form norm}
\label{sec:evolved-energy-rate}

The asymmetric form comparison also gives a rate for the
energy norm. Its regular argument is the limiting quadratic
state. The exact state enters through the operator norm
controlled by its third correlated-energy moment.

\begin{proposition}
\label{prop:evolved-energy-rate}
Use the energies and direct rotation of
Section~\ref{sec:evolved-energy}. Let $\xi_N(t)$ be normalized
exact excitation states and let $\xi_\infty(t)$ be a normalized
quadratic reference evolution. Suppose that, on a compact interval $I$,
\begin{equation}\label{eq:energy-rate-hypotheses}
	\sup_{N,t\in I}\|\widehat F_N(t)^{3/2}\xi_N(t)\|
	+\sup_{t\in I}\|M^2A_{\infty,t}\xi_\infty(t)\|<\infty.
\end{equation}
For the chosen scalar phases, set
\[
	\delta_N(I)=\sup_{t\in I}
	\|e^{i\theta_N(t)}\xi_N(t)-\xi_\infty(t)\|.
\]
Then
\begin{equation}\label{eq:evolved-energy-rate-general}
	\sup_{t\in I}\|Q_N(t)^{1/2}
	(e^{i\theta_N(t)}\xi_N(t)-\mathcal R_N(t)\xi_\infty(t))\|
	\le C_I\bigl(\delta_N(I)^{1/2}+N^{-1/4}\bigr).
\end{equation}
Here $Q_N(t)$ retains its actual sector forms above $N$.
The constant depends on the bounds in
\eqref{eq:energy-rate-hypotheses} and the model bounds on $I$,
but not on $N$.

Define two finite positive energy-weighted measures using the
same operator $Q_N(t)$:
\begin{gather}
	u_N(t)=e^{C_t}\xi_N(t),\qquad
	\widetilde u_N(t)=e^{C_t}\mathcal R_N(t)\xi_\infty(t),
	\label{eq:energy-rate-comparison-vectors}\\
	\nu_{N,t}(g)=
	\langle Q_N(t)^{1/2}u_N(t),
	g(Q_N(t)/N^2)Q_N(t)^{1/2}u_N(t)\rangle,\notag\\
	\widetilde\nu_{N,t}(g)=
	\langle Q_N(t)^{1/2}\widetilde u_N(t),
	g(Q_N(t)/N^2)Q_N(t)^{1/2}\widetilde u_N(t)\rangle.
	\label{eq:energy-rate-comparison-measures}
\end{gather}
With the total variation norm defined by the supremum over
bounded Borel functions $g$ with $\|g\|_\infty\le1$,
\begin{equation}\label{eq:evolved-spectral-transfer-rate}
	\sup_{t\in I}\|\nu_{N,t}-\widetilde\nu_{N,t}\|_{\rm TV}
	\le C_I\bigl(\delta_N(I)^{1/2}+N^{-1/4}\bigr).
\end{equation}
The estimate compares the two measures at the same finite $N$.
\end{proposition}
\begin{proof}
Fix $t\in I$ and suppress the time argument. Work in
$P_{\infty,t}\cF$ and put
\[
	w_N=\mathcal R_N^*e^{i\theta_N}\xi_N,\qquad
	\psi=\xi_\infty,\qquad v_N=\Pi_N\psi,\qquad e_N=w_N-v_N.
\]
Both $w_N$ and $e_N$ lie in the physical sectors.
The rotation bound and $A_{\infty,t}\ge M$ give
\begin{equation}\label{eq:energy-rate-hilbert}
	\|e_N\|\le\delta_N(I)+C_IN^{-1}+C_I(N+1)^{-3}.
\end{equation}
Indeed, $\|(\mathcal R_N-1)\psi\|\le C_IN^{-1}\|M\psi\|$
and $\|(1-\Pi_N)\psi\|\le(N+1)^{-3}\|M^3\psi\|$.

Write $F=\mathcal F_N(t)$, $F_0=F_\infty(t)$ and
$\overline Q_N=\mathcal R_N^*Q_N(t)\mathcal R_N$.
In this notation the direct sum defining $F$ is
\[
	F=(\mathcal R_N^*\widehat F_N(t)\mathcal R_N)_{\le N}
	\oplus(K_{P_{\infty,t}}+b_IM)_{>N}.
\]
The subscripts denote restrictions to the indicated number
sectors. The physical block is invariant and contains $w_N$.
Hence powers of $F$ acting on $w_N$ are determined by
the restriction to this block.
Recall that the actual sector energy is
$Q_N(t)=K_{P_{N,t}}+\mathcal V_{N,P_{N,t}}+M$.
Its first two terms are nonnegative and preserve particle
number. Thus $Q_N(t)\ge M$, as in \eqref{eq:comparison},
and $Q_N(t)$ strongly commutes with $M$.
The rotation commutes with $M$, so $\overline Q_N\ge M$
and the same strong commutation holds after rotation.
On the physical block, $F\asymp\overline Q_N$ and
$F\ge c_I I$. Functional calculus on this invariant block gives
\[
	\|Fw_N\|\le c_I^{-1/2}\|F^{3/2}w_N\|
	=c_I^{-1/2}\|\widehat F_N(t)^{3/2}\xi_N(t)\|\le C_I.
\]
In particular, $w_N\in D(F)$.
Also $D(F_0)=D(A_{\infty,t})$ and
$\|F_0\psi\|\le C_I$, since its quadratic remainder
is bounded from $D(M)$ to $\cF$.
The kinetic comparison following
\eqref{eq:evolved-rotation-bounds} gives, as forms,
\[
	A_{\infty,t}\le
	\mathcal R_N^*A_{N,t}\mathcal R_N+C_IN^{-1}M
	\le(1+C_IN^{-1})\overline Q_N.
\]
Since $F_0\asymp A_{\infty,t}$, these closed-form
comparisons give
\[
	D(F^{1/2})\cap\Pi_NP_{\infty,t}\cF
	\subset D(F_0^{1/2})=D(A_{\infty,t}^{1/2}).
\]
For the regular vector $v_N$, strong commutation of $M$
and $A_{\infty,t}$ gives
\[
	\|\overline Q_N^{1/2}v_N\|^2
	\le C_I\|M^{1/2}A_{\infty,t}^{1/2}v_N\|^2<\infty.
\]
Thus $v_N\in D(F^{1/2})$, and every form below is defined.

The regularity of $\psi$ bounds all right-hand norms of
$v_N$ in \eqref{eq:evolved-form-difference}. For example,
strong commutation with particle number gives
\[
	\|M\overline Q_N^{1/2}v_N\|^2
	=\overline Q_N[Mv_N]
	\le C_I\|M^{3/2}A_{\infty,t}^{1/2}v_N\|^2
	\le C_I\|M^2A_{\infty,t}\psi\|^2.
\]
These identities hold sector by sector. The finite
right-hand norm and closedness give the stated weighted domains.
Consequently
\begin{equation}\label{eq:energy-rate-form-error}
	|(F-F_0)[e_N,v_N]|
	\le C_IN^{-1/4}\|\overline Q_N^{1/2}e_N\|.
\end{equation}
The number-preserving part of $F_0$ has no matrix element
between $e_N$ and $(1-\Pi_N)\psi$. The three number
diagonals of its quadratic remainder $S$ give
$\|M^{-1/2}SM^{-1/2}\|_{\mathrm{op}}\le C_I$, by the
block estimate in \eqref{eq:quadratic-weighted-coefficients}.
Consequently
\begin{align*}
	|F_0[e_N,(1-\Pi_N)\psi]|
	&\le C_I\|M^{1/2}e_N\|
	\|M^{1/2}(1-\Pi_N)\psi\|\\
	&\le C_I(N+1)^{-1}\|\overline Q_N^{1/2}e_N\|
	\|M^{3/2}\psi\|\\
	&\le C_I(N+1)^{-1}\|\overline Q_N^{1/2}e_N\|.
\end{align*}
This keeps the coupling across the physical number cutoff.

Now use $e_N=w_N-v_N$. Taking real parts in the form identity,
and placing the full operator only on $w_N$ or $\psi$, gives
\begin{align}
	F[e_N]
	&=\operatorname{Re}\langle e_N,Fw_N\rangle
	-\operatorname{Re}F[e_N,v_N]\notag\\
	&\le(\|Fw_N\|+\|F_0\psi\|)\|e_N\|
	+C_IN^{-1/4}\|\overline Q_N^{1/2}e_N\|.
	\label{eq:energy-rate-polarization}
\end{align}
Set $X_N=\|F^{1/2}e_N\|$ and use coercivity. Young's
inequality in \eqref{eq:energy-rate-polarization} yields
\[
	X_N^2\le C_I\|e_N\|+C_IN^{-1/4}X_N,\qquad
	\|\overline Q_N^{1/2}e_N\|
	\le C_I\bigl(\|e_N\|^{1/2}+N^{-1/4}\bigr).
\]
Strong commutation of $M$ and $A_{\infty,t}$ also
bounds the omitted tail in the actual energy:
\begin{align*}
	\|\overline Q_N^{1/2}(1-\Pi_N)\psi\|^2
	&\le C_I\|M^{1/2}A_{\infty,t}^{1/2}(1-\Pi_N)\psi\|^2\\
	&\le C_I(N+1)^{-3}\|M^2A_{\infty,t}^{1/2}\psi\|^2.
\end{align*}
Together with \eqref{eq:energy-rate-hilbert}, this proves
\eqref{eq:evolved-energy-rate-general}, uniformly on $I$.

Proposition~\ref{prop:full-graph} bounds $e^{C_t}$ in the
$Q_N(t)^{1/2}$ norm, uniformly on $I$.
Apply this bound directly to
$e^{i\theta_N}\xi_N-\mathcal R_N\xi_\infty$.
Above the physical
number cutoff we use the same zero extension of $C_t$ as
in the preceding spectral statements. Both first energies
in \eqref{eq:energy-rate-comparison-measures} are bounded
uniformly: use coercivity for $\xi_N$, and
$\overline Q_N\le C_IMA_{\infty,t}$ for $\psi$.
For every bounded Borel $g$, spectral calculus and
Cauchy--Schwarz give
\begin{align*}
	|\nu_{N,t}(g)-\widetilde\nu_{N,t}(g)|
	&\le\|g\|_\infty
	\|Q_N^{1/2}(e^{i\theta_N}u_N-\widetilde u_N)\|\\
	&\quad\times
	\bigl(\|Q_N^{1/2}u_N\|+\|Q_N^{1/2}\widetilde u_N\|\bigr).
\end{align*}
The measure is invariant under the scalar phase.
Taking the supremum over the unit ball of bounded Borel
functions proves \eqref{eq:evolved-spectral-transfer-rate}.
\end{proof}

\begin{corollary}
\label{cor:prepared-evolved-energy-rate}
For the single-inverse preparations of
Theorems~\ref{thm:common-preparation}
and~\ref{thm:form-reference-preparation}, both left-hand sides
of \eqref{eq:evolved-energy-rate-general}
and~\eqref{eq:evolved-spectral-transfer-rate} are
$O_I(N^{-1/16})$.
For the triple-inverse preparation of
Theorem~\ref{thm:arbitrary-reference-preparation}, set
$\varepsilon_N=\|\chi_N-\chi\|$. The corresponding bound is
\begin{equation}\label{eq:prepared-evolved-energy-rates}
	C_{I,b,\chi}\bigl(N^{-3/56}+\varepsilon_N^{1/2}\bigr).
\end{equation}
If $\chi\in D(M^2)$, it improves to
$C_{I,b,\chi}(N^{-1/16}+\varepsilon_N^{1/2})$.
The prepared family is independent of $I$.
\end{corollary}
\begin{proof}
Theorem~\ref{thm:D}, the bounded vacuum scalar and
Remark~\ref{rem:exact-number-shift} give the first bound in
\eqref{eq:energy-rate-hypotheses}. The weighted inverse
estimates and Lemma~\ref{lem:quadratic-weighted-operator-domain}
give the second.
For the single-inverse families,
Proposition~\ref{prop:prepared-quadratic-comparison} and the
proof of Theorem~\ref{thm:form-reference-preparation} give
$\delta_N(I)\le C_I N^{-1/8}$.
For the triple-inverse family, the regular comparison
$\Pi_Nz_\infty/\|\Pi_Nz_\infty\|$ gives
\[
	\delta_N(I)\le C_{I,b,\chi}
	\bigl(N^{-1/8}+\|z_N-z_\infty\|\bigr),
\]
as proved in Theorem~\ref{thm:arbitrary-reference-preparation}.
Use \eqref{eq:arbitrary-energy-convergence}, $Q_N\ge1$,
and then \eqref{eq:arbitrary-reference-rate} when
$\chi\in D(M^2)$. In the latter case this reads
\[
	\delta_N(I)\le C_{I,b,\chi}
	(N^{-1/8}+N^{-1/4}+\varepsilon_N)
	\le C_{I,b,\chi}(N^{-1/8}+\varepsilon_N).
\]
Thus the $N^{-1/8}$ Hilbert comparison determines the
$N^{-1/16}$ energy rate. Substitution into
Proposition~\ref{prop:evolved-energy-rate} proves all rates.
\end{proof}

The total variation bound concerns the replacement of the
exact state by a regular quadratic reference at the same
finite-$N$ energy operator. The subsequent limit
$\widetilde\nu_{N,t}\to
A_{\infty,t}[\xi_\infty(t)]\delta_0+2c(t)\mu_v$
holds against fixed continuous test functions.
Indeed, $Q_N(t)\ge1$, so each finite-$N$ measure has
zero mass at $\{0\}$. The limiting measure has an atom there
of mass $A_{\infty,t}[\xi_\infty(t)]\ge1$.
The mass discrepancy at zero therefore bounds their total
variation distance from below by one.

\subsection{Uniform limits on compact time intervals}
\label{sec:uniform-evolved-spectrum}

The model bounds used above are uniform on compact time
intervals. To obtain a uniform spectral limit, we must also
control sequences of observation times. The following argument
does this in one fixed excitation space for each convergent
time sequence.

\begin{theorem}
\label{thm:uniform-evolved-energy-spectrum}
For the prepared family of
Theorem~\ref{thm:evolved-first-energy-convergence} and every
compact interval $I$,
\begin{equation}\label{eq:uniform-evolved-first-energy}
	\sup_{t\in I}\|Q_N(t)^{1/2}
	(e^{i\theta_N(t)}\xi_N(t)-\mathcal R_N(t)\xi_\infty(t))\|
	\longrightarrow0.
\end{equation}
Writing $\nu_{N,t}(g)$ for the left-hand side of
\eqref{eq:exact-evolved-microscopic-spectrum}, we also have,
for every fixed $g\in C_b([0,\infty))$,
\begin{equation}\label{eq:uniform-evolved-microscopic-spectrum}
	\sup_{t\in I}\left|\nu_{N,t}(g)
	-A_{\infty,t}[\xi_\infty(t)]g(0)-2c(t)\int g\d\mu_v\right|
	\longrightarrow0.
\end{equation}
The preparation is independent of $I$.
Corollary~\ref{cor:prepared-evolved-energy-rate} supplies rates
for the first limit.
\end{theorem}
\begin{proof}
Enlarge $I$ to contain zero. Consider arbitrary sequences
$N_j\to\infty$ and $t_j\in I$ with $t_j\to t_*$.
Use a star for the limiting condensate and operators at $t_*$:
\[
	q_*=q_{\infty,t_*},\quad P_*=\Gamma(q_*),\quad
	A_*=K_{P_*}+M,\quad \psi_*=\xi_\infty(t_*),\quad
	F_*=P_*\mathcal J_{2,\infty}(t_*)P_*+b_IM.
\]
Choose $b_I$ once, as in Lemma~\ref{lem:evolved-form-comparison}.
The $H^4$ comparison for the condensates and their continuity
give
\[
	\|\phi_{N_j,t_j}-\phi_{\infty,t_*}\|_{H^4}
	\le C_IN_j^{-1}+\|\phi_{\infty,t_j}-\phi_{\infty,t_*}\|_{H^4}
	\longrightarrow0.
\]
Let $S_j$ be the direct rotation from $q_*$ to $q_{N_j,t_j}$,
defined by \eqref{eq:evolved-direct-rotation} with these two
projections, and put $\mathcal S_j=\Gamma(S_j)$.
The rank-two calculation following that equation gives
\[
	\|S_j-1\|_{L^2\to H^2}+\|S_j^*-1\|_{L^2\to H^2}\longrightarrow0.
\]
The weighted identity \eqref{eq:moving-weighted-rotation}, applied
to these condensates, bounds $\mathcal S_j$ uniformly on
$X=D(M^2A_{\rm amb}^{1/2})$. On finite sectors it converges
to the identity in $H^1$. Density and the uniform bound give
strong convergence on $P_*\cF\cap X$.

Pull back $\widehat F_{N_j}(t_j)$ by $\mathcal S_j$ on the
physical sectors and extend it above $N_j$ by $K_{P_*}+b_IM$.
Call the resulting positive direct sum $G_j$. Its form domain is
\[
	\mathcal S_j^*D(\widehat F_{N_j}(t_j)^{1/2})
	\oplus(1-\Pi_{N_j})D(A_*^{1/2}).
\]
The first summand lies below the physical number cutoff.
Both summands are closed and have the same positive lower
bound, uniformly in $j$. In contrast, $F_*$ is the full
limiting form on $P_*\cF$, including every number-changing
block of $S(t_*)$ on the full Fock space.
The resolvent argument below uses finite-sector sources and
tests the form difference on the physical block.

For $\overline Q_j=\mathcal S_j^*Q_{N_j}(t_j)\mathcal S_j$,
use the actual sector forms of $Q_{N_j}(t_j)$ also above
$N_j$, as in Theorem~\ref{thm:evolved-first-energy-convergence}.
On the physical block, $G_j\asymp\overline Q_j$ with
uniform constants, by unitary conjugation of the corresponding
comparison for $\widehat F_{N_j}(t_j)$.
On all sectors the kinetic comparison and $Q\le C_IMA$ give
$\overline Q_j\le C_IMA_*$ as a form inequality.
These energies preserve particle number, and the rotations
commute with $M$. Thus $M$ strongly commutes with
$\overline Q_j$ and $A_*$. In particular,
\begin{equation}\label{eq:variable-time-weighted-domain}
	\|M\overline Q_j^{1/2}v\|^2
	=\overline Q_j[Mv]\le C_I\|M^{3/2}A_*^{1/2}v\|^2
	\le C_I\|M^2A_*v\|^2.
\end{equation}
This follows sector by sector and then by monotone convergence
of the nonnegative sums.
We claim
\begin{equation}\label{eq:variable-time-forms}
	G_j\longrightarrow F_*\ \text{in the strong resolvent sense},
	\qquad G_j[\Pi_{N_j}v]\longrightarrow F_*[v]
	\quad(v\in D(M^2A_*)).
\end{equation}

To compare the operators at varying times, write $\mathcal J_{2,\infty}(t)=\cK+S(t)$, with
$S(t)=\dGamma(h_t)+\mathcal P^*(k_t)+\mathcal P(k_t)$
as in Lemma~\ref{lem:evolved-form-comparison}.
Let $E_n$ denote the projection onto particle number $n$.
In the one-particle block, $E_1\dot J(t)E_1$ is $\dot h_t$;
in the vacuum-to-two-particle block, $E_2\dot J(t)\Omega$
is a fixed normalization constant times $\dot k_t$.
Equation~\eqref{eq:quadratic-weighted-coefficients} therefore gives
\[
	\|\dot h_t f\|\le C_I\|f\|,\qquad
	\|\dot k_t\|_2\le C_I.
\]
For the first inequality integrate the strong derivative on
each fixed $f$, and then take the supremum over $\|f\|=1$.
For the second, integrate in the two-particle Hilbert space.
Both block derivatives exist by the strong $C^1$ property
used in Lemma~\ref{lem:quadratic-weighted-operator-domain}.
We obtain
$\|h_t-h_s\|_{\mathrm{op}}+\|k_t-k_s\|_2\le C_I|t-s|$.
The kinetic operator has no time dependence.
Let $T_j=R_{N_j}(t_j)^*S_j$. This unitary maps $q_*$ to
$q_{\infty,t_j}$, and $T_j-1$, $T_j^*-1$ tend to zero
as maps $L^2\to H^2$.
To check the kinetic part on the fixed space, let $D=-\Delta$
and $B_j=D(T_j-1)$. On $H^2$,
\[
	T_j^*DT_j-D=T_j^*B_j+B_j^*,\qquad
	\|T_j^*DT_j-D\|_{\mathrm{op}}\le2\|B_j\|_{\mathrm{op}}\longrightarrow0.
\]
Here $T_j$ and $T_j^*$ preserve $H^2$; the right side
is the bounded extension of the difference. Summing this
one-particle operator on each sector gives, for $u,v\in P_*\cF$,
\begin{align*}
	&|K_{P_{\infty,t_j}}[\Gamma(T_j)u,\Gamma(T_j)v]-K_{P_*}[u,v]|\\
	&\qquad\le2\|B_j\|_{\mathrm{op}}\|M^{1/2}u\|\|M^{1/2}v\|.
\end{align*}
Both compressed forms here equal their ambient kinetic forms
on the indicated vectors. The preceding coefficient bounds and the
unitary conjugation estimates in
Lemma~\ref{lem:evolved-form-comparison} give the same
bound for its quadratic part.

Conjugate \eqref{eq:evolved-form-difference} at time $t_j$
by $\Gamma(T_j)$ and add this $o(1)M$ form. Since all
rotations commute with $M$, for physical $u,v$ we obtain
\begin{align}\label{eq:variable-time-form-error}
	|(G_j-F_*)[u,v]|\le C_I\|\overline Q_j^{1/2}u\|
	\bigl\{&N_j^{-1/4}(\|M\overline Q_j^{1/2}v\|+\|M^{5/2}v\|)\notag\\
	&+N_j^{-1/2}(\|A_*v\|+\|M^{3/2}v\|)
	+\omega_j\|M^{1/2}v\|\bigr\},
\end{align}
where $\omega_j\to0$. The operator comparison of the
compressed kinetic energies controls the $A_*$ norm here.

To prove the resolvent claim, fix a finite-sector
$\chi\in D(A_*)$ and set $y=F_*^{-1}\chi$.
The weighted inverse argument gives $y\in D(M^2A_*)$.
For large $j$, $x_j=G_j^{-1}\chi$ belongs to the physical
block. With $v_j=\Pi_{N_j}y$ and $e_j=x_j-v_j$, its
weak equation is
\[
	G_j[u,e_j]=(F_*-G_j)[u,v_j]+F_*[u,(1-\Pi_{N_j})y].
\]
Since $u=\Pi_{N_j}u$, the number-preserving part vanishes:
\[
	(K_{P_*}+b_IM+S_0(t_*))[u,(1-\Pi_{N_j})y]=0.
\]
Only the pair-annihilation blocks cross from above the cutoff
to the physical sectors. On $D(M^2A_*)$ their exact sum is
\begin{equation}\label{eq:variable-time-cutoff-blocks}
	\Pi_{N_j}F_*(1-\Pi_{N_j})
	=E_{N_j-1}F_*E_{N_j+1}+E_{N_j}F_*E_{N_j+2}.
\end{equation}
The number-form bound for these two blocks gives
\begin{align*}
	|F_*[u,(1-\Pi_{N_j})y]|
	&\le C_I\|M^{1/2}u\|\|M^{1/2}(1-\Pi_{N_j})y\|\\
	&\le C_I(N_j+1)^{-1}\|\overline Q_j^{1/2}u\|\|M^{3/2}y\|.
\end{align*}
Equation~\eqref{eq:variable-time-weighted-domain} bounds the
first weighted energy norm of $v_j$ in
\eqref{eq:variable-time-form-error}. The other norms are
bounded by $\|M^2A_*y\|$, since $A_*\ge M$ and these
operators strongly commute.
Testing with $e_j$ gives
$\|\overline Q_j^{1/2}e_j\|\le C_{I,\chi}(N_j^{-1/4}+\omega_j)$.
The dense class of sources and the common positive lower
bound prove strong inverse, hence strong resolvent, convergence
\cite{ReedSimonI}. For the form assertion, take
$u=v_j=\Pi_{N_j}v$ in \eqref{eq:variable-time-form-error}.
The weighted bounds give
\[
	|(G_j-F_*)[v_j]|
	\le C_I(N_j^{-1/4}+\omega_j)\|M^2A_*v\|^2\longrightarrow0.
\]
Moreover $F_*\le C_IA_*$ as forms, so
\[
	\|F_*^{1/2}(v-v_j)\|
	\le C_I(N_j+1)^{-2}\|M^2A_*^{1/2}v\|\longrightarrow0.
\]
Thus $F_*[v_j]\to F_*[v]$, which proves the second assertion
in \eqref{eq:variable-time-forms}.

Put $w_j=\mathcal S_j^*e^{i\theta_{N_j}(t_j)}\xi_{N_j}(t_j)$.
The uniform Hilbert-norm comparison in
Proposition~\ref{prop:prepared-quadratic-comparison}, the
continuity of $\xi_\infty$, and $\mathcal S_j\to1$ give
$w_j\to\psi_*$ in norm. This vector is supported below
$N_j$: the exact state is physical and $\mathcal S_j$
preserves particle number. On that invariant block, $G_j$
is exactly a unitary conjugate of $\widehat F_{N_j}(t_j)$.
Functional calculus therefore gives the identity
\[
	G_j^{3/2}w_j=\mathcal S_j^*e^{i\theta_{N_j}(t_j)}
	\widehat F_{N_j}(t_j)^{3/2}\xi_{N_j}(t_j).
\]
Theorem~\ref{thm:D}, the bounded vacuum scalar and
Remark~\ref{rem:exact-number-shift} now give
\[
	\sup_j\|G_j^{3/2}w_j\|<\infty,\qquad
	\|G_j^{1/2}\mathbf1_{\{G_j>L\}}w_j\|\le C_I L^{-1}.
\]
For $f_R(\lambda)=\min(\lambda^3,R)$ on the common positive
spectral half-line, $f_R-R$ vanishes at infinity. Strong
resolvent convergence hence gives strong convergence of
$f_R(G_j)$. It also applies to these varying vectors, since
\[
	f_R(G_j)w_j-f_R(F_*)\psi_*
	=f_R(G_j)(w_j-\psi_*)+(f_R(G_j)-f_R(F_*))\psi_*
	\longrightarrow0.
\]
Use the resulting expectation bound and let $R\to\infty$.
They place $\psi_*$ in $D(F_*^{3/2})$ with the same bound.
Choose $0\le\vartheta\le1$ continuous, equal to one on $[0,1]$
and zero on $[2,\infty)$, and put
$h_L(\lambda)=\sqrt\lambda\,\vartheta(\lambda/L)$.
For fixed $L$, functional calculus and $w_j\to\psi_*$ give
$h_L(G_j)w_j\longrightarrow h_L(F_*)\psi_*$.
The two omitted tails are bounded by $C_I/L$. Hence
\begin{equation}\label{eq:variable-time-square-root-limit}
	\limsup_{j\to\infty}
	\|G_j^{1/2}w_j-F_*^{1/2}\psi_*\|\le 2C_I/L.
\end{equation}
Letting $L\to\infty$ proves the required strong convergence.
Also $\psi_*\in D(M^2A_*)$ by
Lemma~\ref{lem:quadratic-weighted-operator-domain}.
Set $z_j=G_j^{1/2}\Pi_{N_j}\psi_*$. Form consistency gives
$\|z_j\|^2\to\|F_*^{1/2}\psi_*\|^2$. For every $f$,
\[
	\langle z_j,G_j^{-1/2}f\rangle
	=\langle\Pi_{N_j}\psi_*,f\rangle
	\longrightarrow\langle\psi_*,f\rangle.
\]
The common positive lower bound gives
$G_j^{-1/2}f\to F_*^{-1/2}f$. This range is dense, so
$z_j\rightharpoonup F_*^{1/2}\psi_*$.
Convergence of the norms makes this convergence strong. Subtraction,
coercivity and the tail estimate
\[
	\|\overline Q_j^{1/2}(1-\Pi_{N_j})\psi_*\|
	\le C_I(N_j+1)^{-3/2}\|M^2A_*^{1/2}\psi_*\|
\]
prove
\begin{equation}\label{eq:variable-time-energy-comparison}
	\|Q_{N_j}(t_j)^{1/2}
	(e^{i\theta_{N_j}(t_j)}\xi_{N_j}(t_j)-\mathcal S_j\psi_*)\|
	\longrightarrow0.
\end{equation}

The graph continuity of $\xi_\infty(t)$ implies continuity
in $X$. The rotations above are uniformly bounded on $X$.
More explicitly, $\mathcal S_j^*\mathcal R_{N_j}(t_j)=\Gamma(T_j^*)$
and
\[
	\|\Gamma(T_j^*)\xi_\infty(t_j)-\psi_*\|_X
	\le C_I\|\xi_\infty(t_j)-\psi_*\|_X
	+\|(\Gamma(T_j^*)-1)\psi_*\|_X\longrightarrow0.
\]
Apply the uniformly bounded $\mathcal S_j$ to obtain
\[
	\delta_j=\mathcal R_{N_j}(t_j)\xi_\infty(t_j)-\mathcal S_j\psi_*
	\longrightarrow0\quad\text{in }X.
\]
Both terms lie in the same actual excitation space.
There $Q_{N_j}(t_j)\le C_IMA_{N_j,t_j}$, and the
compressed kinetic form equals the ambient kinetic form.
The weighted ambient form consequently gives
\[
	\|Q_{N_j}(t_j)^{1/2}\delta_j\|
	\le C_I\|M^{1/2}A_{\rm amb}^{1/2}\delta_j\|
	\le C_I\|\delta_j\|_X\longrightarrow0.
\]
Together with \eqref{eq:variable-time-energy-comparison}, this
proves the desired energy limit along every convergent time
sequence. If \eqref{eq:uniform-evolved-first-energy} failed,
a sequence with error bounded away from zero would have
such a subsequence by compactness of $I$. This contradiction
proves the uniform energy limit.

For the spectrum, the same sequences satisfy
\begin{gather*}
	p_{N_j,t_j}\to p_{\infty,t_*},\qquad
	\sigma_{N_j,t_j}\to\sigma_{\infty,t_*}
	\quad\text{in Hilbert--Schmidt norm},\\
	\epsilon_j(\|\nabla p_{N_j,t_j}\|_2+
	\|\nabla\sigma_{N_j,t_j}\|_2)\longrightarrow0.
\end{gather*}
Indeed, the compact-time bounds and time continuity in
\cite[Proposition A.2 and Lemmas A.3--A.4]{COS} give
\[
	\|p_{N_j,t_j}-p_{\infty,t_*}\|_2
	\le C_IN_j^{-1/2}+\|p_{\infty,t_j}-p_{\infty,t_*}\|_2
	\longrightarrow0.
\]
The same inequality holds for $\sigma$, while the displayed
gradient expression is at most $C_IN_j^{-1/4}$.
Apply Theorem~\ref{thm:moving-dressed-spectrum} with particle
numbers $N_j$, backgrounds $\phi_{N_j,t_j}$, kernels
$p_{N_j,t_j},\sigma_{N_j,t_j}$ and fixed reference $\psi_*$.
Its direct rotation is $\mathcal S_j$.
For a fixed finite-sector reference, the source approximation
\eqref{eq:moving-source-approximation} applies to this sequence.
At fixed approximation index $k$, its derivative errors are
\[
	\epsilon_j(\|\nabla p_{N_j,t_j}\|_2+
	\|\nabla\sigma_{N_j,t_j}\|_2)+C_k\epsilon_j\longrightarrow0.
\]
The other constants depend on the bounded $H^4$ background
norms and Hilbert--Schmidt kernel norms. The single-pair limits
hold as $N_j\to\infty$. First take $j\to\infty$ at fixed $k$,
then $k\to\infty$. The phase mapping bound
\eqref{eq:moving-phase-map} and the weighted rotation bound
extend the result by density to $\psi_*\in X$.
This proves the limiting measure for
$e^{C_{t_j}}\mathcal S_j\psi_*$ with the actual operator
$Q_{N_j}(t_j)$.
Recall that $u_N(t)=e^{C_t}\xi_N(t)$.
For the transfer set $a_j=e^{i\theta_{N_j}(t_j)}u_{N_j}(t_j)$,
$b_j=e^{C_{t_j}}\mathcal S_j\psi_*$ and $Q_j=Q_{N_j}(t_j)$.
Equation~\eqref{eq:variable-time-energy-comparison} and the
uniform full-graph bound for $e^{C_{t_j}}$ give
$\|Q_j^{1/2}(a_j-b_j)\|\to0$.
Both energy form norms are uniformly bounded: for $b_j$
use $\overline Q_j\le C_IMA_*$ and the full-graph bound,
and for $a_j$ use this convergence. Directly,
\begin{align*}
	&|\langle Q_j^{1/2}a_j,g(Q_j/N_j^2)Q_j^{1/2}a_j\rangle
	-\langle Q_j^{1/2}b_j,g(Q_j/N_j^2)Q_j^{1/2}b_j\rangle|\\
	&\qquad\le\|g\|_\infty\|Q_j^{1/2}(a_j-b_j)\|
	(\|Q_j^{1/2}a_j\|+\|Q_j^{1/2}b_j\|)\longrightarrow0.
\end{align*}
The operator is the same actual $Q_j$ in both terms;
the scalar phase disappears from the quadratic form. Thus
\[
	\nu_{N_j,t_j}(g)\longrightarrow A_*[\psi_*]g(0)
	+2c(t_*)\int g\d\mu_v.
\]
Finally, $A_{\infty,t}[\xi_\infty(t)]$ is the ambient
kinetic-plus-number form on this excitation vector, hence
is continuous by graph continuity. The contact is continuous
as well. The same compactness argument proves
\eqref{eq:uniform-evolved-microscopic-spectrum}.
\end{proof}

\begin{corollary}
\label{cor:uniform-positive-microscopic-moments}
For every compact interval $I$ and $\rho>1$, there are
$c_{I,\rho}>0$ and $N_I$ such that
\begin{equation}\label{eq:uniform-raw-energy-lower}
	\inf_{t\in I}\langle u_N(t),Q_N(t)^\rho u_N(t)\rangle
	\ge c_{I,\rho}N^{2\rho-2}\qquad(N\ge N_I).
\end{equation}
These are extended nonnegative spectral moments. For the
same prepared initial family, the third physical depletion
moment and sixth collective moments remain bounded uniformly
in $N$ and $t\in I$ as in Theorem~\ref{thm:D}.
\end{corollary}
\begin{proof}
Choose $0\le h\le1$, continuous and supported in
$[\lambda_-,\lambda_+]\subset(0,\infty)$, with
$\mu_v(h)>0$. Such a test exists since $\mu_v$ is nonzero
and has no atom at zero.
Corollary~\ref{cor:contact-all-finite-times} gives
$c_I=\min_Ic>0$. More quantitatively,
Proposition~\ref{prop:contact-energy-coercivity} gives
$c_I\ge\kappa_I/E_I$, where
$E_I=\sup_{t\in I}\|(\cK+M)^{1/2}\xi_\infty(t)\|^2$.
The uniform spectral limit yields
$\nu_{N,t}(h)\ge c_I\mu_v(h)$ for all $t\in I$ and
$N\ge N_I$. Thus
\[
	\langle u_N(t),Q_N(t)^\rho u_N(t)\rangle
	\ge N^{2\rho-2}\lambda_-^{\rho-1}\nu_{N,t}(h)
	\ge c_I\lambda_-^{\rho-1}\mu_v(h)N^{2\rho-2}.
\]
The physical-moment bounds are unchanged applications of
Theorem~\ref{thm:D}, with constants depending on $I$.
\end{proof}

These estimates make the first-energy spectral limit and
the lower bounds for higher excitation-energy moments uniform
on compact time intervals.

\subsection{Moments of the excitation interaction energy}
\label{sec:interaction-moments}

The interaction contribution admits a stronger comparison than
the full excitation energy. It preserves excitation number and is
bounded on each number sector. The weighted first-energy estimate therefore controls
this contribution. Throughout this subsection,
$\mathcal V_N(t)=\mathcal V_{N,P_t}$ denotes the positive
pair-interaction term in the excitation energy $Q_N(t)$.

\begin{proposition}
\label{prop:interaction-moment-transfer}
Let $v\in C_c^\infty(\mathbb R^3)$ be nonnegative and radial.
Let $\phi_t$ solve the modified GP equation with an $N$-independent
normalized initial datum in $H^6(\mathbb R^3)$. Fix a sufficiently
small quadratic scattering radius $\ell\in(0,1/2)$, independent of $N$.
Use the cubic phase \eqref{spec:eq:actual-phase} at this background,
with $m=N^{-1/2}$ and the original COS cutoff $0\le\Theta_N\le1$.
It equals one for $n\le N^{1/2}/2+10$ and zero for
$n\ge N^{1/2}-10$, for sufficiently large $N$.
Fix a compact time interval $I$ and $s\in I$.
Choose a fixed $b_I$ such that $F_t=\mathfrak J_t+b_IM$
satisfies $c_IQ_t\le F_t\le C_IQ_t$ uniformly in $N,t\in I$.
Let $\xi_N(s)\in D(F_s^{3/2})$,
$\xi_N(t)=U_{\rm ex}(t,s)\xi_N(s)$ and
$u_N(t)=e^{C_t}\xi_N(t)$.
For every $1\le p\le2$,
\begin{equation}\label{eq:interaction-moment-upper}
	\sup_{t\in I}N^{2-2p}\|\mathcal V_N(t)^{p/2}u_N(t)\|^2\le C_{I,p}\mathscr E_{3,N}(s;\xi_N(s)).
\end{equation}
More generally, fix $t,N$ and suppose that
$\|M Q_N(t)^{1/2}u\|+\|M Q_N(t)^{1/2}z\|\le B$.
For $1\le p<2$,
\begin{align}
	&\left|N^{2-2p}\|\mathcal V_N(t)^{p/2}u\|^2-N^{2-2p}\|\mathcal V_N(t)^{p/2}z\|^2\right|\notag\\
	&\qquad\le C_{p,B}\|Q_N(t)^{1/2}(u-z)\|^{2-p}.
	\label{eq:interaction-moment-transfer}
\end{align}
These statements include membership in the displayed domains.
\end{proposition}
\begin{proof}
On sector $n$, positivity and boundedness of $v$ give
\[
	0\le\mathcal V_{N,P_t,n}\le \tfrac12\|v\|_\infty N^2 n(n-1).
\]
The compression is positive, and it commutes with $M$.
Their joint spectral calculus therefore gives
\begin{equation}\label{eq:interaction-scaled-power}
	N^{2-2p}\|\mathcal V_N^{p/2}w\|^2\le C_p\|M^{p-1}\mathcal V_N^{1/2}w\|^2,\qquad 1\le p\le2 .
\end{equation}
For $p=1$ the inequality is equality with constant one.
For $p>1$ it follows by bounding $\lambda^{p-1}$ with
$C_pN^{2p-2}m^{2p-2}$ on the joint spectrum.
The form comparison is applied separately on each number sector:
\begin{align*}
	\|M\mathcal V_N^{1/2}w\|^2&=\sum_{n\ge0}(n+1)^2\mathcal V_{N,P_t,n}[w_n]\\
	&\le\sum_{n\ge0}(n+1)^2Q_{N,n}[w_n]=\|M Q_N^{1/2}w\|^2 .
\end{align*}
Together with $M^{p-1}\le M$, this bounds the right-hand
side of \eqref{eq:interaction-scaled-power}.
Proposition~\ref{prop:full-graph} at weight $r=1$ and
\eqref{eq:exact-third-moment} prove
\eqref{eq:interaction-moment-upper}.
For the domain assertion, take $w^{(L)}=\mathbf1_{\{\cN\le L\}}w$
with $w\in D(MQ_N^{1/2})$. These truncations converge in
that graph norm. The finite-sector estimates applied to
$w^{(L)}-w^{(K)}$ make $\mathcal V_N^{p/2}w^{(L)}$ Cauchy
at fixed $N$. Closedness of $\mathcal V_N^{p/2}$ proves
the assertion.

For $1<p<2$, apply H\"older's inequality to the spectral
measure of $M$ on $\mathcal V_N^{1/2}w$. Equation
\eqref{eq:interaction-scaled-power} yields
\begin{equation}\label{eq:interaction-interpolation}
	N^{1-p}\|\mathcal V_N^{p/2}w\|\le C_p\|\mathcal V_N^{1/2}w\|^{2-p}\|M\mathcal V_N^{1/2}w\|^{p-1}.
\end{equation}
The endpoint $p=1$ uses only its first factor.
For $w=u-z$, the first factor is bounded by the first-energy
error and the second by $B$. The scaled norms of $u,z$
are bounded by $C_pB$. Factoring the difference of their
squares proves \eqref{eq:interaction-moment-transfer}.
The spectral interpolation uses the commuting pair
$(M,\mathcal V_N)$ \cite{ReedSimonI}.
\end{proof}

Let $f$ be the zero-energy scattering solution and $w=1-f$,
with the convention of \cite[Section 1]{COS}.
For $1\le p\le2$, define
\begin{equation}\label{eq:interaction-scattering-moments}
	b_p(v)=\int_{\mathbb R^3}v(x)^p w(x)^2\d x .
\end{equation}
These finite constants measure the interaction part of the
short-distance pair profile. In particular $b_1(v)$ is the
potential contribution to $D_v$.
They involve multiplication by $v$, whereas the full
excitation-energy measure $\mu_v$ involves $-2\Delta+v$.
If $v\ne0$, the maximum principle gives $f>0$, and
$-2\Delta w=vf$ implies
\[
	w(x)=\frac1{8\pi}\int\frac{v(y)f(y)}{|x-y|}\d y>0 .
\]
Thus $b_p(v)>0$ for every $1\le p\le2$.

\begin{lemma}
\label{lem:moving-interaction-moments}
Let $v\in C_c^\infty(\mathbb R^3)$ be nonzero, nonnegative and radial.
Let $\phi_N,\phi$ be normalized, with $\phi_N\to\phi$ in
$H^4(\mathbb R^3)$. Put $q_N=1-|\phi_N\rangle\langle\phi_N|$,
$q=1-|\phi\rangle\langle\phi|$, $P_N=\Gamma(q_N)$ and $P=\Gamma(q)$.
Set $m=N^{-1/2}$, $\epsilon=N^{-3/4}$ and
$\kappa_N=k_N^{\mathrm{Neu}}$.
Fix $\ell\in(0,1/2)$ and use the COS phase
\eqref{spec:eq:actual-phase} at $\phi_N$.
Its real cutoff $0\le\Theta_N\le1$ is one below
$N^{1/2}/2+10$ and zero above $N^{1/2}-10$.
Let $\mathcal R_N:P\cF\to P_N\cF$ be the direct rotation
in \eqref{eq:evolved-direct-rotation}.
For $\psi\in P\cF\cap D(M^2A_{\rm amb}^{1/2})$, put
$z_{N,\psi}=e^{C_N}\mathcal R_N\psi$.
For every fixed $1\le p\le2$,
\begin{equation}\label{eq:moving-interaction-moments}
	N^{2-2p}\|\mathcal V_{N,P_N}^{p/2}z_{N,\psi}\|^2\longrightarrow 2c_{\phi,B}[\psi]b_p(v).
\end{equation}
The convergence is uniform on compact subsets of this
reference space. It also holds when the reference converges
in its graph norm.
\end{lemma}
\begin{proof}
All constants below are uniform for the stated $H^4$-bounded
backgrounds. We first record a bound for approximation.
The Sobolev estimate \eqref{spec:eq:energy-comparison} on
sector $n$ gives $Q_{N,P_N,n}\le C(n+1)A_{P_N,n}$ as forms.
For $\zeta=P_N\zeta$, the form of $A_{P_N}$ is the
restriction of the form of $A_{\rm amb}=\cK+M$.
The number operator commutes with $P_N$ and has the same
sector weights in both spaces.
Together with Proposition~\ref{prop:full-graph} and
\eqref{eq:interaction-scaled-power}, this implies
\begin{align}
	N^{1-p}\|\mathcal V_{N,P_N}^{p/2}e^{C_N}\zeta\|
	&\le C_p\|M Q_{N,P_N}^{1/2}\zeta\|\notag\\
	&\le C_p\|M^{3/2}A_{\rm amb}^{1/2}\zeta\|
	\le C_p\|M^2A_{\rm amb}^{1/2}\zeta\|.
	\label{eq:interaction-reference-bound}
\end{align}
The rotations are uniformly bounded on the last space, as
proved after \eqref{eq:moving-phase-map}. We can therefore
start with the projected finite tensor core used in
Theorem~\ref{thm:moving-dressed-spectrum}.

Write $\psi_N=\mathcal R_N\psi$ on this core.
The phase reduction \eqref{eq:moving-phase-reduction} also
holds with one number weight in its remainder:
\begin{equation}\label{eq:interaction-phase-reduction}
	e^{C_N}\psi_N=\psi_N+\varepsilon_{\rm s}P_NG_N\psi_N+r_N,\qquad \|M Q_{N,P_N}^{1/2}r_N\|\longrightarrow0 .
\end{equation}
Here $\varepsilon_{\rm s}\in\{1,-1\}$ is the common sign of
the raising term in the actual phase.
To estimate the number weight, choose $k$
so that every $\psi_N$ has at most $k$ particles. On the
first two insertions, all sectors have at most $k+6$
particles. The block estimates
\eqref{spec:eq:phase-L2}--\eqref{spec:eq:phase-H1},
extended to the dressed kernels in
\eqref{spec:eq:dressed-kernel-estimates}, give
\begin{align*}
	\|C_N\zeta\|&\le C_k\epsilon\|\zeta\|,\\
	\|A_{P_N}^{1/2}C_N\zeta\|&\le C_k\bigl(\epsilon\|A_{P_N}^{1/2}\zeta\|+\|\zeta\|\bigr)
\end{align*}
on these sectors, with constants independent of $N$.
The input norms $\|\psi_N\|+\|A_{P_N}^{1/2}\psi_N\|$
are uniformly bounded. Applying the two estimates
successively gives
\[
	\|C_N^2\psi_N\|\le C_\psi\epsilon^2,\qquad
	\|A_{P_N}^{1/2}C_N^2\psi_N\|\le C_\psi\epsilon .
\]
Since $M\le k+7$ and $Q_{N,P_N}\le C_k A_{P_N}$ on
the output sectors, $\|M Q_{N,P_N}^{1/2}C_N^2\psi_N\|
\le C_\psi\epsilon$.
Proposition~\ref{prop:full-graph} at $r=1$ now gives
\begin{align*}
	&\left\|M Q_{N,P_N}^{1/2}
	\int_0^1(1-a)e^{aC_N}C_N^2\psi_N\d a\right\|\\
	&\qquad\le C\int_0^1(1-a)\|M Q_{N,P_N}^{1/2}C_N^2\psi_N\|\d a
	\le C_\psi\epsilon .
\end{align*}
This is a Bochner integral in the weighted form domain;
for each fixed $N$, the block bounds and the finite number
cutoff make $C_N$ bounded on that graph space.
Its exponential series therefore justifies the Taylor
identity there. The displayed bound is uniform in $N$.
For the first insertion, the adjoint estimate in the proof
of \eqref{spec:eq:dressed-reduction} gives first energy
$O_\psi(\epsilon+t_N)$. Its output has at most $k+3$
particles, so the extra number weight is bounded.
For large $N$ the cutoff equals one on these sectors.
The generalized-field coefficients differ from one by
$O_k(N^{-1})$, while the first energy of the raising term
is bounded. Consequently,
\[
	\|M Q_{N,P_N}^{1/2}(C_N\psi_N-\varepsilon_{\rm s}P_NG_N\psi_N)\|
	\le C_\psi(\epsilon+t_N+N^{-1}).
\]
Combining the last two bounds proves
\eqref{eq:interaction-phase-reduction}, with the explicit
remainder bound $C_\psi(\epsilon+t_N+N^{-1})$.
Equation~\eqref{eq:interaction-scaled-power} makes $r_N$
negligible in every scaled norm under consideration.
The regular term $\psi_N$ is negligible as well:
the fixed-sector trace estimate gives
$\mathcal V_{N,P_N}[\psi_N]\le C_\psi/N$, and $M$ is bounded
on its sectors.

Approximate the raising vector by $P_NH_N^{(j)}$ as in
\eqref{eq:moving-source-approximation}. The error has finitely
many number sectors, independently of $N,j$. On these
sectors, \eqref{eq:interaction-scaled-power} bounds its
scaled $p$ norm by its energy form norm times a fixed
constant. Thus the same source approximation is valid here.
After fixing $j$, every contribution has the form
$g_{N,a}=\kappa_N(x_i-x_j)b_N(X_n)$ in
Lemma~\ref{lem:moving-projected-pair}.
Its diagonal amplitude is denoted by $a_a(R,Y)$.

We give the pair calculation for the interaction operator.
Let $S_{N,a}$ be the pair dilation and let
$E_{N,a}=S_{N,a}P_{N,n}S_{N,a}^*$. On the full dilated
space define the bounded positive operator
\[
	D_{N,a}=N^{-2}S_{N,a}P_{N,n}\mathcal V_{N,n}^{\rm full}P_{N,n}S_{N,a}^* .
\]
Its norm is at most $C_n$. Before compression, each
unmatched scaled potential converges strongly to zero,
by dominated convergence. The matching uncompressed
potential is multiplication by $v(z)$.
Since $E_{N,a}$ converges strongly to the spectator projection
$P_{{\rm sp},a}$, it follows that
\begin{equation}\label{eq:interaction-pair-operator-limit}
	D_{N,a}\longrightarrow v(z)P_{{\rm sp},a}\quad\hbox{strongly}.
\end{equation}
Put $h_{N,a}=N S_{N,a}P_{N,n}g_{N,a}$.
The normalization is
\[
	\|D_{N,a}^{p/2}h_{N,a}\|=N^{1-p}\|\mathcal V_{N,P_N,n}^{p/2}P_{N,n}g_{N,a}\|.
\]
The matching potential and the local convergence of the
Neumann profile give
\begin{equation}\label{eq:interaction-pair-vector}
	D_{N,a}h_{N,a}\longrightarrow a_a\otimes(vw)\quad\hbox{in }L^2.
\end{equation}
We prove \eqref{eq:interaction-pair-vector} directly on the
interaction-weighted vector. Before projection the matching product is
$v(z)w_{N^{1/2}}^{\rm Neu}(z)b_N(R,z/N,Y)$.
It converges in $L^2$ because $v$ has compact support.
The unmatched terms tend to zero by the estimates proving
\eqref{eq:moving-pair-vectors}.
For the removed part, that proof gives
$\|(1-P_{N,n})g_{N,a}\|=O(N^{-3/2})$.
The potential norm on sector $n$ is $O_n(N^2)$, so its
contribution after division by $N$ is $O_n(N^{-1/2})$.
The remaining output projection fixes the limiting vector.
This proves \eqref{eq:interaction-pair-vector}.

The same matching calculation at form level gives
\begin{equation}\label{eq:interaction-pair-first}
	\mathcal V_{N,P_N}[P_{N,n}g_{N,a}]\longrightarrow b_1(v)\|a_a\|^2 .
\end{equation}
The unmatched forms vanish as in
Lemma~\ref{lem:moving-projected-pair}, and removing the
projection error changes the form by $o(1)$.
For distinct pairs, at least one factor in each potential
pairing is unmatched. Cauchy--Schwarz then makes each
mixed first-energy term tend to zero.
This treats $p=1$, including finite sums over pairs.

For $1<p\le2$, choose a continuous cutoff $\chi_\eta$ which
vanishes below $\eta$ and equals one above $2\eta$.
On the common bounded spectral interval of $D_{N,a}$,
the function $\lambda^{p/2-1}\chi_\eta(\lambda)$ is
bounded and continuous. Strong functional calculus and
\eqref{eq:interaction-pair-vector} give
\[
	\chi_\eta(D_{N,a})D_{N,a}^{p/2}h_{N,a}\longrightarrow a_a\otimes\bigl(\chi_\eta(v)v^{p/2}w\bigr).
\]
Different pair dilations of these fixed $L^2$ profiles
are asymptotically orthogonal by
Lemma~\ref{spec:lem:pair-orthogonality}.
The omitted part satisfies
\[
	\|(1-\chi_\eta(D_{N,a}))D_{N,a}^{p/2}h_{N,a}\|^2\le(2\eta)^{p-1}\langle h_{N,a},D_{N,a}h_{N,a}\rangle .
\]
The right-hand side is uniformly bounded times
$\eta^{p-1}$ by \eqref{eq:interaction-pair-first}.
There are finitely many pairs, so this also controls their
sum. Let $N\to\infty$ and then $\eta\downarrow0$.
The scaled $p$ moment is therefore
$b_p(v)$ times the total squared diagonal amplitude.

The ordered-pair formula \eqref{spec:eq:dressed-sector}
adds creation and annihilation sources in each output
sector before taking their squared norm.
For the approximants used in
\eqref{eq:moving-source-approximation}, the total squared
amplitude is $2\|\phi^{(j)}P B^{(j)}\chi^{(j)}\|_{L^2_x(\cF)}^2$.
These source vectors converge to $\phi B\psi$ by the
creation and annihilation bounds, as verified in the proof
of Theorem~\ref{thm:moving-dressed-spectrum}.
Let $j\to\infty$ using that convergence and the first-energy
source approximation. The coefficient becomes $2c_{\phi,B}[\psi]$.
Equation~\eqref{eq:interaction-reference-bound} then extends
the result to the reference space. The same bound gives
uniform Lipschitz control of the scaled norms as functions
of $\psi$ in the reference graph norm. The limiting norm
is continuous by the contact bound on $D(M^{1/2})$.
A finite net in a compact set reduces uniform convergence
to convergence at finitely many core approximants, with
a uniform error given by \eqref{eq:interaction-reference-bound}.
This also proves the assertion for convergent references.
Only the bounded map from the reference graph space to
the scaled observable norm is needed in this argument.
\end{proof}

The interaction measure requires only a first-energy input.
This weaker domain is useful for the general initial class
in Theorem~\ref{thm:admissible-initial-class}.
Define the finite positive measure
\begin{equation}\label{eq:interaction-limit-measure}
	\mu_v^{\rm int}(B)=\int_{\mathbb R^3}\mathbf1_B(v(x))v(x)w(x)^2\d x,
	\qquad B\subset[0,\infty)\ \hbox{Borel}.
\end{equation}
Its support is contained in $[0,\|v\|_\infty]$, its mass
is $b_1(v)$, and $\mu_v^{\rm int}(\{0\})=0$.
It is the weighted distribution of the potential values,
whereas $\mu_v$ is defined by the relative scattering operator.

\begin{lemma}
\label{lem:moving-interaction-spectrum}
Let $v\in C_c^\infty(\mathbb R^3)$ be nonzero, nonnegative and radial.
Let $\phi_N,\phi$ be normalized, with $\phi_N\to\phi$ in
$H^4(\mathbb R^3)$. Put $q_N=1-|\phi_N\rangle\langle\phi_N|$,
$q=1-|\phi\rangle\langle\phi|$, $P_N=\Gamma(q_N)$ and $P=\Gamma(q)$.
Set $m=N^{-1/2}$, $\epsilon=N^{-3/4}$ and
$\kappa_N=k_N^{\mathrm{Neu}}$.
Fix $\ell\in(0,1/2)$ and use the COS phase
\eqref{spec:eq:actual-phase} at $\phi_N$.
Its real cutoff $0\le\Theta_N\le1$ is one below
$N^{1/2}/2+10$ and zero above $N^{1/2}-10$.
Let $\mathcal R_N:P\cF\to P_N\cF$ be the direct rotation
in \eqref{eq:evolved-direct-rotation}. Set
$Y_P=P\cF\cap D(M^{1/2}A_{\rm amb}^{1/2})$.
For every $\psi\in Y_P$ and $g\in C_b([0,\infty))$,
\begin{align}\label{eq:moving-interaction-spectrum}
	&\left\langle\mathcal V_{N,P_N}^{1/2}e^{C_N}\mathcal R_N\psi,
	g(\mathcal V_{N,P_N}/N^2)
	\mathcal V_{N,P_N}^{1/2}e^{C_N}\mathcal R_N\psi\right\rangle\notag\\
	&\qquad\longrightarrow 2c_{\phi,B}[\psi]\int g\d\mu_v^{\rm int}.
\end{align}
For each fixed test, the convergence is uniform on compact
subsets of $Y_P$. It also holds for references converging
in this graph norm.
\end{lemma}
\begin{proof}
Write $\mathcal V_N=\mathcal V_{N,P_N}$ and
$Q_N=Q_{N,P_N}$ in this proof. The form comparison and
Proposition~\ref{prop:full-graph} give
\begin{equation}\label{eq:interaction-weaker-input}
	\|\mathcal V_N^{1/2}e^{C_N}\mathcal R_N\zeta\|
	\le\|Q_N^{1/2}e^{C_N}\mathcal R_N\zeta\|
	\le C\|M^{1/2}A_{\rm amb}^{1/2}\zeta\|.
\end{equation}
Here $Q_N\le CMA_{P_N}$ as forms, and the direct rotations
are uniformly bounded on the displayed weighted form space.
The actual phase is extended by the identity above $N$;
the energy and interaction keep their actual sector forms.

We record the domain facts used here. On each number sector,
$M$ is scalar, so it strongly commutes with $A_{\rm amb}$.
Their joint spectral calculus defines the closed product
$M^{1/2}A_{\rm amb}^{1/2}$.
For a direct rotation $\mathcal R_N$, its kinetic form
difference is bounded by $\delta_N M$, with $\delta_N\to0$,
as proved after \eqref{eq:moving-phase-map}.
The rotation commutes with $M$. Applying the form difference
to $M^{1/2}\zeta$ gives
\begin{align*}
	&\left|\|M^{1/2}A_{\rm amb}^{1/2}\mathcal R_N\zeta\|^2
	-\|M^{1/2}A_{\rm amb}^{1/2}\zeta\|^2\right|\\
	&\qquad\le\delta_N\|M\zeta\|^2
	\le\delta_N\|M^{1/2}A_{\rm amb}^{1/2}\zeta\|^2.
\end{align*}
The calculation can first be made with a finite number
cutoff and then passed to the closed graph norm.
This gives the two-sided graph comparison; the same holds
for the inverse rotation. Strong convergence on finite
smooth sectors follows by expanding the finite tensor
products of the one-particle rotation, whose difference
from the identity tends to zero as a map $L^2\to H^4$.
Density and the uniform graph bound give strong convergence
on $Y$.

For vectors $u,z$ in the potential form domain, set
$J_{N,g}(u)=\langle\mathcal V_N^{1/2}u,
g(\mathcal V_N/N^2)\mathcal V_N^{1/2}u\rangle$.
The bounded functional calculus gives
\begin{equation}\label{eq:interaction-test-comparison}
	|J_{N,g}(u)-J_{N,g}(z)|
	\le\|g\|_\infty\|\mathcal V_N^{1/2}(u-z)\|
	\bigl(\|\mathcal V_N^{1/2}u\|+\|\mathcal V_N^{1/2}z\|\bigr).
\end{equation}
Thus the phase reduction and source approximation in the
proof of Lemma~\ref{lem:moving-interaction-moments} apply
to this test. On the projected finite tensor core, the
regular input has vanishing potential form. Its phase
remainder and the source approximation errors tend to zero
in that form norm. It remains to compute the test on
finitely many projected pair vectors.

Fix a sector and a pair $a$. Use $D_{N,a},h_{N,a},a_a$
from \eqref{eq:interaction-pair-operator-limit}--
\eqref{eq:interaction-pair-first}. Those estimates state
\begin{align*}
	D_{N,a}&\longrightarrow v(z)P_{{\rm sp},a}\quad\hbox{strongly},
	\qquad 0\le D_{N,a}\le C_n,\\
	D_{N,a}h_{N,a}&\longrightarrow a_a\otimes(vw),\\
	\|D_{N,a}^{1/2}h_{N,a}\|^2&\longrightarrow b_1(v)\|a_a\|^2.
\end{align*}
The strong operator limit uses only bounded products.
Indeed, let
$V^{\rm dil}_{N,a}=N^{-2}S_{N,a}\mathcal V_{N,n}^{\rm full}S_{N,a}^*$
and $E_{N,a}=S_{N,a}P_{N,n}S_{N,a}^*$.
The operators $V^{\rm dil}_{N,a}$ are uniformly bounded
on this fixed sector. The matching potential is $v(z)$,
and each unmatched multiplication operator tends strongly
to zero by dominated convergence. Thus
$V^{\rm dil}_{N,a}\to v(z)$ strongly and
$E_{N,a}\to P_{{\rm sp},a}$ strongly.
Writing $E_N,E,V_N,V$ for these four operators, the identity
\[
	E_NV_NE_N-EVE
	=E_NV_N(E_N-E)+E_N(V_N-V)E+(E_N-E)VE
\]
shows strong convergence on every fixed vector.
The spectator projection commutes with $v(z)$.
This proves the first displayed limit.
The other two displayed limits are the separate vector
and form estimates already proved in
\eqref{eq:interaction-pair-vector} and
\eqref{eq:interaction-pair-first}.
We claim that
\begin{equation}\label{eq:interaction-pair-form-vector}
	D_{N,a}^{1/2}h_{N,a}\longrightarrow
	a_a\otimes(\sqrt v\,w)\quad\hbox{in }L^2.
\end{equation}
Choose a continuous $0\le\chi_\eta\le1$ that is zero
below $\eta$ and one above $2\eta$.
On the common bounded spectral interval, the function
$\lambda^{-1/2}\chi_\eta(\lambda)$ extends continuously
by zero at the origin. Strong functional calculus
\cite{ReedSimonI}, applied to the identity
\[
	\chi_\eta(D_{N,a})D_{N,a}^{1/2}h_{N,a}
	=F_\eta(D_{N,a})D_{N,a}h_{N,a},
	\qquad F_\eta(\lambda)=\lambda^{-1/2}\chi_\eta(\lambda),
	\quad F_\eta(0)=0,
\]
therefore gives
\[
	\chi_\eta(D_{N,a})D_{N,a}^{1/2}h_{N,a}
	\longrightarrow a_a\otimes(\chi_\eta(v)\sqrt v\,w).
\]
Since $(1-\chi_\eta)^2\le1-\chi_\eta^2$, the omitted
part satisfies
\begin{align*}
	&\limsup_{N\to\infty}
	\|(1-\chi_\eta(D_{N,a}))D_{N,a}^{1/2}h_{N,a}\|^2\\
	&\quad\le\lim_{N\to\infty}\left(
	\|D_{N,a}^{1/2}h_{N,a}\|^2
	-\|\chi_\eta(D_{N,a})D_{N,a}^{1/2}h_{N,a}\|^2\right)\\
	&\quad= b_1(v)\|a_a\|^2
	-\|a_a\otimes(\chi_\eta(v)\sqrt v\,w)\|^2\\
	&\quad=\|a_a\|^2\int(1-\chi_\eta(v)^2)vw^2\d x.
\end{align*}
Both scalar terms in the second line have limits already
established above: the first by the total potential form
estimate, and the second by the cutoff vector convergence.
In particular the total limit is exactly
$\|a_a\otimes(\sqrt v\,w)\|^2$.
The right-hand side tends to zero as $\eta\downarrow0$.
The same is true for the cutoff error of the limiting
vector. This proves \eqref{eq:interaction-pair-form-vector},
including control at zero potential energy.

For a bounded continuous $g$, this vector convergence
and the bounded strong convergence of $D_{N,a}$ imply
\[
	g(D_{N,a})D_{N,a}^{1/2}h_{N,a}
	\longrightarrow a_a\otimes(g(v)\sqrt v\,w).
\]
The spectator projection fixes the limiting vectors.
Undoing the dilation, each diagonal pair contribution
is therefore $\|a_a\|^2\int g(v)vw^2$.
Put $\zeta_a=a_a\otimes(\sqrt v\,w)$ and
$\zeta_{a,g}=a_a\otimes(g(v)\sqrt v\,w)$.
The two convergences just proved are exactly
\begin{align*}
	x_{N,a}:=\mathcal V_N^{1/2}P_{N,n}g_{N,a}
	&=S_{N,a}^*\zeta_a+o_{L^2}(1),\\
	y_{N,a}:=g(\mathcal V_N/N^2)\mathcal V_N^{1/2}P_{N,n}g_{N,a}
	&=S_{N,a}^*\zeta_{a,g}+o_{L^2}(1).
\end{align*}
In particular these vectors have uniformly bounded norms.
For $a\ne b$, the difference between the tested cross term
$\langle x_{N,a},y_{N,b}\rangle$ and
$\langle S_{N,a}^*\zeta_a,S_{N,b}^*\zeta_{b,g}\rangle$
has absolute value at most
\[
	\|x_{N,a}-S_{N,a}^*\zeta_a\|\,\|y_{N,b}\|
	+\|\zeta_a\|\,\|y_{N,b}-S_{N,b}^*\zeta_{b,g}\|
	\longrightarrow0.
\]
The remaining inner product tends to zero by
Lemma~\ref{spec:lem:pair-orthogonality}, including pairs
with one common particle. Both energy weights and the
test are included in $x_{N,a},y_{N,b}$.
Contributions with the same pair are added before taking
the inner product. The creation and annihilation sources
thus have the same coherent coefficient as in
Lemma~\ref{lem:moving-interaction-moments}.
Removing the finite tensor approximation gives
$2c_{\phi,B}[\psi]\int g\d\mu_v^{\rm int}$ on the core.

Projected finite sums of smooth tensor products are dense
in $Y_P$: truncate particle number, approximate in each
sector's $H^1$ norm, and apply $P$.
To justify the last step, the formula
$q=1-|\phi\rangle\langle\phi|$ and $\phi\in H^1$
give, on sector $n$,
\[
	\|A_{{\rm amb},n}^{1/2}P_n\zeta_n\|^2
	\le C_\phi\|A_{{\rm amb},n}^{1/2}\zeta_n\|^2.
\]
Each differentiated factor uses
$\nabla(qf)=\nabla f-\nabla\phi\langle\phi,f\rangle$;
the other projection factors are $L^2$ contractions.
Summing the derivative bounds gives at most
$2\cK_n[\zeta_n]+2n\|\nabla\phi\|^2\|\zeta_n\|^2$,
and the number term is unchanged or decreased.
The constant is independent of $n$.
Multiplication by the sector weight $n+1$ proves
boundedness of $P$ on $Y$.
Estimates \eqref{eq:interaction-weaker-input} and
\eqref{eq:interaction-test-comparison} extend the limit
to $Y_P$. The contact is continuous on $D(M^{1/2})$,
and hence on $Y_P$. The same estimates give uniform
continuity on each bounded subset of $Y_P$.
A finite net proves compact-set uniformity and the
assertion for graph-convergent references.
\end{proof}

\begin{theorem}
\label{thm:exact-interaction-moments}
For the prepared pure-state families in
Corollary~\ref{cor:prepared-evolved-energy-rate}, write
$u_N(t)=e^{C_t}\xi_N(t)$ and use the contact $c(t)$ of
Theorem~\ref{thm:uniform-evolved-energy-spectrum}.
For every compact interval $I$ and every fixed $1\le p<2$,
\begin{equation}\label{eq:exact-interaction-moments}
	\sup_{t\in I}\left|N^{2-2p}\|\mathcal V_N(t)^{p/2}u_N(t)\|^2-2c(t)b_p(v)\right|\longrightarrow0 .
\end{equation}
At $p=2$ one has the uniform upper bound
\eqref{eq:interaction-moment-upper} and
\begin{equation}\label{eq:exact-interaction-second-lower}
	\liminf_{N\to\infty}\inf_{t\in I}\left(N^{-2}\|\mathcal V_N(t)u_N(t)\|^2-2c(t)b_2(v)\right)\ge0 .
\end{equation}
If $v$ is nonzero, the second interaction-energy moment
is consequently of order $N^2$, uniformly on $I$.
\end{theorem}
\begin{proof}
Put $\psi(t)=\xi_\infty(t)$ and
$\widetilde u_N(t)=e^{C_t}\mathcal R_N(t)\psi(t)$.
The exact third-energy estimate and the phase bound give
\[
	\sup_{N,t\in I}\|M Q_N(t)^{1/2}u_N(t)\|<\infty .
\]
The weighted regularity of $\psi(t)$, used in
\eqref{eq:energy-rate-hypotheses}, and the proof of
\eqref{eq:interaction-reference-bound} give the same bound
for $\widetilde u_N(t)$.
By Proposition~\ref{prop:full-graph} and
\eqref{eq:uniform-evolved-first-energy},
\[
	d_N(I):=\sup_{t\in I}\|Q_N(t)^{1/2}(e^{i\theta_N(t)}u_N(t)-\widetilde u_N(t))\|\longrightarrow0 .
\]
Equation~\eqref{eq:interaction-moment-transfer} bounds the
scaled $p$-moment difference by $C_{I,p}d_N(I)^{2-p}$.

Apply Lemma~\ref{lem:moving-interaction-moments} along an
arbitrary sequence $N_j\to\infty$, $t_j\to t_*\in I$.
The backgrounds converge in $H^4$. The reference path is
continuous in $D(M^2A_{\rm amb})$, and hence in
$D(M^2A_{\rm amb}^{1/2})$, by
Lemma~\ref{lem:quadratic-weighted-operator-domain}.
The direct rotations to the fixed space at $t_*$ are
uniformly bounded and converge strongly on this latter
space, as in the proof of
Theorem~\ref{thm:uniform-evolved-energy-spectrum}.
Thus the pulled-back references converge in its graph norm.
The lemma gives the required limit for
$\widetilde u_{N_j}(t_j)$.
Continuity of the contact and compactness of $I$ now prove
\eqref{eq:exact-interaction-moments}.
The rate $C_{I,p}d_N(I)^{2-p}$ compares two finite-$N$
states.

For the endpoint define
$m_{p,N}(t)=N^{2-2p}\|\mathcal V_N(t)^{p/2}u_N(t)\|^2$.
H\"older's inequality for the spectral measure of
$\mathcal V_N(t)$ gives
\[
	m_{p,N}(t)\le m_{1,N}(t)^{2-p}m_{2,N}(t)^{p-1},\qquad 1<p<2 .
\]
If $v\ne0$, then $b_1(v)>0$ and the scattering length is
positive. The normalized reference $\psi(t)$ has finite
kinetic energy and finite mean excitation number.
Proposition~\ref{prop:contact-positive-kinetic} therefore
gives $c(t)>0$. Its continuity on the compact interval
then yields $\min_{t\in I}c(t)>0$.
The uniform limits for $m_1,m_p$ therefore give
\[
	\liminf_{N\to\infty}\inf_{t\in I}\left(m_{2,N}(t)-2c(t)\left(\frac{b_p(v)}{b_1(v)^{2-p}}\right)^{1/(p-1)}\right)\ge0 .
\]
Let $p\uparrow2$; dominated convergence on the support of
$v$ gives $b_p(v)\to b_2(v)$. This proves
\eqref{eq:exact-interaction-second-lower}. If $v=0$,
all statements are immediate.
The upper bound and the positive minimum of the contact
give the asserted order $N^2$.
\end{proof}

For $1\le p<2$, the positive exponent $2-p$ in
\eqref{eq:interaction-moment-transfer} transfers the moment
limit to the exact evolved state. At $p=2$, the estimates
give an upper bound and the stated asymptotic lower bound.

\subsection{Energy measurement probabilities}
\label{sec:evolved-energy-probability}

The density in Proposition~\ref{prop:radial-microscopic-density}
also determines the probabilities of observing an excitation
energy of order $N^2$. These concern a spectral measurement
of the excitation energy $Q_N(t)$ in $u_N(t)=e^{C_t}\xi_N(t)$.
Its observable on the symmetric $N$-particle space is
\begin{equation}\label{eq:many-body-excitation-observable}
	\mathscr Q_N(t)=\mathcal U_{N,t}^*e^{B_t}
	Q_N(t)e^{-B_t}\mathcal U_{N,t},
\end{equation}
where $Q_N(t)$ is restricted to the physical excitation space.
This unitary conjugation also defines its self-adjoint domain.
For the reconstructed state in Theorem~\ref{thm:D},
unitary spectral calculus \cite{ReedSimonI} gives
\[
	\langle u_N(t),\mathbf1_B(Q_N(t)/N^2)u_N(t)\rangle
	=\langle\Psi_N(t),
	\mathbf1_B(\mathscr Q_N(t)/N^2)\Psi_N(t)\rangle .
\]
Thus the probabilities below refer to this explicitly
reconstructed observable in the exact many-body state.

\begin{corollary}
\label{cor:evolved-energy-probability}
For the prepared families of
Theorems~\ref{thm:common-preparation},
\ref{thm:form-reference-preparation}
and~\ref{thm:arbitrary-reference-preparation}, or any family in
Theorem~\ref{thm:admissible-initial-class}, let
\begin{equation}\label{eq:evolved-energy-probability}
	\mathbb P_{N,t}(B)=
	\langle u_N(t),\mathbf1_B(Q_N(t)/N^2)u_N(t)\rangle,\qquad
	m_v(B)=\int_B\frac{\rho_v(\lambda)}{\lambda}\d\lambda.
\end{equation}
For the latter class use $c(t)=c_\psi(t)$.
If $B=[b,d]$ with $0<b<d<\infty$, or $B=[b,\infty)$
with $b>0$, then
\begin{equation}\label{eq:uniform-rare-energy-probability}
	\sup_{t\in I}\left|N^2\mathbb P_{N,t}(B)-2c(t)m_v(B)\right|
	\longrightarrow0
\end{equation}
on every compact time interval $I$. Here $m_v(B)>0$,
and the probabilities are bounded above and below by
positive multiples of $N^{-2}$ for all large $N$, uniformly
in $t\in I$. The constants may depend on $I$ and $B$.

For a fixed finite window $B=[b,d]$, the conditional
distribution satisfies
\begin{equation}\label{eq:conditional-microscopic-energy}
	\sup_{\substack{t\in I\\x\in[b,d]}}
	\left|
	\frac{\mathbb P_{N,t}([b,x])}{\mathbb P_{N,t}([b,d])}
	-\frac{\displaystyle\int_b^x\rho_v(\lambda)\lambda^{-1}\d\lambda}
	{\displaystyle\int_b^d\rho_v(\lambda)\lambda^{-1}\d\lambda}
	\right|\longrightarrow0.
\end{equation}
Thus its limiting shape depends on $v$ and the window,
while the time-dependent contact determines the probability
of entering that window.
At the lower edge of the microscopic scale,
\begin{equation}\label{eq:evolved-threshold-probability}
	\lim_{b\downarrow0}\limsup_{N\to\infty}
	\sup_{t\in I}
	\left|b^{1/2}N^2\mathbb P_{N,t}([b,\infty))
	-16\sqrt2\,a^2c(t)\right|=0.
\end{equation}
The order of limits in this formula is part of the assertion.
\end{corollary}
\begin{proof}
Spectral calculus on the same positive operator gives
\[
	N^2\mathbb P_{N,t}(B)
	=\int_B\lambda^{-1}\d\nu_{N,t}(\lambda).
\]
For each stated window, the function
$h_B(\lambda)=\mathbf1_B(\lambda)/\lambda$ is bounded,
vanishes near zero, and has only the indicated endpoints
as possible discontinuities. Choose bounded continuous
upper and lower approximations that agree with $h_B$
outside small neighborhoods of those endpoints.
Their values at zero are zero. The measures $\nu_{N,t}$
have uniformly bounded mass by the same spectral theorem
with the test function $1$. The limiting error is bounded
by a constant times the $\mu_v$ mass of those neighborhoods,
uniformly on $I$, because $c(t)$ is bounded there.
That mass tends to zero by
Proposition~\ref{prop:radial-microscopic-density}.
The uniform continuous-test limit in
Theorem~\ref{thm:uniform-evolved-energy-spectrum}, or
Theorem~\ref{thm:admissible-initial-class} for the general class, therefore
proves \eqref{eq:uniform-rare-energy-probability}.
For the unbounded window, the same approximations equal
$1/\lambda$ above the lower endpoint neighborhood; no
unbounded test function is used.

The density is positive except at isolated points, so
$0<m_v(B)<\infty$. Moreover,
$0<\min_Ic\le\max_Ic<\infty$ by
Corollary~\ref{cor:contact-all-finite-times} for the prepared families,
and by Theorem~\ref{thm:admissible-initial-class} for the general class.
These facts give the uniform upper and lower probability
bounds and ensure that the conditional denominator is
positive for all sufficiently large $N$.

Apply the same endpoint approximation to $[b,x]$ for
each fixed $x\in[b,d]$. At $x=b$, the corresponding
rescaled singleton mass tends to zero, by continuous
majorants supported near $b$ and the absence of a limiting
atom there. Division by the uniformly positive rescaled
denominator gives convergence in
\eqref{eq:conditional-microscopic-energy} at each fixed $x$,
uniformly in $t$. The limiting distribution function is
continuous. A finite partition of $[b,d]$ on which its
increments are small, followed by monotonicity of all the
conditional distribution functions, makes the convergence
uniform also in $x$.

Finally, \eqref{eq:uniform-rare-energy-probability} for
$[b,\infty)$ gives, for each fixed $b>0$,
$N^2\mathbb P_{N,t}([b,\infty))\to2c(t)m_v(b)$
uniformly on $I$. Equation~\eqref{eq:radial-threshold-intensity}
implies $b^{1/2}m_v(b)\to8\sqrt2\,a^2$.
Taking the limits in the stated order proves
\eqref{eq:evolved-threshold-probability}.
\end{proof}

The quantitative state replacement from
Proposition~\ref{prop:evolved-energy-rate} also has a direct
probability interpretation. Define $\widetilde{\mathbb P}_{N,t}$
using $\widetilde u_N(t)$ from
\eqref{eq:energy-rate-comparison-vectors} and the same $Q_N(t)$.
For every $b>0$,
\begin{equation}\label{eq:rare-probability-transfer-rate}
	\sup_{\substack{t\in I\\B\subset[b,\infty)\ {\rm Borel}}}
	|\mathbb P_{N,t}(B)-\widetilde{\mathbb P}_{N,t}(B)|
	\le\frac{C_I}{bN^2}
	\bigl(\delta_N(I)^{1/2}+N^{-1/4}\bigr).
\end{equation}
Indeed, $\|\mathbf1_B/\lambda\|_\infty\le b^{-1}$,
so this follows directly from
\eqref{eq:evolved-spectral-transfer-rate}.
This compares the two measurement laws at the same finite $N$.

\subsection{Contact energy and resolvent convergence}

\begin{lemma}\label{spec:lem:trace}
Uniformly in $r\in\mathbb R^3$ and for $u\in C_c^\infty(\mathbb R^6)$,
\[
	\int|u(x,x-r)|^2\d x
	\le C\ip u{(1-\Delta_x)(1-\Delta_y)u}.
\]
It follows on Fock space that, as forms on $D(A)$,
\begin{equation}\label{spec:eq:potential-A2}
	\cV_N\le\frac C N A^2,
	\qquad \norm{\cV_N^{1/2}A^{-1}}^2\le C/N.
\end{equation}
\end{lemma}
\begin{proof}
Take the Fourier transform of the trace in $x$, with total momentum
$P=p+q$. Its phase depending on $r$ has modulus one.
Cauchy--Schwarz leaves the factor
\[
	\int\frac{\d p}{(1+|p|^2)(1+|P-p|^2)}
	\le\norm{(1+|p|^2)^{-1}}_{L^2(\mathbb R^3)}^2<\infty,
\]
uniformly in $P$. Integrating the other weighted factor over $P$
proves the assertion, with the harmless Fourier-normalization
constant included in $C$. Apply it also with Hilbert-space-valued
spectator coordinates, then integrate $r$ against $N^2v(Nr)$.
Since this potential has $L^1$ norm $N^{-1}\norm v_1$, on sector
$n$ one obtains
\[
	\cV_{N,n}\le\frac C N\sum_{i<j}(1-\Delta_i)(1-\Delta_j)
	\le\frac C{2N}\left(\sum_i(1-\Delta_i)\right)^2
	\le\frac C N A_n^2.
\]
The constants are independent of $n$. Closure and direct summation
give the operator-graph bound \eqref{spec:eq:potential-A2}.
\end{proof}

\begin{proposition}\label{spec:prop:inverse}
The model \eqref{spec:eq:phase} satisfies
\begin{equation}\label{spec:eq:inverse}
	\norm{Q_N^{-1}-A^{-1}}\le C/N,
	\qquad
	\norm{\mathscr H_N^{-1}-A^{-1}}
	\le C(N^{-1}+\epsilon^{2/3}).
\end{equation}
The second estimate requires only the $L^2$ sector estimate
\eqref{spec:eq:phase-L2}. In particular,
$\mathscr H_N\to A$ in norm resolvent.
\end{proposition}
\begin{proof}
Work first on a fixed sector, where the potential is bounded.
The positive resolvent formula gives
\[
	0\le A_n^{-1}-Q_{N,n}^{-1}
	\le A_n^{-1}\cV_{N,n}A_n^{-1}\le C/N.
\]
For example the middle difference is
$A_n^{-1}\cV_{N,n}^{1/2}
(1+\cV_{N,n}^{1/2}A_n^{-1}\cV_{N,n}^{1/2})^{-1}
\cV_{N,n}^{1/2}A_n^{-1}$, whose central factor is a contraction.
The bound is uniform in $n$, so direct summation proves the first
estimate.

Let $P_J=\mathbf1_{\{\cN\le J\}}$. Duhamel's formula and unitarity
imply
\[
	\norm{(U_N-1)A^{-1}P_J}
	\le\norm{C_NA^{-1}P_J}
	\le C\epsilon\norm{M^{3/2}A^{-1}P_J}
	\le C\epsilon\sqrt{J+1}.
\]
On the complementary input sectors, use $A\ge M$ to obtain
$\norm{(U_N-1)A^{-1}(1-P_J)}\le2/(J+1)$.
Choose an integer $J$ comparable to $\epsilon^{-2/3}$,
independently of the cutoff $L_N$. Thus
\begin{equation}\label{spec:eq:unitary-inverse}
	\norm{(U_N-1)A^{-1}}\le C\epsilon^{2/3}.
\end{equation}
Since $\mathscr H_N^{-1}=U_N^*Q_N^{-1}U_N$, expand
$U_N^*A^{-1}U_N-A^{-1}$ into two terms and use
\eqref{spec:eq:unitary-inverse} and its adjoint. This proves the second
bound. All these positive operators are at least one; convergence
of their inverses in norm is norm-resolvent convergence, equivalently
by applying $x\mapsto x/(1+x)$ to the inverses at the resolvent
point $-1$.
\end{proof}

\begin{proposition}\label{spec:prop:projected}
Let $q=1-|\phi\rangle\langle\phi|$ and $P=\Gamma(q)$. Denote by
$A_P,Q_{N,P}$ the self-adjoint operators of the restricted forms
on $P\cF$. Let $C_{N,P}=PC_NP$ on that space and $U_{N,P}=e^{C_{N,P}}$.
Then
\[
	\norm{Q_{N,P}^{-1}-A_P^{-1}}\le C_\phi/N,
	\qquad
	\norm{(U_{N,P}^*Q_{N,P}U_{N,P})^{-1}-A_P^{-1}}
	\le C_\phi(N^{-1}+\epsilon^{2/3}).
\]
The microscopic profile is given in
Theorem~\ref{spec:thm:projected-profile}.
\end{proposition}
\begin{proof}
On sector $n$, $P_n=q^{\otimes n}$ preserves $D(A_n)$ and
$[A_n,P_n]$ is bounded by $c_\phi n$. Indeed,
$[-\Delta,q]$ is a sum of two rank-one bounded operators when
$\phi\in H^2$, and the remaining coordinate projections are
contractions. The off-diagonal part of $A_n$ relative to $P_n$
is therefore bounded. Subtracting it gives a self-adjoint
block-diagonal operator on $D(A_n)$. Its restriction to $P_n$
is precisely $A_{P,n}$, with domain $P_nD(A_n)$. For a vector $u$
in this domain,
\[
	\norm{A_nu}\le\norm{A_{P,n}u}+c_\phi n\norm u
	\le(1+c_\phi)\norm{A_{P,n}u},
\]
since $A_{P,n}\ge n+1$. Thus Lemma~\ref{spec:lem:trace} restricts to
$\cV_{N,P,n}\le C_\phi N^{-1}A_{P,n}^2$, uniformly in $n$.
Apply the sectorwise positive resolvent formula and sum sectors.
The compressed phase retains \eqref{spec:eq:phase-L2}, because $P$
commutes with particle number and is a contraction in $L^2$.
Also $A_P\ge M$ and preserves particle number. The proof of
\eqref{spec:eq:unitary-inverse} therefore applies verbatim with $A_P$.
\end{proof}

\begin{corollary}
\label{spec:cor:dressed-inverse}
For the phases of Theorem~\ref{spec:thm:dressed-profile}, including
the extended actual phase of
Theorem~\ref{spec:thm:actual-COS-profile},
\[
	\norm{(\mathscr H_N^{\mathrm d})^{-1}-A_\Pi^{-1}}
	\le C(N^{-1}+\epsilon^{2/3}).
\]
Here $C$ may depend on the uniform Hilbert--Schmidt bounds of
the dressing kernels.
\end{corollary}
\begin{proof}
The $L^2$ monomial bound in the proof of
Theorem~\ref{spec:thm:dressed-profile} gives
$\norm{(\Pi C_N^{\mathrm d}\Pi)M^{-3/2}}\le C\epsilon$.
It also holds for the one-sided cutoff and generalized fields.
The low/high number split in
\eqref{spec:eq:unitary-inverse} therefore applies on $\Pi\cF$.
Combine it with the inverse estimate for $Q_{N,\Pi}$ in
Propositions~\ref{spec:prop:inverse}--\ref{spec:prop:projected}.
\end{proof}

\begin{corollary}\label{spec:cor:limits}
For a one-particle $h$ as in Theorem~\ref{spec:thm:profile},
\[
	\ip h{\mathscr H_Nh}_{\mathrm{form}}
	\longrightarrow q_1[h]+2D_v\norm{\phi h}^2,
	\qquad \mathscr H_N\longrightarrow A
	\quad\hbox{in norm resolvent}.
\]
For $\phi h\ne0$, the energy limit on the constant sequence
$h$ exceeds the energy of the norm-resolvent limit by
$2D_v\norm{\phi h}^2$.
\end{corollary}
\begin{proof}
Use $F=1$ in Theorem~\ref{spec:thm:profile} and Proposition~\ref{spec:prop:inverse}.
For a concrete recovery sequence take $h_N=U_N^*h$. It converges
to $h$ in $L^2$ because $\norm{(U_N^*-1)h}\le\norm{C_Nh}=O(\epsilon)$.
Its $\mathscr H_N$ energy is exactly $Q_N[h]=q_1[h]$.
This recovery sequence attains $q_1[h]$, while the constant
sequence retains the positive contact contribution.
\end{proof}

\subsection{The energy distribution at fixed scale}
\label{sec:evolved-energy-defect}

The microscopic contribution is also visible as a loss of
convergence of the first moment. At fixed energy, the
measurement law has a limit determined by the quadratic
reference state. Its mean omits a positive contribution
carried by rare events at energy of order $N^2$.
The following statement identifies both limits for the same
exactly evolved states.

\begin{proposition}
\label{prop:evolved-energy-defect}
Use any of the prepared families in
Theorems~\ref{thm:common-preparation},
\ref{thm:form-reference-preparation}
and~\ref{thm:arbitrary-reference-preparation}.
Write $\psi(t)=\xi_\infty(t)$ and
$E_\infty(t)=A_{\infty,t}[\psi(t)]$.
For every $h\in C_b([1,\infty))$ and compact interval $I$,
\begin{equation}\label{eq:unscaled-energy-law}
	\sup_{t\in I}\left|
	\langle u_N(t),h(Q_N(t))u_N(t)\rangle
	-\langle\psi(t),h(A_{\infty,t})\psi(t)\rangle
	\right|\longrightarrow0.
\end{equation}
For each $0<p<1$, the same conclusion holds for $h(x)=x^p$:
\begin{equation}\label{eq:sublinear-energy-moments}
	\sup_{t\in I}\left|
	\langle u_N(t),Q_N(t)^p u_N(t)\rangle
	-\langle\psi(t),A_{\infty,t}^p\psi(t)\rangle
	\right|\longrightarrow0.
\end{equation}
At the first moment, however,
\begin{equation}\label{eq:evolved-first-energy-defect}
	\sup_{t\in I}\left|
	Q_N(t)[u_N(t)]-E_\infty(t)-2D_vc(t)
	\right|\longrightarrow0.
\end{equation}
Here $D_v=\mu_v([0,\infty))>0$, and $c(t)>0$.

For a positive operator $B$, write $(B-R)_+$ for the
spectral function $\max(B-R,0)$. The contribution omitted
by the limiting measurement law is
\begin{equation}\label{eq:evolved-escaping-energy}
	\lim_{R\to\infty}\limsup_{N\to\infty}\sup_{t\in I}
	\left|\langle u_N(t),(Q_N(t)-R)_+u_N(t)\rangle
	-2D_vc(t)\right|=0.
\end{equation}
It is accounted for by the microscopic energy scale:
\begin{equation}\label{eq:evolved-microscopic-defect}
	\lim_{b\downarrow0}\limsup_{N\to\infty}\sup_{t\in I}
	\left|
	\langle Q_N(t)^{1/2}u_N(t),
	\mathbf1_{[bN^2,\infty)}(Q_N(t))Q_N(t)^{1/2}u_N(t)\rangle
	-2D_vc(t)\right|=0.
\end{equation}
Both statements use the indicated order of limits.
\end{proposition}
\begin{proof}
The operators $Q_N(t)$ and $\mathscr H_N(t)$ act on
$P_{N,t}\mathcal F$, and $A_{\infty,t}$ acts on
$P_{\infty,t}\mathcal F$. We retain the actual sector forms
above $N$.
The vectors $\xi_N(t)$ remain in the physical sectors.
Set
\[
	\mathscr H_N(t)=e^{-C_t}Q_N(t)e^{C_t},\qquad
	\widehat{\mathscr H}_N(t)=
	\mathcal R_N(t)^*\mathscr H_N(t)\mathcal R_N(t).
\]
Corollary~\ref{spec:cor:dressed-inverse}, with
$\epsilon=N^{-3/4}$, gives
\[
	\sup_{t\in I}
	\|\mathscr H_N(t)^{-1}-A_{N,t}^{-1}\|
	\le C_I N^{-1/2}.
\]
Its proof requires only a uniform $H^2$ bound on the
condensate and uniform Hilbert--Schmidt bounds on the
dressing kernels. These bounds hold on $I$.
The rotation estimate \eqref{eq:evolved-rotation-bounds}
also gives the form bound
\[
	\pm\bigl(\mathcal R_N^*A_{N,t}\mathcal R_N-A_{\infty,t}\bigr)
	\le C_I N^{-1}M\le C_I N^{-1}A_{\infty,t}.
\]
Put $a_N=C_I/N<1$. The inverse reverses the order of
positive forms, so
\[
	(1+a_N)^{-1}A_{\infty,t}^{-1}
	\le(\mathcal R_N^*A_{N,t}\mathcal R_N)^{-1}
	\le(1-a_N)^{-1}A_{\infty,t}^{-1}.
\]
The inverse difference is consequently bounded in norm by
$a_N/(1-a_N)$, since $A_{\infty,t}\ge1$.
Hence
\begin{equation}\label{eq:evolved-raw-inverse-limit}
	\sup_{t\in I}
	\|\widehat{\mathscr H}_N(t)^{-1}-A_{\infty,t}^{-1}\|
	\le C_I N^{-1/2}.
\end{equation}
Since $Q_N(t)$ and $A_{j,t}$ are at least one, all these
inverse operators are positive contractions.
Polynomial approximation on $[0,1]$ therefore proves
uniform norm convergence of their continuous functional
calculi \cite{ReedSimonI}. In particular it applies to
$h(B)$ whenever $h$ is continuous and has a limit at
infinity, by using $h(1/x)$ and its continuous value at
$x=0$.

The Hilbert-norm comparison gives
\[
	\sup_{t\in I}
	\|\mathcal R_N(t)^*e^{i\theta_N(t)}\xi_N(t)-\psi(t)\|
	\longrightarrow0.
\]
Unitary conjugation and the preceding functional calculus
then prove \eqref{eq:unscaled-energy-law} for such tests.
Theorem~\ref{thm:uniform-evolved-energy-spectrum}, applied
to the microscopic test $g=1$, gives
\eqref{eq:evolved-first-energy-defect} and
$\sup_{N\ {\rm large},\,t\in I}Q_N(t)[u_N(t)]<\infty$.
Also $\sup_I E_\infty<\infty$.
Thus the two unscaled probability measures have tails
above $R$ bounded by $C_I/R$. Choose a continuous
$0\le\chi_R\le1$ equal to one on $[1,R]$ and zero
above $2R$. The test $h\chi_R$ has a limit at infinity.
Replacing $h$ by $h\chi_R$ gives two truncation errors whose sum
is at most $2C_I\|h\|_\infty/R$. First let $N\to\infty$
and then $R\to\infty$ to prove
\eqref{eq:unscaled-energy-law} for every bounded
continuous $h$.

For $0<p<1$, the uniform first-energy bound gives
\[
	\langle u_N,Q_N^p\mathbf1_{(R,\infty)}(Q_N)u_N\rangle
	\le R^{p-1}Q_N[u_N]\le C_I R^{p-1}.
\]
The same estimate holds for $A_{\infty,t}$ and $\psi(t)$.
Apply \eqref{eq:unscaled-energy-law} to
$h_R(x)=\min(x^p,R^p)$, and then let $R\to\infty$.
This proves \eqref{eq:sublinear-energy-moments}.

The identity
$(Q_N-R)_+=Q_N-\min(Q_N,R)$ and the two established
limits give, for each fixed $R$,
\[
	\langle u_N,(Q_N-R)_+u_N\rangle
	\longrightarrow 2D_vc(t)
	+\langle\psi(t),(A_{\infty,t}-R)_+\psi(t)\rangle
\]
uniformly on $I$. By
Lemma~\ref{lem:quadratic-weighted-operator-domain},
$\sup_I\|A_{\infty,t}\psi(t)\|<\infty$. Spectral calculus gives
\[
	\langle\psi(t),(A_{\infty,t}-R)_+\psi(t)\rangle
	\le R^{-1}\|A_{\infty,t}\psi(t)\|^2\le C_I/R.
\]
This proves \eqref{eq:evolved-escaping-energy}.

For fixed $b>0$, the expectation in
\eqref{eq:evolved-microscopic-defect} is
$\nu_{N,t}([b,\infty))$.
The continuous upper and lower approximation used in
Corollary~\ref{cor:evolved-energy-probability} applies to
this indicator. The endpoint has zero $\mu_v$ mass by
Proposition~\ref{prop:radial-microscopic-density}.
Consequently,
\[
	\sup_{t\in I}
	|\nu_{N,t}([b,\infty))-2c(t)\mu_v([b,\infty))|
	\longrightarrow0.
\]
There is no atom of $\mu_v$ at zero, and its mass is
$D_v$. Letting $b\downarrow0$ proves
\eqref{eq:evolved-microscopic-defect}.
\end{proof}

The unscaled laws therefore fail to be uniformly integrable
at their first moment, although the excitation-energy moments of
each order $0<p<1$ converge to the corresponding
$A_{\infty,t}$ moment in $\psi(t)$.
Corollary~\ref{cor:uniform-positive-microscopic-moments}
gives divergence of every higher excitation-energy moment as
$N\to\infty$. For the same prepared many-body states,
Theorem~\ref{thm:D} bounds the third depletion and sixth
collective moments uniformly in $N$.

\subsection{Statistical mixtures and thermal references}
\label{sec:mixed-preparation}

The preparation also applies to a density matrix with
infinitely many occupied states. A uniform bound on the
preparation map is needed before summing the pure-state
estimates. Throughout this subsection, $\|\cdot\|_{\mathfrak S_1}$ and
$\|\cdot\|_{\mathrm{HS}}$ denote the trace and Hilbert--Schmidt norms.
For a positive observable $L$, its expectation means
$\operatorname{Tr}(L\omega)=\|L^{1/2}\omega^{1/2}\|_{\mathrm{HS}}^2$,
with the extended value allowed.
All reference density matrices act on the common initial
space $P\cF$.

\begin{lemma}
\label{lem:uniform-prepared-source}
Let $v\in C_c^\infty(\mathbb R^3)$ be nonzero, nonnegative and
radial. Fix an $N$-independent normalized
$\phi\in C_c^\infty(\mathbb R^3)$ and a sufficiently small fixed $\ell\in(0,1/2)$.
Let $q=1-|\phi\rangle\langle\phi|$, $P=\Gamma(q)$, and use
the Neumann pair kernel $\eta_N$ and its limit
$\eta_\infty$ in \eqref{spec:eq:actual-eta-limit}.
Let $C_N=\mathcal C_N^{\mathrm{COS}}$ be the phase
\eqref{spec:eq:actual-phase}, with $m=N^{-1/2}$.
Its real cutoff $0\le\Theta_N\le1$ equals one for
$n\le N^{1/2}/2+10$ and zero for $n\ge N^{1/2}-10$.
Extend $C_N$ by zero above sector $N$.
Use the centered initial energies
$\widehat F_{N,b}=\mathfrak J_N-\mathfrak J_N[\Omega]I+bM$
on $\Pi_NP\cF$ and $F_{\infty,b}=\mathcal J_{2,\infty}+bM$
on $P\cF$, where $\Pi_N=\mathbf1_{\{\cN\le N\}}$.
Choose a fixed $b$ so that, uniformly in $N$,
$c_bQ_N\le\widehat F_{N,b},F_{\rm ex}(0)\le C_bQ_N$
on the physical space and $c_bA\le F_{\infty,b}\le C_bA$.
Writing $H_b=K_P+bM$ and $S_\infty=F_{\infty,b}-H_b$, require
\[
	\|M^rS_\infty H_b^{-1}M^{-r}\|<\tfrac12,
	\qquad r=0,1,2.
\]
All these conditions hold for sufficiently large $b$, independently of $N$.
Put $T_N=\widehat F_{N,b}^{-1}\Pi_N$ and $T=F_{\infty,b}^{-1}$;
$T_N$ is zero on $(1-\Pi_N)P\cF$.
Put
\[
	S_N=T_N^3,\qquad S=T^3,\qquad
	V_N(t)=e^{i\theta_N(t)}U_{\rm ex}(t,0)S_N,\qquad
	V(t)=\mathcal U_{2,\infty}(t,0)S.
\]
These maps satisfy
\begin{align}
	\sup_{t\in I}\|V_N(t)-V(t)\|_{\mathrm{op}}
	&\le C_{I,b}N^{-3/28},\label{eq:prepared-source-hilbert}\\
	\sup_{t\in I}
	\|Q_N(t)^{1/2}(V_N(t)-\mathcal R_N(t)V(t))\|_{\mathrm{op}}
	&\le C_{I,b}N^{-3/56}.\label{eq:prepared-source-energy}
\end{align}
The first comparison uses the ambient Fock space. The second
is on $P_{N,t}\cF$, using the actual sector forms of
$Q_N(t)$ above $N$ and the unitary
$\mathcal R_N(t):P_{\infty,t}\cF\to P_{N,t}\cF$.
The constants are uniform over unit reference vectors.
\end{lemma}
\begin{proof}
The inverse and propagation estimates give
\begin{equation}\label{eq:prepared-source-domains}
	\sup_{N,t\in I}
	\|\widehat F_N(t)^{3/2}V_N(t)\|_{\mathrm{op}}
	+\sup_{t\in I}\|M^2A_{\infty,t}V(t)\|_{\mathrm{op}}
	\le C_{I,b}.
\end{equation}
For the first bound, spectral calculus gives
$\|\widehat F_{N,b}^{3/2}S_N\|\le C_b$ at time zero.
Theorem~\ref{thm:D}, the bounded scalar shifts and
Remark~\ref{rem:exact-number-shift} propagate this bound.
The second follows from
\eqref{eq:triple-inverse-regularity} and
Lemma~\ref{lem:quadratic-weighted-operator-domain}.
The comparison of compressed and ambient kinetic
operator domains is the one used in
Theorem~\ref{thm:form-reference-preparation}.

For $f\in P\cF$, take the unnormalized comparison vector
$v_N=\Pi_NSf$. It satisfies
\[
	\|\cK v_N\|+\|M^3v_N\|\le C_b\|f\|,\qquad
	\|S_Nf-v_N\|\le C_bN^{-3/28}\|f\|.
\]
The second bound follows from
\eqref{eq:triple-inverse-operator-rate} and the number
tail of $Sf$.
Indeed, the proof of \cite[Theorem 2.2]{COS} bounds the
squared comparison error for an unnormalized input $v$
by $C_I N^{-1/4}(\|\cK v\|^2+\|M^3v\|^2)$.
The proof of its Theorem 2.3 is homogeneous as well.
Apply these estimates to $v_N$, as in the proof of
Proposition~\ref{prop:prepared-quadratic-comparison}.
Removing the cubic phases gives an error bounded by
$C_I N^{-3/4}\|M^{3/2}v_N\|$, with the corresponding
number bound after quadratic evolution.
Unitarity controls the difference from $S_Nf$.
The resulting error is
$C_{I,b}(N^{-1/8}+N^{-3/28})\|f\|$.
This proves \eqref{eq:prepared-source-hilbert}, with
the same scalar phase for all $f$.

For the energy estimate use the form polarization in
Proposition~\ref{prop:evolved-energy-rate}, before
normalizing either vector. Put
\[
	w_N=\mathcal R_N^*V_N(t)f,\qquad
	\psi=V(t)f,\qquad e_N=w_N-\Pi_N\psi.
\]
The rotation and number-tail estimates give
$\|e_N\|\le C_{I,b}N^{-3/28}\|f\|$.
In \eqref{eq:energy-rate-polarization}, the full-operator
norms of $w_N$ and $\psi$ are at most $C_{I,b}\|f\|$,
by \eqref{eq:prepared-source-domains}.
The asymmetric form error and the coupling across
sector $N$ are bounded by
$C_{I,b}N^{-1/4}\|f\|\|\overline Q_N^{1/2}e_N\|$.
Coercivity and Young's inequality therefore give
\[
	\|\overline Q_N^{1/2}e_N\|^2
	\le C_{I,b}(N^{-3/28}+N^{-1/2})\|f\|^2.
\]
The omitted number tail has energy norm at most
$C_{I,b}(N+1)^{-3/2}\|f\|$.
Taking square roots proves \eqref{eq:prepared-source-energy}.
\end{proof}

\begin{theorem}
\label{thm:mixed-reference-preparation}
Let $v\in C_c^\infty(\mathbb R^3)$ be nonzero, nonnegative and
radial. Fix an $N$-independent normalized
$\phi\in C_c^\infty(\mathbb R^3)$ and a sufficiently small fixed $\ell\in(0,1/2)$.
Let $q=1-|\phi\rangle\langle\phi|$, $P=\Gamma(q)$, and use
the Neumann pair kernel $\eta_N$ and its limit
$\eta_\infty$ in \eqref{spec:eq:actual-eta-limit}.
Let $C_N=\mathcal C_N^{\mathrm{COS}}$ be the phase
\eqref{spec:eq:actual-phase}, with $m=N^{-1/2}$.
Its real cutoff $0\le\Theta_N\le1$ equals one for
$n\le N^{1/2}/2+10$ and zero for $n\ge N^{1/2}-10$.
Extend $C_N$ by zero above sector $N$.
Use the centered initial energies
$\widehat F_{N,b}=\mathfrak J_N-\mathfrak J_N[\Omega]I+bM$
on $\Pi_NP\cF$ and $F_{\infty,b}=\mathcal J_{2,\infty}+bM$
on $P\cF$, where $\Pi_N=\mathbf1_{\{\cN\le N\}}$.
Choose a fixed $b$ so that, uniformly in $N$,
$c_bQ_N\le\widehat F_{N,b},F_{\rm ex}(0)\le C_bQ_N$
on the physical space and $c_bA\le F_{\infty,b}\le C_bA$.
Writing $H_b=K_P+bM$ and $S_\infty=F_{\infty,b}-H_b$, require
\[
	\|M^rS_\infty H_b^{-1}M^{-r}\|<\tfrac12,
	\qquad r=0,1,2.
\]
All these conditions hold for sufficiently large $b$, independently of $N$.
Put $T_N=\widehat F_{N,b}^{-1}\Pi_N$ and $T=F_{\infty,b}^{-1}$;
$T_N$ is zero on $(1-\Pi_N)P\cF$.
Put $S_N=T_N^3$ and $S=T^3$. Let
$\varrho_N,\varrho$ be positive trace-class operators on
$P\cF$ with trace one and
$r_N=\|\varrho_N-\varrho\|_{\mathfrak S_1}\to0$.
Define, for all sufficiently large $N$,
\begin{gather}
	Z_N=\operatorname{Tr}(S_N\varrho_NS_N),\qquad
	Z=\operatorname{Tr}(S\varrho S)>0,\notag\\
	\omega_N(0)=Z_N^{-1}S_N\varrho_NS_N,\qquad
	\omega_\infty(0)=Z^{-1}S\varrho S.
	\label{eq:mixed-preparation}
\end{gather}
Then
\begin{gather}
	|Z_N-Z|\le C_b(N^{-3/28}+r_N),\notag\\
	\|Q_N(0)^{1/2}(\omega_N(0)-\omega_\infty(0))
	Q_N(0)^{1/2}\|_{\mathfrak S_1}
	\le C_{b,\varrho}(N^{-3/28}+r_N),\label{eq:mixed-initial-energy}\\
	\sup_N\operatorname{Tr}(F_{\rm ex}(0)^3\omega_N(0))
	<\infty.\label{eq:mixed-initial-third}
\end{gather}
The supremum is over the sufficiently large $N$ for which
the preparation is defined.

Evolve $\omega_N(0)$ by $U_{\rm ex}$ and
$\omega_\infty(0)$ by $\mathcal U_{2,\infty}$.
Set
\[
	\omega_N^{\rm raw}(t)=e^{C_t}\omega_N(t)e^{-C_t},\qquad
	\widetilde\omega_N^{\rm raw}(t)=
	e^{C_t}\mathcal R_N(t)\omega_\infty(t)
	\mathcal R_N(t)^*e^{-C_t}.
\]
On every compact $I$,
\begin{equation}\label{eq:mixed-evolved-energy-trace}
	\sup_{t\in I}
	\|Q_N(t)^{1/2}
	(\omega_N^{\rm raw}(t)-\widetilde\omega_N^{\rm raw}(t))
	Q_N(t)^{1/2}\|_{\mathfrak S_1}
	\le C_{I,b,\varrho}(N^{-3/56}+r_N).
\end{equation}
All sandwiched operators in these formulas are trace class.
The physical reconstructions satisfy the third depletion
and sixth collective-moment bounds
\eqref{eq:physical-depletion-third}--\eqref{eq:physical-sixth-moment},
with expectations replaced by traces.

Let $B_{t,x}$ denote the limiting dressed field in
\eqref{spec:eq:dressed-contact}, and define
\begin{align}
	E_\varrho(t)&=\operatorname{Tr}(A_{\infty,t}\omega_\infty(t)),\notag\\
	c_\varrho(t)&=
	Z^{-1}\int|\phi_{\infty,t}(x)|^2
	\|B_{t,x}V(t)\varrho^{1/2}\|_{\mathrm{HS}}^2\d x.
	\label{eq:mixed-contact}
\end{align}
These functions are continuous, and $\min_I c_\varrho>0$.
For every fixed $g\in C_b([0,\infty))$,
\begin{align}
	\sup_{t\in I}\biggl|
	&\operatorname{Tr}\!\left[
	g(Q_N(t)/N^2)Q_N(t)^{1/2}\omega_N^{\rm raw}(t)Q_N(t)^{1/2}
	\right]\notag\\
	&-E_\varrho(t)g(0)-2c_\varrho(t)\int g\d\mu_v
	\biggr|\longrightarrow0.
	\label{eq:mixed-microscopic-spectrum}
\end{align}
The fixed-window probabilities, conditional distributions,
and sequential threshold limit of
Corollary~\ref{cor:evolved-energy-probability} hold with
$c$ replaced by $c_\varrho$.
Proposition~\ref{prop:evolved-energy-defect} holds with
the limiting spectral law
$\operatorname{Tr}(h(A_{\infty,t})\omega_\infty(t))$ and
mean contribution $2D_vc_\varrho(t)$.
The excitation-energy lower bounds of
Corollary~\ref{cor:uniform-positive-microscopic-moments}
hold in trace form. The preparation is independent of $I$.
The constants may depend on the fixed $\varrho$ through
$Z^{-1}$.
\end{theorem}
\begin{proof}
The bounded maps $S_N,S$ satisfy
$\|S_N-S\|\le C_bN^{-3/28}$, and $S$ is injective.
In particular, $Z=\|S\varrho^{1/2}\|_{\mathrm{HS}}^2>0$.
For bounded maps $L_N,L$ and density matrices
$\varrho_N,\varrho$, the trace-ideal inequality
\cite{SimonTrace} gives
\begin{align}
	\|L_N\varrho_NL_N^*-L\varrho L^*\|_{\mathfrak S_1}
	&\le \|L_N\|^2 r_N
	+(\|L_N\|+\|L\|)\|L_N-L\|.
	\label{eq:mixed-trace-ideal}
\end{align}
This follows by first replacing $\varrho_N$ by $\varrho$
and then expanding the difference of the two products.
Apply it to $S_N,S$ to control $Z_N-Z$.
For large $N$, $Z_N\ge Z/2$.
Consequently,
$|Z_N^{-1}-Z^{-1}|\le 2Z^{-2}|Z_N-Z|$.
Apply it again to $Q_N(0)^{1/2}S_N$ and
$Q_N(0)^{1/2}S$, using
\eqref{eq:triple-inverse-operator-rate}.
Division by $Z_N,Z$ proves \eqref{eq:mixed-initial-energy}.
The constants may depend on $Z^{-1}$.

The positive trace convention gives
\[
	\operatorname{Tr}(F_{\rm ex}(0)^3\omega_N(0))
	=Z_N^{-1}\|F_{\rm ex}(0)^{3/2}S_N\varrho_N^{1/2}\|_{\mathrm{HS}}^2.
\]
The operator before $\varrho_N^{1/2}$ is uniformly
bounded, by the same spectral and scalar-shift argument
as in \eqref{eq:prepared-source-domains}.
Since $\|\varrho_N^{1/2}\|_{\mathrm{HS}}=1$, this proves
\eqref{eq:mixed-initial-third}.
Summing the nonnegative quadratic estimates in
Theorems~\ref{thm:D} and~\ref{thm:physical-sixth-moment}
over the columns of $S_N\varrho_N^{1/2}$ proves the
physical bounds. Finite column sums and monotone
convergence justify the unbounded observables.

At time $t$ apply \eqref{eq:mixed-trace-ideal} to
\[
	L_N(t)=Q_N(t)^{1/2}e^{C_t}V_N(t),\qquad
	L(t)=Q_N(t)^{1/2}e^{C_t}\mathcal R_N(t)V(t).
\]
Both have uniformly bounded operator norms on $I$.
This follows from \eqref{eq:prepared-source-domains},
coercivity, $Q_N\le C_IMA_{N,t}$, and the full-graph
bound for $e^{C_t}$ in Proposition~\ref{prop:full-graph}.
The same graph bound and
\eqref{eq:prepared-source-energy} give
$\sup_I\|L_N-L\|\le C_{I,b}N^{-3/56}$.
Their products with the reference density matrices
are the energy-weighted density matrices before division by
$Z_N,Z$. The scalar phase cancels.
In particular, the trace-class operators are defined by
these bounded-map factorizations.
Normalization now proves \eqref{eq:mixed-evolved-energy-trace}.

To pass to the final spectrum, write
$\varrho=\sum_j\lambda_j|f_j\rangle\langle f_j|$,
where $\lambda_j>0$, $\sum_j\lambda_j=1$ and
the $f_j$ are orthonormal.
For each fixed $j$, $Sf_j\ne0$ and
$V(t)f_j$ is the quadratic evolution of a regular
prepared vector, before normalization.
Its uniform microscopic spectral limit follows from
Theorem~\ref{thm:uniform-evolved-energy-spectrum}
and the energy comparison for this fixed source.
Thus every finite sum has the asserted limit.
The omitted energy-weighted trace is bounded by
\[
	Z^{-1}\|g\|_\infty\sup_{t\in I}\|L(t)\|^2
	\sum_{j>J}\lambda_j.
\]
The same tail bound holds for the limiting kinetic and
contact terms, by \eqref{eq:prepared-source-domains}.
Let $N\to\infty$ at fixed $J$, then $J\to\infty$.
The energy trace estimate transfers the result to
$\omega_N^{\rm raw}$, proving
\eqref{eq:mixed-microscopic-spectrum}.

Each nonzero $V(t)f_j$ has finite kinetic form energy,
so its contact is strictly positive by
Proposition~\ref{prop:contact-positive-kinetic}.
For the continuity assertion, the number-form bound
following \eqref{spec:eq:dressed-contact} gives
\[
	\sup_{t\in I}\left(
	A_{\infty,t}[V(t)f_j]+c_{B_t}[V(t)f_j]\right)
	\le C_{I,b}\qquad\text{for every }j.
\]
Here $\|f_j\|=1$, and
\eqref{eq:prepared-source-domains} controls both terms.
Each summand is continuous by graph continuity and
continuity of the dressed contact. The tails of their
weighted sums are at most
$C_{I,b}Z^{-1}\sum_{j>J}\lambda_j$, uniformly on $I$.
Uniform convergence of these sums proves continuity of
$E_\varrho,c_\varrho$.
Hence $c_\varrho(t)>0$ and $\min_Ic_\varrho>0$.
For each fixed positive window $B$,
\[
	N^2\operatorname{Tr}\!\left[
	\mathbf1_B(Q_N(t)/N^2)\omega_N^{\rm raw}(t)\right]
	=\nu_{N,t}^{\varrho}(\mathbf1_B/\lambda),
\]
where $\nu_{N,t}^{\varrho}$ is the measure on the
left-hand side of \eqref{eq:mixed-microscopic-spectrum}.
The test is bounded and vanishes near zero.
Continuous upper and lower approximations at its
endpoints have limiting error controlled by
$\max_Ic_\varrho$ and the $\mu_v$ mass near those endpoints.
That mass tends to zero. The probability limit follows,
and $\min_Ic_\varrho>0$ controls the conditional denominator.
The finite-partition argument of
Corollary~\ref{cor:evolved-energy-probability} gives its
uniform conditional distribution and the same sequential
threshold limit.

For the unscaled law, apply
Proposition~\ref{prop:evolved-energy-defect} to each
fixed source and then use the same finite-rank argument.
For bounded tests, ordinary trace errors are controlled
by the energy trace norm: if $Y=Q_N^{1/2}XQ_N^{1/2}$,
then $X=Q_N^{-1/2}YQ_N^{-1/2}$ and
$\|X\|_{\mathfrak S_1}\le\|Y\|_{\mathfrak S_1}$ because $Q_N\ge1$.
The first-energy bound controls fractional-moment tails.
Also
\[
	\operatorname{Tr}(A_{\infty,t}^2\omega_\infty(t))
	=Z^{-1}\|A_{\infty,t}V(t)\varrho^{1/2}\|_{\mathrm{HS}}^2
	\le C_{I,b,\varrho}.
\]
This supplies the uniform tail estimate needed for the
sequential mean-energy contribution.
Finally, a fixed positive microscopic window gives
the higher excitation-energy lower bounds exactly as in
Corollary~\ref{cor:uniform-positive-microscopic-moments}.
\end{proof}

A concrete reference is a free Bose Gibbs state on the
excitation Fock space \cite[Chapter 17]{DerezinskiGerard}.
Let $h\ge\kappa>0$ on $qL^2$ and suppose
$e^{-\beta h}$ is trace class for some fixed $\beta>0$.
The occupation-number basis gives
\begin{equation}\label{eq:thermal-reference-state}
	\varrho_\beta=\mathcal Z_\beta^{-1}
	\Gamma(e^{-\beta h}),\qquad
	\mathcal Z_\beta=\prod_j(1-e^{-\beta\varepsilon_j})^{-1}<\infty,
\end{equation}
where $\varepsilon_j$ are the eigenvalues of $h$.
The product is finite because
$-\log(1-x)\le x/(1-e^{-\beta\kappa})$
for $0\le x\le e^{-\beta\kappa}$.
The same product with $e^{-\beta\varepsilon_j}$ replaced
by $e^{\tau-\beta\varepsilon_j}$ proves
$\operatorname{Tr}(e^{\tau\cN}\varrho_\beta)<\infty$
for $0<\tau<\beta\kappa$.
For example, the form compression of
$-\Delta+|x|^2+1$ to $qL^2$ satisfies these conditions
for every fixed $\beta>0$. The min--max principle bounds
its heat trace by that of the three-dimensional harmonic
oscillator.

Taking $\varrho_N=\varrho_\beta$ in
\eqref{eq:mixed-preparation} therefore gives a prepared
thermal reference with the same moment and spectral
conclusions. The reference temperature and trap are fixed,
giving excitation populations of order one.
Positive-temperature evolution with a macroscopic
thermal cloud has been studied in the mean-field regime
\cite{DCS2024}.

\subsection{Energy statistics from independent preparations}
\label{sec:energy-sampling}

The spectral measure also determines the energy recorded
over many independent preparations. Measurement distributions
of quantum gases contain information beyond their mean;
for example, particle-number statistics after a quench
have been studied in one-dimensional gases \cite{HR2024}.
Non-Gaussian work fluctuations and algebraic tails also
arise after an interaction quench \cite{PPG2019}.
Here the observable is the excitation energy $Q_N(t)$.
We consider one measurement on each independently prepared
copy at the same time $t$.

Let $\sigma_N(t)$ be either
$|u_N(t)\rangle\langle u_N(t)|$ from
Proposition~\ref{prop:evolved-energy-defect}, or the
mixed state $\omega_N^{\rm raw}(t)$ from
Theorem~\ref{thm:mixed-reference-preparation}.
Write $\pi_{N,t}$ for its unscaled energy measurement law.
Use $E(t)=E_\infty(t)$ and $c(t)$ in the pure case,
and $E(t)=E_\varrho(t)$ and $c(t)=c_\varrho(t)$ in
the mixed case. The corresponding limiting law of
$A_{\infty,t}$ is denoted by $\pi_{\infty,t}$.
Thus $\int x\d\pi_{\infty,t}(x)=E(t)$.
Proposition~\ref{prop:radial-microscopic-density} identifies
the entire positive-energy measure:
$\mu_v(\d\lambda)=\rho_v(\lambda)\d\lambda$ on $(0,\infty)$.
There is no singular continuous part.

\begin{theorem}
\label{thm:independent-energy-sampling}
Let $X_{N,t}^{(1)},\ldots,X_{N,t}^{(m_N)}$ be independent
random variables with law $\pi_{N,t}$, where
$m_N\to\infty$ is an integer sequence. Define
\[
	\overline X_{N,t}=\frac1{m_N}
	\sum_{j=1}^{m_N}X_{N,t}^{(j)},\qquad
	\tau_N=\frac{m_N}{N^2}.
\]
On every compact time interval $I$, the following hold.
If $\tau_N\to0$, then for every $\epsilon>0$,
\begin{equation}\label{eq:energy-sampling-few}
	\sup_{t\in I}
	\mathbb P\bigl(|\overline X_{N,t}-E(t)|>\epsilon\bigr)
	\longrightarrow0.
\end{equation}
If $\tau_N\to\tau\in(0,\infty)$, the sample mean converges
in law to a nonnegative random variable $Y_{\tau,t}$ with
\begin{equation}\label{eq:energy-sampling-laplace}
	\mathbb E e^{-sY_{\tau,t}}
	=\exp\!\left\{-sE(t)-2\tau c(t)
	\int_0^\infty\frac{1-e^{-s\lambda/\tau}}{\lambda}
	\d\mu_v(\lambda)\right\},\qquad s\ge0.
\end{equation}
Convergence is uniform in $t\in I$ for each bounded
continuous test function. The limiting law is infinitely
divisible, with deterministic contribution $E(t)$ and
L\'evy measure
\begin{equation}\label{eq:energy-sampling-levy}
	\Lambda_{\tau,t}(\d x)
	=2\tau c(t)\frac{\rho_v(\tau x)}{x}\d x,\qquad x>0.
\end{equation}
Its mean and variance are
\begin{equation}\label{eq:energy-sampling-limit-moments}
	\mathbb E Y_{\tau,t}=E(t)+2D_vc(t),\qquad
	\operatorname{Var}(Y_{\tau,t})
	=\frac{2c(t)}{\tau}\|v\|_2^2>0.
\end{equation}
If $\tau_N\to\infty$, then
\begin{equation}\label{eq:energy-sampling-many}
	\sup_{t\in I}\mathbb E
	\bigl|\overline X_{N,t}-E(t)-2D_vc(t)\bigr|
	\longrightarrow0.
\end{equation}
For every sequence $m_N$, the mean of $\overline X_{N,t}$
converges uniformly to $E(t)+2D_vc(t)$.
Equation~\eqref{eq:energy-sampling-limit-moments} gives the
moments of the limiting law.
\end{theorem}
\begin{proof}
Let $\nu_{N,t}$ denote the energy-weighted law at scale
$N^2$. The pure and mixed spectral theorems give
\begin{equation}\label{eq:energy-sampling-input}
	\nu_{N,t}(g)\longrightarrow
	E(t)g(0)+2c(t)\mu_v(g),\qquad
	g\in C_b([0,\infty)),
\end{equation}
uniformly on $I$. In particular,
$\int x\d\pi_{N,t}(x)=\nu_{N,t}(1)$ is uniformly
bounded for large $N$.
The measures $\nu_{N,t}$ are also uniformly tight in
this limit. To see this, use continuous cutoffs equal
to one above a large fixed microscopic energy.
Their limiting integrals are controlled by the tail
of the finite measure $\mu_v$, uniformly on $I$.
We will also use the unscaled weak convergence to
$\pi_{\infty,t}$ and
$\sup_I\int x^2\d\pi_{\infty,t}(x)<\infty$,
proved in Proposition~\ref{prop:evolved-energy-defect}
and Theorem~\ref{thm:mixed-reference-preparation}.

For fixed $s>0$, put
\begin{align}
	a_{N,t}(s)&=m_N\int(1-e^{-sx/m_N})\d\pi_{N,t}(x)
	=\nu_{N,t}(g_{\tau_N,s}),\notag\\
	g_{\alpha,s}(\lambda)&=
	\alpha\frac{1-e^{-s\lambda/\alpha}}{\lambda},
	\qquad g_{\alpha,s}(0)=s.
	\label{eq:energy-sampling-test}
\end{align}
These tests satisfy $0\le g_{\alpha,s}\le s$.
Independence gives the exact identity
\[
	\mathbb E e^{-s\overline X_{N,t}}
	=\left(1-\frac{a_{N,t}(s)}{m_N}\right)^{m_N}.
\]
Since $a_{N,t}(s)$ is bounded uniformly on $I$,
the logarithm of the right-hand side is
$-a_{N,t}(s)+O_I(m_N^{-1})$.
This identity also follows by functional calculus
for the commuting observables $Q_N(t)$ on the
different tensor factors of $\sigma_N(t)^{\otimes m_N}$.

Suppose first that $\tau_N\to\tau\in(0,\infty)$.
The tests $g_{\tau_N,s}$ converge uniformly on each
compact energy interval to $g_{\tau,s}$.
For $h_N=g_{\tau_N,s}$ and $h=g_{\tau,s}$, the
replacement is controlled explicitly by
\begin{equation}\label{eq:energy-sampling-moving-test}
	\sup_{t\in I}|\nu_{N,t}(h_N-h)|
	\le C_I\sup_{0\le\lambda\le L}|h_N(\lambda)-h(\lambda)|
	+2s\sup_{t\in I}\nu_{N,t}((L,\infty)).
\end{equation}
First take $N\to\infty$ at fixed $L$, then $L\to\infty$.
The local uniform convergence and the uniform
asymptotic tightness make both terms vanish.
We may thus use the fixed test $g_{\tau,s}$ in
\eqref{eq:energy-sampling-input}.
This proves \eqref{eq:energy-sampling-laplace}.
The jump measure is the pushforward
\[
	\Lambda_{\tau,t}(B)=2\tau c(t)
	\int_0^\infty\mathbf1_B(\lambda/\tau)
	\frac{\d\mu_v(\lambda)}{\lambda}.
\]
Absolute continuity of $\mu_v$ and $\lambda=\tau x$
give \eqref{eq:energy-sampling-levy}. This measure satisfies
\[
	\int_0^\infty x\d\Lambda_{\tau,t}(x)=2c(t)D_v,\qquad
	\int_0^\infty x^2\d\Lambda_{\tau,t}(x)
	=\frac{2c(t)}{\tau}\int_0^\infty\lambda\d\mu_v(\lambda)
	=\frac{2c(t)}{\tau}\|v\|_2^2.
\]
It is therefore a L\'evy measure with finite first
and second moments. The L\'evy--Khintchine formula
\cite[Theorem 1.2]{Bertoin2000}, or a sum of Poisson
jumps with a lower cutoff followed by monotone
convergence, constructs the stated law.
The omitted small jumps tend to zero in $L^2$,
since both their mean and their variance tend to zero.
The Poisson mean and variance formulas then give
\eqref{eq:energy-sampling-limit-moments}.
The nonzero variance and the lower support bound
$Y_{\tau,t}\ge E(t)$ show that this law is not Gaussian.

If $\tau_N\to\infty$, then $g_{\tau_N,s}\to s$
uniformly on compact energy intervals.
Apply \eqref{eq:energy-sampling-moving-test} with
$h_N=g_{\tau_N,s}$ and $h=s$. It gives
$a_{N,t}(s)\to s(E(t)+2D_vc(t))$ uniformly on $I$.
The limiting Laplace transform is that of the
constant $E(t)+2D_vc(t)$.

It remains to consider $\tau_N\to0$.
The test $g_{\tau_N,s}$ develops a boundary layer at zero.
We control this layer using the unscaled law together with
\eqref{eq:energy-sampling-input}.
Choose a continuous $0\le\chi_R\le1$, equal to one
on $[1,R]$ and zero above $2R$. Define
\[
	b_R(t)=\int x\chi_R(x)\d\pi_{\infty,t}(x).
\]
The cutoff transition contributes only to the same tail:
\[
	0\le E(t)-b_R(t)
	\le\int_{(R,\infty)}x\d\pi_{\infty,t}(x)
	\le R^{-1}\int x^2\d\pi_{\infty,t}(x)\le C_I/R.
\]
On the support of $\chi_R$,
$m_N(1-e^{-sx/m_N})\to sx$ uniformly.
Consequently the contribution with $\chi_R$ to
$a_{N,t}(s)$ tends uniformly to $sb_R(t)$.

For fixed $b>0$, the contribution from $x>bN^2$
is at most
\[
	m_N\pi_{N,t}((bN^2,\infty))
	\le \frac{\tau_N}{b}\nu_{N,t}(1)\longrightarrow0.
\]
For $N$ large enough that $2R<bN^2$, the remaining
contribution is nonnegative and bounded by
\[
	s\left(\nu_{N,t}([0,b])
	-\int x\chi_R(x)\d\pi_{N,t}(x)\right).
\]
Here $[0,b]$ refers to $\lambda=x/N^2$; the support
of $\chi_R$ lies in $x\le bN^2$.
Choose continuous upper and lower approximations to
$\mathbf1_{[0,b]}$ that both equal one near zero
and differ only within a $\delta$ neighborhood of $b$.
Their limiting gap is bounded by
$2\max_Ic\,\mu_v([b-\delta,b+\delta])$ for $0<\delta<b$.
This tends to zero with $\delta$. Hence, uniformly on $I$,
$\nu_{N,t}([0,b])\to E(t)+2c(t)\mu_v([0,b])$.
It follows that
\[
	\limsup_{N\to\infty}\sup_{t\in I}
	|a_{N,t}(s)-sE(t)|
	\le \frac{C_{I,s}}{R}
	+2s\max_Ic\,\mu_v([0,b]).
\]
Let $R\to\infty$ and then $b\downarrow0$.
Since $\mu_v(\{0\})=0$, this proves
$a_{N,t}(s)\to sE(t)$ uniformly on $I$.

To pass from Laplace transforms to the stated
uniform limits, note that the sample means are
nonnegative and have uniformly bounded first moments.
Their laws are therefore tight. Any sequence
$t_N\in I$ has a subsequence converging to some
$t_*\in I$. The Laplace limits and continuity of
$E,c$ identify every weak subsequential limit.
Uniqueness follows, for example, by polynomial
approximation after the change of variable $y=e^{-x}$.
The limiting laws themselves depend weakly
continuously on $t$. If convergence for a fixed bounded
continuous test were not uniform, a sequence with an
error bounded away from zero would yield such a
subsequence $t_N\to t_*$, a contradiction.
In the deterministic cases the same argument gives
convergence in probability to the continuous functions
$E(t)$ and $E(t)+2D_vc(t)$, respectively.

Finally, linearity of expectation gives
$\mathbb E\overline X_{N,t}=\int x\d\pi_{N,t}(x)
\to E(t)+2D_vc(t)$.
In the case $\tau_N\to\infty$, put
$a(t)=E(t)+2D_vc(t)$. The identity
\[
	\mathbb E|\overline X_{N,t}-a(t)|
	=\mathbb E\overline X_{N,t}+a(t)
	-2\mathbb E\min(\overline X_{N,t},a(t))
\]
proves the claim along every sequence $t_N\to t_*$:
replacing $a(t_N)$ by $a(t_*)$ changes the minimum
by at most $|a(t_N)-a(t_*)|$, and the resulting
test is bounded and continuous. Compactness of $I$ yields
\eqref{eq:energy-sampling-many}.
\end{proof}

The limiting jump measure has infinite mass near zero.
Indeed, the threshold density gives
\[
	\frac{\d\Lambda_{\tau,t}}{\d x}
	=8\sqrt{2\tau}\,a^2c(t)x^{-3/2}
	+O_{I,\tau}(x^{-1/2}),\qquad x\downarrow0.
\]
Its total energy is finite, as the first integral
in the proof shows. Thus at order $N^2$ repetitions,
the sample mean retains nonzero statistical
fluctuations determined by the scattering measure.
With $m_N\to\infty$ but $m_N/N^2\to0$, its typical
value omits the contact contribution even though
its expectation includes it. With $m_N/N^2\to\infty$,
the sample mean converges to the full limiting mean.
The limit describes independent preparations at a fixed
observation time.

The physical collective observables of the same prepared
families have a different sampling behavior. Let
$Z_{O,N,t}^{(j)}$ be independent outcomes of
$\mathcal X_{O,N}(t)$ in the physical reconstruction,
and let $\overline Z_{O,N,t}$ be their sample mean.
Theorem~\ref{thm:physical-sixth-moment} and its mixed
extension give a uniform centered sixth moment.
H\"older's inequality and independence therefore imply
\begin{equation}\label{eq:collective-sampling-concentration}
	\sup_{t\in I}\mathbb E
	\left|\overline Z_{O,N,t}
	-\mathbb E Z_{O,N,t}^{(1)}\right|^2
	=\frac1{m_N}\sup_{t\in I}
	\operatorname{Var}(Z_{O,N,t}^{(1)})
	\le\frac{C_I\|O\|^2}{m_N}.
\end{equation}
The constant may depend on the fixed preparation.
Thus these averages concentrate for every $m_N\to\infty$,
including the regime where the excitation-energy sample mean
has the nondegenerate limit \eqref{eq:energy-sampling-laplace}.
The two observables are measured on separate ensembles.

\subsection{Initial states with a uniform third energy moment}
\label{sec:admissible-initial-class}

The inverse-energy constructions give explicit initial states.
The qualitative energy and spectral limits require only norm
convergence and a uniform third correlated-energy moment.
We now characterize the possible limits of such families.

We first give the form-domain argument used below. All spaces
in the lemma are fixed; in the time-dependent application,
the direct rotations identify them with one excitation space.
Fractional powers and strong resolvent convergence are used
in their spectral-theorem sense \cite{ReedSimonI}.

\begin{lemma}
\label{lem:third-energy-completion}
Let $A\ge M\ge1$ be strongly commuting positive operators.
Let $\Pi_j\to1$ be orthogonal projections commuting with
$A,M$, and let $G_j,F\ge c>0$ be self-adjoint.
Assume that $\Pi_j$ reduces $G_j$, that $G_j\to F$ in the
strong resolvent sense, and that
\begin{equation}\label{eq:completion-form-consistency}
	G_j[\Pi_jv]\longrightarrow F[v]\quad(v\in D(M^2A)),
	\qquad \|M^2AF^{-3}f\|\le C\|f\|.
\end{equation}
Let $Q_j$ be positive operators strongly commuting with $M$
and reduced by $\Pi_j$. Suppose, with uniform constants, that
\begin{gather}\label{eq:completion-coercivity}
	A\le C Q_j,\qquad Q_j\le C MA,\qquad
	Q_j\le C G_j\ \text{on }\operatorname{Ran}\Pi_j,\\
	\|M Q_j^{1/2}z\|\le C\|G_j^{3/2}z\|,
	\qquad z\in\operatorname{Ran}\Pi_j\cap D(G_j^{3/2}).\notag
\end{gather}
The inequalities in the first line are form inequalities.
Put $R_j=G_j^{-3/2}\Pi_j$ and $R=F^{-3/2}$. Then
\begin{equation}\label{eq:completion-inverse-energy}
	\|MA^{1/2}Rf\|\le C\|f\|,\qquad
	\|Q_j^{1/2}(R_j-R)f\|\longrightarrow0
	\quad(f\in\mathcal H).
\end{equation}
If $x_j=\Pi_jx_j\to x$ in norm and
$\sup_j\|G_j^{3/2}x_j\|<\infty$, then
\begin{equation}\label{eq:completion-state-energy}
	x\in D(F^{3/2}),\qquad
	\|F^{3/2}x\|^2\le\liminf_j\|G_j^{3/2}x_j\|^2,\qquad
	\|Q_j^{1/2}(x_j-x)\|\longrightarrow0.
\end{equation}
\end{lemma}
\begin{proof}
The common lower bound and functional calculus give
$R_jf\to Rf$ in norm. The weighted coercivity and strong
commutation in \eqref{eq:completion-coercivity} imply
\[
	\|MA^{1/2}R_jf\|\le C\|M Q_j^{1/2}R_jf\|
	\le C\|\Pi_jf\|.
\]
Indeed, $R_jf$ belongs to the physical $G_j^{3/2}$ domain,
since $G_j^{3/2}R_jf=\Pi_jf$. The coercivity assumption
therefore supplies its weighted $Q_j$ form norm.
For $E_L=\mathbf1_{\{M\le L\}}$, apply $A\le C Q_j$ to
$E_LMR_jf\in D(Q_j^{1/2})$. Strong commutation gives
\[
	\|A^{1/2}E_LMR_jf\|
	\le C\|Q_j^{1/2}E_LMR_jf\|
	\le C\|M Q_j^{1/2}R_jf\|.
\]
Monotone convergence in the joint spectral calculus of
$A,M$ proves the displayed weighted bound.
Weak closedness of the graph of $MA^{1/2}$ gives the
first bound in \eqref{eq:completion-inverse-energy}.
Consequently,
\[
	\sup_j\bigl(\|Q_j^{1/2}R_j\|_{\mathrm{op}}
	+\|Q_j^{1/2}R\|_{\mathrm{op}}\bigr)<\infty.
\]
For the second term, the bound just proved places $Rf$
in $D(M^{1/2}A^{1/2})$. The form inequality $Q_j\le C MA$
then applies directly.

It suffices to prove convergence for the dense set of sources
$f=F^{-3/2}h$. For these sources, $v=Rf=F^{-3}h$
belongs to $D(M^2A)$. Functional calculus gives
\[
	G_j^{1/2}R_jf=G_j^{-1}\Pi_jf
	\longrightarrow F^{-1}f=F^{1/2}v.
\]
The consistency assumption gives convergence of the norms
of $G_j^{1/2}\Pi_jv$. For every $k\in\mathcal H$,
\[
	\langle G_j^{1/2}\Pi_jv,G_j^{-1/2}k\rangle
	=\langle\Pi_jv,k\rangle\longrightarrow\langle v,k\rangle.
\]
Since $G_j^{-1/2}k\to F^{-1/2}k$ and the range of
$F^{-1/2}$ is dense, every weak limit is $F^{1/2}v$.
Convergence of the norms therefore gives strong convergence.
Subtracting and using physical-block coercivity yields
$\|Q_j^{1/2}(R_jf-\Pi_jv)\|\longrightarrow0$.
The remaining tail tends to zero because
\[
	\|Q_j^{1/2}(1-\Pi_j)v\|
	\le C\|M^{1/2}A^{1/2}(1-\Pi_j)v\|\longrightarrow0.
\]
Here $\Pi_j$ commutes with $A,M$, so strong convergence
of $\Pi_j$ also holds in this graph norm for each fixed $v$.
The uniform operator bounds extend this conclusion to all
sources by density.

For $x_j\to x$, apply strong resolvent convergence to
$\min(\lambda^3,L)$ on the common positive spectral half-line.
This function differs from a constant by a continuous
function vanishing at infinity. Its expectations converge.
Letting $L\to\infty$ proves the domain and lower bound in
\eqref{eq:completion-state-energy}. Set $f=F^{3/2}x$
and $e_j=x_j-R_jf$. Then $e_j\to0$ and
$G_j^{3/2}R_jf=\Pi_jf$ gives
$\sup_j\|G_j^{3/2}e_j\|<\infty$. Spectral interpolation
for this same operator gives
\[
	\|Q_j^{1/2}e_j\|
	\le C\|G_j^{1/2}e_j\|
	\le C\|e_j\|^{2/3}\|G_j^{3/2}e_j\|^{1/3}
	\longrightarrow0.
\]
Together with \eqref{eq:completion-inverse-energy}, this
proves the last assertion.
\end{proof}

Return to a common, $N$-independent normalized initial condensate
$\phi_0\in C_c^\infty(\mathbb R^3)$ and a nonzero, nonnegative
radial potential $v\in C_c^\infty(\mathbb R^3)$.
Use the quadratic and cubic phases of
Section~\ref{sec:microscopic-spectrum}, with $m=N^{-1/2}$.
Write
\[
	H_N=\widehat F_{N,b},\qquad H=F_{\infty,b},\qquad
	A=K_P+M.
\]
The physical operator $H_N$ acts below particle number $N$.
Choose $b$ as in Theorem~\ref{thm:arbitrary-reference-preparation}.
As before, its full-space extension is
$H_N\oplus(K_P+bM)_{>N}$, whereas $Q_N$ retains its
actual sector forms on all sectors.

\begin{theorem}
\label{thm:admissible-initial-class}
Let $v\in C_c^\infty(\mathbb R^3)$ be nonzero, nonnegative and
radial. Fix an $N$-independent normalized
$\phi\in C_c^\infty(\mathbb R^3)$ and a sufficiently small fixed $\ell\in(0,1/2)$.
Let $q=1-|\phi\rangle\langle\phi|$, $P=\Gamma(q)$, and use
the Neumann pair kernel $\eta_N$ and its limit
$\eta_\infty$ in \eqref{spec:eq:actual-eta-limit}.
Let $C_N=\mathcal C_N^{\mathrm{COS}}$ be the phase
\eqref{spec:eq:actual-phase}, with $m=N^{-1/2}$.
Its real cutoff $0\le\Theta_N\le1$ equals one for
$n\le N^{1/2}/2+10$ and zero for $n\ge N^{1/2}-10$.
Extend $C_N$ by zero above sector $N$.
Use the centered initial energies
$\widehat F_{N,b}=\mathfrak J_N-\mathfrak J_N[\Omega]I+bM$
on $\Pi_NP\cF$ and $F_{\infty,b}=\mathcal J_{2,\infty}+bM$
on $P\cF$, where $\Pi_N=\mathbf1_{\{\cN\le N\}}$.
Choose a fixed $b$ so that, uniformly in $N$,
$c_bQ_N\le\widehat F_{N,b},F_{\rm ex}(0)\le C_bQ_N$
on the physical space and $c_bA\le F_{\infty,b}\le C_bA$.
Writing $H_b=K_P+bM$ and $S_\infty=F_{\infty,b}-H_b$, require
\[
	\|M^rS_\infty H_b^{-1}M^{-r}\|<\tfrac12,
	\qquad r=0,1,2.
\]
All these conditions hold for sufficiently large $b$, independently of $N$.
Write $H_N=\widehat F_{N,b}$ and $H=F_{\infty,b}$.
A normalized vector $\psi\in P\cF$
is the norm limit of normalized physical states
$z_N\in\Pi_NP\cF$ satisfying
\begin{equation}\label{eq:admissible-initial-bound}
	z_N\to\psi,\qquad
	\sup_{N\ge N_0}\|H_N^{3/2}z_N\|<\infty
\end{equation}
if and only if $\psi\in D(H^{3/2})$.
For each such $\psi$, the states
\begin{equation}\label{eq:admissible-recovery}
	z_N^{\rm r}=
	\frac{H_N^{-3/2}\Pi_N H^{3/2}\psi}
	{\|H_N^{-3/2}\Pi_N H^{3/2}\psi\|}
\end{equation}
are defined for all sufficiently large $N$ and satisfy
\begin{equation}\label{eq:admissible-optimal-energy}
	z_N^{\rm r}\to\psi,\qquad
	\|H_N^{3/2}z_N^{\rm r}\|^2\longrightarrow
	\|H^{3/2}\psi\|^2.
\end{equation}
Every family in \eqref{eq:admissible-initial-bound}, whether
or not it is given by \eqref{eq:admissible-recovery}, obeys
\[
	\|Q_N^{1/2}(z_N-\psi)\|\longrightarrow0.
\]

For any such family, set
$\xi_N(t)=U_{\rm ex}(t,0)z_N$ and
$\psi(t)=\mathcal U_{2,\infty}(t,0)\psi$.
For every compact time interval $I$, the third depletion
moment and sixth collective moments in Theorem~\ref{thm:D}
are bounded uniformly in $N$ and $t\in I$.
Moreover,
\begin{equation}\label{eq:admissible-evolved-energy}
	\sup_{t\in I}\|Q_N(t)^{1/2}
	(e^{i\theta_N(t)}\xi_N(t)-\mathcal R_N(t)\psi(t))\|
	\longrightarrow0.
\end{equation}
With $u_N(t)=e^{C_t}\xi_N(t)$ and the contact
$c_\psi(t)$ defined by \eqref{eq:evolved-contact} with
$\xi_\infty(t)=\psi(t)$, every fixed
$g\in C_b([0,\infty))$ satisfies
\begin{align}\label{eq:admissible-evolved-spectrum}
	\sup_{t\in I}\bigg|&
	\langle Q_N(t)^{1/2}u_N(t),
	g(Q_N(t)/N^2)Q_N(t)^{1/2}u_N(t)\rangle\notag\\
	&-A_{\infty,t}[\psi(t)]g(0)
	-2c_\psi(t)\int g\d\mu_v\bigg|\longrightarrow0.
\end{align}
The contact is continuous and $\min_{t\in I}c_\psi(t)>0$.
The lower bound \eqref{eq:uniform-raw-energy-lower} therefore
holds for this family for every $\rho>1$.
\end{theorem}
\begin{proof}
Apply Lemma~\ref{lem:third-energy-completion} at time zero.
Strong resolvent convergence follows from
Proposition~\ref{prop:initial-energy-resolvent}.
Lemma~\ref{lem:common-form-comparison} gives form consistency,
and \eqref{eq:triple-inverse-regularity} gives the bound on
$H^{-3}$. The form comparisons and
Lemma~\ref{lem:exact-moment-coercivity} give
\eqref{eq:completion-coercivity}; the bounded scalar shift
is treated as in Theorem~\ref{thm:arbitrary-reference-preparation}.
Thus \eqref{eq:admissible-initial-bound} implies
$\psi\in D(H^{3/2})$ and the initial first-energy convergence.
Conversely, put $f=H^{3/2}\psi$. Strong resolvent convergence
gives $H_N^{-3/2}\Pi_Nf\to\psi$. Its norm tends to one,
so the normalization is eventually well defined. Moreover,
\[
	\|H_N^{3/2}z_N^{\rm r}\|^2
	=\frac{\|\Pi_Nf\|^2}{\|H_N^{-3/2}\Pi_Nf\|^2}
	\longrightarrow\|f\|^2.
\]
This proves sufficiency and attains the lower bound of
\eqref{eq:completion-state-energy}.

The initial bound is equivalent to a bound for
$F_{\rm ex}(0)^{3/2}$ by the bounded scalar shift.
Theorem~\ref{thm:D} and Remark~\ref{rem:exact-number-shift}
therefore give the physical moments and, with one choice of
$b_I$ on each interval,
\begin{equation}\label{eq:admissible-third-propagation}
	\sup_{N,t\in I}\|\widehat F_N(t)^{3/2}\xi_N(t)\|<\infty.
\end{equation}

We first obtain convergence in the ambient Hilbert norm.
The range of $H^{-3}$ is dense. Choose $f_k$ so that
$H^{-3}f_k\to\psi$. Lemma~\ref{lem:uniform-prepared-source}
uses the same scalar phase for all these sources and gives
\[
	\sup_{t\in I}
	\|e^{i\theta_N(t)}U_{\rm ex}(t,0)H_N^{-3}\Pi_Nf_k
	-\mathcal U_{2,\infty}(t,0)H^{-3}f_k\|
	\longrightarrow0
\]
at fixed $k$. Unitarity and $z_N\to\psi$ show that the
limsup of the corresponding error for $z_N,\psi$ is at
most $2\|\psi-H^{-3}f_k\|$. Letting $k\to\infty$ gives
\begin{equation}\label{eq:admissible-hilbert-limit}
	\sup_{t\in I}\|e^{i\theta_N(t)}\xi_N(t)-\psi(t)\|
	\longrightarrow0.
\end{equation}

We next record the regularity of the limiting trajectory.
Write
$F_\infty(t)=P_{\infty,t}\mathcal J_{2,\infty}(t)
P_{\infty,t}+b_IM$.
At each fixed time, pull back the physical energy by
$\mathcal R_N(t)$. Lemma~\ref{lem:evolved-form-comparison}
supplies its strong resolvent limit $F_\infty(t)$.
Let $B_I$ denote the finite supremum in
\eqref{eq:admissible-third-propagation}.
For each fixed $t\in I$, Hilbert convergence, the unitary
pullback and spectral lower semicontinuity give
\[
	\|F_\infty(t)^{3/2}\psi(t)\|
	\le\liminf_{N\to\infty}
	\|\widehat F_N(t)^{3/2}\xi_N(t)\|\le B_I.
\]
The same $B_I$ bounds every $t$, so taking the supremum
proves the uniform limiting bound. This step uses the
uniform exact propagation estimate.
The constants in Lemma~\ref{lem:third-energy-completion}
depend only on the common form and weighted coercivity
constants. These are uniform on $I$.
Its inverse estimate gives the uniform bound for
$MA_{\infty,t}^{1/2}\psi(t)$.
The weighted inverse argument with weight zero also gives
$D(F_\infty(t))=D(A_{\infty,t})$ and
$\|A_{\infty,t}F_\infty(t)^{-1}\|\le C_I$.
Comparison with the ambient kinetic operator, as in
Lemma~\ref{lem:quadratic-weighted-operator-domain}, therefore gives
\begin{equation}\label{eq:admissible-limit-regularity}
	\sup_{t\in I}\bigl(
	\|MA_{\rm amb}^{1/2}\psi(t)\|
	+\|A_{\rm amb}\psi(t)\|\bigr)<\infty.
\end{equation}
For the weaker space $Y=D(M^{1/2}A_{\rm amb}^{1/2})$,
joint spectral calculus and H\"older's inequality give
\begin{equation}\label{eq:admissible-weighted-interpolation}
	\|M^{1/2}A_{\rm amb}^{1/2}e\|
	\le\|MA_{\rm amb}^{1/2}e\|^{1/2}
	\|A_{\rm amb}e\|^{1/4}\|e\|^{1/4}.
\end{equation}
Hilbert-norm continuity of $\psi(t)$ and
\eqref{eq:admissible-limit-regularity} imply continuity in $Y$.

Now let $N_j\to\infty$ and $t_j\to t_*\in I$.
Use the direct rotations $\mathcal S_j$ and the fixed-space
operators $G_j,\overline Q_j,A_*,F_*$ from the proof of
Theorem~\ref{thm:uniform-evolved-energy-spectrum}.
The form convergence \eqref{eq:variable-time-forms} was proved
independently of the initial states. Together with
\eqref{eq:variable-time-weighted-domain}, the weighted
inverse estimates and exact coercivity, it verifies the
assumptions of Lemma~\ref{lem:third-energy-completion}.
In particular $A_*\le C_I\overline Q_j$: the direct
rotation changes the kinetic form by $o(1)M$, and
the actual energy contains $M$.
The vectors
\[
	w_j=\mathcal S_j^*e^{i\theta_{N_j}(t_j)}\xi_{N_j}(t_j)
	\longrightarrow\psi_*:=\psi(t_*)
\]
are physical and have uniformly bounded $G_j^{3/2}$ norm.
The lemma gives
$\|\overline Q_j^{1/2}(w_j-\psi_*)\|\longrightarrow0$.
The Fock rotations are uniformly bounded on $Y$ and converge
strongly there. Indeed, they commute with $M$, their kinetic
form differences are $o(1)M$, and convergence holds first
on finitely many smooth sectors and then by density.
Using the continuity in $Y$ just proved, we obtain
\[
	\|Q_{N_j}(t_j)^{1/2}
	(\mathcal R_{N_j}(t_j)\psi(t_j)-\mathcal S_j\psi_*)\|
	\longrightarrow0.
\]
Here the actual form bound $Q_{N_j}(t_j)\le C_I MA_{N_j,t_j}$
controls the difference in the same excitation space.
These two limits prove \eqref{eq:admissible-evolved-energy}
along every convergent time sequence. Compactness of $I$
gives the stated uniform limit.

For the spectral limit, the actual cubic phase permits a
weaker input domain than the general dressed phase in
Theorem~\ref{thm:moving-dressed-spectrum}.
Proposition~\ref{prop:full-graph} and
$\overline Q_j\le C_I MA_*$ give
\begin{equation}\label{eq:actual-phase-weaker-input}
	\|Q_{N_j}(t_j)^{1/2}e^{C_{t_j}}\mathcal S_jv\|
	\le C_I\|M^{1/2}A_*^{1/2}v\|.
\end{equation}
The phase is extended by the identity above $N_j$;
the energy there keeps its actual sector forms.
Projected finite sums of smooth product vectors are dense
in $D(M^{1/2}A_*^{1/2})$.
For each such vector, Theorem~\ref{thm:moving-dressed-spectrum}
applies along this time sequence, with the kernel convergence
already verified in the proof of
Theorem~\ref{thm:uniform-evolved-energy-spectrum}.
Estimate \eqref{eq:actual-phase-weaker-input} and the
first-energy polarization bound extend its limit to this
weaker space. The limiting kinetic form and contact are
continuous there; the contact is bounded by $C_I\|M^{1/2}v\|^2$.
We may therefore take $v=\psi_*$.
The full-graph bound transfers the same limit from
$e^{C_{t_j}}\mathcal S_j\psi_*$ to
$e^{C_{t_j}}\xi_{N_j}(t_j)$, using the energy comparison above.
Both tests use the same actual operator $Q_{N_j}(t_j)$.
The continuity in $Y$ and that of the limiting kernels
make $A_{\infty,t}[\psi(t)]$ and $c_\psi(t)$ continuous.
Compactness proves \eqref{eq:admissible-evolved-spectrum}.

The vector $\psi(t)$ is normalized and has finite kinetic
energy and finite mean excitation number.
Proposition~\ref{prop:contact-positive-kinetic} gives
$c_\psi(t)>0$. Its positive minimum on $I$ and a fixed
positive microscopic test give
\eqref{eq:uniform-raw-energy-lower}, exactly as in
Corollary~\ref{cor:uniform-positive-microscopic-moments}.
\end{proof}

The initial-domain criterion is necessary and sufficient within
norm-convergent families. Under the overlap condition in
Theorem~\ref{thm:vacuum-third-obstruction}, the unprepared vacuum
lies outside $D(H^{3/2})$. The inverse constructions give
quantitative examples in the admissible class.

\subsection{Correlated energy and microscopic measurements}
\label{sec:physical-energy-comparison}

The two energy estimates can be expressed on the same physical
space. The observable $\mathscr Q_N(t)$ was defined in
\eqref{eq:many-body-excitation-observable}. Using the reconstruction
$\mathcal W_{N,t}$ from \eqref{eq:physical-reconstruction}, define
\begin{equation}\label{eq:physical-correlated-energy}
	\mathscr F_N(t)=\mathcal W_{N,t}F_{\rm ex}(t)\mathcal W_{N,t}^*,
	\qquad D(\mathscr F_N(t))=\mathcal W_{N,t}D(F_{\rm ex}(t)).
\end{equation}
On each compact interval $I$ we use one positive number correction
$b_I$ in $F_{\rm ex}(t)$. Both observables act on the symmetric
$N$-particle space. Their higher moments distinguish the energy
of the correlated state from its microscopic excitation energy.

\begin{proposition}\label{prop:physical-energy-comparison}
Let $v\in C_c^\infty(\mathbb R^3)$ be nonzero, nonnegative and
radial. Fix an $N$-independent normalized
$\phi\in C_c^\infty(\mathbb R^3)$ and a sufficiently small fixed $\ell\in(0,1/2)$.
Let $q=1-|\phi\rangle\langle\phi|$, $P=\Gamma(q)$, and use
the Neumann pair kernel $\eta_N$ and its limit
$\eta_\infty$ in \eqref{spec:eq:actual-eta-limit}.
Let $C_N=\mathcal C_N^{\mathrm{COS}}$ be the phase
\eqref{spec:eq:actual-phase}, with $m=N^{-1/2}$.
Its real cutoff $0\le\Theta_N\le1$ equals one for
$n\le N^{1/2}/2+10$ and zero for $n\ge N^{1/2}-10$.
Extend $C_N$ by zero above sector $N$.
Use the centered initial energies
$\widehat F_{N,b}=\mathfrak J_N-\mathfrak J_N[\Omega]I+bM$
on $\Pi_NP\cF$ and $F_{\infty,b}=\mathcal J_{2,\infty}+bM$
on $P\cF$, where $\Pi_N=\mathbf1_{\{\cN\le N\}}$.
Choose a fixed $b$ so that, uniformly in $N$,
$c_bQ_N\le\widehat F_{N,b},F_{\rm ex}(0)\le C_bQ_N$
on the physical space and $c_bA\le F_{\infty,b}\le C_bA$.
Writing $H_b=K_P+bM$ and $S_\infty=F_{\infty,b}-H_b$, require
\[
	\|M^rS_\infty H_b^{-1}M^{-r}\|<\tfrac12,
	\qquad r=0,1,2.
\]
All these conditions hold for sufficiently large $b$, independently of $N$.
Let $z_N\in\Pi_NP\cF$ and $\psi\in P\cF$ be normalized, with
\[
	z_N\to\psi\quad\text{in norm},\qquad
	\sup_{N\ge N_0}\|\widehat F_{N,b}^{3/2}z_N\|<\infty.
\]
Set $\xi_N(t)=U_{\rm ex}(t,0)z_N$ and
$\psi(t)=\mathcal U_{2,\infty}(t,0)\psi$.
Let $\Psi_N(t)=\mathcal W_{N,t}\xi_N(t)$.
On every compact time interval $I$, the positive forms satisfy
\begin{equation}\label{eq:physical-energy-form-equivalence}
	c_I\mathscr Q_N(t)\le\mathscr F_N(t)\le C_I\mathscr Q_N(t),
	\qquad t\in I,
\end{equation}
with constants independent of $N$. Moreover,
\[
	\sup_{N\ge N_0,\,t\in I}
	\langle\Psi_N(t),\mathscr F_N(t)^3\Psi_N(t)\rangle<\infty.
\]
Fix $B=[b,d]$ with $0<b<d<\infty$ and $p\ge0$. Then
\begin{align}\label{eq:physical-window-moments}
	\sup_{t\in I}\bigg|&N^{2-2p}
	\langle\Psi_N(t),\mathscr Q_N(t)^p
	\mathbf1_B(\mathscr Q_N(t)/N^2)\Psi_N(t)\rangle\notag\\
	&-2c_\psi(t)\int_B\lambda^{p-1}\d\mu_v(\lambda)\bigg|
	\longrightarrow0.
\end{align}
In particular, the probability of this window is of order
$N^{-2}$, while its contribution to the $p$th moment is of
order $N^{2p-2}$. For $p>1$, there are $a_{I,p,B}>0$ and
$N_{I,p,B}$ such that
\begin{equation}\label{eq:physical-window-operator-lower}
	\inf_{t\in I}\left\|
	\mathscr Q_N(t)^{p/2}\mathbf1_B(\mathscr Q_N(t)/N^2)
	\mathscr F_N(t)^{-3/2}\right\|_{\mathrm{op}}
	\ge a_{I,p,B}N^{p-1},\qquad N\ge N_{I,p,B}.
\end{equation}
The third depletion moment and sixth collective moments of
these same states satisfy Theorem~\ref{thm:D}.
\end{proposition}
\begin{proof}
At fixed $t,N$, put $T=\mathcal U_{N,t}^*e^{B_t}$,
so $\mathcal W_{N,t}=Te^{C_t}$. All operators below are
restricted to the physical excitation space before conjugation.
Proposition~\ref{prop:full-graph} gives
\[
	\|Q_N(t)^{1/2}e^{\pm C_t}u\|
	\le C_I\|Q_N(t)^{1/2}u\|.
\]
Applying these estimates in both directions yields
$Q_N(t)[e^{-C_t}u]\asymp Q_N(t)[u]$.
Their constants are uniform on $I$. Since
$F_{\rm ex}(t)\asymp Q_N(t)$, their form domains agree and
\[
	\mathscr F_N(t)[Tu]
	=F_{\rm ex}(t)[e^{-C_t}u]
	\asymp Q_N(t)[u]=\mathscr Q_N(t)[Tu].
\]
This proves \eqref{eq:physical-energy-form-equivalence} on
the common form domain $T D(Q_N(t)^{1/2})$.
Since $Q_N(t)\ge1$, it also gives $\mathscr F_N(t)\ge c_I>0$.
Unitary functional calculus \cite{ReedSimonI} gives
\[
	\|\mathscr F_N(t)^{3/2}\Psi_N(t)\|^2
	=\|F_{\rm ex}(t)^{3/2}\xi_N(t)\|^2.
\]
Theorem~\ref{thm:exact-third-moment} and
Remark~\ref{rem:exact-number-shift} bound this uniformly
for the unchanged initial family.

Write $\nu_{N,t}$ for the energy-weighted measure in
\eqref{eq:admissible-evolved-spectrum}. The same unitary
calculus and $u_N(t)=e^{C_t}\xi_N(t)$ give the exact identity
\[
	N^{2-2p}\langle\Psi_N(t),\mathscr Q_N(t)^p
	\mathbf1_B(\mathscr Q_N(t)/N^2)\Psi_N(t)\rangle
	=\int_B\lambda^{p-1}\d\nu_{N,t}(\lambda).
\]
For $0<\delta<\min(b/2,(d-b)/2)$, choose continuous cutoffs
$0\le\chi_\delta^-\le\mathbf1_B\le\chi_\delta^+\le1$.
Take $\chi_\delta^-=1$ on $[b+\delta,d-\delta]$, supported in $B$,
and $\chi_\delta^+=1$ on $B$, supported in $[b-\delta,d+\delta]$.
The tests $f_\delta^\pm(\lambda)=\lambda^{p-1}\chi_\delta^\pm(\lambda)$,
extended by zero near zero, belong to $C_c((0,\infty))$,
also when $p=0$. They satisfy
\[
	0\le f_\delta^+-f_\delta^-
	\le C_{p,B}\mathbf1_{E_\delta},\qquad
	E_\delta=[b-\delta,b+\delta]\cup[d-\delta,d+\delta].
\]
First fix $\delta$ and apply \eqref{eq:admissible-evolved-spectrum}
to both tests. Positivity then gives
\begin{align*}
	\limsup_{N\to\infty}\sup_{t\in I}
	\left|\int_B\lambda^{p-1}\d\nu_{N,t}
	-2c_\psi(t)\int_B\lambda^{p-1}\d\mu_v\right|
	\le C_{I,p,B}\mu_v(E_\delta).
\end{align*}
Now let $\delta\downarrow0$.
Proposition~\ref{prop:radial-microscopic-density} gives
$\mu_v(\{b,d\})=0$, so the right-hand side tends to zero.
This proves \eqref{eq:physical-window-moments}.
The integral in its limit is strictly positive: the radial
density is positive except at isolated points, and $b<d$.
Also $\min_Ic_\psi>0$. This proves both order statements.

For \eqref{eq:physical-window-operator-lower}, set
$w_N(t)=\mathscr F_N(t)^{3/2}\Psi_N(t)$.
These vectors have uniformly bounded norm, say at most $K_I>0$.
They are nonzero, and $\mathscr F_N(t)^{-3/2}w_N(t)=\Psi_N(t)$.
Applying $\mathscr Q_N(t)^{p/2}\mathbf1_B(\mathscr Q_N(t)/N^2)$
to $\mathscr F_N(t)^{-3/2}w_N(t)$ gives the window
moment in \eqref{eq:physical-window-moments} as its squared norm.
Its lower bound is $c_{I,p,B}N^{2p-2}$ for all sufficiently
large $N$, uniformly in $t$. Division by $K_I$ proves the
operator bound. The window operator
$\mathscr Q_N(t)^{p/2}\mathbf1_B(\mathscr Q_N(t)/N^2)$
has norm at most $(dN^2)^{p/2}$. Its product with
$\mathscr F_N(t)^{-3/2}$ is therefore bounded at every fixed $N$.
The physical fluctuation bounds follow from
Theorem~\ref{thm:physical-sixth-moment}.
\end{proof}

The first-energy form comparison is uniform in $N$.
Equation~\eqref{eq:physical-window-operator-lower} quantifies
the growth of the corresponding higher-energy comparison.
The correlated third moment controls depletion and collective
fluctuations throughout the same evolution.

The same initial class also determines the interaction-energy
measure. The weighted first-energy comparison suffices for
bounded tests. The exact second-moment upper bound then
controls the higher-energy tail.

\begin{theorem}
\label{thm:admissible-interaction-spectrum}
Let $v\in C_c^\infty(\mathbb R^3)$ be nonzero, nonnegative and
radial. Fix an $N$-independent normalized
$\phi\in C_c^\infty(\mathbb R^3)$ and a sufficiently small fixed $\ell\in(0,1/2)$.
Let $q=1-|\phi\rangle\langle\phi|$, $P=\Gamma(q)$, and use
the Neumann pair kernel $\eta_N$ and its limit
$\eta_\infty$ in \eqref{spec:eq:actual-eta-limit}.
Let $C_N=\mathcal C_N^{\mathrm{COS}}$ be the phase
\eqref{spec:eq:actual-phase}, with $m=N^{-1/2}$.
Its real cutoff $0\le\Theta_N\le1$ equals one for
$n\le N^{1/2}/2+10$ and zero for $n\ge N^{1/2}-10$.
Extend $C_N$ by zero above sector $N$.
Use the centered initial energies
$\widehat F_{N,b}=\mathfrak J_N-\mathfrak J_N[\Omega]I+bM$
on $\Pi_NP\cF$ and $F_{\infty,b}=\mathcal J_{2,\infty}+bM$
on $P\cF$, where $\Pi_N=\mathbf1_{\{\cN\le N\}}$.
Choose a fixed $b$ so that, uniformly in $N$,
$c_bQ_N\le\widehat F_{N,b},F_{\rm ex}(0)\le C_bQ_N$
on the physical space and $c_bA\le F_{\infty,b}\le C_bA$.
Writing $H_b=K_P+bM$ and $S_\infty=F_{\infty,b}-H_b$, require
\[
	\|M^rS_\infty H_b^{-1}M^{-r}\|<\tfrac12,
	\qquad r=0,1,2.
\]
All these conditions hold for sufficiently large $b$, independently of $N$.
Let $z_N\in\Pi_NP\cF$ and $\psi\in P\cF$ be normalized, with
\[
	z_N\to\psi\quad\text{in norm},\qquad
	\sup_{N\ge N_0}\|\widehat F_{N,b}^{3/2}z_N\|<\infty.
\]
Set $\xi_N(t)=U_{\rm ex}(t,0)z_N$ and
$\psi(t)=\mathcal U_{2,\infty}(t,0)\psi$.
Put $u_N(t)=e^{C_t}\xi_N(t)$. Define
\begin{equation}\label{eq:admissible-interaction-measure}
	\nu_{N,t}^{\rm int}(B)=\left\langle\mathcal V_N(t)^{1/2}u_N(t),
	\mathbf1_B(\mathcal V_N(t)/N^2)
	\mathcal V_N(t)^{1/2}u_N(t)\right\rangle.
\end{equation}
For every compact time interval $I$ and each fixed
$g\in C_b([0,\infty))$,
\begin{equation}\label{eq:admissible-interaction-spectrum}
	\sup_{t\in I}\left|\int g\d\nu_{N,t}^{\rm int}
	-2c_\psi(t)\int g\d\mu_v^{\rm int}\right|\longrightarrow0.
\end{equation}
For every fixed $1\le p<2$,
\begin{equation}\label{eq:admissible-interaction-moments}
	\sup_{t\in I}\left|
	N^{2-2p}\|\mathcal V_N(t)^{p/2}u_N(t)\|^2
	-2c_\psi(t)b_p(v)\right|\longrightarrow0.
\end{equation}
The second moments satisfy
\begin{gather}
	\sup_{N,t\in I}N^{-2}\|\mathcal V_N(t)u_N(t)\|^2<\infty,
	\label{eq:admissible-interaction-second-upper}\\
	\liminf_{N\to\infty}\inf_{t\in I}
	\left(N^{-2}\|\mathcal V_N(t)u_N(t)\|^2
	-2c_\psi(t)b_2(v)\right)\ge0.
	\label{eq:admissible-interaction-second-lower}
\end{gather}
In particular, for $v\ne0$ they are of order $N^2$
uniformly on $I$.
\end{theorem}
\begin{proof}
Theorem~\ref{thm:admissible-initial-class} gives a uniform
third correlated-energy bound. Proposition~
\ref{prop:interaction-moment-transfer} therefore gives
\eqref{eq:admissible-interaction-second-upper} and
membership in the displayed potential domains.
Set $\widetilde u_N(t)=e^{C_t}\mathcal R_N(t)\psi(t)$.
The full-graph phase bound and
\eqref{eq:admissible-evolved-energy} imply
\[
	\sup_{t\in I}
	\|\mathcal V_N(t)^{1/2}
	(e^{i\theta_N(t)}u_N(t)-\widetilde u_N(t))\|
	\longrightarrow0.
\]
Both potential form norms are uniformly bounded.
For the reference this follows from
\eqref{eq:interaction-weaker-input} and
\eqref{eq:admissible-limit-regularity}.
Equation~\eqref{eq:interaction-test-comparison} thus
transfers every fixed bounded test between these two
states, using the same actual operator $\mathcal V_N(t)$.

Let $N_j\to\infty$ and $t_j\to t_*\in I$.
The backgrounds converge in $H^4$.
The proof of Theorem~\ref{thm:admissible-initial-class}
shows that the reference path is continuous in
$Y=D(M^{1/2}A_{\rm amb}^{1/2})$.
Let $\mathcal S_j$ be the direct rotation from
$P_{\infty,t_*}\cF$ to $P_{N_j,t_j}\cF$.
These rotations and $\mathcal R_{N_j}(t_j)$ are uniformly
bounded and converge strongly to the identity on the
ambient space $Y$. Hence
\[
	\chi_j=\mathcal S_j^*\mathcal R_{N_j}(t_j)\psi(t_j)
	\longrightarrow\psi(t_*)\quad\hbox{in }Y,
	\qquad \chi_j\in P_{\infty,t_*}\cF.
\]
The reference vector under the phase is exactly
$\mathcal S_j\chi_j$.
Lemma~\ref{lem:moving-interaction-spectrum} now gives
the limit of each tested expectation along this sequence.
Continuity of $c_\psi(t)$ and compactness of $I$ prove
\eqref{eq:admissible-interaction-spectrum}.
Indeed, failure of uniform convergence would provide
a sequence of observation times with a fixed positive
error. A convergent subsequence satisfies all the moving
background and graph-convergence assumptions just checked,
which forces that error to tend to zero.

The case $p=1$ is the test $g=1$.
For $1<p<2$, the spectral theorem gives
\begin{align*}
	\int\lambda^{p-1}\d\nu_{N,t}^{\rm int}(\lambda)
	&=N^{2-2p}\|\mathcal V_N(t)^{p/2}u_N(t)\|^2,\\
	\int\lambda\d\nu_{N,t}^{\rm int}(\lambda)
	&=N^{-2}\|\mathcal V_N(t)u_N(t)\|^2\le C_I.
\end{align*}
Hence
\begin{equation}\label{eq:interaction-moment-uniform-tail}
	\sup_{N,t\in I}\int_{\lambda>R}
	\lambda^{p-1}\d\nu_{N,t}^{\rm int}(\lambda)
	\le C_I R^{p-2}\longrightarrow0.
\end{equation}
Apply \eqref{eq:admissible-interaction-spectrum} to
$g_R(\lambda)=\min(\lambda^{p-1},R^{p-1})$.
The limiting measure has bounded support. Letting first
$N\to\infty$ and then $R\to\infty$ proves
\eqref{eq:admissible-interaction-moments}, since
$\int\lambda^{p-1}\d\mu_v^{\rm int}=b_p(v)$.

For the endpoint, fix $R\ge\|v\|_\infty$ and use
$g_R(\lambda)=\min(\lambda,R)$.
The second scaled moment is at least
$\int g_R\d\nu_{N,t}^{\rm int}$. Its uniform limit is
$2c_\psi(t)b_2(v)$, which proves
\eqref{eq:admissible-interaction-second-lower}.
For nonzero $v$, the contact has a positive minimum on
$I$ and $b_2(v)>0$. Together with the upper bound, this
gives the asserted order $N^2$.
\end{proof}

The interaction-moment limits therefore hold for every
norm-convergent family with bounded initial third correlated
energy. The tail estimate \eqref{eq:interaction-moment-uniform-tail}
transfers unbounded tests of order $1\le p<2$.

\subsection{Low-energy states of the initial correlated energy}
\label{sec:correlated-ground-state}

The third-energy hypothesis can be verified by a variational
choice of the initial state. We use the same positive correlated
energy $H_N=\widehat F_{N,b}$ and its limit $H=F_{\infty,b}$.
The fixed term $bM$ is part of these operators in the initial
excitation representation.
The positive self-adjoint operator $H_N$ acts on $\Pi_NP\cF$.
Its spectral projections are extended by zero on the complement.

\begin{proposition}
\label{prop:correlated-norm-resolvent}
Let $v\in C_c^\infty(\mathbb R^3)$ be nonzero, nonnegative and
radial. Fix an $N$-independent normalized
$\phi\in C_c^\infty(\mathbb R^3)$ and a sufficiently small fixed $\ell\in(0,1/2)$.
Let $q=1-|\phi\rangle\langle\phi|$, $P=\Gamma(q)$, and use
the Neumann pair kernel $\eta_N$ and its limit
$\eta_\infty$ in \eqref{spec:eq:actual-eta-limit}.
Let $C_N=\mathcal C_N^{\mathrm{COS}}$ be the phase
\eqref{spec:eq:actual-phase}, with $m=N^{-1/2}$.
Its real cutoff $0\le\Theta_N\le1$ equals one for
$n\le N^{1/2}/2+10$ and zero for $n\ge N^{1/2}-10$.
Extend $C_N$ by zero above sector $N$.
Use the centered initial energies
$\widehat F_{N,b}=\mathfrak J_N-\mathfrak J_N[\Omega]I+bM$
on $\Pi_NP\cF$ and $F_{\infty,b}=\mathcal J_{2,\infty}+bM$
on $P\cF$, where $\Pi_N=\mathbf1_{\{\cN\le N\}}$.
Choose a fixed $b$ so that, uniformly in $N$,
$c_bQ_N\le\widehat F_{N,b},F_{\rm ex}(0)\le C_bQ_N$
on the physical space and $c_bA\le F_{\infty,b}\le C_bA$.
Writing $H_b=K_P+bM$ and $S_\infty=F_{\infty,b}-H_b$, require
\[
	\|M^rS_\infty H_b^{-1}M^{-r}\|<\tfrac12,
	\qquad r=0,1,2.
\]
All these conditions hold for sufficiently large $b$, independently of $N$.
Write $H_N=\widehat F_{N,b}$ and $H=F_{\infty,b}$.
Let $\widetilde H_N=H_N\oplus(K_P+bM)_{>N}$ on $P\cF$.
Then
\begin{equation}\label{eq:correlated-norm-inverse}
	\|\widetilde H_N^{-1}-H^{-1}\|_{\mathrm{op}}
	\le C_bN^{-1/28}.
\end{equation}
In particular, the convergence is in norm resolvent.
If $0<a<d<\infty$ and $a,d\notin\sigma(H)$, put
\[
	E_N=\mathbf1_{(a,d)}(H_N)\Pi_N,\qquad
	E=\mathbf1_{(a,d)}(H).
\]
Then
\begin{equation}\label{eq:correlated-spectral-projection}
	\|E_N-E\|_{\mathrm{op}}\le C_{a,d,b}N^{-3/28}.
\end{equation}
If $E$ has finite rank, $E_N$ has the same rank for all
sufficiently large $N$.
\end{proposition}
\begin{proof}
Set
\[
	T_N=H_N^{-1}\Pi_N,\qquad T=H^{-1},\qquad
	S_N=T_N^3,\qquad S=T^3.
\]
These are bounded nonnegative self-adjoint operators.
Equation~\eqref{eq:triple-inverse-operator-rate} and $Q_N\ge1$
give
\begin{equation}\label{eq:correlated-cube-norm}
	\|S_N-S\|_{\mathrm{op}}\le C_bN^{-3/28}.
\end{equation}
For bounded nonnegative $X,Y$ and $0<\alpha<1$, functional
calculus \cite{ReedSimonI} gives
\[
	X^\alpha=\frac{\sin(\pi\alpha)}{\pi}
	\int_0^\infty t^{\alpha-1}X(t+X)^{-1}\d t.
\]
The resolvent identity bounds the difference of the two
integrands, without the scalar weight, by
$\min(2,\|X-Y\|/t)$. Splitting the integral at
$t=\|X-Y\|$ proves
\[
	\|X^\alpha-Y^\alpha\|\le C_\alpha\|X-Y\|^\alpha.
\]
The zero-difference case is immediate. Apply this estimate
with $X=S_N$, $Y=S$ and $\alpha=1/3$.
On the same full space, write
\[
	B_N=(K_P+bM)^{-1}(1-\Pi_N),\qquad
	\widetilde H_N^{-1}=T_N+B_N,\qquad
	\|B_N\|\le[b(N+2)]^{-1}.
\]
The triangle inequality now gives
\[
	\|\widetilde H_N^{-1}-H^{-1}\|
	\le\|T_N-T\|+\|B_N\|
	\le C_bN^{-1/28}+[b(N+2)]^{-1}.
\]
The first term already compares $T_N$ and $T$ on the
whole space, including their number-changing blocks.
This proves \eqref{eq:correlated-norm-inverse}.
For each fixed $\zeta\in\rho(H)$, $1-\zeta H^{-1}$ is
invertible. Norm convergence makes $1-\zeta\widetilde H_N^{-1}$
invertible for large $N$, and
\[
	(\widetilde H_N-\zeta)^{-1}
	=\widetilde H_N^{-1}(1-\zeta\widetilde H_N^{-1})^{-1}
	\longrightarrow H^{-1}(1-\zeta H^{-1})^{-1}
\]
in operator norm. This is the norm-resolvent conclusion.

For the spectral projection, use the cubes directly.
The interval $(a,d)$ corresponds to $(d^{-3},a^{-3})$
in the spectra of $S_N,S$. Its endpoints lie outside
$\sigma(S)$, and it is separated from zero.
A fixed contour around this part of the spectrum remains
in the resolvent set of $S_N$ for large $N$.
The Riesz projection formula and the resolvent identity
\cite[Chapter IV, Theorem 3.16]{KatoBook} give
$\|E_N-E\| \le C_{a,d,b}\|S_N-S\|$.
The zero eigenvalue of $S_N$ above $N$ is outside the contour.
Thus this is precisely the physical projection in the
statement. Orthogonal projections at distance less than one
have the same finite rank: each projection is injective on
the range of the other. This completes the proof.
\end{proof}

The limiting quadratic ground state is supplied by the
Bogoliubov diagonalization theorem \cite[Theorems 1--2]{NNS2016}.
The new comparison above transfers its isolated lowest
eigenvalue to the exact initial correlated energy.

\begin{theorem}
\label{thm:correlated-ground-state}
Let $v\in C_c^\infty(\mathbb R^3)$ be nonzero, nonnegative and
radial. Fix an $N$-independent normalized
$\phi\in C_c^\infty(\mathbb R^3)$ and a sufficiently small fixed $\ell\in(0,1/2)$.
Let $q=1-|\phi\rangle\langle\phi|$, $P=\Gamma(q)$, and use
the Neumann pair kernel $\eta_N$ and its limit
$\eta_\infty$ in \eqref{spec:eq:actual-eta-limit}.
Let $C_N=\mathcal C_N^{\mathrm{COS}}$ be the phase
\eqref{spec:eq:actual-phase}, with $m=N^{-1/2}$.
Its real cutoff $0\le\Theta_N\le1$ equals one for
$n\le N^{1/2}/2+10$ and zero for $n\ge N^{1/2}-10$.
Extend $C_N$ by zero above sector $N$.
Use the centered initial energies
$\widehat F_{N,b}=\mathfrak J_N-\mathfrak J_N[\Omega]I+bM$
on $\Pi_NP\cF$ and $F_{\infty,b}=\mathcal J_{2,\infty}+bM$
on $P\cF$, where $\Pi_N=\mathbf1_{\{\cN\le N\}}$.
Write $H_N=\widehat F_{N,b}$ and $H=F_{\infty,b}$.
Choose the fixed $b$ sufficiently large. The limiting energy $H$ has
a unique normalized ground state $\psi_b$, up to phase,
with energy $e_b>0$ and a positive spectral gap.
For all sufficiently large $N$, the physical energy $H_N$
also has a unique ground state $z_{N,b}$ with energy $e_{N,b}$.
Its gap is bounded below independently of $N$.
Choose the phase so that $\langle\psi_b,z_{N,b}\rangle>0$.
Then
\begin{gather}\label{eq:correlated-ground-convergence}
	|e_{N,b}-e_b|+\|z_{N,b}-\psi_b\|
	+\|Q_N^{1/2}(z_{N,b}-\psi_b)\|
	\le C_bN^{-3/28},\\
	\|H_N^{3/2}z_{N,b}\|^2=e_{N,b}^3\longrightarrow e_b^3,
	\qquad \psi_b\in D(M^2A).\notag
\end{gather}
This family satisfies the physical moment and microscopic
spectral conclusions of Theorem~\ref{thm:admissible-initial-class}.
More precisely, for
$\xi_{N,b}(t)=U_{\rm ex}(t,0)z_{N,b}$ and
$\psi_b(t)=\mathcal U_{2,\infty}(t,0)\psi_b$,
\begin{equation}\label{eq:correlated-ground-evolved-energy}
	\sup_{t\in I}\|Q_N(t)^{1/2}
	(e^{i\theta_N(t)}\xi_{N,b}(t)-\mathcal R_N(t)\psi_b(t))\|
	\le C_{I,b}N^{-3/56}
\end{equation}
on every compact interval $I$. The initial minimization is
independent of $I$.
\end{theorem}
\begin{proof}
Write the quadratic part of $H$ on the one-particle
excitation space $\mathfrak h=qL^2$ as
$H=b+\dGamma(h_b)+\mathcal P$.
Here $h_b=q(-\Delta)q+b+h_{\rm reg}$, with $h_{\rm reg}$ bounded
self-adjoint. Use the pairing convention
\[
	\mathcal P=\frac12\sum_{i,j}
	\bigl(\langle J^*k f_i,f_j\rangle a(f_i)a(f_j)
	+\overline{\langle J^*k f_i,f_j\rangle}
	a^*(f_i)a^*(f_j)\bigr).
\]
The $f_i$ form an orthonormal basis of $\mathfrak h$,
$J:\mathfrak h\to\mathfrak h^*$ is the canonical antiunitary,
and $k:\mathfrak h\to\mathfrak h^*$ is Hilbert--Schmidt
with $k^*=J^*kJ^*$.
These coefficient bounds follow from
\eqref{eq:common-quadratic-weight} and the quadratic
decomposition in Lemma~\ref{lem:evolved-form-comparison}.
The pairing coefficient uses the factor $1/2$ displayed above.

Take $m_b=b-\|h_{\rm reg}\|>\|k\|$ and set
$\mathcal A_b=\left(\begin{smallmatrix}h_b&k^*\\
k&Jh_bJ^*\end{smallmatrix}\right)$.
Then
\[
	\mathcal A_b\ge(m_b-\|k\|)I,\qquad
	\|h_b^{-1/2}J^*kh_b^{-1/2}\|\le\|k\|/m_b<1.
\]
Both this latter operator and $kh_b^{-1/2}$ are
Hilbert--Schmidt. Theorems 1--2 of \cite{NNS2016}
give a bounded invertible Bogoliubov map $\mathcal V_b$
and its Fock implementer $\mathbb U_b$, with
\[
	\mathcal V_b\mathcal A_b\mathcal V_b^*
	=\operatorname{diag}(\omega_b,J\omega_bJ^*),\qquad
	\mathbb U_bH\mathbb U_b^*=e_b+\dGamma(\omega_b).
\]
The first identity implies
\[
	\omega_b\ge
	(m_b-\|k\|)\|\mathcal V_b^{-1}\|^{-2}
	=:\gamma_b>0.
\]
Thus $\psi_b=\mathbb U_b^*\Omega$ is the unique ground
state, and the rest of $\sigma(H)$ lies in
$[e_b+\gamma_b,\infty)$. The common lower bound
$H\ge c_bI>0$ gives $e_b>0$.
Also $\psi_b=e_b^3H^{-3}\psi_b$, so
\eqref{eq:triple-inverse-regularity} gives $\psi_b\in D(M^2A)$.

The top of $\sigma(S)$ is the simple eigenvalue $e_b^{-3}$;
the remaining spectrum is at most $(e_b+\gamma_b)^{-3}$.
Norm convergence \eqref{eq:correlated-cube-norm} keeps
this eigenvalue isolated. The contour argument of
Proposition~\ref{prop:correlated-norm-resolvent} gives a
rank-one projection for $S_N$ at distance
$O_b(N^{-3/28})$ from $|\psi_b\rangle\langle\psi_b|$.
For bounded self-adjoint operators the spectral distance
is at most their norm difference, by the resolvent
Neumann series \cite{KatoBook}.
Hence this eigenvalue is the top of $\sigma(S_N)$ and
differs from $e_b^{-3}$ by $O_b(N^{-3/28})$.
It is nonzero and its eigenvector is physical.
Applying the scalar function $\lambda^{-1/3}$ in a fixed
neighborhood of $e_b^{-3}$ gives the claimed energy rate.
The rest of $\sigma(H_N)$ is separated from $e_{N,b}$ by,
for example, $\gamma_b/2$ for large $N$.
Normalizing the projection of $\psi_b$ gives the stated
phase and Hilbert-norm rate.

The initial first-energy rate uses the cube estimate. The eigenvalue
identities give
\[
	z_{N,b}-\psi_b
	=e_{N,b}^3(S_N-S)z_{N,b}
	+S(e_{N,b}^3z_{N,b}-e_b^3\psi_b).
\]
The first term is controlled by
\eqref{eq:triple-inverse-operator-rate}.
For the second, apply the form inequality to the vector $Sf$:
\[
	\|Q_N^{1/2}Sf\|
	\le C\|M^{1/2}A^{1/2}Sf\|
	\le C\|M^2ASf\|\le C_b\|f\|.
\]
The last step is \eqref{eq:triple-inverse-regularity}.
The middle inequality uses only the joint spectral calculus
of $A,M\ge1$.
This proves \eqref{eq:correlated-ground-convergence}.
The third initial energy is exactly $e_{N,b}^3$.

Finally, put $f_N=e_{N,b}^3z_{N,b}$ and $f=e_b^3\psi_b$.
Then $z_{N,b}=S_Nf_N$, $\psi_b=Sf$ and
$\|f_N-f\|\le C_bN^{-3/28}$.
Apply Lemma~\ref{lem:uniform-prepared-source} to these
unnormalized sources. Its uniform energy estimate controls
$(V_N(t)-\mathcal R_N(t)V(t))f_N$.
For the remaining term, the direct-rotation form comparison
and \eqref{eq:prepared-source-domains} give
\[
	\|Q_N(t)^{1/2}\mathcal R_N(t)V(t)f\|
	\le C_I\|M^{1/2}A_{\infty,t}^{1/2}V(t)f\|
	\le C_I\|M^2A_{\infty,t}V(t)f\|\le C_{I,b}\|f\|.
\]
Here $Q_N(t)\le C_IMA_{N,t}$ is evaluated on
$\mathcal R_N(t)V(t)f$. The rotation commutes with $M$,
and its kinetic form differs from $A_{\infty,t}$ by
$O_I(N^{-1})M$, as in \eqref{eq:evolved-rotation-bounds}.
This yields \eqref{eq:correlated-ground-evolved-energy}.
The uniform third initial moment also verifies all
hypotheses of Theorem~\ref{thm:admissible-initial-class}.
\end{proof}

The minimizers furnish a family whose initial energy condition
follows directly from the variational prescription. Their finite-$N$ evolution is the exact
many-body evolution after physical reconstruction.
The quadratic diagonalization and spectral perturbation tools
are standard \cite{NNS2016,KatoBook}. Their role here is to
select initial states for the joint moment and microscopic
spectral conclusions. The parameter $b$ and the initial
excitation representation remain part of this prescription.

\section{Cubic commutators and source estimates}
\label{sec:cubic-estimates}

The comparison principle requires a remainder estimate with one
number weight. We derive it from the spatial structure of the cubic
kernels. The time term and degree tail are treated first, followed
by the kinetic and potential commutators and their couplings to
bounded sources. These estimates retain the original number
coefficients, projections and cutoffs.

\subsection{The cubic time term and the degree tail}
\label{sec:time-and-tail}

We begin the verification of the hypotheses of
Theorem~\ref{thm:inverse-comparison} with the cubic time term
and the exact tail in kernel degree. The quadratic time term
$L_B$ will be treated in Section~\ref{sec:BS-quadratic-time}.
All tails are retained in the exact generator.

\subsubsection{Contractions of cubic terms}

Besides normal-ordered three-field monomials with a
three-variable Hilbert--Schmidt kernel, consider
\[
	T_F=\int F(x,y)a_x^*a_y^*a_x\d x\d y,\qquad \alpha(F)=\max\{\|F\|_2,\mathop{\rm ess\,sup}_x \|F(x,\cdot)\|_2,\mathop{\rm ess\,sup}_y\|F(\cdot,y)\|_2\}.
\]
The input coordinate $x$ is retained. This defines a
bounded one-to-two particle map with norm at most
$\alpha(F)$. Its kernel contains a Dirac delta in the retained
coordinate.
Its Wick lift has block bound
$C\alpha(F)(n+1)^{3/2}$.

\begin{lemma}\label{lem:cubic-contractions}
Let $T_1,T_2$ belong to the two cubic term classes just
defined or their adjoints, with kernel bounds
$\alpha_1,\alpha_2$. Their canonical commutator satisfies
\[
	\|[T_1,T_2]z_n\| \le C\alpha_1\alpha_2(n+1)^2\|z_n\|.
\]
In particular, for $D_1,D_2\in\{C,X=\nabla C,Y=\nabla^2C\}$,
including their fixed number-gauge derivatives,
\begin{equation}\label{eq:actual-cubic-contractions}
	\|M^r[D_1,D_2]z\| \le C_{r,I}\epsilon^2\|M^{r+2}E_Nz\|, \qquad E_N=\mathbf1_{M\le\Lambda_N}, \quad\Lambda_N\le C\sqrt N .
\end{equation}
The projection $E_N$ can be chosen once to include
all fixed shifts occurring here.
\end{lemma}
\begin{proof}
The uncontracted six-field terms cancel. Between two
cubic terms with Hilbert--Schmidt kernels, every partial contraction
is a Hilbert--Schmidt tensor contraction of norm at
most $\alpha_1\alpha_2$; complete contractions have
the same scalar bound. At most four fields remain.
For a mixed pair, write $K$ for the three-variable
Hilbert--Schmidt kernel, with any creation and annihilation
assignment. There are three possible contraction kernels:
\begin{align*}
	H_1(x,y,u,v)&=F(x,y)K(x,u,v),\\
	H_2(x;u,v)&=\int F(x,y)K(y,u,v)\d y,\\
	H_3(x,u)&=\int F(x,y)K(x,y,u)\d y .
\end{align*}
They satisfy
\begin{align}
	\|H_1\|_2^2
	&\le\alpha(F)^2\|K\|_2^2,\notag\\
	\mathop{\rm ess\,sup}_x\|H_2(x;\cdot,\cdot)\|_2^2
	&\le\alpha(F)^2\|K\|_2^2,\notag\\
	\|H_3\|_2^2
	&\le\alpha(F)^2\|K\|_2^2 .
	\label{eq:mixed-cubic-kernel-bounds}
\end{align}
For the first line integrate $y$ first. For the second,
apply Cauchy--Schwarz in $y$ at fixed $x,u,v$, then
integrate $u,v$. For the third, apply the same inequality
at fixed $x,u$, then integrate $x,u$.
These arguments use only the full $L^2$ norm of $K$.

A single contraction at the retained coordinate gives $H_1$.
A single contraction at the other coordinate gives $H_2$;
the retained $x$ remains on both sides of the particle map.
This is a direct integral in $x$. Its fibers are pair
creation maps, Hilbert--Schmidt one-particle maps, or their
adjoints, according to the two remaining field operators.
Two contractions give $H_3$. A retained-coordinate cubic
term has at most two creation or two annihilation operators,
so no threefold contraction with it is possible.
Let the cubic $T_K$ have $p$ creation
and $q=3-p$ annihilation operators. Before combining
identical terms, the contraction counts are
\[
	\begin{array}{c|c|c|c}
	(p,q)&T_FT_K,\ r=1&T_KT_F,\ r=1&T_KT_F,\ r=2\\ \hline
	(3,0)&3H_1&0&0\\
	(2,1)&2H_1&H_1+H_2&0\\
	(1,2)&H_1&2H_1+2H_2&2H_3\\
	(0,3)&0&3H_1+3H_2&6H_3
	\end{array}
\]
Here an entry counts kernels of the indicated type, with
permuted coordinates. The zero-contraction terms cancel. The $H_1$ maps
act from $q$ to $p+1$ active particles. The $H_2$ fibers
act from $q-1$ to $p$ particles, with one retained coordinate
on each side. The $H_3$ maps act from $q-1$ to $p$ particles.
These are bounded maps by \eqref{eq:mixed-cubic-kernel-bounds}.
The identity
$[T_F^*,T_K]=-[T_F,T_K^*]^*$ covers the adjoint of
the retained-coordinate term, using the same table with
$(p,q)$ interchanged. It reverses the input and output
spaces without taking a partial trace.

For two retained-coordinate cubic terms, the first
commutator is explicitly
\[
	[T_F,T_G]=\int\{F(x,y)G(z,x)-G(x,y)F(z,x)\} a_x^*a_y^*a_z^*a_z\d x\d y\d z .
\]
For each $z$ its $(x,y)$ kernel has $L^2$ norm at most
$2\alpha(F)\alpha(G)$.
The negative of $[T_F,T_G^*]$ is the sum of
\begin{align*}
	&\int F(x,y)\overline{G(y,w)}
	a_y^*a_x^*a_wa_x\d x\d y\d w,\\
	&\int F(x,y)\overline{G(z,x)}
	a_z^*a_y^*a_za_x\d x\d y\d z,\\
	&\int H(x,z)a_z^*a_x^*a_za_x\d x\d z,\qquad
	H(x,z)=\int F(x,y)\overline{G(z,y)}\d y,\\
	&\dGamma(h)+\dGamma(B),\qquad
	h(x)=\int F(x,y)\overline{G(x,y)}\d y,\quad
	B(y,x)=F(x,y)\overline{G(y,x)} .
\end{align*}
For the first two terms, the retained coordinates are
$x$ and $z$, respectively. Their fiber bounds follow from
\begin{align*}
	\int|F(x,y)G(y,w)|^2\d y\d w
	&\le\alpha(F)^2\alpha(G)^2,\\
	\int|F(x,y)G(z,x)|^2\d x\d y
	&\le\alpha(F)^2\alpha(G)^2 .
\end{align*}
Both $H$ and $h$ have $L^\infty$ norm at most
$\alpha(F)\alpha(G)$ by row Cauchy--Schwarz.
For $B$, that inequality applied once to a row and once
to a column gives
\[
	\max\left\{\mathop{\rm ess\,sup}_y\int|B(y,x)|\d x,
	\mathop{\rm ess\,sup}_x\int|B(y,x)|\d y\right\}
	\le\alpha(F)\alpha(G).
\]
Schur's test \cite{Schur} therefore bounds the remaining
one-particle map. The same row calculation bounds the
$(x,y)$ fiber in $[T_F,T_G]$ at fixed $z$.
Every map just listed is bounded between its active particle
spaces. Its Wick lift has at most four field operators and
therefore a factor at most $C(n+1)^2$.
The commutation identities can first be verified on finite
linear combinations of bounded tensor kernels. Approximate
$K$ in $L^2$. For $F,G$, first truncate their support and
values, then average on finite product partitions.
Jensen's inequality preserves the row, column and full
$L^2$ bounds at the averaging step. Refining the partitions
and removing the truncations gives almost-everywhere
convergence along a subsequence and convergence in $L^2$.
The retained-coordinate maps and their adjoints converge
strongly: first use bounded compactly supported input
functions, then use the common operator bound and density.
The displayed estimates then extend the identities to the
stated kernel classes. The two remaining retained-coordinate
commutators follow from
$[T_F^*,T_G]=-[T_G,T_F^*]$ and
$[T_F^*,T_G^*]=-[T_F,T_G]^*$.
For an adjoint block, the factor $(n+1)^2$ is unchanged
up to a fixed constant, since the particle-number shift is
bounded. This completes the canonical assertion.

For the actual phase, $F(x,y)=\kappa_N(x-y)a(y)$
has $\alpha(F)\le C\epsilon$ for $a\in H^2$.
Its dressed kernels
$\kappa_N(x-y)a(y)h(x,z)$ have Hilbert--Schmidt
norm at most $C\epsilon$.
The two derivatives needed for $X,Y$ were checked
in Lemma~\ref{lem:phase-second-regularity}.
Expanding $q=1-|\phi\rangle\langle\phi|$ on the three active coordinates gives two
types of finite-rank correction. A projection on a retained coordinate
gives a cubic term with a Hilbert--Schmidt kernel. A projection on the
newly created coordinate gives $F'(x,y)=k(x)u(y)$, with
$k(x)=\int F(x,z)\bar v(z)\d z$.
For the latter, the row, column and full $L^2$ estimates follow
from $\|k\|_\infty\le\alpha(F)\|v\|_2$,
$\|k\|_2\le\|F\|_2\|v\|_2$ and $u\in L^2\cap L^\infty$.
Repeated insertions remain within these two classes, so the
required finite expansion introduces no spectator factor.

Collect the original generalized-field square roots
and the original $\Theta$ on the right, writing a
cubic term as $T_a b_a(M)$ with shift $d_a\in\{\pm1,\pm3\}$.
On all active blocks, including the corners of $\Theta$,
\[
	|b_a(n)|\le C,\qquad |b_a(n+d)-b_a(n)|\le C_d N^{-1/2}.
\]
The cutoff is Lipschitz, and the square roots have
first differences $O(N^{-1})$ away from the physical
upper boundary. The active region is below that
boundary for all sufficiently large $N$.
The coefficient in the product part of the commutator is
\[
	b_a(n+d_b)b_b(n)-b_b(n+d_a)b_a(n)=O(N^{-1/2}).
\]
Its block norm is therefore at most
$C N^{-1/2}\epsilon^2(n+1)^3$, which is bounded
by $C\epsilon^2(n+1)^2$ on the particle-number cutoff range.
The matched part uses the canonical estimate.
Each ordered product has particle-number shift at most six.
Enlarge the original active range by six sectors when choosing
$E_N$; its upper endpoint is still at most $C\sqrt N$.
This includes the input, intermediate and output blocks.
On every nonzero block $n\to m$,
$(m+1)^r/(n+1)^r\le7^{|r|}$.
Summing the finitely many shifts by Cauchy--Schwarz gives
\eqref{eq:actual-cubic-contractions}, with $E_N$ on both sides.
Number-gauge derivatives only multiply cubic terms by their shifts.
Use the ordinary bosonic commutation relations
\cite{DerezinskiGerard} before restoring the cutoff,
and absorb the finitely many small $N$ separately.
Only first coefficient differences have been used.
\end{proof}

\begin{proposition}\label{prop:time-remainder-sharp}
The exact cubic time remainder
$R_{\rm time}=-ie^{-C}\nabla(e^C)+iX$ satisfies
\begin{equation}\label{eq:time-remainder-sharp}
	\|M^rTz\|\le C_{r,I}N^{-3/2}\|M^{r+2}E_Nz\|, \qquad T=R_{\rm time},\nabla R_{\rm time},i[M,R_{\rm time}] .
\end{equation}
It satisfies the one-weight form hypothesis
\eqref{eq:one-weight-static}, and its contribution to
the original comparison obeys
\begin{equation}\label{eq:time-remainder-dynamics}
	\left\|\int_s^tU_{\rm ex}(t,r)R_{\rm time}(r) U_3(r,s)z\d r\right\| \le C_{I,s}N^{-5/4}\|F_3(s)^{3/2}z\|.
\end{equation}
\end{proposition}
\begin{proof}
Use the exact identity
\[
	R_{\rm time} =-i\int_0^1(1-a)e^{-aC}[X,C]e^{aC}\d a .
\]
The number-weighted phase bounds and
\eqref{eq:actual-cubic-contractions} prove the
undifferentiated estimate. Its middle derivative is
$\nabla[X,C]=[Y,C]$, since $[X,X]=0$.
The derivatives of the exterior phases insert $X$
through their exact Duhamel formula. The block bound
$C\epsilon(n+1)^{3/2}$ makes $X$ uniformly bounded
on all fixed number-weighted spaces in the cutoff region.
The number derivative is handled by inserting
$[M,X]$ or $[M,C]$, including in the exterior phases.
These have the same bounds. The support remains
two-sided in $E_N$.

For the form estimate choose $r=-3/2$ in
\eqref{eq:time-remainder-sharp} and pair to obtain
\[
	|T[u,v]|\le C\epsilon^2 \|M^{3/2}E_Nu\|\|M^{1/2}E_Nv\| \le C\epsilon^2 \|M Q^{1/2}E_Nu\|\|Q^{1/2}E_Nv\|.
\]
For the evolution estimate, use
$M^2E_N\le\Lambda_N^{1/2}M^{3/2}$,
Corollary~\ref{cor:Taylor-three-halves} and unitarity
of the left flow. This proves \eqref{eq:time-remainder-dynamics}
directly, including the reversed time direction.
\end{proof}

\subsubsection{The tail in total kernel degree}

\begin{lemma}\label{lem:exact-degree-tail}
Group the original quadratic expansion as in
Proposition~\ref{prop:GP-weight}, retaining its scalar,
linear and scattering cancellations inside the groups.
For total kernel degree $J>J_*=\lfloor c_0\sqrt N\rfloor$,
the sum of the exact fourth Taylor remainders satisfies
\eqref{eq:geometric-tail-static}.
\end{lemma}
\begin{proof}
The geometric majorant \eqref{eq:degree-majorant}
and the differentiated majorant proved in
Proposition~\ref{prop:full-second-form} give, for
the cancelled source group $S_J$,
\[
	|T_J[u,v]|\le C(1+J)^b\theta^{\max(J-b_0,0)} \|Q^{1/2}u\|\|Q^{1/2}v\|, \quad T_J=S_J,\nabla S_J,i[M,S_J],\quad\theta<1 .
\]
The fourth commutator is a finite sum of words
$C^j S_J C^k$, $j+k=4$. Full-energy graph
bounds control these words and the exterior phase
in its exact integral remainder by a constant
independent of $J,N$. A time or number derivative
uses the same graph bounds for the inserted
$X$ or $[M,C]$ and the displayed bound for $T_J$.
Finally
$\sum_{J>c_0\sqrt N}(1+J)^b\theta^{\max(J-b_0,0)}
\le C_K N^{-K}$ for every fixed $K$.
This proves the bound for the exact tail.
\end{proof}

\subsection{Fixed-degree kernels and source recursions}
\label{sec:source-foundations}

The remaining static estimates use two ingredients: extraction
of a fixed-degree kernel and a recursion for its source
commutators. Here degree counts the fields in a normal-ordered
monomial.
Normal ordering and contractions use Wick's theorem \cite{Wick};
our Fock-space normalization follows \cite[Section 9.4]{DerezinskiGerard}.
The energy-weighted extraction and recursion estimates below are
proved for the kernels of the present model.

\subsubsection{Extracting a kernel at the correct number argument}

Work at a fixed time on $\mathfrak h=q_tL^2(\mathbb R^3)$.
For a separately symmetric kernel $K:\mathfrak h^{\otimes_s q}
\to\mathfrak h^{\otimes_s p}$ use the Wick convention
without a factor $1/(p!q!)$. With this convention, the sector
formula for Wick quantization \cite[Definition 9.46]{DerezinskiGerard} gives its
$n$ to $m=n+p-q$ block as
\begin{equation}\label{eq:Wick-lift-block}
	\mathcal W_{p,q}(K)\big|_n =\frac{\sqrt{n!\,m!}}{(n-q)!} \operatorname{Sym}_m(K\otimes I_{n-q}) \operatorname{Sym}_n,\quad n\ge q .
\end{equation}
It vanishes for $n<q$. In particular its smallest
$q$ to $p$ block is $\sqrt{p!q!}K$.
The identity includes scalar, pure creation and pure
annihilation kernels.

\begin{lemma}
\label{lem:fixed-degree-extraction}
Let $R$ be a normal-ordered polynomial of degree at most
$d$, with all its number-dependent coefficients on the right:
\[
	R=\sum_{p+q\le d}\mathcal W_{p,q}(K_{pq}(\cN)).
\]
Fix a target input number $n$. To extract $K_{pq}(n)$,
replace every right coefficient $f(k)$ in the expression
for $R$ by $f(k+n-q)$ and take the $q$ to $p$ block;
denote the result by $R^{[n,q]}_{pq}$. Then
\begin{equation}\label{eq:triangular-kernel-extraction}
	\frac{R^{[n,q]}_{pq}}{\sqrt{p!q!}} =K_{pq}(n)+ \sum_{r=1}^{\min(p,q)}\frac1{r!} \operatorname{Sym}_p (K_{p-r,q-r}(n)\otimes I_r)\operatorname{Sym}_q .
\end{equation}
Let $\mathcal E=A_P$ or $Q_{N,P}$.
If all fixed-sector blocks required by this recursion
have $\mathcal E$-form norm at most $\lambda$,
then the extracted kernels have form norm at most
$C_d\lambda$. Their general blocks satisfy
\begin{equation}\label{eq:energy-kernel-lift}
	|\mathcal W_{p,q}(K_{pq}(n))[u_m,v_n]| \le C_d\lambda(n+1)^{(p+q)/2-1} \|\mathcal E_m^{1/2}u_m\| \|\mathcal E_n^{1/2}v_n\|.
\end{equation}
If the fixed-sector bounds are instead ordinary
operator bounds, extraction and lifting give
\begin{equation}\label{eq:ordinary-kernel-lift}
	\|R v_n\|\le C_d\lambda(n+1)^{d/2}\|v_n\|.
\end{equation}
The latter assertion also applies to polynomials taking
values in an auxiliary Hilbert source space, or acting
on its tensor product with Fock space. The source
coordinates do not count as fields.
\end{lemma}
\begin{proof}
Formula \eqref{eq:Wick-lift-block} follows from the
creation and annihilation normalization on symmetric
sectors. Among the terms of $R$, only kernels with
indices $(p-r,q-r)$ can contribute to its $q$ to
$p$ block. Their normalization in that block is
$\sqrt{p!q!}/r!$. Here the right coefficient always acts on the total input
particle number, before annihilation. In particular,
$f(\cN+n-q)\Pi_q=f(n)\Pi_q$.
For $a-b=p-q$ and $b\le q$, the shifted block is therefore
\[
	\Pi_p\mathcal W_{a,b}(K_{ab}(\cN+n-q))\Pi_q
	=\frac{\sqrt{p!q!}}{(q-b)!}
	\operatorname{Sym}_p(K_{ab}(n)\otimes I_{q-b})
	\operatorname{Sym}_q .
\]
Set $a=p-r$, $b=q-r$ to obtain
\eqref{eq:triangular-kernel-extraction}.
This is an auxiliary shift used to read the target kernel
from a small block. Recursively subtract the lower-degree
kernels to complete the triangular extraction.

For the form version, write an extracted kernel as
$K=\mathcal E_a^{1/2}B\mathcal E_b^{1/2}$ in the
sense of forms, where $B$ is bounded between the active
particle spaces. For $r$ spectator coordinates, its pairing is
\[
	\left\langle(\mathcal E_a^{1/2}\otimes I_r)u,
	(B\otimes I_r)(\mathcal E_b^{1/2}\otimes I_r)v\right\rangle .
\]
Its bound is unchanged because $\|B\otimes I_r\|=\|B\|$.
On the partially symmetric particle space,
$\mathcal E_a\otimes I_r\le\mathcal E_{a+r}$ as forms:
the omitted kinetic terms, mass $r$, and, for $Q$,
all omitted pair potentials are nonnegative.
The total energy commutes with symmetrization, so
symmetrization is a contraction in its form norm.
Thus every subtraction in the triangular recursion has
a bound independent of $N$. Induction in $p+q$ gives
the fixed-kernel estimate. This argument also proves
the estimate for Hilbert-valued spectator functions.

For symmetric $u_m$ and $p\le\min(d,m)$, permutation
symmetry gives exactly
\begin{align*}
	\langle u_m,(\mathcal E_p\otimes I_{m-p})u_m\rangle
	={}&\frac p m K_{P,m}[u_m]+(p+1)\|u_m\|^2\\
	&+\mathbf1_{\{\mathcal E=Q\}}
	\frac{p(p-1)}{m(m-1)}V_{P,m}[u_m]\\
	\le{}&\frac{C_d}{m+1}\mathcal E_m[u_m].
\end{align*}
Terms with zero denominator are zero; for $m=p=0$
use $\mathcal E_0=1$ directly. This includes $p=0$:
the mass term still supplies the factor $(m+1)^{-1}$.
Apply the fixed-kernel form estimate to
\eqref{eq:Wick-lift-block} and use this identity
on both arguments. The factorial prefactor is bounded
by $C_d(n+1)^{(p+q)/2}$, and $m+1,n+1$ are
comparable for the fixed shift $p-q$.
The two energy fractions give
\eqref{eq:energy-kernel-lift}.
Without energy weights the same calculation gives
\eqref{eq:ordinary-kernel-lift}.

We also record the passage from blocks to forms on Fock space.
If the hypotheses hold uniformly at every target number $n$,
then, for each fixed real $s$,
\begin{equation}\label{eq:fixed-degree-weighted-form}
	|R[u,v]|\le C_{d,s}\lambda
	\|M^{d/2-1-s}\mathcal E^{1/2}u\|
	\|M^s\mathcal E^{1/2}v\| .
\end{equation}
To prove this, fix a shift $h=p-q$. Put $m=n+h$ in
\eqref{eq:energy-kernel-lift} and sum over $n$.
The ratio $(m+1)/(n+1)$ lies between $(d+1)^{-1}$
and $d+1$ on every nonzero block. Cauchy--Schwarz
therefore gives the displayed bound for this shift.
There are at most $2d+1$ shifts. Summing them changes
only the constant. The same calculation applies with a
particle-number cutoff and with Hilbert-valued spectators.

All these form identities initially hold on finite sums of
symmetric tensors with factors in $q_tC_c^\infty(\mathbb R^3)$.
To check density, take $f_j\in C_c^\infty$ tending to
$f\in q_tH^1$ in $H^1$. Then $q_tf_j\to q_tf=f$
in $H^1$, since the finite-rank projection is bounded there.
Finite tensor sums, followed by symmetrization, give the
same approximation in each particle sector.
The pair potential is a bounded multiplier at fixed $N$
and particle number, so this approximation is also in the
$Q$ form norm.
Particle-number truncation and sectorwise approximation
give density also in every displayed weighted form norm.
Since $M$ and $\mathcal E$ strongly commute, their weighted
products are defined by the joint spectral calculus.
Equation \eqref{eq:fixed-degree-weighted-form} supplies
the unique continuous extension of each pairing.
The preceding form estimates provide the constants uniform in $N$.

For a Hilbert source space $\mathcal S$, replace the
kernel by a map
$\mathfrak h^{\otimes_s q}\to
\mathcal S\otimes\mathfrak h^{\otimes_s p}$, or by
a map between two such tensor products.
The source factor is left untouched in
\eqref{eq:Wick-lift-block} and in the triangular
subtractions. The ordinary operator proof is unchanged.
Finite-dimensional compression of the active and
source variables, followed by the uniform bounds,
gives the result for general bounded kernels.
\end{proof}

\begin{remark}\label{rem:fixed-degree-scope}
The constants depend on the fixed maximal degree $d$.
For normalized quadratic chains, the Leibniz estimate below
controls dependence on the total kernel degree $J$.
The coefficient shift in \eqref{eq:triangular-kernel-extraction}
evaluates the extracted kernel at the target particle number $n$.
\end{remark}

\subsubsection{A source derivative with one unbounded input map}

Let $\mathcal S$ be a source Hilbert space and let
$D$ be an energy map from Fock space to
$\mathcal S\otimes\cF$. The kinetic and potential
examples are
\[
	D_Kz=(\nabla_xa_xz)_x,\qquad D_Vz=(\sqrt{W_N(x-y)/2}\,a_xa_yz)_{x,y}.
\]
Their squared norms are $K_P[z]$ and $V_P[z]$.
We call these auxiliary operator-valued maps source maps; their
free spatial variables are the source coordinates. This terminology
refers to the factorization of the energy forms.
For a canonical cubic term $T$, set
$T_{\rm out}=I_{\mathcal S}\otimes T$ and
$\delta_TD=DT-T_{\rm out}D$.
In this subsection $T$ contains its spatial projections
but no number-dependent cutoff coefficient.

\begin{lemma}\label{lem:source-recursion}
Suppose, for each of finitely many allowed cubic terms,
\[
	\delta_TD=P_1(T)+A_1(T)D.
\]
Assume that $P_1$ has degree at most $a$
and fixed-sector operator norm at most $C$. Assume also
that $A_1$ has degree at most one and fixed-sector
source-operator norm at most $C\epsilon$. The cubic terms have fixed-sector
norm at most $C\epsilon$. Here $a$ is fixed.
Then every ordered source commutator of length
$1\le j\le4$ has a decomposition $S_j=P_j+A_jD$ with
\begin{align}
	\deg P_j&\le a+j-1,&
	\deg A_j&\le j,\notag\\
	\|S_jv_n\|&\le C_j\left\{
	\epsilon^{j-1}(n+1)^{(a+j-1)/2}\|v_n\|
	+\epsilon^j(n+1)^{j/2}\|Dv_n\|\right\}.
	\label{eq:abstract-source-recursion}
\end{align}
All source compositions keep the original operator order.
\end{lemma}
\begin{proof}
The rectangular product rule gives
\begin{align}
	P_{j+1}&=P_jT-T_{\rm out}P_j+A_jP_1(T),\notag\\
	A_{j+1}&=A_jT_{\rm out}-T_{\rm out}A_j+A_jA_1(T).
	\label{eq:ordered-source-recursion}
\end{align}
The final $D$ acts on the original input throughout.
In each commutator the zero-contraction terms cancel,
because $T$ is the identity on the source coordinates.
Commuting with a cubic term therefore raises
normal degree by at most one. The two remaining
products have the asserted degrees as well.
If $a_j,b_j$ denote the required fixed-sector norms
of $P_j,A_j$, every product in
\eqref{eq:ordered-source-recursion} is a composition
of bounded fixed-sector maps and
\[
	a_{j+1}\le C_j(\epsilon a_j+b_j),\qquad b_{j+1}\le C_j\epsilon b_j .
\]
Starting with $a_1\le C$, $b_1\le C\epsilon$
gives $a_j\le C_j\epsilon^{j-1}$ and
$b_j\le C_j\epsilon^j$.
Apply the source version of
Lemma~\ref{lem:fixed-degree-extraction} to these
fixed-degree polynomials. This proves the estimate.
\end{proof}

For an anti-self-adjoint canonical $C_{\rm fr}$, put
$D_j=\delta_{C_{\rm fr}}^jD$ and
$C_{{\rm fr},{\rm out}}=I_{\mathcal S}\otimes C_{\rm fr}$.
Adjunction gives
$(D_1)^*=D^*C_{{\rm fr},{\rm out}}-C_{\rm fr}D^*$,
so the exact source-product identity is
\begin{equation}\label{eq:fourth-source-product}
	\mathcal D_{C_{\rm fr}}^4(D^*D) =D_4^*D+4D_3^*D_1+6D_2^*D_2+4D_1^*D_3+D^*D_4 .
\end{equation}
This is an identity on the form core, extended whenever
the source estimates make the pairings continuous.
The coefficient $6$ and its positive sign concern
the full product identity; integrations by parts in
commutator algebra can redistribute these middle terms.
For non-anti-self-adjoint cubic terms the adjoint
cubic term must be kept explicitly.

The hypotheses of Lemma~\ref{lem:source-recursion}
must be checked from the spatial kernels for each
choice of $D$. In particular inserting the actual
$b_a(\cN)$ inside $T$ before using this lemma would
destroy the zero-contraction cancellation and its
degree count. We retain those coefficients through
the exact block rule
\[
	(T_ab_a(\cN))(T_bb_b(\cN))\big|_n =T_aT_b\,b_a(n+d_b)b_b(n)\big|_n ,
\]
where $d_b$ is the number shift of $T_b$.
Bounded coefficients with first differences
$O(N^{-1/2})$ have products along any fixed-length
path differing from their value frozen at $n$
by $O(N^{-1/2})$, by telescoping.
This observation concerns coefficient size only.
The operator words multiplying the differences
still have to be estimated, including all their
endpoint terms.

\subsection{Spatial estimates for the cubic phase}
\label{sec:spatial-phase}

The support and spatial norms of the cubic kernel supply the
small factors needed in the commutator estimates. We retain the
original cutoff and projected generalized fields, and set
$\epsilon=N^{-3/4}$, $\delta=N^{-5/4}$ and
$E_N=\mathbf1_{M\le\Lambda_N}$, with a fixed enlargement
of $\Lambda_N\le C\sqrt N$ when necessary.
The pointwise bound for $\kappa=\kappa_N$ and its
support in $|x|\le N^{-1/2}$ imply
\begin{equation}\label{eq:spatial-kappa-norms}
	\|\kappa\|_2\le C\epsilon,\quad \|\kappa\|_{6/5}\le C\delta,\quad \|\kappa\|_1\le CN^{-3/2},\quad \|\kappa\|_{3/2}\le CN^{-1}.
\end{equation}
Indeed, integrate $C\sqrt N$ on $|x|\le N^{-1}$
and $CN^{-1/2}|x|^{-1}$ on the remaining support.
In particular
$\|\kappa\|_1\|\kappa\|_{3/2}\le C\delta^2$
and $\delta^2N=\epsilon^2$.

\begin{lemma}\label{lem:paired-cubic-smallness}
Let $f\in L^\infty$ and $F(x,y)=\kappa(x-y)f(y)$.
On the canonical Fock space set
\[
	P_B=\int F(x,y)a_x^*a_y^*B_x\d x\d y .
\]
Either $B_x=\int h(x,z)a_z\d z$ with
$\|h\|_{\mathrm{op}}\le C$, including $h=I$, or
$B_x=\int h(x,z)a_z^*\d z$ with
$\|h\|_{\mathrm{HS}}\le C$ and
$\|\nabla_zh\|_{\mathrm{HS}}\le C\sqrt N$.
For $\nu=n+1$ and $\psi_n$ in the kinetic form domain,
\begin{align}
	\|P_B\psi_n\|
	&\le C\{\epsilon\sqrt\nu\|\psi_n\|
	+\delta\nu\|K^{1/2}\psi_n\|\},\notag\\
	\|P_B^*\psi_n\|
	&\le C\delta\nu\|K^{1/2}\psi_n\|.
	\label{eq:paired-cubic-smallness}
\end{align}
The constants include $\|f\|_\infty$ and the stated pair-kernel norms.
\end{lemma}
\begin{proof}
Write
\[
	I_B=\int\|B_x\psi_n\|^2\d x,\qquad D_B=\int\|\nabla_ya_yB_x\psi_n\|^2\d x\d y .
\]
The annihilation and creation bounds give
\begin{equation}\label{eq:B-source-norms}
	I_B\le C\nu\|\psi_n\|^2,\qquad D_B\le C\{\nu\|K^{1/2}\psi_n\|^2+N\|\psi_n\|^2\}.
\end{equation}
For annihilation, commute $B_x$ through $a_y$.
For creation the additional commutator is $h(x,y)$,
whose derivative contributes at most
$C\|\nabla_yh\|_{\mathrm{HS}}^2\|\psi_n\|^2$.

In $P_B^*P_B$, normal order only the two outer
annihilators against the two outer creators, leaving
$B_x^*,B_z$ at the ends. The term without a contraction is
\[
	\int\overline{F(x,y)}F(z,w) \langle a_za_wB_x\psi_n,a_xa_yB_z\psi_n\rangle \d x\d y\d z\d w .
\]
Cauchy--Schwarz, followed by interchange of $x,z$,
bounds its absolute value by
\[
	\int\d x\d z \left\|\int\overline{F(x,y)}a_xa_yB_z\psi_n\d y\right\|^2 \le C\delta^2\nu D_B .
\]
Here use the Hilbert-valued $L^{6/5}$--$L^6$
H\"older and Sobolev inequalities in $y$, then
integrate the extra annihilator in $x$.

There are four single contractions. The contraction
$x=z$ is
$\int\|\int\overline{F(x,y)}a_yB_x\psi_n\d y\|^2\d x$,
bounded by $C\delta^2D_B$ in the same way.
The contraction $x=w$ is
\[
	\int\overline{F(x,y)}F(z,x) \langle a_zB_x\psi_n,a_yB_z\psi_n\rangle \d x\d y\d z .
\]
Apply weighted Cauchy--Schwarz to
$|F(x,y)F(z,x)|$. In the first square, integration
in $y$ leaves a coefficient bounded by
$C\|\kappa\|_1|\kappa(z-x)|$. In the second,
integration in $x$ leaves
$C(|\kappa|*|\kappa|)(z-y)$.
Their $L^{3/2}$ norms in the remaining annihilation
variable are at most
$C\|\kappa\|_1\|\kappa\|_{3/2}\le C\delta^2$.
The $L^{3/2}$--$L^3$ inequality and Sobolev
therefore bound both squares by $C\delta^2D_B$.
The $y=z$ contraction is its adjoint.
For $y=w$ the coefficient is
$H(x,z)=\int\overline{F(x,y)}F(z,y)\d y$.
Since $|H(x,z)|=|H(z,x)|$, Cauchy--Schwarz
reduces to $\int|H(x,z)|\|a_zB_x\psi_n\|^2\d x\d z$.
Young's inequality gives
$\sup_x\|H(x,\cdot)\|_{3/2}\le C\delta^2$,
so the same Sobolev argument applies.

The first double contraction is bounded by
$\sup_x\int|F(x,y)|^2\d y\,I_B\le C\epsilon^2I_B$.
For the second, the kernel
$\overline{F(x,y)}F(y,x)$ has both Schur integrals
bounded by $C\epsilon^2$. Apply Schur's estimate
to the Hilbert-valued function $x\mapsto B_x\psi_n$.
Consequently
\[
	\|P_B\psi_n\|^2 \le C\{\epsilon^2I_B+\delta^2\nu D_B\}.
\]
Equations \eqref{eq:B-source-norms} and
$\delta^2N=\epsilon^2$ prove the first assertion.

For the adjoint, set
$d_x=\int\overline{F(x,y)}a_ya_x\psi_n\d y$.
Direct H\"older--Sobolev estimation gives
$\int\|d_x\|^2\d x\le
C\delta^2\nu\|K^{1/2}\psi_n\|^2$.
The map $(d_x)_x\mapsto\int B_x^*d_x\d x$
has norm at most $C\sqrt\nu$ between the relevant
sectors, by the bound for $I_B$ and Hilbert-space
adjunction. This proves the second assertion.
All computations can first be made with truncated
kernels on the smooth finite-particle core.
The displayed bounds permit passage to the original
kernels and closure in the input kinetic form norm.
\end{proof}

\begin{proposition}
\label{prop:phase-spatial-smallness}
For $D=C,X=\nabla_tC,Y=\nabla_t^2C$, and for
any fixed number of their number-gauge derivatives,
\begin{align}
	\|M^rDz\|&\le C_{r,I}\{
	\epsilon\|M^{r+1/2}E_Nz\|
	+\delta\|M^{r+1}A_P^{1/2}E_Nz\|\},
	\label{eq:phase-spatial-smallness}\\
	\|M^rDz\|+\|M^r(e^{aC}-I)z\|
	&\le C_{r,I}\epsilon\|M^rQ_{N,P}^{1/2}z\|,
	\qquad |a|\le1 .
	\label{eq:phase-form-to-L2}
\end{align}
The estimates also hold for the individual projected
canonical cubic terms and their allowed coefficient
and connection replacements on the particle-number cutoff range.
\end{proposition}
\begin{proof}
For the annihilation field with kernel $\gamma=I+p$, use the
bounded-operator case of
Lemma~\ref{lem:paired-cubic-smallness}; for $\sigma$
use its creation case. The hyperbolic-kernel bounds
and their first two ordinary time derivatives were
verified in Lemma~\ref{lem:phase-second-regularity}.
Their gradient Hilbert--Schmidt bound remains
$C_I\sqrt N$, as in \eqref{eq:B-source-norms}.
The amplitudes $\phi,\dot\phi,\ddot\phi$ lie in
$H^2\subset L^\infty$.
On each number block, the generalized-field factors
and the original $\Theta$ collect into bounded
scalars. The output projection is an $L^2$
contraction. The input excitation vector has the
same kinetic quadratic form as its ambient vector.
Thus neither operation adds a number weight.

It remains to verify the connection terms. Consider a connection $R=|u\rangle\langle v|$ on an active field operator,
with $u,v$ uniformly bounded in $H^2$.
When the connection acts on the field $B$, it preserves
the required operator or Hilbert--Schmidt class and its gradient bound.
The outer creation operators contribute $a^*(u)$ times a quadratic
field, whose kernel is
\[
	k(x)h(x,z),\quad k=\kappa*(f\bar v),\qquad\hbox{or}\qquad f(y)\int\kappa(x-y)\bar v(x)h(x,z)\d x .
\]
Young's inequality gives the bound $CN^{-3/2}$ in operator
norm for the annihilation case and in Hilbert--Schmidt norm
for the creation case. When $h=I$, we retain the multiplication
or convolution operator and use its operator norm. The cubic block and its adjoint consequently have norm at most
$CN^{-3/2}\nu^{3/2}$. Since $A_{P,n}\ge\nu$ and
$N^{-3/2}\le\delta$, this contribution is absorbed by
$\delta\nu A_{P,n}^{1/2}$. A second insertion either acts on
a smooth finite-rank vector or composes a bounded finite-rank
operator with the quadratic kernel. Both operations preserve
the estimates, completing the check of the connection terms in $X,Y$.

Sum the finite-band blocks with the fixed weight
$M^r$ to obtain \eqref{eq:phase-spatial-smallness}.
Number-gauge differentiation multiplies a block
by its fixed particle-number shift. Since $Q_{N,P}\ge A_P\ge M$
and $\delta\Lambda_N\le C\epsilon$,
the estimate for $D$ in \eqref{eq:phase-form-to-L2}
follows. For the exponential use
$e^{aC}-I=\int_0^a e^{bC}C\d b$ and the already
proved weighted phase bounds.
\end{proof}

\begin{corollary}\label{cor:two-phase-factors}
Let $D_1,D_2$ belong to the families in
Proposition~\ref{prop:phase-spatial-smallness}.
Then
\begin{equation}\label{eq:two-phase-factors}
	\|M^rD_1e^{aC}D_2z\| \le C_{r,I}\delta\|M^{r+1}Q_{N,P}^{1/2}z\|, \qquad |a|\le1 .
\end{equation}
The same conclusion holds for a word of any fixed
length containing at least two such factors,
with phases inserted between or at the ends;
the constant may depend on the length.
In particular, for $j=0,1,2$,
\begin{equation}\label{eq:phase-linear-tail}
	\|M^r\nabla_t^j(e^{C_t}-I-C_t)z\| \le C_{r,I}\delta\|M^{r+1}Q_{N,P}^{1/2}z\|.
\end{equation}
\end{corollary}
\begin{proof}
Abbreviate $Q=Q_{N,P}$. Apply
\eqref{eq:phase-spatial-smallness} at the left end.
The phase preserves each fixed number-weighted
$L^2$ and $Q$ form norm. The ordinary cubic
block bound for $D_2$ and its uniform weighted
$Q$ graph bound then give
\begin{align*}
	\|M^rD_1e^{aC}D_2z\|
	&\le C_r\{\epsilon\|M^{r+1/2}e^{aC}D_2z\|
	+\delta\|M^{r+1}Q^{1/2}e^{aC}D_2z\|\}\\
	&\le C_r\{\epsilon^2\|M^{r+2}E_Nz\|
	+\delta\|M^{r+1}Q^{1/2}z\|\}.
\end{align*}
The first term is absorbed by the second because
$Q\ge M$ and $\epsilon^2\Lambda_N^{1/2}\le C\delta$.
For a longer word use its two rightmost $D$
factors, and the uniform weighted $L^2$ bounds
for all remaining left factors. A phase at the
right end is bounded in the weighted form norm.
This proves \eqref{eq:two-phase-factors}.

For $j=0$, use the exact second-order exponential
remainder. For $j=1$, differentiate by the
Duhamel formula and subtract $X$.
Split each integrand as
$(e^{(1-a)C}-I)Xe^{aC}+X(e^{aC}-I)$,
then expand each phase difference by its exact
first-order integral. Each word has two factors.
For $j=2$, the single-$Y$ integral is treated
in the same way after subtraction of $Y$;
the other ordered integrals have two $X$ factors.
The argument takes place in the parallel coordinates,
so it includes the moving-projection connection.
\end{proof}

These maps act from the energy form domain to $L^2$.
Products containing $G_0$ between phase factors are estimated
as forms below.

\subsection{Canonical kinetic sources}
\label{sec:kinetic-source-endpoints}

We first estimate the kinetic sources with canonical fields
and constant number coefficients. The spatial projections and
the original kernels are retained. The number-dependent
coefficients will subsequently be restored using their exact
block shifts.

\begin{proposition}\label{prop:kinetic-source-decomposition}
For each projected canonical cubic term $T$
of $C$, its adjoint, or one ordinary or covariant
coefficient derivative, the map
$D_K=(\nabla_za_z)_z$ satisfies
\begin{equation}\label{eq:kinetic-first-source-decomposition}
	\delta_TD_K=P_1(T)+A_1(T)D_K .
\end{equation}
The source polynomial $P_1$ has degree at most two
and uniformly bounded fixed-sector operator norm.
The source-to-source polynomial $A_1$ has degree
at most one and fixed-sector norm at most
$C_I\epsilon$. Consequently each ordered
canonical source commutator of length $1\le j\le4$
satisfies
\begin{equation}\label{eq:kinetic-higher-source}
	\|S_jv_n\|\le C_{j,I}\{ \epsilon^{j-1}\nu^{(j+1)/2}\|v_n\| +\epsilon^j\nu^{j/2}\|K_{P,n}^{1/2}v_n\|\}, \qquad \nu=n+1 .
\end{equation}
\end{proposition}
\begin{proof}
First omit the projections. For the identity
raising cubic term $T_0$ used in the proof of
Proposition~\ref{prop:sharp-kinetic-block},
put $f_z(y)=\kappa(z-y)\phi(y)$. The exact
first source is
\begin{align*}
	\nabla_z[a_z,T_0]
	={}&a^*(f_z)\nabla_za_z
	+a^*(\nabla_zf_z)a_z\\
	&+\dGamma\bigl(\nabla_z[
	\phi(z)\kappa(\cdot-z)]\bigr).
\end{align*}
Take the first term as $A_1D_K$.
Its fibre coefficient satisfies
$\sup_z\|f_z\|_2\le C\epsilon$, so $A_1$
has the required fixed-sector bound.
For the second term,
$\sup_z\|\nabla_zf_z\|_2\le C$ and the
creation bound followed by integration of $a_z$
give a bounded fixed-sector source map.
For the density term, put
$m_z(x)=\nabla_z[\phi(z)\kappa(x-z)]$.
The gradient-kernel and amplitude bounds give
$\sup_x\int|m_z(x)|^2\d z\le C$.
On a fixed $n$ sector,
$|\sum_{i=1}^n m_z(x_i)|^2
\le n\sum_i|m_z(x_i)|^2$ proves its boundedness.
Both terms therefore belong to $P_1$.

For $T_0^*$, the annihilator commutator is
$a(f_z)a_z$, with the corresponding conjugations.
After differentiation, $a(f_z)\nabla_za_z$
is $A_1D_K$. The remaining source
$a(\nabla_zf_z)a_z$ has a bounded
two-particle-to-source kernel by the same
uniform $L^2$ bound on $\nabla_zf_z$.
Thus it is a degree-two $P_1$.

For the dressed cubic terms $T_h^\pm$ with
$h=p,\sigma$, all derivatives in
$\nabla_z[a_z,T_h^\pm]$ fall on explicit kernels.
The displayed commutators in the proof of
Proposition~\ref{prop:sharp-kinetic-block}
list their two terms and, for the pure creation
cubic term, the additional pair-creation term.
Their differentiated kernels have joint
$L^2$ norm bounded by
\[
	C_I\{\|\nabla\kappa\|_2 +\|\kappa\|_2(1+\|\nabla h\|_{\mathrm{HS}})\} \le C_I .
\]
If the derivative acts on $\kappa$, integrate the other amplitude
variable and apply the row or column $L^2$ bound of $h$.
If it acts on $h$, first integrate $|\kappa|^2$.
These give bounded quadratic source maps on
each fixed sector, including pair creators.
Set $A_1=0$ for these cubic terms.
The adjoint of the two-creation one-annihilation
cubic term has a pair-annihilation source;
the purely annihilating adjoint has zero source.
These are direct commutator computations.

Now restore $q$ on each active coordinate. For an excitation
input, with $d$ the particle-number shift of $T$, the exact identity is
\begin{equation}\label{eq:source-projection-identity}
	[a_z,T^q]v_n =P_{n+d-1}\bigl([a_z,T] -\phi(z)[a(\phi),T]\bigr)v_n .
\end{equation}
It follows from $a_zP=P_{\rm rest}a(q_z)$.
The remaining-particle projection is independent of $z$
and therefore has zero derivative.
The correction after differentiation is the
bounded quadratic source
$-\nabla\phi(z)[a(\phi),T]$, of fixed-sector
size $C_I\epsilon$. The retained input-gradient
part can still be written $A_1D_K$, since
$D_Kv_n$ has all remaining particles in the
excitation space. Projection preserves the $C_I\epsilon$ bound
for its fibre creation or annihilation vector. All other projections on remaining coordinates are $L^2$
contractions on the bounded source maps.

Ordinary time differentiation preserves these estimates because
the differentiated amplitudes and hyperbolic kernels satisfy the
same bounds. For a connection $R=|u\rangle\langle v|$, acting on the
annihilation operator of the identity term gives a dressed kernel with bounded
$\nabla R$. The creation operator at the retained coordinate produces a
three-variable kernel controlled by $\|\kappa\|_2$,
$\|\nabla\kappa\|_2$ and the smooth finite-rank vectors.

The other creation operator requires a convolution estimate. Its fibre is $u(y)(\kappa*(\phi\bar v))(x)$. The norm of this fibre and of its
derivative in the retained coordinate are bounded by
$C_I\epsilon$. This follows from $\|\kappa\|_1$, with the
derivative transferred to the smooth amplitude. Any resulting contribution to $A_1$ therefore
satisfies the required bound. Hilbert--Schmidt composition
estimates cover the dressed fields, completing the three active
cases; the spectator connection terms cancel.

We have verified every assumption of
Lemma~\ref{lem:source-recursion} with $a=2$.
Applying it gives \eqref{eq:kinetic-higher-source}.
Only a first spatial derivative of $\kappa$
has been used. All subsequent source recursions
leave the final $D_K$ acting on the input.
\end{proof}

\begin{corollary}
\label{cor:kinetic-fourth-endpoints}
Let $C_{\rm fr}$ be an anti-self-adjoint sum
of the projected canonical cubic terms with bounded
constant number coefficients. On the particle-number cutoff range,
\begin{equation}\label{eq:kinetic-fourth-source}
	\|\delta_{C_{\rm fr}}^4D_K\,v_n\| \le C_I\epsilon^2\nu\|A_{P,n}^{1/2}v_n\|.
\end{equation}
Thus, with $S_4=\delta_{C_{\rm fr}}^4D_K$,
\begin{equation}\label{eq:kinetic-fourth-endpoints}
	|(D_K^*S_4+S_4^*D_K)[E_Nu,E_Nv]| \le C_I\epsilon^2 \|MA_P^{1/2}E_Nu\|\|A_P^{1/2}E_Nv\|.
\end{equation}
The bound is stable under one covariant time
derivative and any fixed number of number-gauge
derivatives.
\end{corollary}
\begin{proof}
At $j=4$, \eqref{eq:kinetic-higher-source}
and $A_{P,n}\ge\nu$ give
$C_I(\epsilon^3+\epsilon^4)\nu^2
\|A_{P,n}^{1/2}v_n\|$.
Use $\epsilon\nu\le C N^{-1/4}$ on the
particle-number cutoff range to obtain
\eqref{eq:kinetic-fourth-source}.
Pair with $D_K$, whose norm is the kinetic
form norm, and sum the finitely many number
shifts. Their weights $\nu$ are comparable,
which allows the full $M$ weight on either side.

Time differentiation of the cubic terms was
included in Proposition~\ref{prop:kinetic-source-decomposition}.
The additional derivative of $D_K$ in parallel
coordinates is a bounded linear source $E_K$:
the spatial derivative of the one-body connection
is a smooth finite-rank bounded operator.
Its $j$-fold canonical source commutator has
degree at most $j+1$ and fixed-sector norm
$C_{j,I}\epsilon^j$, by the same product
recursion or Lemma~\ref{lem:fixed-degree-extraction}.
For $j=4$ its block bound is
$C_I\epsilon^4\nu^{5/2}$.
After $A_P\ge M$ and the particle-number cutoff,
this is smaller than \eqref{eq:kinetic-fourth-source}.
An outside $E_K$ is bounded by
$C_I\sqrt\nu\|v_n\|\le C_I\|A_{P,n}^{1/2}v_n\|$.
Number-gauge derivatives multiply each term by its fixed particle-number
shift.
\end{proof}

The preceding corollary controls the two endpoint
terms of \eqref{eq:fourth-source-product} for the
canonical coefficients. The middle source products
and the errors from restoring the actual
number-dependent coefficients remain separate.

\subsection{Canonical pair-potential sources}
\label{sec:potential-source-endpoints}

Put $\rho=N^{-1}$, $\tau=N^{-1/2}$, and
$D=D_V=(\sqrt{W_N(x-y)/2}\,a_xa_y)_{x,y}$.
The source space is $L^2(\mathbb R^6)$, and
$V_P=D^*D$. Throughout this subsection, $Q=Q_{N,P}$.
As in Section~\ref{sec:kinetic-source-endpoints},
the cubic terms keep their spatial projections and
have constant number coefficients.

\begin{lemma}
\label{lem:potential-source-decomposition}
For every allowed canonical cubic term,
\begin{equation}\label{eq:potential-first-source-decomposition}
	S_1(T)=\delta_TD=P_1(T)+A_1(T)D,\qquad P_1=L+F+R .
\end{equation}
Here $L$ has degree one, $F,R$ degree three,
and $A_1$ degree one. On the particle-number cutoff range,
\begin{align}
	\|Lv_n\|&\le C_I\sqrt\nu\|v_n\|,&
	\|Rv_n\|&\le C_I\delta\nu^{3/2}\|v_n\|,\notag\\
	\|Fv_n\|&\le C_I\rho\nu\|Q_n^{1/2}v_n\|,&
	\|A_1w_n\|&\le C_I\epsilon\sqrt\nu\|w_n\|,
	\label{eq:potential-source-parts}
\end{align}
where $w_n$ is an arbitrary source-valued vector.
In addition $P_1$ has uniformly bounded
fixed-sector norm. In particular
\begin{equation}\label{eq:potential-cubic-source}
	S_1=L+B_3,\qquad B_3=F+R+A_1D,\qquad \|B_3v_n\|\le C_I\epsilon\sqrt\nu \|Q_n^{1/2}v_n\|.
\end{equation}
\end{lemma}
\begin{proof}
Before projection, set $D_x(T)=[a_x,T]$ and
$E_{xy}(T)=[a_x,[a_y,T]]$.
Equation \eqref{eq:pair-comm-identity} gives exactly
\[
	S_1(T)=\sqrt{W_N/2} \{D_x(T)a_y+D_y(T)a_x+E_{xy}(T)\}.
\]
For $T_0$ the first-source formulas already proved
give the source-fibre operator
$A_1=a^*(f_x)+a^*(f_y)$, with
$f_x(z)=\kappa(x-z)\phi(z)$.
The remaining terms are the linear contact source
$L=\sqrt{W_N/2}\,E_{xy}(T_0)$ and the two density
sources
\[
	\sqrt{W_N(x-y)/2}\, \phi(x)\dGamma(\kappa(\cdot-x))a_y
\]
and its $x,y$ interchange. These comprise $F$.
For $T_0^*$ use the corresponding annihilation
fibres for $A_1$; both $P_1$ and $E_{xy}$ vanish.
For $p,\sigma$ cubic terms take $A_1=0$, put
all linear double contractions in $L$, and
all normal cubic terms in $R$.
The dressed adjoint with two annihilators in
$D_x$ contributes only to $R$; the pure
annihilation cubic term contributes zero.

The bound for $A_1$ follows from
$\sup_x\|f_x\|_2\le C_I\epsilon$.
For $L$ use $\int W_N|\kappa|^2\le C$
for the identity term. The complete dressed
double-contraction list and its weighted kernel
estimates were given in the proof of
Proposition~\ref{prop:full-graph}; all its
one-field source maps have norm at most $C_I$.
Lifting gives the factor $\sqrt\nu$.
For $R$, the unweighted dressed $D_x$
has a quadratic kernel with row
Hilbert--Schmidt norm at most $C_I\epsilon$,
uniformly in $x$. This includes the extra
pair kernel $\kappa(z-w)\phi(w)h(z,x)$.
Integrating $W_N(x-y)$ in $x$ supplies
$\|W_N\|_1\le C/N$.
Thus the fixed-sector source norm is at most
$C_I\epsilon N^{-1/2}=C_I\delta$.
Lemma~\ref{lem:fixed-degree-extraction} gives the
asserted cubic lifting bound.

For $F$, the sector Cauchy--Schwarz and Hardy
calculation in \eqref{eq:potential-density-source}
gives $C_I\rho\nu\|K^{1/2}v_n\|$.
It uses precisely \eqref{eq:single-W-hardy}
with the fixed support radius of the potential.
It also has a bounded fixed-sector norm:
use instead
$\|\kappa\|_\infty^2\|W_N\|_1\le C$
after the sector Cauchy--Schwarz inequality.

The exact pair-annihilation projection identity gives
\[
	[a_xa_y,T^q]v_n =P_{\rm rest}(q\otimes q)[a_xa_y,T]v_n .
\]
The extra contribution to $S_1$ is therefore
$P_{\rm rest}\sqrt{W_N/2}(q\otimes q-I)
[a_xa_y,T]v_n$.
For every rank-one summand in $q\otimes q-I$,
\[
	\sup_y\int W_N(x-y)|\phi(x)|^2\d x\le C_I/N
\]
shows that
$\sqrt{W_N}(q\otimes q-I)$ has operator norm
$C_I\tau$ on the pair source space.
The unweighted $[aa,T]$ has fixed-sector
source norm $C_I\epsilon$: use the norm
of the pair-annihilation map on fixed sectors
and the cubic block bound on its two ordered terms.
Consequently the correction has size $C_I\delta$.
Its linear and cubic components are separated
by Lemma~\ref{lem:fixed-degree-extraction} and
added to $L,R$ respectively.
Remaining-particle projections are $L^2$
contractions. The input-gradient estimate for
$F$ can equivalently be lifted from its fixed
active-coordinate energy map; inserting the
active projections uses their uniform $H^1$
bound and retains $C_I\rho\nu Q_n^{1/2}$.
Projecting the remaining field coordinate of the fibre of
$A_1$ preserves its norm bound.

This proves \eqref{eq:potential-source-parts}
and the fixed-sector assertion for $P_1$.
Finally $\rho\sqrt\nu\le C\epsilon$,
$\delta\nu\le C\epsilon\sqrt\nu$ and $Q\ge M$
on the particle-number cutoff range prove
\eqref{eq:potential-cubic-source}.
\end{proof}

\begin{proposition}
\label{prop:potential-higher-sources}
Every ordered canonical source commutator satisfies
\begin{equation}\label{eq:potential-higher-source}
	\|S_jv_n\|\le C_{j,I}\{ \epsilon^{j-1}\nu^{(j+2)/2}\|v_n\| +\epsilon^j\nu^{j/2}\|V_{P,n}^{1/2}v_n\|\}, \qquad 1\le j\le4 .
\end{equation}
The second source also satisfies the stronger bound
\begin{equation}\label{eq:potential-second-source}
	\|S_2(T,U)v_n\|\le C_I\epsilon\sqrt\nu\|Q_n^{1/2}v_n\|.
\end{equation}
For an additional cubic term $T'$,
\begin{equation}\label{eq:potential-second-source-word}
	\|S_2(T,U)T'v_n\| \le C_I\epsilon^2\nu^2\|Q_n^{1/2}v_n\|.
\end{equation}
The latter bound also allows one more cubic term
at the right. All assertions remain valid under
one covariant derivative and fixed number-gauge
derivatives.
\end{proposition}
\begin{proof}
Lemma~\ref{lem:potential-source-decomposition}
verifies the hypotheses of
Lemma~\ref{lem:source-recursion} with $a=3$.
It gives \eqref{eq:potential-higher-source},
including the decomposition $S_j=P_j+A_jD$
with degrees $j+2,j$ and fixed-sector norms
$C_j\epsilon^{j-1},C_j\epsilon^j$.

For \eqref{eq:potential-second-source}, the
degree-one $L$ has commutator of degree at most
two and fixed-sector norm $C_I\epsilon$.
Its lift is $C_I\epsilon\nu\|v_n\|$.
For the density source use its sharper energy
bound, the sharp $Q$ block bound for $U$,
and the ordinary bound for $U_{\rm out}$:
\[
	\|\delta_UF\,v_n\| \le C_I\{\rho\epsilon\nu^{5/2}+\rho\nu\} \|Q_n^{1/2}v_n\|.
\]
The two ordered $R$ terms are bounded by
$C_I\delta\epsilon\nu^3\|v_n\|$.
The commutator $\delta_UA_1$ has degree at most
two and fixed-sector norm $C_I\epsilon^2$;
its product with $D$ is bounded by
$C_I\epsilon^2\nu\|Q_n^{1/2}v_n\|$.
The remaining product $A_1S_1(U)$ satisfies
\[
	C_I\{\epsilon\nu\|v_n\| +\epsilon^2\nu\|Q_n^{1/2}v_n\|\}.
\]
These terms are exactly the source product rule
$\delta_U(P_1+A_1D)=\delta_UP_1+
(\delta_UA_1)D+A_1S_1(U)$.
Use $Q\ge M$, $\rho\nu^2\le C$,
$\rho\sqrt\nu/\epsilon\le C$,
$\delta\nu^2\le C$, and $\epsilon\sqrt\nu\le C$
to obtain \eqref{eq:potential-second-source}.
It controls the entire source and its entire
square, including lower-field contractions.

To estimate a right cubic term, use the
decomposition underlying
\eqref{eq:potential-higher-source}.
At $j=2$ combine its two terms with
\eqref{eq:phase-spatial-smallness} and the
sharp $Q$ graph bound. This gives
\begin{align*}
	\|S_2(T,U)T'v_n\|
	\le C_I\{&
	\epsilon^2\nu^{5/2}\|v_n\|
	+\epsilon\delta\nu^3\|Q_n^{1/2}v_n\|\\
	&+\epsilon^3\nu^{5/2}\|Q_n^{1/2}v_n\|
	+\epsilon^2\nu^{3/2}\|v_n\|\}.
\end{align*}
Using $\delta\nu/\epsilon\le C$ and
$\epsilon\sqrt\nu\le C$ proves
\eqref{eq:potential-second-source-word}.
The uniform weighted $Q$ graph bound for
one further cubic term preserves its right side.

We specify the derivative checks. Ordinary
coefficient derivatives preserve the row,
column and amplitude-envelope bounds in the
proof of Proposition~\ref{prop:full-graph}.
Hence the linear, dressed and density parts
retain respectively their $1,\delta,\rho$
estimates. A connection on a remaining field
coordinate is a smooth finite-rank insertion:
use its $L^2$ bound for the ordinary source
maps and its $H^1$ bound for the density
energy map. On a removed source coordinate,
use instead the $C_I\tau$ bound for
$\sqrt{W_N}(R\otimes I)$ or
$\sqrt{W_N}(I\otimes R)$, applied to the
unweighted fixed-sector commutator of size
$C_I\epsilon$. These contributions have
size $C_I\delta$ and join $L,R$.
The derivative of the pair source projection
has the same $C_I\tau$ bound, since its
rank-one vectors remain uniformly bounded
in $L^2\cap L^\infty$.

In parallel coordinates the additional
$E=\nabla_tD$ is a bounded quadratic source,
of fixed-sector norm $C_I\tau$; it is the pair map with a connection acting on one removed coordinate. The same weighted rank-one
estimate proves this statement. Its $j$-fold
canonical commutator has degree at most $j+2$
and fixed-sector norm $C_{j,I}\tau\epsilon^j$.
Thus its general block is at most
$C_{j,I}\tau\epsilon^j\nu^{(j+2)/2}$.
At $j=0,2,4$, or after the additional right
cubic terms just considered, $Q\ge M$ and the
scaling bounds on the cutoff range bound it by the corresponding
displayed estimate. This includes derivatives
of both the energy source and $T$.
Number-gauge derivatives again give fixed particle-number shifts.
\end{proof}

For an anti-self-adjoint canonical $C_{\rm fr}$,
put $S_j=\delta_{C_{\rm fr}}^jD$. The preceding
results in particular give
\begin{align}
	\|S_4v_n\|&\le C_I\epsilon^2\nu\|Q_n^{1/2}v_n\|,
	\label{eq:potential-fourth-source}\\
	|(D^*S_4+S_4^*D)[E_Nu,E_Nv]|
	+|\langle S_2E_Nu,S_2E_Nv\rangle|
	&\le C_I\epsilon^2
	\|MQ^{1/2}E_Nu\|\|Q^{1/2}E_Nv\|.
	\label{eq:potential-endpoints-square}
\end{align}
Indeed, at $j=4$ absorb
$\epsilon\nu^{3/2}\le C$ in
\eqref{eq:potential-higher-source}.
Pair with $D$ and apply
\eqref{eq:potential-second-source} to the
second-source square. Finite bandwidth transfers
the weight between the two arguments.
These estimates include one covariant and
fixed number-gauge derivatives by the proposition.
The products of first and third sources in
\eqref{eq:fourth-source-product} still require
their own estimates, as do the actual number
coefficients. We estimate these products below.

\subsection{The fourth potential commutator}
\label{sec:potential-fourth}

We restore the number-dependent coefficients and estimate the
fourth commutator of the potential. All norms and forms are
restricted to the original particle-number cutoff range, with a fixed enlargement
to accommodate at most four cubic factors. The constants are
uniform in the input particle number on this range.

\subsubsection{Errors from number-dependent coefficients}

\begin{lemma}\label{lem:potential-coefficient-error}
Write the actual phase, with real number coefficients,
as
\[
	C=\sum_a\{T_ab_a(\cN)-T_a^*b_a(\cN-d_a)\}, \qquad d_a\in\{1,3\}.
\]
For a target input number $n$, let
$C_{{\rm fr},n}=\sum_a b_a(n)(T_a-T_a^*)$.
Define $\mathfrak F_V$ by taking the $n$th input
column of $\mathcal D_{C_{{\rm fr},n}}^4V_P$
on the particle-number cutoff range. Then
\begin{equation}\label{eq:potential-coefficient-error}
	\mathcal D_C^4V_P=\mathfrak F_V+\mathcal E_V,\qquad |\mathcal E_V[u,v]| \le C_I\epsilon^2\|MQ^{1/2}E_Nu\|\|Q^{1/2}E_Nv\|.
\end{equation}
The same bound holds for $\nabla_t\mathcal E_V$
and $i[M,\mathcal E_V]$. Only boundedness and
the $O(\tau)$ first differences of the original
$b_a$ are used.
\end{lemma}
\begin{proof}
Set $B_1(T)=[V_P,T]$ and
$B_2(T,U)=[[V_P,T],U]$, and use $S_j$ for
the pair sources. The source product rule gives,
with the adjoint cubic terms kept explicitly,
\begin{align}
	B_1(T)&=D^*S_1(T)-S_1(T^*)^*D,\notag\\
	B_2(T,U)&=D^*S_2(T,U)+S_2(T^*,U^*)^*D\notag\\
	&\quad-S_1(U^*)^*S_1(T)-S_1(T^*)^*S_1(U).
	\label{eq:potential-B2-identity}
\end{align}
For a raising cubic term, the lowering source
$S_1(T^*)$ has no linear contact term.
Indeed its annihilator commutator is a pure
pair annihilator; for the identity family
it is $a(f_x)a_x$, and for the dressed family
its pair kernel has row norm $C_I\epsilon$.
The fibre bound and, respectively, integration
of $W_N$ give
$C_I\epsilon\sqrt\nu Q_n^{1/2}$ after lifting.
The projection correction has size
$C_I\delta\nu^{3/2}$ by the pair projection
identity and is absorbed in the same estimate.
The purely annihilating cubic term has zero source.
It follows that
\begin{equation}\label{eq:potential-oriented-first}
	|B_1(T)[u_m,v_n]| \le C_I\{\epsilon\sqrt\nu\|Q_m^{1/2}u_m\| \|Q_n^{1/2}v_n\| +\sqrt\nu\|Q_m^{1/2}u_m\|\|v_n\|\}.
\end{equation}
For $T^*$ the unweighted argument is on the left.

We record the word estimates used in the rearrangement.
For any choices of directions, the sharp graph bound,
\eqref{eq:phase-spatial-smallness} and
\eqref{eq:two-phase-factors} give
\begin{align}
	\|S_1(T)Uv_n\|&\le C_I\epsilon\sqrt\nu
	\|Q_n^{1/2}v_n\|,\notag\\
	\|Q^{1/2}U_1U_2v_n\|
	&\le C_I\{\epsilon^2\nu^3\|Q_n^{1/2}v_n\|
	+\epsilon\nu^2\|v_n\|
	+\delta\nu^{3/2}\|Q_n^{1/2}v_n\|\},\notag\\
	\|S_1(T)U_1U_2v_n\|
	&\le C_I(\epsilon^3\nu^{7/2}
	+\epsilon^2\nu^2+\delta\nu^{3/2})
	\|Q_n^{1/2}v_n\|.
	\label{eq:potential-source-words}
\end{align}
For the first line, substitution in
$\|S_1(T)w\|\le C_I(\epsilon\sqrt\nu
\|Q^{1/2}w\|+\sqrt\nu\|w\|)$ leaves
$\epsilon^2\nu^2Q^{1/2}$,
$\epsilon\nu\|v_n\|$ and
$\delta\nu^{3/2}Q^{1/2}$; each has the claimed
bound on the particle-number cutoff range.
To get the second line, apply the sharp graph
bound to the left cubic term, use the graph
bound on the first resulting term and the
spatial $L^2$ bound on the second.
The last line follows by inserting the second
line and the two-factor $L^2$ bound into the
first-source estimate. This proves all three estimates.

Let $h$ be any right coefficient with $|h|\le C\tau$.
For $l+r=2$, every form
$W_lB_2(T,U)W_rh(\cN)$ satisfies
the right side of \eqref{eq:potential-coefficient-error}.
For the source products in
\eqref{eq:potential-B2-identity}, use the first
line of \eqref{eq:potential-source-words}
when one cubic term is on each side.
When both are on one side, use its last line
and $\|S_1v_n\|\le C_I\|Q_n^{1/2}v_n\|$.
The relevant scaling inequalities are
\[
	\tau\epsilon\nu^{5/2}\le C,\qquad \tau\nu\le C,\qquad \tau\delta\sqrt\nu\le C\epsilon^2 .
\]
For $D^*S_2$, if a cubic term is to the right
use \eqref{eq:potential-second-source-word};
if neither is to the right, combine the
second line of \eqref{eq:potential-source-words}
on the left with
\eqref{eq:potential-second-source} on the right.
The same scaling bounds and $\tau\nu\le C$ apply.
The adjoint term is handled with the sides reversed.

For a product of three cubic terms $W_3$, the graph
iteration and the spatial estimate give
\[
	\|Q^{1/2}W_3v_n\|\le C_I\epsilon^2\nu^3 \|Q_n^{1/2}v_n\|,\qquad \|W_3v_n\|\le C_I\delta\nu\|Q_n^{1/2}v_n\|.
\]
Equation \eqref{eq:potential-oriented-first} therefore gives the desired
form bound for the ordering $B_1(T)W_3h(\cN)$. For
$T^*$, the order is reversed. The required scaling
inequality is
$\tau(\epsilon^3\nu^{7/2}
+\delta\nu^{3/2})\le C\epsilon^2\nu$.
For the other positions, use $UB_1(T)=B_1(T)U-B_2(T,U)$ to move the first
commutator to the corresponding end of the product.
Every error retains its other two cubic terms
and obeys the preceding bound, with $h$ at the far right.

Since $V_P$ commutes with number functions,
the first commutator with the actual phase
starts with $B_1(T)$ times its coefficient.
Expand the remaining three commutators and
put all number coefficients on the right
using the exact particle-number shifts.
Each coefficient is a product of four $b_a$
at arguments differing from $n$ by fixed integers.
Its difference from the product of $b_a(n)$
is $O(\tau)$, by telescoping first differences.
This includes the original negative-adjoint
shift $b_a(n-d_a)$.
The common products recombine exactly to
$\mathcal D_{C_{{\rm fr},n}}^4V_P$.
Every difference multiplies one of the source
words already bounded. This proves
\eqref{eq:potential-coefficient-error}.

We also estimate the connection terms obtained by differentiating this
ordering.
Let $E=\nabla_tD$ and $R_\pm=\delta_{T^{(*)}}E$.
For a rank-one connection $Z=|u\rangle\langle v|$,
\[
	E=\sqrt{W_N/2}\{(Za)_xa_y+a_x(Za)_y\},\qquad \|Ev_n\|\le C_I\tau\nu\|v_n\|.
\]
The raising product bound gives
$\|R_+v_n\|\le C_I\tau\epsilon\nu^{5/2}\|v_n\|$.
For the lowering direction write
$F_x=[a_x,T^*]$ and $F_v=\int\bar v(x)F_x\d x$.
The four terms in $R_-$ are
$\sqrt{W_N/2}$ times
\[
	u(x)F_va_y,\quad u(x)a(v)F_y,\quad F_xu(y)a(v),\quad a_xu(y)F_v .
\]
The two-particle map for $F$ has norm
$C_I\epsilon$; integrating $W_N|u|^2$
then gives
$\|R_-v_n\|\le C_I\tau\epsilon\nu^{3/2}\|v_n\|$.
The pure annihilation case is zero.
These calculations also cover smooth insertions on the active field
operators of the cubic terms:
on a removed coordinate the weighted rank-one map has norm of order
$\tau$; on a remaining coordinate it composes with the
bounded two-particle map.
Now
\[
	[\nabla_tV_P,T] =E^*S_1(T)+D^*R_+-R_-^*D-S_1(T^*)^*E .
\]
With $W_3h$ at the corresponding end of the product, the
form coefficient is bounded by
\[
	C_I\tau^2\{ \epsilon^3\nu^4+\delta\nu^2+ \epsilon\delta\nu^{7/2} +\epsilon\delta\nu^{5/2}\} \le C_I\epsilon^2\nu .
\]
The differentiated $B_2$ estimates follow
from the displayed source identity, the
derivative bounds in
Proposition~\ref{prop:potential-higher-sources},
and $\|Ev_n\|\le C_I\|Q_n^{1/2}v_n\|$.
All other derivatives replace one already
estimated cubic term. The right number
coefficients are time independent.
This proves the covariant assertion; number
derivatives only insert fixed particle-number shifts.
\end{proof}

\subsubsection{The density pairing and its five-field extension}

Let $\mathcal N_d$ denote extraction of the
normal-ordered component of degree $d$.
Split the linear source as $L=L_c+L_s$,
where $L_s$ is the source-projection correction
of fixed-sector size $C_I\delta$.
The original double contractions, with projections on the remaining field
coordinates, form $L_c$.

\begin{lemma}\label{lem:potential-density-pairing}
For the density source $F$ of
Lemma~\ref{lem:potential-source-decomposition},
the fixed-degree kernels obey
\begin{align}
	|\mathcal N_4(L_c^*F)[u,v]|
	&\le C_I\epsilon^2
	\|A_P^{1/2}u\|\|A_P^{1/2}v\|,\notag\\
	|\mathcal N_5(L_c^*A_1F)[u,v]|
	&\le C_I\epsilon^3
	\|A_P^{1/2}u\|\|A_P^{1/2}v\|.
	\label{eq:potential-density-pairing}
\end{align}
These estimates allow Hilbert-valued spectator
variables, the adjoint terms, and one covariant
coefficient replacement.
\end{lemma}
\begin{proof}
It is enough to consider one unprojected density
term
\[
	\sqrt{W_N(x-y)}\phi(x) \int\kappa(z-x)a_z^*a_za_y\d z .
\]
Write one linear source as
$\sqrt{W_N(x-y)}\int e_{xy}(t)a_t^\sharp\d t$.
Set
\[
	J_e(t,y)=\int W_N(x-y)|\phi(x)e_{xy}(t)|\d x .
\]
For an annihilation source we shall use
$\|J_e\|_{\mathrm{op}}\le C_I\tau$; for a
creation source use
$\|J_e\|_{\mathrm{HS}}\le C_I\tau$.
In the normal four-field pairing, the density
coordinate $z$ occurs on both sides. Uniformly
in $x$, H\"older--Sobolev gives
\[
	\int|\kappa(z-x)f(z)g(z)|\d z \le C\rho\|\nabla f\|_2\|\nabla g\|_2 .
\]
For an annihilation source the remaining
pairing is
\[
	C\rho\int J_e(t,y)U(t)V(y)\d t\d y ,
\]
where $U,V$ are the $z$-gradient norms of
the two wavefunctions. For a creation source
it is
\[
	C\rho\|\nabla u\|_2\int J_e(t,y)V(t,y)\d t\d y .
\]
More explicitly, for the annihilation source set
$U(t)=\|\nabla_z u(t,\cdot)\|_2$ and
$V(y)=\|\nabla_z v(\cdot,y)\|_2$.
For the creation source set
$V(t,y)=\|\nabla_z v(t,\cdot,y)\|_2$.
The normal-ordered kernels then satisfy
\begin{align}
	|\mathcal N_4(L_c^*F)[u,v]|
	&\le C\rho\|J_e\|_{\mathrm{op}}\|U\|_2\|V\|_2
	&&\text{for an annihilation source},\notag\\
	|\mathcal N_4(L_c^*F)[u,v]|
	&\le C\rho\|J_e\|_{\mathrm{HS}}\|\nabla u\|_2\|V\|_2
	&&\text{for a creation source}.
	\label{eq:density-pairing-marginals}
\end{align}
Fixed-sector normalization factors are absorbed into $C$.
The two required kernel norms therefore give
$C_I\rho\tau=C_I\epsilon^2$.
Hilbert-valued variables are integrated in the marginal norms.

For the double contractions, put $k_0=W_N|\kappa|$. Then
$\|k_0\|_1\le C\tau$,
$\|W_N\|_1\le C\rho$, and
$\|\kappa\|_\infty\le C\sqrt N$.
For the identity source,
\[
	e_{xy}(t)=\kappa(x-y) \{\phi(x)\delta(t-y)+\phi(y)\delta(t-x)\}.
\]
With $\Phi=|\phi|$, its positive kernel is the sum
\[
	J_e(t,y)=m(y)\delta(t-y)+k_0(t-y)\Phi(t)\Phi(y),\qquad m(y)=\int k_0(x-y)\Phi(x)^2\d x .
\]
Thus $\|m\|_\infty\le\|\Phi\|_\infty^2\|k_0\|_1$.
Young's inequality gives the same bound for the second
operator, and hence $\|J_e\|_{\mathrm{op}}\le C_I\tau$.
The first two dressed terms are
$\kappa(x-y)\phi(y)h(x,t)$ and
$\kappa(x-y)\phi(x)h(y,t)$.
Their positive kernels are
\[
	\Phi(y)\int k_0(x-y)\Phi(x)|h(x,t)|\d x,\qquad m(y)|h(y,t)|.
\]
Apply the $L^2$ convolution bound in $y$ to the first,
then integrate in $t$:
\[
	\int\left\|\int k_0(x-\cdot)\Phi(x)|h(x,t)|\d x\right\|_2^2\d t\le\|k_0\|_1^2\int\|\Phi h(\cdot,t)\|_2^2\d t.
\]
The multiplier bound applies to
the second. Both Hilbert--Schmidt norms are at most
$C_I\|k_0\|_1\|h\|_{\mathrm{HS}}\le C_I\tau$.
For the four additional creation terms the
needed bounds are
\[
	\begin{array}{c|c}
	e_{xy}(t)&\|J_e\|_{\mathrm{HS}}\\ \hline
	\kappa(y-t)\phi(t)h(y,x)&C_I\epsilon\\
	\kappa(x-t)\phi(t)h(x,y)&C_I\epsilon\\
	\kappa(t-y)\phi(y)h(t,x)&C_I\tau\\
	\kappa(t-x)\phi(x)h(t,y)&C_I\tau
	\end{array}
\]
Write these positive kernels as $J_1,\ldots,J_4$
in the order of the table, and put $K=|\kappa|$.
For the first, define
\[
	b(y)=\int W_N(x-y)\Phi(x)|h(y,x)|\d x .
\]
The envelope \eqref{eq:profile-envelope} gives
\[
	\|b\|_2\le C_I N\|W_N\|_1\sum_\ell\|u_\ell\|_2\|v_\ell\|_\infty\le C_I.
\]
Since $J_1(t,y)=K(y-t)\Phi(t)b(y)$, Tonelli's theorem yields
$\|J_1\|_{\mathrm{HS}}\le\|\Phi\|_\infty\|K\|_2\|b\|_2
\le C_I\epsilon$.
For the second, set $r=x-y$. Minkowski's inequality gives
\begin{align*}
	\|J_2\|_{\mathrm{HS}}
	&\le C_I N\sum_\ell\int W_N(r)
	\|\Phi(t)\Phi(y+r)u_\ell(y+r)v_\ell(y)K(y+r-t)\|_{L^2_{t,y}}\d r\\
	&\le C_I N\|W_N\|_1\|K\|_2
	\sum_\ell\|u_\ell\|_\infty\|v_\ell\|_2
	\le C_I\epsilon .
\end{align*}
For the last two kernels use, respectively,
\begin{align*}
	J_3(t,y)&=K(t-y)\Phi(y)\int W_N(x-y)\Phi(x)|h(t,x)|\d x,\\
	J_4(t,y)&=|h(t,y)|\int W_N(x-y)\Phi(x)^2K(t-x)\d x .
\end{align*}
The $L^2$ convolution bound for $J_3$ and the multiplier
bound for $J_4$ both give
\[
	\|J_3\|_{\mathrm{HS}}+\|J_4\|_{\mathrm{HS}}\le C_I\|K\|_\infty\|W_N\|_1\|h\|_{\mathrm{HS}}\le C_I\tau.
\]
Here $\|W_N\|_1=N^{-1}\|v\|_1$ and
$\|K\|_\infty\le C\sqrt N$. Together with
$\epsilon\le\tau$, these estimates cover all six
dressed double contractions.

For the five-field assertion there is one
additional free coordinate $w$ from the
fibre field $A_1$. Its coefficient
$a_{xy}(w)$ satisfies
$\sup_{x,y}\|a_{xy}\|_2\le C_I\epsilon$.
In the normal five-field part it has no
contraction with either adjacent source.
For a wavefunction $f(z,w,\Xi)$ on that side, put
$f_{\mathrm{marg}}(z,\Xi)=\|f(z,\cdot,\Xi)\|_{L^2_w}$.
Cauchy--Schwarz and the Sobolev chain rule give
\[
	\left|\int a_{xy}(w)f(z,w,\Xi)\d w\right|\le C_I\epsilon f_{\mathrm{marg}}(z,\Xi),\qquad |\nabla_z f_{\mathrm{marg}}(z,\Xi)|\le\|\nabla_zf(z,\cdot,\Xi)\|_{L^2_w}.
\]
The second inequality follows first for smooth functions
from differentiation of $(\|f\|_{L^2_w}^2+\eta^2)^{1/2}$,
and then by $H^1$ approximation and $\eta\downarrow0$.
The density coordinate $z$ remains on both sides.
Using \eqref{eq:density-pairing-marginals} with the marginal
therefore gives the factor $\epsilon\rho\tau=\epsilon^3$.
This argument covers either creation or annihilation of
the additional field.

Remaining-field projections or smooth
connection insertions move to the external
fixed-particle wavefunctions and preserve
their $H^1$ norms. Terms on a removed source
coordinate have already been isolated in
$L_s,R$ by the weighted rank-one estimate.
Ordinary time replacements preserve the
kernel envelope and fibre bounds just used.
Taking Hilbert-valued marginal norms proves
the spectator assertion as well.
The estimates extend by density from smooth fixed-particle
vectors to the corresponding kinetic form spaces.
\end{proof}

\subsubsection{Closing the six-field part}

\begin{theorem}
\label{thm:actual-potential-fourth}
For the original cubic phase,
\begin{equation}\label{eq:actual-potential-fourth}
	|T[u,v]|\le C_I\epsilon^2 \|MQ^{1/2}E_Nu\|\|Q^{1/2}E_Nv\|,\qquad T\in\{\mathcal D_C^4V_P,\, \nabla_t\mathcal D_C^4V_P,\, i[M,\mathcal D_C^4V_P]\}.
\end{equation}
The corresponding exact integral remainder
\[
	R_V=\frac1{3!}\int_0^1(1-a)^3 e^{-aC}\mathcal D_C^4V_Pe^{aC}\d a
\]
and its same derivatives have this form bound.
Its contribution to the exact-versus-Taylor
Duhamel integral on $D(F_3(s)^{3/2})$
is at most $C_{I,s}N^{-5/4}\|F_3(s)^{3/2}z\|$.
This is the bound for the potential contribution.
\end{theorem}
\begin{proof}
First fix a matching anti-self-adjoint
$C_{\rm fr}$ and suppress its subscript.
The source Leibniz rule also writes
\eqref{eq:fourth-source-product} as
\[
	\mathcal D_C^4(D^*D) =D^*S_4+S_4^*D +4\mathcal D_C^2(S_1^*S_1)-2S_2^*S_2 .
\]
The endpoint and square terms are bounded by
\eqref{eq:potential-endpoints-square}.
Use $S_1=L+B_3$.
Set $S=B_3^*B_3$ as a source pairing.
Equation \eqref{eq:potential-cubic-source} and its finite
particle-number bandwidth give
\[
	|S[u,v]|\le C_I\epsilon^2
	\|MQ^{1/2}u\|\|Q^{1/2}v\| .
\]
The three forms $C^jSC^k$, $j+k=2$, satisfy the same
bound: move $C^j$ to the first argument and use the
uniform graph bounds of $C$ in the two displayed norms.
This directly bounds $SC^2-2CSC+C^2S$.
All blocks here lie in the original active range enlarged
by the fixed shifts in these words.
The polynomial $L^*L$ has degree at most two
and fixed-sector norm $C_I$. To extract kernels of degree
at most six below, the input sector is at most six.
Every ordered product used for $L^*L$ and $L^*B_3$
then stays in sectors at most $24$, since each cubic
factor shifts number by at most three and $L$ by one.
This fixed range also contains the lower blocks used
in the triangular recursion. Coefficients frozen at the
target number remain bounded independently of that number.
For sufficiently large $N$ this range lies within the phase
cutoff; the finitely many other $N$ can be absorbed in the
constant. On these fixed sectors, $C$ has ordinary norm
$C_I\epsilon$. Thus each term of
\[
	\mathcal D_C^2(L^*L)
	=L^*LC^2-2CL^*LC+C^2L^*L
\]
has ordinary norm $C_I\epsilon^2$.
The zero-contraction terms cancel at each commutator,
so its degree is at most four. Kernel extraction and
\eqref{eq:fixed-degree-weighted-form} give the target
bound, including the scalar and lower-degree terms.

Only $\mathcal D_C^2(L^*B_3+B_3^*L)$
requires further work. We first check its fixed-sector
form norm. The canonical block estimate in the proof of
Proposition~\ref{prop:full-graph} gives
\[
	\|Q^{1/2}Cz\|\le C_I\{\epsilon\|Q^{1/2}z\|+\|z\|\}
\]
on these fixed sectors. Applying this estimate to $Cz$
and using $\|Cz\|\le C_I\epsilon\|z\|$ gives
$\|Q^{1/2}C^2z\|\le C_I\epsilon\|Q^{1/2}z\|$.
Together with \eqref{eq:potential-cubic-source}, this yields
\[
	\begin{array}{c|c|c|c}
	(j,k)&\|LC^j\|&\|B_3C^kQ^{-1/2}\|
	&\text{form bound for }C^jL^*B_3C^k\\ \hline
	(0,2)&C_I&C_I\epsilon^2&C_I\epsilon^2\\
	(1,1)&C_I\epsilon&C_I\epsilon&C_I\epsilon^2\\
	(2,0)&C_I\epsilon^2&C_I\epsilon&C_I\epsilon^3
	\end{array}
\]
Here $\sqrt\nu\le5$ in the source bound, and each
inverse square root belongs to its input sector.
It maps that sector into its $Q$ form domain; the graph
estimate makes every subsequent $C$ a bounded map between
the corresponding form domains. On the form core,
\[
	\langle u,C^jL^*B_3C^kv\rangle
	=(-1)^j\langle LC^ju,B_3C^kv\rangle .
\]
The two norms in the table bound this source pairing by
the last entry times $\|Q^{1/2}u\|\|Q^{1/2}v\|$,
using $\|u\|\le\|Q^{1/2}u\|$. Density gives its unique
continuous extension. Interchanging $u,v$ treats the
adjoint words. Their sum has
form norm $C_I\epsilon^2$ and degree at most six.
Triangular extraction therefore controls
its degree-zero, two and four parts by
the target bound. We estimate the remaining
six-field part explicitly.

Both cubic terms are bounded on fixed sectors, so the terms
with $R$ have ordinary norm $C_I\delta\epsilon^2$ there.
The preceding word argument gives the same form bound for
the terms with $L_s$. Extracting and lifting the six-field
part then yields $C_I\delta\epsilon^2\nu^2$, which is at
most $C_I\epsilon^2\nu$.

For the density term, set $B_F=\mathcal N_4(L_c^*F+F^*L_c)$.
The first estimate in \eqref{eq:potential-density-pairing},
lifted to the particle-number cutoff range, gives the required bound for
$B_F$, and the weighted graph estimates preserve it under
the full second commutator. On fixed sectors, this commutator
has form norm $C_I\epsilon^2$ and degree at most six.
Subtracting its extracted part of degree at most four therefore
leaves a six-field part with the same target bound. The
lower-degree contacts in $L_c^*F$ do not contribute, since
two canonical commutators cannot raise them to six fields.

It remains to bound
\[
	\mathcal H_6=\mathcal N_6\mathcal D_C^2 (L_c^*A_1D+D^*A_1^*L_c).
\]
Set $B=A_1^*L_c$, a bounded quadratic
source map, possibly with a scalar contact.
The source maps $B,\delta_CB,\delta_C^2B$
have fixed-sector norms
$C_I\epsilon,C_I\epsilon^2,C_I\epsilon^3$
and degrees at most two, three, four.
The exact rectangular identity is
\[
	\mathcal D_C^2(B^*D) =(\delta_C^2B)^*D +2(\delta_CB)^*S_1+B^*S_2 .
\]
Use
$S_2=\delta_CL+\delta_CF+\delta_CR+
(\delta_CA_1)D+A_1S_1$.
All the following bounds are on fixed sectors.
The first term has form norm $C_I\epsilon^3$.
In the middle term, $(\delta_CB)^*L$
has degree at most four, whereas
$(\delta_CB)^*B_3$ has form norm
$C_I\epsilon^3$.
In the last term, $B^*\delta_CL$ and
$B^*A_1L$ have degree at most four.
The term $B^*\delta_CR$ has ordinary norm
$C_I\delta\epsilon^2\le C_I\epsilon^3$.
Both $B^*(\delta_CA_1)D$ and
$B^*A_1B_3$ have form norm
$C_I\epsilon^3$. The source-operator
$\delta_CA_1$ has degree at most two and
fixed-sector norm $C_I\epsilon^2$,
so these are legitimate bounded-source
compositions.

For the final term use another exact identity
$B^*\delta_CF =\mathcal D_C(B^*F)-(\delta_CB)^*F$.
The second term has form norm
$C_I\epsilon^2\rho\le C_I\epsilon^3$.
The polynomial $B^*F=L_c^*A_1F$ has
degree at most five. Its degree-three
and degree-one parts cannot produce degree
six after one commutator, so exactly
\[
	\mathcal N_6\mathcal D_C(B^*F) =\mathcal N_6\mathcal D_C \bigl(\mathcal N_5(L_c^*A_1F)\bigr).
\]
The second estimate in
\eqref{eq:potential-density-pairing}
gives fixed-sector form norm
$C_I\epsilon^3$ to the inner polynomial.
We give the form-domain argument for this commutator.
On sectors $n\le d$, Sobolev's inequality in one pair
coordinate and $\|W_N\|_{3/2}=\|v\|_{3/2}$ give
\begin{equation}\label{eq:fixed-sector-potential-comparison}
	\mathcal V_N[z_n]=\|Dz_n\|^2\le C_v(n-1)K[z_n],\qquad A_{P,n}\le Q_n\le C_dA_{P,n}.
\end{equation}
For $n=0,1$ the potential is zero. These constants are
independent of $N$. Let $\mathcal V_{\le d}$ be the
direct sum of these kinetic form spaces, with norm
$\|A_P^{1/2}z\|$. The cubic graph bound maps
$\mathcal V_{\le d}$ continuously to
$\mathcal V_{\le d+3}$, uniformly in $N$.
Write $S=\mathcal N_5(L_c^*A_1F)$.
The density estimate defines $S$ as a continuous
sesquilinear form on these spaces. Since $C^*=-C$,
its commutator is the continuous form
\begin{equation}\label{eq:density-form-commutator}
	[S,C][u,v]=S[u,Cv]+S[Cu,v],\qquad |[S,C][u,v]|\le C_{I,d}\epsilon^3\|A_P^{1/2}u\|\|A_P^{1/2}v\|.
\end{equation}
Each additional cubic factor enlarges the fixed sector
range by at most three.

At fixed $N$ and on a finite sector range, $W_N$ is
bounded and all source products above are bounded maps
between the indicated Hilbert spaces. Their rectangular
identities and normal ordering can therefore be checked
on smooth vectors and extended in $L^2$. The uniform
form estimates then extend to $\mathcal V_{\le d}$.
In particular, its degree-six identity is
\[
	\mathcal N_6(B^*\delta_CF)=\mathcal N_6[S,C]-\mathcal N_6((\delta_CB)^*F).
\]
The second term has the $C_I\epsilon^2\rho$ bound
given above. Lemma~\ref{lem:fixed-degree-extraction}
applies on its fixed finite range of sectors and preserves
$C_I\epsilon^3$ for the six-field kernels. The adjoint terms have
the same estimates. Consequently
\[
	|\mathcal H_6[u_p,v_q]| \le C_I\epsilon^3 \|Q_p^{1/2}u_p\|\|Q_q^{1/2}v_q\|, \qquad p+q=6 .
\]
On $E_N$ and its fixed enlargements,
$\nu\le C\sqrt N$ by Section~\ref{sec:spatial-phase}.
Thus $\epsilon\nu\le C N^{-1/4}$, and energy lifting gives
$C_I\epsilon^3\nu^2\le C_I\epsilon^2\nu$.
This completes the canonical
estimate while retaining every lower-field
contraction. Restore the entire coefficient
error by Lemma~\ref{lem:potential-coefficient-error}.

For one covariant derivative, all the source
and density-pairing estimates were proved
with the required replacements.
When $D$ in the last displayed rectangular
identities is differentiated, the additional
terms are second commutators of $B^*E$
and its adjoint. Their fixed-sector norm
is $C_I\tau\epsilon^3$ and degree at most
six, so extraction and lifting are smaller
than the bound just obtained.
Replacing one cubic term in the commutator by
$X$ is covered by polarization of the
same word estimates. Number derivatives
insert only the fixed total particle-number shifts.
This proves \eqref{eq:actual-potential-fourth}.

The weighted phase graph bounds, including
their first derivative, pass this estimate
to the exact integral remainder $R_V$.
It has the one-weight static bound required
in Theorem~\ref{thm:inverse-comparison},
with one fixed-degree summand.
Apply that theorem's inverse-energy
integration-by-parts estimate to this
summand, using the same $F_{\rm ex}^{-1}$
and the same $F_3^{3/2}$ propagation.
This proves the stated bound for its
Duhamel contribution.
\end{proof}

\subsection{The second kinetic source}
\label{sec:kinetic-second-source}

The square term in \eqref{eq:fourth-source-product} requires a
sharper estimate for the second kinetic source than the general
recursion in Section~\ref{sec:kinetic-source-endpoints} provides.
We derive it from the kernel maps below. Throughout, $A=A_P$,
$\nu=n+1\le C\sqrt N$, $\rho=N^{-1}$,
$\epsilon=N^{-3/4}$ and $\delta=N^{-5/4}$.
Write $T_0,T_p,T_\sigma$ for the projected
raising cubic terms, and $S_a^\pm=\delta_{T_a^{(*)}}D_K$.
The superscripts refer to the direction of the
original cubic term.

\subsubsection{Kernel maps and source lifting}

Besides \eqref{eq:spatial-kappa-norms}, we use
nonnegative radial majorants $k\ge|\kappa|$ and
$g\ge|\nabla\kappa|+|\kappa|$ with
\begin{align}
	\|k\|_2&\le C\epsilon,&
	\|k\|_1&\le C\epsilon^2,&
	\|k\|_{6/5}&\le C\delta,&
	\|k\|_{3/2}&\le C\rho,\notag\\
	\|k\|_6&\le C,&
	\|g\|_2&\le C,&
	\|g\|_1&\le C\rho,&
	\|g\|_{6/5}&\le C\epsilon,\notag\\
	\|kg\|_1&\le C\epsilon,&
	\|kg\|_{6/5}&\le C N^{-1/2}.
	\label{eq:kinetic-majorants}
\end{align}
These follow by integrating the same pointwise
bounds on the inner ball and outer annulus;
for $g$ the singular upper bound is
$CN^{-1/2}|x|^{-2}$, with its regular inner
bound $CN^{3/2}$.

Define the two-particle contraction
\[
	(F_0\psi)(x)=\int\kappa(x-y)\bar\phi(y)\psi(x,y)\d y, \qquad F_p=p^{\mathsf T}F_0 ,
\]
with complex conjugations fixed by the annihilator convention.
The estimates are invariant under these conjugations. Direct Cauchy--Schwarz
and H\"older--Sobolev give, for $a=0,p$,
\begin{align}
	\|F_a\|_{L^2\to L^2}&\le C_I\epsilon,&
	\|F_a\|_{H^1\to L^2}&\le C_I\delta,&
	\|F_a\|_{H^1\to H^1}&\le C_I\epsilon .
	\label{eq:kinetic-pair-maps}
\end{align}
For the derivative of $F_0$, the derivative
of $\kappa$ uses $\|g\|_{6/5}$ and an input
$y$ derivative; the derivative of the input
uses $\|\kappa\|_2$. For $F_p$, use also
the bounded integral operator $\nabla p^{\mathsf T}$.
These bounds extend with spectator variables
and with active $q$ projections.

We specify the auxiliary dressed-source bounds.
Before projection, write
$S_p^+(z)=\dGamma(L_z)$ and
$S_\sigma^+(z)=a^{*2}(H_z)$, where, up to
symmetrization of creation coordinates,
\begin{align*}
	L_z(x,y)&=\nabla_z[\kappa(z-x)\phi(x)p(z,y)]
	+\nabla_z[\kappa(x-z)\phi(z)p(x,y)],\\
	H_z(x,y)&=\nabla_z[\kappa(z-x)\phi(x)\sigma(z,y)]
	+\nabla_z[\kappa(x-z)\phi(z)\sigma(x,y)]\\
	&\quad+\kappa(x-y)\phi(y)\nabla_z\sigma(x,z).
\end{align*}
The following are distinct kernel maps:
\[
	(\mathcal A_Lf)(z,y)=\int\overline{L_z(x,y)}f(x)\d x, \qquad (\mathcal B_J\psi)(z)=\int J_z(x,y)\psi(x,y)\d x\d y .
\]
For $J=L,H$ the required bounds are
\begin{align}
	\|L\|_2+\|H\|_2+
	\sup_z(\|L_z\|_2+\|H_z\|_2)&\le C_I,\notag\\
	\|\mathcal A_L\|+\|\mathcal B_L\|&\le C_I\rho,&
	\|\mathcal B_H\|&\le C_I\epsilon,\notag\\
	\sup_x\bigl(\|L(\cdot;x,\cdot)\|_2+
	\|H(\cdot;x,\cdot)\|_2\bigr)&\le C_I,&
	\sup_y\|L(\cdot;\cdot,y)\|_2&\le C_I .
	\label{eq:kinetic-dressed-maps}
\end{align}
All operator norms on the second line are between
the indicated unweighted $L^2$ spaces.
The three-creation kernel $t$ of $T_\sigma$
has $\|t\|_2\le C_I\epsilon$, and each
one-coordinate slice has $L^2$ norm
$C_I\epsilon$. The pair kernel $J_z$ of $[a_z,T_\sigma]$
has the same global and fixed-$z$ bounds.

We verify these statements, including the slice estimates,
by first recording the kernel bounds. For $h=p,\sigma$,
the row and column $L^2$ norms are bounded, and their
first-derivative counterparts are at most $C_I\sqrt N$.
For the initial kernel, direct integration of
$|\nabla(Nw_\ell(Nx))|\le CN^2/(1+N^2|x|^2)$ gives this
estimate; projection adds only bounded finite-rank corrections.
In the hyperbolic series, a spatial derivative acts at an end
factor. Multiplying its derivative row by the remaining bounded
operators therefore gives the same estimate for $h$.
Young's inequality applied to $\nabla(Nw_\ell(N\cdot))$
also yields $\|\nabla_jh\|_{\mathrm{op}}\le C_I$.

The series for $p$ and $r=\sigma-\eta$ contain an additional
Hilbert--Schmidt factor, which improves their derivative bound
to $\|\nabla p\|_{\mathrm{HS}}
+\|\nabla r\|_{\mathrm{HS}}\le C_I$.
This agrees with \cite[Lemma A.3]{COS}, while
$\nabla\eta,\nabla\sigma$ retain Hilbert--Schmidt size
$C_I\sqrt N$.

Since $p$ has at least two kernel factors,
its row-column representation further gives
\[
	\|p\|_\infty\le C_I,\quad \|\nabla_jp\|_\infty\le C_I\sqrt N,\quad \sup_y\|\nabla_1p(\cdot,y)\|_2\le C_I .
\]
For the last bound use the derivative operator
on the first factor and the last factor's column
norm, before summing the series.
Now insert the displayed formulas for $L,H$.
A derivative on $\kappa$ uses $\|g\|_2$ for
global or slice norms; a derivative on $h$
uses $\epsilon\sqrt N\le1$.
For $\mathcal A_L,\mathcal B_L$, use the
Hilbert-valued convolution norm $\|g\|_1=\rho$,
or $\|k\|_1\sqrt N=\rho$.
The first two groups of $\mathcal B_H$ obey
the same bound. Its third group is $\nabla\sigma^{\mathsf T}$ applied to $F_0$,
with norm at most $C_I\epsilon$.
For the fixed last-coordinate slice of $L$
use the last displayed derivative column bound.
For $t$ and $J$, integrate $|\kappa|^2$ first
and then use the row or column bound of $\sigma$.
This proves \eqref{eq:kinetic-dressed-maps}
and all the asserted slice estimates.
The same proofs allow one ordinary time
derivative; the series gains only a linear
factor in its degree.

\begin{lemma}\label{lem:cubic-source-criteria}
Use the Wick convention of
Lemma~\ref{lem:fixed-degree-extraction}.
The following conditions each imply
\begin{equation}\label{eq:cubic-source-target}
	\|Rv_n\|_{\mathcal S\otimes\cF} \le C_I\epsilon\sqrt\nu\|A_n^{1/2}v_n\|.
\end{equation}
First, for $R(z)=a^{*3}(r_z)$, let
$\alpha=\|r\|_2$ and let $\beta$ be the norm
of its three-input contraction
$L^2_{\rm sym}(\mathbb R^9)\to L^2_z$.
It suffices that $\alpha\le C_I\epsilon$,
$\beta\le C_I\epsilon^2$.
Second, for
$R(z)=\int r_z(i,j;k)a_i^*a_j^*a_k$,
let $\alpha=\|r\|_2$ and let
$\mathcal A_r\psi(z,k)=\int\bar r_z(i,j;k)
\psi(i,j)\d i\d j$.
It suffices that $\alpha\le C_I\epsilon$
and either $\|\mathcal A_r\|_{L^2\to L^2}
\le C_I\epsilon^2$ or
$\|\mathcal A_r\|_{H^1\to L^2}\le C_I\delta$.
Third, for $R(z)=\int r_z(k;i,j)a_k^*a_ia_j$,
it suffices that its two-input map to
$L^2_{z,k}$ has $L^2$ norm $C_I\epsilon^2$
or $H^1$-to-$L^2$ norm $C_I\delta$.
Any cubic source with ordinary kernel norm
$C_I\epsilon^2$ also satisfies
\eqref{eq:cubic-source-target}.
\end{lemma}
\begin{proof}
For the first type, full normal ordering gives
\[
	\|Rv_n\|^2\le [\beta^2(n)_3+ \alpha^2\{9(n)_2+18n+6\}]\|v_n\|^2 .
\]
The one- and two-contraction kernels are
positive partial traces of $|r\rangle\langle r|$
with trace $\alpha^2$. For the zero-contraction
term use the three-input map and Cauchy--Schwarz.
For the second type the same calculation gives
coefficients $1,4,2$ and
\[
	\|Rv_n\|^2\le [\beta^2(n)_3+\alpha^2\{4(n)_2+2n\}]\|v_n\|^2 .
\]
The four-field kernel may be an index
rearrangement of a partial trace; its
Hilbert--Schmidt norm is still at most
$\alpha^2$.
If the map has only an $H^1$ bound $\beta_1$,
the six-field term is instead bounded by
$C\beta_1^2\nu^2\|A_n^{1/2}v_n\|^2$:
permutation symmetry allocates at most
$2/n$ of the kinetic energy to its two inputs.
The lower contractions retain their
$C\alpha^2\nu^2\|v_n\|^2$ bound.

For the third type the only terms in $R^*R$
have six and four fields. Equivalently lift
the two-input-to-one-output source kernel
directly. Its $H^1$ bound gives
$C\beta_1^2\nu^2\|A_n^{1/2}v_n\|^2$;
its $L^2$ bound gives
$\beta^2\{(n)_3+(n)_2\}\|v_n\|^2$.
This allows identity kernels represented by Dirac delta distributions and
a first derivative on the input of that map.
Use $A_n\ge\nu$,
$\delta^2\nu\le C\epsilon^2$ and
$\epsilon^2\nu\le C$ in these inequalities.
The final assertion follows by ordinary
degree-three lifting and the same scaling bounds.
\end{proof}

\subsubsection{Raising combinations}

\begin{lemma}\label{lem:kinetic-raising-second}
For $a,b\in\{0,p,\sigma\}$,
$[S_a^+,T_b]$ satisfies
\eqref{eq:cubic-source-target}, with the
actual spatial projections and one
ordinary or covariant coefficient replacement.
\end{lemma}
\begin{proof}
We first give all unprojected kernel estimates;
the internal projections are checked below.
For $a=0,p$, let $K_a(i,j;l)$ be the symmetric
two-creation one-annihilation kernel of $T_a$.
The pure creation combinations are
\begin{align*}
	[S_\sigma^+(z),T_a]&=
	-2a^{*3}\!\left(\operatorname{Sym}
	\int K_a(i,j;l)H_z(l,k)\d l\right),\\
	[S_p^+(z),T_\sigma]&=
	3a^{*3}\!\left(\operatorname{Sym}
	\int L_z(i,l)t(l,j,k)\d l\right).
\end{align*}
Their global kernel norms are $C_I\epsilon$.
The full contraction maps have norms
$C_I\epsilon^2$ and $C_I\rho\epsilon$,
respectively, by composing $F_a$ with
$\mathcal B_H$, or the two-input contraction
of $t$ with $\mathcal B_L$.
For $[S_0^+,T_\sigma]$ the four kernel groups
are, suppressing bounded amplitudes and
finite permutations,
\[
	g(z-x)J_z(y,w),\quad k(z-x)H_z(y,w),\quad \phi(z)g(x-z)t(x,y,w),\quad \nabla\phi(z)k(x-z)t(x,y,w).
\]
Their global norms are $C_I\epsilon$.
Their full contraction norms are respectively
$C_I\rho\epsilon$, $C_I\epsilon^2$, $C_I\rho\epsilon$ and $C_I\epsilon^3$.
Use the fixed-coordinate bounds from
\eqref{eq:kinetic-dressed-maps}, followed
by the indicated $g$ or $k$ convolution.
The first case of Lemma~\ref{lem:cubic-source-criteria}
applies. The combination
$[S_\sigma^+,T_\sigma]$ is zero.

For $[S_p^+,T_0]$, contraction with the creation or annihilation
operator at the retained coordinate gives
\[
	L_z(i,k)\kappa(k-j)\phi(j),\qquad -\kappa(i-j)\phi(j)L_z(i,k),
\]
symmetrized in $i,j$.
Both have global norm $C_I\epsilon$.
The second contraction map is $\mathcal A_LF_0$
and has norm $C_I\rho\epsilon$.
For the first, insert the two explicit
groups of $L$. The $g$ terms use the pointwise
bound of $p$ and two Schur convolution
bounds, giving $C_I\rho\epsilon^2$.
The derivative-$p$ terms give
$C_I\sqrt N\,\|k\|_1^2=C_I\rho\epsilon^2$.
The other creation operator gives the density source
\[
	\int F_z(i,j)a_i^*a_j^*a_j\d i\d j,\qquad F_z(i,j)=\int L_z(i,l)\phi(l)\kappa(j-l)\d l .
\]
The fixed-$l$ column bound of $L$ and
$\|k\|_1$ give
$\sup_j\|F(\cdot;\cdot,j)\|_2\le C_I\epsilon^2$.
Its direct sector bound is
$C_I\epsilon^2\nu^{3/2}\|v_n\|$.
For $[S_p^+,T_p]$, the first two kernels
compose with a bounded $p$ operator.
The density kernel becomes $F_z(i,j)p(j,k)$;
its global norm is $C_I\epsilon^2$, by
the row bound of $p$ and Young's inequality.
These are the second and small-kernel
cases of Lemma~\ref{lem:cubic-source-criteria}.

For $[S_0^+,T_p]$, contractions with the two creation operators
give the groups
$g\otimes J_p$, $k\otimes L_p$, and the
density derivative acting on a creation
coordinate of $K_p$. Their global norms
are $C_I\epsilon$ and their
$\mathcal A_r$ norms are, respectively,
$C_I\rho\epsilon,C_I\epsilon^2,
C_I\rho\epsilon$ by the same slice
and convolution estimates.
The contractions with the annihilation operator contain
$a^{*2}(v_z)a_z$ and
$a^{*2}(w_z)\nabla_za_z$, where
\[
	v_z(i,j)=\int K_p(i,j;l)\nabla\kappa(z-l)\phi(l)\d l, \quad w_z(i,j)=\int K_p(i,j;l)\kappa(z-l)\phi(l)\d l .
\]
Since each fixed-$l$ slice of $K_p$ has norm
$C_I\epsilon$, their uniform slice norms
are $C_I\rho\epsilon$ and $C_I\epsilon^3$.
Their source bounds are
$C_I\rho\epsilon\nu^{3/2}\|v_n\|$ and
$C_I\epsilon^3\nu\|K_P^{1/2}v_n\|$,
both sufficient.
The remaining density kernel is
$\phi(z)g(k-z)K_p(i,j;k)$.
Its global norm is $C_I\epsilon$, while
its $\mathcal A_r$ map has $H^1$ norm
$C_I\delta$: first apply $F_p$ using
\eqref{eq:kinetic-pair-maps}, then use
$\|g\|_2\|\phi\|_\infty$.

We give the identity--identity calculation
separately. With $f_z(x)=\kappa(z-x)\phi(x)$,
direct CCR gives
\begin{align}
	[[a_z,T_0],T_0]
	={}&a^{*2}(w_z)a_z\notag\\
	&+\phi(z)\int\kappa(z-y)\phi(y)\kappa(z-x)
	a_x^*a_y^*a_x\d x\d y\notag\\
	&+\phi(z)\int\kappa(x-y)\phi(y)\kappa(y-z)
	a_x^*a_y^*a_x\d x\d y ,
	\label{eq:identity-second-CCR}
\end{align}
where
\[
	w_z(x,y)=\operatorname{Sym}_{xy} \{\kappa(z-x)\phi(x)\kappa(z-y)\phi(y) -\kappa(x-y)\phi(y)\kappa(z-x)\phi(x)\}.
\]
In the density commutator, the creation and annihilation contributions at
the retained coordinate cancel. Only the multiplier at the newly created
coordinate remains.
Differentiate \eqref{eq:identity-second-CCR}
only once in $z$.
For every differentiated pair kernel $W_z$
in its first term,
\begin{align*}
	\sup_z\|W_z\|_2&\le C_I\epsilon,\\
	|\langle W_z,\psi\rangle|&\le C_I\epsilon^2\|\psi\|_{H^1},\\
	\left\|x\mapsto\int W_z(x,y)h(y)\d y\right\|_2
	&\le C_I\delta\|h\|_{H^1}.
\end{align*}
For a star kernel $g(z-x)k(z-y)$ these
are H\"older and Sobolev using
$g\in L^{6/5}\cap L^2$,
$k\in L^2\cap L^{6/5}$.
For a path kernel $g(z-x)k(x-y)$ the
two-input bound first applies $F_0:H^1\to H^1$
and then the $g$ Sobolev duality.
Its one-input bound uses
$\|k*h\|_\infty\le C\delta\|h\|_{H^1}$
in one direction and Young's inequality
with $gh\in L^{3/2}$ in the other.
The bounded amplitudes are $H^1$ multipliers.
Full pair normal ordering, with coefficients
$1,4,2$, now gives
\[
	\|a^{*2}(W_z)a_zv_n\|_{L^2_z}^2 \le C_I\{\epsilon^4\nu^2\|A_n^{1/2}v_n\|^2 +\delta^2\nu\|A_n^{1/2}v_n\|^2 +\epsilon^2\nu\|v_n\|^2\}.
\]
The undifferentiated $w_z$ has slice norm
$C_I\epsilon^2$, so
$a^{*2}(w_z)\nabla_za_z$ has weighted bound $C_I\epsilon^2\nu K_P^{1/2}$.

The two density terms after differentiation
are bounded by sources with kernels
\[
	F_A=g(z-y)k(z-x),\quad F_B=k(z-y)g(z-x),\quad F_C=k(x-y)g(y-z)
\]
in $\int F(z;x,y)a_x^*a_y^*a_x\d x\d y$.
For each fixed retained coordinate $x$,
\[
	\left\|\int F(\cdot;x,y)h(y)\d y\right\|_2 \le C\delta\|h\|_{H^1}.
\]
For $F_A,F_B$, use the $g$ or $k$ dual
Sobolev bound, then the other kernel's
$L^2$ norm; for $F_C$ use $kh\in L^1$
and $g\in L^2$. The squared integrals are at most $C\epsilon^2$ when either the
retained coordinate or the newly created coordinate is fixed.
Order only the two outer fields in the
square of this density source.
The uncontracted term and the diagonal contraction at the retained
coordinate are bounded by $C\delta^2\nu^2 A_n$ and $C\delta^2\nu A_n$,
respectively.
The other diagonal contraction is bounded by $C\epsilon^2(n)_2\|v_n\|^2$.
For the two crossed contractions, weighted
Cauchy--Schwarz gives
\[
	P_1(u,x)=\int|F(z;u,x)|\int|F(z;x,y)|\d y\d z,\qquad P_2(u,y)=\int|F(z;x,y)F(z;u,x)|\d z\d x .
\]
For $F_A$, both are bounded by $C\epsilon^2$:
use $P_1\le C\rho\,k*(kg)$,
$\|k\|_6\|kg\|_{6/5}\le CN^{-1/2}$,
and for $P_2$ use $\|kg\|_1\|k*g\|_\infty$.
For $F_B$, $\sup_u\|P_1(u,\cdot)\|_{3/2}
\le C\epsilon^2\delta$ follows from
$P_1\le C\epsilon^2(kg)*g$; $P_2$ has
the same $C\epsilon^2$ bound.
For $F_C$, $P_1\le C\epsilon^2k(u-x)$,
while $P_2\le C(k(g*g))*k(u-y)$ is
bounded by $C\epsilon^2$.
Use Sobolev for the $L^{3/2}$ kernels,
and $A_n\ge\nu$ for the bounded kernels.
The two double contractions have
$C\epsilon^2$ multiplication or Schur
bounds. Thus the full square is at most
$C_I\epsilon^2\nu\|A_n^{1/2}v_n\|^2$.
Background derivatives replace a $g$ by
$k$ and are covered by the same majorants.

We finish by including the projections in the source commutator.
All uncontracted field coordinates carry exterior $q$
projections. Internal contractions are
$q=I-|\phi\rangle\langle\phi|$.
For the dressed $H$ term, both creation variables
already lie in the range of $q$. The internal
$q$ projection is therefore absorbed there.
For $L$ and $K_0,K_p$ there are two rank-one
corrections: $(L_z\phi)$ times a contraction of a creation variable of
$K_a$ against a smooth vector, and
$(K_a\phi)$ times $\phi^*L_z$.
Here $L_z\phi=0$ in the annihilation coordinate
of the actual $p$ kernel.
The other $p$ contraction has size
$C_I\rho$ in $L^2_{z,k}$ and
$K_0\phi$ has pair norm $C_I\epsilon$,
giving an ordinary $C_I\rho\epsilon$
cubic correction; $K_p\phi=0$.
In the reverse order the two factors
have norms $C_I$ and $C_I\epsilon^2$.
These use $\|\mathcal A_L\|\le C_I\rho$
and convolution by $k$ against a smooth vector.

For the identity--identity case,
$L_0(z)\phi$ has global norm $C_I$;
contracting the other creation variable of $K_0$ against a
smooth vector gives a multiplier plus a convolution operator, with norm
$C_I\epsilon^2$.
The first correction is consequently a
source $a^*(u_z)\dGamma(B)$ of size
$C_I\epsilon^2\nu^{3/2}$.
In the second correction $K_0\phi$
has norm $C_I\epsilon$ and $H^1$ dual
norm $C_I\delta$. The source $\phi^*L_0(z)$
consists of local $a_z,\nabla a_z$ with
$C_I\epsilon^2$ coefficients and an
ordinary global $L^2$ annihilation kernel
of size $C_I$. For its derivative coefficient
use
$(\nabla\kappa)*|\phi|^2=\kappa*\nabla|\phi|^2$,
so it also has size $C_I\epsilon^2$.
The local parts satisfy the pair estimates
above; the other part satisfies the
second criterion of
Lemma~\ref{lem:cubic-source-criteria}.

Finally define $H_{a,f}=[a(f),T_a]$.
The removed-coordinate projection
adds $-\nabla\phi(z)H_{a,\phi}$ to $S_a$.
For $a=p$ its one-body kernel has
Hilbert--Schmidt size $C_I\epsilon^2$,
by $\|k\|_1$ and bounded smooth amplitudes.
For $a=0$ it is a multiplier of size
$C_I\epsilon^2$ plus an operator of norm
$C_I\epsilon^2$ and Hilbert--Schmidt
norm $C_I\epsilon$.
Its commutator with $T_0$ has ordinary
cubic kernels of size $C_I\epsilon^2$
and density fibres of size $C_I\epsilon^3$;
the multipliers at the retained coordinate cancel, and the multiplier at
the newly created coordinate has size $C_I\epsilon^2$.
With a dressed cubic term the same composition
bounds apply. For $a=\sigma$ its pair
kernel has size $C_I\epsilon$, and composition
with $F_b$, $b=0,p$, gives $C_I\epsilon^2$.
All additional rank-one internal contractions
are bounded by these same maps.

The same estimates hold after a time derivative of the background or a
kernel. Any contraction that vanishes by orthogonality must be
differentiated as a whole. The remaining external
connections are bounded in $H^1$ on the inputs and in $L^2$
on the outputs. For density sources without a Hilbert--Schmidt kernel, the newly created
field is estimated with the retained coordinate fixed. The corresponding
$H^1$ bound is $C\delta$. A retained density
variable instead gives an ordinary kernel with global norm
$C_I\epsilon$ and two-input $H^1$ bound
$C_I\delta$. At $\nabla a_z$, the estimate follows from
the bounded map $\nabla Z$ together with $\sqrt n\|v_n\|$.
This covers the covariant derivatives in all raising cases.
\end{proof}

\subsubsection{Lowering and mixed combinations}

\begin{proposition}\label{prop:kinetic-second-source}
Every ordered canonical second source
$S_2(T,U)=\delta_U\delta_TD_K$ satisfies
\begin{equation}\label{eq:kinetic-second-source}
	\|S_2(T,U)v_n\|\le C_I\epsilon\sqrt\nu \|A_n^{1/2}v_n\|.
\end{equation}
Both ordered words $S_1(T)U$ and
$U_{\rm out}S_1(T)$ have the same bound.
The complete actual source
$S_2=\delta_C^2D_K$ also has this bound,
including the original number coefficients.
All statements allow one covariant and
fixed number-gauge derivatives.
\end{proposition}
\begin{proof}
By \eqref{eq:kinetic-pair-maps}, the lowering
first sources obey
$\|S_a^-v_n\|\le C_I\epsilon\sqrt\nu
\|A_n^{1/2}v_n\|$; $S_\sigma^-=0$.
The active two-input kinetic fraction proves
this directly from $\nabla F_a$.
The source-projection correction is
$-\nabla\phi(z)F_{a,\phi}$ and has the
same bound by the unweighted pair map.
Consequently the sharp cubic-operator graph
bound and the ordinary block bound give
\[
	\|S_a^-Uv_n\|+\|U_{\rm out}S_a^-v_n\| \le C_I\{\epsilon^2\nu^2\|A_n^{1/2}v_n\| +\epsilon\nu\|v_n\|\} \le C_I\epsilon\sqrt\nu\|A_n^{1/2}v_n\|.
\]
For every raising first source, the spatial
bound \eqref{eq:phase-spatial-smallness}
on a right cubic term gives instead
\[
	\|S_a^+Uv_n\| \le C_I\{\epsilon^2\nu^2\|A_n^{1/2}v_n\| +\epsilon\nu\|v_n\| +\delta\nu^{3/2}\|A_n^{1/2}v_n\|\} \le C_I\epsilon\sqrt\nu\|A_n^{1/2}v_n\|.
\]
Here use $\epsilon\nu^{3/2}\le C$,
$\delta\nu/\epsilon\le C$ and $A_n\ge\nu$.
Lemma~\ref{lem:kinetic-raising-second}
then gives the reverse word for every
raising $U$.

For a pure three-creator $T_\sigma$,
its full canonical Wick formula gives
\[
	[T_\sigma^*,T_\sigma] =9\mathcal W_{2,2}(\Gamma_1) +18\mathcal W_{1,1}(\Gamma_2)+6\|t\|_2^2\ge0 ,
\]
where $\Gamma_j$ are positive partial traces.
Hence $\|T_\sigma^*\xi\|\le\|T_\sigma\xi\|$,
also for source-valued $\xi$.
The already estimated reverse raising word
therefore controls $T_\sigma^*S_a^+$.
Apply this positivity to the canonical pure creator
before introducing the number cutoff.

Next take $a=0,p$ and the pairing source
$S_\sigma^+(z)=a^{*2}(H_z)$.
Its commutator with $T_a^*$ is exactly
\begin{align}
	[T_a^*,S_\sigma^+(z)]
	={}&4\int\overline{K_a(i,j;k)}H_z(i,v)
	a_k^*a_v^*a_j\d i\d j\d k\d v
	\notag\\
	&+2\int\overline{K_a(i,j;k)}H_z(i,j)
	a_k^*\d i\d j\d k .
	\label{eq:kinetic-mixed-pair-CCR}
\end{align}
The linear kernel is $2F_aH_z$ and
has norm $C_I\epsilon$.
For $a=0$, the two cubic contributions involving the
retained coordinate are
$\kappa(k-j)\bar\phi(j)H_z(k,v)$ and
the retained-density kernel
\[
	G_z(j,v)=\int\kappa(i-j)\bar\phi(i)H_z(i,v)\d i .
\]
The first has global norm $C_I\epsilon$
and $\mathcal A_r$ norm $C_I\epsilon^2$:
use the fixed-$k$ slice bound of $H$
then convolution by $k$.
For the second,
$\sup_j\|G(\cdot;j,\cdot)\|_2\le C_I\epsilon^2$,
so direct density lifting suffices.
For $a=p$, compose the first direction
with $p$; the second becomes
$\bar p(j,k)G_z(j,v)$ with global norm
$C_I\epsilon^2$.
The projected $H_z$ absorbs internal $q$
contractions, while exterior projections remain on the uncontracted variables.
This proves the required commutator bound;
the other ordered word was already estimated.

It remains to treat $[T_a^*,S_b^+]$ with
$a,b\in\{0,p\}$. For $b=p$, the exact
identity is
\begin{align}
	[T_a^*,\dGamma(L_z)]
	={}&2\int\overline{K_a(i,j;k)}L_z(i,l)
	a_k^*a_ja_l\d i\d j\d k\d l
	\notag\\
	&-\int L_z(k,l)\overline{K_a(i,j;l)}
	a_k^*a_ia_j\d i\d j\d k\d l .
	\label{eq:kinetic-mixed-density-CCR}
\end{align}
There is no linear contact.
For $a=0$ the first line has a convolution
direction $\kappa(k-j)\bar\phi(j)L_z(k,l)$
and a fibre direction
$G_z(k,l)=\int\kappa(i-k)\bar\phi(i)L_z(i,l)\d i$.
Their two-input maps have $L^2$ norm
$C_I\epsilon^2$, using the fixed-$k$
slice bound of $L$ and $\|k\|_1$.
For $a=p$ compose on the output with $p$.
The second line is $\mathcal C_LF_a$,
where $\mathcal C_Lf(z,k)=\int L_z(k,l)f(l)\d l$.
Its $H^1$ norm is $C_I\delta$ by the
global $L^2$ bound of $L$ and
\eqref{eq:kinetic-pair-maps}.
Apply the third case of
Lemma~\ref{lem:cubic-source-criteria}.

For $b=0$, use the explicit first source
from Proposition~\ref{prop:kinetic-source-decomposition}.
We give the two-input bounds for every
term with $T_0^*$.
In $[T_0^*,a^*(\nabla f_z)]a_z$,
the first contribution at the retained coordinate is multiplication by a
kernel bounded by $k*g(x-z)$.
Since $\|k*g\|_3\le\|k\|_{6/5}\|g\|_2
\le C\delta$, its $H^1$-to-$L^2$ norm
is $C_I\delta$. The other direction is
\[
	\psi\longmapsto g(z-x)\phi(x)\int\kappa(x-y)\bar\phi(y)\psi(y,z)\d y .
\]
At fixed $z$, the integral is bounded
in $L^\infty_x$ by $C_I\delta$ times
the input $y$-gradient norm; then use
$g\in L^2$.
In $[T_0^*,a^*(f_z)]\nabla_za_z$,
the corresponding multiplier and integral
operator have norms at most
$C_I\|k\|_2^2=C_I\epsilon^2$.
The one derivative is already on the input.
The other two terms are
$(\nabla f_z)(x)(F_0\psi)(z)$ and
$f_z(x)\nabla_z(F_0\psi)(z)$.
Their $H^1$ norms are $C_I\delta$ and
$C_I\epsilon^2$, by
\eqref{eq:kinetic-pair-maps}.

For the first density source, the creation and annihilation contributions
at the retained coordinate cancel, leaving
\[
	\psi\longmapsto \phi(z)\int\kappa(x-y)\bar\phi(y)g(y-z)\psi(x,y)\d y .
\]
Its fixed-$x$ map is the $F_C$ estimate
proved in Lemma~\ref{lem:kinetic-raising-second},
of $H^1$ norm $C_I\delta$.
A background derivative is smaller.
With $T_p^*$, the first two contractions compose with $p$ on the
output,
and the next two use $F_p$ in place of $F_0$.
The density commutator adds
$(p^{\mathsf T}m_z-m_zp^{\mathsf T})F_0$,
where $m_z(x)=\phi(z)g(x-z)$.
The map $f(x)\mapsto m_z(x)f(x)$ is
bounded from $L^2_x$ to $L^2_{z,x}$,
since $\sup_x\int|m_z(x)|^2\d z\le C_I$.
Composing with the $C_I\delta$ pair map
proves the same bound.

Internal projections in these mixed terms
give two types of rank-one correction.
The first is a one-body contraction $B_a$
of $K_a$, of operator norm $C_I\epsilon^2$,
times the annihilation source $\phi^*L_b(z)$.
For $b=p$ that source has global norm
$C_I\rho$; for $b=0$ its ordinary kernel
has size $C_I$ and its local $a_z,\nabla a_z$
coefficients have size $C_I\epsilon^2$,
as verified in the preceding lemma.
The resulting source is bounded by $C_I(\epsilon^2\nu^{3/2}\|v_n\|
+\epsilon^4\nu\|A_n^{1/2}v_n\|)$.
The second correction is $L_b(z)\phi$,
of global norm $C_I$, times the pair
functional $\phi^*F_a$, of $H^1$ norm
$C_I\delta$. Both satisfy the third
source criterion. The removed-coordinate
term $-\nabla\phi(z)H_{b,\phi}$ has
one-body operator norm $C_I\epsilon^2$;
its two ordered products with $T_a^*$ are bounded by
$C_I\epsilon^3\nu^{5/2}\|v_n\|$, which also suffices. External projections
act on the two $H^1$ inputs or on the
$L^2$ output.
These checks complete all canonical directions.

All the stated maps allow one time
replacement by the already proved kernel
bounds. For an input or output connection,
compose the two-input map with the
$H^1$ or $L^2$ finite-rank operator.
Local $\nabla a_z$ terms use $\nabla Z$.
The differentiated source projection is
estimated as above with the differentiated
smooth vectors. This proves the derivative
assertion for the canonical words.

Finally put $C=\sum_\alpha T_\alpha b_\alpha(\cN)$
with the exact right coefficients, including
their adjoint shifts. Since $D_K$ lowers
the remaining number by one, its first
actual source is exactly
\begin{align}
	\delta_CD_K&=\sum_\alpha S_1(T_\alpha)b_\alpha(\cN)+E_1,
	\notag\\
	E_1&=\sum_\alpha(T_\alpha)_{\rm out}D_K
	\{b_\alpha(\cN)-b_\alpha(\cN-1)\}.
	\label{eq:actual-kinetic-source-coeff}
\end{align}
The original first difference has size $C\tau$,
so
$\|E_1v_n\|\le C_I\tau\epsilon\nu^{3/2}
\|K_P^{1/2}v_n\|\le
C_I\epsilon\sqrt\nu\|A_n^{1/2}v_n\|$.
Differentiate once more by $\delta_C$.
The first sum gives the two canonical
ordered source words with their possibly
different exact shifted coefficients.
Both words have been separately bounded.
The additional term $E_1C-C_{\rm out}E_1$ is bounded by
$C_I(\epsilon^2\nu^2\|A_n^{1/2}v_n\|
+\epsilon\nu\|v_n\|)$, which also suffices.
One time derivative inserts an estimated
cubic term or the bounded connection source
$\nabla_tD_K$; first differences are
time independent. Number derivatives
only multiply by fixed particle-number shifts.
This proves the full actual assertion.
\end{proof}

In particular every complete product of
two such second sources, including the
actual $S_2^*S_2$ and its indicated derivatives,
satisfies the asymmetric form bound
\[
	C_I\epsilon^2 \|MA_P^{1/2}E_Nu\|\|A_P^{1/2}E_Nv\|.
\]
It includes all lower-field contractions
and the number-coefficient source terms
in \eqref{eq:actual-kinetic-source-coeff}.

\subsection{The fourth kinetic commutator}
\label{sec:kinetic-fourth}

\subsubsection{Restoring the original coefficients}

\begin{lemma}\label{lem:kinetic-coefficient-error}
With the columnwise matching convention of
Lemma~\ref{lem:potential-coefficient-error},
\[
	\mathcal D_C^4K_P=\mathfrak F_K+\mathcal E_K, \qquad \mathfrak F_K\big|_n =\mathcal D_{C_{{\rm fr},n}}^4K_P\big|_n .
\]
The form $\mathcal E_K$, its first covariant
derivative and its number commutator obey
\begin{equation}\label{eq:kinetic-coefficient-error}
	|\mathcal E_K[u,v]| \le C_I\epsilon^2 \|MA_P^{1/2}E_Nu\|\|A_P^{1/2}E_Nv\|.
\end{equation}
\end{lemma}
\begin{proof}
The rectangular identity
\eqref{eq:potential-B2-identity} holds with
$D=D_K$, $V_P=K_P$, and the corresponding
kinetic sources. We verify the estimates
needed to use the same rearrangement.
The first-source bounds and the bounds for products with two cubic terms
are given by \eqref{eq:potential-source-words}, with $Q$
replaced by $A=A_P$. The proof uses the sharp kinetic graph
bound, the spatial smallness estimate and the first-source bound from
Proposition~\ref{prop:sharp-kinetic-block}.
For a second source with a right cubic term
the kinetic recurrence is slightly stronger.
Equation \eqref{eq:kinetic-higher-source}
at $j=2$ gives
\begin{align*}
	\|S_2(T,U)T'v_n\|\le C_I\{&
	\epsilon^2\nu^2\|v_n\|
	+\epsilon\delta\nu^{5/2}\|A_n^{1/2}v_n\|\\
	&+\epsilon^3\nu^{5/2}\|A_n^{1/2}v_n\|
	+\epsilon^2\nu^{3/2}\|v_n\|\}
	\le C_I\epsilon^2\nu^{3/2}\|A_n^{1/2}v_n\|.
\end{align*}
Use $\delta\nu/\epsilon\le C$,
$\epsilon\nu\le C$ and $A_n\ge\nu$.
One more right cubic term is allowed by
uniform weighted graph boundedness.
The direct second-source bound is
\eqref{eq:kinetic-second-source}.
For $l+r=2$ and $|h|\le C\tau$, every word
$W_lB_2(T,U)W_rh(\cN)$ has the bound
\eqref{eq:kinetic-coefficient-error}:
use the source-word bounds on the two
source products and the two second-source
bounds on its endpoint pairings.
The scaling estimates are those displayed
in Lemma~\ref{lem:potential-coefficient-error},
with a smaller weight for right multiplication by a cubic term.

The lowering first kinetic source has
the $\epsilon\sqrt\nu$ energy bound
proved in Proposition~\ref{prop:kinetic-second-source}.
Consequently $B_1(T)=[K_P,T]$ has the
oriented estimate
\eqref{eq:potential-oriented-first}
with $Q=A$. The graph and $L^2$ bounds for products of three cubic
terms apply when the first commutator is at the appropriate end of the
product. They give the coefficient $\tau(\epsilon^3\nu^{7/2}+\delta\nu^{3/2})
\le C\epsilon^2\nu$.
Reordering to those endpoints retains
only the already controlled $B_2$ terms.
Since $K_P$ commutes with number functions,
the exact fourfold expansion and first-difference
telescoping from the potential proof now
give \eqref{eq:kinetic-coefficient-error}.

The additional kinetic connection is
$\nabla_tK_P=P\,\dGamma([-\Delta,Z])P$.
Its one-body kernel is bounded finite rank,
with vectors $\phi,\dot\phi,\Delta\phi,
\Delta\dot\phi$ uniformly controlled by
the prescribed background regularity.
Its commutator with a cubic term has block
norm $C_I\epsilon\nu^{3/2}$.
For this ordering, its coefficient is bounded by $C_I\tau\epsilon\delta\nu^2$,
which is smaller than $\epsilon^2\nu$.
In the differentiated $B_2$ formula,
use the proved source derivatives and
the bounded linear source $\nabla_tD_K$.
This preserves the same endpoint and
reordering estimates. All other derivatives
replace an already estimated cubic term;
number differentiation inserts a fixed particle-number shift.
\end{proof}

\subsubsection{Reduction to a bounded six-field kernel}

Fix one matching anti-self-adjoint $C_{\rm fr}$
and suppress the subscript. Use
$S_j=P_j+A_jD_K$ from
Proposition~\ref{prop:kinetic-source-decomposition}.
The rectangular recurrence gives the ordered
formula
\begin{align}
	P_2&=\delta_CP_1+A_1P_1,&
	A_2&=\delta_CA_1+A_1^2,\notag\\
	P_3&=\delta_C^2P_1+A_1\delta_CP_1
	+2(\delta_CA_1)P_1+A_1^2P_1 .
	\label{eq:kinetic-P3-ordered}
\end{align}
Here $\delta_C$ acts on a source map or
a source-to-source map with the corresponding
domain and range. The product $(\delta_CA_1)P_1$ retains
this order.
The degrees of $P_1,P_3,A_1,A_3$ are
at most two, four, one and three,
and their fixed-sector norms are
$C_I,C_I\epsilon^2,C_I\epsilon,
C_I\epsilon^3$.

Split $P_1=P_1^++P_1^-$ according to
the raising or lowering original cubic term.
The lowering source estimate and
$\|A_1^-w_n\|\le C_I\epsilon\sqrt\nu\|w_n\|$
imply
$\|P_1^-v_n\|\le C_I\epsilon\sqrt\nu
\|A_n^{1/2}v_n\|$.
Consequently the following source products
have the listed balanced form constants:
\[
	\begin{array}{c|c}
	(P_1^-)^*P_3&C_I\epsilon^3\nu^2\\
	P_1^*A_3D_K&C_I\epsilon^3\nu^2\\
	(A_1D_K)^*P_3&C_I\epsilon^3\nu^2\\
	(A_1D_K)^*A_3D_K&C_I\epsilon^4\nu^2
	\end{array}
\]
Use ordinary fixed-degree lifting and
$\|v_n\|\le\nu^{-1/2}\|A_n^{1/2}v_n\|$
on the unweighted side of each product.
Every entry is at most $C_I\epsilon^2\nu$.
The same holds for the adjoints.
Equation \eqref{eq:fourth-source-product},
the fourth-source endpoint estimate and
Proposition~\ref{prop:kinetic-second-source}
therefore leave only
\[
	H_6=\mathcal N_6\{(P_1^+)^*P_3+P_3^*P_1^+\}.
\]
Indeed the full bounded-source product has
fixed-sector norm $C_I\epsilon^2$ and
degree at most six. Its degree-zero, two
and four parts have the desired bound by
triangular extraction and energy lifting.
After restoring the coefficient error,
the exact decomposition is
\begin{equation}\label{eq:kinetic-six-reduction}
	\mathcal D_{C_{\rm actual}}^4K_P =L_{\rm controlled}+4H_6^{\rm col},
\end{equation}
where the column convention is that of
Lemma~\ref{lem:kinetic-coefficient-error}.
$L_{\rm controlled}$ has the target
$\epsilon^2MA_P$ bound, including the
indicated derivatives. The remaining $H_6$
contains no derivative acting directly on
an input wavefunction; its derivatives are
in the original source coefficients.

\subsubsection{Positive kernel bounds and the spatial gain}

\begin{lemma}\label{lem:kinetic-six-spatial}
For every six-field kernel in $H_6$,
\begin{equation}\label{eq:kinetic-six-spatial}
	|(H_6)_{pq}[u_p,v_q]| \le C_I N^{-2}\|A_{P,p}^{1/2}u_p\| \|A_{P,q}^{1/2}v_q\|, \qquad p+q=6 .
\end{equation}
The same estimate holds after one covariant
coefficient replacement.
\end{lemma}
\begin{proof}
We first construct the positive kernel bound
needed below. A normally ordered field
operator with nonnegative scalar kernel
preserves nonnegative wavefunctions in
the coordinate representation. The same
is true with an explicit source coordinate.
For a product $W$ of such operators,
each Wick contraction term $W_\alpha$
has nonnegative kernel and, on fixed sectors,
\[
	|\langle u,W_\alpha v\rangle| \le\langle|u|,W_\alpha|v|\rangle \le\langle|u|,W|v|\rangle \le\|W\|\,\|u\|\,\|v\|.
\]
The order here is the order of positivity-preserving integral maps.
Apply it on the minimal input/output
block of each chosen Wick term; the
normalization is its fixed factorial.

Put $k=|\kappa|$, $g=|\nabla\kappa|$ and $\Phi=|\phi|$.
Replace each coefficient in $C,A_1,P_1$ by its absolute
value before multiplication or contraction. Denote the resulting
positive maps by $C_+,A_+,P_+$, with the adjoint terms of
$C$ included with positive signs.
The identity cubic uses the one-to-two map
\[
	f(x)\longmapsto k(x-y)\Phi(y)f(x),\qquad \|k(x-y)\Phi(y)f(x)\|_2\le\|\Phi\|_\infty\|k\|_2\|f\|_2.
\]
For a dressed cubic, the three-variable coefficient obeys
\[
	\int k(x-y)^2\Phi(y)^2|h(x,z)|^2\d x\d y\d z\le\|\Phi\|_\infty^2\|k\|_2^2\|h\|_{\mathrm{HS}}^2.
\]
Its adjoint has the same norm. The fibre maps of $A_+$ use
$\sup_z\|k(z-\cdot)\Phi\|_2\le C_I\epsilon$.
Thus $C_+$ and $A_+$ have norm $C_{I,d}\epsilon$ on
each fixed sector range $n\le d$.

The identity pieces of $P_+$ are controlled by
\[
	\sup_z\int g(z-x)^2\Phi(x)^2\d x+\sup_x\int\Phi(z)^2g(x-z)^2\d z\le2\|\Phi\|_\infty^2\|g\|_2^2.
\]
The first term bounds $a^*(g_z\Phi)a_z$ and its
annihilation counterpart after integrating in the source
coordinate $z$. For the density term, use
$|\sum_{i=1}^n\Phi(z)g(x_i-z)|^2
\le n\sum_i\Phi(z)^2g(x_i-z)^2$ and then the second term.
The dressed coefficients have, up to permutations,
the following squared integral bounds:
\begin{align*}
	\|g(z-x)\Phi(x)h(z,y)\|_2^2
	+\|g(x-z)\Phi(z)h(x,y)\|_2^2
	&\le C_I\|g\|_2^2\|h\|_{\mathrm{HS}}^2,\\
	\|k(z-x)\Phi(x)\nabla_zh(z,y)\|_2^2
	+\|k(x-y)\Phi(y)\nabla_zh(x,z)\|_2^2
	&\le C_I\|k\|_2^2\|\nabla h\|_{\mathrm{HS}}^2.
\end{align*}
These bounds follow by integrating $x$ or $y$ first.
They give $\|P_+\|\le C_{I,d}$, since
$\|\nabla h\|_{\mathrm{HS}}\le C_I\sqrt N$ and
$\epsilon\sqrt N\le1$. Terms with background derivatives,
$\nabla p$, $\nabla r$ or the source projection have
norm $C_{I,d}\epsilon$ by their bounded derivative
Hilbert--Schmidt norms and the same integrals.

Internal $q$ contractions are retained.
Their absolute kernels are bounded by
$Q_+=I+|\phi|\,\langle|\phi|,\cdot\rangle$.
Here $\|Q_+\|=2$ and
$Q_+^2=I+3|\phi|\,\langle|\phi|,\cdot\rangle$.
Insert $Q_+$ on each projected field argument;
the contraction of two such arguments is dominated
by $Q_+^2$. Composition with these bounded maps
preserves the preceding row and Hilbert--Schmidt bounds.
Their number is bounded by the fixed length of the word.
We include these insertions in $C_+,A_+,P_+$.

Write $C_{+,\mathrm{out}}=I_{L^2_z}\otimes C_+$.
For the source endomorphism $A_1$, both sides of the
rectangular commutator act on the source space:
\[
	\delta_CA_1=A_1C_{\mathrm{out}}-C_{\mathrm{out}}A_1.
\]
Consequently,
\[
	A_1\delta_CP_1+2(\delta_CA_1)P_1=A_1P_1C+(-1+2)A_1C_{\mathrm{out}}P_1-2C_{\mathrm{out}}A_1P_1.
\]
Expanding \eqref{eq:kinetic-P3-ordered} before taking
absolute values gives seven ordered products:
\begin{align}
	P_3={}&P_1C^2-2C_{\mathrm{out}}P_1C+C_{\mathrm{out}}^2P_1
	+A_1P_1C+A_1C_{\mathrm{out}}P_1\notag\\
	&-2C_{\mathrm{out}}A_1P_1+A_1^2P_1.
	\label{eq:kinetic-P3-seven-words}
\end{align}
Their positive majorant is
\begin{align}
	\mathscr W_+={}&P_+C_+^2+2C_{+,\mathrm{out}}P_+C_+
	+C_{+,\mathrm{out}}^2P_+\notag\\
	&+A_+P_+C_++A_+C_{+,\mathrm{out}}P_+
	+2C_{+,\mathrm{out}}A_+P_++A_+^2P_+.
	\label{eq:kinetic-positive-seven-words}
\end{align}
For an input sector $n\le d$, all intermediate sectors
lie below $d+8$. Each product has one bounded $P_+$
factor and two factors of size $C_{I,d}\epsilon$.
It follows that $\|\mathscr W_+\|\le C_{I,d}\epsilon^2$.

Let $\Pi_3$ be the sum of its four-field Wick kernels,
with their positive combinatorial coefficients. Each
contraction in the signed formula is dominated by the
corresponding contraction in \eqref{eq:kinetic-positive-seven-words}.
On the minimal block from $q$ to $p$ particles, $p+q=4$,
the associated Wick operator has kernel
$\sqrt{p!q!}(\Pi_3)_{pq}$. It is a nonnegative summand
of the same block of $\mathscr W_+$. The positive-map
inequality above therefore gives
\begin{equation}\label{eq:kinetic-positive-source}
	|(P_3^{(4)})_{pq}(z;Y,X)| \le(\Pi_3)_{pq}(z;Y,X),\qquad \|(\Pi_3)_{pq}\|_{L^2_{\rm sym}(\mathbb R^{3q}) \to L^2_z\otimes L^2_{\rm sym}(\mathbb R^{3p})} \le C_I\epsilon^2,\quad p+q=4 .
\end{equation}
These are ambient spaces, so absolute values
of excitation wavefunctions are permitted.
The norm is taken in the full source space. Dirac delta
kernels represent identity maps in the field operators.
For the integral kernels, first restrict the coordinates
to balls and truncate the nonnegative coefficients.
The preceding row and squared integral bounds remain valid.
Keep identity kernels as identity maps. Wick expansion is
then justified on bounded compactly supported vectors.
Monotone convergence gives each nonnegative contraction
pairing in the limit. Its uniform norm bound extends the
map to the full $L^2$ source space.

Three spatial inequalities supply the gain.
With $g=|\nabla\kappa|$ here,
\begin{align}
	\int g(z-x)|f(x)|\d x
	&\le C\epsilon\|\nabla f\|_2,\notag\\
	\int|\kappa(z-x)f(x)|\d x
	&\le C\delta\|\nabla f\|_2,\notag\\
	\int g(z-x)|f(x)h(x)|\d x
	&\le C\tau\|\nabla f\|_2\|\nabla h\|_2,
	\qquad \tau=N^{-1/2}.
	\label{eq:kinetic-Hardy-pairings}
\end{align}
The first two follow from the $L^{6/5}$ kernel bounds
and $\|f\|_6\le C\|\nabla f\|_2$.
For the third, $g(r)\le C\tau|r|^{-2}$ gives
\[
	\int g(z-x)|f(x)h(x)|\d x\le C\tau\left\|\frac{f}{|\cdot-z|}\right\|_2\left\|\frac{h}{|\cdot-z|}\right\|_2\le C\tau\|\nabla f\|_2\|\nabla h\|_2.
\]
These are the three-dimensional Sobolev and Hardy inequalities
\cite{LiebLoss}. Hilbert-valued spectators are handled by
marginal $L^2$ norms and their weak gradient bound, as in
the proof of Lemma~\ref{lem:potential-density-pairing}.

The six-field component pairs a
two-field $P_1^+$ with a four-field $P_3$
without contractions between their field
coordinates. They share only the source
coordinate $z$. Up to bounded amplitudes,
conjugations and spatial components,
the parts of $P_1^+$ without an additional small factor have
the following eight forms:
\[
	\begin{array}{c|l|c}
	&\text{coefficient and fields}
	&\text{additional pairing factor}\\ \hline
	U&g(z-x)\phi(x)a_x^*a_z& C_I\epsilon\\
	W&\phi(z)g(x-z)a_x^*a_x& C_I\tau\\
	D_1&g(z-x)\phi(x)p(z,y)a_x^*a_y& C_I\epsilon\\
	D_2&g(x-z)\phi(z)p(x,y)a_x^*a_y& C_I\tau\\
	H_1&g(z-x)\phi(x)\sigma(z,y)a_x^*a_y^*&C_I\epsilon\\
	H_2&g(x-z)\phi(z)\sigma(x,y)a_x^*a_y^*&C_I\epsilon\\
	J_1&\kappa(z-x)\phi(x)\partial_z\eta(z,y)a_x^*a_y^*
	&C_I\delta\sqrt N\\
	J_3&\kappa(x-y)\phi(y)\partial_z\eta(x,z)a_x^*a_y^*
	&C_I\delta\sqrt N
	\end{array}
\]
Every last-column factor is at most
$C_I\tau$. The first-source formulas identify these terms
directly. In $P_1(T_0)$, the derivative of $\kappa$ gives
$U$ and $W$. In $L_z$ of Section~\ref{sec:kinetic-second-source},
it gives $D_1,D_2$; derivatives of $p$ or $\phi$ give
the smaller class. In $H_z$ it gives $H_1,H_2$.
Writing $\sigma=\eta+r$ in the two differentiated
hyperbolic kernels gives $J_1,J_3$, with the derivatives
of $r$ again in the smaller class. Projections contribute
the source correction already estimated above.
We verify the pairings so that
the source norm in
\eqref{eq:kinetic-positive-source} is
used only after integration in $z$.

For $U$, after adjunction the external
wavefunctions contain $u(z,Y)$ and $v(x,X)$.
Apply the first inequality in
\eqref{eq:kinetic-Hardy-pairings} in $x$
to the latter, yielding
$\epsilon\|\nabla_xv(\cdot,X)\|_2$.
The remaining source pairing is controlled
by \eqref{eq:kinetic-positive-source}.
For $W$, both wavefunctions retain $x$.
Apply the third inequality to their
$x$ factors, yielding
$\tau U(Y)V(X)$ from one derivative
on each side. The source test vector
is $|\phi(z)|U(Y)$, which belongs to $L^2$.

For $D_1$ first integrate the input $x$
by Sobolev. The other test vector is
$U_p(z,Y)=\int|p(z,y)u(y,Y)|\d y$,
of $L^2$ norm at most $C_I\|u\|$
by the Hilbert--Schmidt norm of $p$.
For $D_2$ use
$U_p(x,Y)=\int|p(x,y)u(y,Y)|\d y$.
Its $x$-gradient norm is at most
$C_I\|u\|$, since
$\|\nabla p\|_{\mathrm{HS}}\le C_I$ and
$|\nabla|p||\le|\nabla p|$ in the weak sense.
Pair this vector with $v(x,X)$ by Hardy.
The remaining source weight is $|\phi(z)|$.

For $H_1$, both outside fields occur in
$v(x,y,X)$. Cauchy--Schwarz in $y$ leaves
$w_\sigma(z)=\|\sigma(z,\cdot)\|_2$.
Apply the first inequality in
\eqref{eq:kinetic-Hardy-pairings} in $x$
to the $y$ marginal of $v$.
The source test vector is
$w_\sigma(z)u(Y)$ and
$\|w_\sigma\|_2=\|\sigma\|_{\mathrm{HS}}\le C_I$.
For $H_2$, use the uniform row bound
of $\sigma(x,\cdot)$ in $y$, then the
same Sobolev inequality in $x$.
The source weight is now $|\phi(z)|$;
the derivative acts on the marginal
of $v$.

For $J_1$ repeat $H_1$ with the second
Sobolev inequality and
$w(z)=\|\partial_z\eta(z,\cdot)\|_2$.
Its $L^2$ norm is at most $C_I\sqrt N$,
giving $\delta\sqrt N=\epsilon$.
For $J_3$, first integrate $y$ by the
same $k$ Sobolev bound, leaving
$\delta\|\nabla_yv(x,\cdot,X)\|_2$.
Then integrate $x$ by Cauchy--Schwarz
and retain
$w(z)=\|\partial_z\eta(\cdot,z)\|_2$,
again of size $C_I\sqrt N$.

We collect the resulting pairings. Write
$U_0(Y)=\|u(\cdot,Y)\|_2$,
$U_1(Y)=\|\nabla_xu(\cdot,Y)\|_2$ and
$V_1(X)=\|\nabla_xv(\cdot,X)\|_2$.
For two extra coordinates of $v$, use
$V_{1,x}(X)=\|\nabla_xv(\cdot,\cdot,X)\|_2$ and
$V_{1,y}(X)=\|\nabla_yv(\cdot,\cdot,X)\|_2$.
In each row the absolute pairing is bounded by
\begin{equation}\label{eq:kinetic-source-test-pairing}
	\beta\langle F,\Pi_3G\rangle_{L^2_{z,Y}}\le C_I\beta\epsilon^2\|F\|_{L^2_{z,Y}}\|G\|_{L^2_X},
\end{equation}
with the following nonnegative tests:
\[
	\begin{array}{c|c|c|c}
	&F(z,Y)&G(X)&\beta\\ \hline
	U&|u(z,Y)|&V_1(X)&C_I\epsilon\\
	W&\Phi(z)U_1(Y)&V_1(X)&C_I\tau\\
	D_1&U_p(z,Y)&V_1(X)&C_I\epsilon\\
	D_2&\Phi(z)U_0(Y)&V_1(X)&C_I\tau\\
	H_1&w_\sigma(z)|u(Y)|&V_{1,x}(X)&C_I\epsilon\\
	H_2&\Phi(z)|u(Y)|&V_{1,x}(X)&C_I\epsilon\\
	J_1&\|\partial_z\eta(z,\cdot)\|_2|u(Y)|&V_{1,x}(X)&C_I\delta\\
	J_3&\|\partial_z\eta(\cdot,z)\|_2|u(Y)|&V_{1,y}(X)&C_I\delta
	\end{array}
\]
For $D_2$, the factor $\|\nabla p\|_{\mathrm{HS}}$ is
included in $C_I$. The first six $F$ norms are bounded
by $C_I\|A_{P,p}^{1/2}u\|$; the last two have the
additional factor $\sqrt N$. All $G$ norms are bounded
by $\|A_{P,q}^{1/2}v\|$. The remaining coordinates
$X,Y$ are symmetric separately, so these are vectors
in the spaces of \eqref{eq:kinetic-positive-source}.
Source-coordinate identity kernels are included in that
bounded map. The pairing uses its full $L^2_{z,Y}$ norm.
Thus every coefficient is at most
$C_I\max(\epsilon,\tau,\delta\sqrt N)\epsilon^2
=C_I\tau\epsilon^2=C_IN^{-2}$.
The omitted parts of $P_1^+$ contain
background derivatives, $\nabla p,\nabla r$,
or source-projection corrections.
Their fixed-sector norm is
$C_I\epsilon$, so their product with
$P_3$ has norm $C_I\epsilon^3\le C_IN^{-2}$.
Triangular extraction controls its
six-field part with the same constant.
This proves \eqref{eq:kinetic-six-spatial}.

Uncontracted field projections move to
the external fixed-particle wavefunctions
and preserve $H^1$; internal projections
were retained in the positive majorant.
Ordinary time differentiation preserves
all row, envelope, Hilbert--Schmidt
and smooth-amplitude bounds just used.
An internal $\dot q$ is a smooth
finite-rank insertion and has the same
bounded positive majorant construction.
External covariant insertions preserve
the external $H^1$ norms.
The source connection contributes a
bounded initial source of size
$C_I\epsilon$ after its commutator
with a cubic term, and joins the smaller
class. Thus the same proof includes
one covariant replacement.

The displayed estimates first hold for smooth fixed-particle
vectors. They bound the resulting sesquilinear forms on
$D(A_{P,p}^{1/2})\times D(A_{P,q}^{1/2})$, so density
gives their unique continuous extensions. The projections
$q$ and their smooth coefficient replacements are bounded
on $H^1$ and preserve this extension. After energy lifting,
the finite number of particle-number shifts are summed by
Cauchy--Schwarz; on each such shift the $M$ weights are
comparable. This gives the weighted form used below.
\end{proof}

\begin{theorem}
\label{thm:actual-kinetic-fourth}
For the original cubic phase,
\begin{equation}\label{eq:actual-kinetic-fourth}
	|T[u,v]|\le C_I\epsilon^2 \|MA_P^{1/2}E_Nu\|\|A_P^{1/2}E_Nv\|, \quad T\in\{\mathcal D_C^4K_P,\, \nabla_t\mathcal D_C^4K_P,\, i[M,\mathcal D_C^4K_P]\}.
\end{equation}
The exact integral remainder $R_K$ of the conjugation expansion
and the same derivatives
satisfy this estimate, also with $A_P$
replaced by $Q_{N,P}$. Its Duhamel
contribution on $D(F_3(s)^{3/2})$ is bounded
by $C_{I,s}N^{-5/4}\|F_3(s)^{3/2}z\|$.
\end{theorem}
\begin{proof}
Energy lifting in
Lemma~\ref{lem:fixed-degree-extraction}
changes the six-field bound
\eqref{eq:kinetic-six-spatial} to
$C_IN^{-2}\nu^2$.
Since $\nu\le C\sqrt N$, this is at most
$C_I\epsilon^2\nu$.
Combine with the exact reduction
\eqref{eq:kinetic-six-reduction} and
the full coefficient-error estimate.
Each component was proved with its
covariant replacements; number differentiation
only inserts a fixed particle-number shift.
For the kinetic connection appearing
outside the source products, its bounded
finite-rank one-body kernel was retained
in Lemma~\ref{lem:kinetic-coefficient-error}.
Equivalently its canonical fourfold
commutator has degree at most six and
fixed-sector norm $C_I\epsilon^4$,
which is smaller after energy lifting.

The original weighted phase graph bounds
and their first derivative pass the
estimate to
\[
	R_K=\frac1{3!}\int_0^1(1-a)^3 e^{-aC}\mathcal D_C^4K_Pe^{aC}\d a .
\]
Finally apply the one-summand inverse-energy
estimate of Theorem~\ref{thm:inverse-comparison},
as for the potential remainder.
The inverse on the left is still the
full exact energy inverse, and the right
trajectory is the same Taylor evolution with
the proved $F_3^{3/2}$ domain propagation.
This proves the asserted contribution;
the other BS generator groups require
their separate static estimates.
\end{proof}

\subsection{Coupling energy sources to bounded sources}
\label{sec:bounded-source-couplings}

The remaining generator groups involve a coupling between an
energy source and a bounded source. The estimate below treats
these couplings in a common form. Throughout this subsection $Q=Q_{N,P}$,
$D=D_V=\sqrt{W_N/2}\,a_xa_y$,
$\epsilon=N^{-3/4}$, $\delta=N^{-5/4}$,
$\tau=N^{-1/2}$ and $\nu=n+1$.
All sector bounds are on the original particle-number cutoff range
enlarged by the finitely many particle-number shifts in the displayed words.
In particular $\nu\le C\sqrt N$.

Let $F$ be a normally ordered source-valued polynomial of
degree at most $d$, whose fixed-sector operator norms, and
those of one covariant replacement, are bounded by $C_I\beta$.
We use only the two cases
\begin{equation}\label{eq:bounded-source-class}
	(d,\beta)=(1,1)\ \hbox{or}\ (2,\tau), \qquad H_F=D^*F+F^*D .
\end{equation}
The fixed sectors here include all sectors required for
degree extraction after four commutators.
Thus $\|Fv_n\|\le C_I\beta\nu^{d/2}\|v_n\|$.
Each component with fixed particle-number shift of $D^*F$ or $F^*D$ may have a bounded
right coefficient $g(\cN)$; its first differences, and
those of its required time replacement, are $O(\tau)$.
In an adjoint its exact shifted argument is retained.

\begin{proposition}\label{prop:bounded-source-coupling}
For every coupling in \eqref{eq:bounded-source-class},
including these right coefficients, the actual fourth
commutator satisfies
\begin{equation}\label{eq:bounded-source-fourth}
	|B[u,v]|\le C_I\epsilon^2 \|MQ^{1/2}E_N'u\|\|Q^{1/2}E_N'v\|, \quad B\in\{\mathcal D_C^4H_F,\, \nabla_t\mathcal D_C^4H_F,\, i[M,\mathcal D_C^4H_F]\}.
\end{equation}
Here $E_N'$ enlarges the original particle-number cutoff projection by
the range of particle-number shifts of $H_F$ and a fixed constant.
The exact integral remainder of its conjugation expansion has the same bounds.
Its contribution to the comparison in
Theorem~\ref{thm:inverse-comparison} is at most
$C_{I,s}N^{-5/4}\|F_3(s)^{3/2}z\|$.
\end{proposition}
\begin{proof}
Write $W_j$ for an ordered word of $j$ canonical cubic terms,
allowing adjoints and one covariant replacement.
The previously proved graph and spatial estimates give
\begin{align}
	\|W_jv_n\|&\le C_I\|v_n\|,&
	\|Q^{1/2}W_jv_n\|&\le C_I\|Q_n^{1/2}v_n\|,\notag\\
	\|W_jv_n\|&\le C_I\delta\nu\|Q_n^{1/2}v_n\|,
	&&j=2,3,4,\notag\\
	\|Tv_n\|&\le C_I\{\epsilon\sqrt\nu\|v_n\|
	+\delta\nu\|Q_n^{1/2}v_n\|\},\notag\\
	\|Q^{1/2}W_3v_n\|&\le C_I\epsilon^2\nu^3
	\|Q_n^{1/2}v_n\|.
	\label{eq:coupling-word-bounds}
\end{align}
The first line uses $\epsilon\nu^{3/2}\le C$ and
is asserted only on the particle-number cutoff range.
The last line was proved in
Lemma~\ref{lem:potential-coefficient-error};
it also follows by one more sharp graph iteration in
\eqref{eq:potential-source-words}.

First fix a matching anti-self-adjoint constant-coefficient
phase. Its iterated source commutators
$F_j=\delta_C^jF$ have degree at most $d+j$ and fixed-sector
norm $C_I\beta\epsilon^j$.
This follows either by the finite Wick expansion or
Lemma~\ref{lem:fixed-degree-extraction}: on fixed sectors
each ordered product has one factor $\epsilon$ per cubic term,
and the top uncontracted terms cancel at each commutator.
By Proposition~\ref{prop:potential-higher-sources},
$D_j=P_j+A_jD$, where for $j\ge1$ the degrees of $P_j,A_j$
are at most $j+2,j$ and their fixed-sector norms are
$C_I\epsilon^{j-1},C_I\epsilon^j$.

Apply the exact fourth source Leibniz rule to $D^*F$.
For $j\ge1$, the products $P_j^*F_{4-j}$ have degree
at most $d+6$ and fixed-sector norm $C_I\beta\epsilon^3$.
Extraction and lifting with $Q\ge M$ give the form constant
$C_I\beta\epsilon^3\nu^{d/2+2}$.
The products containing $A_jD$, and the $j=0$ endpoint,
have the smaller constant
$C_I\beta\epsilon^4\nu^{(d+3)/2}$.
Add the adjoint terms. Consequently the matching column obeys
\begin{equation}\label{eq:coupling-canonical-fourth}
	|\mathcal D_{C_{\rm fr}}^4H_F[u_m,v_n]| \le C_I\beta\epsilon^3\nu^{d/2+2} \|Q_m^{1/2}u_m\|\|Q_n^{1/2}v_n\| \le C_I\epsilon^2\nu \|Q_m^{1/2}u_m\|\|Q_n^{1/2}v_n\|.
\end{equation}
The last step uses $\epsilon\nu^{3/2}\le C$ for $d=1$,
and $\tau\epsilon\nu^2\le C$ for $d=2$.
Lower-degree contractions are part of the same extraction.

We next estimate the coefficient errors, including those of bare
$H_F$, with its possible particle-number shifts.
Let $|h(n)|\le C_I\tau$ and consider
$W_jH_FW_kh(\cN)$, $j+k=4$.
If $j,k\ge1$ and the $F$ side has at least two cubic terms,
use the second line of \eqref{eq:coupling-word-bounds}
there and the uniform graph bound on the $D$ side.
Before multiplying by $\tau$ the form constant is
$C_I\beta\delta\nu^{d/2+1}$.
If the $F$ side has only one cubic term, the $D$ side has three.
The last two lines of \eqref{eq:coupling-word-bounds}
give instead
\[
	C_I\beta\{\epsilon^3\nu^{(d+6)/2} +\epsilon^2\delta\nu^{(d+8)/2}\}.
\]
Interchanging the two sides gives the bound for the other source pairing.
All these constants, multiplied by $\tau$, are bounded by
$C_I\epsilon^2\nu$, since
\begin{equation}\label{eq:coupling-budgets}
	\frac{\tau\beta\delta}{\epsilon^2}\nu^{d/2}\le C,\qquad \tau\beta\epsilon\nu^{(d+4)/2}\le C,\qquad \tau\beta\delta\nu^{(d+6)/2}\le C.
\end{equation}
For $d=1$ all three are bounded even at $\nu=C\sqrt N$;
for $d=2,\beta=\tau$ they gain at least $N^{-1/4}$.

For an endpoint put $S_T=\delta_TD$ and $F_T=\delta_TF$.
The rectangular commutator gives exactly
\begin{equation}\label{eq:coupling-first-identity}
	[H_F,T]=D^*F_T-F_{T^*}^*D +F^*S_T-S_{T^*}^*F.
\end{equation}
We have $\|F_Tv_n\|\le
C_I\beta\epsilon\nu^{(d+1)/2}\|v_n\|$.
Also \eqref{eq:potential-cubic-source} and
\eqref{eq:coupling-word-bounds} imply, for every direction,
\[
	\|S_TW_3v_n\| \le C_I\{\epsilon^3\nu^{7/2}+\delta\nu^{3/2}\} \|Q_n^{1/2}v_n\|.
\]
Thus the four terms of $[H_F,T]W_3$ have combined form constant
\[
	C_I\beta\{\epsilon^3\nu^{(d+6)/2} +\delta\nu^{(d+2)/2} +\epsilon\delta\nu^{(d+3)/2}\}.
\]
For the first term of \eqref{eq:coupling-first-identity}
use the $L^2$ bound on $W_3$; for the second use its refined
graph bound. For the third use the displayed source-word
bound; for the fourth use
$\|S_{T^*}u\|\le C_I(\epsilon\sqrt\nu\|Q^{1/2}u\|
+\sqrt\nu\|u\|)$ and the $L^2$ bound on $FW_3$.
This accounts also for a raising source in the fourth term.
After multiplication by $\tau$, the first two scaling bounds are
in \eqref{eq:coupling-budgets}; the remaining one is
$\tau\beta\delta\epsilon^{-1}\nu^{(d+1)/2}\le C$.
Finally the identities
$H_FW_4=T_1H_FW_3+[H_F,T_1]W_3$
and its right-endpoint adjoint reduce both endpoints to
the estimated middle positions and their retained commutators.
The same argument applies separately to a non-self-adjoint
component with fixed particle-number shift of $D^*F$ or $F^*D$.

In every term of the actual fourth commutator collect all
number functions at the far right, with their exact particle-number shifts.
Their difference from the matching constant column is
$O(\tau)$ by telescoping first differences. This includes
both $[g,T]$ and $[b,H_F]$ errors. Each multiplies one of
the words just estimated; the common products recombine to
\eqref{eq:coupling-canonical-fourth}.
Only first differences are used at the corners of the
original cutoff. Summing the finite particle-number bands proves
the undifferentiated statement.

For a covariant derivative, replacements of $F$, of the
cubic terms and of their coefficients have the same bounds.
The extra source derivative $E=\nabla_tD$ must also be retained.
As proved above, it is a degree-two bounded source of
fixed-sector norm $C_I\tau$.
Hence $E^*F+F^*E$ has degree at most $d+2$ and fixed-sector
norm $\gamma=C_I\tau\beta$.
Its canonical fourth commutator has degree at most $d+6$
and norm $C_I\gamma\epsilon^4$, which suffices by lifting.
For its actual coefficient errors, at least one of the two
products of cubic terms has length at least two. Use its $\delta\nu$
bound, the other word's uniform $L^2$ bound, and $Q\ge M$
on the remaining unweighted side.
The form constant is $C_I\tau\gamma\delta\nu^{(d+3)/2}$,
and its ratio to $\epsilon^2\nu$ is at most
$C_I\tau^2\beta\delta\epsilon^{-2}\nu^{(d+1)/2}\le C_I$.
This proves the connection estimate.
Number differentiation multiplies each component by its
fixed particle-number shift. The weighted phase bounds pass all three
estimates to the exact integral remainder.
Apply Theorem~\ref{thm:inverse-comparison} to this one summand
using the full exact inverse and the original Taylor evolution.
\end{proof}

\subsubsection{Two concrete source classes}

For the bare cubic interaction in \cite[(5.3)]{BS}, move the
generalized-field square root to the right. Its canonical
part is $D^*F_1$ and its adjoint, with
\begin{equation}\label{eq:bare-cubic-source}
	F_1(x,y)=\sqrt{2N}\sqrt{W_N(x-y)}\,\phi(y)a_x, \qquad \|F_1v_n\|^2\le C_In\|v_n\|^2.
\end{equation}
Indeed $N\|W_N\|_1\le C$ and $\phi$ is uniformly bounded.
The same calculation covers a time derivative of a coefficient and the
smooth finite-rank connection on an active field operator.
Thus $(d,\beta)=(1,1)$ applies with the original right coefficient.

The second class consists of normally ordered quadratic sources multiplied
by $\sqrt{W_N(x-y)}$. At least one field operator contains a kernel
$h=p,\sigma$ or one of its required replacements. Any field at a
source coordinate without such a kernel must be an annihilation operator. Examples are
$\sqrt{W_N}\,a^*(h_y)a_x$,
$\sqrt{W_N}\,a_xa(h_y)$ and sources with both fields smoothed by kernels. When only one field is
smoothed, use the uniform row norm and $\|W_N\|_1\le C\tau^2$. When both
are smoothed, the scalar integral is bounded by
\[
	\int W_N(x-y)\|h_x\|_2^2\|h_y\|_2^2\d x\d y \le \|W_N\|_1\sup_x\|h_x\|_2^2\|h\|_{\mathrm{HS}}^2.
\]
The ordinary creation/annihilation sector bounds therefore
give the class $(2,\tau)$.
The argument works with two different kernels by the same
row and Hilbert--Schmidt bounds.
A bare $a_x^*$ produces a delta output; the source class
therefore includes creators with smoothed output coordinates.
The scalar sources $\sqrt{W_N}h(x,y)$ produced by normal
ordering enter the scattering and scalar cancellations below.

\section{Expansion of the full fluctuation generator}
\label{sec:model-expansion}

We apply the source estimates to the full quadratically transformed
generator. Normalized quadratic chains provide summable bounds for
its conjugation series. We then treat the interaction, time-dependent,
one-body and pair terms, keeping the scalar convention and scattering
cancellations explicit. Their combination verifies the remainder
hypotheses of Theorem~\ref{thm:inverse-comparison} and completes the
stronger comparison in Theorem~\ref{thm:C}. The decomposition and
the estimate for each contribution are collected in
Section~\ref{sec:full-model-assembly}.

\subsection{Normalized quadratic chains}
\label{sec:normalized-source-chains}

For the quadratic conjugation series, we formulate the source
estimate directly in terms of its bandwidth. Let $F_J$ have bandwidth and construction shifts
bounded by $L_J\le c(J+1)$.
On the required sectors assume, for every ordered list
of cubic terms and for $0\le j\le4$,
\begin{equation}\label{eq:chain-source-hypothesis}
	\|\delta_{T_j}\cdots\delta_{T_1}F_Jv_n\| \le a_J\beta\epsilon^j (\nu+L_J)^{(d+j)/2}\|v_n\|, \qquad (d,\beta)=(1,1)\ \hbox{or}\ (2,\tau).
\end{equation}
Require the same estimate for one covariant replacement.
The original right coefficients are retained; their
shifted first differences are bounded by a fixed polynomial
in $J$ times $\tau$, with the original scalar majorant.
Here $d$ denotes the base degree in the estimate;
the additional fields belong to the normalized chains.

\begin{proposition}\label{prop:normalized-source-coupling}
Let $0\le J\le J_*=\lfloor c_0\sqrt N\rfloor$, and let
$F_J$ have number bandwidth and construction shifts bounded by
$L_J\le c(J+1)$. Put $\nu=n+1$, $\tau=N^{-1/2}$ and
$\epsilon=N^{-3/4}$. For every ordered list of canonical cubic
terms and $0\le j\le4$, assume
\[
	\|\delta_{T_j}\cdots\delta_{T_1}F_Jv_n\|
	\le a_J\beta\epsilon^j(\nu+L_J)^{(d+j)/2}\|v_n\|,
	\qquad (d,\beta)=(1,1)\ \text{or}\ (2,\tau).
\]
Require this inequality also when one coefficient or cubic term
is replaced by its covariant derivative. Keep the original right
number coefficients. Their absolute values are bounded by their
scalar majorant, and their shifted first differences are bounded
by that majorant times $C(1+J)^k\tau$, for fixed $C,k$.
These bounds hold on the particle-number range enlarged by $L_J$.
For $D=D_V$ or $D=D_K$, the actual fourth commutator of
$D^*F_J+F_J^*D$, including the specified right coefficients,
and its covariant and number derivatives satisfy
\begin{equation}\label{eq:normalized-source-fourth}
	|B_J[u,v]|\le C_Ia_J(1+J)^b\epsilon^2 \|MQ^{1/2}E_Ju\|\|Q^{1/2}E_Jv\|.
\end{equation}
Here $b$ is fixed, and $E_J$ enlarges the original particle-number cutoff support by $O(J+1)$. The same conclusion holds for the
exact integral remainder of the conjugation expansion. In particular, if $a_J$
has all polynomial moments summable, these contributions
sum to the $N^{-5/4}$ comparison bound.
\end{proposition}
\begin{proof}
For $D_V$ repeat the source Leibniz formula without
extracting kernels of $F_J$.
Pair $P_j$ directly with $\delta_C^{4-j}F_J$.
Their sector norm product is at most
$C_Ia_J\beta\epsilon^3(\nu+L_J)^{d/2+3}$.
Using $Q\ge M$ on the two sides and
$\nu+L_J\le(1+L_J)\nu$ gives the first constant in
\eqref{eq:coupling-canonical-fourth}, times a fixed
polynomial in $J$.
The $A_jD$ terms and the endpoint are smaller.
The coefficient-error proof uses only the $j=0,1$
instances of \eqref{eq:chain-source-hypothesis},
the word estimates \eqref{eq:coupling-word-bounds}
and the exact identity \eqref{eq:coupling-first-identity}.
It therefore has the same scaling bounds with polynomial
bandwidth losses. Summation over the allowed particle-number bands introduces only another
fixed polynomial factor.

For $D_K$, Proposition~\ref{prop:kinetic-source-decomposition}
gives the stronger bound
\[
	\|\delta^jD_Kv_n\| \le C_I\{\epsilon^{j-1}\nu^{(j+1)/2}\|v_n\| +\epsilon^j\nu^{j/2}\|K_P^{1/2}v_n\|\}.
\]
Its first source has the same two-term bound used in
the preceding endpoint proof. Thus all canonical and
coefficient-error estimates remain valid with $Q\ge K_P+M$.
The additional connection
$E_K=\nabla_tD_K$ is a bounded linear source of
fixed-sector norm $C_I$, from the smooth finite-rank kernel
$\nabla Z$. Its product with $F_J$ has base degree $d+1$
and norm constant $C_Ia_J\beta$.
Distribute four commutators between these two sources.
The canonical bound follows directly from
\eqref{eq:chain-source-hypothesis}.
For a coefficient-error word place the $\delta\nu$
estimate on a side with at least two cubic terms, exactly
as in the last paragraph of the preceding proof.
After division by $\epsilon^2\nu$ the scaling bound is now
$\tau\beta\delta\epsilon^{-2}\nu^{d/2}\le C$,
the first inequality in \eqref{eq:coupling-budgets}.
This also proves the derivative assertion.
Number derivatives and exact remainders follow as before.
Finally $\operatorname{supp}E_J$ lies below
$\Lambda+C(J+1)$. Theorem~\ref{thm:inverse-comparison}
absorbs the additional fixed polynomial and sums the terms.
\end{proof}

\subsubsection{Verifying the chain hypothesis}

Let $\mathcal Q(h)$ be an ordinary quadratic field
operator, allowing a scalar contact only when its trace
is controlled. Let $\alpha_h$ bound its fixed-sector norm.
For a normal pair or one-body kernel one may use its
Hilbert--Schmidt norm; in reverse normal order include
$\int|h(x,x)|\d x$ as well.
For $s\le4$, a mixed $s$-fold commutator has degree at most
$s+2$ and fixed-sector norm $C_{I,s}\alpha_h\epsilon^s$.
The finite-degree lifting lemma, applied at degree at most
six, therefore gives
\begin{equation}\label{eq:normalized-quadratic-factor}
	\|\delta^s(N^{-1}\mathcal Q(h))v_n\| \le C_{I,s}\alpha_h\epsilon^s \frac{\nu}{N}\nu^{s/2}\|v_n\|, \qquad 0\le s\le4.
\end{equation}
This follows from the finite ordered-word expansion on
fixed sectors.
For the kernel products $\eta_\natural^{(m)}$ in
\cite[Lemma 3.2]{BS}, one has
$\alpha_h\le C\|\eta\|_{\mathrm{HS}}^m$.
In reverse normal order that lemma requires $m\ge2$;
in this case the needed diagonal integral follows from
\[
	\int|\eta_\natural^{(m)}(x,x)|\d x \le \|\eta\|_{\mathrm{HS}}\|\eta_\natural^{(m-1)}\|_{\mathrm{HS}} \le \|\eta\|_{\mathrm{HS}}^m.
\]
This trace estimate uses $m\ge2$.
A time or projection replacement expands in at most $m$
positions and loses at most one small kernel factor.

For a bounded source $F_0$, insert
at most $c(J+1)$ factors
$R_i=N^{-1}\mathcal Q(h_i)$ on its input Fock space or the Fock component
of its output, in any fixed prescribed order.
On $J\le J_*$, every intermediate particle number is
at most $C\sqrt N$.
For sufficiently large $N$ these sectors stay below the
physical upper boundary $N$, so the ordinary CCR expansion
is valid there. The finitely many smaller $N$ are absorbed
in a constant using the original fixed-$N$ bounds.

Apply at most four commutators by the Leibniz rule.
There are only polynomially many assignments, at most
$(cJ+c)^j$ for $j$ commutators.
Factors on which no commutator acts retain their uniform bound
$c_i\alpha_{h_i}$ on the enlarged particle-number cutoff range. At
most four factors receive a commutator. Only these factors carry the
additional number weights from \eqref{eq:normalized-quadratic-factor}.
The same estimate holds on the Fock component of the output space.
Thus each such chain satisfies
\begin{equation}\label{eq:normalized-chain-leibniz}
	\|\delta^j(R_{1,\rm out}\cdots F_0\cdots R_p)v_n\| \le C_{I,j}(1+J)^{b_j} \Big(\prod_i c_i\alpha_{h_i}\Big) \beta\epsilon^j(\nu+L_J)^{(d+j)/2}\|v_n\|.
\end{equation}
The unchanged factors retain the constants of
\cite[Lemma 2.4]{BS}. At most four factors acquire new constants.
Their contribution multiplies the geometric majorant
by a fixed polynomial. Thus the resulting bound has the form
$C(1+J)^b\theta^{\max(J-b_0,0)}$, $\theta<1$.
The geometric ratio $\theta$ and the kernel ball are unchanged.

\subsubsection{Ordered endpoints and scalar contacts}

To recover the operators in \cite[(2.10)--(2.12)]{BS},
move the endpoint generalized-field square roots
to the right with their particle-number shifts.
The remaining ordered fields in each integrated pair
form one ordinary $\mathcal Q(h_i)$.
Each kernel $h_i$ couples only the two field operators in
its pair, as in the proof of \cite[Lemma 2.4]{BS}.
Hence the explicit $N^{-p}$ supplies one $N^{-1}$ per
quadratic factor. A $\Pi^{(1)}$ operator has one further
terminal linear field; its adjoint has that terminal on
the other side.

On the enlarged low-$J$ range, the normalized number factors
and endpoint square roots are uniformly bounded. Telescoping the differences across the $O(J+1)$ factors and
shifts gives at most a polynomial in $J$ times
$N^{-1}$ relative to the original majorant. Positive shifts contribute factors at most
$1+C(J+1)/N$, whose product is bounded by
$\exp(C(J+1)^2/N)$. This bound is uniform for
$J\le c_0\sqrt N$, where the square-root arguments also stay
away from zero. Thus the original number functions satisfy the
coefficient bounds of Proposition~\ref{prop:normalized-source-coupling}
without being removed from the expression.

The source-coordinate estimates involve two types of endpoints.
The first is $\sqrt N\sqrt{W_N(x-y)}\phi(y)$ times either
$a_x$ or a smoothed linear field. The second is
$\sqrt{W_N(x-y)}$ times a normally ordered quadratic expression.
In this expression, every creation operator is smoothed and
there is at most one bare annihilator.
For a smoothed linear field at fixed $x$ its $s$-fold
commutator has fixed-sector norm
$C_{I,s}\epsilon^s\|h_x\|_2$ and degree at most $s+1$.
For $a_x$ the corresponding statement is for the
$L_x^2$-valued source and has constant $C_{I,s}\epsilon^s$:
first use the actual canonical kernel's $L^2$ source
bound and then its ordered commutators.
The finite-degree lemma is used only on these local
factors of degree at most five.

Quadratic normalized factors may lie between the endpoints.
When a bare annihilator is on the left, use the exact identity
\begin{equation}\label{eq:chain-bare-reordering}
	a_xR_1\cdots R_p =R_1\cdots R_pa_x+ \sum_{i=1}^pR_1\cdots R_{i-1}[a_x,R_i] R_{i+1}\cdots R_p.
\end{equation}
Each new commutator is a smoothed linear field with its
retained $N^{-1}$. There are $p+1$ terms.
When the bare annihilator is at the right, apply the
fixed-coordinate bound on the smoothed field first, then
integrate using
$\sup_y\int W_N(x-y)\|h_x\|_2^2\d x\le C_I\tau^2$.
Two smoothed endpoints use the row and Hilbert--Schmidt
estimate following \eqref{eq:bare-cubic-source}.
Insert \eqref{eq:normalized-quadratic-factor} on the
intermediate factors and distribute at most four
commutators. These calculations prove
\eqref{eq:chain-source-hypothesis} for the stated structures,
including their derivatives, with only polynomial
losses in $J$.

If a bare annihilator passes a smoothed creator, keep the
scalar contact $h_y(x)$ and its original normalization.
One useful separate case is a scalar source $s_N$ with
$\|s_N\|_2\le C_I\sqrt N$ and a retained normalized factor:
\begin{equation}\label{eq:normalized-large-contact}
	F_{\rm contact}(x,y)=s_N(x,y)N^{-1}\mathcal Q(h).
\end{equation}
Equation \eqref{eq:normalized-quadratic-factor} shows that
this is in the $(d,\beta)=(2,\tau)$ class with the additional
factor $\alpha_h$, including four commutators.
Further normalized factors are handled by
\eqref{eq:normalized-chain-leibniz}.
For example $\sqrt{W_N}\eta(x,y)$ and
$N\sqrt{W_N}\phi(x)\phi(y)$ have source norm at most
$C_I\sqrt N$: use $\|W_N\|_1=O(N^{-1})$,
$|\eta(x,y)|\le C_IN|\phi(x)\phi(y)|$ on the potential
support, and the projected smooth corrections.
The same statement holds for their first replacements.
Terms with an explicit normalized factor obey this source
bound. The remaining scalar and pair terms enter the
scattering cancellations below.

\subsubsection{The exact high-degree tail}

Consider a BS group after the required linear, scattering and scalar
cancellations. Assume that its order-$J$ summands
satisfy the geometric $Q$ form bound below, together
with their first covariant and number derivatives:
\[
	C_I(1+J)^b\theta^{\max(J-b_0,0)} \|Q^{1/2}u\|\|Q^{1/2}v\|, \qquad 0<\theta<1.
\]
On the tail $J>J_*$ the original phase and its first
derivative are uniformly bounded on the $Q$ graph.
Every fourfold ordered word and every exact phase
conjugation therefore has the same majorant up to a
fixed constant. As in Lemma~\ref{lem:exact-degree-tail},
the sum is, for every fixed $K>0$,
\begin{equation}\label{eq:chain-exact-tail}
	|B_{>J_*}[u,v]|\le C_{I,K}N^{-K} \|Q^{1/2}u\|\|Q^{1/2}v\|, \quad B=R,\nabla_tR,i[M,R].
\end{equation}
This estimate is on the exact tail of the original
generator.
The inverse-energy argument in
Theorem~\ref{thm:inverse-comparison} gives an
$N^{-K}\|F_3(s)^{3/2}z\|$ time integral without a number
support assumption for this tail.
We next verify the geometric premise for each combined
generator group using these source and tail estimates.

\subsection{The transformed quartic interaction}
\label{sec:BS-quartic}

Let $B=B(\eta)$ be the original generalized quadratic
phase, $k$ its unprojected correlation kernel,
$\eta=(q\otimes q)k$, and $\mu=\eta-k$.
Keep the scalar and pair convention of \cite[(5.115)]{BS}:
\begin{align}
	E^{(4)}_{\rm BS}
	&=e^{-B}V_Pe^B-V_P-\sigma_4 I-P_k,\notag\\
	\sigma_4&=\frac12\int W_N(x-y)|k(x,y)|^2\d x\d y,\qquad
	P_k=\frac12\int W_N(x-y)
	\{k(x,y)b_x^*b_y^*+\mathrm{h.c.}\}\d x\d y .
	\label{eq:BS-quartic-definition}
\end{align}
All equalities below are on the compressed common form domain.

\begin{lemma}
\label{lem:bounded-chain-products}
Suppose the canonical part $S_J$, with its right number
coefficients collected separately, satisfies
\begin{equation}\label{eq:bounded-product-hypothesis}
	\|\delta_{T_j}\cdots\delta_{T_1}S_Jv_n\| \le a_J\gamma\epsilon^j(\nu+L_J)^{(q+j)/2}\|v_n\|, \quad j\le4,\qquad (q,\gamma)=(2,1),(3,\tau),(4,\tau^2).
\end{equation}
Require \eqref{eq:bounded-product-hypothesis} also when one
coefficient or cubic term is replaced by its covariant derivative.
Here $\nu=n+1$, $\tau=N^{-1/2}$ and $\epsilon=N^{-3/4}$.
Assume $0\le J\le\lfloor c_0\sqrt N\rfloor$ and that both
the number bandwidth and the construction shifts are at most
$L_J\le c(J+1)$. Keep the right number coefficients on this
enlarged range. Their absolute values are bounded by their scalar
majorant, and their shifted first differences by that majorant
times $C(1+J)^k\tau$, for fixed $C,k$.
Then its actual fourth commutator, its two required
derivatives and its exact conjugation remainder obey
\eqref{eq:normalized-source-fourth}, with a fixed
additional polynomial in $J$.
\end{lemma}
\begin{proof}
The $j=4$ bound and $Q\ge M$ on both sides give the
canonical form constant
$C_Ia_J\gamma\epsilon^4\nu^{q/2+1}$, up to bandwidth
polynomials. Each actual coefficient error is an
$O(\tau)$ right coefficient times $W_lS_JW_r$,
where $l+r=4$. A side with at least two cubic terms uses
the $\delta\nu$ bound in \eqref{eq:coupling-word-bounds}.
The other side uses its uniform unweighted norm and
$\|u_m\|\le(m+1)^{-1/2}\|Q_m^{1/2}u_m\|$.
The resulting form constant is
$C_Ia_J\tau\gamma\delta\nu^{(q+1)/2}$.
Relative to $\epsilon^2\nu$, the two scaling bounds are
\[
	\gamma\epsilon^2\nu^{q/2}\le C,\qquad \tau\gamma\delta\epsilon^{-2}\nu^{(q-1)/2}\le C.
\]
They hold in all three listed cases.
The same ordered estimates include replacements, and
number-gauge differentiation adds only a bandwidth
factor. Weighted phase conjugation gives the remainder.
\end{proof}

\subsubsection{Exact source decomposition}

Write $a^*(\eta_x)=\int\eta(x,z)a_z^*\d z$ and define
\[
	M^0_{xy}=a^*(\eta_y)a_x+a^*(\eta_x)a_y,\qquad P_s(x,y)=e^{-sB}b_x^*b_y^*e^{sB},\qquad H_r(x,y)=e^{-rB}[M^0_{xy},B]e^{rB}.
\]
The ordinary CCR, with all field projections retained, give the exact integral form of \cite[(5.117)]{BS}:
\begin{equation}\label{eq:BS-quartic-integral}
	\begin{split}
	e^{-B}V_Pe^B-V_P
	&=\frac12\int W_N(x-y)\int_0^1 P_s(x,y)\\
	&\quad\times
	\left\{\eta(x,y)+M^0_{xy}
	+\int_0^sH_r(x,y)\d r\right\}
	\d s\d x\d y+\mathrm{h.c.}
	\end{split}
\end{equation}
Here the scalar contact follows from
$a_xa^*(\eta_y)=\eta(x,y)+a^*(\eta_y)a_x$.
With $\eta^{(2)}=\eta\bar\eta$, the inner commutator is
\[
	[M^0_{xy},B] =2b^*(\eta_x)b^*(\eta_y) +b(\eta^{(2)}_y)b_x+b(\eta^{(2)}_x)b_y,
\]
where the annihilation kernels follow the alternating
conjugation convention of \cite[(3.8)]{BS}.
This can also be checked by applying the mixed-field
commutator to each of the two terms of $M^0_{xy}$.
In particular the composed kernel for a complex symmetric
$\eta$ is $\eta\bar\eta$.
Expanding the exact integrals gives the four oriented
families in \cite[(5.118)]{BS}, each with its adjoint.
The one-integral coefficient is $1/(n+k+1)$; the
double-integral coefficient is
$1/[(m+r+1)(n+k+m+r+2)]$, in addition to the original
factorials and commutator signs.

Put $A_j(x)=\delta_B^j b_x$, with $\delta_BX=[X,B]$;
using the opposite BS commutator convention changes only
the displayed series signs.
For each monomial in \cite[Lemma 3.2]{BS}, the source
$\sqrt{W_N(x-y)}A_n(y)A_k(x)$ decomposes into
\begin{equation}\label{eq:BS-pair-source-classes}
	\mathcal U_{\rm out}D_V,\qquad F_2,\qquad s_\eta(x,y)\mathcal U,\qquad s_\eta=\sqrt{W_N}\eta .
\end{equation}
Numerical normalizations such as $\sqrt2$ are included
in the coefficients. The symbols $\mathcal U$ are the
original ordered normalized quadratic words, including
scalar contacts with their $N^{-1}$ factors.
The source $F_2$ satisfies
\eqref{eq:chain-source-hypothesis} with $(d,\beta)=(2,\tau)$.
The last source has $\|s_\eta\|_2\le C_I\sqrt N$.
Each term retains its right number coefficients,
conjugation and coordinate order.

Part (v) of that lemma implies that an unsmoothed
terminal field is always an annihilator.
If both terminals are unsmoothed, move the intervening
normalized factors to the output Fock space using
\eqref{eq:chain-bare-reordering}.
The zero-commutator term is $\mathcal U_{\rm out}D_V$;
every other term has at least one field smoothed by a kernel and is of
type $F_2$.
If $n+k>0$, this zero-commutator term retains at least
one normalized quadratic factor. Indeed the kernel-degree
identity in \cite[Lemma 3.2(iii)]{BS} excludes a term
with neither a terminal kernel nor an internal kernel
at positive total degree. Contraction of an internal
factor preserves its normalization.

If a terminal is smoothed, its normal quadratic part
is covered by Section~\ref{sec:normalized-source-chains}.
A scalar terminal contact is
$\sqrt{W_N}\eta_\natural^{(\ell)}(x,y)$, possibly with
extra $N^{-1}$. For $\ell\ge2$, the row, column and
Hilbert--Schmidt bounds give
\[
	\|\sqrt{W_N}\eta_\natural^{(\ell)}\|_2 \le C_I\tau\|\eta\|_{\mathrm{HS}}^{\max(\ell-2,0)}.
\]
One can see the integrability directly by retaining the
$L_x^2$ norm of the first row factor and the uniform
bound on the last one when integrating $W_N(x-y)$.
Such a scalar is allowed as a degree-zero part of $F_2$.
At $\ell=1$ retain $s_\eta\mathcal U$ unless a factor
$N^{-1}$ permits \eqref{eq:normalized-large-contact}.
Finally reverse-normal global quadratic factors have
at least two kernels by \cite[Lemma 3.2(vi)]{BS}, so
their traces have precisely the bounds already proved.
Only the two explicit terminal field operators are reordered.
At most polynomially many source corrections occur;
normalized factors between smoothed terminals can stay
in their original order.
Leibniz distribution of four cubic commutators and one
time replacement therefore adds only fixed polynomial
losses to the original geometric majorant.

\begin{theorem}\label{thm:BS-quartic-remainder}
The exact fourth-order conjugation remainder of
$E^{(4)}_{\rm BS}$ satisfies the low-degree static
hypothesis \eqref{eq:one-weight-static}, with all
polynomial moments of its coefficients summable, and
the exact high-degree hypothesis
\eqref{eq:geometric-tail-static}.
The assertions include its first covariant and number
derivatives. Its individual time-integrated contribution
is bounded by
$C_{I,s}N^{-5/4}\|F_3(s)^{3/2}z\|$.
\end{theorem}
\begin{proof}
The two terms in $M^0_{xy}$ are $\tau$-small normal
quadratic sources after multiplication by $\sqrt{W_N}$.
The two-smoothed-creator term of $H_r$ and its
$\eta^{(2)}$-smoothed terms have the same property after
their generalized transformation is expanded.
They always retain a smoothed terminal, and any normal
contact contains at least two kernel factors.
Pair them with the three types in
\eqref{eq:BS-pair-source-classes}.
The $D_V$ type yields $D_V^*\mathcal U^*F_2$ and its
adjoint, covered by Proposition~\ref{prop:normalized-source-coupling}.
The two $F_2$ types yield the base class $(q,\gamma)=(4,\tau^2)$.
Pairing $F_2$ with $s_\eta\mathcal U$ yields $(2,1)$
because $\sqrt N\tau=1$.
Both are covered by Lemma~\ref{lem:bounded-chain-products}.
The pairings are inner products in the source space.
This accounts for the second, third and fourth oriented
families and all their lower-field terms.

The first family pairs the outer source with $s_\eta$.
Its $n=k=0$ bare term is $P_\eta$.
Subtracting the original $P_k$ in
\eqref{eq:BS-quartic-definition} leaves $P_\mu$.
The finite-rank projection formula for $\mu$, whose
profiles have bounded $L^2$ and $L^\infty$ norms, gives
\begin{equation}\label{eq:BS-small-projection-source}
	\|\sqrt{W_N}\mu\|_2\le C_I\tau .
\end{equation}
Thus $P_\mu$ is a small bounded-source coupling.
For positive degree, the bare $D_V$ type retains a
normalized quadratic factor. Combine its first factor
with $s_\eta$ as in
\eqref{eq:normalized-large-contact}, and use the
normalized-source proposition.
Pairing $F_2$ directly with $s_\eta$ again gives $(2,1)$.

It remains to pair two large scalar sources.
Their scalar product is at most the original degree
majorant times $C_IN$.
If there is a normalized factor, combine $N$ with its
$N^{-1}$, leaving a bounded quadratic base operator;
other factors remain normalized. This is again $(2,1)$,
also when that quadratic factor has become a scalar trace.
If no normalized factor remains, the term is
$c_J(t)Ng_J(\cN)$.
On the low-degree range its smooth BS number function
has $|g_J'|\le C_J/N$, $|g_J''|\le C_J/N^2$.
The constants have only polynomial degree losses:
products of positive shifted factors are bounded by
$\exp(C(J+1)^2/N)$ as above.
Use the exact identity
\begin{equation}\label{eq:BS-large-number-scalar}
	c_JNg_J(\cN)=c_JNg_J(0)I+c_J\cN h_J(\cN), \qquad h_J(n)=N\int_0^1 g_J'(sn)\d s .
\end{equation}
The coefficient $h_J$ is bounded and has first differences
$C_J/N$. The second term is the $(2,1)$ one-body base
class with a right number coefficient.
The first is a scalar identity, whose fourth commutator
and derivative of that commutator are exactly zero.
This argument uses a second derivative of a smooth BS
number function; the original piecewise linear $\Theta$
is still used only through its first differences.
The original subtraction $\sigma_4I$ stays unchanged.

The kernel estimates remain valid under ordinary time
differentiation. Connections on the remaining field coordinates
are covered by the bounded finite-rank terms in the source
classes, and the removed-coordinate contribution $\nabla_tD_V$
was included explicitly in
Proposition~\ref{prop:bounded-source-coupling}.
Number differentiation adds at most $O(J+1)$. These polynomial
losses are absorbed by the original factorials and kernel-degree
count within the small kernel range of
\cite[Lemmas 2.4, 3.2 and (5.118)--(5.126)]{BS}, leaving
coefficients with every fixed polynomial moment summable.
For the exact high-degree tail use the geometric $Q$ form
estimates for this already subtracted BS group and
\eqref{eq:chain-exact-tail}.
The bare pair and scalar terms remain the explicit extractions
in \eqref{eq:BS-quartic-definition}.
Finally apply Theorem~\ref{thm:inverse-comparison}.
The extracted $P_k$ will still require the original
scattering combination with the kinetic and pair groups.
\end{proof}

\subsection{The quadratic time term}
\label{sec:BS-quadratic-time}

The time term of the quadratic transformation is separate
from the cubic time remainder in
Proposition~\ref{prop:time-remainder-sharp}.
In the chosen parallel trivialization the compressed
logarithmic derivative is
\begin{equation}\label{eq:BS-time-integral}
	L_B=-i\int_0^1e^{-sB}\dot Be^{sB}\d s,\qquad \dot B=\frac12\int \{\dot\eta(x,y)b_x^*b_y^* -\overline{\dot\eta(x,y)}b_xb_y\}\d x\d y .
\end{equation}
Here the kernel derivative includes the finite-rank
corrections in its covariant version.
The compressed identity was justified in
Section~\ref{sec:second-regularity}.
The prescribed background regularity gives uniform
Hilbert--Schmidt bounds on the first two required kernel
derivatives.

Writing $\delta_BX=[X,B]$, the homogeneous term with
$J$ undifferentiated $\eta$ kernels is exactly
\begin{equation}\label{eq:BS-time-homogeneous}
	\begin{split}
	L_{B,J}
	&=-\frac{i}{(J+1)!}\delta_B^J\dot B\\
	&=-\frac i2\sum_{n+k=J}
	\frac1{n!k!(J+1)}
	\int\dot\eta(x,y)
	\delta_B^n(b_x^*)\delta_B^k(b_y^*)\d x\d y
	+\mathrm{h.c.}
	\end{split}
\end{equation}
This is the adjoint-complete version of
\cite[(5.128)--(5.129)]{BS}. The basic Hilbert--Schmidt
kernel $h=\dot\eta$ carries its own HS bound alongside the
$J$ small kernel factors.

\begin{lemma}
\label{lem:HS-pair-chains}
Let $h$ be a symmetric Hilbert--Schmidt pair kernel with
a uniformly bounded first covariant derivative.
The low-degree expansion of $e^{-sB}P(h)e^{sB}$,
$0\le s\le1$, where
\[
	P(h)=\frac12\int\{h(x,y)b_x^*b_y^* +\mathrm{h.c.}\}\d x\d y,
\]
has the base block bounds
\eqref{eq:bounded-product-hypothesis} with $(q,\gamma)=(2,1)$,
uniformly in $s$, after its right number coefficients are
collected separately.
The coefficients have all fixed polynomial moments
summable. The assertion holds also for the anti-self-adjoint
pair combination multiplied by $-i$, and for integrals
in $s$ with bounded weights.
\end{lemma}
\begin{proof}
Apply \cite[Lemma 3.2]{BS} to the two explicit endpoints
and keep every normalization and coefficient.
The remaining structure consists of normalized factors
$R_i=N^{-1}\mathcal Q(h_i)$ and two terminal fields
linked by $h$.
Move only the first terminal past the factors between
it and the second terminal. At each step
\begin{equation}\label{eq:HS-terminal-reordering}
	a^\#(f)R_i=R_i a^\#(f)+N^{-1}\mathcal L_i(f),
\end{equation}
where $\mathcal L_i(f)$ is a sum of at most two linear
fields, with kernels obtained by applying $h_i$, its
transpose or its adjoint. Their norms are bounded by
the same $\alpha_{h_i}\|f\|$.
A scalar contact in $R_i$ has zero commutator.
The global normalized factors retain their original order.
At the end one has a normalized word multiplying an
ordinary quadratic kernel $h_{\rm eff}$, or a scalar trace.
The effective kernel has Hilbert--Schmidt norm bounded
by $\|h\|_{\mathrm{HS}}$ times the original internal kernel norms.
A scalar terminal contact contains at least one additional
$\eta$ kernel, since the two original unsmoothed pair
endpoints have the same direction and cannot contract.
Its trace is bounded by the product of two
Hilbert--Schmidt norms.

Repeated commutation of terminal fields produces a sum of terms.
Estimate this sum with the normalization factors retained.
On the low-degree range, a term retaining the quadratic factor is bounded
by $C\alpha_{h_i}(\nu+O(J+1))/N$. A term from the commutator is bounded by
$C\alpha_{h_i}/N$ and contains one additional kernel composition in
$h_{\rm eff}$.
Their sum at each factor is bounded by
\begin{equation}\label{eq:HS-branch-budget}
	C\alpha_{h_i}\frac{\nu+O(J+1)}{N} \le C'\alpha_{h_i}N^{-1/2}.
\end{equation}
For $N\ge N_0$ this is at most $\alpha_{h_i}$.
Distributing at most four cubic commutators affects at most four factors. Equation
\eqref{eq:normalized-quadratic-factor} gives their extra
$\epsilon^s(\nu+L_J)^{s/2}$ factors; their constants and the number of possible positions contribute only a
fixed polynomial in $J$.
For the other factors use the sum
\eqref{eq:HS-branch-budget}.
The original geometric majorant and small-kernel range are preserved.
The finite set $N<N_0$, $J\le J_*(N)$ is absorbed in the
constant using the original finite-$N$ bounds.

The remaining ordinary quadratic operator has degree two,
and its first four canonical commutators have fixed-sector
norms $C_{I,j}\epsilon^j\|h_{\rm eff}\|_{\mathrm{HS}}$.
Scalar traces satisfy the weaker degree-two bound as well.
Combine this with the normalized-factor estimates and
the original factorial count. This proves
\eqref{eq:bounded-product-hypothesis} with $(2,1)$.
The right number functions have the same polynomial
shift bounds as before; they are treated as coefficients
in Lemma~\ref{lem:bounded-chain-products}.

Under covariant differentiation, the derivative falls either
on $h$ or on one of the $J$ kernel factors. The latter
contribution has $J$ possible positions and loses only one
small kernel factor, while preserving the row, Hilbert--Schmidt
and summed contraction estimates. The same kernel classes already include
the finite-rank connection terms. Hence differentiation keeps
the geometric ratio unchanged and introduces only a fixed
polynomial loss. Finally, the weights $s^J$ and their integrals
have magnitude at most one, giving the remaining assertions.
\end{proof}

\begin{theorem}\label{thm:BS-quadratic-time-remainder}
The exact fourth-order conjugation remainder of $L_B$, including its
covariant and number derivatives, satisfies
\eqref{eq:one-weight-static} on low degrees and
\eqref{eq:geometric-tail-static} on the exact tail.
Its individual contribution to the original comparison
is at most $C_{I,s}N^{-5/4}\|F_3(s)^{3/2}z\|$.
\end{theorem}
\begin{proof}
Apply Lemma~\ref{lem:HS-pair-chains} with $h=\dot\eta$
to \eqref{eq:BS-time-homogeneous}, then
Lemma~\ref{lem:bounded-chain-products}.
A time derivative uses $\ddot\eta$ for the basic kernel
or $\dot\eta$ at one of the $J$ other positions, exactly
as accounted for in that lemma.
The low-degree canonical and coefficient-error form
constants are respectively
$a_J\epsilon^4\nu^2$ and
$a_J\tau\delta\nu^{3/2}$, both at most
$C_Ia_J\epsilon^2\nu$.
Number differentiation adds a fixed polynomial bandwidth
factor. The geometric number-form bounds established
in \cite[(5.127)--(5.129)]{BS} control the high-degree
tail by \eqref{eq:chain-exact-tail}.
Exact phase conjugation and
Theorem~\ref{thm:inverse-comparison} now give the claim,
using the same full exact inverse and the same Taylor
trajectory. The cubic-transform time term remains the
separate summand proved earlier.
\end{proof}

\subsection{Number and one-body terms}
\label{sec:BS-regular-groups}

For a bounded one-body operator $h$ put
$T_h=e^{-B}\dGamma(h)e^B$ and $T_n=e^{-B}\cN e^B$.
For self-adjoint $h$, number preservation implies
\begin{equation}\label{eq:BS-onebody-conjugation}
	T_h=\dGamma(h)+\int_0^1e^{-sB}P(k_h)e^{sB}\d s,\qquad k_h=h\eta+\eta h^{\mathsf T}.
\end{equation}
Indeed $\dGamma(h)$ commutes with the generalized-field
number square roots; the CCR then give
$[\dGamma(h),B]=P(k_h)$.
The transpose is the kernel transpose, with the original
complex conjugations retained. In particular
$\|k_h\|_{\mathrm{HS}}\le2\|h\|_{\mathrm{op}}\|\eta\|_{\mathrm{HS}}$.
For a non-self-adjoint $h$ split it into its self-adjoint
real and imaginary parts.

Lemma~\ref{lem:HS-pair-chains} applies to the integral.
The bare $\dGamma(h)$ has fixed-sector norm
$C\|h\|_{\mathrm{op}}$ and its first four cubic commutators
are controlled by finite-degree extraction.
Thus $T_h$ has the base $(2,1)$ bounds.
The constant depends on the operator norm of $h$.
If $\nabla_t h$ is bounded, the differentiated pair kernel
is the sum
\[
	(\nabla_t h)\eta+h\nabla_t\eta +(\nabla_t\eta)h^{\mathsf T} +\eta(\nabla_t h)^{\mathsf T};
\]
all four terms have uniform Hilbert--Schmidt bounds.
Taking $h=I$ gives the same assertion for $T_n$.

For two such operators, retain the displayed normalization
in $N^{-1}T_hT_g$.
On low total degree $J$, the ordered Leibniz rule gives
\begin{equation}\label{eq:BS-onebody-product}
	\|\delta^j(N^{-1}(T_hT_g)_J)v_n\| \le a_JN^{-1}\epsilon^j (\nu+L_J)^{(4+j)/2}\|v_n\|,\qquad j\le4.
\end{equation}
The two geometric series convolve with only a polynomial
loss in $J$. Coefficients retain their intermediate
particle-number shifts. Since $N^{-1}=\tau^2$, this is the
$(4,\tau^2)$ case of Lemma~\ref{lem:bounded-chain-products},
including its derivative replacements.
For the high-degree tail, the original weighted
number estimates give the form bound
$C_Ia_JN^{-1}\|Mu\|\|Mv\|$.
On the physical space $\cN\le N$,
$N^{-1}M^2\le C M\le C Q$, so this too is a geometric
$Q$ form bound. This step is needed before applying
\eqref{eq:chain-exact-tail} to the product.

We now identify the original groups.
With the fixed scattering radius $\ell$ of $B$, set
\[
	\begin{split}
	\alpha_N&=\tfrac12\langle\phi,
	[N^3v(N\cdot)(1-2f_\ell(N\cdot))*|\phi|^2]\phi\rangle,\\
	\beta_N&=\tfrac12\langle\phi,
	[N^3v(N\cdot)*|\phi|^2]\phi\rangle .
	\end{split}
\]
These scalars and their first time derivatives are
uniformly bounded by the $L^1$ scaling of $N^3v(N\cdot)$,
$0\le f_\ell\le1$, and the background bounds.
The original expression in \cite[(5.3)]{BS} gives exactly
\begin{equation}\label{eq:BS-number-polynomial}
	L^{(0)}=(N\alpha_N-\beta_N)I +(-\alpha_N-\beta_N+\beta_N/N)\cN +(\beta_N/N)\cN^2 .
\end{equation}
Consequently its transformed non-scalar part is
\begin{equation}\label{eq:BS-T0}
	T_0=(-\alpha_N-\beta_N+\beta_N/N)T_n +(\beta_N/N)T_n^2 .
\end{equation}
The original scalar convention is unchanged.
Its fourth cubic commutator and its derivative vanish.

For the two regular one-body terms in $L^{(2)}$ set
\[
	U_N=N^3v(N\cdot)*|\phi|^2,\qquad J_N(x,y)=N^3v(N(x-y))\phi(x)\overline{\phi(y)},\qquad h_N=q(U_N+J_N)q .
\]
Multiplication by $U_N$ and the convolution representation
of $J_N$ give $\|h_N\|_{\mathrm{op}}\le C_I$.
Time replacement and the finite-rank derivative of $q$
give $\|\nabla_t h_N\|_{\mathrm{op}}\le C_I$.
The exact generalized-field ordering is
$b_x^*b_y=a_x^*a_y(N-\cN+1)/N$.
Its $+N^{-1}$ shift combines with the explicit
$-N^{-1}a_x^*a_y$ in the original expression. Hence
\begin{equation}\label{eq:BS-T2}
	L_{\rm reg}^{(2)}=\dGamma(h_N)(1-\cN/N),\qquad T_2=e^{-B}L_{\rm reg}^{(2)}e^B =T_{h_N}-N^{-1}T_{h_N}T_n .
\end{equation}
The factors on the right commute because their unconjugated
operators commute.

\begin{proposition}\label{prop:BS-regular-remainder}
The fourth-order conjugation remainders of $T_0$ and $T_2$,
and their first covariant and number derivatives,
satisfy \eqref{eq:one-weight-static} and
\eqref{eq:geometric-tail-static}. Each has the
$N^{-5/4}$ time-integral bound of
Theorem~\ref{thm:inverse-comparison}.
\end{proposition}
\begin{proof}
Apply \eqref{eq:BS-onebody-conjugation} and
\eqref{eq:BS-onebody-product} to
\eqref{eq:BS-T0}--\eqref{eq:BS-T2}, followed by
Lemma~\ref{lem:bounded-chain-products}.
The low-degree coefficients have all polynomial moments
summable, and the preceding weighted number argument
gives the exact high-degree premise.
All derivatives were estimated through their explicit
kernel and scalar replacements.
The weighted phase bounds and
Theorem~\ref{thm:inverse-comparison} finish the proof.
\end{proof}

\subsection{Linear and cubic terms}
\label{sec:BS-linear-cubic}

The linear and cubic groups share leading linear terms whose
signs are fixed by the original generator. We keep these terms
together until their cancellation is made explicit. With the
fixed radius $\ell$ used by $B$, set
\[
	g=[N^3v(N\cdot)w_\ell(N\cdot)*|\phi|^2]\phi,\qquad \widetilde g=[N^3v(N\cdot)*|\phi|^2]\phi .
\]
Their $L^2$ norms and first covariant replacements are
uniformly bounded by convolution and the background
$L^2\cap L^\infty$ bounds. Active $q$ projections are
understood in the original generalized fields.
Define, with the alternating kernel convention of BS,
\begin{equation}\label{eq:BS-linear-principal}
	\mathcal L_g=b(\cosh_\eta g)+b^*(\sinh_\eta\bar g), \qquad D_g=e^{-B}b(g)e^B-\mathcal L_g .
\end{equation}
The hyperbolic expression is the principal part, and $D_g$
is the exact correction.
Equations (5.3) and (5.45) of \cite{BS} give
\begin{align}
	e^{-B}L^{(1)}e^B
	&=\sqrt N(\mathcal L_g+\mathcal L_g^*)+E^{(1)},\notag\\
	E^{(1)}
	&=\sqrt N(D_g+D_g^*)
	-\tau e^{-B}\{(\cN+1)b(\widetilde g)
	+\mathrm{h.c.}\}e^B .
	\label{eq:BS-linear-error}
\end{align}

\subsubsection{The error of a single transformed field}

Part (iv) of \cite[Lemma 3.2]{BS} singles out, at degree
$J$, the term with no normalized quadratic factor.
For $\delta_B=[\,\cdot\,,B]$, it is
$\gamma_J(\cN)b^\#(g_J)$, where
\begin{equation}\label{eq:BS-principal-number}
	\begin{split}
	\gamma_{2r}(n)&=(1-n/N)^r(1+(1-n)/N)^r,\\
	\gamma_{2r+1}(n)&=(1-n/N)^{r+1}(1+(1-n)/N)^r .
	\end{split}
\end{equation}
Even terms annihilate and odd terms create;
$g_J$ is the specified alternating kernel product and
$\|g_J\|\le\|\eta\|_{\mathrm{HS}}^J\|g\|$.
The coefficient in this formula is on the left.
Moving it to the right uses the exact particle-number shifts.
Replacing $\gamma_J$ by one and summing with $1/J!$
gives precisely \eqref{eq:BS-linear-principal}.
All other terms contain at least one normalized
quadratic field factor.

The first two derivatives of $\gamma_J$ on the enlarged
low-degree range have bounds polynomial in $J$ times
$N^{-1}$ and $N^{-2}$, respectively.
Use the exact identity, including the vacuum,
\begin{equation}\label{eq:BS-linear-coefficient-split}
	\sqrt N(\gamma_J(n)-1)=\tau\{n u_J(n)+v_J\},\qquad u_J(n)=N\int_0^1\gamma_J'(sn)\d s,\qquad v_J=N(\gamma_J(0)-1).
\end{equation}
The error in this principal term is therefore a
$\tau$-small base cubic term and a
$\tau$-small linear term, with
right coefficients whose first differences are
polynomial in $J$ times $N^{-1}$.
For every nonprincipal term retain one normalization:
\[
	\sqrt N\,N^{-1}\mathcal Q(h)b^\#(g_J) =\tau\mathcal Q(h)b^\#(g_J).
\]
The other factors remain normalized.
If the selected factor contracts, its original $N^{-1}$
is still retained. Reordering the single terminal uses
\eqref{eq:HS-terminal-reordering} and the sum over the kernel expansion
\eqref{eq:HS-branch-budget}.
It follows, for the canonical cores with coefficients
separate, that
\begin{equation}\label{eq:BS-linear-blocks}
	\|\delta^j(\sqrt N D_g)_Jv_n\| \le a_J\tau\epsilon^j (\nu+L_J)^{(3+j)/2}\|v_n\|,\qquad j\le4 .
\end{equation}
Here and below the notation means the sum of the
coefficient-separated block estimates.
The coefficients $a_J$ have all polynomial moments
summable on the original small kernel range.
The linear part of
\eqref{eq:BS-linear-coefficient-split} is covered using
$\nu\ge1$.

A transformed single field itself has base degree one
and constant one by the same chain proof without the
principal subtraction. Multiplying it by $T_n+I$ from
Section~\ref{sec:BS-regular-groups} and by $\tau$ gives
the $(3,\tau)$ class. This treats the second term in
\eqref{eq:BS-linear-error}.
Actual derivatives replace $g,\eta,q$ and the coefficient
profiles, with at most a polynomial degree loss.
Lemma~\ref{lem:bounded-chain-products} therefore controls
the low-degree fourth remainder of $E^{(1)}$.
For the tail use the estimates for the already subtracted
$D_g$ in \cite[(5.49)--(5.58)]{BS}: the external $\sqrt N$
is paired with that error's $N^{-1/2}$ gain.
They give the required geometric $Q$ form majorant.
The other term is of size $\tau M^{3/2}\le C M$ on
the physical space and has the same geometric bound.
Consequently $E^{(1)}$ satisfies both static hypotheses
of Theorem~\ref{thm:inverse-comparison}, including their
derivatives. Its extracted large linear term remains
to be cancelled below.

\subsubsection{Three exact families for the cubic interaction}

The original cubic interaction is
\[
	L^{(3)}=\sqrt N\int W_N(x-y)\phi(y)b_x^*a_y^*a_x \d x\d y+\mathrm{h.c.}
\]
The mixed-field identity is
\[
	[a_y^*a_x,B]=b_y^*b^*(\eta_x)+b(\eta_y)b_x .
\]
Integrate it for $e^{-B}a_y^*a_xe^B$ and expand the
outer transformed $b_x^*$ separately.
With $A_j(x)=\delta_B^j b_x$, the resulting three
oriented families, all integrated in $x,y$, are
\begin{align}
	\mathcal A&=\sqrt N\sum_{n\ge0}\frac1{n!}
	\int W_N\phi(y)A_n(x)^*a_y^*a_x,\notag\\
	\mathcal B&=\sqrt N\sum_{n,k,r\ge0}
	\frac1{n!k!r!(k+r+1)}
	\int W_N\phi(y)A_n(x)^*A_k(y)^*
	\delta_B^r b^*(\eta_x),\notag\\
	\mathcal C&=\sqrt N\sum_{n,k,r\ge0}
	\frac1{n!k!r!(k+r+1)}
	\int W_N\phi(y)A_n(x)^*
	\delta_B^k b(\eta_y)A_r(x).
	\label{eq:BS-cubic-families}
\end{align}
The full operator is their sum plus their adjoints.
Their total kernel degrees are $n,n+k+r+1,n+k+r+1$.
These formulas follow directly from the exact density
identity, and keep its coefficients and field order.

The estimates use two bounded source classes. The first consists
of $\sqrt N\sqrt{W_N(x-y)}\phi(y)$ times either a bare annihilator or a transformed linear field whose
creation operators are smoothed by kernels. Equation \eqref{eq:bare-cubic-source} and the row bounds
place these terms in the base class $(d,\beta)=(1,1)$,
including $\delta_B^r b^*(\eta_x)$. The second consists of
the $\tau$-small quadratic sources in
\eqref{eq:BS-pair-source-classes}, which contain at least one field smoothed by a kernel and have
annihilation operators for all bare fields.

Their products belong to the class $(q,\gamma)=(3,\tau)$.
When the first source is paired with a bare $D_V$, we instead
apply Proposition~\ref{prop:normalized-source-coupling}.
Throughout, the normalized chains retain their original order.

In $\mathcal C$ put $A_n(x)$ on the left state.
The first field in the right source $\delta_B^k b(\eta_y)A_r(x)$ is always
smoothed by a kernel. If the second field is bare, it is an annihilation
operator. If both fields are smoothed, their contact contains at least two
kernel factors. Thus the entire third family,
including $k=0$, is of class $(3,\tau)$.

\subsubsection{The unique large contacts and their cancellation}

The first family in \eqref{eq:BS-cubic-families} pairs
$\sqrt{W_N}a_yA_n(x)$ with
$\sqrt N\sqrt{W_N}\phi(y)a_x$.
The second uses the pair of sources
\[
	\sqrt{W_N}A_k(y)A_n(x),\qquad \sqrt N\sqrt{W_N}\phi(y)\delta_B^r b^*(\eta_x).
\]
The bare $D_V$ and small $F_2$ types are covered by the
two source estimates just stated.
An unnormalized large contact $\sqrt{W_N}\eta$ can
occur only at
\[
	\mathcal A:\ n=1,\qquad \mathcal B:\ n=1,\ k=0,\ r\ge0 .
\]
Uniqueness follows by keeping the left source $A_k(y)A_n(x)$
in its original order. A left-terminal creator from the principal
$k\ge1$ term is already normally ordered with a bare right
annihilator. An annihilating left terminal at $k\ge1$ instead
contains at least two kernel factors, as does a contact between
two smoothed terminals. In $\mathcal A$, the bare left terminal
$a_y$ therefore permits an order-one contact only when $n=1$.
Every other first-kernel contact retains a normalization and
is controlled by \eqref{eq:normalized-large-contact}.

Put $c(n)=1-n/N$.
The exact principal part of the first generalized
commutator is
$\delta_Bb_x^*=b(\eta_x)c(\cN)$.
Its remaining terms contain a normalized quadratic factor.
The contact with $a_y^*$ in $\mathcal A$ is therefore
\begin{equation}\label{eq:BS-cubic-first-contact}
	\sqrt N\int W_N(x-y)\phi(y)\overline{\eta(x,y)} b_xc(\cN)\d x\d y .
\end{equation}
For $\mathcal B$ use the exact relation
\[
	[b(\eta_x),b_y^*] =c(\cN)\overline{\eta(x,y)} -N^{-1}a_y^*a(\eta_x).
\]
Since $c(\cN)b_y^*=b_y^*c(\cN+1)$, its large contact is
\begin{equation}\label{eq:BS-cubic-second-contact}
	\sqrt N\sum_{r\ge0}\frac1{(r+1)!} \int W_N(x-y)\phi(y)\overline{\eta(x,y)} c(\cN)c(\cN+1)\delta_B^r b^*(\eta_x)\d x\d y .
\end{equation}
The shift in the second coefficient is essential.
The normal products and the displayed $N^{-1}$ error
remain controlled. For instance put their bare $a_y^*$
or $b_y^*$ on the left state, giving the bounded linear
source $\sqrt N\sqrt{W_N}\phi(y)a_y$ or its generalized
version. The other source is
$\sqrt{W_N}a(\eta_x)a_x$ in the first family, or a quadratic source with two fields smoothed by $\eta_x$ in
the second. Both fields depend on $x$. To obtain the
$\tau$ bounds, first integrate $y$ using
$\|W_N\|_1$. Then apply the uniform row bound and the integrated
squared row norm of $\eta$.
This justifies the orientation of these source estimates.

In \eqref{eq:BS-cubic-first-contact} and
\eqref{eq:BS-cubic-second-contact} put $\eta=k+\mu$.
The finite-rank profile of $\mu$ and
$\|W_N\|_1=O(N^{-1})$ give
\[
	\left\|\sqrt N\int W_N(x-y)\phi(y) \overline{\mu(x,y)}\d y\right\|_2 \le C_I\tau.
\]
These terms are consequently small linear or cubic
chain classes, including their derivatives.
The $k$ part has the exact identity
\[
	\sqrt N\int W_N(x-y)\phi(y)\overline{k(x,y)}\d y =-\sqrt N\,\overline{g(x)}.
\]
Writing $c(\cN)=1-\cN/N$ in the first contact gives
the main term $-\sqrt N b(g)$ and a $(3,\tau)$ error.
For the second contact use the unique principal terms
of $\delta_B^r b^*(\eta_x)$.
Every combined number coefficient, including
$c(\cN)c(\cN+1)$, splits into one plus an error by
\eqref{eq:BS-linear-coefficient-split}; it has the same
polynomial first- and second-derivative bounds.
Every nonprincipal term retains $N^{-1}$.
After the external $\sqrt N$ all these errors satisfy
\eqref{eq:BS-linear-blocks}.
The remaining alternating series, with its exact
$1/(r+1)!$, is
$-\sqrt N(\mathcal L_g-b(g))$.
Indeed applying the principal linear transformation to
$b^*(\eta\bar g)$ supplies the powers numbered $r+1$;
the even and odd subsequences are precisely the
cosh and sinh series with the zeroth annihilation term
removed. This identifies the principal series.
The two contacts therefore give $-\sqrt N\mathcal L_g$.

\begin{theorem}\label{thm:BS-linear-cubic-remainder}
With the original BS definitions,
\begin{equation}\label{eq:BS-linear-cubic-cancellation}
	e^{-B}L^{(3)}e^B =-\sqrt N(\mathcal L_g+\mathcal L_g^*)+E^{(3)},\qquad e^{-B}(L^{(1)}+L^{(3)})e^B=E^{(1)}+E^{(3)}.
\end{equation}
The fourth-order conjugation remainder of each error $E^{(1)},E^{(3)}$
satisfies both static hypotheses of
Theorem~\ref{thm:inverse-comparison}, including the first
covariant and number derivatives. Their combined
time integral is bounded by
$C_{I,s}N^{-5/4}\|F_3(s)^{3/2}z\|$.
\end{theorem}
\begin{proof}
The preceding contact calculation proves the exact
first identity with the definition in \cite[(5.98)]{BS};
\eqref{eq:BS-linear-error} then gives the second identity.
Every low-degree term left in the three cubic families
is either a bounded linear-source coupling with $D_V$,
a $(3,\tau)$ bounded product, or an error of the form
\eqref{eq:BS-linear-blocks}.
Proposition~\ref{prop:normalized-source-coupling} and
Lemma~\ref{lem:bounded-chain-products} give the
one-weight static estimates, with the original
coefficients and all particle-number shifts.
Reordering involves a fixed number of terminal field operators.
The summed contraction estimate
\eqref{eq:HS-branch-budget} is used before dropping any
$N^{-1}$, so the original geometric ratio is preserved.
Derivatives replace $\phi,\eta,\mu,g,q$ and their
coefficient profiles; all the row and source estimates
have already been proved with these replacements.
Number differentiation adds only a particle-number shift factor.

The high-degree estimate uses the original bounds
in \cite[(5.99)--(5.114)]{BS} for $E^{(3)}$ after
the large linear term has been subtracted, and the
corresponding linear-error bounds established above.
They provide the geometric $Q$ form majorants needed
for \eqref{eq:chain-exact-tail}.
Theorem~\ref{thm:inverse-comparison} completes the argument.
\end{proof}

\subsection{Singular pairs and scattering cancellations}
\label{sec:BS-scattering}

\subsubsection{Two singular pair kernels}

For $\chi_N=1$ or $\chi_N=f_\ell(N\cdot)$ define
\begin{equation}\label{eq:BS-singular-pair}
	\begin{split}
	h_\chi(x,y)&=NW_N(x-y)\chi_N(x-y)\phi(x)\phi(y),\qquad
	P_\chi=P(h_\chi),\\
	\sigma_\chi&=\operatorname{Re}\int\overline{h_\chi(x,y)}
	k(x,y)\d x\d y .
	\end{split}
\end{equation}
The radius $\ell$ is fixed, and $\chi_N$ is real,
symmetric, independent of time and between zero and one.
Keep the BS scalar convention and put
\begin{equation}\label{eq:BS-singular-correction}
	E_\chi(s)=e^{-sB}P_\chi e^{sB}-P_\chi-s\sigma_\chi I, \qquad 0\le s\le1 .
\end{equation}
We estimate the combined correction $E_\chi(s)$.

\begin{proposition}\label{prop:BS-singular-correction}
Uniformly in $s\in[0,1]$, the fourth-order conjugation
remainder of $E_\chi(s)$ satisfies
\eqref{eq:one-weight-static} and
\eqref{eq:geometric-tail-static}, including its covariant
and number derivatives. The same holds for its average
in $s$.
\end{proposition}
\begin{proof}
Use the scalar source
$s_\chi=N\sqrt{W_N}\chi_N\phi(x)\phi(y)$.
The scaling of $W_N$ gives
\begin{equation}\label{eq:BS-singular-source}
	\|s_\chi\|_2+\|\partial_t s_\chi\|_2\le C_I\sqrt N, \qquad |\langle s_\chi,\sqrt{W_N}\mu\rangle|\le C_I .
\end{equation}
The second assertion follows from
\eqref{eq:BS-small-projection-source}.
Projected finite-rank replacements satisfy the same
estimates.
Expand the two transformed fields in
$e^{-sB}P_\chi e^{sB}$ and apply the source classification
\eqref{eq:BS-pair-source-classes}.
Its zero-degree bare term is removed exactly by $P_\chi$.
At positive degree a bare $D_V$ source retains at least
one normalized quadratic factor. Combine that factor
with $s_\chi$ as in
\eqref{eq:normalized-large-contact}, then use the
bounded-source coupling proposition.
The small $F_2$ source paired with $s_\chi$ is the
$(2,1)$ bounded-product class.
Two large scalar sources give an $O(N)$ coefficient;
a retained normalized factor absorbs this $N$ and
again leaves the $(2,1)$ class.
All additional factors keep their original normalization,
coefficient order and adjoints.

A large scalar contact without a normalized factor can
only have total degree one. A higher-degree terminal
contact contains at least two kernel factors and is
already in the small-source class.
The exact degree-one scalar is
\begin{equation}\label{eq:BS-pair-vacuum-coefficient}
	s\,\sigma_{\eta,\chi}g_N(\cN),\qquad \sigma_{\eta,\chi}=\operatorname{Re} \int\overline{h_\chi}\eta,\qquad g_N(n)=\frac{(N-n)(N-n-1)}{N^2}.
\end{equation}
This is the vacuum scalar of the generalized pair
commutator; both endpoint square roots remain present.
Equation \eqref{eq:BS-singular-source} gives
$|\sigma_{\eta,\chi}-\sigma_\chi|\le C_I$ and
$|\sigma_{\eta,\chi}|\le C_IN$.
Subtracting the original scalar yields the exact identity
\begin{equation}\label{eq:BS-pair-number-polynomial}
	s(\sigma_{\eta,\chi}-\sigma_\chi)I +s\frac{\sigma_{\eta,\chi}}N \left\{-(2\cN+1)+\frac{\cN^2+\cN}{N}\right\}.
\end{equation}
The scalar commutator is zero.
The other terms have base classes $(2,1)$ and $(4,\tau^2)$.
The identity includes the vacuum and the physical upper
number boundary.

Proposition~\ref{prop:normalized-source-coupling} and
Lemma~\ref{lem:bounded-chain-products} now cover all low-degree
terms. This remains true after differentiation: time derivatives
act on $\phi,\eta,\mu,q$ and preserve
\eqref{eq:BS-singular-source}, while number differentiation
inserts the particle-number shift. The resulting series therefore retains
the original geometric ratio, with only fixed polynomial losses.
For the high-degree tail use the absolute kernel
estimates in \cite[(5.91)--(5.97)]{BS}.
These bounds remain valid when $W_N$ is replaced by $W_N\chi_N$,
since the weight is nonnegative and $\chi_N\le1$.
The only large scalar has already been isolated in
\eqref{eq:BS-pair-vacuum-coefficient}; replacing $\eta$
by the original $k$ in the subtraction has only the
bounded difference in \eqref{eq:BS-singular-source}.
Thus the same geometric $Q$ form estimate holds for
both choices of $\chi_N$, including derivatives.
Apply \eqref{eq:chain-exact-tail}.
All bounds are uniform in the auxiliary parameter $s$,
so its average has the same estimates.
\end{proof}

\subsubsection{The full symmetric kinetic kernel}

Put $a_N(r)=Nw_\ell(Nr)$ and $f_N(r)=f_\ell(Nr)$.
For sufficiently large $N$ the support of the potential
lies inside the microscopic Neumann ball.
The Neumann scattering equation \cite[Eq.~(4.1)]{BS} and
the constant extension give, in distributions,
\begin{equation}\label{eq:BS-scaled-scattering}
	2\Delta a_N =2N^3\lambda_\ell f_N\mathbf1_{|r|\le\ell} -NW_Nf_N,\qquad N^3|\lambda_\ell|\le C_\ell .
\end{equation}
There is no boundary measure: the Neumann normal
derivative matches the constant extension.
The finitely many smaller $N$ can be absorbed in the
constants of the original fixed-$N$ estimates.
The same profile bounds imply
$\|a_N\|_1+\|a_N\|_2+\||r|\nabla a_N\|_2\le C_\ell$.
The quadratic correlation radius $\ell$ is fixed.

For $k(x,y)=-a_N(x-y)\phi(x)\phi(y)$, both Laplacians
must be kept together:
\begin{equation}\label{eq:BS-symmetric-laplacian}
	\begin{split}
	(-\Delta_x-\Delta_y)k
	={}&2(\Delta a_N)(x-y)\phi(x)\phi(y)\\
	&+2\nabla a_N(x-y)\cdot
	[\nabla\phi(x)\phi(y)-\phi(x)\nabla\phi(y)]\\
	&+a_N(x-y)[\Delta\phi(x)\phi(y)+\phi(x)\Delta\phi(y)] .
	\end{split}
\end{equation}
The bracket in the second line vanishes on the diagonal.
Write it as
$[\nabla\phi(x)-\nabla\phi(y)]\phi(y)
-[\phi(x)-\phi(y)]\nabla\phi(y)$.
The mean value formula bounds it by
\[
	|x-y|\{\|\nabla^2\phi\|_\infty|\phi(y)| +\|\nabla\phi\|_\infty|\nabla\phi(y)|\}.
\]
Consequently the second line has a uniform
Hilbert--Schmidt bound using $\||r|\nabla a_N\|_2$.
The last line uses $\|a_N\|_2$, $\|\phi\|_\infty$ and
$\|\Delta\phi\|_2$.
For a time derivative apply the same mean value
formula to the differentiated diagonal-vanishing
bracket. The bounds $\dot\phi\in H^4\subset W^{2,\infty}$
and $\phi\in H^6$ suffice.

The projection correction also has controlled second
derivatives. With $\rho=|\phi|^2$,
$A_N=a_N*\rho$ and $d_N=\int\rho A_N$, the exact
formula is
\begin{equation}\label{eq:BS-projection-profile}
	\mu(x,y)=\phi(x)\phi(y)[A_N(x)+A_N(y)-d_N].
\end{equation}
For $|\alpha|\le2$ transfer derivatives to the density:
$\partial^\alpha A_N=a_N*\partial^\alpha\rho$.
Young's inequality gives the required $L^\infty$ bounds,
and $\phi A_N$ has uniformly bounded $H^2$ norm.
Thus $\mu$ is a finite sum of tensor products of
uniformly bounded $H^2$ vectors and
\[
	\|\mu\|_{\mathrm{HS}} +\|(\Delta_x+\Delta_y)\mu\|_{\mathrm{HS}}\le C_I.
\]
Their first time derivatives have the same bounds.
This explicitly retains the failure of $q$ to commute
with the spatial Laplacian.

Combining these facts gives
\begin{equation}\label{eq:BS-kinetic-pair-decomposition}
	H_N:=(-\Delta_x-\Delta_y)\eta=-h_f+r_H,\qquad h_f=NW_N f_N\phi\otimes\phi,\qquad \|r_H\|_{\mathrm{HS}}+\|\partial_t r_H\|_{\mathrm{HS}}\le C_I.
\end{equation}
The eigenvalue term in \eqref{eq:BS-scaled-scattering}
has bounded Hilbert--Schmidt norm, as do the background
and projection terms just checked.
Moreover $\|h_f\|_{\mathrm{op}}\le C_I$ by its convolution
$L^1$ bound, so $H_N$ is uniformly bounded as an
integral operator. Projecting both field coordinates replaces the kernel by
$(q\otimes q)H_N$.
The difference $(q\otimes q-I)h_f$, and each connection
insertion, is a composition of a bounded operator with
a smooth finite-rank operator. Its Hilbert--Schmidt
norm is bounded. Hence the covariant version of
\eqref{eq:BS-kinetic-pair-decomposition} has the same
remainder bound.

As compressed forms $[K_P,B]=P(H_N)$, and therefore
\begin{equation}\label{eq:BS-kinetic-duhamel}
	e^{-B}K_Pe^B-K_P =P(H_N)+\int_0^1 \{e^{-sB}P(H_N)e^{sB}-P(H_N)\}\d s .
\end{equation}
Proposition~\ref{prop:BS-singular-correction} treats
the $-h_f$ part of the integrand.
Lemma~\ref{lem:HS-pair-chains} treats the entire
$r_H$ part and its conjugation.
For that regular pair the same proof also gives a
geometric number-form tail: the estimates for the
basic Hilbert--Schmidt kernel in the time-term proof
apply with $r_H$ in place of $\dot\eta$.
This calculation uses generalized pairs throughout.

\subsubsection{Original scalars and bare pair cancellation}

The auxiliary $s$ integral in
\eqref{eq:BS-kinetic-duhamel} gives the scalar coefficient
\begin{equation}\label{eq:BS-kinetic-scalar}
	\tfrac12\operatorname{Re}\langle H_N,\eta\rangle_{\mathrm{HS}} =\|\nabla_x\eta\|_{\mathrm{HS}}^2 .
\end{equation}
Symmetry equates the two gradient norms.
The original scalar is instead
$\sigma_K=\|\nabla_xk\|_{\mathrm{HS}}^2$.
Their difference is bounded: in the cross term use
\[
	|\langle\nabla_x k,\nabla_x\mu\rangle| =|\langle k,-\Delta_x\mu\rangle| \le\|k\|_{\mathrm{HS}}\|\Delta_x\mu\|_{\mathrm{HS}}\le C_I,
\]
and add $\|\nabla_x\mu\|_{\mathrm{HS}}^2$.
The differentiated identity is bounded by the same
integration by parts with $k$ or $\mu$ replaced.
The $\eta$-to-$k$ scalar changes in the singular-pair
proposition also have the bound
\eqref{eq:BS-singular-source}.
These scalar observations preserve the original
generator convention; their fourth commutators vanish.

Let $h_\Delta=2N^3(\Delta w_\ell)(N(x-y))\phi(x)\phi(y)$.
The preceding symmetric calculation shows that
$H_N-h_\Delta$ is uniformly Hilbert--Schmidt, including
one covariant derivative.
The original bare quadratic pair is $h_1=NW_N\phi\otimes\phi$;
the quartic group extracted the pair kernel $W_Nk$.
Their exact scattering combination is
\begin{equation}\label{eq:BS-bare-pair-cancellation}
	h_\Delta+h_1+W_Nk =2N^3\lambda_\ell f_N\mathbf1_{|x-y|\le\ell} \phi(x)\phi(y).
\end{equation}
This is \cite[(5.130)]{BS}.
The right side and its first covariant derivative have
uniformly bounded Hilbert--Schmidt norms.

\begin{theorem}\label{thm:BS-scattering-remainder}
Put $\sigma_2=\operatorname{Re}\int\overline{h_1}k$.
The original scattering group
\begin{equation}\label{eq:BS-scattering-group}
	S_{\rm scat}=e^{-B}K_Pe^B-K_P +e^{-B}P(h_1)e^B+P(W_Nk) -(\sigma_K+\sigma_2)I
\end{equation}
has an exact fourth-order conjugation remainder satisfying
\eqref{eq:one-weight-static} and
\eqref{eq:geometric-tail-static}, including their
covariant and number derivatives.
Its time-integrated contribution has the
$N^{-5/4}$ bound of
Theorem~\ref{thm:inverse-comparison}.
\end{theorem}
\begin{proof}
For the transformed corrections use
\eqref{eq:BS-kinetic-duhamel},
Proposition~\ref{prop:BS-singular-correction} with both
$\chi_N=f_N$ and $\chi_N=1$, and the regular
Hilbert--Schmidt pair-chain estimate.
Their bare pair terms combine by
\eqref{eq:BS-bare-pair-cancellation} and
\eqref{eq:BS-kinetic-pair-decomposition} to a bounded
Hilbert--Schmidt pair.
Their scalar differences are exactly those retained
above and have zero fourth commutator.
Every low-degree non-scalar piece is therefore already
in a proved source or bounded-product class.
The geometric high-degree estimate uses these
subtracted transformed corrections, or equivalently
the original BS estimates after the scattering and
scalar cancellations. Apply
\eqref{eq:chain-exact-tail} and then the inverse-energy
comparison theorem. This controls the group with the
original scalars and the original two evolutions.
\end{proof}

\subsection{The full comparison of the two evolutions}
\label{sec:full-model-assembly}

We can now combine the estimates for all groups of the
compressed generator $\mathfrak G=P(G_N-c_N)P$ and apply the
inverse-energy comparison principle. With the definitions above,
the exact decomposition is
\begin{equation}\label{eq:full-model-groups}
	\mathfrak G=K_P+V_P+L_B+T_0+T_2 +E^{(1)}+E^{(3)}+E^{(4)}_{\rm BS} +S_{\rm scat}.
\end{equation}
Indeed \eqref{eq:BS-linear-cubic-cancellation} cancels
the two original large linear terms;
\eqref{eq:BS-quartic-definition} contributes $V_P$,
$E^{(4)}_{\rm BS}$ and the extracted pair $P(W_Nk)$.
The latter belongs to \eqref{eq:BS-scattering-group},
together with the kinetic correction and the transformed
bare quadratic pair. The regular number and one-body
parts are \eqref{eq:BS-T0} and \eqref{eq:BS-T2}.
The scalar identity needed for \eqref{eq:full-model-groups} is
\begin{equation}\label{eq:full-model-scalar}
	c_N=(N\alpha_N-\beta_N)+\sigma_K+\sigma_2+\sigma_4 .
\end{equation}
Substituting their definitions gives exactly
\cite[(4.25)]{BS}: its first integral is
$N\alpha_N-\beta_N$; its other three integrals are
$\sigma_K,\sigma_4,\sigma_2$.
The cubic phase contributes its own time term to the exact
generator $\mathfrak J$, with the estimate proved above.

\begin{theorem}
\label{thm:full-model-comparison}
Let $v\in C_c^\infty(\mathbb R^3)$ be nonnegative and radial,
and use the modified GP background with an $N$-independent normalized
initial condensate wave function in $H^6$.
Use the fixed small radius and kernel range of the
original quadratic transformation, and the original
cubic phase with $m=N^{-1/2}$, its cutoff $\Theta$,
generalized fields, $p,\sigma$ kernels and active projections.
Let $U_{\rm ex}$ and $U_3$ be the two evolutions defined
in this paper, with the same connection.
For every compact time interval $I$, fixed $s\in I$
and sufficiently large fixed $b_I$, set
$F_3(s)=\mathfrak J^{[3]}(s)+b_IM$.
Then
\begin{equation}\label{eq:full-model-comparison}
	\|(U_{\rm ex}(t,s)-U_3(t,s))z\| \le C_{I,s}N^{-5/4}\|F_3(s)^{3/2}z\|, \qquad t\in I,\quad z\in D(F_3(s)^{3/2}).
\end{equation}
The constants are independent of $N$.
In particular the normalized preparations from the seventh
power of the resolvent in \eqref{eq:Taylor-resolvent-data}, and the positive inverse
data of Remark~\ref{rem:inverse-preparation}, have a
uniform $O_{I,s}(N^{-5/4})$ error.
\end{theorem}
\begin{proof}
By linearity the exact generator difference is the
sum of the exact fourth integral remainders of
\eqref{eq:full-model-groups}, plus the cubic time
remainder of Proposition~\ref{prop:time-remainder-sharp}.
The necessary estimates have the following sources:
\[
	\begin{array}{c|l}
	\text{group}&\text{proved remainder estimate}\\ \hline
	K_P&\text{Theorem }\ref{thm:actual-kinetic-fourth}\\
	V_P&\text{Theorem }\ref{thm:actual-potential-fourth}\\
	L_B&\text{Theorem }\ref{thm:BS-quadratic-time-remainder}\\
	T_0,\ T_2&\text{Proposition }\ref{prop:BS-regular-remainder}\\
	E^{(1)},\ E^{(3)}&\text{Theorem }\ref{thm:BS-linear-cubic-remainder}\\
	E^{(4)}_{\rm BS}&\text{Theorem }\ref{thm:BS-quartic-remainder}\\
	S_{\rm scat}&\text{Theorem }\ref{thm:BS-scattering-remainder}
	\end{array}
\]
The bare kinetic term appears once; its quadratic
conjugation correction is contained in $S_{\rm scat}$.
The same scattering combination contains the pair
extracted from the quartic group.

To combine the estimates, group the series by total kernel
degree $J$ and set $J_*=\lfloor c_0\sqrt N\rfloor$, assigning
the finite bare-energy and cubic-time contributions to a fixed
degree. For $J\le J_*$, the preceding proofs establish
\eqref{eq:one-weight-static}, with every polynomial moment
of $a_J$ summable, also after covariant and number
differentiation. By Remark~\ref{rem:enlarged-support},
the support is the original particle-number cutoff support enlarged by
$O(J+1)$. The source estimates supply the required bilinear
bound with the number weight on the specified side, including
all adjoint orientations.
For $J>J_*$, the geometrically summable BS groups, with
their prescribed subtractions, give
\eqref{eq:geometric-tail-static} through
\eqref{eq:chain-exact-tail}. These exact tails remain in the
generator. Uniform convergence of both the kernel series and
their differentiated series also establishes the required $C^1$
form regularity, completing the verification of the hypotheses.

All hypotheses of Theorem~\ref{thm:inverse-comparison}
are now verified for the model.
The small factor is
$\epsilon^2\Lambda_N^{1/2}=O(N^{-5/4})$:
the inverse of the full positive exact energy removes
half the number weight, and the original cutoff supplies
the remaining $\Lambda_N^{1/2}$.
The required right-trajectory bound is the already
proved propagation of $F_3^{3/2}$ in
Corollary~\ref{cor:Taylor-three-halves}.
Applying that theorem proves
\eqref{eq:full-model-comparison}, including $t<s$.
Both changes of coordinates are unitary.
Finally Proposition~\ref{prop:Taylor-resolvent-preparation}
and Remark~\ref{rem:inverse-preparation} give the stated
uniformly prepared classes.
\end{proof}

\section{Discussion and outlook}
\label{sec:discussion}

The main results describe two energy scales of the same exact
many-body evolution. The third moment of the positive correlated
energy controls the third depletion moment and sixth moments
of collective fluctuations. For norm-convergent pure-state
families satisfying the initial moment bound, the excitation-energy
measure also resolves a nonzero contribution at scale $N^2$.
The common initial class is characterized in
Theorem~\ref{thm:admissible-initial-class}.

The distinction between these observables has a quantitative
meaning. A fixed positive microscopic window with nonzero limiting
weight has probability of order $N^{-2}$, while its contribution
to the first excitation-energy moment is of order one.
The full microscopic measure uses the relative Hamiltonian
$-2\Delta+v$. Its contact coefficient evolves with the limiting
Bogoliubov state. The radial density and conditional window law
are given in Proposition~\ref{prop:radial-microscopic-density}
and Corollary~\ref{cor:evolved-energy-probability}.
The threshold coefficient depends on the scattering length;
the density at a fixed positive rescaled energy retains the
potential. The threshold limit is taken after $N\to\infty$.

Proposition~\ref{prop:evolved-energy-defect} identifies the
unscaled measurement law and the positive excess in its first
moment. Higher excitation-energy moments have growing lower
bounds, although the depletion and collective moments remain
bounded on every compact time interval.
Theorem~\ref{thm:independent-energy-sampling} describes the
resulting sample mean over order $N^2$ independent preparations.
The L\'evy--Khintchine representation is standard
\cite{Bertoin2000}; the relative scattering measure and the
contact determine its parameters.

For the interaction energy,
Theorem~\ref{thm:admissible-interaction-spectrum} identifies
the scaled moments for $1\le p<2$ on the same initial class.
Their coefficients involve $2c_\psi(t)\int v^p(1-f)^2$.
At order two, the exact evolution has an $O(N^2)$ upper bound
and an asymptotic lower bound. The separate coefficient
limit in Theorem~\ref{thm:spec:reference-second-moment}
concerns a prescribed cubic transform of a reference in
$D(M^2A_P)$.

The model estimates extend the correlation framework of
\cite{BS,COS}. Fixed-kernel number bounds first control the
cancelled groups on a common condensate neighborhood.
Their real polynomial structure then gives the two time
derivatives needed for form-space propagation
\cite{Kisynski,BLP}. The weighted coercivity follows from
the inverse-commutator identity. These steps keep the
microscopic cancellations and the particle-number weights
throughout the exact evolution.
The finite cubic comparison in Theorem~\ref{thm:C}
is a further application of this energy analysis.
Its reference generator retains the interaction energy,
and its two rates use different initial spectral norms.

The preparation results give explicit admissible families.
One positive form equation applies to the weighted form class;
three equations admit arbitrary Fock-space references.
Pure quasi-free references, statistical mixtures and free
Bose Gibbs references are treated in their respective
preparation theorems. The prepared state supplies the
initial correlated-energy moment.
The minimizers in Theorem~\ref{thm:correlated-ground-state}
belong to the positive initial correlated energy, including
its fixed number correction $bM$.
Their selection uses the quadratic diagonalization of
\cite{NNS2016}.

The initial domain is fixed by the correlation representation.
Under the overlap condition of
Theorem~\ref{thm:vacuum-third-obstruction}, the unprepared
vacuum cannot be approached in norm by a family with a
uniform third correlated-energy moment.
The finite-excitation extension is given in
Corollary~\ref{cor:finite-excitation-obstruction}.
These results make the role of the preparation explicit.

Further questions concern unbounded one-particle observables,
higher correlated-energy moments and low-energy trapped initial
states. Verifying the initial third-moment condition for trapped
states would connect the present theorem to release evolution.
The higher-order mean-field expansions of \cite{BPPS} and
the equilibrium exponential estimates of \cite{NR,BBR}
also suggest corresponding questions at the GP correlation scale.

\section*{Declarations}

\subsection*{Funding}
The authors declare that no funds, grants, or other support were received during the preparation of this manuscript.

\subsection*{Competing interests}
The authors have no relevant financial or non-financial interests to disclose.

\subsection*{Authors' contributions}
All authors contributed to the conceptualization, mathematical analysis, proof verification, and writing of the manuscript. All authors contributed equally to this work. All authors read and approved the final manuscript.

\subsection*{Data availability}
Data sharing is not applicable because no datasets were generated or analyzed in this theoretical study.

\subsection*{Ethical Approval/Human and Animal Rights}
This article does not contain any studies with human participants or animals performed by any of the authors.


\clearpage
\begin{thebibliography}{99}
\bibitem{Gross1961} E. P. Gross,
\emph{Structure of a quantized vortex in boson systems},
Nuovo Cim. \textbf{20} (1961), 454--477.
\href{https://doi.org/10.1007/bf02731494}{doi:10.1007/bf02731494}.

\bibitem{Pitaevskii1961} L. P. Pitaevskii,
\emph{Vortex lines in an imperfect Bose gas},
Sov. Phys. JETP \textbf{13} (1961), 451--454.
\href{https://jetp.ras.ru/cgi-bin/dn/e_013_02_0451.pdf}{Publisher archive}.

\bibitem{BS} C. Brennecke and B. Schlein,
\emph{Gross--Pitaevskii dynamics for Bose--Einstein condensates},
Anal. PDE \textbf{12} (2019), 1513--1596.
\href{https://doi.org/10.2140/apde.2019.12.1513}{doi:10.2140/apde.2019.12.1513}.
Equation and theorem numbering follows \href{https://arxiv.org/abs/1702.05625v4}{arXiv:1702.05625v4}.

\bibitem{COS} C. Caraci, J. Oldenburg and B. Schlein,
\emph{Quantum fluctuations of many-body dynamics around the Gross--Pitaevskii equation},
Ann. Inst. H. Poincar\'e C Anal. Non Lin\'eaire \textbf{42} (2025), 1313--1374.
\href{https://doi.org/10.4171/aihpc/132}{doi:10.4171/aihpc/132}.
Equation and theorem numbering follows \href{https://arxiv.org/abs/2308.11687v2}{arXiv:2308.11687v2}.

\bibitem{ESY2010} L. Erd\H{o}s, B. Schlein and H.-T. Yau,
\emph{Derivation of the Gross--Pitaevskii equation for the dynamics of Bose--Einstein condensate},
Ann. Math. \textbf{172} (2010), 291--370.
\href{https://doi.org/10.4007/annals.2010.172.291}{doi:10.4007/annals.2010.172.291}.

\bibitem{Anderson1995} M. H. Anderson, J. R. Ensher, M. R. Matthews, C. E. Wieman and E. A. Cornell,
\emph{Observation of Bose--Einstein condensation in a dilute atomic vapor},
Science \textbf{269} (1995), 198--201.
\href{https://doi.org/10.1126/science.269.5221.198}{doi:10.1126/science.269.5221.198}.

\bibitem{Davis1995} K. B. Davis, M.-O. Mewes, M. R. Andrews, N. J. van Druten, D. S. Durfee, D. M. Kurn and W. Ketterle,
\emph{Bose--Einstein condensation in a gas of sodium atoms},
Phys. Rev. Lett. \textbf{75} (1995), 3969--3973.
\href{https://doi.org/10.1103/physrevlett.75.3969}{doi:10.1103/physrevlett.75.3969}.

\bibitem{DGPS} F. Dalfovo, S. Giorgini, L. P. Pitaevskii and S. Stringari,
\emph{Theory of Bose--Einstein condensation in trapped gases},
Rev. Mod. Phys. \textbf{71} (1999), 463--512.
\href{https://doi.org/10.1103/revmodphys.71.463}{doi:10.1103/revmodphys.71.463}.

\bibitem{LSY2000} E. H. Lieb, R. Seiringer and J. Yngvason,
\emph{Bosons in a trap: a rigorous derivation of the Gross--Pitaevskii energy functional},
Phys. Rev. A \textbf{61} (2000), 043602.
\href{https://doi.org/10.1103/physreva.61.043602}{doi:10.1103/physreva.61.043602}.

\bibitem{LS2002} E. H. Lieb and R. Seiringer,
\emph{Proof of Bose--Einstein condensation for dilute trapped gases},
Phys. Rev. Lett. \textbf{88} (2002), 170409.
\href{https://doi.org/10.1103/physrevlett.88.170409}{doi:10.1103/physrevlett.88.170409}.

\bibitem{BBCS2018} C. Boccato, C. Brennecke, S. Cenatiempo and B. Schlein,
\emph{Complete Bose--Einstein condensation in the Gross--Pitaevskii regime},
Commun. Math. Phys. \textbf{359} (2018), 975--1026.
\href{https://doi.org/10.1007/s00220-017-3016-5}{doi:10.1007/s00220-017-3016-5}.

\bibitem{BBCS2020} C. Boccato, C. Brennecke, S. Cenatiempo and B. Schlein,
\emph{Optimal rate for Bose--Einstein condensation in the Gross--Pitaevskii regime},
Commun. Math. Phys. \textbf{376} (2020), 1311--1395.
\href{https://doi.org/10.1007/s00220-019-03555-9}{doi:10.1007/s00220-019-03555-9}.

\bibitem{ESY2006} L. Erd\H{o}s, B. Schlein and H.-T. Yau,
\emph{Derivation of the Gross--Pitaevskii hierarchy for the dynamics of Bose--Einstein condensate},
Commun. Pure Appl. Math. \textbf{59} (2006), 1659--1741.
\href{https://doi.org/10.1002/cpa.20123}{doi:10.1002/cpa.20123}.

\bibitem{BOS2015} N. Benedikter, G. de Oliveira and B. Schlein,
\emph{Quantitative derivation of the Gross--Pitaevskii equation},
Commun. Pure Appl. Math. \textbf{68} (2015), 1399--1482.
\href{https://doi.org/10.1002/cpa.21542}{doi:10.1002/cpa.21542}.

\bibitem{BK2025} C. Brennecke and W. Kroschinsky,
\emph{Derivation of the Gross--Pitaevskii dynamics through renormalized excitation number operators},
Forum Math. Sigma \textbf{13} (2025), e107.
\href{https://doi.org/10.1017/fms.2025.10073}{doi:10.1017/fms.2025.10073}.

\bibitem{Bogoliubov1947} N. N. Bogoliubov,
\emph{On the theory of superfluidity},
J. Phys. (USSR) \textbf{11} (1947), 23--32.
\href{https://ufn.ru/pdf/jphysussr/1947/11_1/3jphysussr19471101.pdf}{Publisher archive}.

\bibitem{Shale} D. Shale,
\emph{Linear symmetries of free boson fields},
Trans. Amer. Math. Soc. \textbf{103} (1962), 149--167.
\href{https://doi.org/10.1090/s0002-9947-1962-0137504-6}{doi:10.1090/s0002-9947-1962-0137504-6}.

\bibitem{NNS2016} P. T. Nam, M. Napi\'orkowski and J. P. Solovej,
\emph{Diagonalization of bosonic quadratic {H}amiltonians by {B}ogoliubov transformations},
J. Funct. Anal. \textbf{270} (2016), 4340--4368.
\href{https://doi.org/10.1016/j.jfa.2015.12.007}{doi:10.1016/j.jfa.2015.12.007}.
The cited numbering refers to \href{https://arxiv.org/abs/1508.07321v2}{arXiv:1508.07321v2}.

\bibitem{Derezinski2017} J. Derezi\'nski,
\emph{Bosonic quadratic Hamiltonians},
J. Math. Phys. \textbf{58} (2017), 121101.
\href{https://doi.org/10.1063/1.5017931}{doi:10.1063/1.5017931}.
The cited equation numbering refers to \href{https://arxiv.org/abs/1608.03289v2}{arXiv:1608.03289v2}.

\bibitem{Seiringer2011} R. Seiringer,
\emph{The excitation spectrum for weakly interacting bosons},
Commun. Math. Phys. \textbf{306} (2011), 565--578.
\href{https://doi.org/10.1007/s00220-011-1261-6}{doi:10.1007/s00220-011-1261-6}.

\bibitem{GrechSeiringer2013} P. Grech and R. Seiringer,
\emph{The excitation spectrum for weakly interacting bosons in a trap},
Commun. Math. Phys. \textbf{322} (2013), 559--591.
\href{https://doi.org/10.1007/s00220-013-1736-8}{doi:10.1007/s00220-013-1736-8}.

\bibitem{BBCS2019} C. Boccato, C. Brennecke, S. Cenatiempo and B. Schlein,
\emph{Bogoliubov theory in the Gross--Pitaevskii limit},
Acta Math. \textbf{222} (2019), 219--335.
\href{https://doi.org/10.4310/acta.2019.v222.n2.a1}{doi:10.4310/acta.2019.v222.n2.a1}.

\bibitem{BSSSpectrum} C. Brennecke, B. Schlein and S. Schraven,
\emph{Bogoliubov theory for trapped bosons in the Gross--Pitaevskii regime},
Ann. Henri Poincar\'e \textbf{23} (2022), 1583--1658.
\href{https://doi.org/10.1007/s00023-021-01151-z}{doi:10.1007/s00023-021-01151-z}.

\bibitem{NamTriay} P. T. Nam and A. Triay,
\emph{Bogoliubov excitation spectrum of trapped Bose gases in the Gross--Pitaevskii regime},
J. Math. Pures Appl. \textbf{176} (2023), 18--101.
\href{https://doi.org/10.1016/j.matpur.2023.06.002}{doi:10.1016/j.matpur.2023.06.002}.

\bibitem{WernerCastin2012} F. Werner and Y. Castin,
\emph{General relations for quantum gases in two and three dimensions. II. Bosons and mixtures},
Phys. Rev. A \textbf{86} (2012), 053633.
\href{https://doi.org/10.1103/PhysRevA.86.053633}{doi:10.1103/PhysRevA.86.053633}.
The cited table follows \href{https://arxiv.org/abs/1210.1784v3}{arXiv:1210.1784v3}.

\bibitem{BKP2011} E. Braaten, D. Kang and L. Platter,
\emph{Universal relations for identical bosons from three-body physics},
Phys. Rev. Lett. \textbf{106} (2011), 153005.
\href{https://doi.org/10.1103/PhysRevLett.106.153005}{doi:10.1103/PhysRevLett.106.153005}.

\bibitem{HofmannZwerger2017} J. Hofmann and W. Zwerger,
\emph{Deep inelastic scattering on ultracold gases},
Phys. Rev. X \textbf{7} (2017), 011022.
\href{https://doi.org/10.1103/PhysRevX.7.011022}{doi:10.1103/PhysRevX.7.011022}.

\bibitem{Hepp1974} K. Hepp,
\emph{The classical limit for quantum mechanical correlation functions},
Commun. Math. Phys. \textbf{35} (1974), 265--277.
\href{https://doi.org/10.1007/bf01646348}{doi:10.1007/bf01646348}.

\bibitem{GinibreVelo1979} J. Ginibre and G. Velo,
\emph{The classical field limit of scattering theory for non-relativistic many-boson systems. {I}},
Commun. Math. Phys. \textbf{66} (1979), 37--76.
\href{https://doi.org/10.1007/bf01197745}{doi:10.1007/bf01197745}.

\bibitem{RodnianskiSchlein2009} I. Rodnianski and B. Schlein,
\emph{Quantum fluctuations and rate of convergence towards mean field dynamics},
Commun. Math. Phys. \textbf{291} (2009), 31--61.
\href{https://doi.org/10.1007/s00220-009-0867-4}{doi:10.1007/s00220-009-0867-4}.

\bibitem{GMM2010} M. G. Grillakis, M. Machedon and D. Margetis,
\emph{Second-order corrections to mean field evolution of weakly interacting bosons. {I}},
Commun. Math. Phys. \textbf{294} (2010), 273--301.
\href{https://doi.org/10.1007/s00220-009-0933-y}{doi:10.1007/s00220-009-0933-y}.

\bibitem{GMM2011} M. G. Grillakis, M. Machedon and D. Margetis,
\emph{Second-order corrections to mean field evolution of weakly interacting bosons. {II}},
Adv. Math. \textbf{228} (2011), 1788--1815.
\href{https://doi.org/10.1016/j.aim.2011.06.028}{doi:10.1016/j.aim.2011.06.028}.

\bibitem{LNS2015} M. Lewin, P. T. Nam and B. Schlein,
\emph{Fluctuations around {H}artree states in the mean-field regime},
Amer. J. Math. \textbf{137} (2015), 1613--1650.
\href{https://doi.org/10.1353/ajm.2015.0040}{doi:10.1353/ajm.2015.0040}.

\bibitem{NN2017} P. T. Nam and M. Napi\'orkowski,
\emph{Bogoliubov correction to the mean-field dynamics of interacting bosons},
Adv. Theor. Math. Phys. \textbf{21} (2017), 683--738.
\href{https://doi.org/10.4310/atmp.2017.v21.n3.a4}{doi:10.4310/atmp.2017.v21.n3.a4}.

\bibitem{BCS2017} C. Boccato, S. Cenatiempo and B. Schlein,
\emph{Quantum many-body fluctuations around nonlinear {S}chr\"odinger dynamics},
Ann. Henri Poincar\'e \textbf{18} (2017), 113--191.
\href{https://doi.org/10.1007/s00023-016-0513-6}{doi:10.1007/s00023-016-0513-6}.

\bibitem{BNNS} C. Brennecke, P. T. Nam, M. Napi\'orkowski and B. Schlein,
\emph{Fluctuations of $N$-particle quantum dynamics around the nonlinear Schr\"odinger equation},
Ann. Inst. H. Poincar\'e C Anal. Non Lin\'eaire \textbf{36} (2019), 1201--1235.
\href{https://doi.org/10.1016/j.anihpc.2018.10.007}{doi:10.1016/j.anihpc.2018.10.007}.

\bibitem{DCS2024} A. Deuchert, M. Caporaletti and B. Schlein,
\emph{Dynamics of mean-field bosons at positive temperature},
Ann. Inst. H. Poincar\'e C Anal. Non Lin\'eaire \textbf{41} (2024), 995--1054.
\href{https://doi.org/10.4171/AIHPC/93}{doi:10.4171/AIHPC/93}.

\bibitem{HR2024} D. X. Horv\'ath and C. Rylands,
\emph{Full counting statistics of charge in quenched quantum gases},
Phys. Rev. A \textbf{109} (2024), 043302.
\href{https://doi.org/10.1103/PhysRevA.109.043302}{doi:10.1103/PhysRevA.109.043302}.

\bibitem{PPG2019} G. Perfetto, L. Piroli and A. Gambassi,
\emph{Quench action and large deviations: Work statistics in the one-dimensional Bose gas},
Phys. Rev. E \textbf{100} (2019), 032114.
\href{https://doi.org/10.1103/PhysRevE.100.032114}{doi:10.1103/PhysRevE.100.032114}.

\bibitem{Bertoin2000} J. Bertoin,
\emph{Subordinators, L\'evy processes with no negative jumps and branching processes},
MaPhySto Lecture Notes \textbf{8} (2000).
\href{https://webdoc.sub.gwdg.de/ebook/e/2002/maphysto/publications/mps-ln/2000/8.pdf}{MaPhySto MPS-LN 2000-8}.

\bibitem{BAKS} G. Ben Arous, K. Kirkpatrick and B. Schlein,
\emph{A central limit theorem in many-body quantum dynamics},
Commun. Math. Phys. \textbf{321} (2013), 371--417.
\href{https://doi.org/10.1007/s00220-013-1722-1}{doi:10.1007/s00220-013-1722-1}.

\bibitem{BSSCLT} S. Buchholz, C. Saffirio and B. Schlein,
\emph{Multivariate central limit theorem in quantum dynamics},
J. Stat. Phys. \textbf{154} (2014), 113--152.
\href{https://doi.org/10.1007/s10955-013-0897-3}{doi:10.1007/s10955-013-0897-3}.

\bibitem{RSCLT} S. Rademacher and B. Schlein,
\emph{Central limit theorem for Bose--Einstein condensates},
J. Math. Phys. \textbf{60} (2019), 071902.
\href{https://doi.org/10.1063/1.5094348}{doi:10.1063/1.5094348}.

\bibitem{NR} P. T. Nam and S. Rademacher,
\emph{Exponential bounds of the condensation for dilute Bose gases},
Trans. Amer. Math. Soc. \textbf{378} (2025), 3229--3278.
\href{https://doi.org/10.1090/tran/9335}{doi:10.1090/tran/9335}.

\bibitem{BBR} N. Behrmann, C. Brennecke and S. Rademacher,
\emph{Exponential control of excitations for trapped BEC in the Gross--Pitaevskii regime},
Lett. Math. Phys. \textbf{115} (2025), 91.
\href{https://doi.org/10.1007/s11005-025-01986-1}{doi:10.1007/s11005-025-01986-1}.

\bibitem{Kisynski} J. Kisy\'nski,
\emph{Sur les op\'erateurs de Green des probl\`emes de Cauchy abstraits},
Studia Math. \textbf{23} (1964), 285--328.
\href{https://doi.org/10.4064/sm-23-3-285-328}{doi:10.4064/sm-23-3-285-328}.

\bibitem{BLP} A. Balmaseda, D. Lonigro and J. M. P\'erez-Pardo,
\emph{On the Schr\"odinger equation for time-dependent Hamiltonians with a constant form domain},
Mathematics \textbf{10} (2022), 218.
\href{https://doi.org/10.3390/math10020218}{doi:10.3390/math10020218}.
Equation and theorem numbering follows \href{https://arxiv.org/abs/2112.11063v2}{arXiv:2112.11063v2}.

\bibitem{BPPS} L. Bo{\ss}mann, S. Petrat, P. Pickl and A. Soffer,
\emph{Beyond Bogoliubov dynamics},
Pure Appl. Anal. \textbf{3} (2021), 677--726.
\href{https://doi.org/10.2140/paa.2021.3.677}{doi:10.2140/paa.2021.3.677}.

\bibitem{BPPaS2020} L. Bo{\ss}mann, N. Pavlovi\'c, P. Pickl and A. Soffer,
\emph{Higher order corrections to the mean-field description of the dynamics of interacting bosons},
J. Stat. Phys. \textbf{178} (2020), 1362--1396.
\href{https://doi.org/10.1007/s10955-020-02500-8}{doi:10.1007/s10955-020-02500-8}.

\bibitem{NRS2026} P. T. Nam, S. Rademacher and A. Soffer,
\emph{Dispersive estimates and long-time validity for Bogoliubov dynamics of interacting Bose gases},
\href{https://arxiv.org/abs/2511.12748v3}{arXiv:2511.12748v3} (2026).

\bibitem{DerezinskiGerard} J. Derezi\'nski and C. G\'erard,
\emph{Mathematics of Quantization and Quantum Fields},
Cambridge Monographs in Mathematical Physics, Cambridge University Press, Cambridge (2023), open-access reissue of the 2013 edition.
\href{https://doi.org/10.1017/9781009290876}{doi:10.1017/9781009290876}.

\bibitem{KatoBook} T. Kato,
\emph{Perturbation Theory for Linear Operators},
Classics in Mathematics, Springer, Berlin (1995), reprint of the second edition.
\href{https://doi.org/10.1007/978-3-642-66282-9}{doi:10.1007/978-3-642-66282-9}.

\bibitem{SimonForms} B. Simon,
\emph{Quantum Mechanics for Hamiltonians Defined as Quadratic Forms},
Princeton Series in Physics, Princeton University Press, Princeton (1971);
Princeton Legacy Library reissue (2015).
\href{https://doi.org/10.1515/9781400868834}{doi:10.1515/9781400868834}.

\bibitem{LiebLoss} E. H. Lieb and M. Loss,
\emph{Analysis},
2nd ed., Graduate Studies in Mathematics 14, American Mathematical Society, Providence (2001).
\href{https://doi.org/10.1090/gsm/014}{doi:10.1090/gsm/014}.

\bibitem{Magnus} W. Magnus,
\emph{On the exponential solution of differential equations for a linear operator},
Commun. Pure Appl. Math. \textbf{7} (1954), 649--673.
\href{https://doi.org/10.1002/cpa.3160070404}{doi:10.1002/cpa.3160070404}.

\bibitem{Wilcox} R. M. Wilcox,
\emph{Exponential operators and parameter differentiation in quantum physics},
J. Math. Phys. \textbf{8} (1967), 962--982.
\href{https://doi.org/10.1063/1.1705306}{doi:10.1063/1.1705306}.

\bibitem{KatoTransport} T. Kato,
\emph{On the adiabatic theorem of quantum mechanics},
J. Phys. Soc. Japan \textbf{5} (1950), 435--439.
\href{https://doi.org/10.1143/jpsj.5.435}{doi:10.1143/jpsj.5.435}.

\bibitem{ReedSimonII} M. Reed and B. Simon,
\emph{Methods of Modern Mathematical Physics, Vol. II: Fourier Analysis, Self-Adjointness},
Academic Press, New York (1975).
\href{https://shop.elsevier.com/books/ii-fourier-analysis-self-adjointness/reed/978-0-08-092537-0}{Publisher archive}.

\bibitem{Kato} T. Kato,
\emph{Integration of the equation of evolution in a Banach space},
J. Math. Soc. Japan \textbf{5} (1953), 208--234.
\href{https://doi.org/10.2969/jmsj/00520208}{doi:10.2969/jmsj/00520208}.

\bibitem{SimonTrace} B. Simon,
\emph{Trace Ideals and Their Applications},
2nd ed., Mathematical Surveys and Monographs 120, American Mathematical Society, Providence (2005).
\href{https://doi.org/10.1090/surv/120}{doi:10.1090/surv/120}.

\bibitem{KohldorferBalazs} L. K\"ohldorfer and P. Balazs,
\emph{Wiener pairs of Banach algebras of operator-valued matrices},
J. Math. Anal. Appl. \textbf{549} (2025), 129525.
\href{https://doi.org/10.1016/j.jmaa.2025.129525}{doi:10.1016/j.jmaa.2025.129525}.
Section references use
\href{https://arxiv.org/abs/2407.16416v3}{arXiv:2407.16416v3}.

\bibitem{BerghLofstrom} J. Bergh and J. L\"ofstr\"om,
\emph{Interpolation Spaces: An Introduction},
Grundlehren der mathematischen Wissenschaften 223, Springer, Berlin (1976).
\href{https://doi.org/10.1007/978-3-642-66451-9}{doi:10.1007/978-3-642-66451-9}.

\bibitem{LionsMagenes} J. L. Lions and E. Magenes,
\emph{Non-Homogeneous Boundary Value Problems and Applications, Vol. 1},
Grundlehren der mathematischen Wissenschaften 181, Springer, Berlin (1972).
\href{https://doi.org/10.1007/978-3-642-65161-8}{doi:10.1007/978-3-642-65161-8}.

\bibitem{Cazenave2003} T. Cazenave,
\emph{Semilinear {S}chr\"odinger Equations},
Courant Lecture Notes in Mathematics 10, American Mathematical Society, Providence (2003).
\href{https://doi.org/10.1090/cln/010}{doi:10.1090/cln/010}.

\bibitem{Harris} L. A. Harris,
\emph{A proof of Markov's theorem for polynomials on Banach spaces},
J. Math. Anal. Appl. \textbf{368} (2010), 374--381.
\href{https://doi.org/10.1016/j.jmaa.2010.02.050}{doi:10.1016/j.jmaa.2010.02.050}.

\bibitem{GoleniaJecko2013} S. Gol\'enia and T. Jecko,
\emph{Weighted Mourre's commutator theory, application to Schr\"odinger operators with oscillating potential},
J. Operator Theory \textbf{70} (2013), 109--144.
\href{https://doi.org/10.7900/jot.2011may06.1922}{doi:10.7900/jot.2011may06.1922}.

\bibitem{Teschl2009} G. Teschl,
\emph{Mathematical Methods in Quantum Mechanics:
With Applications to Schr\"odinger Operators},
Graduate Studies in Mathematics \textbf{99},
American Mathematical Society, Providence (2009).
\href{https://www.mat.univie.ac.at/~gerald/ftp/book-schroe/schroe.pdf}{Author's first-edition text}.

\bibitem{ReedSimonI} M. Reed and B. Simon,
\emph{Methods of Modern Mathematical Physics, Vol. I: Functional Analysis},
Revised and enlarged ed., Academic Press, New York (1980).
\href{https://shop.elsevier.com/books/i-functional-analysis/reed/978-0-08-057048-8}{Publisher archive}.

\bibitem{Schur} J. Schur,
\emph{Bemerkungen zur Theorie der beschr\"ankten Bilinearformen mit unendlich vielen Ver\"anderlichen},
J. Reine Angew. Math. \textbf{140} (1911), 1--28.
\href{https://doi.org/10.1515/crll.1911.140.1}{doi:10.1515/crll.1911.140.1}.

\bibitem{Wick} G. C. Wick,
\emph{The evaluation of the collision matrix},
Phys. Rev. \textbf{80} (1950), 268--272.
\href{https://doi.org/10.1103/physrev.80.268}{doi:10.1103/physrev.80.268}.

\bibitem{EMS} L. Erd\H{o}s, A. Michelangeli and B. Schlein,
\emph{Dynamical formation of correlations in a Bose--Einstein condensate},
Commun. Math. Phys. \textbf{289} (2009), 1171--1210.
\href{https://doi.org/10.1007/s00220-009-0828-y}{doi:10.1007/s00220-009-0828-y}.

\bibitem{ReichmannTriay2026} L. Reichmann and A. Triay,
\emph{On the spherical symmetry and finite-range assumptions of the
interaction potential in the low energy study of dilute Bose gases},
\href{https://arxiv.org/abs/2607.00859v2}{arXiv:2607.00859v2} (2026).

\end{thebibliography}
\end{document}